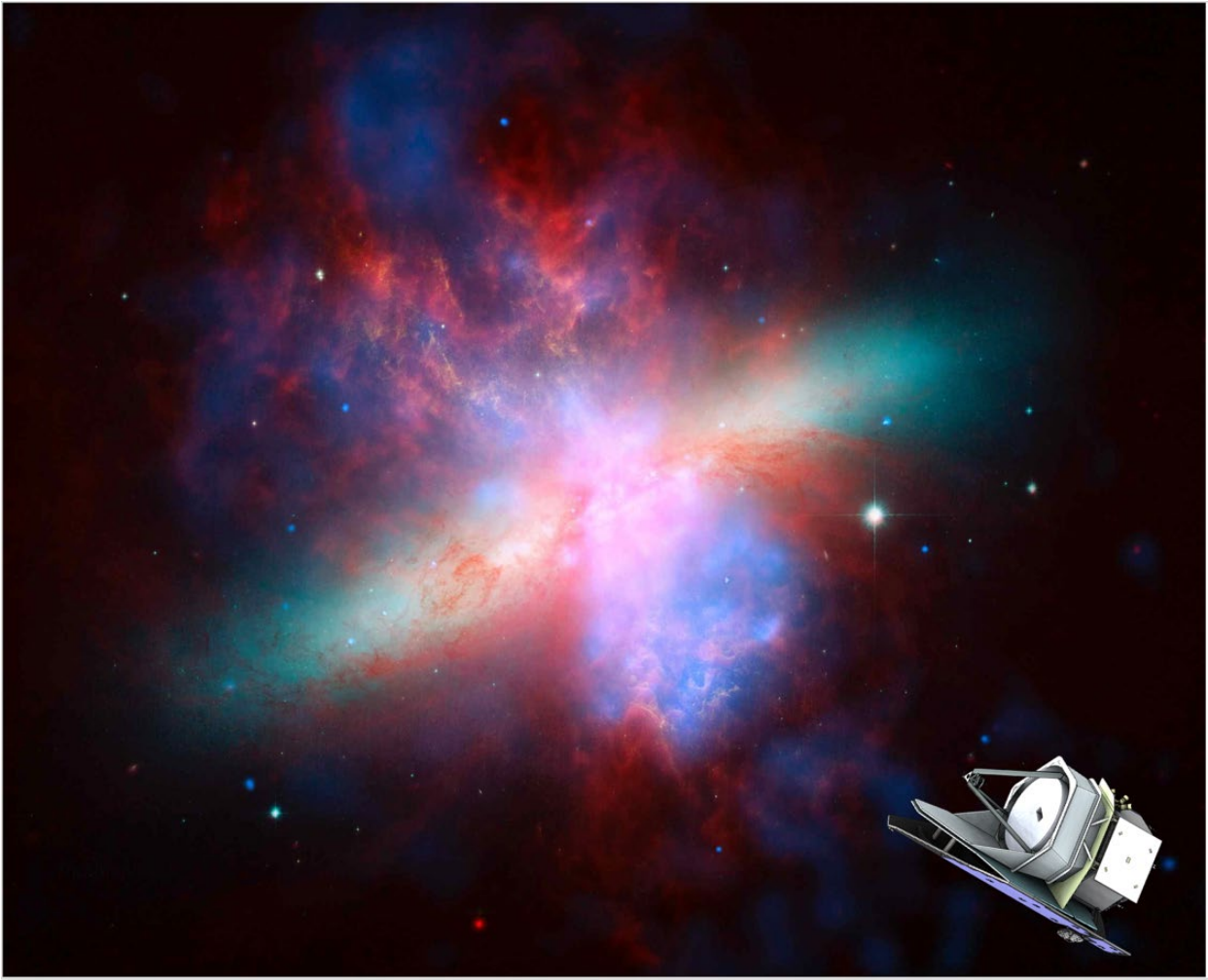

# PRIMA General Observer Science Book

Editors: A. Moullet (National Radio Astronomy Observatory), T. Kataria, D. Lis, S. Unwin, & Y. Hasegawa (Jet Propulsion Laboratory, California Institute of Technology), E. Mills (University of Kansas), C. Battersby (University of Connecticut), A. Roc (Pomona College), M. Meixner (Jet Propulsion Laboratory, California Institute of Technology)

PRIMA (The PRobe for-Infrared Mission for Astrophysics) is a concept for a far-infrared (IR) observatory. PRIMA features a cryogenically cooled 1.8 m diameter telescope and is designed to carry two science instruments enabling ultra-high sensitivity imaging and spectroscopic studies in the 24 to 235 μm wavelength range. The resulting observatory is a powerful survey and discovery machine, with mapping speeds better by 2–4 orders of magnitude with respect to its far-IR predecessors. The bulk of the observing time on PRIMA should be made available to the community through a General Observer (GO) program offering 75% of the mission time over 5 years. In March 2023, the international astronomy community was encouraged to prepare authored contributions articulating scientific cases that are enabled by the telescope's massive sensitivity advance and broad spectral coverage, and that could be performed within the context of GO program. This document, the PRIMA General Observer Science Book, is the edited collection of the 76 received contributions.

# Table of Contents

# Highlights at a Glance

In **5,000 hours**, a PRIMA all-sky survey can reach a depth comparable to Herschel extragalactic deep fields. With a sensitivity 100x better than the Infrared Astronomical Satellite (IRAS) all-sky survey, it will increase the number of detected sources by more than a factor of 100 and will be unsurpassed for decades to come. *(section 66; Wright et al.)*

In **1,200 hours**, PRIMA can measure the D/H isotopic ratio of water in a statistically-significant sample of solar system comets to determine how, when, and in what quantity comets contributed to the Earth's water content– a key measurement for understanding whether water-rich Earthlike planets are common in the Universe. (*section 41; Lis et al.*)

In **500 hours**, PRIMA can map a 200 square degree area of the Small Magellanic Cloud (SMC) and Large Magellanic Cloud (LMC) to a sensitivity that allows separation of their dust emission from the Milky Way foreground and the Cosmic Infrared Background. This will yield the first well-constrained dust templates for galaxies outside of the Milky Way, which is critical to the interpretation of galaxy properties in the early universe. (*section 49; Galliano et al.*)

In **400 hours**, an extragalactic polarization survey can detect 10,000 galaxies out to z=2, revolutionizing our view of magnetic fields and dust properties at redshifts beyond the local universe, where there are currently no polarization measurements. (*section 13; Bethermin et al.*)

In **400 hours**, PRIMA's unprecedented sensitivity can yield a sample of 100 quiescent low-metallicity dwarf galaxies that have not been previously detectable at infrared wavelengths. This sample of R∼100 spectra will break the degeneracy between star formation history and metallicity that currently limits the understanding of dust grain evolution in early stages of galaxy evolution. (*section 26; Galliano et al.*).

In **200 hours**, PRIMA can measure redshifts of hundreds of optically dark galaxies and quantify the prevalence of dust-obscured active galactic nuclei (AGN) before the peak of cosmic star formation by observing the rest-frame mid-infrared emission of z>3 galaxies that is inaccessible to both James Webb Space Telescope (JWST) and Atacama Large Millimeter/submillimeter Array (ALMA). (*section 24; Faisst et al.*).

In **20 minutes**, PRIMA can conduct the first far-infrared dust reverberation mapping studies of one AGN tori. Taking advantage of PRIMA's 5-year long mission, AGN structure can be monitored at the relevant cadence to investigate the relation between supermassive black hole (SMBH) accretion rate and accretion disk clumpiness. (*section 4; Gorjian et al.*).

In **75 hours**, PRIMA can detect the circumgalactic medium (CGM) in emission, one of the six 'discovery areas' highlighted by Astro2020. (*section 38; Tarantino et al.*)

In **40 hours**, a deep imaging survey can characterize a newly detected sub-mJy population of far-infrared galaxies, resolving individual sources to enable a multiwavelength census of one of the final major constituents of the cosmic infrared background. (*section 20; Clements et al.*)

In **30 hours**, PRIMA can determine whether large (>1 µm) crystalline water grains that are undetectable with near-infrared or X-ray observations are the missing reservoir for oxygen in the interstellar medium. (*section 58; Onaka et al.*)

In **5 minutes**, PRIMA can detect transient sources discovered by future CMB experiments, enabling efficient monitoring of hundreds of sources over day to week timescales. (*section 3; Clements et al.*)

In **600 hours**, PRIMA can monitor 2,000 protostars over the mission's lifetime, to determine the frequency and types of bursts that protostars undergo, and ultimately their importance in the stellar mass assembly process and whether the primary mode of stellar mass accretion is steady state or through accretion bursts. (*section 43; Battersby et al.*)

In **200 hours**, PRIMA can directly measure the early appearance and conditions of hydrocarbon dust and metals in the Universe by obtaining high-quality full band rest-mid-infrared spectra of 50 star-forming galaxies out to z=7 and enabling the photometric recovery of polycyclic aromatic hydrocarbon (PAH) emission in thousands more from archival deep field surveys. (*section 21*, Donnelly et al.).

## The PRIMA General Observer Science Book: context and scope

The PRIMA General Observer Science Book is a collection of 76 contributions authored by members of the international astronomy community, each articulating a scientific case that is enabled by PRIMA's massive sensitivity advance and broad spectral coverage. These contributions were encouraged at a science community workshop held in March 2023, as part of the PRIMA team's community outreach activities. The objective of this effort was to capture and characterize the community's interest towards the far-infrared wavelength regime in general, and towards the PRIMA concept in particular.

PRIMA (The PRobe for-Infrared Mission for Astrophysics) is a concept for a far-IR observatory, developed in response to the Announcement of Opportunity (AO) for an Astrophysics Probe Explorer (APEX) issued by NASA in July 2023. PRIMA features a cryogenically cooled 1.8 m diameter telescope and is designed to carry two science instruments enabling both imaging and spectroscopic studies. PRIMAger, provided by a European consortium, is an ultra-sensitive multi-band spectrophotometric imager offering hyperspectral narrow-band imaging (R=10) from 24 to 80 μm, and polarimetric capabilities in 4 broadband filters from 80 to 235 μm. FIRESS, led by a JPL team, is a versatile spectrometer offering several two spectral resolution modes (R∼130 and up to 4400) over from 24 to 235 μm. Both instruments offer unprecedented instantaneous spectral coverage; FIRESS can cover that entire range in two settings while PRIMAger does it naturally via scan mapping. The resulting observatory is a powerful survey and discovery machine, with mapping speeds better by 2–4 orders of magnitude with respect to its far-IR predecessors Herschel and Spitzer. The specific capabilities are tailored to enable the realization of a focused Principal Investigator (PI) team-defined observing program, structured around three scientific themes identified by the Astro2020 Decadal survey as priority areas for a far-IR probe mission:

- Evolution of Galactic Systems: probing the co-evolution of galaxies and their supermassive black holes across cosmic time

- Formation of Planets: tracing the astrochemical signatures of planet formation

- Rise of Dust and Metals: measuring the buildup of heavy elements and interstellar dust from early galaxies to today

The APEX AO specifies that the bulk of the observing time should be made available to the community through a General Observer (GO) program offering 75% of the mission time. With PRIMA, we anticipate >80% observing efficiency, providing over 26,000 hours of observing time for GO programs during the nominal 5-year mission, about twice as many hours as Herschel for General Observer science.

The cases collected in this GO Science Book, while representing just a snapshot of possible proposals that could be submitted to the GO program, already total about 21,000 h of observations, or about 80% of the expected time available for GO observations in the nominal 5-year mission. This demonstrates the community's eagerness for the type of observations enabled by PRIMA. 5000h alone are proposed by Wright et al. (section 66) for a PRIMAger all-sky survey.

To facilitate the collection and production of these community-authored contributions, which realistically represent examples of GO science cases, a widely advertised 2-day virtual workshop was held in March 2023. In addition to presentations on the planned capabilities of the observatory, breakout sessions by scientific area were organized to encourage discussions of scientific questions addressed by the PRIMA instruments, moderated by leaders in the field. PRIMA team members attended the breakout sessions to answer questions about feasibility. At the end of the workshop, attendees were encouraged to submit contributions to the science book following a common template resembling an observing proposal. The PRIMA team distributed fact sheets and presentations on instrument capabilities to prospective authors, providing a snapshot of the mission concept development as of March 2023 (see Figures 1, 2, and 3, and the workshop proceedings https://prima.ipac.caltech.edu/page/2023-workshop). The received contributions were then verified quantitatively, edited, and formatted by a team of editors.

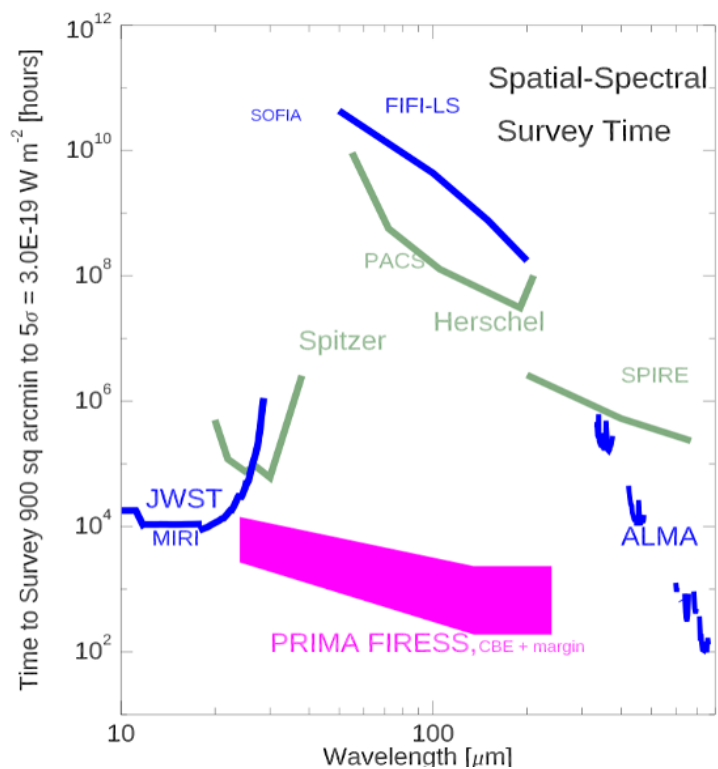

Figure 1: Time to survey to a depth (5 σ) of 3×10⁻¹⁹ W/m² with FIRESS over a quarter square degree and with full spectral coverage. This incorporates sensitivity, spatial coverage, and total instantaneous bandwidth. The figure is for the R~100 FIRESS spectrometer optimized for detection of spectral lines; but the PRIMAger hyperspectral instrument offers a similar advance with R~10 bandpasses optimized for broad dust features and rapid high-fidelity SED extraction and polarization.

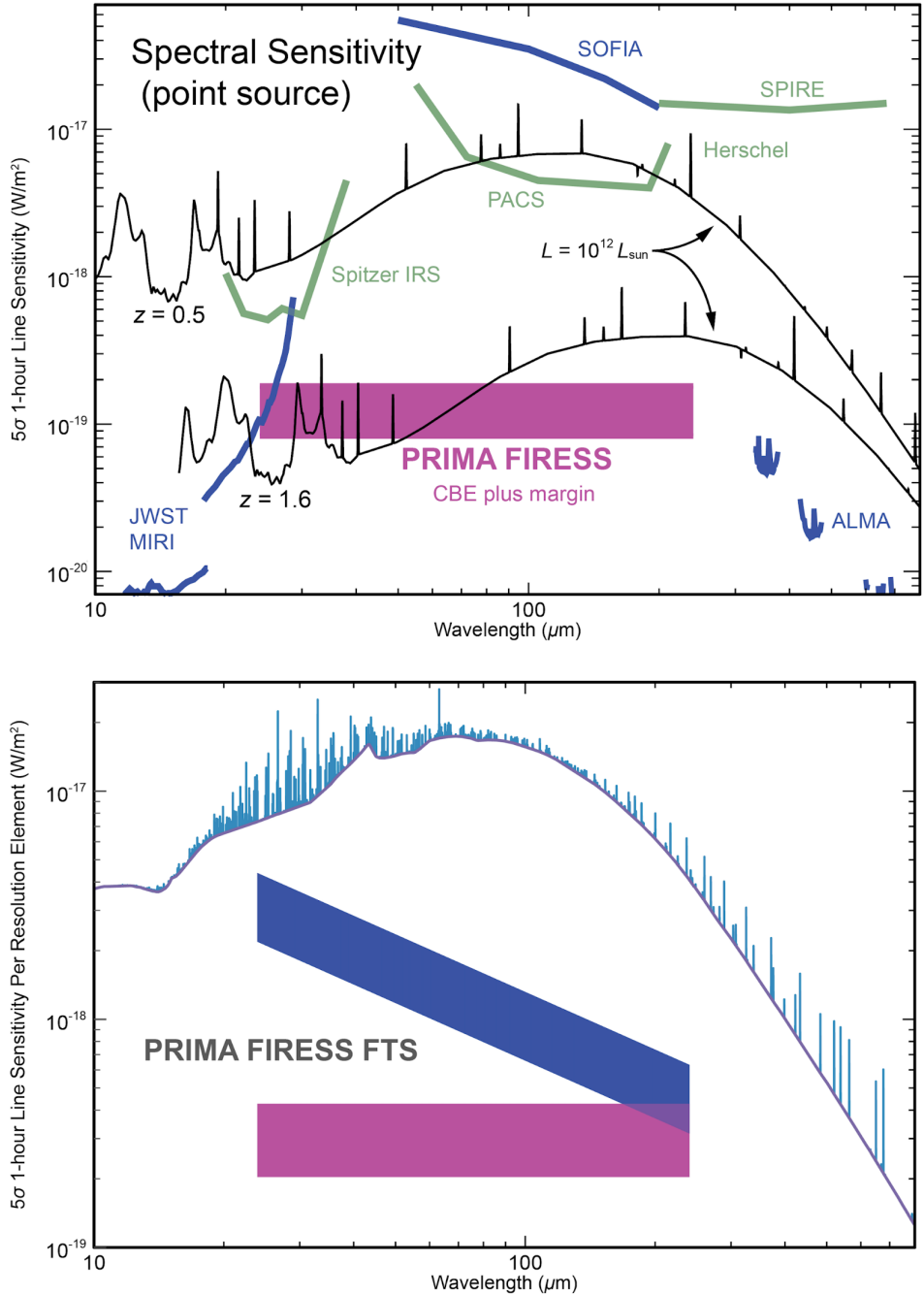

Figure 2: FIRESS pointed spectral line sensitivity used in developing the cases in this book. The instrument sensitivities have evolved slightly but do not materially impact the contributions outlined in this book. Up to date sensitivities will be maintained on the PRIMA website: prima.ipac.caltech.edu. The full 24 to 235 μm range is covered in four logarithmically spaced slit-fed grating modules. The slits are aligned in pairs, so that a source of interest is observed in 2 of the 4 bands simultaneously, and the complete spectrum of a given source requires 2 pointings. **Top** shows the low-resolution mode, with R~100 which is most sensitive. **Bottom** shows the high-resolution mode, which engages a Fourier-transform interferometer to process light en route to the grating modules, this boosts the resolving power up to 4400 at 112 μm and >2000 across everywhere with a modest sensitivity penalty. The blue swath shows the sensitivity when viewing a 1 Jansky source. For comparison, the model disk spectrum is for a 1.5 M$_\odot$ disk around a 4 L$_\odot$ star is plotted.

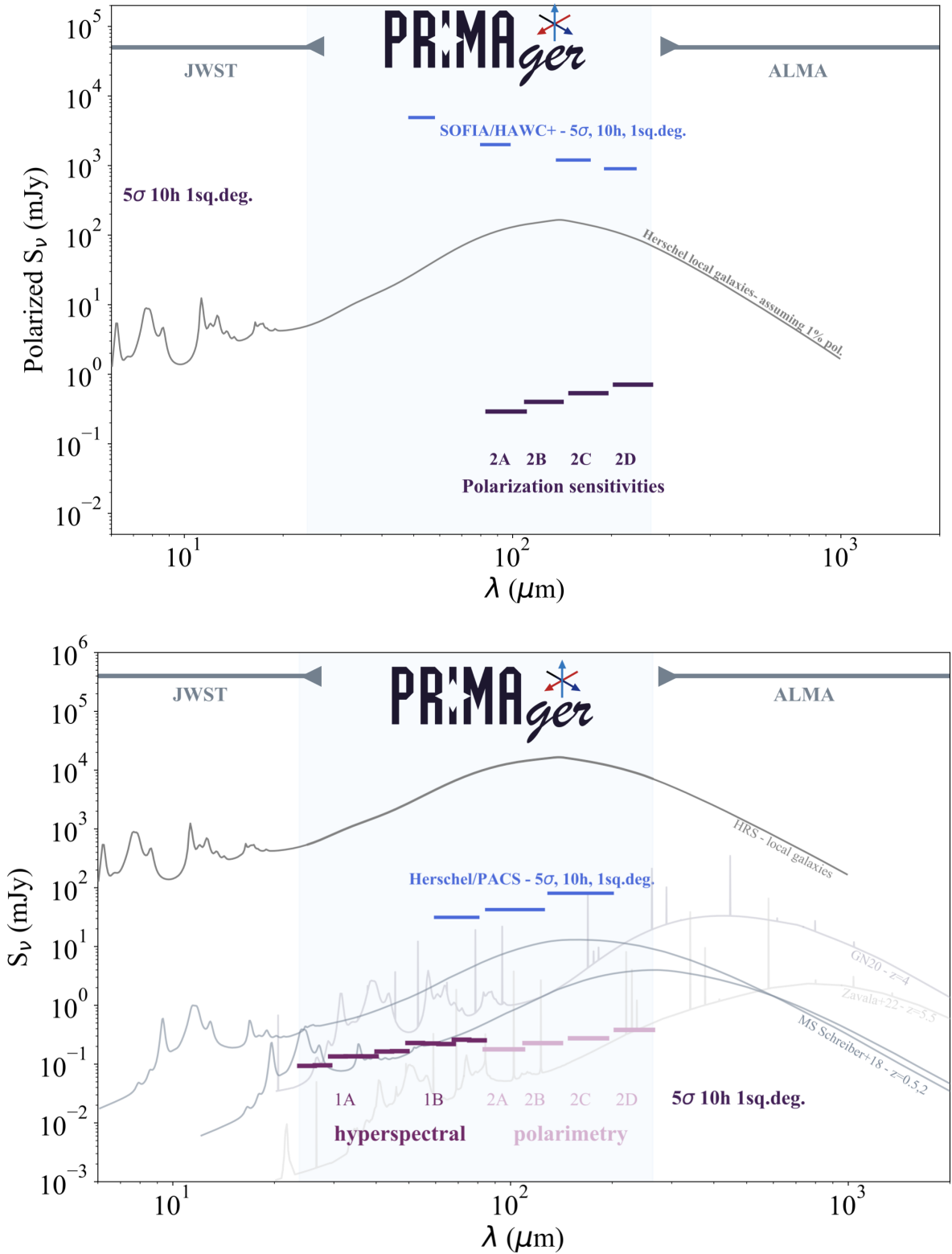

Figure 3: PRIMAger estimated performance used in developing the cases in this book. The instrument sensitivities have evolved slightly but do not materially impact the contributions outlined in this book. Up-to-date sensitivities will be maintained on the PRIMA website: prima.ipac.caltech.edu.

The ensemble of contributions provides an initial view of the broad and exciting science enabled by PRIMA's new far-IR measurement capabilities (See the Highlights at a Glance table.) In aggregate, the collection addresses more than 70% of the scientific questions identified in the Astro2020 survey (see Table 1), and two of the three decadal priority areas: 'New Windows of the Dynamic Universe' and 'Unveiling the Drivers of Galaxy Growth'. All instrument and observing modes are represented in the submitted cases, with similar total time requests for PRIMAger and FIRESS, and a third of programs requesting both instruments. At least seven cases are based on time-domain capabilities, either through relatively rapid response or long-term monitoring. Leveraging the agility of the observatory, blind imaging surveys with PRIMAger are proposed in several cases, with both the hyperspectral capability shortward of 80 μm and the polarimetric capability longward of 80 μm. Figure 4 captures the trade between depth and area for a 3,000-hour fiducial allocation. At one extreme, an all-sky-survey (Wright et al., section 66) could reach comparable depth to that of the pencil-beam Herschel PEP survey (Lutz et al., 2011). On the deep end, PRIMA can reach to below L* at z~2 on multi-square-degree fields. As noted by Bethermin, section 13, the polarization content of such confusion-limited surveys offers a new and exciting probe of the cosmological dusty galaxy population.

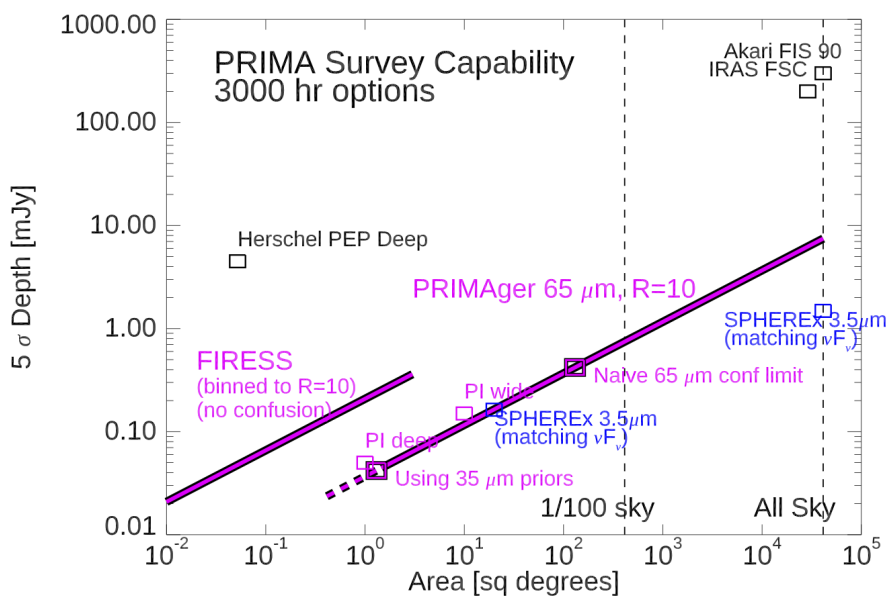

Figure 4: Depth / area trade space for PRIMA surveys, assuming a 3,000-hour fiducial allocation. Heavy curve shows the depth at 65 μm for both PRIMAger and FIRESS (left curve). While source confusion plays a role at PRIMAger's longer wavelengths, the shortest bands can go very deep, and FIRESS offers the 3rd dimension for spectral line-based deconfusion.

As shown in the highlights table, PRIMA GO capabilities produce decades of community legacy value, support large GO programs that leverage large statistical samples to bring far-IR into the era of astronomical big data, and enable smaller GO programs with revolutionary impact. Of course, we fully expect evolution in these proposed programs, as well as the genesis of new ideas in the years before PRIMA's launch. If experience with prior observatories offering a comparable sensitivity advance is any guide, many of PRIMA's most exciting discoveries are yet to be identified.

The 76 contributions received are divided into seven chapters. The first six chapters correspond to scientific areas covering the entire field of Astrophysics and Planetary Science, structured along the scientific areas of the Astro2020 Decadal Survey panels. The last chapter, Technical Notes, includes contributions that do not necessarily fit within the format of an observing proposal, but capture a technique or exploration area enabled by PRIMA's capabilities.

Table 1: Science Panel Questions and discovery areas as defined in the Astro2020 Decadal Survey (Tables 2.1 and 2.2, p 2-49), and the number of PRIMA GO Book Science Cases addressing them. Some science cases address multiple questions.

| Panel | Decadal Question | Citation | Number of Cases in GO Book | Example Case |
|---|---|---|---|---|
| Compact Objects and Energetic Phenomena | What powers the diversity of explosive phenomena across the electromagnetic spectrum? | B-Q2 | 5 | 'The infrared view of energetic cosmological transients: kilonovae, gamma-ray bursts, and fast radio bursts,' Watson et al., section 8, |
| | Why do some compact objects eject material in nearly light-speed jets, and what is that material made of? | B-Q3 | 1 | 'Rapid far-IR spectral timing of X-ray binaries,' Tetarenko et al., section 7 |
| | What seeds supermassive black holes and how do they grow? | B-Q4 | 2 | 'The dust distribution around accreting black holes using reverberation mapping,' Gorjian et al., section 4 |
| | Transforming our View of the Universe by Combining New Information from Light, Particles, and Gravitational Waves | B-DA | 1 | 'Probing heavy element nucleosynthesis in binary neutron star mergers with PRIMA,' Andreoni et al., section 1 |
| Cosmology | What are the properties of dark matter and the dark sector? | C-Q2 | 1 | 'Probing the history of cosmic star formation, black hole growth, and metallicity/dust evolution with line-intensity mapping,' Cheng et al., section 9 |
| | What physics drives the expansion and large-scale evolution of the Universe? | C-Q3 | 1 | 'Search for the radiative decay of the cosmic neutrino background by spectral measurements of the cosmic infrared background,' Takeuchi et al., section 10 |
| | How will measurements of gravitational waves reshape our cosmological view? | C-Q4 | 1 | 'Probing heavy element nucleosynthesis in binary neutron star mergers with PRIMA,' Andreoni et al., section 1 |
| Galaxies | How did the intergalactic medium and the first sources of radiation evolve from cosmic dawn through the epoch of reionization? | D-Q1 | 1 | 'Unveiling rapid and obscured quenching processes in galaxies,' French et al., section 25 |
| | How do gas, metals, and dust flow into, through, and out of galaxies? | D-Q2 | 35 | 'Starbursts under the microscope: how the feedback from a super star cluster impacts its surroundings and its host galaxy,' Goldsmith et al., section 51 |
| | How do supermassive black holes form and how is their growth coupled to the evolution of their host galaxies? | D-Q3 | 11 | 'Coevolution of star formation and black hole accretion at the Cosmic Noon,' Spignolio et al., section 37 |

| Panel | Decadal Question | Citation | Number of Cases in GO Book | Example Case |
|---|---|---|---|---|
| | How do the histories of galaxies and their dark matter halos shape their observable properties? | D-Q4 | 4 | 'Unveiling the evolution of the dust mass function with PRIMA,' Calura et al., section 19 |
| | Mapping the Circumgalactic Medium and Intergalactic Medium in Emission | D-DA | 6 | 'Detecting FIR emission from the circumgalactic medium,' Tarantino et al., section 38 |
| Exoplanets, Astrobiology, and the Solar System | What is the range of planetary system architectures, and is the configuration of the solar system common? | E-Q1 | 1 | 'Protoplanetary disk properties over a wide range of star-forming environments,' Villenave et al., section 64 |
| | What are the properties of individual planets, and which processes lead to planetary diversity? | E-Q2 | 3 | 'Size distribution in the Solar System Kuiper Belt,' Moullet et al., section 42 |
| | How do habitable environments arise and evolve within the context of their planetary systems? | E-Q3 | 1 | 'Comets and the origin of Earth's water,' Lis et al., section 41 |
| ISM and Star and Planet Formation | How do star-forming structures arise from, and interact with, the diffuse ISM? | F-Q1 | 12 | 'Revealing the role of magnetic fields in the formation of giant molecular clouds and star formation in local galaxies,' Lopez-Rodriguez et al., section 55 |
| | What regulates the structure and motions within molecular clouds? | F-Q2 | 9 | 'Investigating the role of magnetic fields in the formation and evolution of striations in interstellar clouds with PRIMA,' Tassis et al., section 63 |
| | How does gas flow from parsec scales down to protostars and their disks? | F-Q3 | 3 | 'How stars gain their mass: uncovering the primary mode of stellar accretion with PRIMA,' Battersby et al., section 43 |
| | Is planet formation fast or slow? | F-Q4 | 3 | 'Probing solid dynamics and evolution in protoplanetary disks using dust mineralogy,' Seo et al., section 61 |
| Stars, the Sun and Stellar Populations | What are the most extreme stars and stellar populations? | G-Q1 | 2 | 'Exploring the magnetized circumstellar environment of evolved stars through dust polarization analysis,' Sabin et al., section 70 |
| | How does multiplicity affect the way a star lives and dies? | G-Q2 | 1 | 'The extraordinary deaths of ordinary stars,' Sahai et al., section 71 |
| | "Industrial Scale" Spectroscopy | G-DA | 1 | 'Identifying extreme star-forming galaxies with the PRIMA mission,' Oliver et al., section 35 |

## Acknowledgement:

A portion of this research was carried out at the Jet Propulsion Laboratory, California Institute of Technology, under a contract with the National Aeronautics and Space Administration (80NM0018D0004).

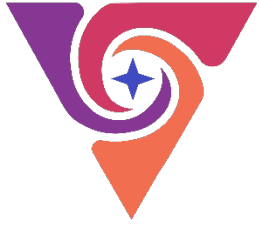

# Compact Objects and Energetic Phenomena

# 1. Probing heavy element nucleosynthesis in binary neutron star mergers with PRIMA

Igor Andreoni (UMD)

Binary neutron star and neutron star – black hole mergers are multi-messenger sources. They generate loud gravitational wave signals detectable by the LIGO-Virgo-KAGRA instruments and can be accompanied by luminous electromagnetic transients. The optical/infrared counterpart to these mergers, called "kilonova", is powered by the decay of newly synthesized heavy elements. The infrared component, in particular, can provide direct evidence of r-process nucleosynthesis in the second and third abundance peaks (A > 120) in the Solar heavy element distribution. PRIMA mid-infrared observations will directly probe the presence of such elements in an uncharted discovery space.

## General Scientific Area:

Time Domain

## Science Justification:

The direct detection of gravitational waves (GWs) changed our way of exploring the dynamic Universe. When the first binary neutron star (NS) merger was discovered in gravitational waves, the event (named GW170817) was found to be accompanied by electromagnetic (EM) radiation in the form of a short-duration gamma-ray burst (GRB), an optical/infrared transient, and a radio afterglow. The multi-messenger discovery [1] has led to hundreds of studies addressing, for example, astrophysics of energetic phenomena, fundamental physics, and cosmology. Combining GW and EM information also provided us with unique insights on the NS equation of state and nuclear physics in extreme conditions.

The finding of a multi-wavelength counterpart was the long-sought confirmation that that short GRBs and fast optical/infrared transients called "kilonovae" can be generated by the coalescence of a binary NS or neutron star-black hole (NSBH) systems (see for example [2] for a review). When the NS is disrupted by tidal forces, a "tail" of neutron-rich material is launched. In addition, a 0.01–0.3 solar-masses disk of post-merger ejecta forms around the NS or black hole (BH) remnant. Along with tidal tails, outflows from these post-merger ejecta can have a sufficiently small electron fraction ($Y_e < 0.3$) to synthesize heavy r-process elements such as lanthanides and actinides. Radioactive decay of those unstable nuclei powers the optical and near-infrared transient.

There are three distinct abundance peaks in the solar heavy element distribution: the first abundance peak spans $70 \leq A \leq 88$, the second peak spans $120 \leq A \leq 140$, lanthanides span A 139 $\leq A \leq 180$, and the third peak spans $180 \leq A \leq 200$. The optical/infrared photometric data suggested heavy-element nucleosynthesis in the ejecta of GW170817, which could however be explained both with and without the third r-process peak [2,3,4]. The Spitzer space observatory has provided us with the most constraining data in the nebular phase, suggesting that elements in the second and third r-process abundance peak were indeed synthesized [5]. Infrared observations are key to probing the presence of the heaviest elements synthesized in mergers including at least one neutron star.

It is possible that other types of phenomena, such as the core-collapse of rapidly rotating massive stars, are responsible for a large fraction of the r-process elements in the Universe [6]. A theoretical debate is yet to be settled on this matter, which can greatly benefit from new observations and larger samples. While time-domain surveys yield many supernovae that can help us understand better the core-collapse scenarios, kilonovae remain elusive because of their intrinsic rarity, rapid evolution, and low luminosity compared to supernovae. The question is open: are NS mergers "the dominant site" or just "a site" for the heavy element production in the Universe?

PRIMA will probe the presence of newly formed heavy r-process elements via the photometric detection and/or spectroscopic characterization of the infrared transient associated with binary neutron star mergers. Simulations [7] show that the nebular phase emission from binary neutron star mergers can reveal signatures of heavy element production in the near infrared and likely in the mid-infrared. Mid-infrared observations of binary NS mergers have yet to be acquired, thus constitute an excellent ground for exploration using PRIMA.

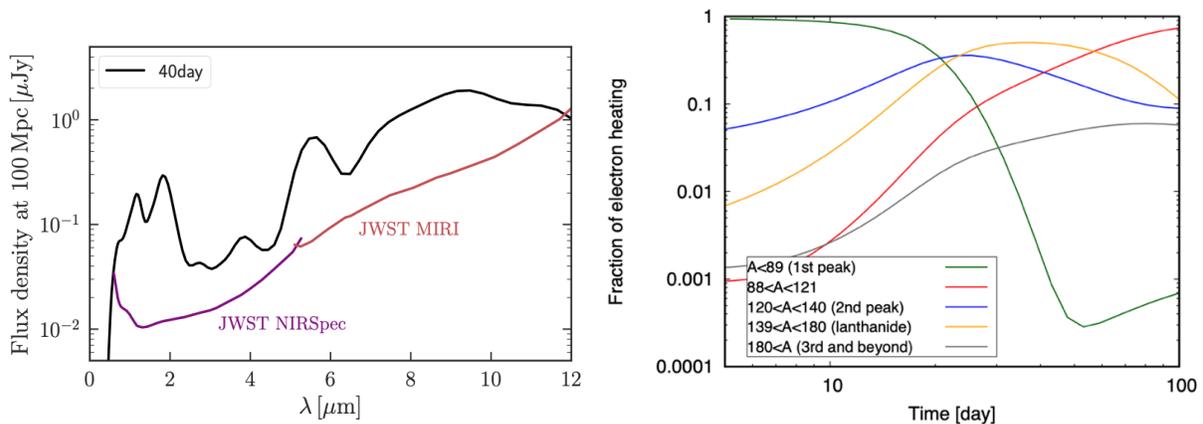

Figure 1: (Left) Spectrum at 40d for the fiducial model at a distance of 100 Mpc (from [7]). Although the plot stops at 12 μm, the emission is expected to be there past 30 μm [8, 10]. This regime is still little explored both theoretically and observationally. (Right) Fraction of electron heating contributed by various sets of elements as a function of time using the solar abundance pattern including the first peak (from [5]). While the first peak dominates at early-time, emission requiring elements in the second and third peak dominates at later times.

### Instruments Required:
- PRIMAger: Small Map (<20'×20'), Hyperspectral band

### Program Size:
- Large (100 + hr of Target of Opportunity observations)

### Approximate Integration Time:
- At least 200h depending on the source distance and phase

## Special Capabilities Needed:

- To observe the target source at its brightest flux level, Target of Opportunity observations should be initiated ∼1 week from their request.

## Synergies with Other Facilities:

PRIMA observations will complement multi-wavelength observations of GW counterparts obtained with other space- and ground-based telescopes sensitive to radiation from gamma-rays to optical, near-infrared, millimeter, and radio bands.

## Description of Observations:

PRIMA will probe the presence of newly formed heavy r-process elements via PRIMAger photometric detection of the infrared transient associated with binary neutron star mergers. In case the signal is exceptionally bright, spectroscopic characterization with FIRESS may also be attempted. This wavelength regime is currently unexplored for compact binary mergers. We expect to observe a continuum signal with very broad lines (see Fig. 1 left and [6,7]). The one known counterpart to a binary neutron star merger, GW170817, was observed in the NIR by Spitzer. At 3.6 μm, GW170817 is not detected to a limit of 23.2mag AB at +43 days and 23.1mag AB at +74 days [5].

Little theoretical modeling is currently available for kilonova nebular spectra in the far infrared. At 40 days from the merger, existing models predict a continuum signal with very broad lines at wavelengths longer than 10 μm [7, 10] on top of a T∼2,000 K blackbody. The continuum is expected to have $L\nu \sim 1 \times 10^{24}$ erg/s/Hz at ∼20 μm, $5 \times 10^{22}$ erg/s/Hz at 40 μm, and $5 \times 10^{21}$ erg/s/Hz around 70 μm, but uncertainties are large (Hotokezaka, private comm.). The spectrum is likely to be different weeks before or after the 40 days mark, therefore observations at multiple epochs are required. Given the above flux densities, we can expect a kilonova at 40Mpc (like GW170817) to have flux densities of 0.52, 0.026, and 0.003 μJy at 20 μm, 40 μm, and 70 μm, respectively, 40 days from the merger, which is challenging to achieve with PRIMAger even with many days of exposure time.

We therefore recommend using PRIMA to observe kilonovae:

- At early phases, when the intrinsic luminosity is expected to be larger across the optical/IR spectrum by at least 1 order of magnitude, starting ∼1 week from the merger. These observations will enable the building of the SED from the optical to the IR in synergy with other observatories for the first time and will have great legacy value.

- That are exceptionally bright because of their proximity. At a late phase of 40 days post-merger, a flux density of ∼8 μJy is expected at 20 μm if the merger occurs 10 Mpc away. This would require ∼200 hr of exposure time.

- That are likely brighter than the current models predict based on deep near-infrared observations obtained during follow-up observations. In fact, kilonovae with higher ejecta mass than GW170817 could generate a brighter signal.

These sources will typically be distant (tens to hundreds of Mpc) and coarsely localized (tens to thousands of square degrees), so we expect PRIMA to characterize well-localized (sub-arcsec) transients found by other telescopes in nearby galaxies via wide-field or galaxy-targeted searches. Event rates are still uncertain: the expected rate for binary neutron star mergers within 60 Mpc is up to ~0.4/year.

We describe here a possible baseline observing plan for PRIMA. When a nearby (D ≲ 60 Mpc) counterpart is located and confirmed spectroscopically, PRIMAger will obtain at least two epochs via Target of Opportunity observations. If possible, the first observation should occur within one week from the merger to maximize the chances of detecting the source. The second epoch should be acquired 3 and weeks after merger. The transient might be bright at early phases, but radiation from newly synthesized elements from the third peak is expected to dominate (see Fig. 1 right and [5]) in the later spectra. If a source is detected 3 weeks after merger, a third epoch should therefore be acquired around 9 weeks from the merger. PRIMA will probe kilonova emission uniquely complementing JWST observations.

# 2. Exploiting the PRIMA-Athena synergies to search for Compton Thick AGN

Luigi Barchiesi (Bologna University – INAF-OAS), Cristian Vignali (Bologna University – INAF-OAS), Francesca Pozzi (Bologna University – INAF-OAS), Laura Bisigello (Padova University – INAF-OAPD), Carlotta Gruppioni (INAF-OAS), Calura Francesco (INAF-OAS)

In the black hole -galaxy co-evolution framework, most of the star-formation (SF) and the black hole (BH) accretion is expected to take place in highly obscured conditions. The large amount of gas and dust absorbs most of the UV-to-soft-X radiation and re-emits it at longer wavelengths in the IR. Thus, obscured Active Galactic Nuclei (AGN) are very difficult to identify in optical and X-ray bands, but shine bright in the IR. Recent works show that the fraction of obscured AGN increases with redshift (up to ∼80% at z∼6, e.g., Vito+18) and that at high-z, due to the host galaxy being compact and gas-rich, the ISM can contribute significantly to the AGN obscuration (Gilli+20 and references therein). The key role played by obscured AGN is also demonstrated by X-ray background (XRB) synthesis models, which predict that a large fraction of the yet-unresolved XRB is due to the most obscured AGN (Compton thick, CT: NH >$10^{24}$ cm$^{-2}$).

We propose to use PRIMAger to perform a deep survey for the same field that will be observed with the ESA Athena X-ray Observatory. The capabilities of PRIMA will allow us to detect ∼6000 of the most obscured sources (60% of all CT-AGN at z<6) that would be otherwise missed by current facilities. The complementarity of the two facilities will allow to cover almost the whole AGN parameter space, with Athena able to detect high-luminosity AGN up to very high-redshift, and PRIMAger detecting the extremely obscured AGN that will be missed by Athena. We estimate that PRIMA will be able to detect ∼17000 AGN (70% of all AGN at z<10), with ∼12000 being also detected with Athena WFI. This will be a significant step forward in the process of placing more substantial constraints on the yet-unresolved XRB and investigating the BH accretion rate evolution up to very high redshift.

## General Scientific Area:

Growth of Galaxies and Black Holes Through Cosmic Time

## Science Justification:

One of the key, still open issues in astrophysics is the role of super-massive black holes (SMBHs) in shaping galaxies and the influence of the active galactic nuclei (AGN) on the star formation (SF) properties. While it is now accepted that the growth in BH mass of the AGN and in stellar mass (M$_*$) of the host galaxy are coupled, the exact mechanism, timescales, and how these influence each other are still a matter of debate. Therefore, their study is fundamental for understanding the evolution of both galaxy and AGN. In the co-evolution scenario (Hopkins+08), on the one hand, the stellar feedback may help in funneling gas to the nuclear region of the galaxy, thus triggering the AGN activity. On the other hand, the AGN feedback heats up and expels the gas, reducing or quenching the SF. Signatures of this interaction have been observed in winds and outflows of cold molecular (e.g., Feruglio+10, Cicone+14), neutral atomic (e.g., Rupke+05) and ionised (e.g., Weymann+91, McKernan+07) gas.

## The obscured accretion phase

Despite the differences between the various co-evolution models, for example regarding the co-evolution triggering mechanism and the involved timescales, all scenarios agree that a key phase is that of obscured accretion. In this phase, most of the BH accretion is expected to take place, however the high quantity of material fueling the SMBH growth has also the effect of hiding the core of the galaxy. Thus, **this first phase of growth takes place in very obscured conditions, making its study extremely challenging**. In fact, very obscured AGN can appear as normal non-active galaxies in the UV-optical band (because the accretion disk emission is totally extincted), as a dusty SF galaxy in the IR, and/or not being detected at all in the X-rays. However, these "missing" AGN are extremely important for explaining the X-ray background (XRB), as the current spectrum of the XRB can only be reproduced by assuming also the presence of a population of very obscured AGN that we are currently able to efficiently detect only at low redshift (e.g., Gilli+07). Finally, a complete census of these obscured accreting AGN is also needed for reconstructing the evolution of the BHAR at high redshift. In fact, the discrepancy between the model prediction and the measurements of the BH accretion rate density (BHAR) could be explained by assuming that current high-redshift surveys are missing a significant population of very obscured sources (Figure 1). **We are seeing hints of this population with IR surveys** (e.g., Delvecchio+14), **although in the high-z Universe (z>3) we must rely almost uniquely on the UV and X-ray bands, that are prone to miss the highly obscured AGN.**

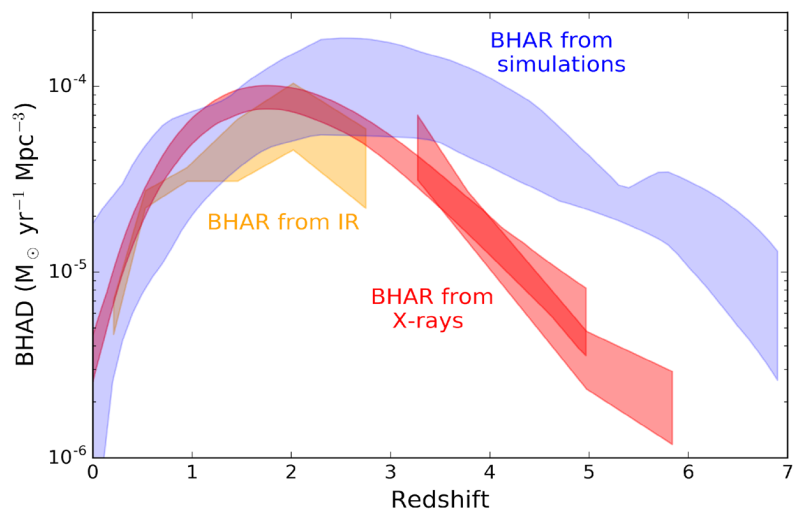

Figure 1: Redshift evolution of the BHAR from Barchiesi+21. The blue area is a theoretical BHAR curve obtained from Volonteri +16, Sijacki+15, and Shankar+13. The red areas are the X-ray-derived BHARs, obtained from observation, from Aird+15, Ueda+14, and Vito+14. The yellow area is the IR-derived BHAR from Delvecchio+14. PRIMA will allow to sample the BHAR at higher z than actual IR survey and will be able to detect very obscured CT-AGN, missed by X-ray surveys, that should contribute to the total BHAR.

A breakthrough in the selection and identification of AGN is expected with the new ESA X-ray observatory Athena, due for launch in the 2030s. Athena, with its combination of angular resolution, field of view and collecting area at ~1 keV, is the ideal instrument to perform X-ray surveys and will be more than two orders of magnitude faster than Chandra and XMM-Newton (Nandra+13). Unfortunately, even this new generation instrument will not be able to probe completely the AGN parameter space: most of the CT-AGN will be detected only at very low-z, and the detection of low-luminosity AGN at high-z will be challenging.

However, the detection of these elusive sources is well within the capabilities of the PRIMA mission. **Deep PRIMAger photometric surveys will be able to detect these sources up to z~6.** Moreover, **with a combined used of Athena observatory we will be able to probe almost the entire AGN population.** Finally, the synergies between the two facilities will help in overcoming the degeneracies between obscured SF galaxies and obscured AGN and to recognize the AGN as such.

## Simulations

We used XRB synthesis models to predict the total number of AGN as a function of intrinsic X-ray luminosity ($L_X$), amount of obscuration ($N_H$) and redshift (see Barchiesi+21). The prediction of the AGN emission are based on the SEDs of more than 500 AGN from the COSMOS field with both X-ray spectra and optical-to-FIR SED-fitting. For each $L_X$-$N_H$-z bin, we estimated the fraction of AGN that will be detected by Athena WFI, taking into account the source density and the instrument capabilities. Beside the source detection, we stated that we can characterize the AGN properties if the fitted luminosity and obscuration, obtained from the X-ray spectral fitting of the simulated spectrum, were within 30% of the input values. Regarding the IR side of our simulations, we used the above-mentioned SED and randomly extracted 20 of them from the corresponding bins. We measured their total (AGN+host) flux densities in all the PRIMAger bands. If the SED had a higher flux than the survey sensitivity (filter-wise), we considered that source as detected. We simulated the deepest pointing of the Athena WFI survey strategy (~1 deg$^2$) and assumed a 100 hr 1 deg$^2$ deep survey for PRIMA in the same field (see justification below).

## PRIMA and Athena capabilities

Our simulations show that more than 17000 AGN (70% of all the AGN at z<10) will be detected by at least one of the PRIMAger filters, while ~15000 (60% of all AGN) will be detected in at least all the first 6 PRIMAger filters (up to λ=45μm). Moreover, ~12000 AGN (~50%) will also have a detection in the X-ray with ATHENA WFI and for ~5000 (~20%) we will be able to characterize (with a 80% accuracy) their properties via X-ray spectral fitting.

As we can see from Figure 2, by exploiting both the X-ray and IR bands we will be able to detect AGN over the whole parameter space: while ATHENA WFI will detect almost all the high-luminosity ($L_X$>10$^{44}$ erg/s) AGN even to z>6 and a significant fraction of the low- and mid-luminosity ones, PRIMAger will cover extremely well the highly-obscured end of the parameter space. From our PRIMAger simulations, we can expect to detect ~6000 CT-AGN at z<6 (~60% of all the CT-AGN). In particular, low- and mid-luminosity CT-AGN ($L_X$ <10$^{43}$ erg/s) are the ones that are missed by current X-ray and UV surveys and that, even with the next-generation X-ray observatory, will be extremely challenging to detect beyond z~1. Moreover, these low- and mid-luminosity AGN should compose the majority of the population of heavily obscured AGN, thus covering this locus of the parameter space that will allow us to place stronger constraints on the BHAR and hopefully to bridge the gap between the predicted and the observed BHAR.

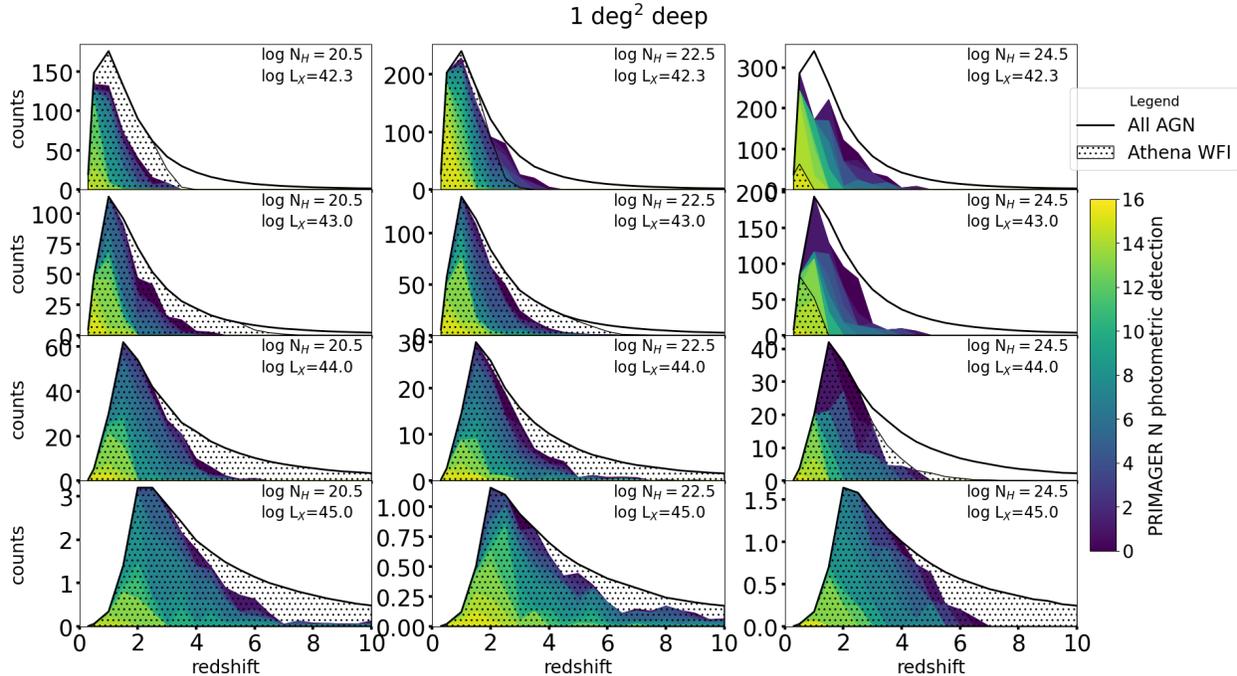

Figure 2: Number of predicted AGN per deg$^2$ and per $\Delta z = 1$ for a PRIMAger deep survey. The black lines are the total number of expected AGN, the dotted area represents the sources that can be detected in the X-rays by Athena WFI. The color code indicates our prediction for the number of PRIMAger photometric filters for which we will have a detection. The columns refer to AGN with different amounts of obscuration ($20 < \log (N_H / cm^{-2}) < 21$; $22 < \log (N_H / cm^{-2}) < 23$; $24.2 < \log (N_H / cm^{-2}) < 25$, from left to right), the rows to different AGN luminosity (from top to bottom, $42.0 < \log ( L_X / erg\ s^{-1} ) < 42.5$, $42.9 < \log ( L_X / erg\ s^{-1} ) < 43.2$, $43.9 < \log ( L_X / erg\ s^{-1} ) < 44.2$ , $44.9 < \log ( L_X / erg\ s^{-1} ) < 45.2$). **The contribution of PRIMAger will be fundamental in detecting the low-luminosity CT-AGN** (third column, first and second row) **that will be missed by Athena due to their extreme obscuration**.

## Instruments Required:

PRIMAger large map (1 square degree), using the hyperspectral band.

## Program Size:

- Large (100 + hr)

## Approximate Integration Time:

(based on calculation from sensitivity curves)

We took into consideration various survey sizes and depths. As we are most interested in the detection of the sources that will be missed by the Athena WFI survey (CT-AGN with $L_X < 10^{43}$ erg/s), we focused on a deep 1 deg$^2$ survey to match the deepest Athena WFI survey area.

With the latest PRIMAger sensitivity curves, we are mostly confusion-limited wavelengths with a 10 hr 1 deg$^2$ survey (with the exception of the first 1A filters). Choosing a 100hr 1deg$^2$ survey allows us to detect a significant number (~50%) of low- and mid- luminosity CT-AGN, while still being just ~3.1 times below the confusion limit. This depth allows us to detect the most-luminous AGN up to z~6. In particular, we will be able to detect ~180 AGN between 4<z<6 in all the first six PRIMAger filters.

With a shallower survey, the number of detected low-luminosity CT-AGN quickly drops. We also took into consideration deeper surveys, but the longer integration time does not provide a significant increase in the number of detected CT-AGN until we reach a ~400–500 hr survey, while being strongly affected by the confusion.

### Special Capabilities Needed:

- None

### Synergies with Other Facilities:

The synergies with the ESA X-ray observatory Athena are extremely important for two reasons. On the one hand, the combined use of both instruments allows us to cover the whole AGN parameter space. In fact, PRIMAger performance in detecting AGN lowers at z>3 and drops for z>4, while ATHENA WFI is capable of detecting even mid-luminosity AGN at z~6 (z>8 for the ones with $L_X > 10^{45}$ erg/s). However, even with its deepest survey, the detection of CT-AGN is extremely challenging for Athena, especially for the low- and mid-luminosity ones that compose the majority of the CT population. These sources will be easily detected by PRIMAger up to z~4.

On the other hand, exploiting both wavelengths helps in recognizing these sources as AGN. At z>2, an X-ray photometric detection with Athena corresponds to a luminosity of $L_X > 10^{42}$ erg/s, usually associated to AGN activity. Moreover, having IR and X-rays coverage can help to discern the AGN and host-galaxy components or to constrain the AGN X-ray spectral properties.

### Description of Observations:

To exploit the synergies between X-ray and IR, PRIMA will need to target the same fields as Athena. At the moment this is not defined yet, although it is known that it will be one of the deep fields with multi-wavelength coverage. Targeting an already extensively studied field (e.g., COSMOS, CDFS) will provide further benefits, such as having already at our disposal a complete multi-λ photometric coverage that will allow us to perform reliable SED-fitting, separating the AGN and host-galaxy contribution, thus allowing us to distinguish between dusty SF-galaxies and obscured AGN.

Moreover, having near-IR coverage, such as the one provided by the JWST COSMOS-Webb survey, could help in delivering priors to deblend the sources, thus allowing us to make the best of the deepest PRIMAger observations.

# 3. PRIMA follow-up of mm-wave transients: from M-stars to GRBs, TDRs and beyond

Dr. Dave Clements (Imperial College London), Dr Mike Peel (Imperial College London)

Today, the time variable far-IR/mm sky is largely unexplored. By the time that PRIMA flies, however, the next generations of ground-based CMB experiments, including Simons Observatory and CMB-S4, will be monitoring large areas of the sky for transient mm sources as a byproduct of their CMB observations, and will be producing regular alerts for newly discovered mm-transients. The first results from the current generation of such experiments have shown they are capable of detecting a wide variety of transient sources in the mm, ranging from stars in our own galaxy to extragalactic sources associated with AGN. These results, together with theoretical predictions for other classes of sources, indicate that future facilities will be able to detect many interesting classes of source including flaring stars of a wide range of spectral types, protostars, gamma ray bursts (GRBs), tidal disruption events (TDEs), neutron star mergers, faint blue optical transients (FBOTs), and supernovae (SNe). There is also the possibility that new classes of mm-variable will be uncovered by these CMB observatories. Predicted rates for these experiments range from 10s (SO) to 1000s (CMB-HD) of transient events over the lifetime of these projects. The addition of far-IR data points for these sources can determine much about their radiation mechanisms and the physics of the underlying source. PRIMA follow-up of representative examples of various mm-wave transients could thus have a powerful impact on our understanding of a wide range of astrophysical phenomena.

## General Scientific Area:

Time Domain

## Science Justification:

Transient phenomena have so far been only poorly studied at mm wavelengths, and studied very little in the far-IR. A wide range of Galactic and extragalactic sources are expected to show transient activity at these wavelengths. These range from luminous distant sources such as gamma ray bursts (GRBs), tidal disruption events (TDEs), neutron star mergers, faint blue optical transients (FBOTs), and supernovae (SNe) to more local phenomena including stellar flares and accreting protostars. Searches for transients that are bright at mm/submm wavelengths have so far been limited by the nature of the surveys being conducted at these wavelengths. Two approaches have been adopted to date: Firstly, serendipitous detections of a wide variety of transients have been made by CMB experiments (e.g., Guns et al., 2021; Naess et al., 2021). These CMB surveys cover a range of areas and cadences, and have discovered transients varying on timescales of minutes to weeks; Secondly, explicit monitoring campaigns have been mounted targeting specific objects such as star-forming regions (e.g., Lee et al., 2021) which have uncovered protostellar sources varying on year-long timescales. The next generation of CMB experiments, such as the Simons Observatory and CMB-S4, represent a significant enhancement to our ability to detect mm/submm transients since they will cover large areas of the sky with a roughly daily cadence, and have large array detectors which in principle can be used to look for variations on timescales shorter than the scanning cadence. Recent calculations (Eftekhari et al., 2022) suggest that these future CMB experiments will detect many tens to hundreds of transient sources over their lifetimes, with the more advanced experiments, such as CMB-S4 and CMB-HD, which will be operating while PRIMA is flying, producing the largest numbers. There are thus likely to be many mm/submm transients for PRIMA to follow-up during its lifetime. Furthermore, other observatories studying the transient sky in other ways, including the VRT-LSST and gravitational wave observatories, may also provide PRIMA with important targets of opportunity where far-IR observations can provide important new information about the nature of these objects.

The addition of far-IR observations to the study of mm/submm transients can provide a range of physical insights into these diverse sources. For the most luminous sources, including GRBs, FBOTs, GW-sources and others, the principal emission mechanism is synchrotron radiation. In these cases, observations in the far-IR can help determine the turn-over frequency of the synchrotron emission and thus the parameters of the shock responsible for the emission. This will provide insights into the interstellar medium around the transient source, including density and magnetic field, as well as a better measurement of the overall energy budget of the shock. The timescales for these events are in the days to weeks region which should make it relatively easy to observe and monitor such sources on a target-of-opportunity basis.

A long-lived sub-class of extragalactic transients are tidal disruption events (TDEs) where a star or other body is disrupted and accreted by a supermassive black hole (e.g., Yuan et al., 2016; Andreoni et al., 2022; Mattila et al., 2018). These events can be very luminous and can have durations lasting days to years. Event rates for such TDEs can be as high as several hundred per year that would be detectable by large area surveys with sensitive mm-telescopes such as ALMA or the LMT, with far-IR fluxes for such sources with peak far-IR fluxes of 10–100 mJy for sources in the nearby universe. In some case the TDEs may take place in heavily dust obscured regions (e.g., Mattila et al., 2018). This leads to dust heating and much of the energy of the TDE being reradiated into the infrared. The infrared emission of a dust obscured TDE can last years. Few events of this nature have so far been studied in detail, but the addition of short-wavelength far-IR emission, in PRIMA's high frequency bands, will add new constraints to the energetics of such objects and our understanding of the circumnuclear dust heated by them.

Galactic mm/submm transients also exist, with two main classes currently known. The first of these are accreting protostars which vary on month to year timescales. These may be the subject of specific monitoring programmes but may also appear as targets of opportunity in the case of significant events. The known mm/submm protostar transients are all quite bright (Lee et al., 2021), with 850 µm fluxes > 5 Jy, so there may be many more less luminous systems that will be identified as transients by future observations by CMB experiments and others. With their long period variability these sources can easily be dealt with as targets of opportunity or specific monitoring projects. The other class of mm/submm Galactic transient are stellar flares, which can have fluxes of 100s of mJy at 220 GHz, but which have short durations of minutes to hours.

These will be more difficult for PRIMA to study as they will need fast response times after initial detection. A specific target of opportunity interrupt process would be needed to allow such targets to be observed if accessible once detected by e.g., a CMB experiment.

The mm/submm transient sky remains very poorly known, so there is also the possibility of novel or unusual sources that may be high priority targets. An example of this type of source is AT2018cow (Ho et al., 2019) – an extreme FBOT with very luminous mm emission also detected in the X-rays. It is thought to be the result of a fast sub-relativistic shock in a dense environment but the origin of this shock remains unclear. Submm observations proved crucial to understanding the nature of this source, identifying the frequency of the spectral turn over, but it may well be that other extreme sources have turnovers at still higher, far-IR frequencies accessible only by PRIMA.

## Instruments Required:

PRIMAger: Small Map (< 20' × 20'); Hyperspectral band (25–80 μm; 12 filters; R=10); Polarimeter band (96, 126, 172, 235 μm; R=4)

## Program Size:

Medium (10–100 hrs)

## Approximate Integration Time:

A sample of 50 transient sources in the far-IR, detected by future CMB experiments and selected to be brighter than 10 mJy (significantly above the confusion limit in most bands) in the far-IR will lead to quite short individual integration times – 5 minutes or less. To follow these sources for 10 epochs of observation then this project will require about 50 hours of observations.

## Special Capabilities Needed:

Monitoring over days to weeks timescales depending on the nature of the transient. Some specific objects (e.g., dust embedded TDEs) may need monitoring over year time scales.

## Synergies with Other Facilities:

Synergy directly with CMB experiments that will provide event triggers, and with other monitoring projects such as VRT-LSST.

## Description of Observations:

Single source photometric observations in all PRIMAGER bands with 5-minute integration time. Observations to be repeated on timescales determined to be appropriate for each specific source by other observatories that can characterize the timescale of variation.

# 4. The dust distribution around accreting black holes using reverberation mapping

Varoujan Gorjian (Jet Propulsion Laboratory, California Institute of Technology), Michael W. Werner (Jet Propulsion Laboratory, California Institute of Technology)

Variability studies are a powerful tool for studying the structures of unresolved sources. One such type of variability study, called reverberation mapping (RM), established that the source of infrared radiation from active galactic nuclei (AGN) was from dust absorption and re-emission by showing that optical brightening and fading was followed by a corresponding (delayed) variation at infrared (IR) wavelengths. Since the IR universally carries a substantial fraction of the total AGN energy output, dust absorption plays a fundamental role in the structure and appearance of all AGN. Also, in the simplest version of how the dust is distributed, a thick dust torus determines whether a viewer sees a broad-lined/unobscured Type I AGN or a narrow-lined/obscured Type II AGN.

Although the dusty torus model has successfully explained many aspects of AGN phenomenology (e.g., polarization), direct observational constraints on its size and structure are still sparse. Dust RM has been done exclusively in the near-IR (1–5 µm) which traces the inner edge of the dust near the dust sublimation radius. Hence extending RM to the mid-IR, especially to the peak of the dust emission between 25 and 30 µm, allows for an examination of the main distribution of the dust around AGN.

RM with PRIMA will focus on variability-monitoring of a sample of low- to high-luminosity AGN, using both the photometer and the spectrometer, to trace the 25–30 µm emission that is reverberated from the UV/optical accretion disk emission which will be monitored by ground-based observatories. It is vital for the variability monitoring to have near continuous cadence of revisiting each source throughout several years. This will restrict sources to be placed near the North and South Ecliptic poles where PRIMA has extensive visibility throughout its orbit around Sun. Also, by being able to use both the spectrometer and the photometer, the cadence will not be compromised by switching campaigns between the instruments.

The expected lag between the optical and the 25–30 µm emission region can be long (months to years), but the nominal five-year duration of the PRIMA mission should allow for reverberation to be detected. Then, by detailed modeling of the response of the dust to the shorter wavelengths, the distribution of the dust around the observed AGN will be revealed and linked to the accretion disk luminosities.

## General Scientific Area:

Time Domain

## Science Justification:

**Broader Context:** Variability studies are a powerful tool for studying the structures of unresolved sources. Such studies established the source of infrared radiation from active galactic nuclei (AGN) by showing that optical brightening and fading was followed by a corresponding (delayed) variation at 2 μm. This reverberation mapping (RM: Peterson 1993) determined that the infrared is thermal re-radiation by dust grains which are heated by the primary continuum emission from the compact accretion disk around a galaxy's supermassive black hole (Clavel, Wamsteker, & Glass 1989, Nelson 1996) (Figure 1).

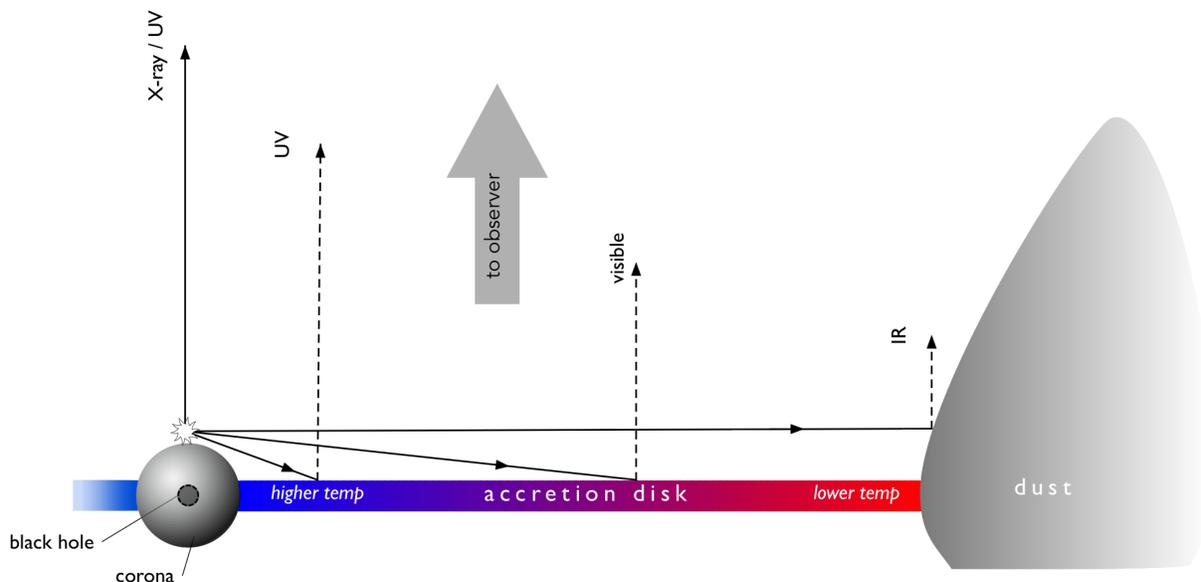

Figure 1: A schematic (not to scale) representation of reverberation mapping. The differing arrival times of differing wavelength (UV vs visible vs IR) correspond to the physical separation at a length of cΔt between those regions where Δt is the difference in arrival times of differing wavelengths.

Since the IR universally carries a substantial fraction of the total AGN energy output (see Figure 2), dust absorption plays a fundamental (not merely incidental) role in the structure and appearance of all AGN. Also, in the simplest version of how the dust is distributed, a thick dust torus determines whether a viewer sees a broad-lined/unobscured Type I AGN or a narrow-lined/obscured Type II AGN.

Although the dusty torus model has successfully explained aspects of AGN phenomenology (e.g., polarization), direct observational constraints on its size and structure are still sparse. In fact, detailed SED modeling has shown that the dust around the supermassive black hole is inhomogeneously distributed (i.e., "clumpy") and has multiple components. (e.g., Nenkova et al. 2002, Schartmann et al. 2008, Hönig and Kishimoto 2010).

But dust RM has been done exclusively in the near-IR (1–5 μm) which traces the inner edge of the dust at and near the dust sublimation radius, and so extending RM to the mid-IR, especially to the peak of the dust emission between 25 and 30 μm, allows for an examination of the main body of the dust around AGN.

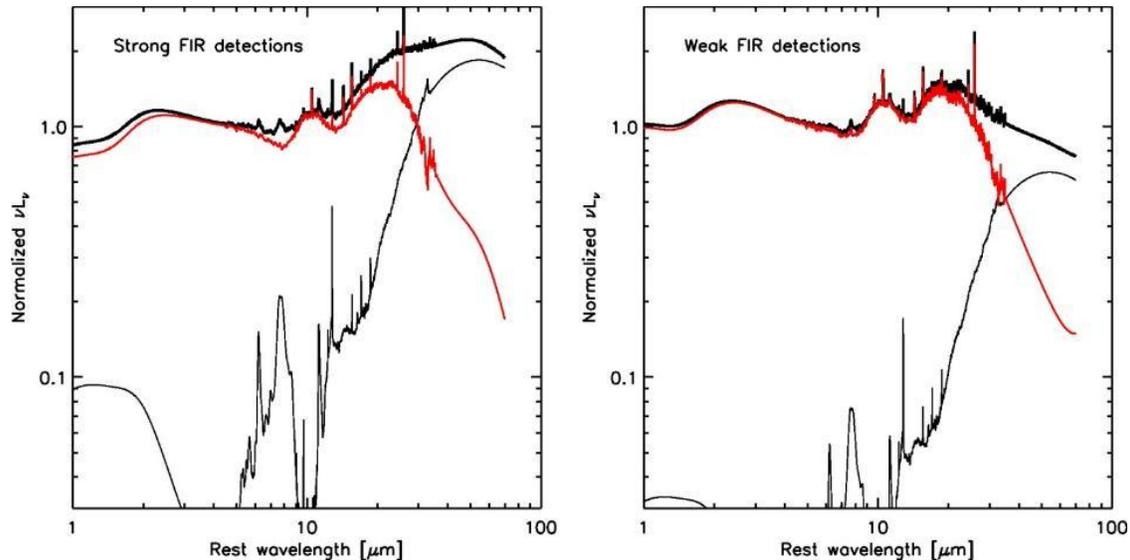

Figure 2: Normalized mean SEDs for strong FIR quasars (left, top curve) and weak FIR quasars (right, top curve). The red SED curves show "intrinsic" AGN SEDs obtained by the subtraction of the scaled mean starburst (ULIRG) spectrum (shown in black) from the mean SEDs. The red AGN curves indicates how IR emission (>1 μm) carries a substantial portion of the AGN's energy and that the peak AGN emission is between 20 and 30 ⬜m. In the PRIMA 25–30 μm wavelength region, the variability will be dominated by the AGN allowing for RM to provide critical insight into the dust distribution (Netzer et al., 2007).

### Science Questions:

It is striking that with the important role that dust plays in the appearance and energetics of AGN, we still do not have a definitive picture of how dust is distributed around the accretion disk of an AGN. RM at the shortest wavelengths (1–5 μm) has shown where the dust sublimation radius is: a few light weeks for low-luminosity AGN to a few light months for quasars (Koshida et al 2014). Interferometry from 8–12 μm has traced dust on much larger scales: tens to hundreds of parsecs (e.g., Lopez-Gonzaga et al. 2016). So, there is a critical gap of knowledge about the distribution of dust between the sublimation radius and the farthest reaches probed by the current capabilities of interferometry.

The key questions that RM with PRIMA can address are:

1. What is the shape of the dust distribution around the accretion disk of a supermassive black hole?

2. How far does that dust extend?

3. Where is the transition between dust heated by the accretion disk and host galaxy dust?

### Need for PRIMA:

The capability of PRIMA to provide high sensitivity observations in the 25–30 μm range provides the very first opportunity to do 25–30 μm RM on nearby AGN. Previously, the Spitzer Space Telescope could observe at those wavelengths, but due to its rotating schedule for its 25 μm camera, it could not provide the needed regular cadence to do the reverberation observations. Currently the James Webb Space Telescope has the needed 25 µm capabilities, but due to its large overhead needed to slew to any target, the hundreds of observations needed to provide the cadence to detect reverberation would dwarf the actual on-target observing time, thus making for a radically inefficient use of the telescope. This imbalance between overhead and observations would be exacerbated by the fact that we would need to monitor a large sample of AGN (>10) to come to a statistically representative model of dust distribution around the accretion disks of AGN. PRIMA with its high sensitivity from 25 to 30 µm, but smaller physical size, will allow for RM programs to repeatedly visit multiple targets without excessive overhead, and get the needed observations and build the needed statistics for a large sample of AGN over the five-year lifetime of the mission.

## Interpretation Methods

The approach for understanding the dust distribution is to take an optical light curve (obtained by other facilities) and combine it with the IR light curve provided by PRIMA. As an example, we show the optical light curve obtained for the AGN Zw229-015 by the Kepler Spacecraft and the 3.6 µm light curve provided by the Spitzer Space Telescope (Figure 3, Guise et al. 2022). As can be seen, the dust reverberated light curve is a time-displaced, smoothed, and widened version of the optical light curve. Each of these changes from the optical light curve to the IR light curve is a function of the dust distribution.

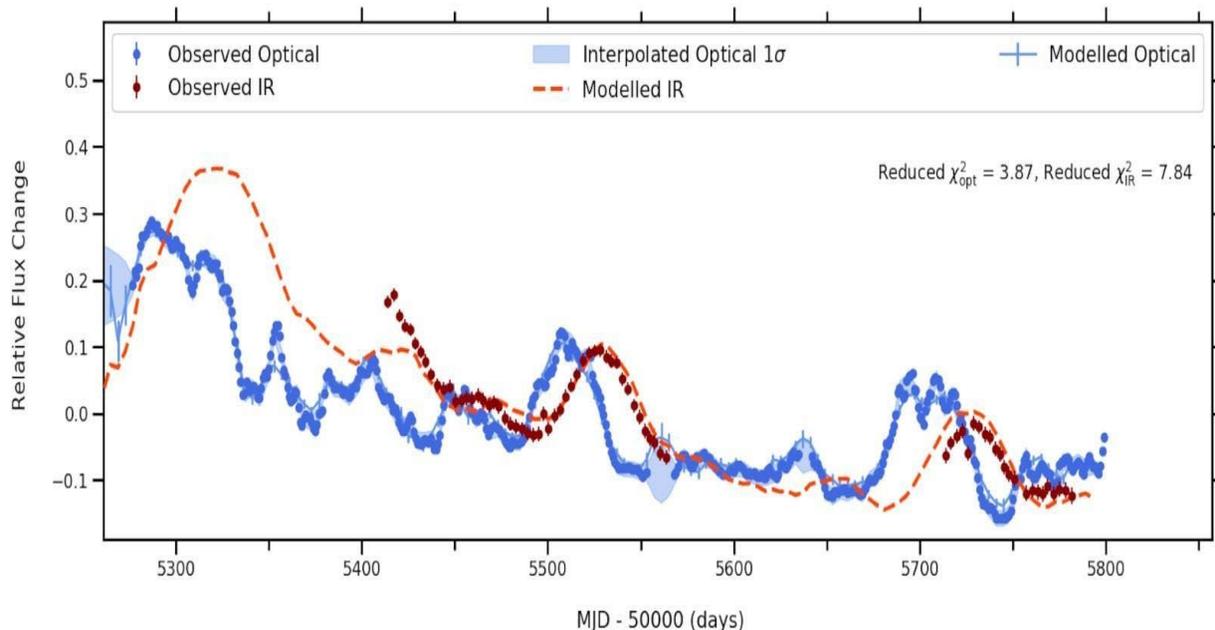

Figure 3: Kepler (dark blue points) and Spitzer (dark red points) observed light curves for the 2010-2011 observing season for Zw229-015, plotted with the interpolated optical (light blue) and interpolated infrared (light red dashes) that corresponded to the highest posterior distribution.

Using MCMC approaches, a model IR light curve is created that best matches a delayed, smoothed, and widened version of the optical light curve. The displacement between their peaks and valleys gives the first order separation of the IR emitting dust from the optical emitting accretion disk, which in this case was found to be ~11 light days.

The light curves are then used to create a dust-cloud distribution model that properly reproduces the reverberated light curves. In the case of the Zw229-015, a distribution of 10,000 dust clouds corresponding to the mean parameters from the MCMC modelling of the Zw229-015 light curves produced the dust distribution represented Figure 4.

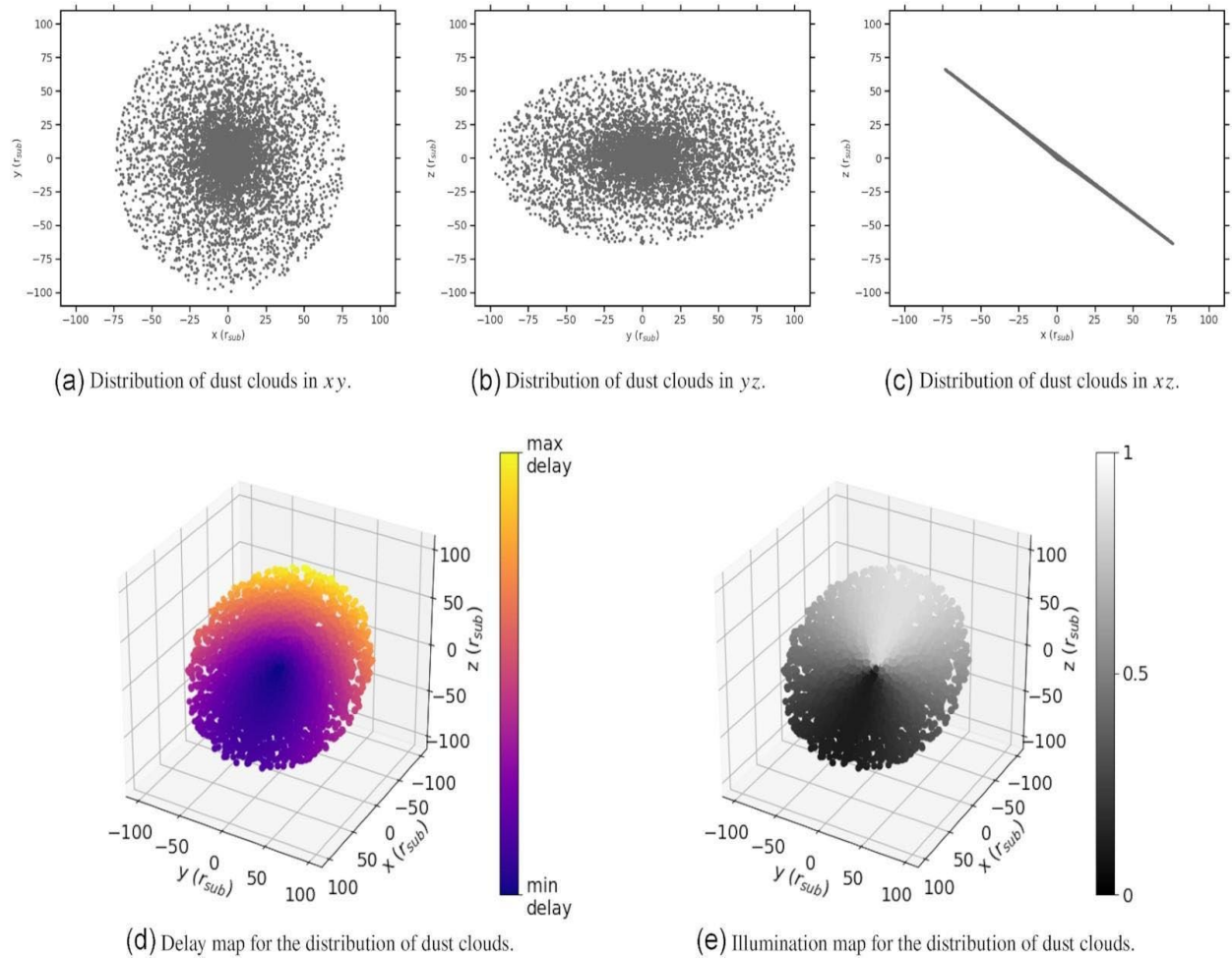

(a) Distribution of dust clouds in $xy$.   (b) Distribution of dust clouds in $yz$.   (c) Distribution of dust clouds in $xz$.

(d) Delay map for the distribution of dust clouds.   (e) Illumination map for the distribution of dust clouds.

Figure 4: Distribution of 10,000 dust clouds surrounding the Zw229-015 accretion disk. These distributions reproduce the delayed infrared light curves and are derived from the mean parameters of the MCMC modeling of the delayed IR light curves vs the optical light curves (Guise et al., 2022).

## Instruments Required:

- PRIMA FIRESS: Small Map Low-res, R~130
- PRIMAger: Small Map (< 20'×20') Hyperspectral band

Since the AGN monitoring will require a uniform cadence, but PRIMA observations will be split between the imager and the spectrometer, we will use both instruments to gain an uninterrupted cadence for our 25-30 μm observations. This will require a good cross calibration between the PRIMAger R~10 Hyperspectral Band and the PRIMA FIRESS Spectrometer Small Map, low res (R~130) mode in the 25–30 µm region. By choosing the FIRESS Small Map option, we come closest to recreating the PRIMAger's maps and thus minimizing the influence of the slit when comparing the resultant images.

## Program Size:

- Large (100 + hrs)

## Approximate Integration Time:

Integration times will be relatively short since the 25 µm flux densities for the sources are known to be in the range of ~10 to 800 mJy. But signal to noise needs to be very high: 100 to 1 to allow enough detection of inflection points to allow good cross correlation with shorter wavelengths. This is an empirically established SNR based on years of the community's experience with RM, but is of greatest necessity as the expected variations in the 25–30 µm region will be smaller than those at 1–5 µm.

So according to the early sensitivity plots for the PRIMAger, this high SNR for our fainter sources (~10 mJy) covered by a small mosaic (10'×10') can be achieved by a ~4-minute integration.

For the same source brightness, area covered, and SNR, FIRESS should achieve the same SNR in approximately the same time since by binning the spectrum from an R=130 to R=10 we gain a factor of ~4 in sensitivity. But since spectrographs tend to have more optical elements and spectral mapping usually takes longer than image mosaicking, we can add a factor of 4 to the exposure time (which will increase sensitivity by a factor of 2) and say that the spectral coverage will achieve the needed SNR in ~16 minutes per target.

In either case the, there are still uncertainties about the exact mosaicking/spectral mapping time, but it is safe to say that even for our fainter sources, we expect to get the required SNR in less than 20 minutes of observing time.

## Special Capabilities Needed:

- None

## Synergies with Other Facilities:

Since RM observations for dust reverberation are dependent on monitoring short wavelength emission from the accretion disk, optical observations are necessary. Fortunately, with the advent of large-scale optical monitoring surveys from the ground (e.g., Rubin) that predate the PRIMA launch, there will not be any need to plan for separate optical observations.

But it is always beneficial to have additional information, and so the AGN RM community has a consortium that gets multiwavelength photometric and spectroscopic monitoring for any AGN that is being monitored for reverberation, based on the availability of the facilities that each member has access to. Members also apply to general observer facilities on the ground and in space to supplement the RM data. AGN that are being monitored from space get a higher priority from the team members. This process has been employed successfully for decades starting with observations by the International Ultraviolet Explorer of six AGN from 1988 to 1996 (https://www.asc.ohio-state.edu/astronomy/agnwatch/) and continuing till today with observations of NGC 5548 and Mrk 817 led by the Hubble Space Telescope (De Rosa et al 2015, and Kara et al 2021) and supplemented by multiple ground and space-based telescopes like the IRTF, Swift, XMM, and others.

## Description of Observations:

A sample of AGN will be identified spanning a range of luminosities that would set differing dust sublimation radii and accretion rates, both of which are critical in setting the structure of the surrounding dust.

Since PRIMA will have a field of regard 30 degrees wide on the sky (Figure 5). This will limit the AGN sample to sources that will be at or close to the North and South Ecliptic Poles because those regions will get longer accessibility throughout the year.

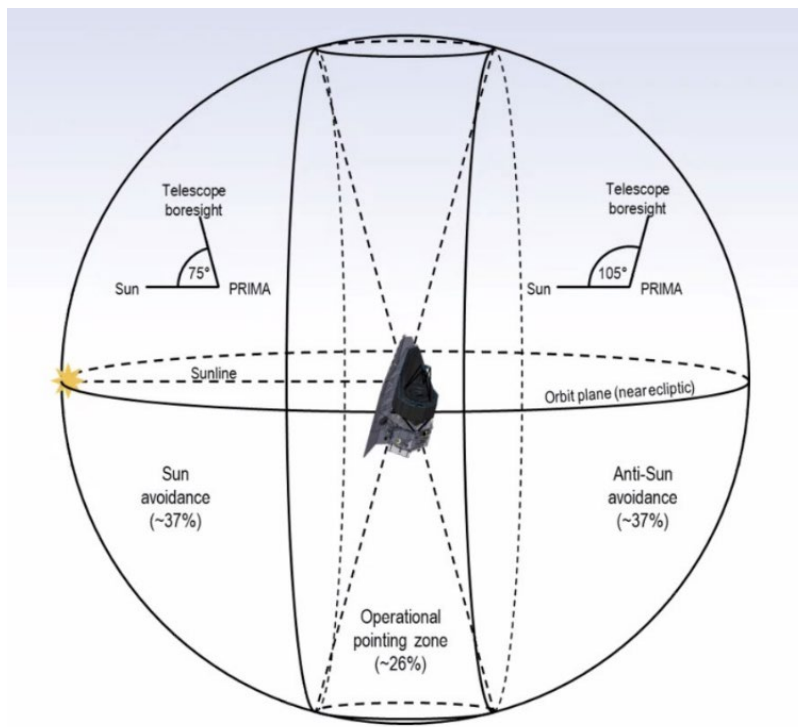

Figure 5: PRIMA field of regard +15 degrees from plane normal to the Sun-Earth line. As the orbit progresses during the year, the field of regard covers the entire sky with the North and South Ecliptic poles getting continuous coverage.

As the Spitzer Space Telescope had a similar restriction, our experience showed that for AGN like Mrk 817 and Zw229-015, we could get over 220 days of continuous observations. These are ideal candidates because they are low luminosity AGN and so their 25–30 μm emitting region will be closer to the accretion disk and so their reverberation timescales will be shorter and so can be easily captured in those 220 days.

For more luminous AGN like quasars, the reverberation timescale would be much longer, and so would need a correspondingly longer continuous timespan. Fortunately, within the continuous viewing zone (CVZ) near the North and South Ecliptic Poles, there are multiple quasars which would lend themselves to long term monitoring lasting for a few years to detect reverberation.

Once the targets have been identified, they would be placed within the observing plan of PRIMA such that each source would be revisited once every 4–5 days as that would cover the expected delay and smoothing of the optical light curve. Since these observations don't have to be taken at an exact moment in time, a buffer of ±1 days will be added to the timing. This provides the PRIMA schedulers enough slack so that these repeated observations can be optimally placed in the schedule to take the least amount of slew-overhead as well as not interrupting critical extended observations.

## Acknowledgement:

A portion of this research was carried out at the Jet Propulsion Laboratory, California Institute of Technology, under a contract with the National Aeronautics and Space Administration (80NM0018D0004).

# 5. Probing the afterglow of relativistic stellar explosions in the far-IR

Anna Ho (Cornell)

Gamma ray bursts (GRBs) are some of the most energetic explosive events in the Universe and as such can be viewed as a potential excellent probe of the early phases of cosmic history. While significant progress has been made both theoretically and observationally over the past decades, the origins (or progenitors) of GRBs are still elusive. In addition, several new classes of transients have been discovered in the past few years that may be related to gamma-ray bursts, but do not produce luminous gamma-ray emission. One example is "luminous fast blue optical transients" (LFBOTs), the prototype of which is AT2018cow (Prentice et al. 2018).

Models currently available in the literature suggest that the peak of GRB afterglow SEDs is located at far-IR wavelengths (e.g., about 100 μm) a few minutes to hours post-explosion. Despite such importance, there is currently no GRB afterglow observation at far-IR wavelengths. PRIMA will be a game changer in high energy astrophysics on this topic.

## General Scientific Area:

Time Domain

## Science Justification:

The afterglow of GRBs—synchrotron radiation from the collision of the shockwave with the ambient medium—is a key diagnostic of the progenitor system. Modeling the afterglow SED as it evolves with time provides insight into the structure, velocity, and magnetization of the relativistic outflow launched by the collapsing star, which in turn constraints the nature of the compact object "engine" and the progenitor star, as well as the physics of the outflow. Afterglow emission can show very complex behavior, particularly at early times (first few hours) when the peak is in the far-IR (see Figure). Detailed multifrequency ALMA observations of GRBs have yielded significant progress in early-afterglow studies by enabling distinct components in the afterglow emission to be separated (Laskar et al. 2018). In addition, millimeter-wave observations of other transients with relativistic outflows such as LFBOTs (Ho et al. 2019) have shown prominent components that may also peak in the far-IR—perhaps also a reverse shock (Margutti et al. 2019) or an interesting electron populations such as a relativistic Maxwellian (Margalit & Quataert 2021).

Models currently available in the literature suggest that the peak of GRB afterglow SEDs is located at far-IR wavelengths (e.g., about 100 μm) a few minutes to hours post-explosion, and that it shifts to longer wavelengths as time goes on (e.g., Perley et al. 2014; see Figure). The time evolution of the SEDs is intimately coupled with the environment and explosions properties, and prompt multiwavelength observations are needed to break the degeneracy between all of these parameters. Therefore, quick (within 10 minutes) follow-up and monitoring observations of GRB afterglows, particularly at far-IR wavelengths, with careful characterization of the SEDs, will be key. Despite such importance, there is currently no GRB afterglow observation at far-IR wavelengths. PRIMA will be a game changer in high energy astrophysics on this topic.

Thus, far-IR observations being conducted by PRIMA can advance our understanding of relativistic outflows from stellar explosions and address the question raised by the recently released Astro2020 decadal survey, "B-Q2. WHAT POWERS THE DIVERSITY OF EXPLOSIVE PHENOMENA ACROSS THE ELECTROMAGNETIC SPECTRUM?"

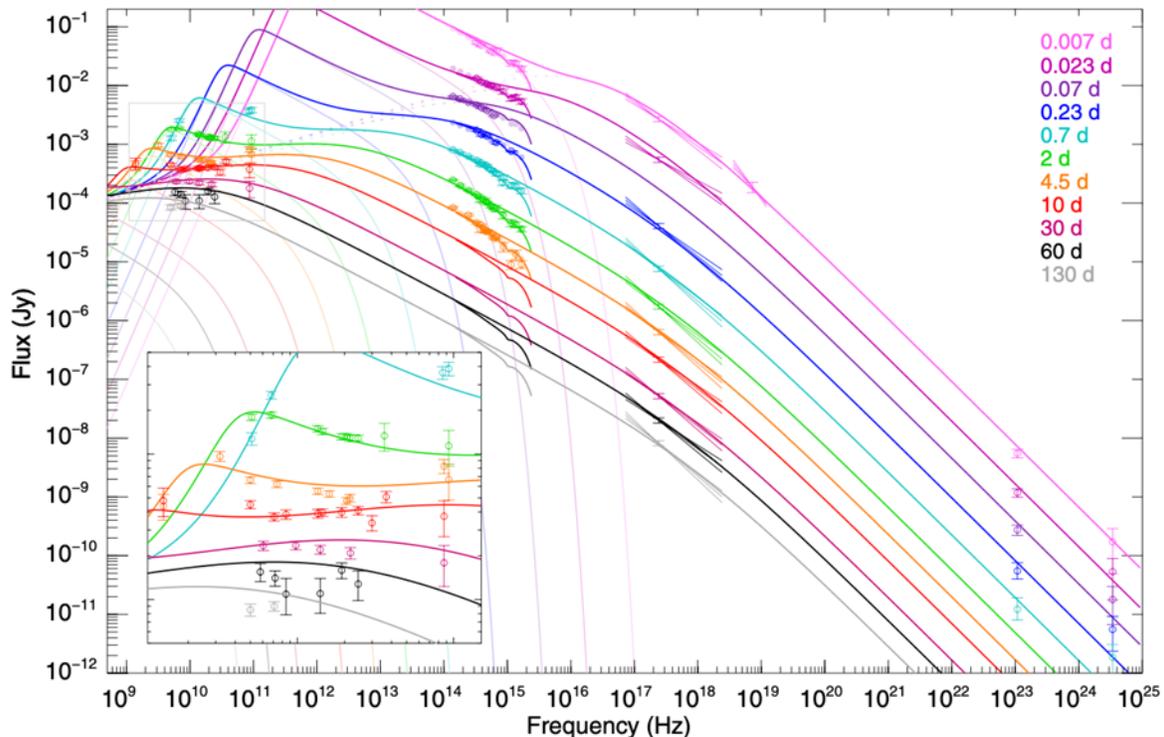

Figure 1: Observations of the afterglow of GRB 130427A. The SEDs originating from reverse shocks take the peak values at far-IR wavelengths ($10^{12}$–$10^{13}$ corresponding to 300-30 µm). Taken from Perley et al. 2014.

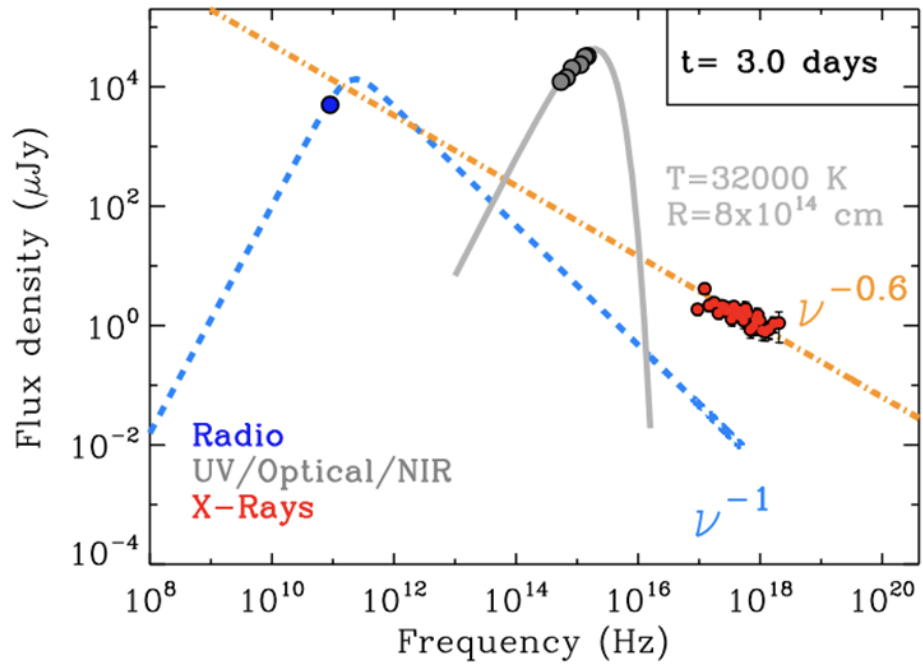

Figure 2: Broadband SED of AT2018cow at 3 days post-explosion (Margutti et al. 2019). The SED of the radio component at this stage is highly uncertain and would greatly benefit from observations in the PRIMA band.

## Instruments Required:

PRIMAGer small maps at all bands

## Program Size:

- Medium (10–100 hr)

## Approximate Integration Time:

(based on calculation from sensitivity curves):

- Integrating for 1 minute gives a 5-σ point-like sensitivity of 860 μJy in Band 2 (100–200 μm, i.e., few × $10^{12}$ Hz).

- Integrating for 18.5 minutes gives 5-σ point-source sensitivity of 200 μJy in Band 2 (100–200 μm, i.e., few × $10^{12}$ Hz).

- Integrating for 1.2 hours gives 5-σ point-source sensitivity of 100 μJy.

- For GRBs, the total request is very modest, 4 min per burst + 30 min = 70 min total.

We request five luminous fast blue optical transients (see below) observations over the course of PRIMA's lifetime, each with integration times of 1 min + 18.5 minutes + 1.2 hours + 1.2 hours = 2.7 hours. The total request is 13.5 hours.

## Special Capabilities Needed:

- None

## Synergies with Other Facilities:

GRB instruments (e.g., Fermi, Swift), submillimeter (ALMA), radio (VLA, ngVLA)

## Description of Observations:

For GRBs we use observations of GRB 130427A (z=0.340; Figure 1) as a model, since it is considered to be a relatively typical burst (Perley et al. 2014). We request four observations on a logarithmic timescale (times given post GRB): 10min, 20min, 40min, and 80min. An integration time of 1 min would enable us to detect a source as bright as ~1 mJy. At the distance of GRB 130427A, the flux densities at these times would be 1000 mJy, 200 mJy, 60 mJy, and 30 mJy. Therefore, we could detect a burst like GRB 130427A a factor of five as distant, or out to z=1.3. In practice, most GRBs have an unknown redshift and poor localization (square degrees), and not all of them have bright reverse shocks (the fraction is not well constrained). Therefore, we would focus on well-localized GRBs that are known to have bright afterglows in the wavelength range relevant to PRIMA. We estimate that we would trigger on ~10 GRBs over the course of the five-year mission (~2/year), and that we would additional obtain deeper (~10 min integration) observations for a subset (say 2–3).

For LFBOTs we use observations of AT2018cow (d=60 Mpc; Figure 2) as a model, since it is the best-observed LFBOT. However, we emphasize that the early emission from LFBOTs is highly uncertain. We request four observations (log timescale, times given post discovery, which will likely be in the X-rays or UV during the PRIMA mission): the first ASAP (minutes to an hour), 4 hours, 17 hours, 3 days. At the distance of 18cow the flux densities at 3d might be 1–10 mJy, so a 1.2 hr integration could detect an event out to d=200 Mpc or even 600 Mpc. Within this volume the expected number in five years is ~10–60. To be conservative we assume the lower end, and take into account the fact that not all of them will be discovered.

# 6. Why far-infrared observations are important for supernovae and supernova remnants?

Mikako Matsuura (Cardiff University), Ilse De Looze (Ghent University), Ori Fox (STScI), Ciska Kemper (E-CSIC / ICREA / IEEC), Florian Kirchschlager (Ghent University), Dan Milisavljevic (University of Purdue), Armin Rest (STScI), Laurance Sabin (UNAM), Arka Sarangi (University of Copenhagen), Roger Wesson (Cardiff University)

Supernovae (SNe) are critical engines that drive the evolution of the interstellar medium (ISM) in galaxies: SNe are considered to be the dominant source of chemical enrichment, a key source of dust creation, and the origin of energetic shocks and outflows, which destroy dust present ISM. Moreover, an instant burst of SN explosion energy can illuminate the surrounding ISM, so that SNe can work as a lighthouse that enables us to investigate the ISM gas and dust in distant galaxies, which normally may not be detected without SN illumination.

Yet, it is not well determined how much of the elements are synthesised, and how much of the elements condensed into dust within the SNe/SNRs and how much of freshly formed SNe dust and existing ISM dust is destroyed by SNe and SNRs. With high-sensitivity imaging and spectroscopic capabilities, PRIMA will be able to measure the true figures of dust mass formed in SNe, with unveiling the dust compositions within.

## General Scientific Area:

Galaxy Ecosystems and Interstellar Medium

## Science Justification:

PRIMA's far-infrared imaging capabilities will detect dust emission, and atomic and molecular line emissions from SNe and SNRs in the Milky Way, local group galaxies and galaxies up to 5 Mpc. With these new exploitations of wavelengths and distant galaxies, PRIMA will solve the following questions, relevant to 'How important are the SNe for ISM evolution of galaxies?'

- Are SNe important sources of dust in cosmic history? Can supernovae form the first dust of the Universe?

- What kind of dust grains are produced by supernovae? Are the grain properties (grain compositions and grain sizes) the same as those in the interstellar medium?

- How do supernovae cool down to become supernova remnants?

- What are the dust properties of the interstellar media in distant galaxies?

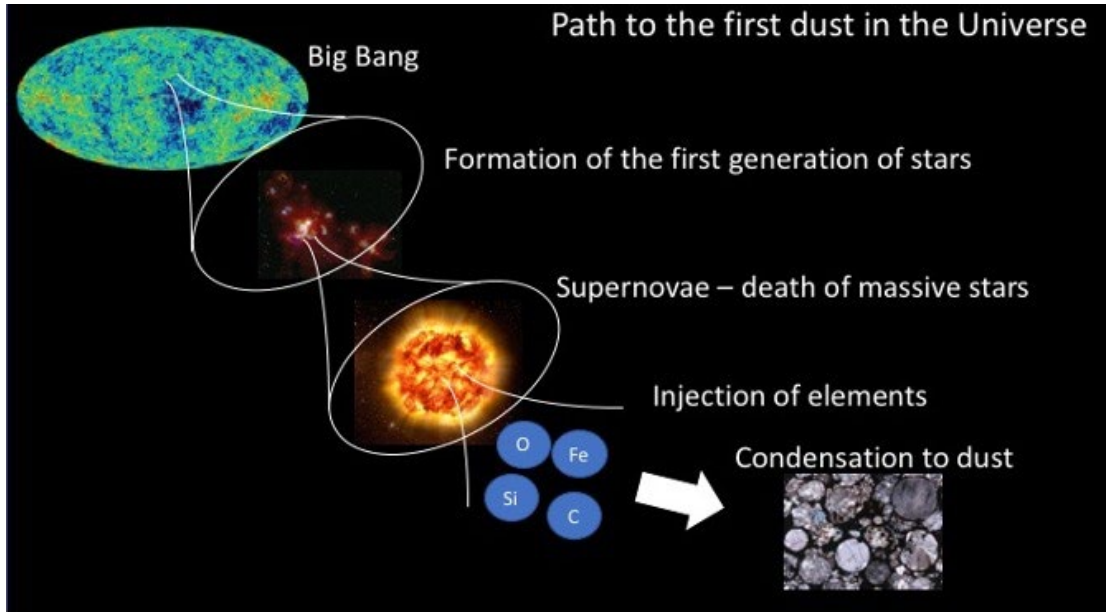

Figure 1: The pathway to form the first dust in the universe: after the Big Bang, the first generation of stars are formed, and high mass stars die as supernovae. These SNe synthesize elements, and some of refractory elements condense into dust grains.

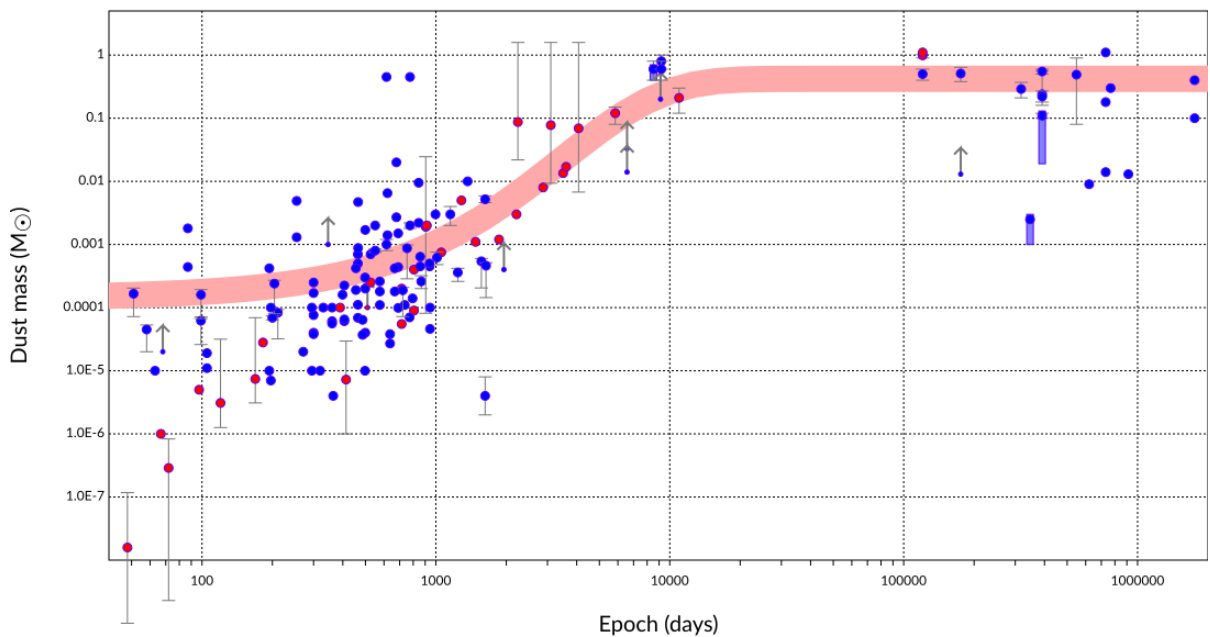

Figure 2: Dust masses measured in SNe, as a function of epoch since the SN explosion (Revised from Niculescu-Duvaz et al. 2022; https://nebulousresearch.org/dustmasses/)

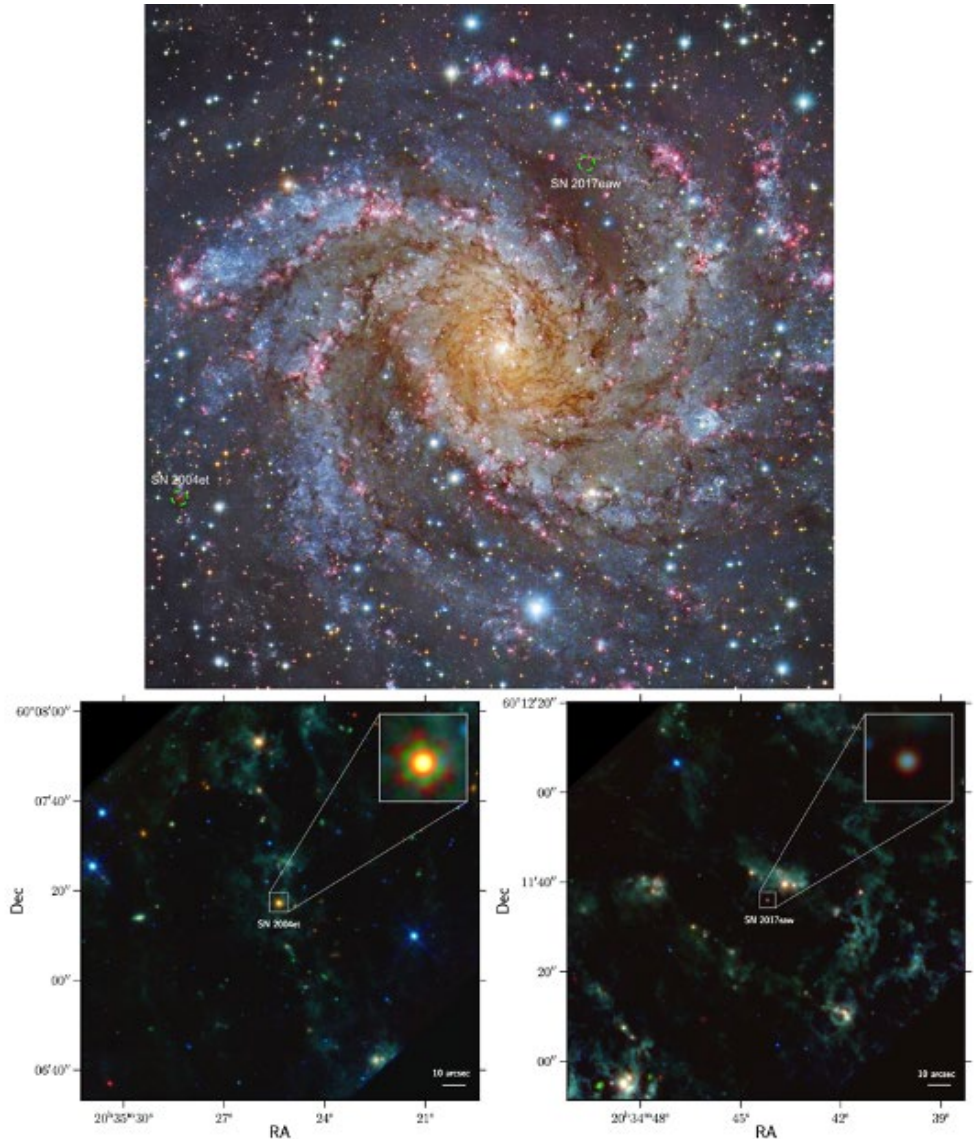

Figure 3: Optical images (top) of the galaxy NGC 6946 with two SNe, and JWST MIRI images (bottom) of SN 2004et and SN 2017eaw (Shahbandeh et al. 2023)

## [1] Dust in Supernovae

### [1.1] Freshly formed dust: what is the true figure of SN dust masses?

It has been proposed that SNe formed the first dust in the Universe, and they are the key source of dust from high-redshift to the Milky Way (e.g., Dwek 1998; Nozawa et al. 2003). Massive stars (>10 Msun) evolve quickly, and die as SNe with a large quantity of heavy elements being ejected from a single star (Fig. 1). If some of these heavy elements can condense into dust grains, SNe can fill the ISM of a galaxy with dust quickly (Laporte et al. 2017; Nanni et al. 2020). Theory predicts that if a SN can produce on average 0.1-1.0 Msun of dust per SN, the dust in high redshift galaxies can be accounted for by SN dust (Dwek & Cherchneff 2011).

The issue of SN dust is the discrepancy of the inferred dust masses in young (<2 years) SNe and older (>10 years) SN remnants. Fig 2 demonstrated the reported dust masses. The reported dust masses of young SNe tend to be in an order of $10^{-6}$ to $10^{-3}$ Msun, while higher dust masses (0.01-0.5 solar masses; e.g., Sugerman et al. 2006) have been reported in older SNe and SN remnants (Gomez et al. 2012, Matsuura et al. 2011, De Looze et al. 2017, Cigan et al. 2019, Chawner et al. 2019). The reasons for this discrepancy are undetermined: either dust masses slowly increase in time over 10 years (Wesson et al. 2015), or dust is optically thick in early days (Dwek et al. 2019), so that inferred dust masses are underestimated. The largest problem is lack of dust measurements in later phases. Young SNe have hotter dust, hence, their SED peaks at near- and mid-infrared, which previously covered by Spitzer, Gemini, and now JWST. While SNe are getting older, gas and dust in SNe gradually cool so that the peak of the dust SEDs shift towards far-infrared wavelengths. Historically, there were only a few far-infrared instruments which are capable of detecting colder and older SN dust, with only a handful of measurements with Herschel. JWST started detecting dust in ~18 years old SNe (Shahbandeh et al. 2023), but is still limited to the warm dust (>100K). Hence, in order to evaluate the true figure of dust mass formed in SNe, far-infrared observations are essential, and the high-sensitivity of PRIMA will provide a unique opportunity.

### [1.2] What types of dust grains are formed?

Although it is getting clearer that SNe do form dust, what types of dust grains being formed remain as an unresolved question. Currently (before JWST spectral observations), only limited examples are known: Cas A and G54.1+0.3 (Rho et al. 2008; Temim et al. 2017; Rho et al. 2018). As found in Fig. 4, dust found in SNRs are far from those found in cosmic silicate dust grains, including MgSiO3, FeO, SiO2, FeS, Al2O3. These dust grains typically have features beyond 25 μm, and PRIMA will be able to identify the characteristics of SN dust.

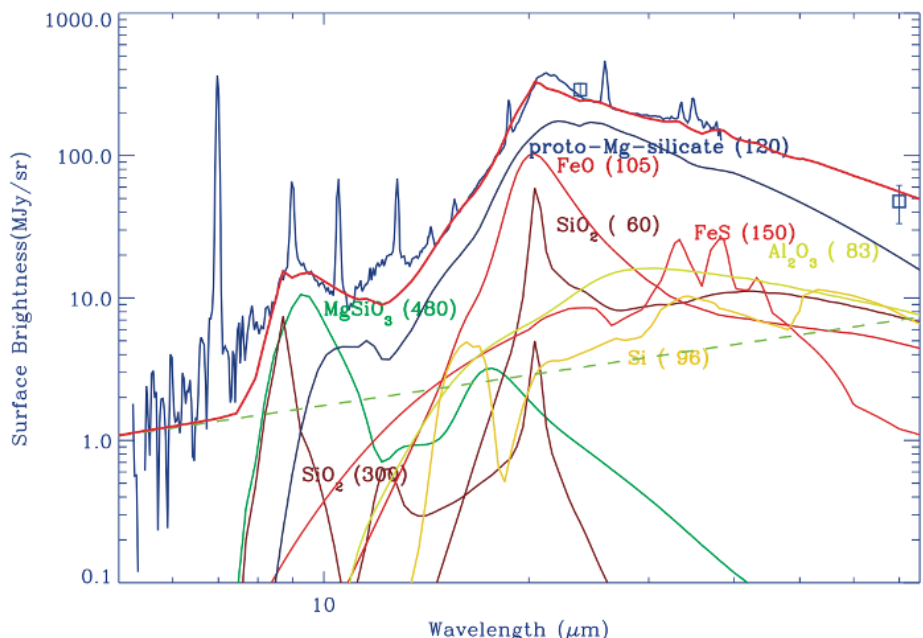

Figure 4: Dust identified in Cas A (Rho et al. 2008).

PRIMA, which has capabilities to detect numerous SNe, will open up the opportunities to investigate what types of dust are formed in which types of SNe. Typically, elemental yields of SNe have the core mass dependence, which would result in different dust compositions (Sarangi et al. 2015), and this will be a new research field to be opened by PRIMA.

The formation of molecules, molecular clusters, and their transition to solid grains of dust, control the chemical budget of SNe from their nebular phase to the remnants. Based on the type of progenitor, theories predict the dominance of C-rich (expected in smaller progenitor stars) or O-rich dust components (expected in larger progenitors). Moreover, the timescales of dust formation for individual species differ based on the densities of the clumps within the ejecta, which had been developed by Rayleigh-Taylor instabilities immediately after SN explosion. The cooling rates of the clumps are defined by elements present in those clumps, affecting dust compositions formed in clumps.

### [1.3] SNe illuminating surrounding Circums and interstellar dust

The interaction of the SN shock with circumstellar environments, which has been formed by the mass loss from the progenitor star, influences the physical and chemical evolution of the SN profusely. In interaction-dominated SNe, typically donated as type IIn SNe, dust is assumed to form in the post-shock gas (behind the SN blast wave) with surrounding circumstellar material.

### [1.4] Survival of Supernova and ISM dust: How much of dust is being destroyed?

Florian: While supernovae can be significant producers of dust, a large fraction of the dust can potentially be destroyed by the reverse shock. Moreover, the forward shock can trigger the destruction of interstellar dust grains. The net dust survival rate is crucial for determining whether or not SNe can contribute significantly to the dust budget in the ISM.

Theoretical studies about the destruction rate of SN formed grains in the SNR (e.g., Silvia et al. 2010, Bocchio et al. 2016, Slavin et al. 2020) or destruction of pre-existing ISM grains by the forward shock (e.g., Bocchio et al. 2014, Slavin et al. 2015, Hu et al. 2019) vary significantly. The dust survival rates of these studies span a broad range between 0 and 99 % which is due to different initial conditions in the SNR, considering different dust processes, and different initial grain size distributions and materials.

Highly resolved MHD simulations have improved our understanding of shocked clumps in the SNR and shocked regions in the ISM. We know that the gas densities, gas-to-dust mass ratios, magnetic field strengths and grain sizes are the main quantities that impact the dust survival rates. Taking all these aspects into account and considering the combined effects of gas sputtering and grain-grain collisions, we are able to better constrain the surviving dust masses (e.g., Kirchschlager et al. 2019, 2022, 2023). However, due to a lack of observational data, key parameters like the grain size or clump gas densities are uncertain, which makes the dust survival rates unclear. PRIMA's wavelength coverage, resolution and sensitivity can help to derive these key parameters and thus shed light on the dust survival rates.

## [2] From SNe to SN remnants: cooling of gas

The explosion energy of $10^{51}$ erg is gradually lost, and SNe can cool down to become SN remnants. In very early phases, the majority of the SNe energy is emitted at UV and optical, but once it reaches the nebular phase after a few years, far-infrared atomic lines are predicted to be the main cooling in SNe and SNRs. The dominant cooling lines are predicted to be [Fe II] 25.99 mm, [Si I] at 68.47 mm and [O I] 63.19 mm (Jerkstrand et al. 2015). However, the detections of far-infrared lines are very limited, and found in SNR Kes 75 with its pulsar age of 450±50 years (Temin et al. 2019) and Cas A (Rho et al. 2008). Hence, detecting these far-infrared dominant cooling lines is crucial to test SNe/SNR physics, and PRIMA is the only telescope that can tackle this physics.

About a few months after the SN explosion, part of the ejecta gas is getting cooler, and molecules are starting to form. Molecules have been proposed to be responsible for the rapid cooling of the gas (Jerkstrand et al. 2015). On top, molecules are indispensable chemical constituents to form dust grains in chemical reactions (Sarangi et al. 2015), Detection of molecules in extragalactic SNe, a few months to a few years after the explosion, will provide the critical evidence for determining the pathways of ejecta gas cooling process, and dust formation.

In order to truly describe the evolution of SNe to SN remnants, and evaluate their impact on surrounding ISM, detecting these cooling lines is important. With PRIMA's sensitivity, these lines could be detected in SNRs in a few Mpc.

### Instruments Required:

- PRIMA FIRESS: PointedHigh-res R~4400@ 112 µm; Pointed Low-res R~130; Large Map Low-res R~130

- PRIMAger: Small Map (< 20′×20′), Large Map (>20′×20′), Hyperspectral band (25–80 µm; 12 filters; R=10)

### Program Size:

- Medium (10-100 hrs)

### Approximate Integration Time:

Brightness estimates for dust-forming supernovae after 1 yr and 10 yrs - SN1987A had a flux at 10 µm of about 4Jy after 415 days and 12Jy after 615 days (Wooden et al), while the estimated dust mass evolution by Wesson et al. (2015). The SED would peak at around 100µm, with a flux of 0.2–0.45 Jy after 10 years. SN2004et, meanwhile, at a distance of about 7.75Mpc, has an SED peaking at about 15 µm, with a flux of 0.75 mJy, at age 18 yrs (Shahbandeh et al 2023).

From the sensitivity information on the PRIMA website, a 10-hour integration would detect a point source with a flux of 200 µJy at 100 µm or 300 µm at ~35 µm. So SN1987A would be detectable at distances of up to 2Mpc after 10 years. A supernova like SN2004et where CSM interaction is providing additional heating would be detectable at age ~20 yr out to about 15Mpc in the shorter wavelength filters.

### Special Capabilities Needed:

- Monitoring over up to the whole mission lifetime might be required, if appropriate targets are found. Young SNe but distant SNe monitoring could last a few years with a few months interval, while older (>5 years old) SNe monitoring would require the whole mission lifetime with a year to two year interval.

- Reasonably quick response time (one month), after ToO trigger.

## Synergies with Other Facilities:

- JWST and ALMA

## Description of Observations:

There are mainly two different observing strategies.

One is to detect newly discovered SNe. First, PRIMA will observe the SNe detection sites after discovery, and keep monitoring with imaging. Once detected and bright enough in far-infrared, SNe will be observed with spectra. These strategies will be used to study dust formation and line cooling.

The second strategy is to observe historically known SNRs. These PRIMA observations include Cas A, Crab, SN 1987A, and monitor dust formation and destruction science Herschel time. The observations require imaging and spectroscopy with a large or a small mapping.

# 7. Rapid far-IR spectral timing of X-ray binaries

Alexandra Tetarenko (Texas Tech University/University of Lethbridge), Poshak Gandhi (University of Southampton), Devraj Pawar (R. J. College, Mumbai-86)

One of the key open questions in relativistic astrophysics is understanding how compact objects act as powerful cosmic engines, gravitationally capturing material and expelling matter in the form of relativistic jets. Stellar-mass compact objects in Galactic X-ray binaries are ideal test-beds for such accretion-driven jet phenomena, as they enter into bright outbursts on timescales of days to months, providing a real-time view of how these jets evolve and interact with their environment. As X-ray binaries are known to produce highly variable emission (on timescales from sub-seconds-hours), time-domain analyses (e.g., Fourier domain techniques, cross-correlations) can be powerful tools to probe the complex physical processes in these systems. In particular, through detecting and characterizing rapid flux variability and time-lags between multi-wavelength signals (expected to be proportional to wavelength), we can measure properties that are difficult, if not impossible, to measure by other means (e.g., size scales, accretion flow geometry, jet speeds/acceleration, energetics, the sequence of events leading to jet launching).

Since X-ray binaries emit across the electromagnetic spectrum, with the in-flowing matter dominating shorter wavelengths and the out-flowing jet dominating longer wavelengths, we need a suite of fast timing capable facilities to take full advantage of time-domain techniques. X-ray satellites were the first to pioneer such studies of X-ray binaries, but the recent invention of new instrumentation and observing techniques have now allowed us to expand these studies into the optical, infrared, and even the longest wavelength radio/sub-mm bands. PRIMA uniquely probes a wavelength range that we have been so far unable to sample (connecting sub-mm [probed by ALMA] and mid-infrared [probed by JWST]) and thus represents an exciting new possibility for characterizing X-ray binary variability as the jet/accretion flow evolves throughout the outburst.

With PRIMA observations of X-ray binaries, our main science goals are to determine how matter propagates from inflow to outflow in these systems, through tracking how accretion flow variations propagate into the jet in different accretion states, and between systems housing different central compact objects (both black holes and neutron stars). We will accomplish these science goals by opening up rapid time-domain science with PRIMA, and conducting detailed time-series analyses, from which we can make accurate measurements of key physical properties in X-ray binaries. To perform rapid time-domain science with PRIMA, we require two main operational capabilities, a rapid sub-second sampling mode and a triggered target of opportunity mode.

The key open questions these PRIMA observations of X-ray binaries will help answer are: (a) How are jets launched and accelerated in black hole systems? (b) What are the initial conditions for the formation of jets in the launching and acceleration region? (c) What role do mass, spin, nature of the accretor, accretion rate, and outburst duty cycle, play in jet production? (d) How do jet properties (energetics, jet speed, jet size scales) compare across systems housing different compact objects? (e) What factors drive jet evolution? (f) How does the energy released by jets compare to other feedback processes, such as star formation?

## General Scientific Area:

Time Domain

## Science Justification:

X-ray Binaries (XRBs) are Galactic binary systems with a stellar-mass compact object (black hole or neutron star) accreting matter from a companion star, where a portion of the accreted material can be ejected in the form of a relativistic jet. The majority of XRBs are transient, i.e., they evolve from periods of minimal activity into a bright outbursting state on timescales of days to months [1]. During an outburst, the XRB source will cycle between accretion states, in which the properties of the accretion flow (geometry, mass accretion rate) and the jet (brightness, morphology, spectral, and temporal properties) emission can vary dramatically.

In XRBs, jet emission dominates at lower electromagnetic frequencies (radio, sub-mm, infrared; [2, 3, 4]), while emission from the accretion flow dominates at higher electromagnetic frequencies (UV, X-ray; [5, 6, 7]; Fig. 1 left). The intermediate optical regime can contain contributions from both inflow and outflow, including jet synchrotron emission and reprocessed thermal emission from the outer accretion disc. These different emission processes will operate on different timescales (jet emission will vary on faster timescales) and show different spectral shapes (reprocessed emission should have a Rayleigh-Jeans $F_\nu \propto \nu^{+2}$ spectrum). While jet emission is produced as a result of synchrotron emission [8, 9], and characterized by a flat-to-inverted optically thick spectrum ($\alpha \geq 0$, where $F_\nu \propto \nu^\alpha$), which transitions to an optically thin spectrum ($\alpha < 0$) at a spectral break. This spectral shape is not static, but rather can be quite dynamic through accretion changes, where the spectral break shifts to lower electromagnetic bands as the accretion rate increases [10, 11, 12].

*Extracting physics from spectral-timing signals:* Since both the jet and accretion flow in XRBs produce highly variable emission (in the spectral and temporal domains), spectral-timing analyses offer new avenues to probe detailed jet and accretion properties [13, 14]. Spectral-timing can probe <u>size scales not accessible with current imaging capabilities</u>. The smallest timescale over which the jet emission significantly varies (measured with Fourier power spectra) provides a direct measure of the jet size scale at different wavebands, enabling geometric measurements of the <u>jet cross-section/opening angle</u> (Fig. 1, right). As the jet propagates downstream from the compact object, optical depth effects cause longer wavelength emission to appear as a delayed version of shorter wavelength emission. Here the delay is directly related to jet speed and geometry. By modelling size-scale constraints and time-lag measurements between emission features at different wave- bands, constraints can be placed on the jet kinetic power and composition (e.g., [15]). Further, by cross-correlating multi-band signals, accreting matter can be tracked from inflow to outflow, directly linking changes in the accretion flow with the jet at different scales.

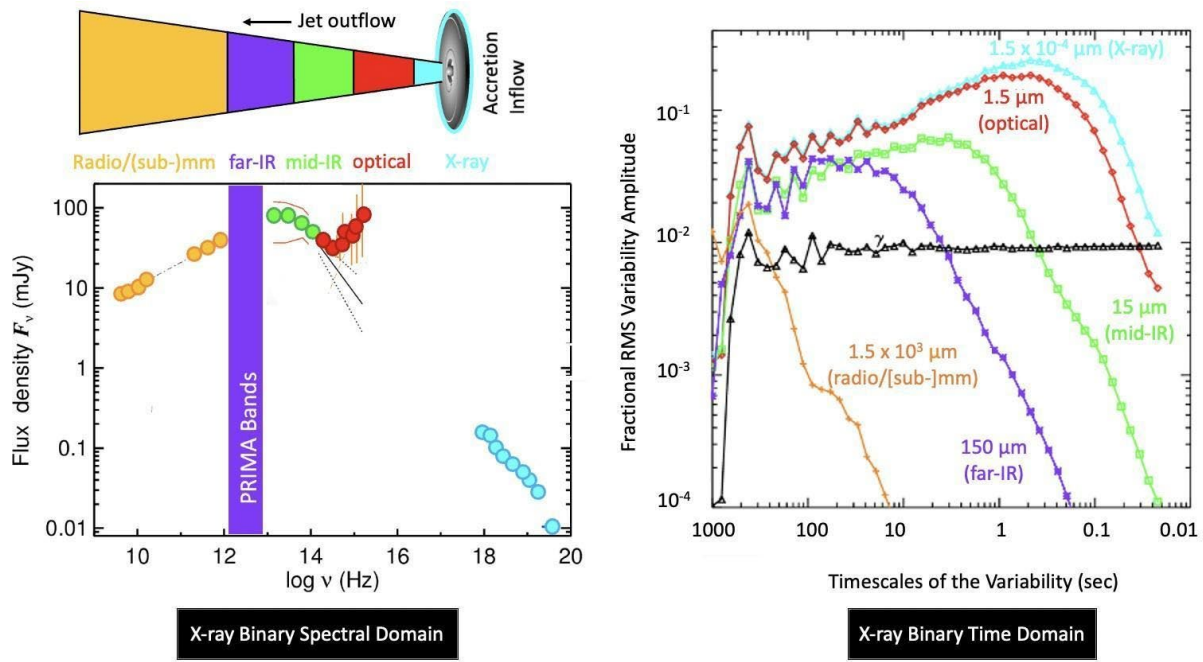

Figure 1: Illustration of the XRB spectral-timing phase space, based on real outbursts of the well-known jet-dominated system GX 339-4 [22, 23]. Left: A schematic of XRB emission regions (top) and a snapshot of a characteristic XRB broad-band spectrum (bottom). Right: Fourier power spectra, characterizing the amplitude of variability on different timescales, from signals observed at five electromagnetic bands (based on a theoretical model, where the black curve represents the matter fluctuations injected into the jet in the simulation). In XRBs, higher (lower) electromagnetic frequencies probe closer to (further from) the black hole and show shorter timescale/larger amplitude (longer timescale/smaller amplitude) variations. The PRIMA bands fill in a 2 order of magnitude gap in our broad-band electromagnetic frequency coverage and are expected to show strong variability on timescales longer than about 0.1 seconds (see purple curve where the power spectrum breaks between ∼1–10 Hz). Linking simultaneous, multi-wavelength observations of XRBs, covering a range of timescales (sub-seconds-hours), with X-ray observations probing the in-flowing matter, is needed to measure how the variability signal propagates through the different emission regions.

Broad-band variability measurements also allow for detailed tests of XRB models. One class of models predict that jet variability is driven by the injection of discrete shells of plasma at the base of the jet with variable speeds [16]. The behavior of these shells is directly linked to the amplitude of X-ray variability (traced by the X-ray power density spectrum; Fig. 1 right cyan curve). As the timescale of the jet variability depends on the shock speed and shell thickness, detecting correlated variability over a wide frequency range could disentangle these parameters. Alternative models suggest variability could originate from cyclo-synchrotron emission produced in a precessing magnetized inner accretion flow [17] or reprocessing of X-ray emission by the outer accretion disc [18]. These models predict very different variability properties (e.g., anti-correlations between optical-IR/X-ray, different spectral components present in the optical-IR bands) when compared to the jet-based model, thus spectral-timing can be used to distinguish between the competing models.

*State of the field:* While X-ray satellites were the first to pioneer timing studies of XRBs, new instrumentation has enabled expansion of these studies into the optical, infrared, and radio (e.g., [13, 15, 19]). These studies detected variability on timescales as short as ∼0.1 seconds in the optical/infrared, and tens of seconds in radio. The variability signals have been found to be highly correlated with X-ray emission. Such discoveries have subsequently allowed a connection to be drawn between variability properties observed in light curves and real jet physics. For example, [15] have measured fundamental jet properties (e.g., jet power, speed, composition, opening angle) through detecting and modeling Fourier spectral properties and time-lags between signals at radio, optical/infrared, and X-ray wavebands. Further, optical/infrared studies [13, 19, 20, 21] have shown that Fourier domain techniques can also probe physical processes in the jet and accretion geometry in the innermost regions near the compact object. These results strongly suggest that variability in the accretion flow is subsequently driving variability in the jet and confirm the diagnostic potential of the time-domain for studying jet/accretion physics.

*PRIMA's role:* There exist a number of limitations to current XRB spectral-timing studies: (1) the far-IR regime has never been sampled; (2) since the majority of studies are one-shot observations, they cannot probe how variability evolves as the spectrum/geometry of emission components change; (3) since studies focusing on lower electromagnetic frequencies have remained largely separate from those focusing on higher electromagnetic frequencies, connecting variability properties across different scales in the jet/inflow has not been possible; and (4) the different XRB emission models [16, 17, 18, 24] have never been thoroughly compared and tested within the time-domain parameter space. PRIMA's Imager instrument can overcome these limitations, providing sampling down to hundreds of milliseconds in twelve simultaneous filters across the far-IR. Further, this instrument should have sufficient sensitivity to observe tens of mJy-level objects, typical of infrared emission levels in outburst accretion states, and the ability for rapid response and flexible scheduling options which are optimal for targeting XRB sources. Further, the far-IR represents the interface between jet-dominated and accretion flow-dominated emission bands, making it the ideal band to disentangle the different variability models with PRIMA time-series measurements.

## Instruments Required:

PRIMAger: Small Map (< 20'×20'); Hyperspectral Band (25–80 μm; 12 filters; R=10)

## Program Size:

- Medium (10–100 hr)

## Approximate Integration Time:

Our required integration time is not primarily dependent on sensitivity, but rather we need to observe long enough to catch several cycles of variability (typically we observe on the order of hours at other wavelengths). Further, the total integration time will define our target brightness threshold. To ensure we can achieve the sensitivity necessary for a timing analysis with PRIMA data we follow [25], where detecting variability features in Fourier space at a certain significance is dependent upon the overall sensitivity achieved in the full observation time. In particular, to achieve a $S_{var}$=10$\sigma$ significance in Fourier space, we require a $S_{total}$=290$\sigma$ detection in the full observation ($S_{var}$=½ $S^2_{total}$ $F^2_{rms}(\lambda\ t)^{-\frac{1}{2}}$; assuming representative XRB values of variability amplitude matching $F_{rms}$=30%, width of variability in Fourier space of $\lambda$ =4 Hz, and exposure time of t=10 hours matching the PRIMA fact sheet). Given a PRIMA point source sensitivity (at 5$\sigma$) of ∼200μJy, this requirement sets the trigger threshold for our target sources to be >12 mJy.

## Special Capabilities Needed:

XRBs show drastic changes in the emission across all wavelengths when they are in an outburst as a result of an unstable accretion rate onto the compact object. During these outbursts (which are usually not predictable) we observe phenomena like periodic and aperiodic variability, bursts on time scales of seconds, correlated spectral variability which is related to accretion dynamics in relativistic gravity, and radiative reprocessing. The following special capabilities are valuable for far-IR spectral-timing observations of XRBs:

*Target of Opportunity and rapid response:* Since XRB outbursts are mostly unpredictable, the instrument time allocation policy should consider anticipated target of opportunity proposals (AToO) and target of opportunity (ToO) proposals, with response times on the order of 24–48 hr.

*High Cadence Monitoring:* Once in outburst the source remains above quiescent flux levels of most instruments for weeks to months, during which multiple pointings at a variable cadence are needed (ranging from the fastest monitoring timescales on the order of a day to slower weekly monitoring timescales; see the observational strategy below for more details).

*Windowed Timing Mode:* Due to the compact nature of XRBs, they show rapid variability (up to kHz in neutron star systems), thus high time resolution photometry using instrument specific bands is the most suitable observation mode. Instruments with imaging CCDs designed for relatively longer integration times achieve this by using a subset of the total pixels on the CCD to enable rapid readout times. For example, such a strategy is achieved through the windowed timing mode (WT) on NASA's Swift/XRT instrument (now renamed as the Neil Gehrels Swift Observatory). A WT mode with variable time resolution will be very helpful for pushing the limits of far-IR timing detections to sampling rates of ∼100 Hz or higher.

*Timing Accuracy:* To complete our science goals, we need to be able to accurately compare timing signals across the PRIMA filters and to multi-wavelength signals obtained from other facilities. Thus, accurate relative and absolute timing accuracy is useful. Timing calibration of the PRIMA instrument can be performed on a system with known timing properties (e.g., this procedure is currently being used for JWST).

## Synergies with Other Facilities:

To maximize science output, we will coordinate simultaneous multi-wavelength coverage with PRIMA. We aim to synchronize PRIMA observations with both ground and space-based facilities: radio (VLA, VLBA, ATCA, LBA), [sub-]mm (JCMT, ALMA, SMA, NOEMA), mid-IR (JWST, VLT HAWK-I), optical (Gemini `Alopeke/Zorro), X-ray (XMM-Newton, NICER, Swift).

## Description of Observations:

We aim to use PRIMA to observe multiple XRB systems in the far-IR bands at high time resolution. These observations will be performed across different timescales and accretion regimes, and on a mixture of black hole and neutron stars, to give us a coherent census of far-IR variability properties in XRBs. To observe XRBs with PRIMA, we will use the PRIMAger hyperspectral imaging and polarimetry instrument. This instrument provides exquisite spectral coverage, as well as the ability to sample on rapid timescales. Our targets are point-like and we understand that point-source mapping with sampling rates of about 10 Hz is feasible. We can compromise on filters vs. speed, if technical constraints demand this, and use of the Pointed Low-res FIRESS is also possible, though we do not discuss this option here until more technical details are known. Sub-array modes to enable higher frame rates could open up new science windows for the brightest targets.

With these capabilities, we can track rapid spectral changes across the far-IR band, as well as correlate time-domain signals between the filters (and with other multi-wavelength signals).

We will (ideally) aim to use the hyperspectral band, observing our target XRB in 12 different filters from 25–80 μm, with monitoring epochs of several hours on-source at a time resolution of 100 milliseconds. We chose the hyperspectral band to maximize the wavelength range that we sample. With the 100 milliseconds time resolution we will characterize variability at Fourier frequencies that exceed the expected break in the Fourier spectra by one order of magnitude (see Fig. 1 right purple curve), while staying well within the instrument's capabilities. We choose several hours of on-source time per epoch to ensure we sample multiple variability cycles and to match other facilities, to compare far-IR variability with variability at other electromagnetic frequencies. We do not expect to spatially resolve the system in these observations.

We request use of the small map mode, where we need to place our target source as well as at least one comparison star (for differential photometry) in our field of view. We follow this procedure to verify that the variability we detect is intrinsic, and not due to instrumental effects, with the additional point source object within our field of view acting as a check source. We do not require absolute flux calibration for a timing analysis, as we are only concerned with relative flux differences.

## Narrative of observational strategy:

*Target Selection:* To select our sample of target sources, we will aim to balance observational constraints while sampling a wide range of the XRB parameter space (i.e., compact object mass and spin, accretion regime). This ensures we can both successfully detect and analyze far-IR variability signals in our sources, as well as determine which XRB properties are the fundamental drivers behind far-IR variability. As such, we adhere to the following selection criteria: (i) target sources must have known jet emission (through a clear radio detection), (ii) target sources must show levels of infrared emission detectable on short timescales with PRIMA (i.e., ≥ 12 mJy, XRBs typically reach these fluxes during outburst; see Fig. 1 left), (iii) target sources must not be in complicated/crowded fields, and (iv) the target source pool must be a mix of black holes and neutron stars.

*Observing Campaigns:* As XRBs evolve over different timescales, showing a mix of accretion state change time-scales (transient XRBs spend on average 28 days in the hard state and 9 days transitioning to the soft state [1]), we define two types of observing campaigns to properly sample the far-IR variability in each target; Short and Long campaigns. For all of the target sources, X-ray monitoring observations will be obtained (via X-ray satellites such as Swift, MAXI, NICER), to track accretion state changes in the targets.

Short campaigns will target those sources showing evolutionary timescales on the order of hours-days. These campaigns will consist of up to a week of observations with a daily cadence.

Long campaigns will target those sources showing evolutionary timescales on the order of weeks-months. These campaigns will consist of 10–20 observations with a cadence depending on the target, taken across multiple months. Here the target specific cadence is defined by the target source evolution time-scales (as determined by radio and X-ray monitoring). This strategy ensures we are able to sample different accretion states and the transition between states during outburst.

*Scheduling/Triggering:* As our target sources are mainly transients, we require triggered ToO observations. Based on our estimated target brightness threshold (defined above) we expect ∼1–2 of these transient sources meeting our target selection criterion to be observable per 6-month period (∼63–86% chance of triggering).

# 8. The infrared view of energetic cosmological transients: kilonovae, gamma-ray bursts, and fast radio bursts

Darach Watson (University of Copenhagen)

To date, the 20–200 μm region is almost entirely unexplored when it comes to energetic cosmological transients, i.e., gamma ray bursts, merging neutron stars (kilonovae), and fast radio bursts. There are two main ways to approach the problem. One is to observe the host galaxies, and the other is to observe the afterglows directly. Neither approach has been explored significantly at the PRIMA wavelengths. The host galaxy observations are similar for all transients, i.e., determine the properties of the hosts and the positions of the transients within the hosts to estimate the population and the environment of the transient as clues to its origin, age, and mechanism. For gamma-ray bursts, observing the afterglow would tie the near-infrared to the radio, to date a blank region in the afterglow spectral energy distribution, would permit the absolute extinctions and hence extinction curves of heavily dust-reddened bursts to be measured unambiguously, providing the best extragalactic dust extinction data, and could enable the detection of ultra-dark burst afterglows. For neutron star mergers, their ejecta, i.e., the kilonovae, are expected to be dominated by high-opacity neutron capture elements. For extremely nearby kilonovae, PRIMA imaging and spectroscopy would be unique in providing information on the mid-infrared emission from these heavy elements, offering the opportunity to identify elements and potentially measure the ejecta mass, since the emission at PRIMA wavelengths is likely to be optically thin at all epochs. PRIMA would also allow issues such as heavy metal dust formation and dust echoes of kilonovae and neutron star merger jets to be addressed. Finally, the emission from fast radio bursts is unknown at such short wavelengths, and sensitive PRIMA imaging could hope to link the radio emission to shorter wavelengths, at least for repeating FRBs.

## General Scientific Area:

Time Domain

## Science Justification:

To date very little has been done in the area of mid to far infrared observations of energetic cosmological, transients, such as gamma-ray bursts (GRBs), neutron star (NS) mergers and kilonovae (KNe), and fast radio bursts (FRBs). There have been several Spitzer programmes on GRB host galaxies and afterglows and one Spitzer observation of the well-observed kilonova AT2017gfo associated with GW170817, leading to important and interesting results, and some observations with the Herschel Space Observatory targeting GRB host galaxies. However, none of these have been very extensive, leading only to a few novel results in a handful of cases. The most intriguing of these was the detection by Spitzer at 4.5 μm of the KN AT2017gfo, but its non-detection at 3.6 μm, indicating potentially strong line emission in the nebular phase of the KN (Villar et al. 2018, Hotokezaka et al. 2022). While both Spitzer and Herschel were modestly successful observing the host galaxies of GRBs, they were limited (due to their sensitivity) to very luminous galaxies at z <∼ 1.5 in most cases because the mean redshift of GRBs is z ∼ 2.6. In terms of afterglow observations, both were limited by sensitivity and by the satellite response time. We expect that JWST will provide significant new insights in the near and mid-infrared regime because of its extreme sensitivity and a response time which can be as good as 48 to 72 hours. So far JWST has indeed observed several GRB afterglows, and there are GO programmes to observe GRB hosts at high spatial resolution, providing important information on the properties of the host galaxies, and the locations of the GRBs within their hosts.

## Science Questions:

Probably the most important questions related to these transients is the nature of their progenitor systems in each case, and how they explode/occur.

For GRBs, both long and short, the basic nature of their progenitor systems is believed to be well-understood: collapsing stripped envelope massive stars (Hjorth et al. 2003, Stanek et al. 2003), and merging neutron stars (Mooley et al. 2018), respectively. However, there are many outstanding questions.

*Long GRBs:* For long GRBs, questions such as the exact nature of the star or binary star that collapsed, how the jet was produced and interacted with the ISM, and the impact of the metallicity and dust environment on the progenitor, as well as the nature of the still somewhat mysterious ultra-dark GRBs. The questions of the nature of the GRB jet, the nature of ultra-dark GRBs, and measurement of a sample of absolute dust extinction curves at high redshift (e.g., Zafar et al. 2011) can be addressed by PRIMA with rapid response observations, since the expected afterglow flux will typically be in the range of a few mJy to a few tenths of a mJy in the first hours to a month after the GRB from 20–200 μm (e.g., Laskar et al. 2023), something that has never been achieved before. JWST will be able to do some of this science at shorter wavelengths, and so may resolve some of the questions of the ultra-dark GRBs. However, there will likely be very few follow-ups with JWST, limiting the sample size, JWST will not cover the range beyond 28 μm, and finally, the typical JWST response time could be ten times longer than the PRIMA response time (72 hours compared to 6–7 hours), leading to a factor of about 30 higher afterglow flux during the early PRIMA observations.

The host galaxies of long GRBs are now fairly well studied, with samples taken using ALMA, Spitzer, Herschel, HST, and ground-based optical/NIR telescopes. The Herschel PACS observations cover the most similar wavelength space to PRIMA and so far fewer than ten galaxies were detected (Hunt et al. 2014, Schady et al. 2014), using heavily biased samples of GRBs, at modest redshift (z ∼ 1). This means that in spite of the extensive datasets at other wavelengths, long GRB host galaxies are poorly sampled in the 20–200 μm region, with virtually no spectroscopy between the NIR and ALMA – one GRB host galaxy has Spitzer IRS spectroscopy at z = 0.1055 (Watson et al. 2011). PRIMA certainly has the power to change this, allowing the silicate dust and PAH features to be detected in hyperspectral imaging for more luminous galaxies at the median GRB redshift and covering the main FIR fine structure cooling lines at redshifts up to $z \sim 2$, something which has so far never been achieved.

*Short GRBs:* For short GRB afterglows some of the same questions apply, but are likely to be more challenging, owing to the fainter afterglows. On the other hand, for the lower redshift subset of short GRBs, this may well be possible, providing insights into the jet mechanism in merging neutron stars. For the associated kilonovae, the real question is sensitivity, and almost all short GRBs will be too distant for PRIMA to have a realistic prospect of detecting them – for kilonovae discovered with gravitational wave detectors in the 10–150 Mpc range, there may be some limited prospect (see below). The host galaxies of short GRBs offer opportunities for study in the same vein as long GRB hosts, except that they are generally lower redshift and therefore more accessible, though in general also less star-forming.

*Kilonovae:* Expectations for the MIR brightness of KNe has been calculated to be about a microJy at 100Mpc at 20–50 days post merger at 10 µm, which effectively rules out late detection of the KN with PRIMA for mergers further than about 10 Mpc, a very rare occurrence even over a decade of future GW observatories. However, the early brightness and the presence of line emission significantly longward of 10 µm is unknown, making exploratory observations a worthwhile endeavor, particularly for the few closest mergers. In principle, low-lying atomic states could produce nebular line emission that might be detectable in hyperspectral imaging or even in low-resolution spectroscopy. Furthermore, the effects of dust scattering of shorter wavelength afterglow light could produce a very IR-dominated emission. The possibility of metallic dust growth in KNe, akin to SN dust growth, is entirely unexplored, and might, in principle give rise to warm dust emission, though, again, such warm emission is unlikely to be detected by PRIMA beyond a few Mpc, a very low likelihood occurrence. Overall, the prospects of detecting KN emission with PRIMA seems remote unless there is a nearby (~10Mpc) merger in its lifetime, or a currently unrecognized spectacularly bright line in the PRIMA bands from an abundant element dominates the cooling. For these reasons, PRIMA responses to GW events should probably be limited to close bursts (~<50Mpc), but should nevertheless be performed because currently our knowledge of these events at these wavelengths is extremely limited. It should also be borne in mind, that we do not know all the properties of r-process element producers, for example, it has been proposed that super-kilonovae may be produced from massive collapsars, with r-process masses 1000 times larger than normal kilonovae, in which case, they could be visible out to >100 Mpc with PRIMA (Siegel et al. 2022).

*Fast radio bursts:* PRIMA, in addition to characterizing the mid and far-infrared properties of FRB host galaxies—an endeavor that is only beginning now even at optical wavelengths—could also contribute to understanding the prompt emission of these enigmatic objects by making a detection of an FRB repeater at a high frequency, placing strong constraints on the emission mechanism.

## Instruments Required:

- PRIMA FIRESS: Pointed Low-res R~130
- PRIMAger: Small Map (< 20'×20'); Large Map (> 20'×20'); Hyperspectral band (25–80 µm; 12 filters; R=10); Polarimeter band (96, 126, 172, 235 µm; R=4)

## Program Size:

- Small (<10 hr)
- Medium (10–100 hr)
- Large (100 + hr)

## Special Capabilities Needed:

Rapid response repointing based on external triggers, ideally as fast as possible, a few hours is definitely much better than a few days. This is not necessary for host galaxy observations.

## Synergies with Other Facilities:

A transient triggering and localization facility is naturally required: either a GRB satellite such as Swift currently, or perhaps THESEUS in the future. Similarly, a gravitational wave facility e.g., LIGO, or the Einstein Telescope or the Cosmic Explorer.

## Description of Observations:

For rapid response follow-up, in all cases, faster is better. A trigger from a satellite or ground-based observatory, probably human-mediated, would allow initial trigger delays as low as 10 minutes in some cases. If PRIMA has a well-designed communications and response programme, e.g., akin to XMM-Newton or Athena, it might be able to get down to response times of 4–6 hours, which would be much more useful than the several day response times of HST and JWST, though several day response times are still useful.

For host galaxy observations, these are better done using carefully curated sample surveys in pointed mode, whether in imaging or spectroscopy.

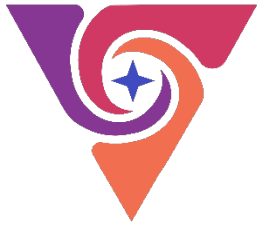

# Cosmology

# 9. Probing the history of cosmic star formation, black hole growth, and metallicity/dust evolution with line-intensity mapping

Yun-Ting Cheng (Jet Propulsion Laboratory, California Institute of Technology)

Line Intensity Mapping (LIM) is a promising technique for probing galaxy evolution and large-scale structure. By measuring spatial fluctuations in integrated spectral line emissions from all sources, LIM traces the 3D structure of the universe through the redshift-frequency relation. LIM is sensitive to information from faint populations below the detection limit of galaxy surveys, providing a crucial tool for studying both astrophysics and cosmology at high redshifts. The FIR spectral coverage by PRIMA provides a unique window for conducting LIM measurements on several emission lines that trace star formation activities, black hole growth, and the dust and metallicity content (e.g., ionized oxygen, silicon, PAH), enabling a comprehensive study of galaxy properties across cosmic time. We propose a 1000-hour observation with PRIMA FIRESS to map a 1000 arcmin$^2$ area for LIM. According to our model forecast, this survey can achieve high-sensitivity detection of the large-scale power spectrum of several lines up to z $\sim$ 5. In conclusion, PRIMA offers an unparalleled opportunity to conduct far-infrared LIM surveys, unveiling the galaxy evolution across cosmic history using information from all galaxies.

## General Scientific Area:

Growth of Galaxies and Black Holes Through Cosmic Time

## Science Justification:

Line intensity mapping (LIM) is an emerging technique that promises to study the galaxy evolution and the large-scale structure throughout cosmic history [1,2]. LIM measures the spatial fluctuations in the integrated emission of a specific spectral line originating from all emitting sources, including faint or diffuse galaxies that remain unresolved in individual detections. By mapping a large sky area with moderate spectral resolution, LIM offers a direct window into the three-dimensional structure of the universe, with the line-of-sight distance inferred from the redshift-frequency relation.

LIM provides a complementary approach to traditional galaxy surveys, as it is sensitive to all sources of emission, and thus accesses the faint populations that remain undetected in individual detections. This is especially crucial for studying the high-redshift universe, where detection-based methods primarily capture the brightest galaxies and leave substantial information from the fainter population missing, hindering the construction of a comprehensive picture of galaxy formation and evolution in the early universe. Additionally, LIM does not require high angular resolution, as it focuses on mapping integrated emissions rather than resolving individual sources, allowing for the coverage of vast sky areas in limited observing time.

LIM promises to enhance our understanding of key questions in both galaxy evolution and cosmology. Many emission lines are tied to star formation activities, enabling us to infer the star formation rate density across cosmic epochs by measuring the total spectral line emission within a given cosmic volume. Furthermore, different lines originate from distinct environments within the ISM or IGM of galaxies, facilitating a comprehensive analysis that encompasses the evolution of galaxy properties such as metallicity, dust content and composition, supernova feedback, black hole accretion, and more [2]. On larger scales, line emission from galaxies traces the underlying density field. The clustering of the line intensity field at high redshift thus provides a unique probe of the expansion history [3], primordial non-Gaussianity in the early universe [4], where traditional galaxy surveys face limitations. Lastly, LIM holds significant promise for shedding light on the Epoch of Reionization (EoR), which is believed to be dominated by intrinsically faint galaxies which are hard to detect individually even with flagship-class facilities. LIM datasets targeting multiple spectral lines can capture their emission in the neutral ISM tracers, star-forming tracers, and indicators of black-hole accretion [5,6].

Presently, ongoing and upcoming LIM experiments primarily target lines such as the HI 21 cm line in the radio [7], CII or CO lines in the mm/sub-mm range [8–15], or atomic/molecular lines in near-infrared wavelengths [16,17]. PRIMA offers a unique frequency coverage in the far-infrared spectrum (24–239 μm), encompassing crucial star-formation tracer lines like ionized neon, oxygen, silicon, $H_2$, and PAH features. Conducting LIM measurements with PRIMA will not only provide constraints on star formation activities across cosmic time but will also offer complementary information when combined with LIM measurements in other wavelengths, creating a comprehensive census of galaxy properties.

We propose a 1000-hour observation with PRIMA FIRESS to map a 1000 arcmin$^2$ area. Using a linear scaling relation between line luminosity and star formation rate [18,19], we can predict the total line intensity as a function of redshift based on the current star formation rate density constraints from observations [20]. The power spectrum of the large-scale clustering of the line intensity field is proportional to the square of the line intensity (modulo a bias factor). On small scales, the power spectrum of the lines will be dominated by the shot noise due to the discreteness of the sources. Fig. 1 illustrates an example of the power spectrum and its sensitivity for [NeIII] (15.55 μm) at z=2 and z=5, respectively, at large scales (k=0.1-1 Mpc/h), where the linear power spectrum dominates over the shot noise and non-linear clustering. We further calculate the total SNR by summing up the SNR in each k bin within k=0.1-1 Mpc/h, yielding a 20σ and 6σ sensitivity in the two cases, respectively.

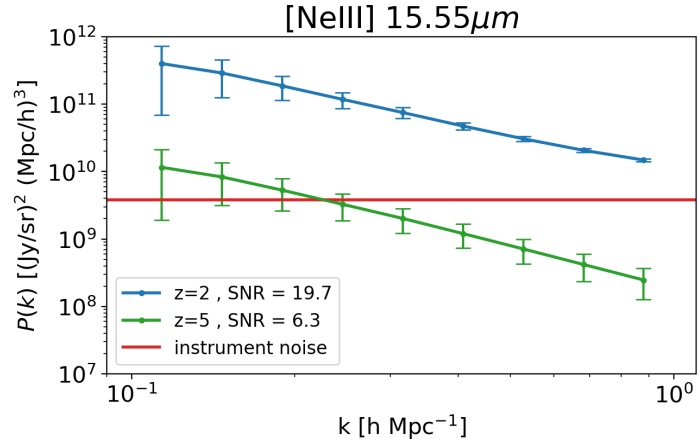

Figure 1: Power spectrum of the [NeIII] 15.55 um line and its uncertainties at z=2 (blue) and z=5 (green), assuming our proposed survey (1000 arcmin² survey with 1000 hours of total integration time). The red line denotes the noise power spectrum level in this survey. The power spectra are calculated with a redshift width Δz/(1+z) = 0.5. We focus on the larger (clustering) scales at k=0.1∼1 h/Mpc, where the power spectrum amplitude is proportional to the square of the line intensity (modulo a bias factor). The total SNR is obtained by summing over all modes within k=0.1∼1 h/Mpc.

In Fig. 2, we present the power spectrum SNR as a function of redshift for 10 lines detectable with PRIMA, and the numerical results are summarized in Table 1. At z=1, all 10 considered lines can be detected at >5σ level. This enables a multi-line census of galaxy properties. For instance, the ratio of [OIV] and [NeII] flux has been shown to trace black hole accretion rates [21]. With LIM measurements of these two lines, we can further constrain the global average of black hole accretion activity. At the peak of cosmic star formation around z=2, almost all considered lines can achieve high-sensitivity detection, facilitating the study of the ISM of galaxies during the cosmic noon. At z > 3, PRIMA can still detect LIM signals from a couple of bright lines, enhancing our understanding of star formation activities in this relatively under-explored territory by traditional galaxy surveys.

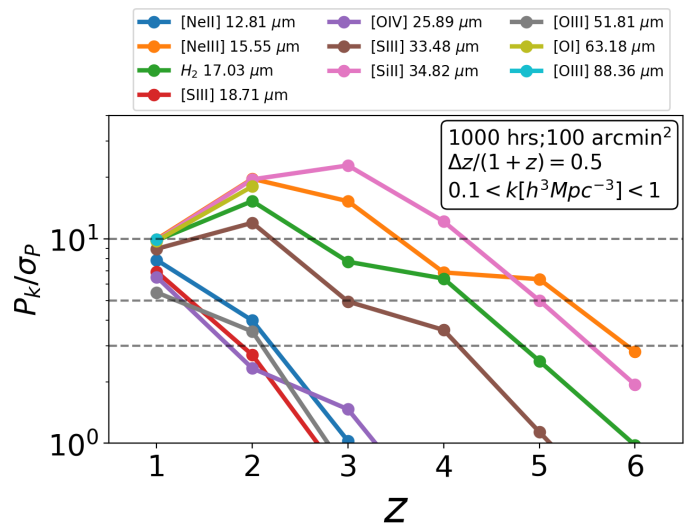

Figure 2: Power spectrum SNR (summing over all modes within k=0.1∼1 h/Mpc) as a function of redshift for 10 lines detectable with our proposed survey (1000 arcmin², 1000 hours total integration time). The power spectra are calculated with a redshift width Δz/(1+z) = 0.5. Gray dashed lines mark SNR levels of 3, 5, and 10, respectively.

Table 1: Power spectrum SNR shown in Fig 2. Cases with SNR > 5 are highlighted in red.

| line name/z | z=1 | z=2 | z=3 | z=4 | z=5 | z=6 |
|---|---|---|---|---|---|---|
| [NeII] 12.81 μm | 7.9 | 4.0 | 1.0 | 0.3 | 0.1 | 0.1 |
| [NeIII] 15.55 μm | 10.0 | 19.7 | 15.3 | 6.9 | 6.4 | 2.8 |
| H2 17.03 μm | 9.8 | 15.3 | 7.7 | 6.4 | 2.5 | 1.0 |
| [SIII] 18.71 μm | 6.9 | 2.7 | 0.6 | 0.5 | 0.1 | 0.1 |
| [OIV] 25.89 μm | 6.5 | 2.3 | 1.5 | 0.4 | 0.3 | 0.1 |
| [SIII] 33.48 μm | 8.9 | 12.0 | 4.9 | 3.6 | 1.1 | 0.4 |
| [SiII] 34.82 μm | 9.8 | 19.5 | 22.9 | 12.2 | 5.0 | 1.9 |
| [OIII] 51.81 μm | 5.5 | 3.5 | 0.7 | — | — | — |
| [OI] 63.18 μm | 19.8 | 18.1 | — | — | — | — |
| [OIII] 88.36 μm | 9.9 | — | — | — | — | — |

In addition to probing the history of galaxy formation and evolution, PRIMA's capability to study density fluctuations on large scales from redshift 1 to 6 through LIM provides crucial cosmological constraints. This is especially important at higher redshifts (z>3), where spectroscopic galaxy surveys are challenging to conduct. LIM offers a unique opportunity to investigate the effects of dark matter and dark energy on the universe's evolution during this cosmic era. This will contribute to addressing one of the main questions in cosmology: What are the properties of dark matter and the dark sector? For instance, by measuring the evolution of the Baryon Acoustic Oscillation scale as a function of redshift, we can probe the expansion rate of the universe, which is a key test for dark energy in the standard cosmological model.

In summary, PRIMA offers a unique opportunity to conduct LIM surveys in the far-infrared range. With our proposed observations, we can detect the large-scale clustering power from multiple spectral lines with high sensitivity. This enables us to unravel the story of star formation activities across cosmic time from all galaxies in the universe, as well as providing constraints on the cosmological model by measuring the clustering of the large-scale structure throughout cosmic history.

### Instruments Required:

PRIMA FIRESS, Large Map Low-res R~130

### Program Size:

- Large (100 + hr)

### Approximate Integration Time:

- 1000 hr

### Special Capabilities Needed:

- None

## Synergies with Other Facilities:

- Roman, Euclid, LSST, SPHEREx.

In addition to the auto-correlations of line emission, cross-correlating the LIM dataset with an external galaxy catalog provides additional information regarding the relationship between the line emission and the external tracers. Furthermore, cross-correlation measurements are more robust against systematics and foreground contaminations in the LIM data. In this regard, galaxy catalogs from upcoming large-scale surveys such as Roman, Euclid, LSST, and SPHEREx are well-suited for synergy with the PRIMA LIM survey.

## Description of Observations:

Uniformly scan a 1000 arcmin$^2$ field with a total integration time of 1000 hours. The observing field will be selected to overlap with future deep galaxy surveys like Roman, enabling cross-correlation analysis.

## Acknowledgement:

A portion of this research was carried out at the Jet Propulsion Laboratory, California Institute of Technology, under a contract with the National Aeronautics and Space Administration (80NM0018D0004).

## 10. Search for the radiative decay of the cosmic neutrino background by spectral measurements of the cosmic infrared background

Yuji Takeuchi (University of Tsukuba, Japan), Shuji Matsuura (Kwansei Gakuin University, Japan), Shinhong Kim (University of Tsukuba, Japan), Takashi Iida (University of Tsukuba, Japan)

We propose to search for a faint yet distinguishable contribution to the spectrum of the cosmic infrared background due to the radiative decay of the cosmic neutrino background. In the standard model of particle physics, neutrino decay is significantly suppressed, and its lifetime is expected to be on the order of $10^{43}$ years. However, some non-standard models suggest the possibility of a much shorter lifetime, ranging from $10^{14}$ to $10^{17}$ years, while observations only provide a lower limit on the neutrino lifetime, which is on the order of $10^{12}$ years. In PRIMA's low-resolution (R~130) mapping mode, a diffuse background analysis with the removal of point sources associated with known galaxies in a wide field (~1 square degree) spectroscopic survey in the 24-240 μm range could enable a search for neutrino decay lifetimes in the range of $10^{15}$ to $10^{16}$ years, achievable with a total observation time of approximately 100 hours in terms of surface brightness sensitivity. Search for neutrino decay at this sensitivity could place strong constraints on several non-standard theories. If observed, it would provide powerful evidence for the contribution of non-standard models to neutrino decay and could also enable a direct detection of the cosmic neutrino background. Additionally, information on the absolute mass of neutrinos, as well as cosmological parameters, could be obtained from the decay photon spectrum.

### General Scientific Area:

Cosmology

### Science Justification:

A few seconds after the birth of the universe, neutrinos became thermally decoupled from other particles, and the neutrinos present at that time have continued to exist almost uniformly in the present universe as the Cosmic Neutrino Background (CvB). Based on measurements of the Cosmic Microwave Background (CMB), the CvB is predicted to have a temperature of 1.95 K and a density of 110 particles per $cm^3$ per generation, and indirect evidence [1] have been provided through CMB measurements. However, direct observation has not been achieved thus far.

Neutrinos have three mass eigenstates, and it is known from oscillation experiments that the flavor eigenstates are a mixture of mass eigenstates. The mass-squared differences between the mass eigenstates have been measured, allowing the heavier neutrino to decay into a lighter neutrino with the emission of a photon ($v_3 \rightarrow v_{1,2} + \gamma$). However, in the Standard Model of particle physics, there is a very strong suppression mechanism for neutrino decay processes, and the lifetime of the heaviest neutrino ($v_3$) is expected to be on the order of $10^{43}$ years [2]. Conversely, the neutrino decay process provides an extremely stringent test of the Standard Model. Some non-standard models [3,4] suggest that this lifetime could be shortened to around $10^{14}$ to $10^{17}$ years. However, the observed $v_3$ lifetime is only limited to a lower bound of around $10^{12}$ years [5] based on observations of cosmic infrared background radiation.

Since neutrino radiative decay is a two-body decay, the energy of the photon observed in the rest frame of the parent particle is monochromatic. If we assume a mass of 50 meV for $v_3$, the energy of the photon would be approximately 25 meV (wavelength of 50 µm). An upper limit of 230 meV for the sum of the masses of neutrinos has been established through cosmic observations [6]. Using the mass-squared differences measured in neutrino oscillation experiments, it is expected that the mass of $v_3$ falls between 50–87 meV, and that the wavelength of the decay photons in the rest frame of $v_3$ is expected to be in the range of 50–87 µm. Given circumstances above, exploring the decay light from CvB in the 50µm wavelength region with a sensitivity exceeding the current lower limit of $10^{12}$ years is of great significance for both cosmology and particle physics.

The wavelength and luminosity of the decay light produced by neutrino decay at redshift z when observed around Earth are affected by the z dependence of the density and decay rate of CvB, as well as the redshift of the decay light. Additionally, they are influenced by the Doppler shift due to the thermal motion of CvB, the temperature of which also depends on z. Figure 1 shows the expected surface brightness distribution of CvB decay light assuming a $v_3$ lifetime of $10^{16}$ years and a mass of 50 meV. The characteristic of the distribution is that it has a sigmoidal-like shape with a sharp rising edge of about 1µm in width around the photon wavelength position in the $v_3$ rest frame. The width of this sharp rising edge is related to the current temperature of CvB, which is 1.95K. The distribution with such a shape is characteristic of the photon distribution resulting from the two-body decay of particles with a certain mass uniformly distributed in the universe, and thus, it can be considered as a decisive and clear evidence of CvB decay. Hence, detecting the sharp rising edge in the CvB radiative decay through spectral measurement is crucial.

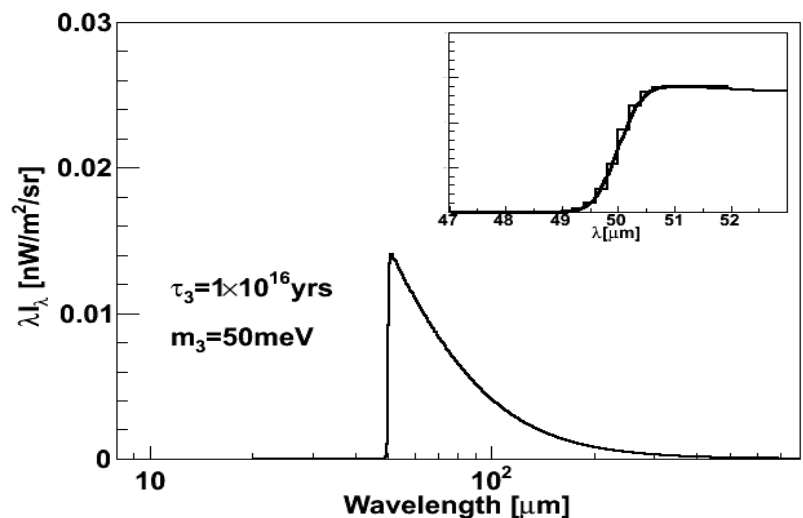

Figure 1: The expected surface brightness of CvB decay light assuming a $v_3$ lifetime of $10^{16}$ years and a mass of 50meV. The inset on the upper right shows a magnified view of the edge.

Observing a wide field of view enables us to assume that the integrated light from line emissions of point sources, such as distant galaxies over various redshifts z, follows a continuous distribution. In contrast, the CvB radiative decay is nearly isotropic, and its spectrum does not depend on the line of sight. Thus, employing a wide-field observation is a strategic approach for conducting a search for the edge in the CvB radiative decay. Our proposal is a search for an area of approximately 1 square degree with low interstellar dust, such as the AKARI Deep Field South (ADF-S), the minimum dust density region in the whole sky, which is one of the best fields for the cosmic infrared background measurement [7], where a known galaxy map is also available. This search will be conducted using a 100-hour observation time. Figure 2 presents a mass-dependent estimate of the lifetime of $v_3$, at which the surface brightness of the edge in the CvB radiative decay spectrum corresponds to a 5σ sensitivity level obtained from a 100-hour observation. This estimation is based on the fields of views of band2 and band3, which cover the range of 50–87 meV, and the 5σ sensitivity value of point-source spectrum measurement at R~130. It is indicated that we can explore the CvB decay for lifetimes ranging from $10^{15}$ to $10^{16}$ years.

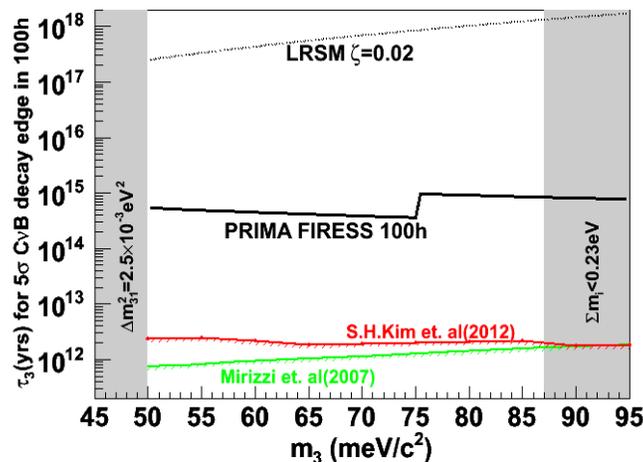

Figure 2: The mass-dependent estimate of the lifetime of $v_3$, at which the surface brightness of the edge in the CvB radiative decay spectrum corresponds to a 5σ sensitivity level obtained from a 100-hour observation.

In summary, we propose to search for the radiative decay of cosmic background neutrinos by observing a region of about 1 sq deg for 100 hours, using the edge of the spectrum as a clear signal to achieve a sensitivity of decay lifetimes in the range of $10^{15}$ to $10^{16}$ years. If neutrino decay is observed with this sensitivity, it would provide compelling evidence for non-standard model contributions in neutrino decay, as well as direct detection of the cosmic neutrino background. Moreover, it would be possible to determine the absolute mass of neutrinos from the edge wavelength of the decay spectrum, and the long-wavelength tail would contain information on the redshift dependence of the cosmological term.

## Instruments Required:

- PRIMA FIRESS: Large Map Low-res, R~130
- ADF-S Field (~1 sq. deg.)

## Program Size:

- Medium (10–100 hr)

**Approximate Integration Time:**

- 100 hours

**Special Capabilities Needed:**

- None

**Description of Observations:**

- In FIRESS mapping mode, we target a region of approximately 1 deg$^2$ within the ADF-S area ($\sim$12 deg$^2$) where there is little interstellar dust influence and precise galaxy mapping information has been obtained through previous research.

- The observation time for the target is 100 hours in total.

- For point sources, two-pointing observations are necessary to cover the 4-band full range. However, for CvB radiative decay, there is no need to specifically focus on conducting two-pointing observations, as the decay can be observed uniformly.

- CvB is almost isotropic and does not vary over time, so all data can be combined into a single spectrum for each wavelength bin, regardless of the line-of-sight direction or the order of the scan.

- By observing a wide field of view at a uniform speed, emission lines from distant galaxies can be integrated at various redshifts, appearing as a continuous spectral component of integrated galaxy light. On the other hand, CvB radiative decay light is almost isotropic, and its spectral component is largely independent of the chosen field of view.

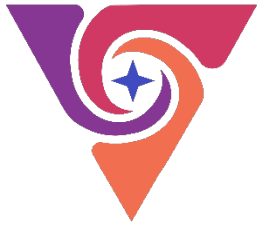

# Galaxies

# 11.   PriViCS: the PRIMA Virgo Cluster Survey

Maarten Baes (Universiteit Gent, Belgium), Simone Bianchi (INAF, Italy); Viviana Casasola (INAF, Italy); Luca Cortese (University of Western Australia, Australia); Ilse De Looze (Universiteit Gent, Belgium); Jacopo Fritz (UNAM Morelia, Mexico); Frédéric Galliano (CEA Paris-Saclay, France); Suzanne C. Madden (CEA Paris-Saclay, France); Angelos Nersesian (Universiteit Gent, Belgium); Matthew Smith (Cardiff University, UK); Manolis Xilouris (IAASARS, Greece)

Clusters of galaxies are unique laboratories for investigating the dependence of galaxy evolution on their environment. The Herschel Virgo Cluster Survey (HeViCS) mapped the central ~84 deg$^2$ region of the Virgo Cluster in five bands between 100 and 500 μm, which resulted in the first detailed view of cold dust in cluster galaxies. Major limitations of the HeViCS survey were the lack of data between 20 and 100 μm and the limited sensitivity of the PACS instrument, resulting in poor constraints on the warmer dust component. We want to use the PRIMAger instrument onboard PRIMA to map the HeViCS regions of the Virgo Cluster in the hyperspectral and polarimeter bands from 25 to 265 μm. Combining the PRIMA and Herschel data with existing multi-wavelength data we will, amongst others, infer the stellar and dust properties of a complete sample of cluster galaxies, investigate the evolutionary processes at work in the cluster environment, map the magnetic field structure of the cold ISM, search for dust emission from the intra-cluster medium, and investigate the evolution of the ISM in galaxies behind the Virgo Cluster.

## General Scientific Area:

Galaxy Ecosystems and Interstellar Medium

## Science Justification:

Clusters of galaxies are unique laboratories for investigating the dependence of galaxy evolution on their environment. We know that in addition to undergoing the morphological segregation effect, cluster galaxies differ significantly from their field counterparts. In particular, they are found to be deficient in atomic gas and their HI distribution is less extended than isolated objects of similar morphological type and luminosity. The lack of neutral gas in cluster galaxies is probably at the origin of the quenching of star-formation, the truncation of the star forming disks, and the higher metal content observed in cluster spirals. Virgo is probably the most thoroughly studied cluster of galaxies. It is a relatively populous system, consisting of ~2000 catalogued members. It contains vast quantities of X-ray emitting gas and shows clear evidence for both substructure and non-virialized motions. The property that sets Virgo aside from other clusters is its proximity: at a distance of ~17 Mpc, it may be studied in a level of detail that is not possible with more distant systems.

In the past decades there have been numerous photometric and spectroscopic studies of the Virgo cluster and its constituent galaxies at all wavelengths, ranging from X-ray to radio (e.g., Boselli et al. 2011, 2018; Auld et al. 2006; Chemin et al. 2006; and many others). Thanks to the Herschel mission, the Virgo cluster was also observed at far-infrared and submm wavelengths. The Herschel Virgo Cluster Survey (HeViCS: Davies et al. 2010) generated FIR/submm images of about 84 deg$^2$ over the Virgo cluster in 5 broadbands from 100 to 500 μm. The survey resulted in a large suite of results on the properties of dust in Virgo Cluster galaxies, including the global properties of large optically or FIR-selected samples (Davies et al. 2012; Auld et al. 2013; Baes et al. 2014; Pappalardo et al. 2016), specific classes of galaxies (Cortese et al. 2010; Grossi et al. 2010; De Looze et al. 2010, 2013; di Serego Alighieri et al. 2013), or detailed properties of interested special cases (Baes et al. 2010; Magrini et al. 2011).

As HeViCS was limited to 5 bands from 100 to 500 μm, the lack of data 25 to 100 μm prevented detailed investigations of the evolving dust properties within galaxies, which depend on the local environments. These include effects of the processing of the dust around star-forming regions, for example, or grain coagulation in dense molecular clouds. Studies involving the widely-varying dust size distributions require the full MIR to submm SEDs to study the evolutionary processes the dust is subjected to within galaxies as well as within the cluster environment. PRIMA is the only facility available to cover this important wavelength regime. Moreover, with PRIMAger we can, for the first time, map the polarized far-infrared emission and thus the structure of interstellar magnetic fields in the cold ISM in a large sample of nearby galaxies.

We plan to map with PRIMAger the ~84 deg$^2$ HeViCS regions of the Virgo Cluster. We will use both the hyperspectral and the polarimeter bands of the PRIMAger instrument, and we will combine the PRIMAger fluxes (and images) with data from the HeViCS survey and the many multiwavelength data (from UV to mm wavelengths).

Specific goals include:

- We will apply global and pixel-per-pixel panchromatic UV-mm SED fitting to infer the stellar and dust properties of a complete sample of Virgo Cluster galaxies, without biases due to a lack of data in the 25–100 μm range.

- We will compare the dust characteristics between cluster and field galaxies. We will search for evidence of stripping of dust from cluster galaxies and interpret the multi-wavelength characteristics of galaxies to investigate which evolutionary processes are at work in the cluster environment.

- We will investigate the magnetic-field structure via polarized dust emission, which traces the cold gas where star formation occurs. We will compare this to measurements of the dynamical structure and SFR maps as derived from IFU observations to study the role of magnetic fields in regulating star formation, in driving outflows, and in fueling galactic nuclei.

- We will search for the first definitive detection of dust emission from the intra-cluster medium and investigate the consequences on the extinction of background sources.

- We will make a census of galaxies behind the Virgo Cluster. The deep surveys of the Virgo Cluster region cover a wide "blind" cosmological volume with excellent multi-wavelength data. We will apply SED fitting of all sources and investigate the evolution of dust properties over the past ~5 Gyr.

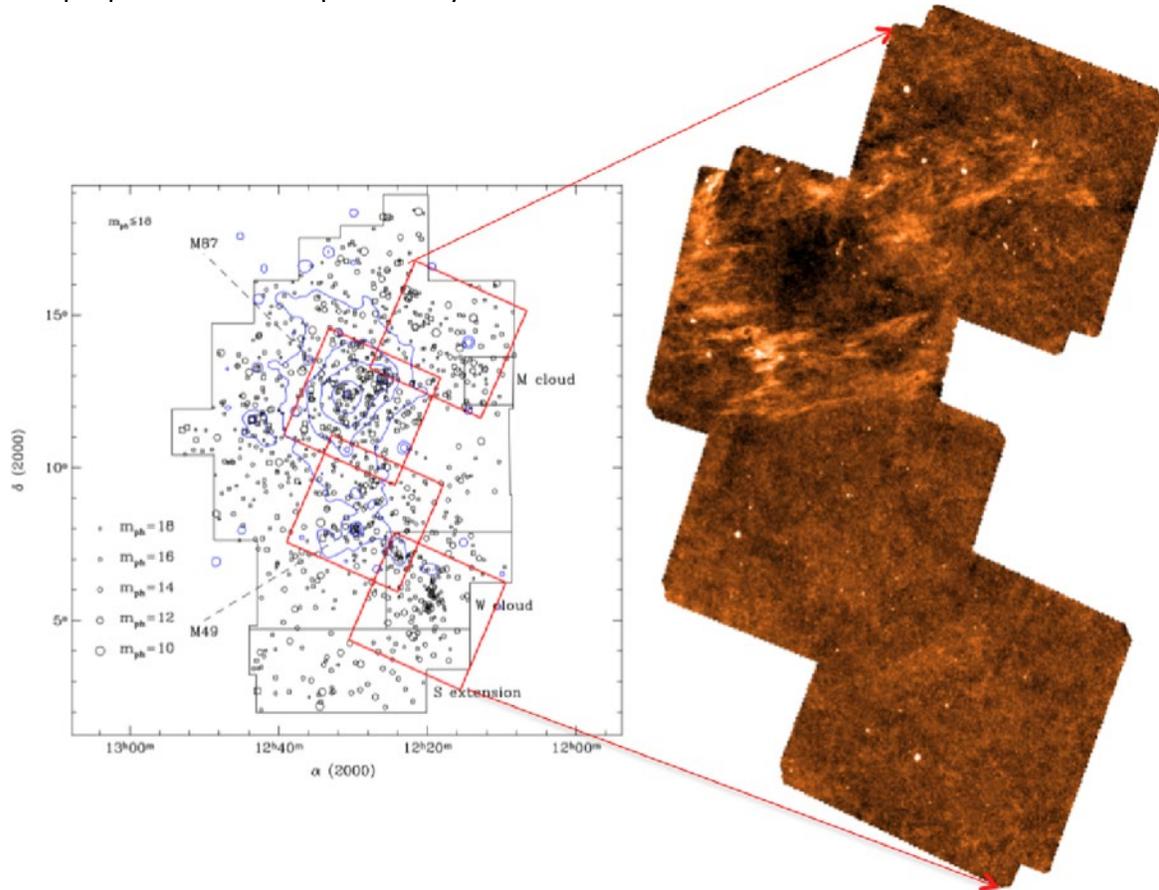

Figure 1: The Virgo Cluster. The left-hand panel shows the broad view of the cluster, with each of the ~2,000 bright member galaxies marked as a circle. The area covered by the Herschel Virgo Cluster Survey, which is the prime target of the PriViCS programme, is indicated by the red boxes. X-ray contours are shown in blue. On the right-hand side are the four fields as observed with Herschel SPIRE at 250 μm. Bright galaxies can be seen as white dots, and the Galactic cirrus is also prominently present. Figure taken from Davies et al. (2012).

## Instruments Required:

PRIMAger:Large Map (> 20'×20'); Hyperspectral band (25–80 μm; 12 filters; R=10); Polarimeter band (96, 126, 172, 235 μm; R=4)

## Program Size:

- Medium (10–100 hr)

## Approximate Integration Time:

- ~84 hours

## Special Capabilities Needed:

- None

## Description of Observations:

We request PRIMAger mapping in the hyperspectral and polarimeter bands. For an integration time of 1h/deg², we should reach 5σ point-source detection limits between ~0.7 and 1 mJy in the hyperspectral PRIMAger bands. A similar observing time gives point-source sensitivities of ~0.6 to 1.6 mJy in the polarimeter bands. This is an enormous gain compared to the HeViCS sensitivity (Auld et al. 2013) and more than sufficient to fully characterize the SED of all SPIRE-detected sources in the HeViCS fields. A 1h/deg² integration time corresponds to a 5σ polarized surface brightness of about 350 kJy sr$^{-1}$ at 170 μm. A representative surface brightness at 160 μm of galaxies at ~0.5 $R_{25}$ is 1 mJy arcsec$^{-2}$ (e.g., Smith et al. 2010; Pohlen et al. 2010; Bendo et al. 2012). Assuming a conservative 1% polarization, this corresponds to a polarized surface brightness of $10^{-5}$ Jy arcsec$^{-2}$, or 420 kJy sr$^{-1}$. With a 1h/deg² integration we will hence be able to reach our goals. To fully cover the 84 deg² HeViCS fields, we require about ~84 hours of observing time.

# 12.   Far-IR mAPPing of CGM Emission (FRAPPE)

James G. Bartlett (Jet Propulsion Laboratory, California Institute of Technology)

What is the make-up of the circumgalactic medium (CGM)? How does material and energy flow between the CGM and its host galaxy? These two fundamental questions about the cosmic ecosystem remain largely unanswered because of a dearth of observational constraints.

We propose a 100-hour program to map emission from dust in the halos of 50 nearby galaxies. Such a program will provide deep multi-band imaging (100 times deeper than Herschel/PACS) over the peak of the thermal dust emission, enabling us to constrain dust temperature and mass as a function of galactic radius. The polarimeter band will enable us to look for structures in the magnetic field associated with outflows.

## General Scientific Area:

Galaxy Ecosystems and Interstellar Medium

## Science Justification:

The circumgalactic medium (CGM) is a key link in the cosmic ecosystem connecting galaxies to the cosmic web. It contains the majority of a galaxy's baryons as a diffuse, multiphase gas, exchanging mass and energy with its host galaxy and the intergalactic medium (IGM). This interaction is critical to galaxy evolution, but the details of exactly how it takes place remain elusive. Key science questions are:

- **What is the make-up of the circumgalactic medium (CGM)?**

- **How does material and energy flow between the CGM and its host galaxy?**

Most of our knowledge about the CGM comes from absorption line studies along the line-of-sight to background sources (Tumlinson et al. 2017). These studies indicate that the CGM is a highly ionized, metal-enriched gas with a complex spatial and thermal structure. We suspect the CGM provides fresh gas for star formation, and we expect that its contents, spatial and thermal structure reflect feedback mechanisms from the host galaxy.

Absorption line studies provide only limited information at specific points through a galaxy halo. Observing the CGM in emission, in contrast, would enable us to map its structure and contents. This has proven difficult because the emission signal in all wavebands is extremely faint. There are only a handful of observations of CGM emission in individual galaxies – in X-rays, ultraviolet, and infrared (IR) (see, e.g., Tüllmann et al. 2006, Li et al. 2018, Kusakabe et al.2022); in particular, Herschel/PACS has been used to trace far-IR (FIR) emission from dust in the inner CGM (close to the host galaxy) for a few nearby galaxies (e.g., Veilleux et al. 2021, Yoon et al. 2021).

Dust in the CGM is an important find, because it must be transported there from star forming regions in the host galaxy. It therefore traces outflows into the CGM and gives us information on feedback. Mapping the dust distribution identifies how mass enters the CGM, and it also displays a historical record of mass accumulated over the history of the host galaxy. In addition, the dust temperature gives information about the thermal structure of the CGM.

**We propose a large GO program that will qualitatively improve our view of the CGM: detailed FIR photometry of the CGM in a sample of 50 nearby galaxies to sensitivities ~100 times deeper than existing observations.** We will map the FIR emission with PRIMAger to ~10–100 times larger radii than current observations. Modeling the dust emission by a modified blackbody, we will map the dust temperature and spectral index in the halo of each galaxy. With this information, we will

- Determine dust masses
- Map the spatial structure in the dust, identifying "break-out" regions that transport mass from the galaxy into the CGM
- Measure temperature structure and dust composition variations (through spectral indices)
- Trace magnetic field lines through measures of the polarized dust emission

## Instruments Required:

- PRIMAger: Large Map (> 20'×20') for 50 galaxies: 2h/galaxy; Hyperspectral band (25–80 µm; 12 filters; R=10) for 50 galaxies: 2h/galaxy; Polarimeter band(96, 126, 172, 235 µm; R=4) for 50 galaxies: 2h/galaxy

## Program Size:

- Large (100 + hr)

## Approximate Integration Time:

~2h/galaxy, based on extrapolation from Herschel/PACS observations and PRIMA improvement over PACS.

## Special Capabilities Needed:

- None

## Synergies with Other Facilities:

Our FIR observations will probe the dust and cooler gas in the CGM. They can be combined with UV and X-ray emission observations that probe metal lines and the hotter gas phases.

## Description of Observations:

Map the halo of each of 50 galaxies for 2 hours in PRIMAger.

## References:

Kusakabe et al. 2022, A&A 660, 44

## Acknowledgement:

A portion of this research was carried out at the Jet Propulsion Laboratory, California Institute of Technology, under a contract with the National Aeronautics and Space Administration (80NM0018D0004).

# 13.  Dust polarization from high-redshift galaxies

Matthieu Béthermin (Strasbourg Observatory/LAM), François Boulanger (ENS); Brandon Hensley (Princeton); Guilaine Lagache (LAM); Enrique Lopez Rodriguez (KIPAC/Stanford University); Marc Sauvage (CEA Saclay)

The emergence of coherent magnetic fields in high-redshift galaxies remains very hard to probe observationally. Thanks to its unprecedented sensitivity, PRIMA will open a new window on the dust polarization measurements in the high-redshift Universe, putting constraints on the magnetic fields of dusty systems. In this white paper, we present the forecast for the combination of a 2×200h survey including a deep (2 deg²) and a wide (20 deg²) field. Such a survey will revolutionize our view of the dust polarization at high redshift with ~10000 detections up to z=2, while no detection has been published so far beyond the nearby Universe.

## General Scientific Area:

Magnetic Fields

## Science Justification:

Even in our era of modern large galaxy surveys, the emergence of magnetic fields in high-redshift galaxies remain extremely hard to study. These magnetic fields can be probed through the polarization of various emission mechanisms. While SKA should open a new window on the polarization of the synchrotron emission, PRIMA will be the perfect instrument to probe the dust polarization. They are complementary, since the dust emission tends to come from the core of star-forming regions, while synchrotron emission is more diffuse. In addition, the dust polarization is also sensitive to local physical conditions (e.g., turbulence) that can affect the degree of depolarization due to changing magnetic field orientations along the line of sight.

- The first measurements of the dust polarization of high-redshift galaxies in the far-infrared will help us to address to address the following questions:

- Do high-redshift galaxies have already coherent large-scale magnetic fields?

- Which properties (inclination, stellar mass, gas density, turbulence, outflows, presence of a starburst and/or AGN, environment) drive the dust polarization?

- Can we probe on-going starburst or AGN activity using the polarized SED as suggested by observations in the local Universe (SALSA, Lopez-Rodriguez et al. 2022)?

- Are there correlations in the polarization orientations of neighboring galaxies?

PRIMA will have polarization measurement capabilities in the PRIMAGer band 2 between 85 and 256 µm. Due to the resolution of PRIMA, all high-z galaxies will be unresolved, and we will be sensitive only to the integrated polarization. The SOFIA/HAWC+ SALSA program (Lopez-Rodriguez et al. 2022) measured the integrated polarization fraction of nearby galaxies and found an average value of 1.3% with no evidence for a dependence to inclination or the starburst nature of a galaxy.

Here, we assume a conservative 1% polarization fraction to estimate the number of expected detections and their properties. We use the SIDES (Simulated Infrared Dusty Extragalactic Sky, Béthermin et al. 2022) semi-empirical simulation modified to include polarization and presented in the technical note about the confusion (see cutouts in Fig. 1). We estimate the number of detections for two different 200h surveys of 2 deg² (deep) and 20 deg² (wide). Using our simulated maps, we checked that our survey will not be affected by the confusion limit in polarization. Since our simulation covers only 2 deg², we renormalize our forecasts by a factor of 10 for the wide survey. Table 1 summarizes the expected number of detections in the various bands. Thousands of detections are expected, while no source has ever been detected in the far-infrared beyond the local Universe. PRIMA would directly open the window with statistical samples.

In Figure 2 we present the redshift and SFR distributions of the detections in each band. The deep survey will provide a large sample of galaxies between z=0 and 1 with a tail extending up to z=3. In contrast, the wide survey will provide most of its detections below z=0.5, but with even more detections. A wide range of SFR will be explored from 1 to 1000 M$_\odot$/yr.

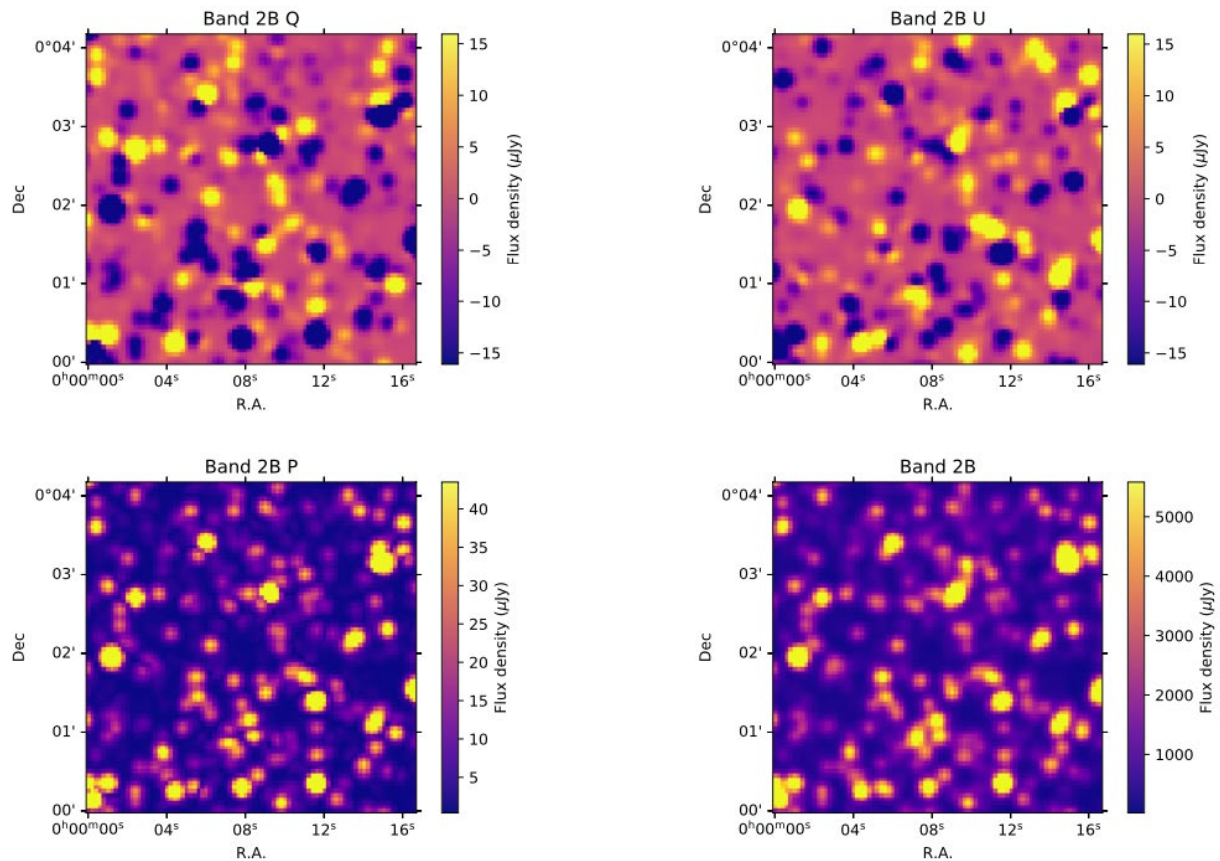

Figure 1: Cutouts of noiseless PRIMAGer band 2B observation in Q (upper left), U (upper right), P (lower left) and intensity (lower right).

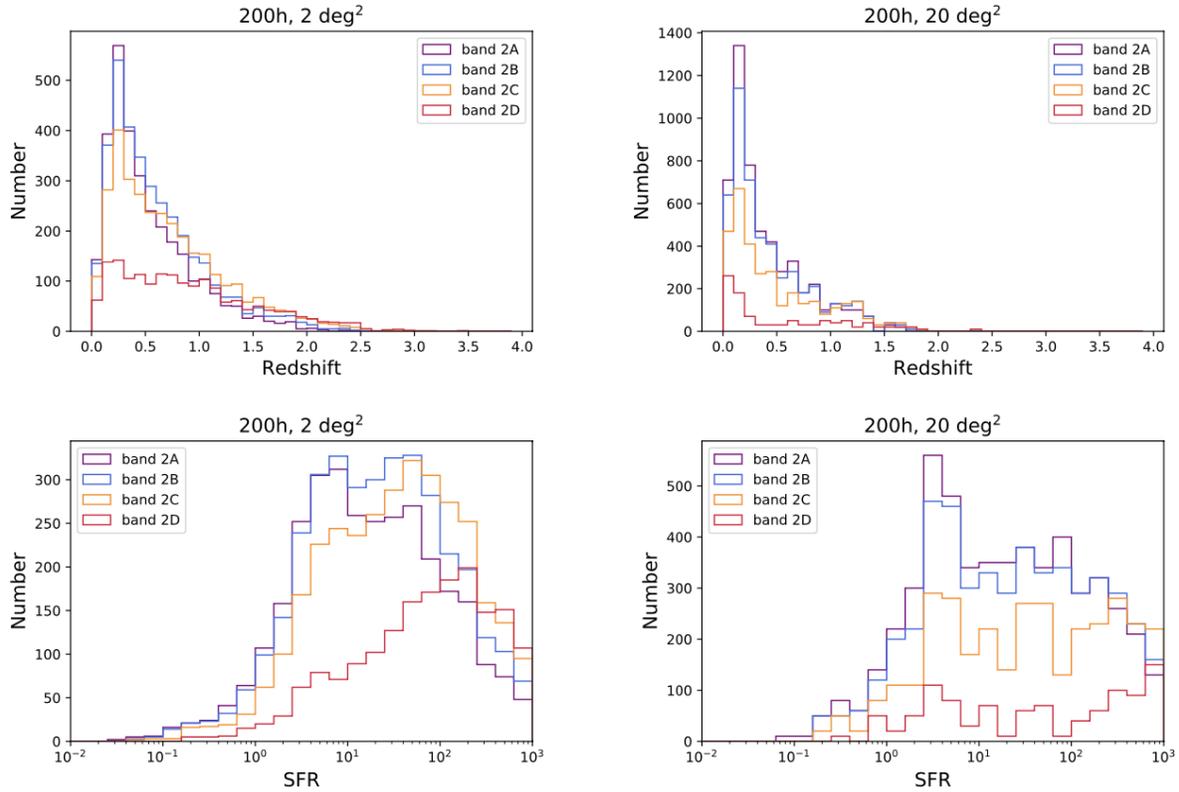

Figure 2: Top panels: expected redshift distribution of detections in a deep (left) and a wide (right) PRIMA survey. Bottom panel: SFR distribution in the same surveys.

By instrument design, such polarization surveys can be performed in parallel with a hyperspectral survey at shorter wavelength in PRIMAGER band 1, providing additional precious data for galaxy evolution studies in intensity. We can thus combine the risky pathfinding polarization observations with safer important intensity surveys, mitigating the risks.

Table 1: Expected number of detections ($N_{det}$) in polarization in the various PRIMAGer bands. The flux limit is given in polarized flux ($P=sqrt(Q^2+U^2)$).

| Band | λ μm | Flux limit mJy | $N_{det}$ | <z> | <SFR> M⊙/yr |
|---|---|---|---|---|---|
| **200h, 2 deg²** | | | | | |
| 2A | 96 | 0.092 | 3102 | 0.53 | 67 |
| 2B | 126 | 0.126 | 3499 | 0.60 | 78 |
| 2C | 235 | 0.224 | 1731 | 0.90 | 174 |
| **200 h, 20 deg²** | | | | | |
| 2A | 96 | 0.291 | 5280 | 0.40 | 89 |
| 2B | 126 | 0.401 | 4890 | 0.43 | 105 |
| 2C | 172 | 0.534 | 3340 | 0.51 | 148 |
| 2D | 235 | 0.709 | 1010 | 0.57 | 226 |

## Instruments used:

PRIMAGer large map (>20′ × 20′) using the polarimetric capability of Band 2. Band 1 hyperspectral data would be obtained as a bonus. An ultra-deep field (small map) could be added but is not necessary.

## Program Size:

Large (100+ hr)

## Approximate Integration Time:

400h

## Special Capabilities Needed:

None

## Synergies with Other Facilities:

The interpretation of PRIMA polarized data will benefit the wealth of data in next-generation multi-lambda deep and wide (10s of deg$^2$ or more) fields by EUCLID and Roman (spectroscopic redshifts, morphology, inclination, stellar masses, SFRs), or SKA (neutral gas content, radio SFRs, presence of a radio AGN, synchrotron polarization).

## Description of Observations:

In the scientific justification, we explored the combination of a deep (2 deg$^2$) and a wide (20 deg$^2$) survey of 200h each.

These surveys will be performed using the polarimetric bands of PRIMAger using the large-map mode. Ideally, if a parallel mode is possible, a hyperspectral scan can be performed at the same time. Depending on the separation of the two field of view, there might be a large overlap between the two footprints allowing a combined analysis.

The most obvious candidate for the deep field is the COSMOS field, which matches the field size and benefits from an extensive multi-wavelength coverage. The choice of the wide field(s) is less trivial. The Euclid deep field in Fornax could be an excellent candidate, but future Roman deep fields could be also very relevant.

## 14. Cluster survey with PRIMA: a survey of the growth of clusters, and critical line mapping to reach the deepest possible view of galaxy formation

Andrew Blain (University of Leicester)

The evolution of galaxies in clusters, transforming into the characteristic early-type population, and polluting the intracluster medium with their emitted gas is an important process is understanding the richest regions of the Universe, and their use as probes of the evolution. Clusters also offer a valuable cosmic telescope to enhance the performance of astronomical tools, especially at far-IR wavelengths where confusion can be severe. In order to make the deepest surveys, it may be necessary to exploit the natural telescope provided by clusters, to reach the faintest objects. Knowing the properties of fields surveyed using PRIMA in a large sample of clusters, building on work conducted by the Herschel Lens Survey, will highlight the most promising fields for this type of "critical line mapping", to highlight the faintest intrinsic objects.

### General Scientific Area:

Growth of Galaxies and Black Holes Through Cosmic Time

### Science Justification:

The astrophysics of galaxy clusters tells us about the evolution of the densest regions of the Universe, and keeps an archaeological record of the winds and outflows from galaxies. Clusters also provide tools to challenge the outputs of models, and a natural telescope to image Mpc-sized regions of the background sky, regions that are inaccessible to telescopes with limited resolution in the absence of this effect.

Galaxies within the clusters can be probed at PRIMA wavelengths, in both colour and spectral properties in order to trace the ongoing level of star-formation and AGN activity as they transform into early-type cluster member galaxies, and as cooling gas continues to rain down and fuel the central cD galaxy in clusters. The degree of dynamical relaxation in clusters can also be probed in terms of ongoing star-formation activity.

Clusters are Mpc in scale, and thus 5×5 arcmin maps provide a good view of the members over a wide range of redshifts from z~0.2 to beyond 1.

High-magnification regions of clusters about an arcmin from the centre and a few arcminute in length can be targeted specifically in very deep observations to see galaxies intrinsically too faint for detection in the field using PRIMA. This approach of "critical line mapping" will provide the ultimate deep field from the PRIMA mission, building on the results of a wider cluster survey.

Acknowledgement: this writeup is based on previous work with Eiichi Egami (University of Arizona). Note that a related case from Egami et al. is also included in this GO book.

## Instruments Required:

- FIRESS small maps at low-res are required for all sources
- FIRESS pointed observations (low res) should be obtained for subsequent critical line mapping
- FIRESS pointed high-res observations on detected sources
- PRIMAGer small maps for context

## Program Size:

- Large (100 + hr)

## Approximate Integration Time:

To reach the confusion limit with PRIMAGER, less than an hour of integration is required per cluster, and there are hundreds to thousands of potential targets, certainly 100 with excellent models available.

For critical line mapping spectral pointings with FIRESS, staring for hours will be required, with the maximum possible depth being determined based on results.

Observations will consist in imaging hundreds of clusters, mapping tens of auspicious high-magnification regions, taking deep spectroscopic observations along the fixed geometry of a slit, placed exactly at the high-magnification position, where the foreground galaxy properties are well-determined in the first stage of the survey, and where the magnification is known to be high.

Follow up of tens-hundreds of specific interesting detected sources for critical line mapping at the highest resolution could also be performed.

## Special Capabilities Needed:

Mapping critical lines requires the slit to be aligned with the high-magnification region in order to take the best advantage of the natural telescope. This will impose constraints on the timing of observations, and may push field choices for deep critical line mapping towards high ecliptic latitude.

## Synergies with Other Facilities:

Existing observations using Herschel and Spitzer provide background, but more clusters are now well-mapped, and the higher sensitivity and extended wavelength coverage of PRIMA ensures that it will be probing new territory.

## Description of Observations:

Maps around known clusters of galaxies, down to a deep level, and deep spectroscopy pointings chosen to probe the most promising single pointings down the maximum confused depth, with the expectation of order an hour of integration per target. The choice of critical line mapping depth will be made iteratively as depth is built up in the survey, and based on the brightness of objects that are discovered in the field.

PRIMAGer maps can be helpful for detailed SED shape determination of hot dust in and behind the cluster, and for mitigating confusion.

# 15. Deep tiers of dusty galaxy surveys to probe representative environments

Andrew Blain (University of Leicester, UK)

The goal of covering a representative volume of the Universe in which galaxies are growing and evolving remains a key goal of observational cosmology. Surveys in small areas using ALMA, and covering large fields into the limit of confusion using Herschel have highlighted much more about the role of dusty in galaxy evolution, but in order to make more progress, it is necessary to obtain spectral information about the properties of the dusty galaxies, and to find their relationships with other populations and their clustering in three dimensions. In order to investigate the most representative galaxies, and compare with samples obtained in the deepest surveys, made with Hubble and Webb, it will be necessary to push down as far as possible into the confusion limit, but over a field of order 10 square degrees in size, to fairly sample a representative volume, which would encompass the full extent of a forming cluster and surrounding filamentary structure, that should be much more extended at redshifts 2–8 than the evolved and virialized structure covers today. The wavelength range of PRIMA allows unique access to the hot dust, AGN diagnostic lines and PAH features that were sampled by Spitzer in large numbers for the first time. When made in a Herschel field, with deep coverage from Hubble and soon Euclid, to provide optical morphology, such as COSMOS, UDS, GOODS-N etc., PRIMA can make a unique contribution to resolving the ratio of dusty and UV-visible galaxies, and split activity from AGN and global star formation throughout the host galaxy.

## General Scientific Area:

Growth of Galaxies and Black Holes Through Cosmic Time

## Science Justification:

The goal to probe galaxy evolution using direct probes of hit components of the interstellar medium has been underlined by ISO, Spitzer and Herschel. The requirement to observe from space has ensured that even missions from the 1990s have not been supplanted in spectral coverage. PRIMA offers to improve dramatically on the progress made using these earlier FIR missions, and to provide imaging and spectral information at moderate resolution, to help address the relative contribution of AGN, with hot dust and low PAH equivalent widths, and possible 9.7-$\mu$m silicate absorption features, against the SED templates that are typical of star-forming regions in the Milky Way, 30 Doradus and well-studied low-redshift galaxies.

By probing the deep spectral images that are possible to obtain using a space-borne telescope, hoping to cover fields with the best supporting data at optical wavelengths, including forthcoming fields imaged by Euclid, the deepest level of a wedding-cake programme can provide the best insight into the properties of evolving galaxies from reionization, through cosmic noon, to the present day.

Covering an area of order 10 square degrees is important in order to include a representative sample of all cosmic environments at moderate to high redshifts, and this will demand multiple coverage passes using both instruments, PRIMAger and FIRESS, to gain the maximum sensitivity possible, good cross calibration and the ideal legacy.

The prospect of mapping redshifted Halpha emission from a wide range of redshifts is also a possible product from the deepest tier of a spectral survey, correlating line features with underlying continuum emission, with colors provided by hyperspectral mode images.

By the time that PRIMA flies, the results of the Euclid survey for the growth of large-scale structure should be known, and so it would be sensible to conduct surveys for evolving large-scale structure using a similar approach, but with fields chosen to match regions highlighted by Euclid, in addition to a representative deep field based on pre-existing data. Such regions may well already be imaged serendipitously in wide-field layers of a coherent PRIMA survey strategy. However, the desire to follow-up Euclid high-density regions down to the deepest level that can be achieved means that this should remain a potential extension to the deepest survey strategy as planning evolves.

## Instruments Required:

- FIRESS large map at low res
- PRIMAGer large maps using the hyperspectral band

## Program Size:

- Large (100 + hr)

## Approximate Integration Time:

With PRIMAGer, of order of 100h, considering that the confusion limit is reached at 60 μm in under 10 hours per square degree. It's likely to require the deepest fields to be located in continuous viewing zones, which introduces a tension with the location of some of the best studied fields, placed equatorially to facilitate access from ground-based sites.

With FIRESS, about 800h would be needed to cover to full area at a $3.10^{-18}$ W.m$^{-2}$ sensitivity. Observing time can be reduced with a strategic area coverage, for example focusing on representative regions. In general, a contiguous field of at least 30 arcmin to 1 degree in size would be required, so that the range of density in large-scale structure can be followed as a function of galaxy colour, projected density and redshift, where spectral lines are detected.

## Special Capabilities Needed:

- None

## Synergies with Other Facilities:

Mostly Euclid in terms of direct uniform coverage, Hubble and Webb at the deepest levels, where Euclid will also overlap, and within existing Herschel coverage. Background information from Planck and WISE coverage.

## Description of Observations:

Uniform survey to the deepest level possible to reach confusion limit. Choice and mix of modes and instruments to be decided based on sensitivity and relative confusion limits. A mix of roll angles, zodiacal foreground conditions, and other systematics would be desirable, with an iterative strategy to conduct the observations to the ideal depth as the performance in space is determined.

# 16.  PRIMA spectroscopy of extreme galaxies

Andrew Blain (University of Leicester)

Powerful dusty galaxies, incorporating galaxy mergers, high-luminosity AGNs, and the overall highest, most dramatic levels of luminosity reached by galaxies over the full history of galaxy formation have been found through a wide range of preceding surveys: using Spitzer, Herschel, Planck, WISE and ground-based submm telescopes. These ultraluminous galaxies should include the most dramatic phases of galaxy evolution, with feedback and fueling processes driven to their most extreme levels, offering insight into the most dramatic and instructive phases of a galaxy's history. PRIMA is in an excellent position to determine much more information about the processes within, via spectroscopy to highlight Ne lines that reveal excitation near to an AGN, silicate absorption that can provide insight into the column density to an AGN, and PAH emission that can indicate the strength of distributed star-formation activity around the galaxy. The continuum colour can also provide insight into the optical depth and heating profile of dust around an extremely luminous region. A PRIMA survey to investigate classes of pre-existing dusty galaxies, which can include all of the categories above, and red QSOs that might be abundant from Vera Rubin Observatory surveys in the run up to PRIMA's launch will all provide insight into the most luminous galaxies. Alongside a pointed spectral image of these targets, the surrounding field will enable a deep survey to be compiled, subject to the useful bias of knowing that there is a very luminous galaxy present, thus providing us with insight into the differences in fields with and without different classes of very-luminous sources.

## General Scientific Area:

Growth of Galaxies and Black Holes Through Cosmic Time

## Science Justification:

Planck ultraluminous galaxies number in the hundreds. With modest resolution, PRIMA is in a good position to locate undetermined examples, and to provide spectra of hundreds of examples. There are many hundreds of ultraluminous galaxies discovered by WISE at 12 μm, and all are suitable to have their AGN properties revealed in moderate-resolution PRIMA spectroscopy. Herschel-detected and ground-based submm detected galaxies, both lensed and unlensed, provide thousands of targets for PRIMA rest frame-mid-IR diagnosis of their interstellar medium properties, and energy source, whether accretion or star-formation.

PRIMA will provide coverage of a very important but ill-determined wavelength range, where the hot dust, possible silicate absorption, characteristic emission lines from species sensitive to high-ionization near intense AGN power, and the relative abundance of the hottest and cooler dust can be discriminated. This offers the potential to determine the AGN content of even the most heavily-obscured systems, and to provide insight into the utility of different AGN diagnostics to apply to less dramatically luminous objects.

Acknowledgements: this case was developed based on discussions with Tanio Diaz-Santos et al. concerning WISE HotDOG ULIRGS, David Clements and other UK FIR Probe team members on Planck sources, Julie Wardlow on submm-selected galaxies

## Instruments Required:

- FIRESS low-res pointed observations for most sources, small maps for sources which are poorly localized
- FIRESS high-res pointed observations for a subset of sources
- PRIMAGER small maps in hyperspectral bands

## Program Size:

- Large (100 + hr)

## Approximate Integration Time:

To confusion limit, approximately an hour per source. Sources with interesting spectra can be followed up at higher resolution for similar times. Expectation is that there would be up to several thousand possible targets from WISE, Herschel, Planck and submm telescopes, including SPT.

## Special Capabilities Needed:

- None

## Description of Observations:

Pointed observations to take spectra of known interesting high-redshift sources. FIRESS can be used for gas line, and PRIMAGER for broad features (PAHs, silicates) as well as the general shape of the mid-IR spectrum. Some of the Herschel and Planck sources might be relatively poorly localized, and require a small map to be made. As this is being done, a deep imaging survey of the regions around these sources is built up for free.

# 17.  Dust in the tails of ram pressure stripped galaxies

Alessandro Boselli (Laboratoire d'Astrophysique de Marseille, France), Laure Ciesla (Laboratoire d'Astrophysique de Marseille, France), Marc Sauvage (AIM-CEA, France)

The evolution of galaxies in rich environments such as clusters and groups can be significantly perturbed during their interaction with the hot intracluster medium (ICM) trapped within the gravitational potential well of the massive host halo (ram pressure stripping). The external pressure exerted by the hot ICM on the galaxy ISM during its high velocity journey within the cluster is able to remove most, if not all of it, producing extended tails of stripped material. Along with the different gas phases (cold atomic and molecular, ionised, hot), the external pressure can remove also dust, thus contributing to the pollution of the ICM. The aim of this project is that of using the superior capabilities of PRIMA to observe this dust component, a crucial ingredient in the energetic balance of the stripped gas. We propose to observe a selected sample of ram pressure stripped tails detected at other frequencies (HI, CO, Halpha, X-rays) using both PRIMAger (in imaging and polarimetric mode) and FIRESS. These data will be used to determine the relative distribution of the dust component with respect to that of the other gas phases, derive its temperature, calculate different gas physical parameters (electron density, photoelectric heating efficiency, gas metallicity), and quantify the strength of the turbulent magnetic fields, all fundamental parameters used to constrain the most recent hydrodynamic simulations of gas stripping in clusters. The ultimate aim of this research is that of understanding the fate of the cold stripped material once mixed with the surrounding hot medium and study under which condition it can collapse into GMC to form new stars.

## General Scientific Area:

Galaxy Ecosystems and Interstellar Medium

## Science Justification:

Galaxies inhabiting rich environments have physical properties significantly different than those in the field. They are principally relaxed systems, composed of evolved stars, with a very reduced amount of gas and dust, while their counterparts in the field are generally gas-rich, star forming rotating systems. All this evidence clearly indicates that the environment in which they reside plays a major role in shaping galaxy evolution. Different physical mechanisms have been proposed in the literature to explain these differences. They can be broadly divided into two major families: the gravitational interactions between the different group/cluster members, and hydrodynamic interactions between the cold ISM of the galaxies and the hot intracluster medium (ICM) trapped within the gravitational potential well of the high-density region (see for a review Boselli & Gavazzi 2006). Among these mechanisms, the one dominant in massive ($M_{cluster} > 10^{14}$

M$_\odot$) local clusters seems to be the ram pressure stripping exerted by the ICM on the ISM of galaxies moving at high velocity ($\sim 1000$ km s$^{-1}$) within it (Boselli et al. 2022). This mechanism is able to remove the gas of the ISM in all its phases, principally the one located in the outer disc of the perturbed galaxies, forming extended tails of stripped material.

Along with the different gas phases of the ISM (cold atomic, molecular, ionised, hot), ram pressure is expected to remove also the interstellar dust, which in normal galaxies is generally well mixed with the gaseous component. Direct observation of dust in the stripped tails of perturbed galaxies, however, is still lacking principally due to the limited sensitivity of the instruments used in the observations. Longobardi et al. (2020a) observed an asymmetric distribution of the dust component in a few galaxies showing an extended tail of gas witnessing an ongoing ram pressure stripping event. Kenney et al. (2015), using high-resolution HST imaging, discovered prominent filaments of dust escaping from the disc of NGC 4921 in the Coma cluster, also suggesting that dust is removed along with the gaseous component. There is also evidence of dust truncated discs in stripped objects in the Virgo cluster (Cortese et al. 2010), or the presence of diffuse dust mixed with the hot ICM still in Virgo (Longobardi et al. 2020b), but the presence of dust associated to the stripped gas at large distances from the galaxy disc is still lacking. The presence of dust in these stripped structures is fundamental since it might contribute to the pollution of the ICM as an alternative origin of the often claimed ejecta of supernovae in the early phase formation of the giant ellipticals inhabiting high-density regions. Its presence is also fundamental for understanding the fate of the stripped material since dust can contribute to the cooling of the stripped gas, necessary for the formation of GMC where star formation can take place, observed in most, but not all, the stripped tails. Indeed, if present, dust would play a major role in the energetic equilibrium of the cold stripped gas once mixed with the surrounding hot ICM thanks to its cooling properties. Its observation is thus fundamental for understanding the physics of the different components in this extreme environment and thus posing strong observational constraints to models and simulations.

The tails of stripped material have been observed in almost all gas phases (cold atomic, molecular, ionised, hot) up to hundred kpc from the stellar disc of the perturbed galaxies (see Fig. 1). Considering a typical gas-to-dust ratio of G/D = 100, we can estimate the expected dust emission from the observed HI column density of a few representative objects (e.g., Chung et al. 2007). The far-IR emission of dust can be detected in continuum. The far-IR surface brightness can be measured using the PRIMAger camera, and used to derive a dust column density in the tail. The emission of the tail will be identified and recognised from that of the confusion-limited background since correlated on large scales and with an expected distribution similar to the one observed in the different gas phases. Using several bands, the data can also be used to infer a typical dust temperature, a critical parameter necessary to constrain models and simulations with the aim of understanding the gas phase transition within the tails. Polarimetry observations are also of great interest since they can be used to quantify the strength of the turbulent magnetic field and compare it to the one that SKA observations will provide for the ordered magnetic field in the radio domain. Recent hydrodynamic simulations have indicated the magnetic field as a crucial parameter in confining the stripped material, thus self-shielding it from the heating due to the surrounding ICM, triggering gas collapse and star formation (Tonnesen & Stone 2014; Ruszkowski et al. 2014). The degree of polarisation observed in radio continuum in similar tails of stripped material is still poorly known since observed in a very few objects, where it ranges between 5% and 30% (NGC 4522, Vollmer et al. 2004).

The spectrometer FIRESS, on the other hand, can be used to measure several IR line emissions and thus to constrain the composition, density and temperature of the stripped ISM. In the spectral domain covered by FIRESS (24–240 μm) there are several emission lines of primordial importance for constraining the properties of the stripped gas. The brightest of them is the [CII] line at 158 μm, which is one of the principal coolers of the ISM in star forming galaxies. Other lines (e.g., [NII] 122 μm and [NII] 205 μm), characterised by different ionising potentials and critical densities for collisions with electrons, can be used to identify the origin of the [CII] emission (ionised gas, molecular gas, atomic gas) in the stripped material, the electron density and the metallicity of the stripped gas, or the photoelectric heating efficiency using the [CII]158 μm+[OI]63 μm/TIR ratio (Croxall et al. 2017, Kewley et al. 2019). Variations of the dust column density and temperature will be used to constrain timescales for dust sputtering once stripped from the cold galaxy ISM and mixed within the hot ICM.

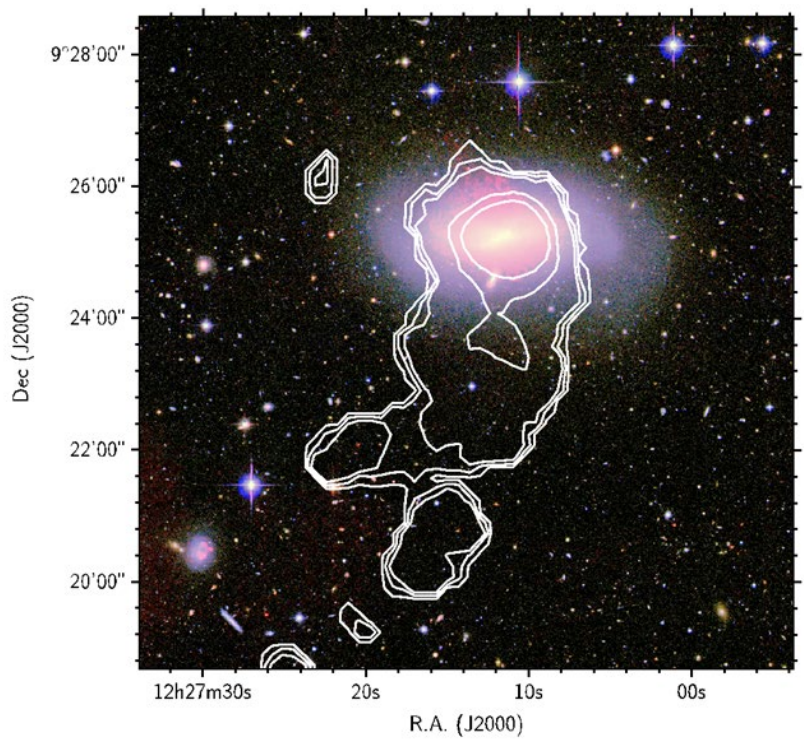

Figure 1: pseudo-colour image of NGC 4424 in the Virgo cluster (1 arcmin = 4.8 kpc), adapted from Boselli et al. (2018a). The red diffuse low surface brightness feature extending in the north direction is ionised gas detected in $H_{alpha}$ during the VESTIGE survey (Boselli et al. 2018b). White contours show the HI gas distribution starting from a column density n(HI)=$10^{19}$ cm$^{-2}$ (from Chung et al. 2009).

## Instruments Required:

- FIRESS low-res small maps
- PRIMAGer small maps in all bands

## Program Size:

- Medium (10–100 hr)

## Approximate Integration Time:

According to the PRIMAger expected sensitivity, a surface brightness of Sigma235 ~ 87 kJy/sr (5σ) appropriate for G/D = 100 can be reached on this area (~ 72 arcmin²) in 400 sec, or Sigma235 ~ 10 kJy/sr (5σ) in 8h if G/D = 1000. Assuming a polarisation of ~ 10% we can also estimate the integration time necessary to measure it with the same S/N than in imaging mode, which is 22h for G/D = 100. These integration times can be significantly reduced adopting a less constraining S/N ratio or rebinning the data on more extended regions (Voronoi binning), which is possible thanks to the very extended nature of these tails.

We can estimate the expected surface brightness of the main emission lines in the FIR spectral domain ([CII]158 μm, [OI]63 μm, [NII]122 μm and 205 μm). Assuming the typical line/TIR ratios of Croxall et al. (2012, 2017) ([CII]158 μm/TIR ~ 0.005, [CII]158 μm/[OI]63 μm ~ 5, [CII]158 μm/[NII]205 ~ 10, and [NII]122/205 μm ~ 1-8) and the S250 vs. STIR ratio of Galametz et al. (2013) (nuS250 ~ 0.3 STIR) we can estimate the expected surface brightness of the main emission lines in the FIR spectral domain ([CII]158 μm, [OI]63 μm, [NII]122 μm and 205 μm) to respectively be $2\times10^{-12}$ W m$^{-2}$ sr$^{-1}$, $4\times10^{-13}$ W m$^{-2}$ sr$^{-1}$, $2\times10^{-13}$ W m$^{-2}$ sr$^{-1}$ and Sigma[NII]122 = $10^{-13}$ W m$^{-2}$ sr$^{-1}$. With emission potentially extended to >1000 beams across, similarly to PRIMAger observations, FIRESS integration times can be reduced using Voronoi binning to under 10h for each target and spectral setting, making these important observations accessible.

## Special Capabilities Needed:

- None

## Synergies with Other Facilities:

**PRIMAger:** Ram pressure stripped tails of gas have fairly extended dimensions (10–100 kpc), corresponding at the distance of the Virgo cluster (16.5 Mpc, where 1 arcsec = 80 pc) to ~ 20 x 20 arcmin. Observations should thus be done using the steering mirror. For a typical column density of n(HI) ~ $10^{19}$ cm$^{-2}$ on a stripped tail of area A ~ 100 kpc x 10 kpc at the distance of the Virgo cluster (see Fig. 1), the total mass of atomic gas is MHI ~ 7.5 x$10^7$ M⊙. Assuming a gas-to-dust ratio G/D = 100, the corresponding Mdust ~ $7.5 \times 10^5$ M⊙. At the same distance, this would correspond to a flux density at 250 μm S250 ~ 350 mJy, thus to a typical surface brightness of Sigma250 ~ 100 kJy/sr, or a factor of 10 lower if G/D = 1000 as claimed by Longobardi et al. (2020a) for these extreme environments.

**FIRESS:** The expected contribution of the main IR emission lines in these extreme environments, where the gas and the dust can be heated by the stellar radiation whenever star forming regions are present, but also by the heat conduction and mixing with the surrounding hot ICM, is totally unknown. A first order estimate can be done assuming that the line emission is comparable to that observed in the typical ISM of star-forming galaxies.

# 18.  Metallicity ($Z_{ISM}$) and Dust Mass ($M_{DUST}$) from PAH bands and fine-structure Line Constraints (MeDALiC)

Denis Burgarella (Laboratoire d'Astrophysique de Marseille, France), Laura Bisigello (Università degli studi di Padova, Italy); Matthieu Bethermin (Observatoire de Strasbourg, France); Matt Bradford (Caltech, USA); Laure Ciesla (Laboratoire d'Astrophysique de Marseille, France); Carlotta Gruppioni (Osservatorio Astronomico di Bologna, Italy)

The basic constituents of today's Universe are: 73% dark energy, 23% dark matter, and 4% in baryons (Peroux & Howk 2020). Metals, i.e., chemical elements heavier than H and He form about 2% of this baryonic matter. They are a key element of many physical processes: they help cool gas clouds, form stars, and planetary systems, and they are the building blocks of life (prebiotic molecules). Understanding how metals formed in the Universe over Cosmic ages is one of the major astrophysical and cosmological goals for humankind.

This objective is a statistical one in essence: we need a large and representative sample of all types of galaxies to be able to constrain the rise. In the next years JWST, SUBARU/PFS, VLT/MOONS will provide unique spectroscopic information on the rise of metals.

However, most metals are in massive and dusty galaxies that only a mid- and far-infrared (IR) telescope in space can reach. The MeDALiC program will make use of the two PRIMA instruments: FIRISS and PRIMAger. PRIMAger will reach intrinsically fainter and more distant galaxies. The PAH bands and the fine line structure lines will bring us this information on the rise of metal and dust in the Universe (e.g., Ciesla et al. 2006).

## General Scientific Area:

Rise of Metals and Dust

## Science Justification:

### Context

After the Big Bang and until the recombination, the Universe only contained hydrogen and helium. When the first stars were born, the first metals and probably the first dust grains started to rise. Until now, it has been quite difficult to trace the rise of metals and dust in the early Universe. JWST, but also new massive spectroscopic surveys like PFS on the Japanese SUBARU telescope, MOONS on the European VLT will open up a new era to decipher the information coming from more distant galaxies, and over very large galaxy samples. Most of the metals and dust grains have been formed in dusty starbursts. We don't understand yet which of the physical processes that could increase the dust mass are responsible for the formation of the bulk of the dust mass.

## Science Question:

What are the first phases of the formation and early evolution of metals and first dust grains in the Universe's lifetime?

## Need for PRIMA:

Measuring the metallicity and the dust mass in dusty galaxies can only be performed from infrared (IR) observations. Previous IR observatories (AKARI, Spitzer, Herschel in space, and ALMA, NOEMA on the ground) lack the sensitivity for the former and the statistics for the latter. PRIMA will bring both.

## Interpretation methods:

From PRIMA spectroscopic and photometric observations, we will use modern tools to perform fitting of the galaxy spectral energy distributions (SEDs with photometry, line fluxes/equivalent widths and full spectra). One of these methods is the CIGALE code (Burgarella et al. 2005, Boquien et al. 2019, and Burgarella et al. 2023, in prep. for spectroscopy). It is important to stress that we will need to derive how much of the IR energy is due to star formation and to any active galactic nuclei (Yang et al. 2020). Another PRIMA program (Mountrichas et al.) demonstrates that we will be able to derive the AGN fraction from PRIMAger data. This will be further checked with the FIRISS spectroscopic sample.

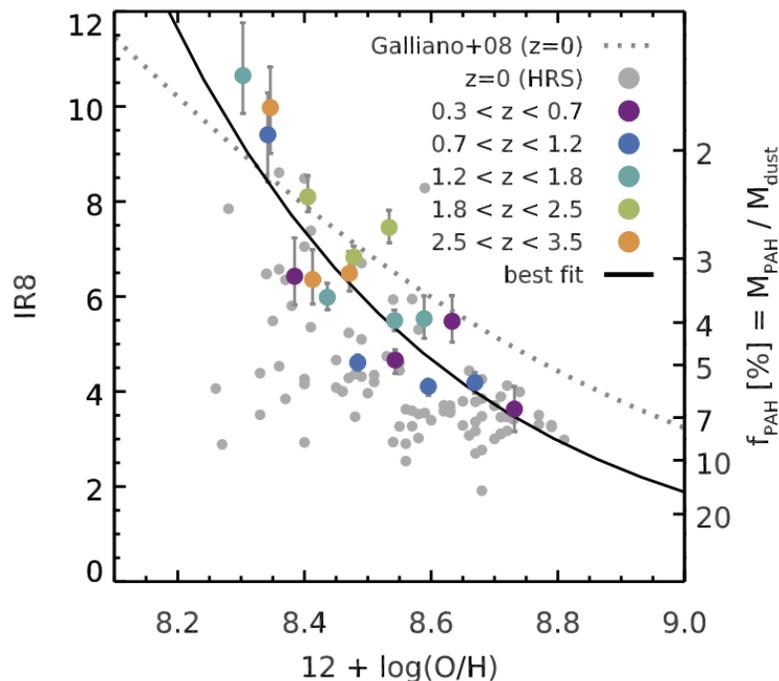

Figure 1: $M_{PAH} / M_{DUST}$ is related to the metallicity as measured by $12 + \log(O/H)$ and would provide a metallicity tracer that will be measured by PRIMAger.

**Link to testable hypotheses:**

JWST/MIRI will provide a test sample at 1<z<3 that could be used to calibrate PAH into $Z_{ISM}$ (see also Schreiber et al. 2018, and Fig. 1 and Fig. 2). Beyond this initial presentation of the MeDALiC program, we will need to build a true test sample. It will be necessary to take into account other constraints from luminosity functions (e.g., Gruppioni et al. 2013) infrared and other relations like the main sequence and the evolution of the dust attenuation with redshift (Burgarella et al. 2013, Bogdanoska & Burgarella 2020), from which we will specifically simulate PRIMAger and FIRISS observations and check how well we can recover the important physical parameters $Z_{ISM}$ and $M_{DUST}$.

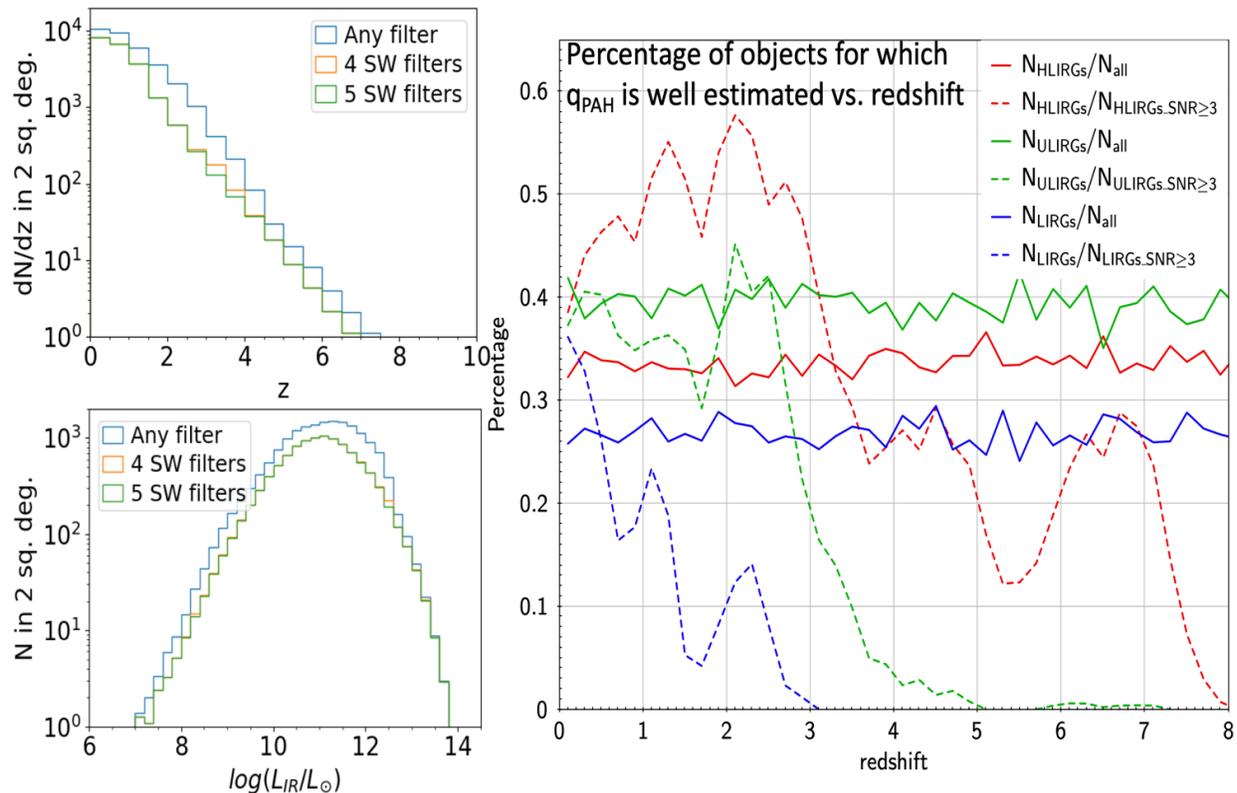

Figure 2: Left: statistics (redshift and LIR) of galaxies that would be detected with PRIMAger (in 10 hr, over 2 sq. deg. at 5 σ), and more specifically, for the 4 and 5 bluest bands that are less affected by confusion. Right: redshift ranges where qPAH ($M_{PAH}$ / $M_{dust}$ is estimated with SNR (qPAH) ≥ 3.

**Instruments Required:**

● PRIMA FIRESS: Medium Map Low-res R~150

● PRIMAger: Large Map (> 20'×20'); Hyperspectral band (25–80 μm; 12 filters; R=10)

**Program Size:**

● Medium (10–100 hr)

**Approximate Integration Time:**

● About 20 hours with PRIMAger; About 50–80 hours with FIRESS

**Special Capabilities Needed:**

- None

**Synergies with Other Facilities:**

To estimate the rest-frame UV/optical properties ($M_{STAR}$, $SFR_{TOT}$) of the galaxies, we will need complementary data from Euclid/Roman in the UV/optical frame, and also ALMA/NOEMA in the sub-mm, ideally in one of the SUBARU/PFS and/or VLT/MOONS deep fields.

**Description of Observations:**

The MeDALiC program will need to observe with PRIMAger a relatively large sample of galaxies over the largest redshift range (Table 1), and up to the sensitivity limits, to estimate the strengths of the PAH bands (separately and all together).

In addition to PRIMAger, a sub-sample of galaxies will be observed with FIRISS. This sub-sample will be observed by PRIMAger as well to calibrate the PRIMAger observations and check how well we can estimate the strengths of the PAH bands, and maybe derive the equivalent widths of the stronger fine structure emission lines. The metallicity for this common sample will be derived using the fine structure emission lines (e.g., Ferkinhoff et al. 2015, Diaz-Santos et al. 2021).

## Narrative of observational strategy:

The MeDALiC program will need to build a statistically representative sample of galaxies (Table 1). We will not know how well the targeted derived parameters ($Z_{ISM}$ and $M_{DUST}$) will be sampled before actually performing the observations. We should therefore rely on known proxies that allow deriving $Z_{ISM}$ and $M_{DUST}$, like the stellar mass $M_{STAR}$ (related to both $Z_{ISM}$ and $M_{DUST}$) or the star formation rate (related to $L_{DUST}$ and therefore to $M_{DUST}$).

Models such as those developed by, e.g., Bisigello et al. (2022), Béthermin et al. (2022) will used to define the minimum size of the samples to be observed to collect at least $N_{galaxies}$ = 10–20 galaxies per parameter bin ($Z_{ISM}$, $M_{DUST}$, $M_{STAR}$, SFR, redshift). This threshold of $N_{galaxies}$ is defined to obtain a minimum signal-to-noise, SNR = 3 assuming a Poisson noise (sensitivities in Fig. 3).

Table 1: Total number of objects detected by PRIMAger in 10 hours at 5σ over 10 sq. deg. Each of the LIR samples will be further cut as best as possible for each of the $Z_{ISM}$, $M_{DUST}$, $M_{STAR}$, SFR, redshift parameters.

| Area (deg$^2$) | LIR < 11 | 11<LIR <12 | 12<LIR <13 | 13 < LIR |
|---|---|---|---|---|
| 10 | 26,355 | 20,895 | 5,840 | 380 |

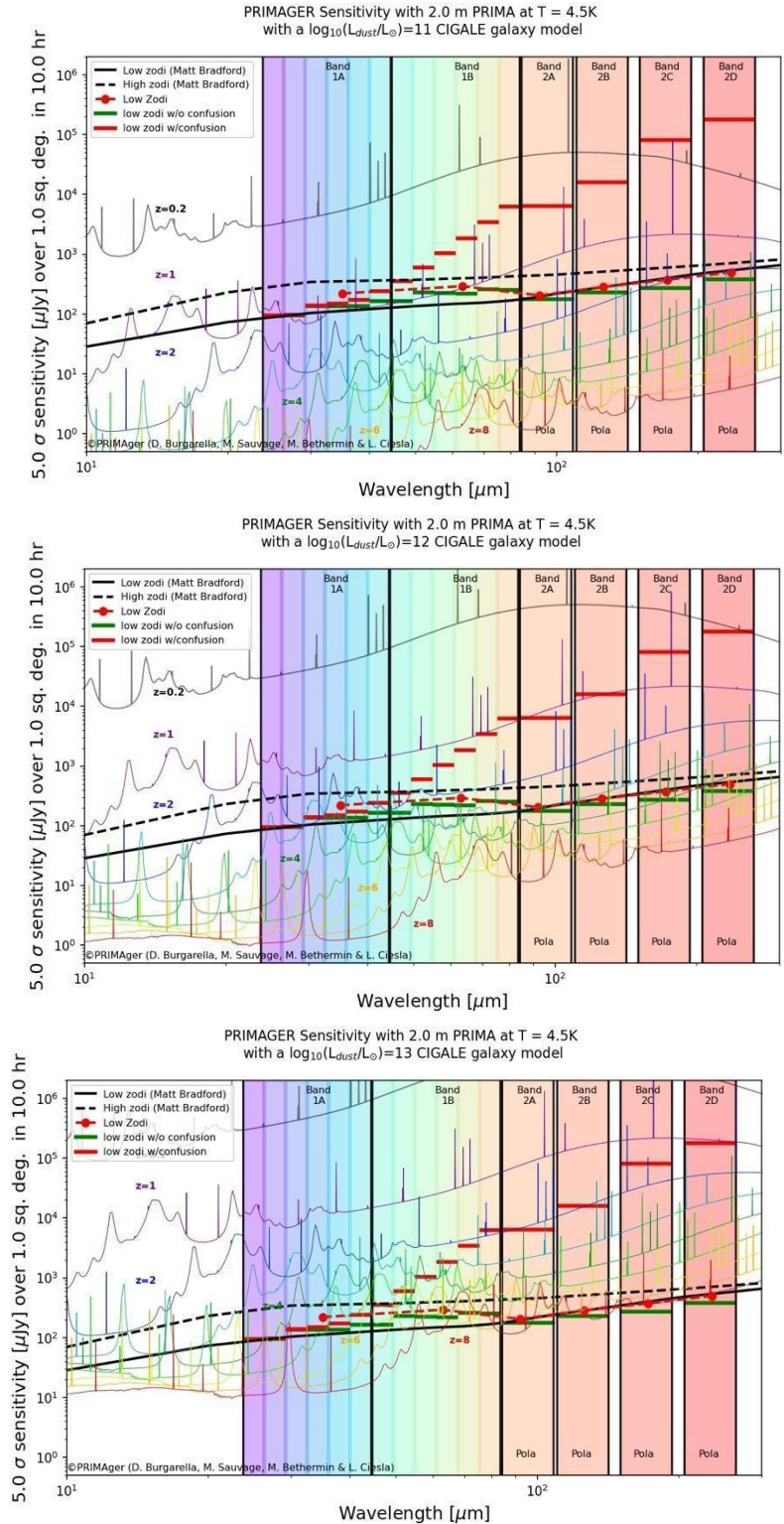

Figure 3: Sensitivity of a 10-hours survey over 10 sq. deg. at 5σ for LIRGs, ULIRGs and HLIRGs as a function of the redshift. For the highest redshifts in the epoch of reionization, only HLIRGs will be detected. Their detectability will define how soon in the Universe's lifetime this MeDALiC program will reach out. We will try to optimize the program to make use of the bands less impacted by confusion at shorter wavelengths.

# 19. Unveiling the evolution of the dust mass function with PRIMA

Francesco Calura (INAF-OAS, Bologna, Italy), Francesca Pozzi (UniBo, Italy), Carlotta Gruppioni (INAF-OAS, Bologna, Italy); Laura Bisigello (UniPd, Italy), Alberto Traina (UniBo, Italy); Ivan Delvecchio (INAF-OAB, Italy), Margherita Talia (UniBo, Italy)

Achieving a complete picture of the evolution of galaxies throughout their history is a primary goal of modern science. To accomplish this ambitious task, a great deal of multi-wavelength surveys have been devoted to assessing the cosmic evolution of the cold gas and of the stellar mass across a significant fraction of the age of the Universe. In this cosmic census, one elusive component is represented by interstellar dust. Dust grains constitute the solid component of the interstellar medium, they are a significant repository of heavy elements and affect the spectral properties of galaxies across a broad range of wavelengths. Achieving a complete census of the cosmic dust budget is of paramount importance to access the yet unknown amount of obscured star formation and to shed light on one galactic component invisible by optical and UV instruments. In this proposal, we plan to exploit PRIMA to perform a survey aimed at estimating the still poorly known dust mass function up to redshift z~3, i.e., across a major fraction of cosmic time. We will probe the evolution of the comoving dust density across the 'cosmic noon' and see if its peak is coincident with the one of unobscured star formation, already accessible with JWST. The synergy with ALMA will be fundamental to cover the IR emission peak and constrain the dust temperature of distant dusty galaxies, enabling a direct assessment of the evolution of the dust mass density throughout a >90% fraction of cosmic history.

## General Scientific Area:

Rise of Metals and Dust

## Science Justification:

One major goal of extragalactic astronomy is to trace in the most complete possible way the evolution of baryonic matter across cosmic time. One fundamental component of galaxies is interstellar dust, representing the solid component of the interstellar medium and affecting the spectral properties of galaxies over a wide range of wavelengths, ranging from the far-infrared to the ultraviolet domain (e.g., Draine 2009). **How did the dust mass budget evolve through cosmic time? Addressing this question is of the utmost importance to constrain one significant component of the cold mass fraction in galaxies, to access obscured star formation and the fraction of heavy elements removed from the gas phase and incorporated into solid grains.**

A fundamental quantity suited for this purpose is the dust mass function (DMF). A thorough estimate of the evolution of the DMF will allow us to reconstruct how the production and destruction of interstellar dust have changed over time in galaxies of various masses.

Thus far, the evolution of the DMF has been the subject of a handful of studies. Dunne et al. (2011) considered a sample of 250 μm-selected sources from the Herschel-ATLAS Science Definition Phase to look at the evolution of galaxies' space density as a function of their dust mass from z=0 up to redshift z∼0.5. Their results supported a sharp increase in the bright end of the DMF across this redshift interval, consistent with a previously determined DMF estimate at z∼2.5 based on a limited sample of SCUBA submillimetre galaxies (Dunne et al. 2003).

The most comprehensive study of the evolution of the DMF was performed by Pozzi et al. (2020), based on a far-IR (160μm) Herschel-selected catalog from the COSMOS field, deriving the DMF from z ∼ 0.2 up to z ∼ 2.5.

The sample of Pozzi et al. consisted of ∼ 5500 sources with flux density >16 mJy and estimated spectroscopic or photometric redshift. For each of these systems, the dust mass Md was derived from the observed flux by assuming a modified black-body relation, valid for a single-temperature (Td) dust component and in the standard optically thin regime (Bianchi 2013). This estimate requires the assumption of a temperature value, which comes from an empirical relation between Td, SFR and redshift out to z∼2 (Magnelli et al. 2012).

Fig. 1 shows the current state of the art regarding estimates of the DMF. The observations performed with Herschel suffer from severe limitations due mostly to its poor sensitivity and resolution, which caused significant confusion noise and allowed us to prove only the bright end of the DMF at 1<z<2. At z>3 the DMF is essentially still unexplored, with the exception of one ALMA-based ongoing study (Traina et al. 2023).

Current cosmological simulations allow one to track the evolution of basic galaxy properties, such as star formation, metal content and dust mass (Parente et al. 2022). At z∼1 the faint end is unconstrained. Some present models underpredict the DMF at z>1, fail to account for the characteristic mass or lack the brightest objects (Hou et al. 2019). Fig. 1 highlights the current poor theoretical understanding of the evolution of the dust mass budget and the need for more surveys at high redshift, to better constrain the shape of the DMF and achieve new estimates at the so-called 'Cosmic Noon', i.e., at 1<∼z<3 where the peak of cosmic star formation is likely to lie. PRIMA is the only instrument that will enable a blind, therefore unbiased IR survey that will overcome the limitations suffered by Herschel. With its unprecedented features, PRIMA will:

1) allow us to improve considerably the characterisation of the DMF and the basic parameters that define its shape, enabling a better estimate of the faint end and the sampling of the characteristic dust mass across a wide redshift range;

2) extend the redshift range where the dust emission in individual galaxies can be probed, allowing us to derive the DMF at z=3 and, in synergy with ALMA, at z>3; and

3) allow us to significantly constrain dust production in galaxy formation models, improving our poor theoretical knowledge of this process and filling the gap in our understanding of dust-obscured galaxies.

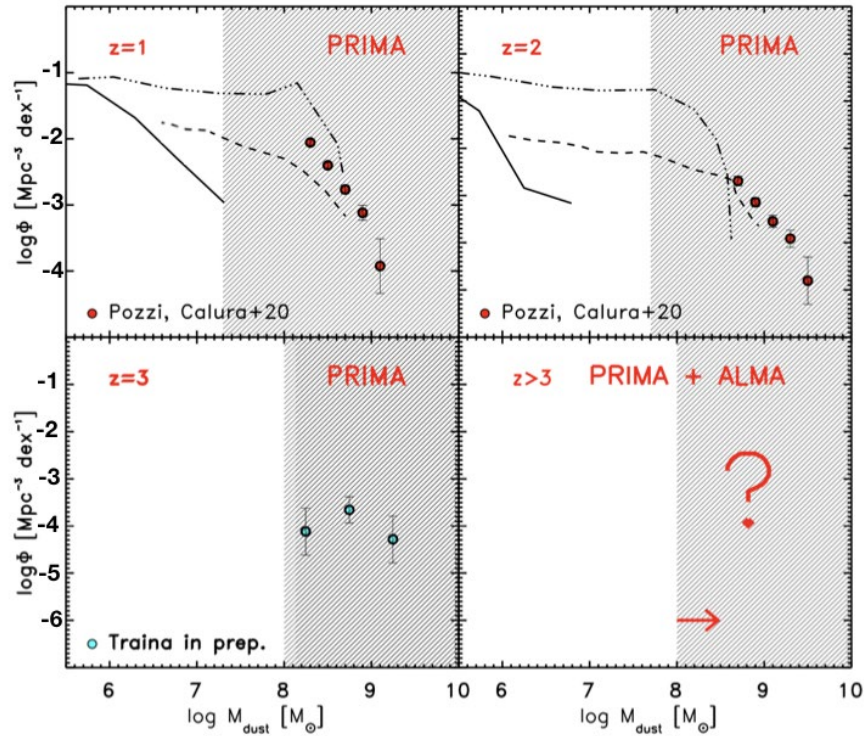

Figure 1: Observed and theoretical Dust Mass Function as a function of redshift. The solid circles are the observational estimates, whereas the black lines are available predictions from cosmological models (solid line: McKinnon et al. 2017; dashed line: Li et al. 2019; Dash-dotted line: Hou et al. 2019). The black Shaded area represents the dust mass regime accessible with PRIMA. PRIMA will enable to constrain significantly the present models, to characterize the shape of the DMF and, in synergy with ALMA, to constrain it at z=3.

## Instruments Required:

PRIMAger:Large Map (2 deg$^2$); Hyperspectral band (25–80 μm; 12 filters; R=10; both 1A and 1B); Polarimeter band (96, 126, 172, 235 μm; R=4; 2A, 2B, 2C, 2D)

## Program Size:

Medium (10–100 hr)

## Approximate Integration Time:

By considering the sensitivities reported in the PRIMAger fact sheets, to reach our goal of S60=0.07 mJy (see the 'Description of Observations' section) over an area of ∼2 deg$^2$ as planned in this survey, we will need an integration time of 25 hours/deg$^2$, i.e., 50 hours in total.

## Special Capabilities Needed:

- None

## Synergies with Other Facilities:

PRIMA will bring significant progress in our understanding of the cosmic evolution of dust production and obscured star formation. The synergy with other instruments will be fundamental in various aspects.

First of all, an open question concerns how the evolution of the dust mass budget is connected with the cosmic star formation history. This fundamental issue will be addressed with PRIMA, in synergy with the James Webb Space Telescope (JWST).

Previous studies evidenced an evolution of the comoving dust density peaking at lower redshift with respect to the cosmic star formation rate, possibly indicating a delay in the production of the bulk of the dust with respect to the main epoch of star formation (Driver et al. 2018, Pozzi et al. 2020). It is possible that such a result is due to the strong limitations in the studies carried out so far, including the already mentioned sensitivity limits of the Herschel instruments. In this regard, PRIMA will allow us to alleviate such limitations significantly. Another fundamental, still open and relevant question is what fraction of the cosmic star formation is obscured and how such a quantity evolves with redshift.

The unobscured cosmic star formation has been accessible for three decades and significant progress in its knowledge was brought by the launch of the Hubble Space Telescope (HST).

Through the detection of the dust emission in distant systems, Herschel enabled access to the brightest and obscured sources up to z~4, allowing us to estimate robustly only the bright end of the IR luminosity function, therefore providing us with a limited view of the amount of obscured SF at high redshift (Gruppioni et al. 2013; 2015). Besides the evolution of the dust mass budget, PRIMA will also allow us to push our current knowledge of the obscured star formation function (Gruppioni et al. 2015) to higher redshift and fainter luminosities. This progress will complement the recent breakthrough results achieved with JWST, which are enabling new estimates of the unobscured cosmic star formation rate in the near-infrared up to z ~ 15 (i.e., Bowens et al, 2023). Thanks to the synergy with JWST, we will address an important physical question and unveil if the growth of the dust mass evolves in lockstep with the cosmic star formation rate or is delayed, providing crucial constraints to current galaxy evolution models (Gioannini et al. 2017; Parente et al. 2022).

One second, fundamental aspect is the synergy of PRIMA with the Atacama Large Millimetre/Sub-millimetre Array (ALMA). Thanks to its unprecedented sensitivity and spatial resolution in the sub-millimetre/millimeter wavelength range, ALMA will allow us to probe the Rayleigh-Jeans regime of the PRIMA-selected sources at z<~3 to constrain the dust temperature and to cover the IR emission peak of higher-redshift sources (z>3) for which, even at the longest wavelength, the PRIMA polarimeter will sample only the short-wavelength side of the peak.

On the other hand, ALMA will need PRIMA for a comprehensive sampling of the FIR SED of these sources, which is crucial to calibrate and test to fainter luminosities fundamental scaling laws, such as the one between L_IR and L'_CO (e. g. Hodge & Da Cunha 2020). In this plan, we will leverage the power of the ALMA archive (Liu et al. 2019) and propose ALMA follow-up observations of pre-targeted PRIMA sources at z>3. The synergy between ALMA and PRIMA will enable a direct assessment of the evolution of the comoving dust mass density up to z>6, i.e., throughout a >90% fraction of cosmic history.

## Description of Observations:

Our aim is to sample the Dust Mass Function ~1 dex fainter than Herschel, in order to accurately probe the faint-end of the DMF up to z~2, discriminate between different DMF predictions and, for the first time, explore the DMF at z>3 (Fig. 1). We plan to perform our survey in the cosmological COSMOS field (of area $\sim$2 deg$^2$) and take advantage of the unique plethora of available multi-wavelength data and robust photometric redshifts (Weaver et al. 2022), including the forthcoming JWST-NIRCAM imaging over its central 0.6 deg$^2$ (Casey et al. 2023).

To reach this goal, we plan a blind survey in the Hyperspectral band 1B_6 (at 60 μm) with a 5σ flux limit of S60=0.07 mJy. Based on predictions obtained with the SPRITZ code (Bisigello et al. 2021), we expect $\sim$250,000 sources to lie above the proposed flux limit.

This limit, obtained with SPRITZ considering various populations of galaxies and AGNs, is sufficient to reach our goal of $S_{170}\sim$1.6 mJy, a factor $\sim$10 lower than the Herschel limit and obtained from the $S_{60}/S_{170}$ flux ratio in the most conservative case. The choice of the 1B$_6$ band is motivated by the fact that it is sensitive to dust emission and that it is not affected by confusion. We will circumvent blending issues affecting PRIMA fluxes by relying upon sub-arcsec resolution priors, including HST and upcoming JWST-NIRCam data in the central COSMOS-Web area. These de-blending techniques (Liu et al. 2018) have been tested and widely employed for extracting accurate Herschel and sub-mm fluxes in the COSMOS field (e.g., Jin et al. 2018).

## 20. Probing the far-IR galaxy population below the confusion limit

Dr. Dave Clements (Imperial College London), Co-authors: Dr Chris Pearson (RAL-Space), Ayushi Parmar (Imperial College London)

Analysis of the deepest Herschel SPIRE observations at 250–500 µm (Pearson et al., in prep; Varnish et al, in prep) have shown that there is an additional bump in the number counts derived from P(D) analysis at fluxes of 300–400 microJy. The nature of these sources is currently unclear, but they appear to represent the final population necessary to account for the cosmic infrared background. Such sources can be detected in Band 2D in integration times of ∼40 hours (5σ over a 1 sq. degree field according to the PRIMAger fact sheet) but will be well below the confusion limit. While such observations will allow statistical studies of this population, with important new constraints coming from the shorter wavelength band 2 channels, a full understanding of this population will require individual sources to be detected for further follow-up by PRIMA and other instruments such as ALMA. The 44 times smaller beams available in Band 1A, as well as the Band 1 hyperspectral imaging capabilities, will allow observations in these bands, if deep enough, to break the confusion limit and, through cross identification processes, determine the positions and far-IR properties of individual sub-mJy far-IR sources. We thus propose deep PRIMAger observations of several fields (20′×20′ − 30′×30′ in size), amounting to a total of 1 sq. deg., with all PRIMAger bands to sensitivities well below the confusion limit. These fields will be chosen to host deep multiwavelength observations from other missions (eg. Euclid, HST, JWST, Herschel, Spitzer, NGRST etc.) to ensure that all sources detected by PRIMA can be properly characterized using photo-z and SED fitting analyses. Followup spectroscopic observations with FIRESS of specific sources and sub-fields are anticipated, but the details of these observations will depend on what is found in the deep imaging. In this way the last significant galaxy population emitting in the far-IR will be fully characterized.

### General Scientific Area:

Growth of Galaxies and Black Holes Through Cosmic Time

### Science Justification:

The Cosmic Infrared Background (CIB) contains the integrated history of star-formation in the universe. Comparison to the integrated UV-Opt-IR shows that roughly half of the energy generated in the universe through star formation is obscured by dust and re-radiated at far-IR wavelengths. A full understanding of the history of galaxy and star formation over cosmic time thus requires observations at these long wavelengths. Observations by Herschel and other mid-IR-to-submm facilities have made significant advances in probing this population over the last 20 years, with up to 70% or more of the CIB resolved into individual sources brighter than the Herschel confusion limit of ∼10s of mJy at 250–500 μm. However, analysis of the deepest Herschel field available (Pearson et al., in prep, Fig. 1) using P(D) analysis to probe below the confusion limit (Varnish et al., in prep, Fig. 2) provides evidence of a second bump in the number counts, indicating the presence of an unexpected population of far-IR sources at flux levels of a few 100 microJy. At 250 μm the counts of this new population peak at fluxes around 500microJy. At still fainter fluxes, the counts drop away, indicating that this sub-mJy population is the last significant source of far-IR emission contributing to the CIB. The nature of this sub-mJy population, however, remains to be determined.

While individual sources responsible for this population might be uncovered by other observatories (Bisigello et al., 2023) a full understanding of this population requires further deep observations in the far-IR, with supporting mid-IR observations to help break the confusion limit and provide SEDs for these objects. PRIMA observations using both the long wavelength PRIMAger channels, to directly detect the far-IR emission of these objects, and the shorter wavelength hyperspectral channels, to provide high resolution observations to break the confusion limit and mid-IR SED points to characterize the sources, are thus required.

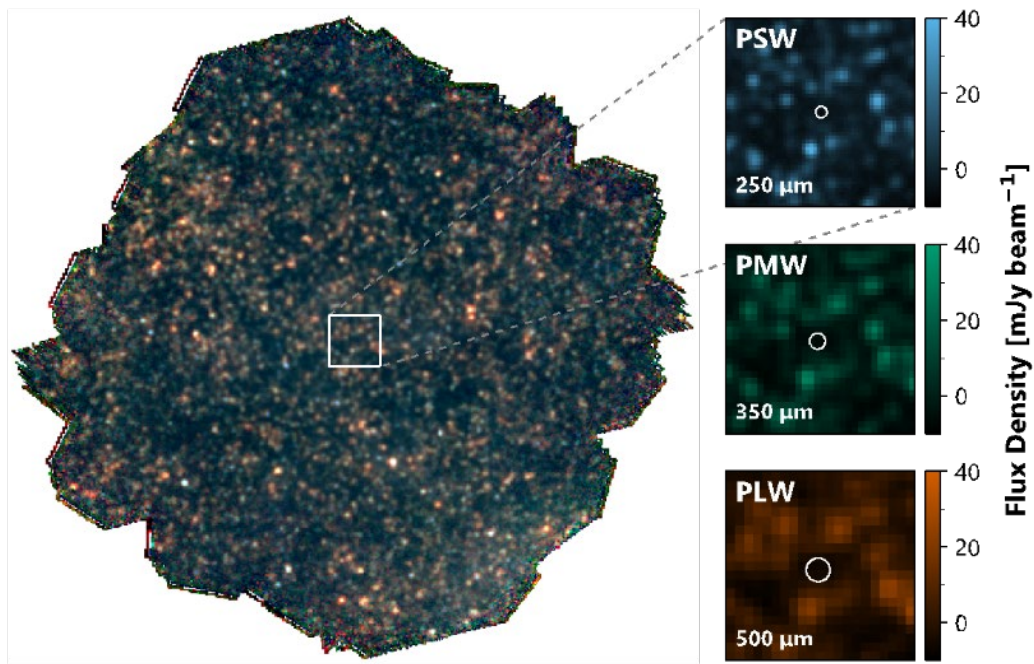

Figure 1: Left: 3 colour image of 250, 350 and 500 μm (blue, green, red respectively) of the SPIRE Deep field. The image is roughly 30' across with the deepest part of the field lying in the central 12'. Right: blow ups of part of the deepest central region.

## Interpretation methods

We will use advanced prior-based cross matching tools and pre-existing data or short wavelength PRIMAger observations to cross identify and determine the fluxes of confused sources. At this stage the most advanced such tool is XID+SED (Hurley et al., 2017; Pearson et al., 2017) but we expect more advanced tools to be developed by the time of launch, with some potentially being specifically designed for PRIMAger use.

We will use SED fitting tools using cross matched fluxes from UV to radio, and including X-ray if available, to determine the physical properties of the sources, including stellar mass, star formation rate, AGN content, photometric redshift, star formation history and more. Our current tool of choice for this is CIGALE but we expect more advanced tools, potentially optimized for PRIMAger, to be available at the time of launch.

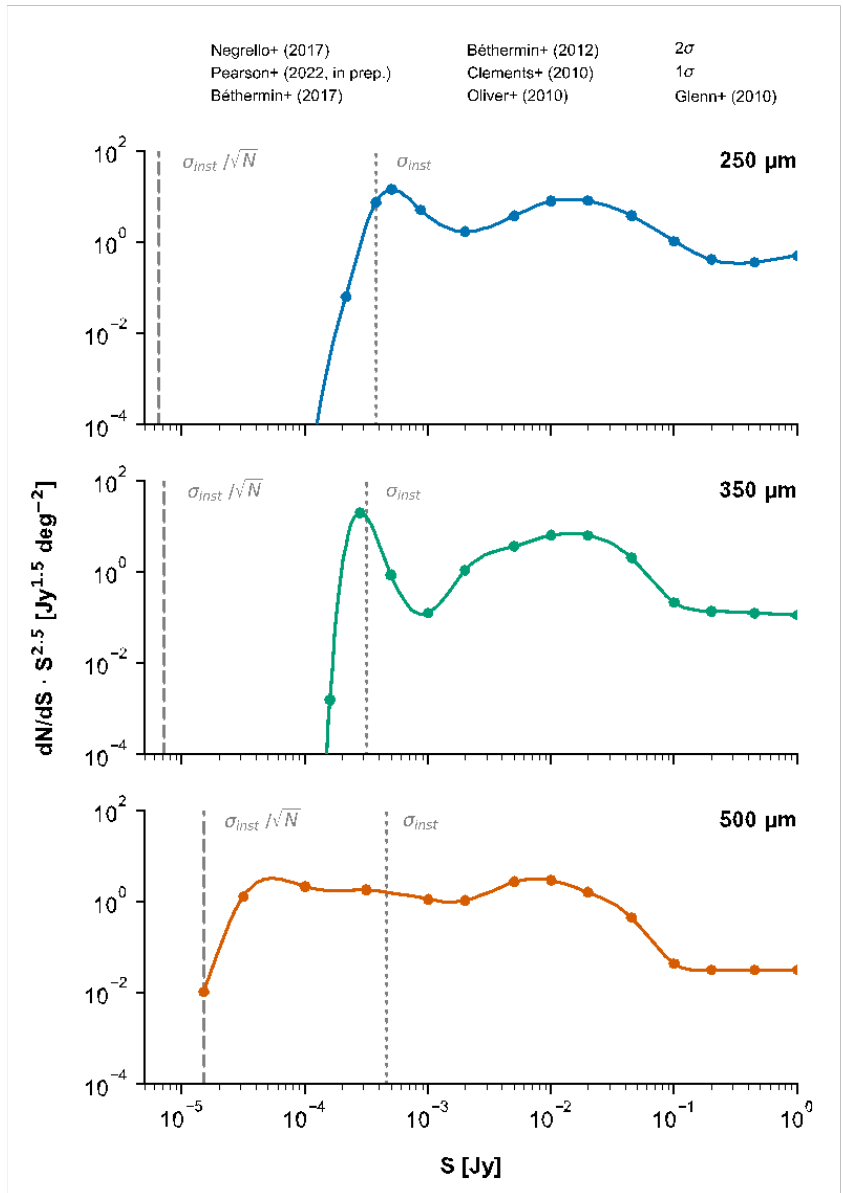

Figure 2: Result of P(D) analysis applied to the SPIRE Deep field observations. The analysis agrees with discrete number counts and existing models at brighter fluxes, but shows an additional population of sub-mJy sources at 250 μm and plausible evidence for similar populations at 500 μm. The dark grey band indicates the 68% confidence internal of the P(D) derived counts, the light grey region the 95% confidence interval.

Follow-up spectroscopy for specific sources of interest or for statistically relevant sub-samples will be proposed subsequent to the completion of the imaging survey. These observations will validate the photometric-based methods for determining redshift and other properties, and provide additional information such as emission line diagnostics to further our understanding of this population.

### Link to testable hypotheses

Test number count models for this population which are currently still in development.

### Instruments Required:

PRIMAGer small maps in both hyperspectral and polarimeter bands, for 40 h total.

### Program Size:

- Medium (10–100 hr)

### Approximate Integration Time:

The source density on the sky means that large samples of these objects can be obtained in each 20'×20' small map, but field-to-field variations, and the need to understand the clustering environment of these sources, require several fields of this size to be observed, amounting to a total area on the sky of 1 sq. deg. We wish to reach a sensitivity of 250 microJy (5σ) so that we can properly characterize the peak of the sub-mJy population, whose counts peak at ∼ 500 microJy. This will require an integration time of 40 hours according to the PRIMAGer information sheet. Since PRIMAGer observes simultaneously in all bands, these observations will also provide the necessary sensitivity at shorter wavelengths.

### Special Capabilities Needed:

- None

### Synergies with Other Facilities:

Selection of fields with existing (or future) deep imaging from UV, and ideally X-ray, through optical to submm and radio will be beneficial though not required.

### Description of Observations:

These observations require deep far- and mid-IR images of the fields using all bands of the PRIMAGer instrument.

# 21. The rise of dust

Grant Donelly (Ritter Astrophysical Research Center, University of Toledo), J.D.T. Smith (Ritter Astrophysical Research Center, University of Toledo)

Emission from PAHs will serve as a critical tracer for understanding dust growth in the early Universe, and PRIMA will shed light on this process for the first time. As a primary sink for the radiation emitted by stars and source of heating in the interstellar medium, dust acts as a thermal regulator within galaxies and thus has a significant impact on the star-formation efficiency and other properties. Recent observational evidence confirms that dust was present within galaxies during the first Gyr after the Big Bang, but little is known about its conditions, origins, and rate of growth. This poses significant problems for models and simulations of the early Universe, which rely upon these observational constraints to predict dust attenuation and emission. We propose a 200-hour program using PRIMAger deep-field imaging together with FIRESS follow-up spectroscopy to survey PAH and nearby continuum emission at redshifts z > 2, for the first time reaching back to the first Gyr after the Big Bang (z > 6).

## General Scientific Area:

Galaxy Ecosystems and Interstellar Medium

## Science Justification:

Dust shrouds the majority of star formation and black hole accretion in the Universe.

The presence of interstellar dust plays a major role in the evolution of galaxies (e.g., Narayanan et al. 2023, Hu et al. 2023), but the nature of dust growth and accumulation in the early Universe remains highly uncertain (see Figure 1). The harsh radiation and frequent supernovae present during this time could be lethal to dust, but observational evidence suggests that some massive galaxies already had significant dust reservoirs as early as several hundred Myr after the Big Bang (Viero et al. 2022, Vistok et al. 2023, Markov et al. 2023). Important physical properties of dust in early galaxies, including grain size distribution and ionization state are unconstrained, but have major implications on the energy balance in the interstellar medium (ISM). How can models and simulations accurately represent the impact of metal and dust enrichment within galaxy environments without an understanding of this important component? Observational constraints on dust characteristics at high redshifts are required to answer these pressing questions.

In the local Universe, thermal emission from polycyclic aromatic hydrocarbons (PAHs), which are highly sensitive to metallicity (Whitcomb et al., 2023), has been a widely used as a probe of conditions in the ISM. PAHs produce a family of prominent spectral features between 3–20 μm, each of which arise to a greater or lesser extent from grains with different qualities like size, temperature, and ionization. PAH emission is readily observable, accounting for as much as 20% of the infrared luminosity of typical galaxies (Smith et al. 2007). As the smallest of dust grains, PAHs are especially sensitive to the production, growth, and processing factors that alter the grain size distribution and overall dust abundance. Given these properties, observations of PAH emission at high redshifts will be a powerful approach to inferring the physical characteristics of dust at different stages within the developmental history of galaxies.

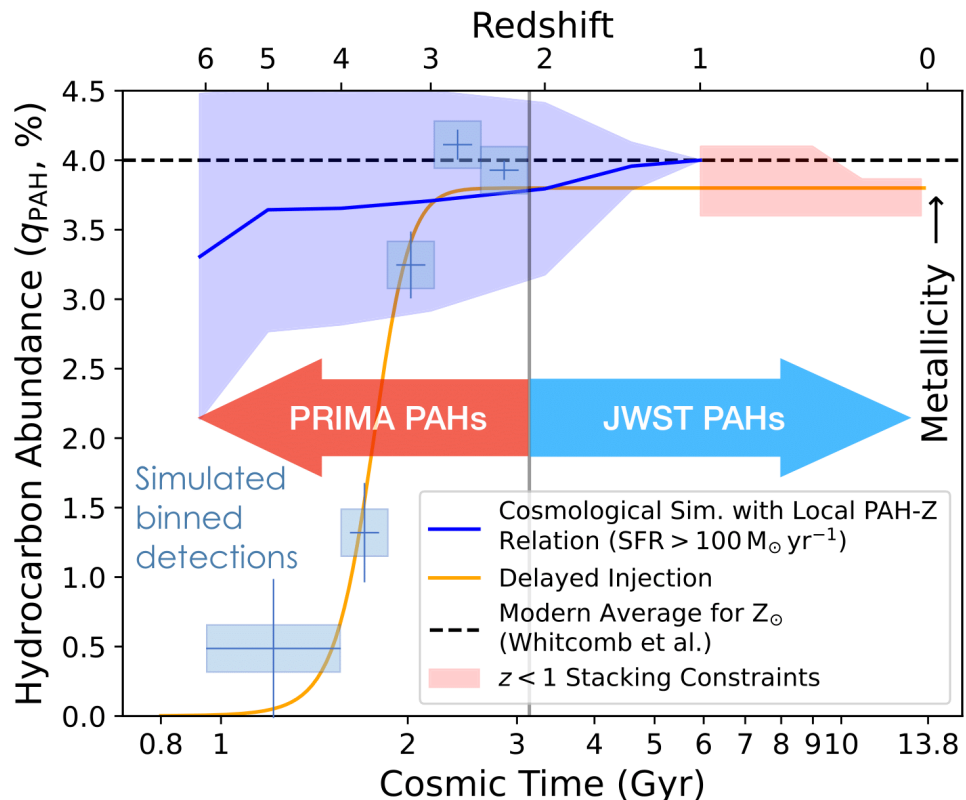

Figure 1: Luminous systems in the early Universe may host hydrocarbon abundances similar to that of local galaxies (black dashed line). Projections from the hydrodynamic simulations of Davé et al. 2019 (SIMBA, blue band), and coarse stacking analyses from Spitzer and Herschel (pink band) are shown. At high redshift, little is known about the emergence of PAHs and whether they will track expected metal abundance (blue) or suffer time-delays due to the stellar evolution injection (orange).

While *Spitzer* has tracked the abundance and conditions of PAHs in the modern Universe, there have been just a handful of detections of PAHs at higher redshifts up to z ~ 4, (Riechers et al. 2014). *JWST* has pushed this even farther, with the current record of PAHs observed in a strongly lensed galaxy at z = 4.2 (Spilker et al. 2023). Where PAH spectra have been observed (z <= 3), the band positions, shapes and strengths appear similar to those of modern galaxies, potentially indicating that processes controlling PAH formation developed much earlier. However, despite JWST's high sensitivity, the red cutoff for *JWST* at 28 μm limits observations of PAHs during the critical first stages of galaxy evolution, as even at z=4.2 only 6% of the power within PAH features is still in the *JWST* spectral window (see Figure 2). Furthermore, the long wavelength sensitivity of JWST has dropped substantially from launch levels >20 μm. Constraining PAH abundance and physical properties at earlier times is not possible with present facilities.

The Astro2020 decadal report tasks far-infrared probes with "*measuring the formation and buildup of galaxies, heavy elements, and interstellar dust from the first galaxies to today.*" PRIMA is uniquely capable of addressing the key related questions: **What are the conditions of dust in the early Universe? When did hydrocarbons emerge in star-forming galaxies?**

PRIMA will be the first facility capable of spectroscopically and photometrically observing PAHs and the processes of dust growth in the first Gyr of the Universe through Cosmic Noon; a transformational time in cosmic history where constraints on dust growth are badly needed. By observing the 7.7 and 11.3 μm bands (rest frame), which account for more than half of PAH power, PRIMA will measure the variation in PAH mass fraction as a function of redshift using the methods of Draine et al. (2021). This will serve as an important indicator for the grain properties and abundance in the early Universe.

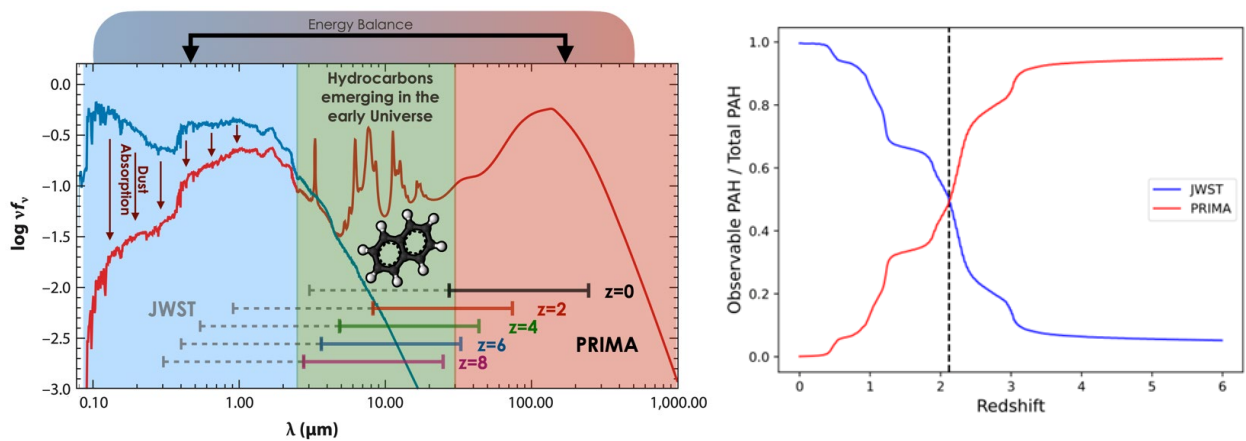

Figure 2: *Left:* The SED of a typical star-forming galaxy, demonstrating the shift of PAH emission into the PRIMA bandpass at z > 2. *Right:* PRIMA's spectral coverage compliments that of JWST. The majority of PAH power shifts into PRIMA at z ∼ 2. By z > 4, PRIMA covers >95% of PAH emission, while JWST bands captures <5%.

## Instruments Required:

- FIRESS low-resolution maps
- PRIMAGer small maps in both bands

## Program Size:

- Large (100 + hr)

## Approximate Integration Time:

- 200 hr

## Special Capabilities Needed:

- None

## Synergies with Other Facilities:

Building on results JWST will deliver at lower redshifts, the PAH mass fraction and dust conditions can be constrained through all of cosmic time back to z > 6 (<1 Gyr). We anticipate producing highly reliable photometric redshifts, but will also use Roman spectroscopy available across the full field for confirmation.

## Description of Observations:

Redshift and PAH band fluxes are recovered directly from PRIMAger hyperspectral SEDs, with full-band FIRESS spectroscopic follow-up for a subset of 10 of the most luminous objects per $\Delta z = 0.5$ redshift bin to precisely retrieve the PAH mass fraction ($q_{PAH}$) and band-ratio dependent properties using (at minimum) the accessible 7.7 and 11.3 µm band fluxes and the rest-frame 20 µm mid-infrared dust continuum (Figure 4). We require detection of both features in the limiting scenario where they are 6 times weaker than in canonical star forming galaxies (corresponding to $q_{PAH}$ = 4.5%). As a bounding case, we require this measurement in a z=4.5 galaxy with log $L/L_\odot$ =13.2. This requires an RMS uncertainty of $3.6\times10^{-20}$ W m$^{-2}$ per spectral resolution element, achievable in 2 hours of integration per module. This sensitivity enables spectroscopy of 50 galaxies in 200 hours.

To measure the variation in mass fraction relative to total dust of PAHs in a sample of luminous star-forming galaxies, we require resolved spectroscopy of the strong PAH bands at 7.7 and 11.3 µm (rest frame). From SED analysis, we identify galaxies in 5 high-redshift bins, based on the hyperspectral surveys. Simulated deep survey yields predict more than 1000 sources detected in bins near z ∼ 2, and more than 1,000 galaxies at z > 4.2, where no PAH emission has ever been measured (Figure 3).

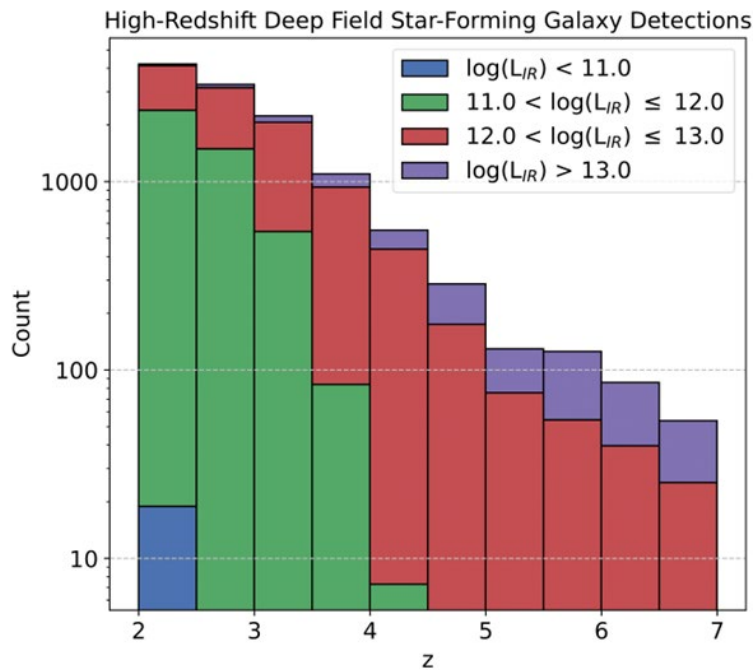

Figure 3: Projections from the *Spritz* model (Bisigello et al, 2021) using current best estimate PRIMAger sensitivities. PRIMA's deep hyperspectral imaging survey will recover PAH band measurements in thousands of sources across a range of luminosities out to z > 7.

FIRESS follow-up spectroscopy will yield high quality, full-band spectra of 50 luminous galaxies; Figure 4 shows two simulated PRIMA/FIRESS detections with realistic noise for two example CIGALE model spectra with low and high PAH abundance.

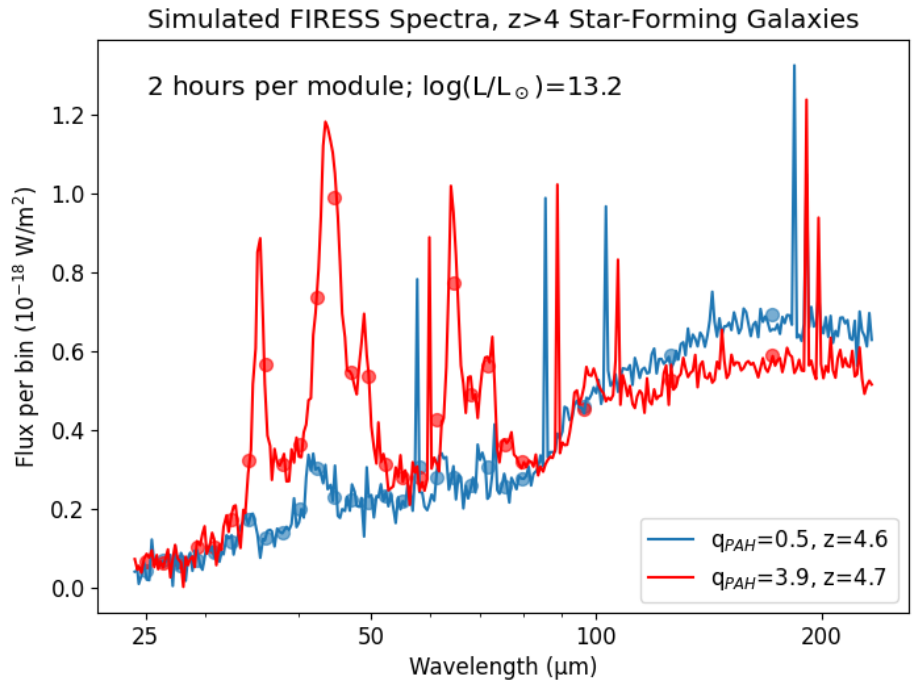

Figure 4: Simulated PRIMA/FIRESS spectra of two luminous galaxies at z > 4.5, with realistic noise model reflecting current best estimate FIRESS sensitivity. The overlaid points illustrate selected R ≥ 10 PRIMAger hyperspectral coverage, which will be available for more than 10,000 galaxies at z > 2.

## 22. Probing the dust properties of z>5 quasars through PRIMA/FIRESS spectroscopy

Eiichi Egami (University of Arizona)

The physical properties of galaxies in the epoch of reionization (EoR; z>5) may be fundamentally different from those of lower-redshift galaxies because dust properties (e.g., composition, grain-size distribution, abundance) could be substantially different at such high redshift. Multiple processes are known for dust production (i.e., AGB stars, core-collapse SNe, and grain growth in the ISM), but it is difficult to quantify their relative contributions. PRIMA/FIRESS will allow us to obtain the rest-frame mid-IR spectra of quasars in the EoR and study their dust properties for the first time. From Herschel studies, we already know that there are ~30 z>5 quasars that are bright enough (>2 mJy at 100 μm) to produce high-quality FIRESS spectra (Tint~1 hr), promising the success of such an observing program.

### General Scientific Area:

Rise of Metals and Dust

### Science Justification:

The physical properties of galaxies in the epoch of reionization (EoR; broadly defined as z>5 here) may be fundamentally different from those of lower-redshift galaxies. This is because basic dust properties (e.g., composition, grain-size distribution, abundance) could be substantially different in galaxies in the EoR. The origin of interstellar dust is a complex subject even today: although it has been known for some time that there are three main dust-production processes (i.e., AGB stars, core-collapse supernovae (CCSNe), and grain growth in the ISM; e.g., Gehrz 1989), it is still difficult to quantify their relative contributions even in the local Universe. In fact, the relative importance of these processes is expected to change as galaxies evolve (e.g., Asano et al. 2013; Ginolfi et al. 2018).

Although some EoR galaxies may be old enough that AGB stars could contribute significantly to the dust production (e.g., the host galaxy of the z=6.4 quasar SDSS J1148+5251; Valiante et al. 2009; Dwek & Cherchneff 2011), this will not be the case with significantly younger EoR galaxies whose ages are too short for the formation of AGB stars (< 200–300 Myr). This suggests that in such young galaxies, CCSNe are expected to play an important role for dust production (e.g., Todini & Ferrara 2001; Nozawa et al. 2003; Schneider et al. 2004; Cherchneff & Dwek 2010; Marassi et al. 2014, 2019).

Furthermore, in some of the most infrared-luminous dusty star-forming galaxies (DSFGs) and quasar hosts in the EoR, their inferred dust masses are so large that these stellar sources alone do not appear to be capable of producing the observed amount of dust in the available time, possibly requiring another ISM-based process such as grain growth in cold, dense molecular gas clouds (Michalowski et al. 2010, Valiante et al. 2011, 2014; Mancini et al. 2015, 2016; Ginolfi et al. 2019; Graziani et al. 2020). Grain growth has been recognized as an important (or even dominant) dust-production process in the local Universe (e.g., Draine 2003), but it should be noted that its effectiveness at high redshift has been seriously questioned by some studies (Ferrara et al. 2016; Ceccarelli et al. 2018).

The main science question is, "What are the properties and origin of the dust surrounding z>5 quasars and therefore those of EoR galaxies in general?" It is not yet clear observationally how the different types of dust-production processes are at work in the EoR, and it is conceivable that the dust properties and therefore the physical properties of many (but not necessarily all) EoR galaxies may be quite different from those observed at lower redshift. With PRIMA/FIRESS, we will be able to probe the dust properties of EoR galaxies for the first time.

Only a sensitive far-IR space mission like PRIMA will allow us to obtain rest-frame mid-IR spectra of galaxies and quasars in the EoR and study their dust properties.

Different types of dust would produce different spectral signatures in the rest-frame mid-IR, and PRIMA/FIRESS will be especially powerful for probing such signatures. Figure 1 shows simulated FIRESS spectra (Tint= 1 hr) of quasars at z=6 using the dust spectral models derived to fit the observed Spitzer IRS spectra of three low-redshift (z=0.07–0.24) PG quasars (Xie et al. 2017). Because of the different mixtures of silicate and graphite dust grains, the mid- IR spectra of quasars show a large variation even at such low redshift. The 9.7/18 μm silicate emission features seen in the figure are particularly important, allowing us to probe the composition (e.g., Draine's "astronomical" silicates vs. Ossenkopf's "cosmic" silicates) and geometry (e.g., clumpiness) of dust surrounding the central AGN (e.g., Hatziminaoglou et al. 2015).

What is important here is that these quasar model spectra are scaled to 5 mJy at 100 μm, which is a typical brightness of ~30 bright quasars at z=5–6.4 observed by Herschel/PACS (Leipski et al. 2014). In other words, the quality of the simulated FIRESS spectra presented here should be highly realistic. Obviously, the actual rest-frame mid-IR spectra of z>5 quasars may be very different from these models, but Figure 1 clearly demonstrates that FIRESS will deliver high-quality spectra for bright z>5 quasar targets, allowing us to discriminate various dust models and to probe the formation process of dust in the EoR.

Many z>5 quasars are also known to reside in vigorously star-forming (SF) host galaxies, which will likely produce PAH features that are detectable by FIRESS. The QSO+SF model in Figure 1 (the red line) predicts that this will likely be the case, allowing us to probe the star-forming properties of z>5 quasar hosts as has been done for PG quasars with Spitzer IRS (e.g., Shi et al. 2007, 2009). In some z>5 quasars, we may also detect strong molecular hydrogen ($H_2$) lines because high-J CO lines are often detected in these quasars, revealing the presence of a significant amount of warm $H_2$ gas. Most notably, the CO (17-16) emission line has been detected with the z=6.4 quasar SDSS J1148+5251 mentioned above, with a corresponding molecular gas temperature of 750 K (Gallerani et al. 2014).

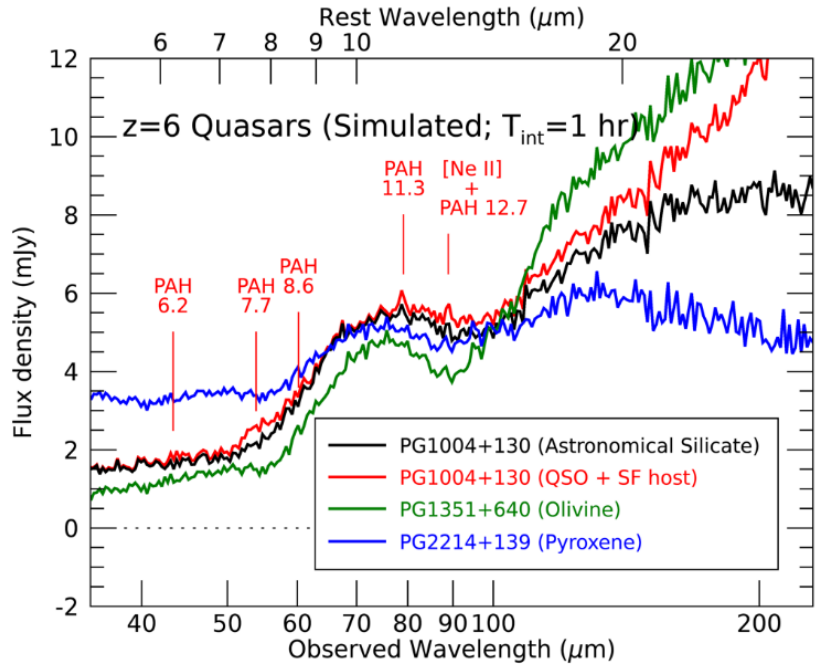

Figure 1: Simulated PRIMA/FIRESS low-resolution (R=130) spectra of quasars at z=6 (Tint=1hr) using the dust spectral models from Xie et al. (2017) derived to fit the observed Spitzer IRS spectra of three low-redshift PG quasars, PG1004+130 (z=0.240; astronomical silicate), PG1351+640 (z=0.087; olivine), and PG2214+139 (z=0.067; pyroxene). The model spectra were scaled to 5 mJy at 100 μm, a typical brightness of ~30 bright quasars at z=5–6.4 observed by Herschel PACS (Leipski et al. 2014). A LIRG spectrum ($L_{IR}=10^{11.75}$ $L_{\odot}$; Rieke et al. 2009) scaled to $L_{IR}=2 \cdot 10^{13}$ $L_{\odot}$ was also added to the PG1004+130 spectrum to illustrate the strengths of the PAH features produced by such a star-forming host galaxy. For such bright z>5 quasars, FIRESS will deliver high-quality rest-frame mid-IR spectra, allowing us to study their dust and star-forming properties in detail. Here, a line sensitivity of $2\times10^{-19}$ W m$^{-2}$ (5σ, 1 hour) was assumed over the whole spectral range.

## Instruments Required:

PRIMA FIRESS Spectrometer: Pointed Low-res R~130

## Program Size:

- Medium (10–100 hr)

## Approximate Integration Time:

- ~1 hr per target for the brightest z>5 quasars, assuming a line sensitivity of ~2×10⁻¹⁹ W m⁻² (5σ, 1 hour, R=130) as currently estimated for FIRESS.

## Special Capabilities Needed:

- None

## Synergies with Other Facilities:

- JWST and ALMA in combination will allow us to obtain the full multi-wavelength view of z>5 quasars from the rest-frame UV to millimeter.

## Description of Observations:

- Pointed FIRESS low-resolution spectroscopy of known z>5 quasars.

# 23. Probing the physical properties of galaxies in the epoch of reionization through gravitational lensing

Eiichi Egami (University of Arizona)

The ALMA dust continuum detections of a few z∼8 galaxies strongly signal that infrared-luminous (i.e., dusty) star-forming galaxies play a much more important role at z>5 than originally thought. Furthermore, recent observational and theoretical studies suggest that the dust temperature of galaxies at z∼8 (and more generally in the EoR) may be significantly higher and similar to (or even hotter than) those of local low-metallicity blue compact galaxies (BCDs). These recent findings make PRIMA a highly critical mission for studying galaxies in the Epoch of Reionization (EoR). Although such high-redshift galaxies are faint, this can be overcome by targeting gravitationally lensed galaxies. **Lensed galaxies with a magnification factor of ∼10x are often found, and observing these galaxies would make the D=2 m PRIMA effectively a D=6.3 m far-IR telescope in space, enabling sciences that were envisaged for the Origins Space Telescope.** There are indeed a dozen hyper-luminous ($L_{IR} \gtrsim 10^{13} L_\odot$) infrared galaxies already known at z=5-7 (mostly gravitationally lensed) that should produce good-quality FIRESS spectra ($T_{int} \sim 10$ hr), and we expect the number of such high-redshift lensed galaxies to increase significantly over the coming decade. This will allow PRIMA/FIRESS to investigate the physical properties of EoR galaxies through various rest-frame mid-IR diagnostics.

## General Scientific Area:

Growth of Galaxies and Black Holes Through Cosmic Time

## Science Justification:

### Broader context

**(1) IR-luminous dusty star-forming galaxies (DSFGs) at z>5:** ALMA dust continuum detections with two Lyman break galaxies (LBGs) at z=8.3, A2744_YD4 at z=8.38 (Laporte et al. 2017) and MACS0416_Y1 at z=8.31 (Tamura et al. 2019), have shown that a significant amount of dust already existed at such high redshift. These results are also consistent with the ALMA stacking analysis of LBGs at z∼7 showing an average infrared luminosity of ∼2・$10^{11}$ $L_\odot$ (Bowler et al. 2018). At the same time, wide-field submillimeter/millimeter surveys (e.g., with Herschel and the South Pole Telescope) have detected more than a dozen hyper-luminous DSFGs ($L_{IR} \gtrsim 10^{13}$ $L_\odot$; mostly gravitationally lensed) at z>5 and up to z=6.9 (Strandet et al. 2017; Marrone et al. 2018). On the theoretical side, there is also a suggestion that giant molecular clouds (GMCs) at high redshift are more compact with a significantly longer dispersal time, possibly absorbing ∼40% of the total UV radiation produced by newly born stars (Sommovigo et al. 2020). All these observational and theoretical studies in recent years strongly suggest that infrared-luminous (i.e., dusty) star-forming galaxies play a much more important role at z>5 than originally thought (e.g.,

Mancini et al. 2015, 2016; Graziani et al. 2020), making PRIMA a highly critical mission for studying galaxies in the Epoch of Reionization (EoR: broadly defined as z>5 here).

**(2) High dust temperature of z~8 galaxies:** Recent observational and theoretical studies also suggest that the dust temperature of galaxies at z~8 (and more generally in the EoR) may be significantly higher and similar to (or even hotter than) those of local low-metallicity blue compact galaxies (BCDs). For example, the ALMA follow-up observations of the z=8.3 LBG MACS0416-Y1 mentioned above by Bakx et al. (2020) indicate that in this galaxy, the dust temperature would have to be > 80 K with a dust emissivity of β < 2. Though extreme, such a high dust temperature is not totally unexpected by theoretical models due to the combination of the intense interstellar radiation field and compact GMC structures expected in such a high-redshift galaxy (e.g., Behrens et al. 2018; Sommovigo et al. 2020). High dust temperature will also result from silicate-rich dust produced by core-collapse SNe (CCSNe), which has a high emission efficiency at 8-60 µm (De Rossi et al. 2018). This would make EoR galaxies more luminous in the rest-frame mid-IR, making PRIMA effective for detecting and studying them.

## Science question

The main science question is, "What are the physical properties of EoR galaxies as compared to those of lower-redshift ones (e.g., high dust temperature)?" If some of the EoR galaxies are affected by a significant amount of dust extinction, PRIMA/FIRESS spectroscopy will be crucial. Also, the rest-frame mid-IR spectral range offers information that is not available at other wavelength ranges, such as dust composition (e.g., strength of the PAH emission) and dust temperature.

## Need for PRIMA

Only a sensitive far-IR space mission like PRIMA will allow us to obtain rest-frame mid-IR spectra of galaxies in the EoR.

## Interpretation methods

**(1) IR-luminous dusty star-forming galaxies (DSFGs) at z>5:** For some of the hyper-luminous lensed DSFGs at z>5 ($L_{IR} \gtrsim 10^{13}$-$10^{14} L_\odot$), their infrared SEDs have been accurately measured by Herschel, and local LIRG SEDs were often found to make good fits (e.g., Rawle et al. 2014). Figure 1a simulates the 10-hr PRIMA/FIRESS spectrum of a DSFG at z=6 using the $L_{IR}=10^{11.75} L_\odot$ LIRG SED template (Rieke et al. 2009) scaled to $L_{IR}=2 \cdot 10^{13} L_\odot$ (the black line). The figure shows that if the target galaxy is this luminous, FIRESS will deliver good-quality spectra that are comparable to those of Spitzer/IRS but at much higher redshift.

Although bright LBGs in the EoR may also have LIRG-like mid-infrared spectra (as implied by their LIRG-like infrared luminosities estimated by Bowler et al. 2018), it has also been suggested (e.g., Lyu et al. 2016; De Rossi et al. 2018) that the SEDs of some high-redshift galaxies and quasar hosts are more similar to that of the local low-metallicity blue compact dwarf (BCD) Haro 11 ($L_{IR}=2 \cdot 10^{11} L_\odot$; Z=1/3 $Z_\odot$). This agreement may be fortuitous in the sense that all it means is that some EoR galaxies have a high dust temperature, which does not necessarily require a low metallicity. With this caveat in mind, we also used the Haro 11 IRS spectrum (e.g., Hunt et al. 2005, Wu et al.

2006) to simulate the FIRESS spectrum of a low-metallicity LBG at z=6 (Figure 1a, the red line; also scaled to $L_{IR}=2 \cdot 10^{13}$ $L_{\odot}$). Note the substantial brightness increase in the rest-frame mid-infrared due to a higher dust temperature. If the galaxy is truly of low-metallicity, FIRESS will see weak PAH features and strong higher ionization lines (e.g., [Ne III] 15.5 μm and [S IV] 10.5 μm) due to a harder UV radiation field as shown in Figure 1a.

(2) High dust temperature of z~8 galaxies: If high dust temperature is prevalent among EoR galaxies, it will substantially increase PRIMA's chance for detecting galaxies at z~8 because higher dust-temperature galaxies are more luminous in the rest-frame mid-infrared. Figure 1b shows simulated FIRESS spectra for z=8.3 galaxies using three spectral models: (1) the $L_{IR}=10^{11.75}$ $L_{\odot}$ LIRG spectral template scaled to $L_{IR}=2 \cdot 10^{13}$ $L_{\odot}$, (2) the observed Haro 11 spectrum scaled to $L_{IR}=3.6 \cdot 10^{12}$ $L_{\odot}$, and (3) the theoretical model by Behrens et al. (2018) that fits the intrinsic (i.e., unlensed) SED of A2744_YD4, also scaled to $L_{IR}=3.6 \cdot 10^{12}$ $L_{\odot}$. The original infrared luminosities of these model/observed spectra are $10^{11.75}$, $2 \cdot 10^{11}$, and $3.6 \cdot 10^{11}$ $L_{\odot}$, so we are essentially boosting these spectra by factors of 36, 18, and 10, respectively, to make them detectable with FIRESS. With such scaling, all three models produce comparably bright mid-infrared spectra (Figure 1b) although the infrared luminosities of the latter two are a factor of ~5 smaller. This is because the SEDs of the latter two models are significantly warmer (i.e., of higher temperature), releasing more luminosity in the mid-infrared (Figure 1c). What this means is that with warm SEDs like those of Haro 11 and the Behrens 2018 model, FIRESS will be able to obtain good-quality spectra for z~8 galaxies for galaxies with infrared luminosities below $10^{13}$ $L_{\odot}$.

### Link to testable hypotheses

**"Are EoR galaxies similar to blue compact dwarf galaxies (BCDs) at low redshift in terms of their physical properties, showing signs of a low metallicity, harder UV radiation field, and warm dust temperature?"**

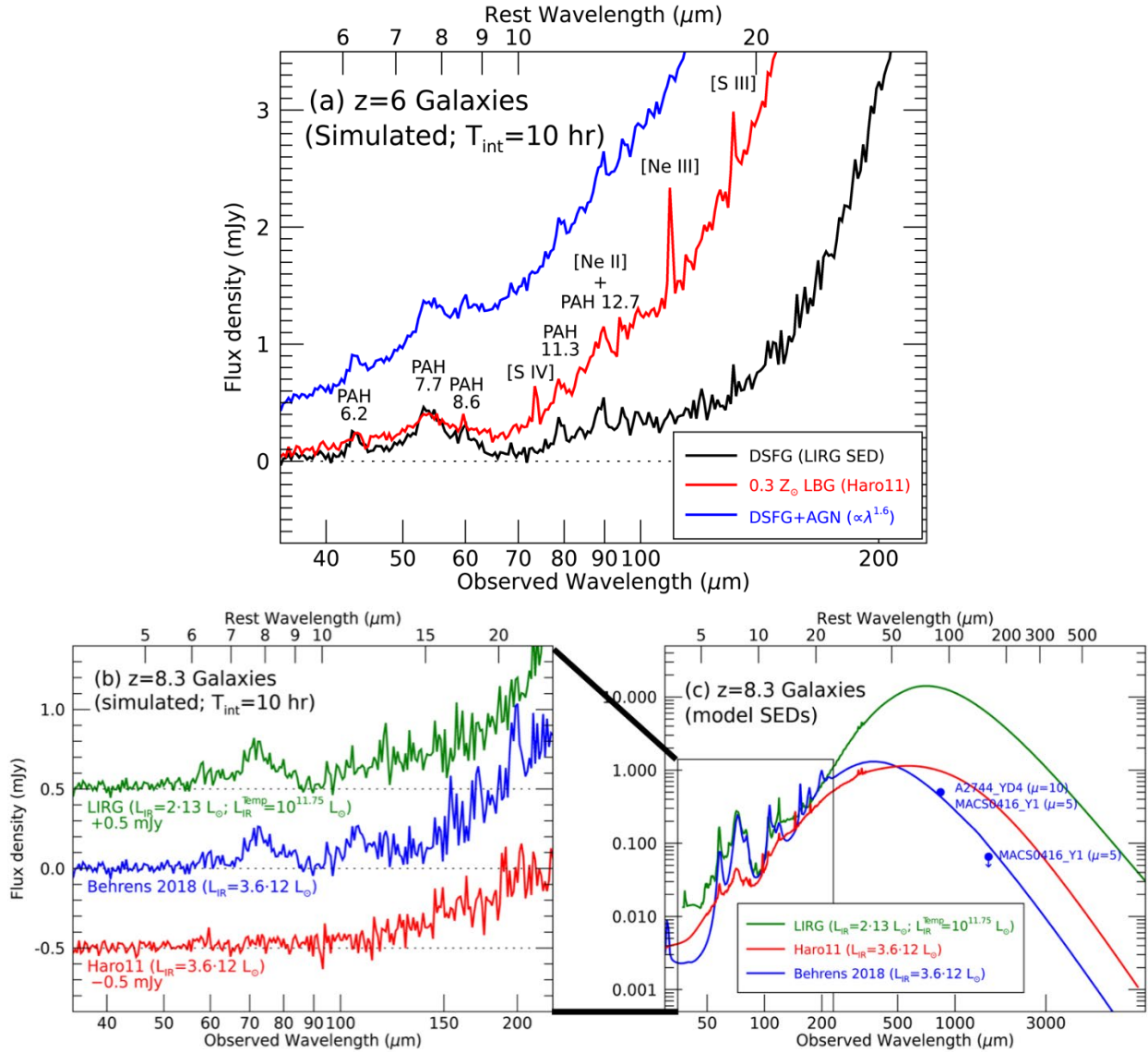

Figure 1: (a) Simulated PRIMA/FIRESS low-resolution (R=130) spectra of star-forming galaxies at z=6 (Tint=10 hr) using the spectral template of a LIRG ($L_{IR}=10^{11.75}$ $L_\odot$; Rieke et al. 2009) as a DSFG (the black line) and the observed Spitzer IRS spectrum of Haro 11 (Hunt et al. 2005; We et al. 2006) as a low-metallicity (0.3 $Z_\odot$) LBG (the red line). Both SEDs are scaled to $L_{IR}=2 \cdot 10^{13}$ $L_\odot$. The former produces a good fit to the observed far-infrared SED of the lensed DSFG at z=5.24 HLSJ0918 (Combes et al. 2012; Rawle et al. 2014) while the latter is taken as a good analog of young low-metallicity galaxies at high redshift (De Rossi et al. 2018). An AGN power-law continuum (e.g., $\propto \lambda^{1.6}$; Mullaney et al. 2011) was also added to the DSFG spectrum to illustrate how it changes the spectrum of a star-forming galaxy (the blue line). The figure clearly shows that FIRESS will deliver good-quality spectra for sufficiently luminous galaxies ($L_{IR}\gtrsim10^{13}$ $L_\odot$; intrinsically or gravitationally lensed); (b) Simulated FIRESS spectra of z=8.3 galaxies using three spectral models: (1) the $L_{IR}=10^{11.75}$ $L_\odot$ LIRG spectral template scaled to $L_{IR}=2 \cdot 10^{13}$ $L_\odot$, (2) the observed Haro 11 spectrum scaled to $L_{IR}=3.6 \cdot 10^{12}$ $L_\odot$, and (3) the theoretical model by Behrens et al. (2018) that fits the intrinsic SED of A2744_YD4, also scaled to $L_{IR}=3.6 \cdot 10^{12}$ $L_\odot$.; (c) The rest-frame far-IR part of the three model SEDs are compared. The Behrens et al. model is broadly consistent with the ALMA measurements (the blue solid circles) of A2744_YD4 (z=8.38) and MACS0416_Y1 (z=8.31) if we scale up their IR luminosities by factors of 10x and 5x, respectively. A line sensitivity of 2×10⁻¹⁹ W m⁻² (5σ, 1 hour) was assumed for the simulation of the FIRESS spectra.

**Instruments Required:**

- PRIMA FIRESS: Pointed Low-res, R~130
- PRIMA FIRESS: Small Map Low-res, R~130

**Program Size:**

- Large (100 + hr)
- For example, a 120-hr program will allow us to obtain FIRESS spectra of almost all the known hyper-luminous IR galaxies ($L_{IR} \gtrsim 10^{13} L_\odot$) currently known at z>5 (12 targets).

**Approximate Integration Time:**

- ~10 hr per target for the brightest z>5 galaxies, assuming a line sensitivity of ~$2\times10^{-19}$ W m$^{-2}$ (5σ, 1 hour, R=130) as currently estimated for FIRESS.

**Special Capabilities Needed:**

- None

**Synergies with Other Facilities:**

- JWST and ALMA in combination will allow us to obtain the full multi-wavelength view of z>5 galaxies from the rest-frame UV to millimeter.
- In combination, various upcoming wide-field surveys (e.g., LSST, Euclid, Roman, and those planned with ground-based submillimeter/millimeter facilities) will likely find a significant number of exceptionally bright lensed galaxies at high redshift, which will serve as great targets for PRIMA/FIRESS.

**Description of Observations:**

- Pointed FIRESS low-resolution spectroscopy of known z>5 galaxies with exceptional brightness.
- FIRESS mapping of lensing clusters can also be considered.
- One way to increase the sensitivity of PRIMA by a large factor is to target gravitationally lensed galaxies. Lensed galaxies with a magnification factor of ~10× are often found in the field (mostly galaxy-lensed systems), and those with a magnification factor of ~20–30× are common in the fields of massive galaxy clusters (cluster-lensed systems). A magnification factor of 10× would directly translate into an increase of the signal-to-noise ratio by a factor of 10x, which is equivalent to increasing the telescope aperture by a factor of $\sqrt{10}$x. Here, note that for a far-infrared telescope of modest size, high-redshift galaxies will appear as diffraction-limited point sources, so the S/N will increase as D$^2$. Therefore, when observing a lensed galaxy with a magnification factor of 10×, the D = 2 m PRIMA would effectively become a D = 6.3 m (=2x$\sqrt{10}$) far-infrared telescope in space, allowing us to conduct many extragalactic studies planned for the Origins Space Telescope (OST).

- **Extensive spectroscopic survey of galaxies in the cosmic noon (z ∼ 1–4):** Although the main emphasis of this science case is on EoR galaxies, spectroscopy of lensed galaxies is also powerful for studying galaxies at lower redshift. In fact, the majority of gravitationally lensed galaxies discovered so far resides in the redshift range of z = 1–4, which is often referred to as the cosmic noon, when galaxies and their central black holes were most active. By targeting lensed galaxies, we will be able to detect faint spectral features that are inaccessible otherwise. Detection of faint high-excitation lines from AGN, for example, will help us investigate the coevolution of star formation and black-hole accretion. We can also use the gain in sensitivity to increase the sample size (or reduce the observing time).

# 24.  PRIDES – The PRIMA Imaging Deep Extragalactic Survey

A. Faisst (Caltech/IPAC), C.-C. Chen (ASIAA), L. Ciesla (LAM), C. Gruppioni (INAF Bologna)

The cosmic evolution of obscured star formation, dust properties and production mechanisms, and the prevalence of dust-obscured AGN out to high redshifts is currently one of the hot topics in astrophysics. While wide-field surveys with Spitzer and Herschel have not reached the necessary depth to observe the mid-IR light of high-redshift (z > 3) galaxies, targeted surveys with ALMA are biased towards UV-detected and less dust obscured galaxies. To progress our understanding of star formation, dust, and AGN activity in the early universe, deep wide-field blind surveys are needed. We propose the PRIMA Imaging Deep Extragalactic Survey (PRIDES), a 1.6 deg$^2$ survey with the main science goals to (i) characterize the optical dark galaxy population in the early Universe, (ii) constrain the obscured star formation and dust properties across cosmic time, and (iii) quantify the prevalence of dust-obscured AGN before the peak of cosmic star formation density. PRIDES is situated on the COSMOS field, one of the best studied extragalactic fields providing deep UV to sub-mm/radio observations from ground-based (including SCUBA2, ALMA, LOFAR) and space-based (including HST, Spitzer, Herschel, and JWST/NIRCam+MIRI over 40% of the PRIDES footprint) facilities. It is designed to observe the mid-IR continuum of typical main-sequence galaxies at z = 3 at stellar masses of >3×10$^{10}$M$_\odot$ (L$_{IR}$ > 10$^{12}$L$_\odot$). The large sky coverage of PRIDES allows the study of the above science goals over a large range of galaxy properties and depending on large-scale structure environments. All in all, PRIDES will provide the first complete and unbiased deep survey of star-formation, dust, and AGN occurrence in the early Universe in addition to a wealth of infrared legacy observations.

## General Scientific Area:

Growth of Galaxies and Black Holes Through Cosmic Time

## Science Justification:

**Motivation.** Studying the spectral energy distribution (SED) of galaxies at infrared (IR) wavelengths is crucial for understanding galaxy formation and evolution: the total IR luminosity probes dust-obscured star formation, and various silicate and polycyclic aromatic hydrocarbon (PAH) spectral features allow us to study dust grain properties and the prevalence of dust-obscured AGN. Recently, wide-field targeted and blind-field surveys with ALMA, such as ALPINE (z = 4–6; LeFèvre et al. 2020), REBELS (z > 6.5; Bouwens et al. 2022), and (Ex-)MORA (z = 3–6; Casey et al. 2021), have shown considerable dust obscuration even in the early Universe at z > 3 (e.g., Gruppioni et al. 2020; Fudamoto et al. 2020, 2021; Talia et al. 2021, Inami et al. 2022), which was surprising. More than 70% of the light of galaxies at z ∼ 3 is reprocessed by dust (i.e., emitted in the IR) and about 20% of galaxies are completely missed in the UV (Bouwens et al. 2009; Franco et al. 2018; Wang et al. 2019). Some of these "optical dark" galaxies are detected in recent JWST imaging, but most are too faint to derive robust redshifts. Unfortunately, all the above ALMA surveys cover only a small area on sky and provide only one data point at IR wavelengths at various redshifts. With currently no constraints on the shape of IR SEDs of z > 3 galaxies (except some constraints from stacking – Schreiber et al. 2017, 2018; Béthermin et al. 2015, 2020), we are completely ignorant about dust temperature, the general dust properties, and obscured AGN contribution, resulting in significant uncertainty in almost any properties derived from the IR SED (Faisst et al. 2017, 2020). Large statistical samples are furthermore needed to study the dust properties and production mechanisms across cosmic time. On another front, JWST recently confirmed AGN via rest-frame UV and optical emission lines in early galaxies out to z ∼ 8 (e.g., Ding et al. 2022; Larson et al. 2023), indicative of accreting massive black holes which are little dust obscured. Dust-obscured AGN remain hidden at UV/optical wavelengths and even in the X-rays (Carroll et al. 2023), and mid-IR observations are crucial for their identification. AGN likely have a significant impact on galaxy evolution, as they are thought to be coupled with the quenching of star formation (Dubois et al. 2013). These studies lead to several questions: *(i) What is the fraction of obscured star formation across cosmic time? (ii) How is dust produced and how do its properties change over time? (iii) What is the prevalence of dust-obscured AGN (missed in rest-UV/optical) and how do they shape galaxy evolution at early stages? (iv) How do the above parameters change as a function of location on the main sequence and in different environments across cosmic time?*

**This proposal.** To study the above questions, we need a full coverage of the IR SED for large samples of galaxies across cosmic time – especially at high redshifts, z > 3, where so far little IR data exist. Samples need to cover various stages of galaxy formation, including different spatial environments. Data from past IR facilities (e.g., Spitzer and Herschel) are not deep enough to study typical main-sequence galaxies in the early Universe except through stacking. Furthermore, at these redshifts, ALMA covers wavelengths redward of the IR SED peak ($\lambda_{rest} \sim 100$ µm) as its transmission drops significantly at higher frequencies covering the mid-IR. On the other hand, JWST does not go red enough. In both cases, the small FoV / beam size makes large-area surveys prohibitively expensive. PRIMA is perfectly suited to fill in the gap in wavelength and sensitivity of current facilities, which would lead to breakthrough results in the study of dust and AGN of galaxies in the early Universe. We propose the PRIMA Imaging Deep Extragalactic Survey (PRIDES), designed to observe the rest-frame mid-IR of typical galaxies out to z=4 over an area of 1.6 deg$^2$, large enough to cover various galaxy properties and environmental regimes. PRIDES is observed at the PRIMAger 25–80 µm wavelength range to a depth of ∼70 µJy and is driven by the following two main science goals:

- Study the obscured SFR density as well as dust properties and production in typical galaxies out to z = 4 (and to higher redshifts for more sub-mm bright galaxies) by constraining the (mid-) IR SED including silicate and PAH spectral features.

- Study the prevalence of dust obscured AGN with strong mid-IR emission and their relation to galaxy physical properties and large-scale environment at high redshifts.

**PRIMA is crucial to pursue the above science goals:** neither JWST nor ALMA can observe the rest-frame mid-IR (silicate features to IR SED peak) of z > 3 galaxies. A large FoV and fast mapping speed is crucial for obtaining a large survey area to (i) cover the Universe's large-scale structure and diversity of environments, (ii) probe a variety of galaxy properties (from main-sequence to starburst galaxies), and (iii) pursue a true flux-limited blind survey to mitigate various biases introduced by targeted surveys (e.g., UV selection bias).

**Field, sensitivity, and time request.** Targeted surveys of galaxies selected at rest-frame UV and optical wavelengths are inherently biased against dusty sources. Only a blind survey over a large area provides a study of the whole population of galaxies with minimal biases. We propose to cover a 1.6 deg$^2$ area centered on the COSMOS field (Scoville et al. 2007), one of the best studied extragalactic fields to-date. It provides a wealth of deep UV to near-IR photometry and spectroscopy, complemented with some of the deepest Spitzer and Herschel observations. The COSMOS-Web JWST cycle 1 program (Casey et al. 2022) observes more than 40% of the PRIDES footprint with JWST/NIRCam 1–5 μm imaging (completion end of 2023). Additional observations at lower frequencies with ALMA, SCUBA-2, and LOFAR are available or being taken. Furthermore, Euclid, Roman, and SPHEREx will cover the COSMOS field in the future in the near-IR. The area covered by PRIDES is comparable to the scale of dark matter density fluctuations (Figure 1), thus ideal to study environmental dependencies and to minimize cosmic variance. The requested sensitivity is driven by the requirement to detect the continuum of typical z > 3 galaxies at M$_*$ ∼ $3 \times 10^{10}$ M$_\odot$ in the mid-IR at ∼70μJy (Figure 2). **In total, we request 200h of observation time.**

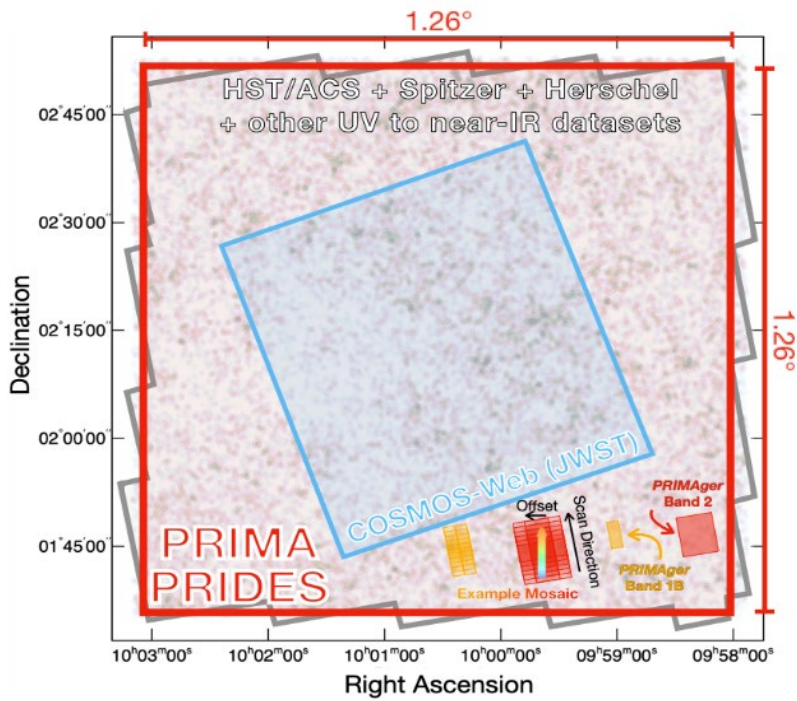

Figure 1: Layout of the 1.6 deg$^2$ PRIDES observations (red) on COSMOS. The COSMOS-Web JWST survey is indicated in blue. The gray outline marks the coverage of the COSMOS HST/ACS observations. COSMOS is observed with a wealth of other facilities from UV to sub-mm. The background cloud shows the large-scale structure of galaxies at 3 < z < 4 (from the Galacticus simulation, Benson et al. 2010). The PRIDES area is optimal to cover various environmental properties including voids and overdensities. PRIMAger footprints for Band 1 (orange) and Band 2 (red) are shown in the lower right.

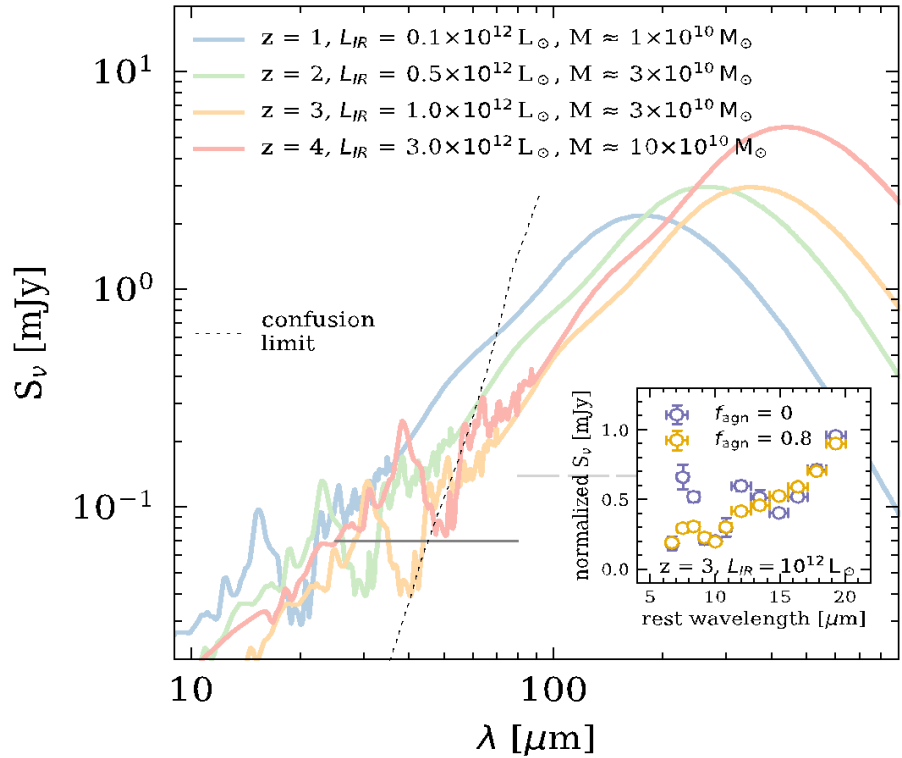

Figure 2: Requested PRIDES sensitivity limits (Band 1, dark gray; Band 2, light gray). The different SEDs show approximate detection limits in total IR luminosity for galaxies at different redshifts. Expected numbers of detections are listed in Table 1. The inset shows simulated photometry in Band 1 of a z = 3 galaxy at for 0% and 80% AGN contribution (from Kirkpatrick et al. 2015) to visualize the measurable changes in the silicate spectral features.

**Science goals.** The main survey goal of PRIDES is to map the mid-IR regime with prominent silicate features ($\lambda_{rest} \sim 10\ \mu m$) of 100s of z > 3 galaxies over a large area to study dust, obscured star formation, and the prevalence of dust-obscured AGN in the early Universe. Imaging makes it possible to study the continuum and broad PAH/silicate emission/absorption through template fitting at R ~ 10 to greater depth than spectroscopy (inset Figure 2). PRIDES will:

- **Measure redshifts of 100s of optical dark galaxies** so far only detected in ALMA and the reddest JWST NIRCam bands by adding important mid-IR coverage. This will be crucial to trace the cosmic dust-obscured star-formation density at z > 3, which currently is uncertain because of unreliable distance measurements.

- **Measure obscured star formation and dust properties** via the abundance of PAH emission and silicate absorption/emission features for the first and most complete census of dust build-up and star formation of the galaxy population at z > 3. The large survey volume will enable to probe a complete sample in terms of galaxy properties and large-scale structure environment at stellar masses of $>3 \times 10^{10} M_\odot$ (see Table 1).

- **Study the prevalence of dust-obscured AGN in the early Universe** by measurements of spectral features and the mid-IR slope to derive the first unbiased census of dust-obscured AGN in typical z > 3 galaxies. Dust-obscured, Compton-thick AGN will be missed by rest-frame UV/optical emission line indicators as used by JWST (and even in X-rays), but they have a significant impact on the evolution of galaxies at early times.

Table 1: Estimated limits and number of galaxies for PRIDES. Number densities are estimated using the current COSMOS2020 catalog (Weaver et al. 2022).

| Redshift | $L_{IR}$ ($L_\odot$) | SFR ($M_\odot yr^{-1}$) | $M_*$ ($M_\odot$) | Number |
|---|---|---|---|---|
| z~1 | >$10^{11}$ | > 10 | >$10^{10}$ | ~18,000 |
| z~2 | >$0.5\times10^{12}$ | > 50 | >$3\times10^{10}$ | ~4,000 |
| z~3 | >$10^{12}$ | > 100 | >$3\times10^{10}$ | ~700 |
| z~4 | >$3\times10^{12}$ | > 300 | >$10^{11}$ | ~30 |

PRIDES will enable a wealth of legacy science in synergy with other facilities at low and high redshifts, including the characterization of extreme sub-mm galaxies (e.g., GN20 at z = 4; Pope et al. 2005) or the observation of IR lines in metal-poor galaxies at low redshifts.

## Instruments Required:

PRIMAGer large map in band 2, 1.26 × 1.26 deg² area

## Program Size:

Large (100 + hr)

## Approximate Integration Time:

Approximately 200h to cover a 1.6 deg² field to a 5σ depth of ~70 μJy. This includes 64% efficiency but no other overheads (which are supposed to be minimal).

## Special Capabilities Needed:

None

## Synergies with Other Facilities:

The COSMOS collaboration has obtained deep JWST NIRCam imaging in F115W, F150W, F277W, and F444W as well as MIRI F777W. PRIDES is designed to cover these JWST observations, which will be one of the main synergies. In addition, COSMOS has been imaged in a wealth of UV to near-IR bands (including Spitzer and Herschel) and provides over 70,000 spectroscopic measurements. ALMA observations (either targeted or blind field) also exist (and some are ongoing) for a large fraction of the COSMOS field. In the future, the COSMOS field will be observed by Euclid (for which it serves as calibration field), Roman, and SPHEREx (which will provide 0.7–5 μm R > 40 spectroscopy at similar resolution as PRIMA). We expect significant synergy with PRIDES and these future surveys. Follow-up of interesting sources observed in PRIDES blind survey with JWST and ALMA as well as other telescopes will be carried out.

## Description of Observations:

The goal of PRIDES is to survey a total area of 1.6 deg², covering the HST/ACS footprint on the COSMOS field (Figure 1), with the PRIMAger instrument to observed typical (main-sequence) z>3 galaxies with total IR luminosities of >10¹²L_⊙ (corresponding of a SFR of >100 M_⊙ yr⁻¹ or a stellar mass of >3 × 10¹⁰M_⊙) in Band 1. As shown in Figure 2, this requirement is satisfied for typical observed 50–100 μm fluxes on the order of 70 μJy (see also Faisst et al. 2020; Talia et al. 2021; Gruppioni et al. 2020; McKinney et al. 2023). Assuming an average sensitivity of 250 μJy (5σ) in 10 hours over 1 deg² (PRIMA factsheet, including 64% efficiency), **we request a total of 200 hours**.

We note that the requested depth is at the confusion limit in the PRIMAger Band 1 (see Béthermin et al. Technical Notes), complementing less deep but more wide-field surveys with PRIMA. We will use state-of-the-art deblending methods (e.g., Béthermin et al. 2015; Weaver et al. 2022), making use of deep near-IR and IR imaging for flux and positional priors to push PRIMA's effective sensitivity limits to the maximum. Specifically, we will use the existing deep UltraVISTA 2.2 μm Ks-band (~25.2 AB, 3σ) and Spitzer 3.6 μm and 4.5 μm (~26.3 AB, 3σ) imaging as well as the blue PRIMAger bands for positional priors. In the COSMOS-Web footprint (covering 40% of the PRIDES footprint), we will in addition use deep JWST/NIRCam F227W and F444W imaging (~28.5 AB, 5σ). Band 2 will be observed simultaneously for free (including polarimetry), but these observations will suffer higher confusion. Deblending may be possible for bluer wavelengths of Band 2. Moreover, depending on the observation strategy (i.e., number of frames per exposure and number of scans across the field) and data reduction (e.g., reducing half of the frames), the impact of confusion could be reduced in Band 2.

Spectroscopy is not requested, as the broad PAH and silicate spectral features and the mid-IR slope can be measured well with model SED fitting at the resolution of the photometry channels (see inset in Figure 2 for an example of a z ~ 3 galaxy at the survey sensitivity limit). Redshifts of dust-obscured galaxies can be confirmed directly using the silicate spectral features using template fitting (resulting in redshifts accuracies comparable with those from narrow-band emission line surveys). If no such features are present, the additional continuum point will greatly enhance the redshift accuracy compared to one single sub-mm observations (in the case of optical dark galaxies).

To cover the requested area, the observations will have to be carried out in mapping mode (scanning in wavelength direction and spatial offsets in perpendicular direction, see Figure 1). We do not constrain the orientation angle to maximize scheduling of the observations. Ideally, the observations should be taken at the same orientation angle to maximize consistency over the field and minimize data reduction efforts (e.g., PSF orientation, deblending, etc.). To achieve a continuous coverage in PRIMAger Band 1A (which covers the smallest FoV), the other filters will overlap, resulting in an increased depth.

# 25. Unveiling rapid and obscured quenching processes in galaxies

K. Decker French (University of Illinois), J.D. Smith (University of Toledo), Adam Smercina (University of Washington)

We propose a PRIMA program to target recently quenched post-starburst galaxies using the pointed low-resolution spectrograph mode. Studying galaxies that have been recently quenched allows us to test the energetic processes that drive galaxies from star formation to quiescence. Post-starburst galaxies from z~0.1 out to z~0.7 have been shown to have large ISM reservoirs, with gas and dust content well below that expected given the low SFRs. Despite the role of AGN feedback in galaxy quenching models, evidence for strong AGN activity in post-starburst galaxies is also lacking. These results raise key questions of whether heavy dust obscuration is hiding star formation and AGN activity, that only mid-IR spectroscopy of large samples can provide.

## General Scientific Area:

Growth of Galaxies and Black Holes Through Cosmic Time

## Science Justification:

Understanding the processes that drive galaxies to evolve from star-forming disks to quiescent ellipticals is key to understanding the current population of galaxies in the Universe. At the root of these evolutionary processes is the fate of the interstellar medium (ISM) and its ability to form stars. Blue star-forming galaxies are rich in molecular gas, dust, PAHs, and ionized gas tracing on-going star formation, while red quiescent galaxies have significantly lower ISM fractions consistent with low or largely absent star formation (e.g., Kennicutt 1998, Davis et al. 2014). Several lines of evidence suggest that star formation alone is not sufficient to deplete the ISM in galaxies and that other processes such as AGN feedback may be required (e.g., Silk & Rees 1998, Springel et al. 2005, Hopkins et al. 2008).

In order to test predictions for how galaxies quench, we can identify **galaxies which have recently quenched** or are in the process of quenching to assess the possible energetic processes at work and their impact on the ISM. The stellar populations of galaxies are dominated by the most recent epochs of star formation, resulting in clear observational signatures for recently quenched galaxies. Galaxies which have experienced a burst of star formation that ended within ~200 Myr show a lack of current strong star formation coupled with strong Balmer absorption from A-star dominant stellar populations (e.g., Dressler & Gunn 1983, Couch & Sharples 1987, Zabludoff et al. 1996). These recently quenched galaxies are often called post-starburst, E+A, or K+A galaxies (see French 2021 for a recent review).

Observations of the cold molecular gas and dust in these recently quenched post-starburst galaxies has shown that substantial ISM reservoirs can persist in the Gyr after a starburst ends (French et al. 2015, Rowlands et al. 2015, Alatalo et al. 2016a, Smercina et al. 2018, Li et al. 2019).

These observations indicate that quenching can occur **without** a complete consumption or removal of the ISM and suggest that feedback processes are delayed after a starburst. Similar results have been observed for higher redshift samples of post-starburst galaxies at z~0.7, where the galaxies contain excess quantities of cold molecular gas for the low current star formation rates (Suess et al. 2017, Bezanson et al. 2022).

However, significant unknowns remain in understanding how galaxies evolve during and after quenching, primarily due to dust obscuration and the lack of unambiguous tracers of AGN activity. Post-starburst galaxies often show total IR luminosities well above what would be predicted from low optical SFR tracers, which may be due to obscured star formation (Baron et al. 2022ab) or intermediate age stellar heating from the recent burst (Hayward et al. 2014, Roebuck et al. 2019). **Mid-IR ionized gas tracers are uniquely powerful SFR tracers in recently quenched galaxies as they trace the unobscured and instantaneous SFR** (Fig 1). Similarly, optical tracers of AGN activity in post-starburst galaxies often show weak, LINER-like emission (Yan et al. 2006, Yang et al. 2006), despite the expectation that AGN feedback may be at work. Mid-IR tracers of highly ionized gas can determine whether post-starburst galaxies lack strong AGN or the AGN are heavily dust obscured.

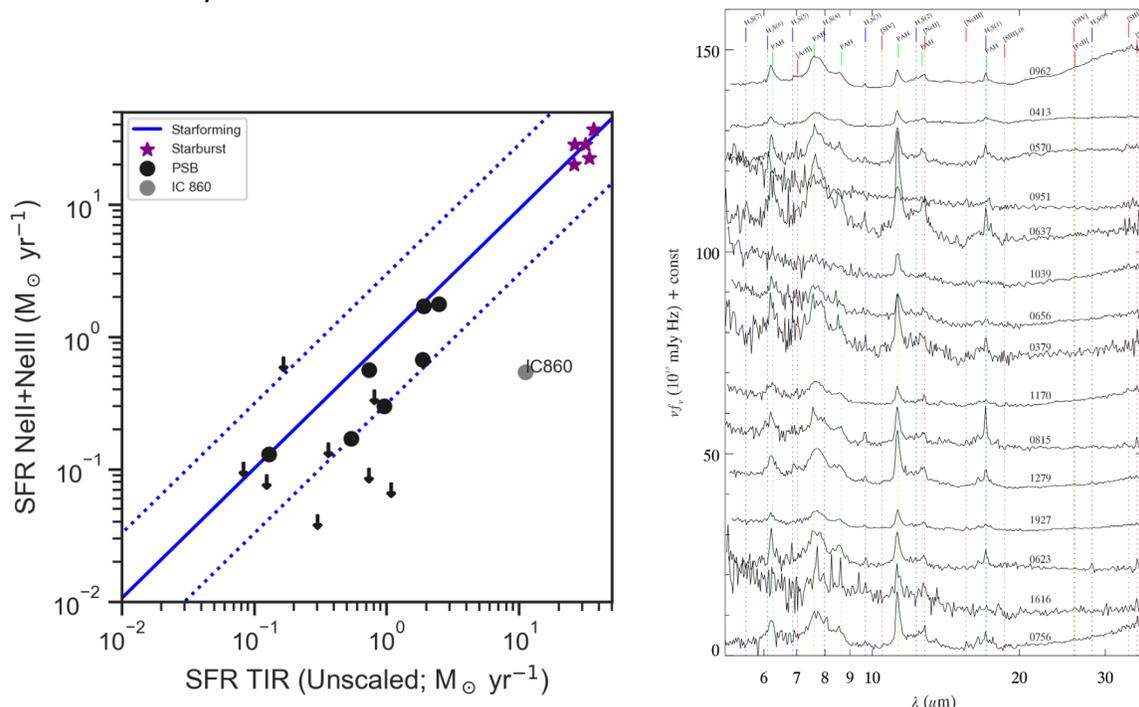

Figure 1: **Left:** (Adapted from Smercina et al. 2018, Luo et al. 2022, French et al. 2023). SFR measured from MIR line tracers vs. SFR from total IR luminosity. Post-starburst galaxies with Ne II+Ne III constraints from Spitzer IRS (Smercina et al. 2018, Luo et al. 2022) lie systematically below star forming galaxies (Ho & Keto 2007) and stacks of starburst galaxies (Stone et al. 2022). PRIMA would provide Ne III measurements for an additional 27 galaxies, including z~0.7 post-starbursts to test whether the total IR luminosity in these galaxies is tracing obscured star formation or heating by intermediate age stellar populations, as well as other tracers of the SFR like CII and other tracers of the ionized gas state to calibrate Ne III as a SFR tracer in these galaxies. **Right:** (From Smercina et al. 2018) Spitzer IRS spectra for 15 post-starburst galaxies showing substantial PAH emission, consistent with the presence of gas and dust in these galaxies, paired with weak or absent ionized gas emission lines. PRIMA would allow for these faint ionized gas emission lines to be detected at a sensitivity ~10x deeper than these IRS spectra and allow for line detections of a larger sample of post-starburst galaxies, including those at higher redshift.

PRIMA provides valuable mid-IR spectroscopy to constrain the star formation, ionized and cold gas, and AGN activity during the phase of galaxy evolution where quenching processes are at work. We outline the lines we aim to detect in our proposed program below.

## Instruments Required:

FIRESS pointed observations at low resolution

## Program Size:

- Medium (10–100 hr)

## Approximate Integration Time:

- 90 hours

## Special Capabilities Needed:

- None

## Synergies with Other Facilities:

ALMA observations of cold molecular gas, HST imaging of unobscured stellar light, JWST observations of shorter wavelength tracers of star formation, AGN activity, and interstellar medium tracers.

## Description of Observations:

The proposed program would observe a sample of 25 post-starburst galaxies at z~0.1 plus an additional two post-starburst galaxies at z~0.7.

The low redshift science case will observe each galaxy for two hours (one hour per pointing to cover the full wavelength range), reaching $2\times10^{-19}$ W/m$^2$ sensitivity. This sensitivity will enable the following science goals:

- Star-Formation and Cold Gas: [CII] 158μm detections down to a SFR~0.3 Msol/yr. We have used the dust SED fits to Herschel observations of post-starburst galaxies from Smercina et al. (2018) to estimate the likely continuum levels, and find the line/continuum ratio to be >10% per R~130 resolution element. Any non-detections would provide useful upper limits, as SFRs < 0.3 Msol/yr are unambiguously quenched.

- Star-Formation and Ionized Gas: [SIII] 33.5 μm, [SiII] 34.8 μm, [NeIII] 36μm, [OIII] 51.8 μm, [OIII] 88.4μm detections to constrain the state of the ionized gas and star formation rates. The SIII, SiII, and NeIII lines were measured for 2–4 post-starburst galaxies each using Spitzer IRS by Smercina et al. (2018, Fig 1), and the proposed observations would provide ~10x better sensitivity for a larger number of objects to enable faint detections.

- AGN: [OIV] 25.9μm, [FeII] 26 μm detections to test for obscured AGN. While these lines will be blended in the proposed low resolution spectra, a lack of strong detection will allow us to rule out the presence of obscured AGN activity.

- Shocks: [OI] 63 μm detections to measure the presence of dust-obscured shocked gas. Many post-starbursts show shocked gas tracers in their optical spectra (Alatalo et al.

2016a), and these observations would allow for shocks in the heavily obscured central regions to be constrained. For the galaxies with shocks, we expect the [OI] 63 μm line to be brighter than the [CII] 158 μm, enabling detection with our proposed sensitivity.

For the z∼0.7 science case, we would target two post-starburst galaxies for 20 hours each (10 hours per pointing to cover the full wavelength range). While the CII line will have redshifted out of observability, we will be able to measure the other lines described above. These observations will allow for detections of the brighter lines seen by Smercina et al. (2018) for low redshift post-starburst galaxies. While the number of possible targets for the z∼0.7 sample is small, these observations are the only way to test whether the large molecular gas content observed in this sample is due to a true suppression of star formation efficiency or whether these galaxies host significant obscured star formation.

## 26. Disentangling the effects of metallicity and star formation activity on the dust properties

Frédéric Galliano (DAp, CEA Paris-Saclay, France), Maarten Baes (Ghent University, Belgium), Simone Bianchi (INAF, Italy), Caroline Bot (Observatoire astronomique de Strasbourg, France), Viviana Casasola (INAF, Italy), Jérémy Chastenet (Ghent University, Belgium), Christopher Clark (STScI, USA), Ilse De Looze (Ghent University, Belgium), Vianney Lebouteiller (DAp, CEA Paris-Saclay, France), Suzanne Madden (DAp, CEA Paris-Saclay, France), Takashi Onaka (Tokyo University, Japan), Lara Pantoni (DAp, CEA Paris-Saclay, France), Monica Relaño Pastor (Granada University, Spain), Marc Sauvage (DAp, CEA Paris-Saclay, France), Matthew Smith (Cardiff University, UK), Tsutomu Takeuchi (Nagoya University, Japan) Manolis Xilouris (Athens Observatory, Greece)

The exceptional sensitivity of PRIMA will allow us to detect, for the first time in the IR, the population of low-surface brightness (quiescent), low-metallicity galaxies. These objects are the key to separate the effects of metallicity and star formation activity on the dust properties that observations of star-forming low-metallicity galaxies cannot. PRIMAger observations of the SED of these objects will thus be a novelty, important to understand the early stages of dust evolution.

### General Scientific Area:

Galaxy Ecosystems and Interstellar Medium

### Science Justification:

Our understanding of dust evolution relies on comparing the grain properties in regions (Galactic or extragalactic) exhibiting different physical conditions (e.g., Galliano et al. 2018). Different objects are then considered as snapshots of galaxy evolution at different stages. The most used environmental parameter to quantify the evolution of a galaxy is its metallicity, Z, as it traces the cumulated elemental enrichment of its InterStellar Medium (ISM). However, different Star Formation Histories (SFH), and thus different evolutionary paths, can lead to the same Z at a given time. Consequently, we often face a degeneracy between the effects of metal enrichment and star formation activity, when attempting to interpret dust evolution trends. This is illustrated by Fig. 1, showing the evolution of the fraction of small amorphous carbon grains as a function of Z and average starlight intensity, $<U>$, in individual galaxies (Galliano et al. 2021). It ambiguously suggests that the evolution of these small grains could be driven either by Z or by the specific Star Formation Rate (sSFR). This is because the low-Z galaxies detected with Herschel are primarily actively star-forming. This selection effect therefore results in a correlation between these two parameters in our sample, and we are unable to understand which one is fundamental. This degeneracy is encountered when looking at other quantities, too.

There is however a population of low sSFR, low-Z galaxies, distinct from those in our sample (e.g., Lara-López et al. 2013). Deriving the dust properties in these objects would thus allow us to break this type of degeneracy, as these sources would necessarily appear as a distinct branch in one of the two panels of Fig. 1 (solution 1 or solution 2). In addition, learning more about these local galaxies is interesting for the interpretation of deep surveys, as they probe the faint end of the luminosity function. These objects are also local analogs to distant primordial galaxies. Herschel was successful in detecting a few very low-Z star-forming objects (essentially I Zw 18 and SBS 0335-052). PRIMA could observe a sample of 100 nearby very low-Z galaxies (1/10 to 1/30 $Z_\odot$; typical size 2'×2'). For instance, the ALFALFA HI21cm survey (Haynes et al. 2018) contains hundreds of galaxies with $M(HI)<10^8$ $M_\odot$ and dozens with $M(HI)<10^7$ $M_\odot$. Among them, several have been well observed in stellar and ionized gas tracers: the 12 sources from the SHIELD sample (Cannon et al.2011); the very low surface density, nearby object, Leo P (Giovanelli et al. 2013); the most metal-poor gas-rich galaxy known to date, AG 198691 (Hirschauer et al. 2016). We also need more statistics in the low-Z/high-sSFR branch. We thus need to explore the whole sSFR range at very low Z.

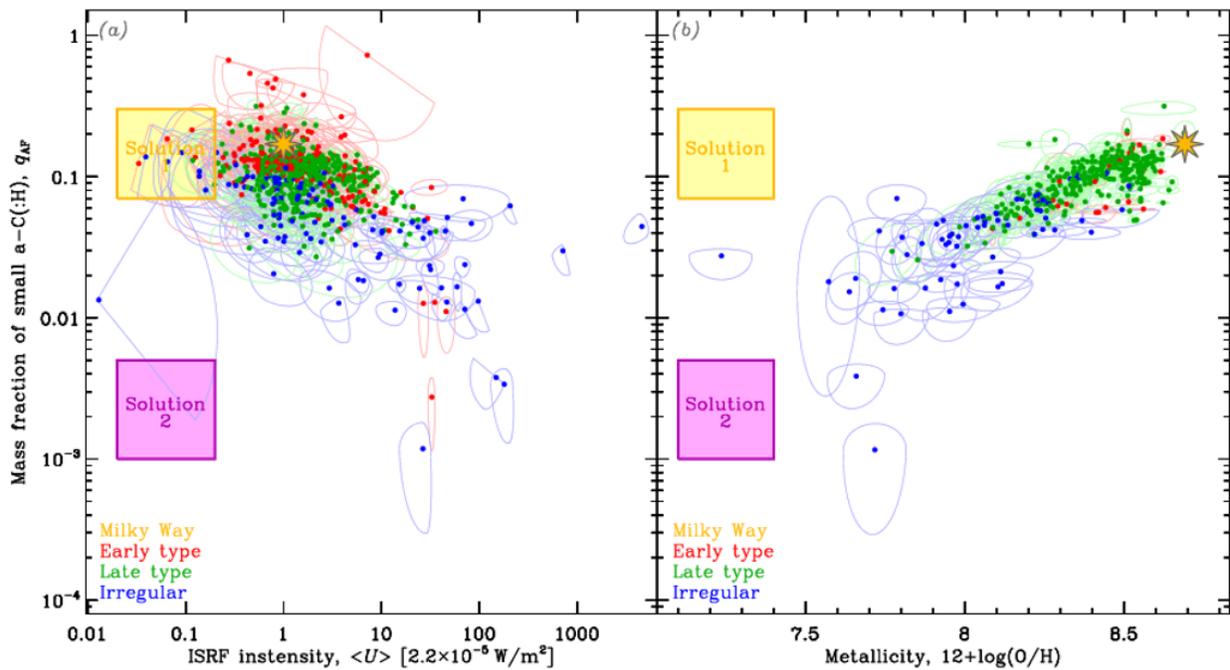

Figure 1: The potential of quiescent very-low-Z galaxies. We show the mass fraction of small amorphous carbon grains as a function of Z and starlight intensity, <U>. Each point corresponds to one galaxy of the Galliano et al. (2021) sample. We have added two hypothetical observations of a quiescent very-low-Z galaxy (solutions 1 and 2). Such observations would break the degeneracy between Z and <U>, as they cannot be consistent with both trends.

## Instruments Required:

- PRIMA FIRESS: Small Map Low-res R~130

- PRIMAger: Small Map(< 20'×20')

**Program Size:**

- Large (100 + hr)

**Approximate Integration Time:**

- 400 hours

**Special Capabilities Needed:**

- None

**Description of Observations:**

The unprecedented sensitivity of PRIMA should allow us to detect these objects at all IR wavelengths. Obtaining the well-sampled IR Spectral Energy Distribution (SED) of these objects would allow us to understand the role of Z and sSFR in shaping the dust properties at early stages.

- Ideally, we want to map this galaxy sample with all PRIMAger bands.

- We also need low-resolution spectral maps, with PRIMA-S, to better constrain the shape of the continuum in different regions.

Mapping 100 of these sources (except at 35 μm) would take less than 150 hours (Table 1), and R≈200 spectral mapping (except in band 1) would take less than 300 hours (Table 2).

- The flux sensitivity has been estimated using the dust model of Jones et al. (2017) with a radiation field intensity of $U$=1 and for $N$(HI)=2×10^{20} H/cm². At first approximation, the emission and the column density both scale with Z. We should thus scale the emission of the model by $Z/Z_{\odot}$. However, we are interested in the emission of an optically thin medium ($A$(V)≈0.1). This $A$(V) will be reached at a $Z_{\odot}/Z$ times higher column density than in the Milky Way. Z therefore cancels out in this (rough) estimate.

- We do not know much about the ISM properties of low surface brightness galaxies. Most of these objects have been detected only through their stellar emission or HI$_{21cm}$.

Table 1: Photometric time estimates. We used the PRIMAger Python time estimator (Feb. 2023; Marc Sauvage, priv. comm.).

| Band [μm] | Fields | Sensitivity [MJy/sr] | Time [hours] |
|---|---|---|---|
| 35 | 100×2'×2' | 0.07 | 8200 |
| 63 | 100×2'×2' | 0.24 | 131 |
| 92 | 100×2'×2' | 0.99 | 2.4 |
| 126 | 100×2'×2' | 1.9 | 2.4 |
| 172 | 100×2'×2' | 2.1 | 2.4 |
| 235 | 100×2'×2' | 1.7 | 2.4 |

Table 2: Spectroscopic time estimates.

| Band [μm] | Fields | Sensitivity [W/m$^2$] | Time [hours] |
|---|---|---|---|
| 32 | 100×2'×2' | 1.8×10$^{-20}$ | 30000 |
| 57 | 100×2'×2' | 8.3×10$^{-20}$ | 288 |
| 101 | 100×2'×2' | 1.1×10$^{-18}$ | <1 |
| 179 | 100×2'×2' | 3.4×10$^{-18}$ | <1 |

## 27. What is the role of HST-dark galaxies in massive galaxy formation and evolution?

Carlotta Gruppioni (INAF – OAS, Italy), M. Bethermin (LAM, France), L. Bisigello (UniPd, Italy), F. Calura (INAF-OAS, Italy), I. Delvecchio (INAF-OAB, Italy), A. Enia (INAF-OAS, Italy), A. Faisst (Caltech, USA), F. Pozzi (UniBo, Italy), G. Rodighiero (UniPD, Italy)

The main goal of this project is to characterize the IR SED and the ISM physics of dusty (ALMA-selected) HST-dark galaxies by observing about 100 HST-dark galaxies at 2<z<6 with PRIMAGER (of the order of 100 are required for a good statistics spread over the whole redshift and luminosity range covered by these galaxies), and a lower-z/brighter subsample with FIRESS, to observe some of the strongest mid-/far-IR fine structure lines. The ALMA HST-dark galaxies, detected in deep ALMA blind surveys and unseen in the deepest UV - near-IR images, often show a faint counterpart in the Spitzer-IRAC images. They are massive ($M*>10^{10}$ $M_\odot$) and dusty SF galaxies, likely the missing progenitors of massive ellipticals, and likely to provide a significant contribution to the obscured star formation rate density (SFRD) at high-z. Given their dusty nature (confirmed by their ALMA detection) and their mid-IR detection, the only way to unveil their nature and study the physics of their ISM is through mid-/far-IR photometric and spectroscopic observations. PRIMA photometric observations with PRIMAGER over the full 24–235 μm range will characterize their still uncovered SEDs between Spitzer/IRAC and ALMA detections, providing accurate photometric redshifts, while spectroscopic observations with FIRESS will be the key to study the physics of their ISM by detecting and measuring emission-lines in the mid- and far-IR domain (not suffering from dust extinction as the optical and UV emission lines). The spectroscopic data for the brighter/lower-z subsample will provide unique insight into the physical conditions (i.e., electron density and temperature, degree of ionization and excitation, chemical composition) of the gas responsible for emitting the lines within the dust-obscured regions of galaxies with intense SF activity or surrounding an AGN. **Unveiling the nature and the main physical properties of these high-z elusive, massive dust obscured galaxies, likely to make an important contribution to the z>2 obscured activity in the Universe, is the key to understand massive-galaxy formation and evolution and to account for the total cosmic SFRD at early epochs.**

### General Scientific Area:

Growth of Galaxies and Black Holes Through Cosmic Time

## Science Justification:

Tracing the history of star formation across cosmic time (or star formation rate density, SFRD) is one of the hottest topics of current galaxy evolution studies. Since stars form in dusty regions, it is crucial to understand how much star formation (SF) activity in the early Universe is obscured by dust and missed by optical/UV selections and what are the objects contributing to this quantity. However, our current knowledge of the SFRD at high z (>3) is mainly based on galaxies detected in the UV rest- frame (e.g., [1]), missing the more massive and obscured star forming galaxies (SFGs; [2]). These systems are likely connected to the formation of massive ellipticals, whose existence at z~2–3 challenges the current galaxy formation theories ([3], [4]). We are therefore still lacking an important piece in the galaxy formation puzzle: unveiling these missing SFGs, understanding their nature (e.g., do they host an AGN? What is their gas density, metallicity, temperature?), and probing the environment around them is the key to understand massive-galaxy formation and evolution and to account for the total cosmic SFRD at early epochs.

## A missing population of "optically invisible" dusty high-z galaxies discovered by ALMA

The Atacama Large Millimetre/submillimetre Array (ALMA) has opened a breach in the wall, revealing serendipitous galaxies in blind surveys and uncovering faint (sub-)mm populations at z>2– 3 (e.g., [5]). An important product of these surveys is the discovery of a new class of "dark" galaxies undetected even in the deepest Hubble Space Telescope (HST) images (down to H~27; [2], [5], [6], [7]). These "HST-dark" galaxies, representing up to 20% of the total ALMA blind samples, are detected in deep Spitzer- IRAC images ([5]; [7]; [8]; see figure 1), even though some remain totally unidentified ([6]; [9]). These results indicate the existence of a prominent population of dusty star-forming galaxies at z >2–3, likely providing a significant contribution to the high-z SFRD ([2]; [6]; [7]). These galaxies apparently lay on the main-sequence of SFGs (MS), and occupy the region below it at z>2 (e.g., [10]; C. Gruppioni et al., in prep.): can they be the missing progenitors of high-z ellipticals, caught in the act of quenching their SF?

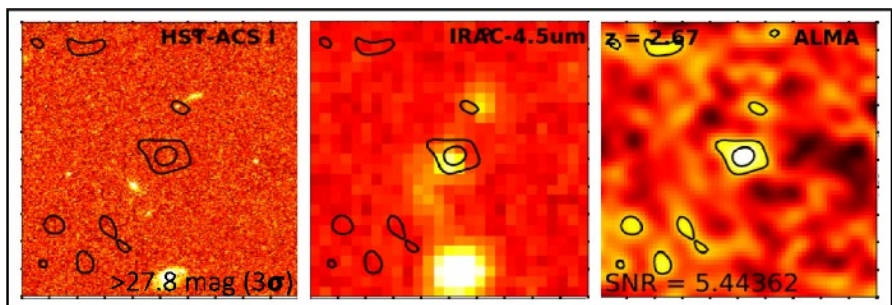

Figure 1: Example cutout of an HST-dark galaxy: ALMA-860 µm 3, 5σ contours on HST/ACS-i, Spitzer/IRAC and ALMA images.

**ALMA selection ensures the presence of dust and confirms their SF nature, while different selections (i.e., H-dropouts: IRAC sources undetected in HST/H-band, [2], or the mid-IR bright/near-IR faint galaxies recently identified by JWST, e.g., [11], [12], [13]), require additional information to exclude dust-poor populations.**

JWST will help in identifying these objects, since it is significantly deeper than Spitzer and redder than HST; however, since JWST is rest-frame optical at the redshift of these galaxies, for a complete characterization of these dusty galaxies we will still need to fill in the wavelength gap between JWST, Spitzer, and ALMA. Recent works based on deep ALMA surveys have set constraints on the obscured SFRD at z>3, showing that the SFRD distribution seems to remain almost constant over the whole 0.5<z<6 range (e.g., [8]), with no significant decrease beyond the cosmic noon (i.e., at z>3), at odds with UV-rest-frame derivations. **However, at z>3 there is no clear consensus also on the IR-based SFRD derivations (see Fig. 2).** In Fig. 2, the estimated contribution to the SFRD from HST-dark galaxies (ALMA selected, e.g., [6], [7], [8]; H- dropouts, [2]; and radio selected [14]) is shown. The uncertainties in the HST-dark contributions are large, due to their small numbers and poorly constrained SEDs. **The ALMA HST-dark galaxies contribute a significant fraction (>17–20%) of the total SFRD at z>3 (at z~5 the contribution is similar to the dust-corrected one from all the UV-selected galaxies; Fig. 2). This means that the dust-obscured SF is still significant up to z~5–6, and massive dusty galaxies might play an important role.**

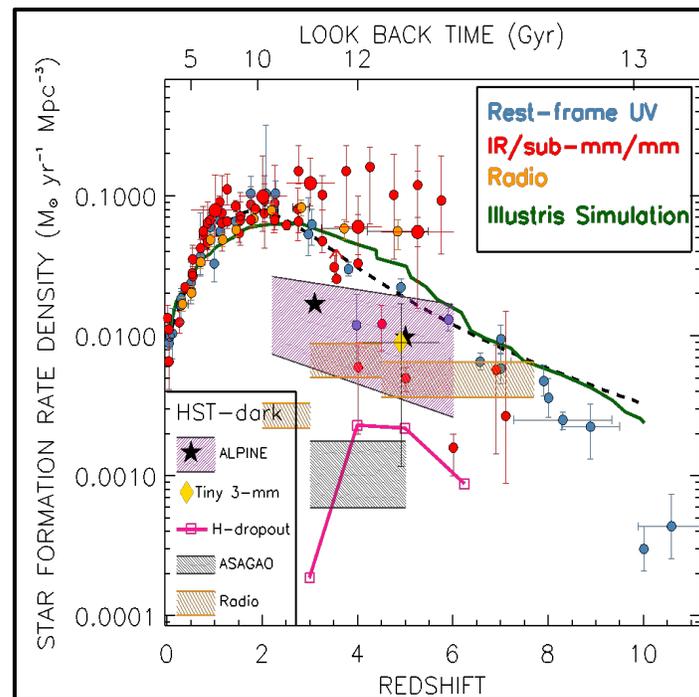

Figure 2: Cosmic SFRD versus z from IR/sub-mm/mm surveys (red circles), radio (orange) and dust-corrected rest-frame UV surveys (blue circles). The black curve is the SFRD by Madau & Dickinson (2014), the green curve is a theoretical simulation ([17]). The estimated contribution from different HST-dark galaxies samples is shown ([2], [6], [7], [8], [14]), the differences likely due to different limits and selections (ALMA, H-dropouts, radio).

The early formation of such a large number of massive, dusty galaxies is not predicted by current galaxy formation models and simulations (e.g., [15], [16]; see Fig.2), which underestimate the density of luminous and massive galaxies at z>2 (see [2]; [17]). The direct implication of such large abundances of massive dusty galaxies at high-z is that our current knowledge of the formation and evolution of massive/luminous galaxies is still far from being complete: **the robust estimate of the main physical properties, the ISM conditions and metallicities of HST-dark galaxies will be crucial for revising the theoretical models considering such a massive and dusty population.**

## Mid- and far-IR photometry and spectroscopy: the key to unveil the nature of HST-dark galaxies

**Sensitive photometric and spectroscopic observations of ALMA-selected HST- dark galaxies in the 24–235 µm range with PRIMAGER and FIRESS will be the key to investigate the role of such high-z dusty galaxies in early galaxy formation.** In fact, while a significant fraction of present-day stars have been produced in such obscured galaxies, we have very limited knowledge about their nature at those redshifts, on which process regulated the SF along the evolution of these galaxies and their metal content at early epochs. Dusty galaxies are the only ones that can tell us the whole story of galaxy and AGN evolution (whose activity peaks at about the same epoch) during that obscured era. So far, mid- and far-IR facilities like ISO, Spitzer and – more recently – Herschel have allowed us to observe the "obscured" side of the Universe at non-local redshifts, mostly in photometry, by measuring total fluxes of galaxies, but without peering into their internal physics. This have been done through mid- and far-IR spectroscopic observations only for local galaxies or, at high redshifts, for few lensed and/or ultraluminous sources not representative of the bulk of the galaxy population. The range covered by PRIMAGER and its narrow-bands will provide a full SED characterization (see Fig. 3) and secure photometric redshifts for a sizable sample of obscured galaxies, allowing to derive accurate estimates of their main physical parameters, and to identify the presence of an AGN. **The FIRESS low-res mode over the whole range of wavelengths is perfect for a blind mapping of the most relevant lines in dust-obscured galaxies, providing the crucial information on the main physical processes occurring in this massive and dusty population at cosmic noon.**

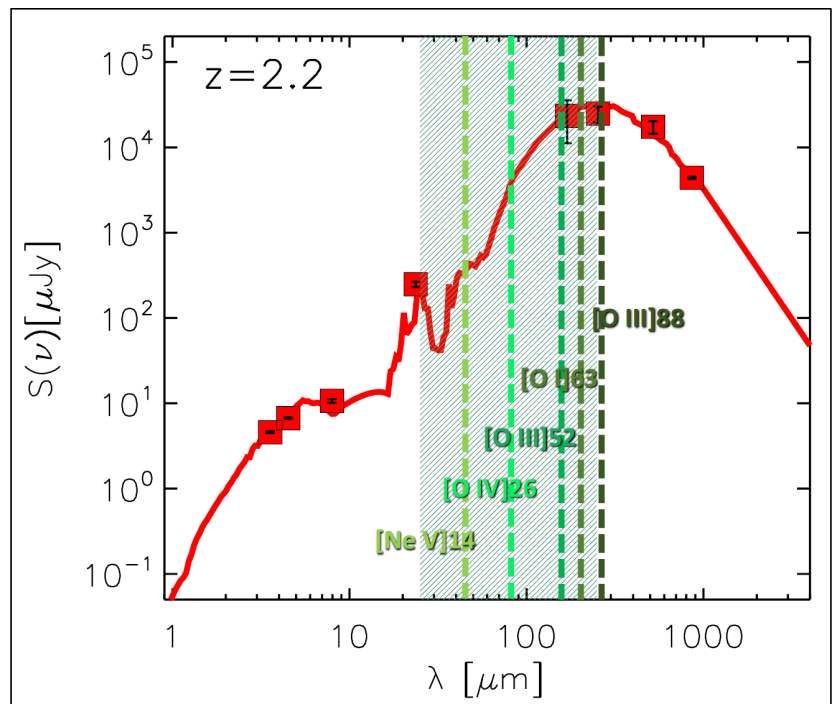

Figure 3: Example SED of one of the brightest and lower-redshift (zphot≃2.2) ALMA HST-dark galaxy (detected also by Herschel and Spitzer/MIPS; see [8]): observed data are shown as red squares, while the red curve shows the best-fit SED. The wavelength range that will be covered by PRIMA is shown as a green area. The main IR lines are shown as vertical dashed lines.

## Why PRIMA?

**PRIMA will be the only facility allowing us to observe the mid- and far-IR continuum and fine structure lines in dust-obscured galaxies invisible in the UV/optical/near-IR bands. In particular, PRIMA will allow us to unveil the nature of high-z obscured galaxies (like the HST-dark population) not observable at optical wavelengths, by sampling their currently unconstrained spectral range without extinction issues.** These galaxies, detected by ALMA in the sub-mm and with a counterpart only in Spitzer/IRAC images, can be observed by JWST in their rest-frame optical range. However, to fully characterize these elusive galaxies and their dust and ISM properties, longer wavelengths observations will be needed. The 24–235 μm range in z>2 galaxies is crucial to characterize the SED and nature of dusty galaxies, to derive accurate photometric redshifts (possibly identifying PAHs within the PRIMAGER narrow-bands), and to obtain important information on the physical conditions of the gas (through mid- and far-IR fine structure lines). A statistically significant sample of HST-dark galaxies (of the order of 100, for a good statistic over the whole redshift and luminosity range) observed with PRIMAGER will be the key to shed light on the still elusive nature of this population. Moreover, for a brightest, lower-z, subsample also further investigation into the ISM physics will be possible with FIRESS. In fact, emission-line intensities and ratios in the mid- and far-IR domain, not suffering from dust extinction as the optical and UV emission lines, provide unique information on the physical conditions (i.e. electron density and temperature, degree of ionization and excitation, chemical composition) of the emitting gas within the dust-obscured regions of galaxies with intense SF activity or surrounding an AGN ([18]; [19]; [20]). The typical electron density of the emitting gas can be identified through IR fine-structure lines, since different IR fine-structure transitions of the same ion have different critical density to collisional de-excitation (e.g., [O III] at 52 and 88 μm). Moreover, the relative strengths of the fine-structure lines in various ionization stages can probe the primary spectrum of the ionizing source, with these line ratios providing information on the age of the ionizing stellar population and on the ionization parameter in single H II regions or in starbursts of short duration. In addition, these line ratios are sensitive to the presence of non-stellar sources of ionizing photons, like the AGN, with some of these lines with higher ionization potential ([Ne V] at 14.32 and 24.32 μm, and [O IV] at 25.89 μm) excited only – or primarily – by an AGN ([21]; [22]). These line diagnostics will not only enable us to identify the primary source of ionization, but also to interpret the relative contribution of different ionization mechanisms (AGN/stellar feedback, shocks) to the observed mid-IR line intensities via comparison with state-of-the-art photoionization models (e.g., [23]). The detection of such lines will allow us to identify the presence of AGN in HST-dark galaxies. Previous studies, in fact, suggest exceptionally high AGN fraction among the ALMA blind surveys (up to 40%; [5]), although recent works seem to indicate lower fractions in HST-dark samples ([24]). Given the highly obscured nature of these galaxies, X-ray observation will not be able to identify AGN even if present in the large majority of them. **Estimating the AGN contribution in HST-dark galaxies will be necessary for understanding the role of AGN in massive galaxies formation (i.e., are they responsible for quenching SF?), and the early build-up of super massive black holes. Fully characterizing the nature of HST-dark galaxies will be crucial for adding such a still missing piece to the galaxy evolution puzzle.**

## Instruments Required:

- PRIMA FIRESS: Pointed Low-res, R~130: 10 brightest sources (log($L_{IR}/L_⊙$)~12.8 at lower-z: 2<z<2.5) in all (4) bands: ≤ 2 hr/source (depending on band) × 2 (2 pointings to cover 4 bands) for a total time of ~40 hr

- PRIMAger: Small Map (< 20'×20') Hyperspectral band and Polarimetry band
    - Pointed observations: 10'x10' for ~100 ALMA-selected HST-dark galaxies (Band1 + Band2).
    - Hyperspectral band: ~50 hr to observe all targets to 5σ (band 1A).
    - Polarimetry band: Simultaneously observe in this band.

## Program Size:

- Medium (10–100 hr)

## Approximate Integration Time:

- About 50 hr to observe ~100 targets (conservatively considering the faintest ones) with PRIMAGER
- About 40 hr to cover the whole wavelength range of FIRESS for 10 brighter (~$10^{12.8}$ $L_⊙$) 2<z<2.5 sources.

## Special Capabilities Needed:

- None

## Synergies with Other Facilities:

These obscured galaxies have been detected by ALMA at 860 or 1100 μm, and by Spitzer/IRAC (the brighter ones also by Herschel at 100–500 μm): the range covered by PRIMAGER, thought crucial for identifying AGN and characterizing the nature of high-z dusty galaxies, is currently unconstrained in their SEDs, since no facilities are observing or will observe at those wavelengths. Spectroscopically HST-dark galaxies could be observed in the sub-mm (e.g., with ALMA), in particular with long integration spectral scans, aiming at detecting one or more CO lines, providing information on the molecular gas. Moreover, information on ionised gas, stellar continuum and consequent diagnostics will be possible with JWST in the mid-IR range (detecting the rest-frame optical continuum with NIRSpec and MIRI), though the lines will be affected by extinction issues. **However, for a complete characterization of these galaxies, for a full comprehension of the different ISM components of and the main physical conditions of these galaxies, it will be necessary to cover the large gap between rest-frame optical and sub-mm by observing the mid- and far-IR range both in photometry and detecting the fine structure lines.**

## Description of Observations:

**Given the elusive and obscured nature of our targets (i.e., without HST counterpart and detected in λ>3–5 μm images), mid- and far-IR photometric and spectroscopic observations over the 24–240 μm range is the only way to obtain crucial information about the nature and**

**physical properties of such a massive and dusty high-z population. While Mid-/Far-IR photometry is key to fully characterise the SEDs of such obscured galaxies and unveil their nature, IR spectroscopy is the only observing mode to capture fine-structure lines providing crucial information about their main physical properties and conditions. Given the range covered by FIRISS, it represents the most efficient configuration for detecting [Ne V] and [O IV] lines, eventually probing the presence of an AGN, and other lines (e.g., [Si II], [O I], [O III]), crucial to unveil the physical and chemical properties of the gas inside this obscured population.** The current lack of mid-/far-IR photometric and spectroscopic information (and the consequently very uncertain physical parameters estimate and correct epoch location due to their non-detection in the optical/near-IR images), makes PRIMAGER photometry and FIRISS spectroscopy the only possible way to study the nature of these galaxies without being affected by dust extinction. We propose to observe 100 HST-dark galaxies selected in ALMA blind fields with photometric redshifts >2 with PRIMAGER over the full wavelength range, and the 10 brightest (in LIR, e.g., $10^{12.8}$ L$_\odot$, see Figure 3) HST-dark galaxies at 2<z<2.5 also with FIRISS in spectroscopy. These latter galaxies constitute the lower-z tail of the HST-dark population distribution, typically ranging from z=2 to z=6–7 and peaking at z=3.5–4: given the nominal sensitivities of FIRESS, this lower-z subsample is the only one observable in a reasonable - though significant - amount of time. With PRIMAGER, we estimate to detect the brighter galaxies at 5σ in most of the 1A/1B bands in less than 30 minutes (longward of band 1B for the faintest/higher-z sources; based on the best-fit SEDs of the ALMA HST-dark galaxies). Since the full range is observed simultaneously, and much shorter times are needed to reach the expected fluxes (or confusion) in Band 2, a conservative estimate of 30 minutes per sources will meet our goal. We would therefore estimate about **50 hr** total to observe 100 HST-dark galaxies with PRIMAGER (10'x10' observations centered on each target, allowing the simultaneous coverage of 4'x4'with Bands 1 and 2).

For FIRESS, by assuming a sensitivity $S_{line}$ (5σ, 1h)= $2\times10^{-19}$ W/m², we obtained the following exposure time for the stronger lines in the considered redshift interval (assuming the local relations between $L_{line}$ and $L_{IR}$ by [25] for $L_{IR} = 10^{12.8}$ L$_\odot$ – the typical value for the brightest HST-dark galaxies at z~2–2.5; the estimates are conservative, since times will be shorter by considering the ratios derived by [26]): with $t_{exp}$ = 2 h/source we expect to be able to detect [Si II]34.5, [O I]63, [O III]88, plus [Ne V] and [O IV] if an AGN will be present. In particular, if the fraction of AGN will be >40% over the 5–40μm range, [O IV]25.9 will be likely observable in about 0.5h and [Ne V]14.3 in about 2h. We would therefore estimate about 40 hr to detect the stronger IR lines in 10 HST-dark galaxies at z=2–2.5 over the full FIRESS spectral range (two pointings per source to cover the four bands).

**This ambitious experiment, expensive in terms of observing time, will be unique in unveiling the physics of a high-z massive dust obscured population, whose existence is currently challenging galaxy formation and evolution models, otherwise impossible to study with UV/optical/near-IR facilities.**

# 28. A Complete view of the ism at the era of peak galaxy growth

Hanae Inami (Hiroshima University, Japan); Amit Vishwas (Cornell University, USA), Luigi Spinoglio (IAPS INAF, Rome, Italy); Juan Antonio Fernández-Ontiveros (CEFCA, Teruel, Spain); Tohru Nagao (Ehime University, Japan); Hideki Umehata (Nagoya University, Japan); Jason Surace (California Institute of Technology, USA)

Understanding the physical and chemical conditions of gas and dust in galaxies is vital for understanding galaxy evolution. At the peak of galaxy growth near z ∼ 2–3, active galaxies are deeply enshrouded in dust, and thus mid- and far-infrared observations are indispensable. PRIMA will bridge the gap between the wavelength range of JWST and ALMA, providing coverage of a suite of polycyclic aromatic hydrocarbon (PAH) features, fine-structure lines, and molecular hydrogen lines, allowing us to explore the interstellar medium (ISM) embedded in the dust. In this PRIMA program we propose to use FIRESS to observe active galaxies detected with ALMA and/or JWST in the distant Universe in order to directly trace the ISM conditions at the dust embedded site of galaxy activity. The redshift range from 3 < z < 4.5 is ideal for PRIMA as it allows coverage of rest-frame wavelengths from 5 to 80 μm, where rich spectral features reside. For galaxies at z < 4, PRIMA will efficiently detect PAH features and emission lines from neon, sulfur, argon, oxygen and silicon for dwarf-like main-sequence galaxies allowing measurement of star formation rates, gas ionization, density, and metallicity. At z > 4, it will be more feasible to observe luminous galaxies to obtain the same set of spectral features to explore their ISM. Simultaneous observations of multiple PAH features, multiple fine-structure lines, and molecular hydrogen lines, which are only available with PRIMA, are the only way to fully reveal the underlying condition of the ISM of active galaxies in the early Universe.

## General Scientific Area:

Galaxy Ecosystems and Interstellar Medium

## Science Justification:

### Context

The physical and chemical conditions of gas and dust are fundamental to understanding galaxy evolution. In particular, emission lines play important roles in constraining properties such as star formation rates, star formation age, ionization, density, and metallicity. Recent ground- and space-based observatories have enabled observations of rest-frame optical emission lines at z > 1 to investigate these key physical parameters for early galaxies. However, due to the dustier nature of active galaxies at high redshifts of at least up to ∼3 (e.g., Gruppioni et al. 2013, Magnelli et al. 2013; or possibly even higher, e.g., Gruppioni et al. 2020, Algera et al. 2023a, Rodighiero et al. 2023), it is likely that we have not yet probed the extent and nature of star formation activity deeply enshrouded in dust. In addition, an unexpectedly large population of heavily obscured AGN have been discovered by ALMA and JWST in recent years (Gilli et al. 2022, Yang et al. 2023).

Although JWST is pushing this limit to higher redshift, its wavelength cutoff at 28 µm still leaves important coolants of the ISM inaccessible. Several PAH bands are sprinkled throughout the rest-frame mid-infrared (mid-IR) bands at 3–20 µm and can directly trace the properties of dust (e.g., Galliano et al. 2008, Stierwalt et al. 2014, Lai et al. 2020). Fine-structure lines are sensitive tracers of the physical conditions of the ISM like the ionization parameter (e.g., Cormier et al. 2015, Inami et al. 2013). In particular, primary elements like neon, sulfur, argon, oxygen, and silicon are available in the mid-IR bands that are less sensitive to large dust columns than optical lines. They provide powerful diagnostics for ongoing energetic processes that accompany star formation, supermassive black hole growth, and feedback mechanisms that regulate the growth of galaxies (e.g., Spinoglio et al. 2017). A probe mission like PRIMA that covers wavelengths from 25–250 µm, which are otherwise inaccessible from Earth, and traces the rest-frame mid-to-far-IR spectra of galaxies is essential to seeing past the dusty veil and understanding the key processes that drive the assembly of massive galaxies in the early Universe.

## Science Questions:

- What are the ISM properties of active galaxies in the early Universe?
- What role does dust play in our understanding of galaxy properties?

## Need for PRIMA:

PRIMA will be the only observatory that fully bridges the wavelength range between JWST and ALMA. This wavelength range covers many key spectral features that trace the physical conditions of the gas and dust. The dust-penetrating power of mid- and far-IR spectroscopy of PRIMA is essential because most active galaxies are known to be dusty at least up to z∼3 and likely beyond. With rest-frame optical observations, the evolution of ISM properties from z∼3 to z=0 has already been suggested (Shapley et al. 2015, Steidel et al. 2016). However, these features are likely from less obscured regions, outside the active regions of galaxies. PRIMA will be able to reveal the hidden aspects of galaxies where most stars are being made and the vicinity of supermassive black holes.

The redshift range of 3 < z < 4.5 will be the sweet spot for PRIMA to follow up galaxies that are detected with ALMA and JWST (Figure 2). The two strongest cooling lines, [C II]158 and [O III]88, are covered by ALMA Bands 7/8 and Bands 9/10, respectively, in this redshift range. On the other hand, JWST can find dusty active galaxies with imaging but its spectroscopy will miss most fine-structure lines, molecular hydrogen lines, and PAH emission for studying the ISM embedded in dust.

PRIMA will be the only observatory that can directly observe the rest-frame wavelength of ∼5–80 µm which hosts rich spectral features that directly trace the physical and chemical conditions of the exact site of galaxy activity.

## Interpretation Methods

The wavelength range between ~5–80 µm in a galaxy's rest-frame spectrum is filled with astrophysically important tracers. The most prominent features in this spectral range arise from PAH emission, especially at 6.2 µm, 7.7 µm, 8.6 µm, 11.2 µm, and 12.7 µm (Tielens 2008). These PAH features can have large equivalent widths (EW~0.5 µm) when star formation is the dominant energy source; conversely when their EWs are < 0.1µm it is likely that an AGN dominates (Armus et al. 2007, Petric et al. 2011). When multiple PAH features are detected, their flux ratios can be used to determine the dust grain size and charge (e.g., Draine & Li 2001).

In addition, PRIMA will cover all of the following fine structure lines for galaxies in 3 < z < 4.5: [Ar II] 7 µm, [Ar III] 8.99, [S IV] 10.5, [Ne II] 12.8, [Ne V] 14.3, [Ne III] 15.6, [S III] 18.7, [Ar III] 22, [Ne V] 24.3, [Fe I] 24, [O IV] 25.9, [Fe II] 26.0, [S III] 33.5, [Si II] 34.8, [Fe I] 35, [Fe II] 35, and [Ne III] 36. Among these fine structure lines, [Ne II] and [Ne III] are often strongest and their line flux ratio is the most sensitive tracer for star formation rate (Ho & Keto 2007, Mordini et al. 2021) and the hardness of the radiation field (e.g., Brandl et al. 2006, Bernard-Salas et al. 2009). This line ratio will facilitate breaking the degeneracy between gas ionization, density, and metallicity when combined with other line ratios, particularly strong lines such as [S III]33.5/[Si II]34.8 (Figure 1 left) that are detectable with PRIMA (cf. Snijder et al. 2007, Inami et al. 2013). Furthermore, when emission lines are present from the same species and the same excitation level, their ratios can be used for density diagnostics. In particular, [S III]33.5 and [S III]18.7 are expected to be as strong as [Ne III], and thus their ratio will be a powerful density tool. Although much weaker (at most ~0.1% of the total IR luminosity, LIR), the coronal line [Ne V] at 14.3 µm and 24.3 µm is a robust indicator of the presence of an AGN. Thus, [Ne V] can be used to search for an AGN and [Ne V]/[Ne II] can be used to calculate the AGN contribution (or its upper limit) to the ISM. In addition, the [O IV] line, which is often brighter than [Ne V], also has a relatively high ionization potential, so [O IV]/[Ne II] can be used to identify an AGN (e.g., Spinoglio et al. 2022, Stone et al. 2022). If shocked gas is present, H2 emission lines may be prominent and detectable with PRIMA (Ogle et al. 2010, Guillard et al. 2012). Mergers are more prominent at high redshifts (e.g., Ventou et al. 2017, Duncan et al. 2019) and thus merger-induced shocks traced by H2 observations of them will open a new window to explore their nature. Combinations of multiple H2 lines will provide estimates of the temperature and mass of warm molecular gas (Higdon et al. 2006, Roussel et al. 2007).

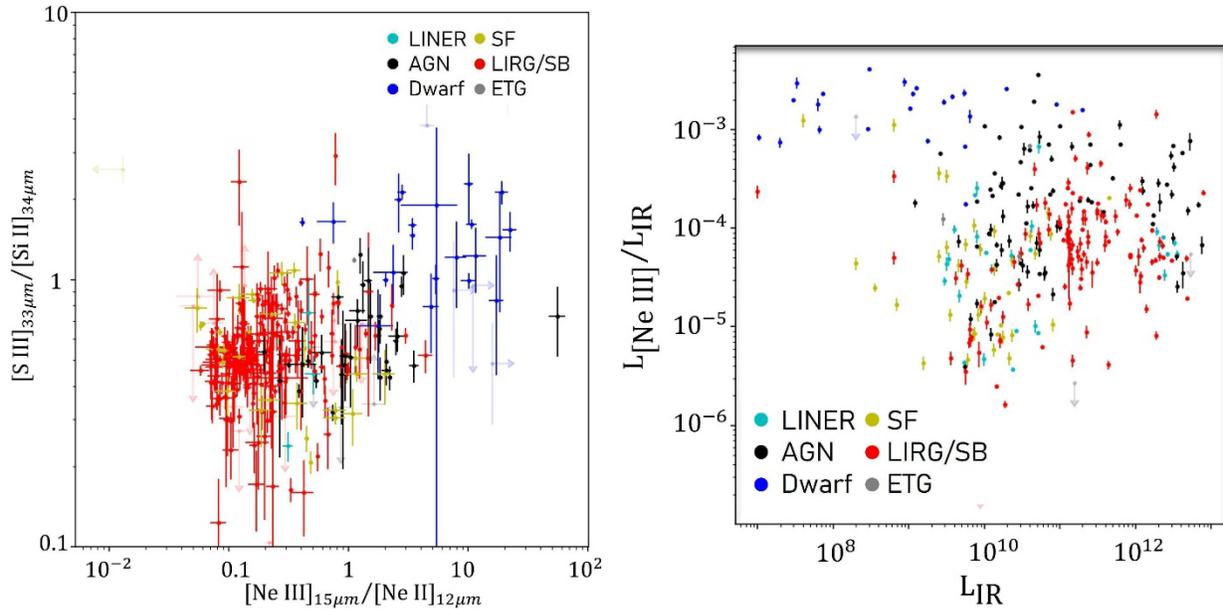

Figure 1: Parameter space of emission lines in the mid-IR that PRIMA can obtain with reliable detections. Left: Line ratio-ratio diagram of the four prominent emission lines that PRIMA can easily detect, [Si II]33, [Si II]34, [Ne III]15, [Ne II]12. Right: L[Ne III]/LIR vs. LIR of galaxies at z=0 from the literature (e.g., Inami et al. 2013, Cormier et al. 2015).

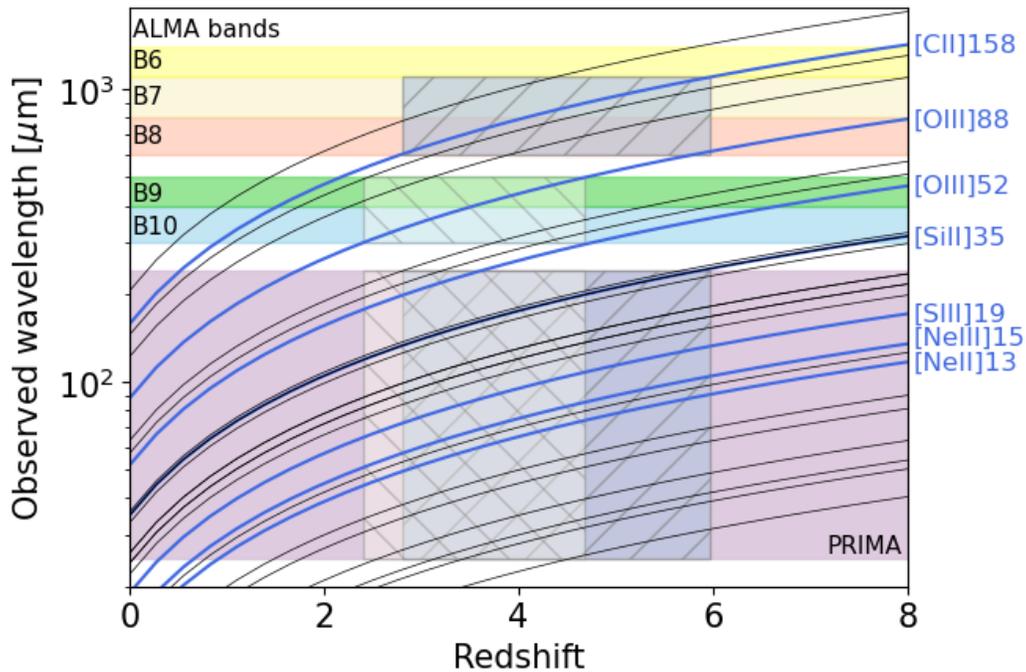

Figure 2: The redshift range that is ideal for PRIMA to follow up ALMA- and JWST-selected samples.

## Instruments Required:

PRIMA FIRESS: Pointed Low-res R~130

## Program Size:

- Medium (10–100 hr)

The program size can vary depending on target selection. This program does not need to be a single program but can be split into multiple programs based on the characteristics of the targets. However, the size will need to be a medium (10–100 hr) or large (100 + hr) program in order to provide adequate sampling statistics.

## Approximate Integration Time:

The observation depths are set by the need to detect the most important fine-structure lines for studying the ISM. At these depths, all the PAH features will be well detected. Based on observations of local galaxies near z=0, the median luminosity ratio of [Ne III] to total IR luminosity is 5e-4 for the whole sample, or 1e-3 for dwarf-like galaxies (Figure 1, right). The [Ne II] line flux is 0.1–10x of [Ne III] and the [S III]18,33 lines and the [Si II] line are ~0.1–1x of [Ne III] depending on the age and metallicity. Given the FIRESS sensitivity of 2e-19 W/m$^2$, in Table 1, we summarize the IR luminosity of dwarf-like galaxies that can be detected with all of these five important lines for the ISM diagnostics with 1, 3, 10, and 20 hour integrations. For more evolved galaxies, ~4x deeper observations will be needed.

Table 1. IR luminosities of galaxies at z=1, 3, and 5 to detect at least [Ne II], [Ne III], [S III]18, [S III]33, and [Si II] (in addition to all PAH features) with the given integration time of 1, 3, 10, and 20 hours. For reference, IR luminosity converted from SFR of main-sequence galaxies with a stellar mass of 10$^{10}$ M$_\odot$ (10$^{11}$ M$_\odot$) is shown in the rightmost column (Speagle et al. 2014). Galaxies above the MS relation will be detected with shorter integration times.

| Integration time [hours] | 1 | 3 | 10 | 20 | LIR of the MS galaxies with 10$^{10}$ M$_\odot$ (10$^{11}$ M$_\odot$) |
|---|---|---|---|---|---|
| z=1 | 2.9e+11 L$_\odot$ | 1.7e+11 L$_\odot$ | 9.2e+10 L$_\odot$ | 6.5e+10 L$_\odot$ | 3.9e+10 L$_\odot$ (1.9e+11 L$_\odot$) |
| z=3 | 4.2e+12 L$_\odot$ | 2.4e+12 L$_\odot$ | 1.3e+12 L$_\odot$ | 9.4e+11 L$_\odot$ | 1.4e+11 L$_\odot$ (8.6e+11 L$_\odot$) |
| z=5 | 1.4e+13 L$_\odot$ | 8.1e+12 L$_\odot$ | 4.4e+12 L$_\odot$ | 3.1e+12 L$_\odot$ | 2.0e+11 L$_\odot$ (1.3+12 L$_\odot$) |

## Special Capabilities Needed:

- None

## Synergies with Other Facilities:

- ALMA, JWST, LMT, SKA, and AtLAST/LST (if realized) will provide the parent sample of targets to be followed up with PRIMA.

At longer wavelengths, ALMA observations increasingly detect the [C II]158μm and [O III]88μm lines for galaxies at z > 4 (e.g., Bethermin et al. 2020, Bouwens et al. 2022). These lines have been studied in galaxies in the local Universe with Herschel, but they are limited to z~0 due to a lack of sensitivity of the onboard spectrometers (e.g., Kennicutt et al. 2011, Diaz-Santos et al. 2017). At 6 < z < 10, the [O III]/[C II] ratios of currently known ``normal'' star-forming galaxies appear to be higher than those of local galaxies. The cause of this difference is still under debate (e.g.,

Harikane et al. 2020, Algera et al. 2023b) due to a lack of observations of other fine-structure lines for these galaxies to interpret the findings. PRIMA can bridge this gap between z~0 and z=6–10, where observing a large sample of [O III] and [C II] lines is either entirely impossible or in sparse redshift windows (around z~3) due to telluric transmission. Even when possible, the low transmission makes high-frequency observations challenging with current facilities. Furthermore, LMT/TolTEC is expected to build up a large sample of distant galaxies, selected based on mm colors, in the near future (Bryan et al. 2018). Optically dark galaxies of this sample, in particular, will benefit from followup by PRIMA. If AtLAST and/or LST become available in the 2030s (Kawabe et al. 2016, Klaassen et al. 2020), their submm-selected high-z galaxies would be perfect candidates for PRIMA. Their spectroscopic capability may also provide large samples of [O III]- and/or [C II]-emitters to follow up.

At shorter wavelengths, JWST will also provide a source of good candidates for PRIMA followup based on imaging surveys. For example, the highest redshifts where the 3.3μm and 6.2μm PAH features can be detected with JWST are z=7 and 3.5, respectively, by MIRI. PRIMA will be sensitive enough to detect the other PAH features at longer wavelengths in order to investigate dust grain properties. Moreover, JWST has started to identify NIRCam-dropout galaxies due to their dusty nature and/or large distance (e.g., McKinney et al. 2023). These sources are likely at z > 4 and are obvious targets to follow up by ALMA. However, since only a small number of spectral features are accessible with ALMA, preventing obtaining a full picture of this interesting and potentially dominant population of galaxies which build up most of the mass in the Universe. PRIMA will be the only observatory that can access their mid-IR spectra.

## Description of Observations:

FIRESS will be used to target galaxies that have already been observed with ALMA, JWST, and PRIMA imaging. Pointed observations with the low-resolution mode of FIRESS will be needed. It is essential to cover the full 25–250 μm observed wavelength range to simultaneously study multiple spectral features. Thus, two spectral settings are required per pointing to cover all four spectral bands.

Using the baseline FIRESS sensitivity of 2e-19 W/m² (5σ, 1 hour) across the band, a practical strategy will be selecting relatively luminous galaxies at z > 4, while observing ``normal'' galaxies on the star-formation main-sequence at z < 4. At z > 4, luminous galaxies (LIR >~ 1e13 $L_\odot$) are perfect targets for PRIMA (< 1–3 hours for each spectral setting to obtain at least the five primary lines). For galaxies at z < 4, on average 10–30 of the normal galaxies can be observed within 50–80 hours in total (~1–5 hours for each spectral setting).

With the low-resolution mode, we note that the 12.7 μm PAH feature can overlap in coverage with the spectral line of [Ne II] at 12.8 μm. In addition, [Ne V]14.3 can be blended with the [Cl II] line at 14.4 μm, as well as [O IV]25.9 and [Fe II]26.0. We can either use [Ne V]24.3, which will not suffer from blending, to identify an AGN or use [Ne V]14.3 and [O IV] in conjunction with PAH emission to avoid misidentifying [Cl II] as [Ne V]14.3 and [Fe II] as [O IV]. The observations undertaken by the Spitzer/IRS spectrometer have provided a significant baseline of extragalactic observations in the local Universe useful for studying the relative strengths of the PAH features and spectral lines, and which will aid in simultaneously fitting these features and spectral lines in the low-resolution mode. Significant advances in these techniques will be made in the near future with the JWST/MIRI IFU, which offers an even higher spectral and spatial resolution, allowing us to refine our existing understanding from IRS spectra, and we will be able to apply these lessons to programs designed for FIRESS (R~130).

## 29. Crystals in the interstellar medium of star-forming galaxies at Cosmic Noon

Francisca Kemper (ICE-CSIC / ICREA / IEEC), Frédéric Galliano (CEA-Saclay), Janet Bowey (Cardiff), Howard Smith (Harvard), Hendrik Linz (MPIA), Henrik Spoon (Cornell), Paul van der Werf (Leiden Observatory)

Silicates are the most common form of dust in the universe. In the interstellar medium of our Milky Way they are found to be amorphous, while stellar ejecta contain significant fractions of crystalline silicates. Crystallization occurs through thermal annealing, while cosmic ray hits are thought to cause the amorphization of crystalline material. Nearby starbursting galaxies are reported to contain crystalline silicates, presumably recently formed by massive, rapidly evolving, stars. We thus hypothesize that measuring the crystalline fraction of silicates in starbursting galaxies helps to constrain galactic properties, such as the star formation rate, the recent star formation history and the cosmic ray fluence. With its spectral coverage from 24 to 264 µm, the FIRESS spectrograph on PRIMA is uniquely capable of observing redshifted crystalline silicate features, which can be found at a rest wavelength from 8 to 72 µm. We propose to measure the crystalline silicate content of a sample of 130 starforming galaxies from 1.5<z<4, bracketing the peak of cosmic star formation, also known as Cosmic Noon, in order to characterize the dust formation and evolution process in these galaxies, related to their star formation activity.

### General Scientific Area:

Rise of Metals and Dust

### Science Justification:

Silicates are the main constituent of the interstellar dust in the Milky Way and most other galaxies. In the Milky Way, they are found to be completely amorphous, with the crystallinity lower than the detection limit of 2% of the silicate mass (Kemper et al., 2004), while stellar ejecta may contain considerable amounts of crystalline silicates. It is thought that amorphitization occurs upon impact of cosmic ray hits, while crystalline silicates are formed through thermal annealing of amorphous silicates, or direct condensation at high temperatures, typically in the dust formation zone around mass-losing stars. Detecting crystallinity thus traces the balance between dust injection by stars in the ISM, the cosmic ray fluence, and dust destruction in supernova shocks, and is a useful tracer of starburst activity (Kemper et al., 2011). Indeed, Spitzer has revealed that many actively starforming galaxies host detectable amounts of crystalline silicates (e.g., Spoon et al., 2006, 2022).

Amorphous silicates show two broad resonances in the infrared at 9.7 and 18 µm due to Si-O stretching and O-Si-O bending modes in the silicate tetraedra. In the spectra of crystalline silicates these modes are split up in a number of narrow and sharply peaked resonances, while resonances at longer wavelengths appear due to lattice (phonon) modes. The longest wavelength feature due to crystalline silicates detected in astronomical objects is the 69 μm forsterite ($Mg_2SiO_4$) feature, which shifts to longer wavelengths (∼72 micron) with the replacement of a small fraction of the Mg by Fe (Molster et al. 2002b). The laboratory measurements of crystalline silicates show a width of the spectral features of the order of 1–2 μm, and thus the spectral resolution of R=130 will resolve and detect a <1 μm wide ∼70 μm feature.

In their recent study, Spoon et al. (2022) catalogued the presence of extragalactic crystalline silicates in more than 700 relatively nearby galaxies using the Spitzer archive. As of yet, extragalactic crystalline silicates have not been reported in JWST observations, although Donnan et al. (2022), assume their presence when interpreting the data for VV114.

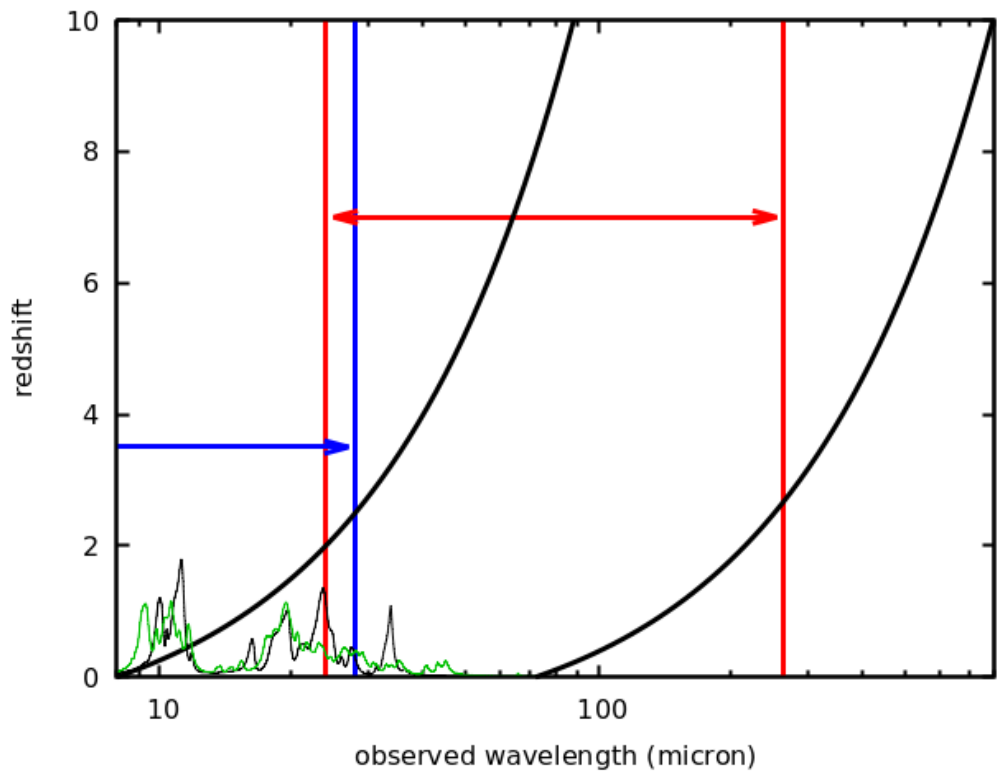

Figure 1: Discovery space of the FIRESS instrument on PRIMA when surveying extragalactic crystalline silicates. As a function of redshift, the wavelength range over which crystalline silicate features can be observed (demarcated by the two black curves, representing rest wavelengths of 8 and 72 μm) is shown. The two red lines show the observable range for the FIRESS instrument, and the blue line shows the long-wavelength limit of the JWST-MIRI instrument. From this plot it is clear that the FIRESS wavelength coverage is ideally suited to detect crystalline silicates at Cosmic Noon (around z=2-3), and that it is uniquely capable of detecting the crystalline silicate features at wavelengths longer than 28 μm at any redshift. Laboratory spectra of crystalline silicate species forsterite (black; Koike et al. 1999) and enstatite (green; Koike et al. 1999) are shown for reference at the bottom of the plot.

JWST has a wavelength coverage up to 28 μm, and is not suitable to detect crystalline silicate features at redshifts z>1.5, despite its superb sensitivity in the mid-infrared. With a wavelength coverage of up to 38 μm for Spitzer, the situation was only marginally better for that observatory. Indeed, even at the rest wavelength, there are several important crystalline silicate features present beyond 28 (or 38) μm, including the strong 33 μm forsterite and 43 μm enstatite features and the diagnostic 69 μm forsterite feature (Molster et al. 2002a). Thus, discovery space for PRIMA, with FIRESS operating from 24 to 235 μm, can be found in exploring the crystallinity of silicates in galaxies in the era of peak star formation ($1 \lesssim z \lesssim 3$; Cosmic Noon).

Since crystalline silicate features are intrinsically much stronger than the features due to amorphous silicates, small amounts, of the order of a few percent of the total silicate mass, can be detected whenever the resonances due to amorphous silicates are visible. The dedicated survey proposed here will complement the data already available in the Spitzer archive. Spoon et al. (2022) have analysed 3335 Spitzer spectra of galaxies, of which 350 have a redshift $1<z<2$ and 204 have a redshift $z>2$. In order to complement this dataset in a meaningful way, and assuming a detection rate of crystalline silicates of 20%, we want to add 500 sources with a redshift of $1.5 \lesssim z \lesssim 4$, to bracket Cosmic Noon, but at the same time to not have too much overlap with what is already in the Spitzer archive.

The observations will allow us to measure the feature strength of crystalline silicates, with several features being present over the wavelength range covered. We can thus measure the occurrence and degree of crystallinity at and before Cosmic Noon, during the peak of star formation, study the mineralogical composition, and relate the measured parameters to characteristics of the galaxies, such as star formation rate, and cosmic ray fluence.

## Instruments Required:

FIRESS pointed observations at low-res (R~130)

## Program Size:

- Large (100 + hr)

## Approximate Integration Time:

- 130 hours on-source

## Special Capabilities Needed:

- None

## Synergies with Other Facilities:

- JWST, possibly MATISSE at ESO (the 10 μm silicate feature).

## Description of Observations:

The fact sheet shows the spectrum of a $10^{12}$ L$_\odot$ galaxy at z=1.6 with respect to the 5-σ spectral line sensitivity is obtained in 1 hour for R~130. The value of $2 \times 10^{-19}$ W/m$^2$ for a 5-σ is met at a rest wavelength of around 30 μm. At longer rest wavelengths, where several important crystalline silicate features are present, the signal-to-noise increases. A z=4 galaxy is a factor of 3 more distant than a z=1.6 galaxy (in luminosity distance, assuming ΛCDM; Wright 2006), requiring 9 times more integration time for the same result. For brighter galaxies with luminosities of at least $10^{13}$ L$_\odot$, integration times become a factor of 3 shorter.

In favor of the science case, the spectral energy distribution of a galaxy continuously rises beyond 30 μm, until it peaks at a rest wavelength of around 80 μm, thus reducing the need for longer integrations, especially over the wavelength range with crystalline silicate features not previously covered by other observatories.

Assuming 1 hour of on-source integration time, which seems to be a reasonable value averaged over the 1.6<z<4 redshift range, 100 galaxies with a luminosity of $10^{13}$ L$_\odot$ galaxies will require 100 hours of observing time. Additionally, we will also add a smaller number (30) of $10^{12}$ L$_\odot$ galaxies with redshifts z<2, yielding an additional 30 hours of on-source observing time.

# 30. Dust temperature in objects at Cosmic noon

Ece Kilerci, Sabancı University, Faculty of Engineering and Natural Sciences, 34956, Istanbul, Turkey, Denis Burgarella, Aix Marseille Univ, CNRS, CNES, LAM Marseille, France

We are in a new era of detecting fainter and further galaxies from cosmic noon to cosmic dawn. Photometric and spectroscopic observations from near-infrared to millimeter wavelengths are essential to obtain accurate measurements of galaxies at redshift 2 and beyond. The combination of already available HST, JWST, Herschel, SCUBA, ALMA observations allowed to cover the most of the spectral energy distributions of variety of galaxies including AGNs and dusty star forming galaxies (DSFGs) in the Cosmic Evolution Survey (COSMOS) field. We propose to map central region of the COSMOS field with PRIMager and FIRESS to fill in the gap between JWST and ALMA especially for higher redshift (z ∼ 2) galaxies. With these observations we will obtain secure spectroscopic redshifts of fainter galaxy populations and fully sampled SEDs in the near-IR to millimeter wavelength range. With the PRIMA observations we will be able to measure the physical properties of these galaxies including dust temperatures, dust mass, IR luminosities, AGN fraction, stellar mass and star formation rate accurately. And we will have a unique collection of high-resolution measurements of PAH and the fine line structure lines that will allow us to measure star formation, dust properties, metallicities and AGN activity in these galaxies as an independent tool. Such a well observed sample of variety of galaxies at cosmic noon will be major sample to compare with higher redshift newly discovered galaxies at cosmic dawn and get further in sight of galaxy evolution through cosmic time.

## General Scientific Area:

Rise of Metals and Dust

## Science Justification:

We are witnessing JWST revealing new galaxy populations from cosmic noon to cosmic dawn (e.g., Finkelstein, S. L., et al. 2023; Yang et al. 2023). Detection of such fainter (at least 2 order of magnitude compared to our previous detection capabilities) galaxy populations at redshift 1 and beyond just started a new era in our understanding of galaxy evolution. In order to understand the physical relation between the gas supply, the star formation, the dust formation, the dust presence, the dust properties, the metallicities, the presence of AGN activity and the effect of AGN activity to the host galaxy for different galaxy populations at different cosmic epochs we need both photometric and spectroscopic observations from near-infrared to millimeter wavelengths. Any gap lack of observations in this wavelength range brings limit for our understanding of galaxy physics. Therefore, it is very important to eliminate this gap especially for galaxies at cosmic noon (z ∼1-3) that we eager to understand.

FIR coverage is essential to obtain new diagnostics from the PRIMA observations such as the PAH and the fine structure lines. Therefore, we want to start adding PRIMA data to the well observed COSMOS-MORA Survey field that already have HST, JWST, Herschel, SCUBA, ALMA observations (Casey et al. 2021). We propose to map the 230 arcmin² COSMOS-MORA Survey field with PRIMA to obtain spectroscopic redshifts and near-IR to millimeter SED coverage of a variety of galaxies including AGNs and DSFGs in this field that contains at least more than 20 DSFGs up to redshift of 3. PRIMA observations will be combined with archival HST, JWST, Herschel, SCUBA, ALMA measurements to derive physical properties variety of galaxies galaxies (star forming, DSFGs, AGNs) at cosmic noon (z ~ 1–3), via detailed SED fitting with the advanced SED modelling capabilities of CIGALE.

Figure 1 shows two example galaxies at cosmic noon. These are 2mm ALMA selected dusty star forming galaxies at z= 2.29 (left panel) and z= 2.47 (right panel). MORA-8 is a cold dust emission dominated DSFG with a photometric redshift estimate. MORA-10 is a composite galaxy with AGN contribution with a spectroscopic redshift measurement. We performed SED analysis with CIGALE by using the available HST to ALMA photometric data to measure the fluxes at PRIMA bands from the best-fitting SED model. As seen from the SED analysis PRIMA fluxes of these galaxies are above 0.2-0.3 mJy. Therefore, they can be detected by PRIMA. With its unique Hyperspectral imaging with 12 continuous filters between 25 and 80 µm, PRIMAger will fill in the GAP between the JWST mid-IR and ALMA wavelength range. As seen in Figure 1a, for a DSFG at z=2.29 PRIMAger observations of Bands 1A, 1B, 2A, 2B, 2C and 2D probes the mid-IR to FIR-IR SED. As a result of these observations physical properties of these galaxies, the PAH emission, metallicities, dust mass, dust temperature, total IR luminosity and star formation rates will be obtained accurately. In less than 1 hour (~40 min), PRIMAger has the following point source sensitivity limits at the 5σ level in a 230 arcmin square area: for Band 1A (0.2 mJy), Band 1B (0.3 mJy), Band 2A (0.2 mJy), Band 2B (0.3 mJy), Band 2C (0.4 mJy), Band 2D (0.5 mJy). As shown in Fig 1a, a DSFG at z= 2.29 can be observed within sensitivity limits of PRIMAger from Band 1A to 2D.

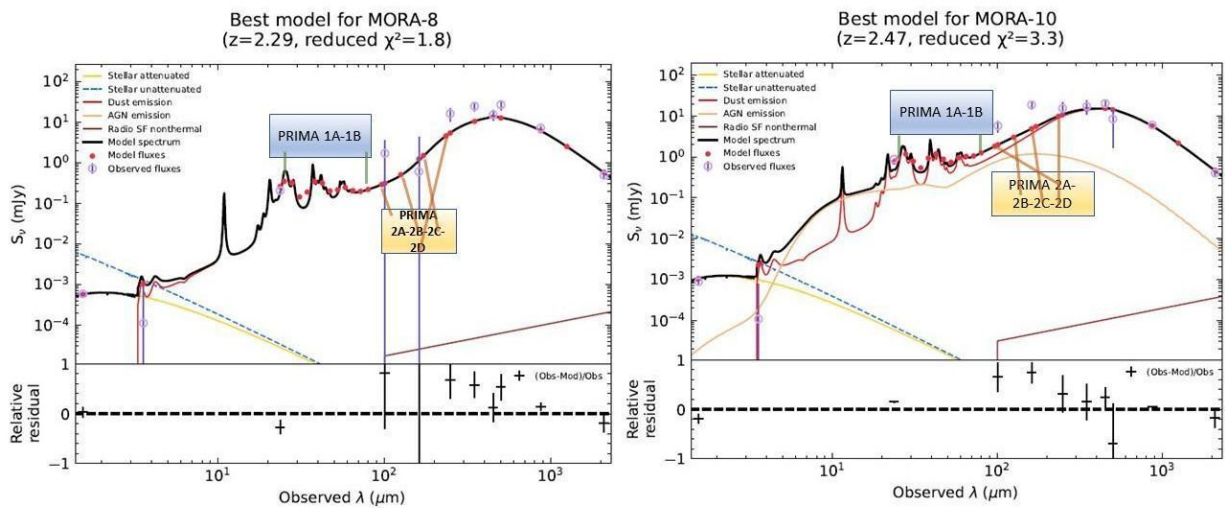

Figure 1: SED analysis of two example galaxies at cosmic noon showing expected PRIMA fluxes from Band1A to Band2D.

PRIMA_FIRESS low resolution mapping mode observations will provide measurements of PAH and the fine line structure lines. Within the sensitivity limits of PRIMA_FIRESS, it would take 416 hours in 230 arcmin$^2$ to detect [OIII] at $4.7 \times 19^{-19}$ Wm$^{-2}$ dept.

## Instruments Required:

- FIRESS Large map at low-resolution (R~130)
- PRIMAGer Large map in hyperspectral band

## Program Size:

- Large (100 + hr)

## Approximate Integration Time:

1 hour for PRIMAger observation of a large map. 416 hours for PRIMA_FIRESS high resolution mode observations; for 2 bands.

## Special Capabilities Needed:

- None

## Synergies with Other Facilities:

- COSMOS-Web JWST and ALMA observations are available in this field.

## Description of Observations:

PRIMAger will observe the 230 arcmin$^2$ COSMOS_Web_MORA_Survey area with all bands. The same field will also be mapped by PRIMA_FIRESS low resolution mapping mode for 416 hours. Full band will be covered in two pointing (each for 208 hours).

# 31. Cosmic evolution of chemical abundances & metallicity

Vianney Lebouteiller (Université Paris-Saclay, Université Paris Cité, CEA, CNRS, AIM, 91191, Gif-sur-Yvette, France), Frédéric Galliano, Suzanne Madden (Université Paris-Saclay, Université Paris Cité, CEA, CNRS, AIM)

Determining accurate ionic and elemental abundances as well as metallicities is paramount to understand the buildup of metals and cosmic dust in galaxies within the context of galaxy evolution but also to constrain and set the proper conditions for interstellar medium (ISM) models and diagnostics. Gas-phase abundances can be determined in various phases, from emission or even absorption spectral lines arising in various wavelength ranges. Ionized gas emission lines have been given particular attention as many transitions arise from star-forming regions in the visible domain. Such diagnostics shift to the mid- and far-infrared (IR) at high-redshift and need to remain accessible through a continuous wavelength sampling which PRIMA can provide.

Rest-frame infrared transitions are also powerful diagnostics that depend little on dust extinction and on electron temperature compared to optical diagnostics. The numerous available IR transitions probe various ionization stages and critical densities, across a wide redshift range including above z=1 for which mid-IR diagnostics shift out of JWST's range.

## General Scientific Area:

Rise of Metals and Dust

## Science Justification:

Determining accurate ionic and elemental abundances as well as metallicities is paramount to understand the buildup of metals and cosmic dust in galaxies but also to constrain and set the proper conditions for interstellar medium (ISM) models and diagnostics. From the galaxy evolution perspective, it is important to study the relationship between metallicity, star-formation rate, and galaxy mass, including in dust-obscured galaxies, active galactic nuclei and possibly even γ-ray burst hosts. Within galaxies of the Local Group including the Magellanic Clouds, metallicity maps of the extraplanar gas or abundance gradients are useful to examine how metals may be ejected from and accreted onto galaxies. On the smallest scales, it is important to probe the dispersal and mixing spatial and time-scales of heavy elements through potential metallicity inhomogeneities/variations in the interstellar medium due to, e.g., local enrichment around nearby HII regions, Wolf-Rayet stars, supernovae, or planetary nebulae, or local dilution through metal-poor gas infall. For the brightest galaxies, some constraints can be put on nucleosynthesis models (production in supernovae Type Ia/II, primary vs. secondary pathways etc...), the gas-phase chemistry, (e.g., in the dense gas associated with ongoing star-formation), as well as dust composition and processing through depletion patterns (e.g., presence and properties of shocks through refractory elements).

The method to determine gas-phase ionic abundances in the infrared (IR) relies on calculating fine-structure level populations, with assumed or self-consistently calculated physical conditions (gas temperature and density), and ideally compared to Hydrogen recombination lines (RL) in the same spectrum. It is also possible to measure the metallicity without Hydrogen, relying instead on the metallicity-dependency of abundance ratios.

Some ionization corrections are then required for the missing ionization stages, usually determined from photoionization grids or specific models, in order to calculate elemental abundances. This technique has been used in the IR with ISO-SWS, Spitzer/IRS, Herschel/PACS, SOFIA/FIFI-LS, and now JWST, with or without direct access to H lines. The IR fine-structure lines have low excitation energies compared to optical collisionally excited lines (CELs) and, as such, their strength is almost independent on the electronic temperature, thereby mitigating known abundance discrepancies between CEL/RL-determined abundances (up to 0.4 dex; Esteban et al. 2002, 2016). Depending on the available tracers, the electron density can also be determined self-consistently by virtue of different critical densities, leading to the most accurate gas-phase abundances possible in any wavelength range if we ignore methods using the notoriously faint radio recombination lines. Finally, total abundances may be calculated by deriving or assuming depletion patterns to account for metals locked in dust grains.

## Why PRIMA?

JWST provides access to optical diagnostics including H recombination lines at high-z but dust obscuration quickly becomes problematic and most IR diagnostics (best for abundance determinations) become unavailable at z>1. We need an observatory to 1) access the diffuse gas not detected with previous IR telescopes in nearby galaxies, 2) add far-IR lines to complement the mid-IR line set for nearby galaxies observed, e.g., with JWST, and 3) enable access to useful mid-IR diagnostics at z≥1. We list below several potential configurations depending on redshift.

**Self-consistent abundance & metallicity measurements at z~0:** Considering the wavelength range of PRIMA, the best option is HI (9-8) at 27.8μm (typical flux ~0.02% of intrinsic, i.e., dust-corrected, Hβ). At z=0, one can calculate the abundances of Fe ([FeII], [FeIII]), N ([NII], [NIII]), O ([OIII], [OIV]) and also P ([PII]), Cl ([Cl III])... Examples of direct determinations in the IR include Lebouteiller et al. (2008) and Rubin et al. (2011).

**Non-self-consistent abundance & metallicity measurements with H recombination lines at z≥0:** For high-z galaxies, HI (9-8) quickly becomes too faint and it is necessary to rely on Huα, Pfα, Brα... from, e.g., JWST, Roman Space Telescope, Euclid... Extinction and aperture corrections are important limitations. Many diagnostics involving mid-IR lines, however, become accessible, with abundances from Ne ([NeII], [NeIII], [NeV]), Ar ([ArII], [ArIII], [ArV]), S ([SIII], [SIV]), as well as those mentioned above for z~0. Examples of such measurements in the IR include Bernard-Salas et al. (2001), Vermeij & van der Hulst (2002), or Wu et al. (2008).

**Metallicity measurements at z≥0:** In case no H lines is available, it is possible to estimate the metallicity using abundances ratios and a proper calibration for the metallicity-dependency, e.g., in the fashion of, e.g., Fernandez-Ontiveros et al. (2017), Pereira-Santaella et al. (2017), or Spinoglio et al. (2021a, 2021b).

### Instruments Required:

FIRESS pointed observations and low-resolution maps

### Program Size:

- Small (<10 hr)

### Approximate Integration Time:

For a given z∼0 source with HI (9-8)∼2×10$^{-18}$ W m$^{-2}$ (estimated from HI 7-6 in Cormier et al. 2015), 5 σ detection in FIRESS low-res mode would be achieved in just about a minute. The sample may include at a minimum the 48 sources from the Dwarf Galaxy Survey observed with Spitzer and Herschel (Madden et al. 2013).

### Special Capabilities Needed:

- None

### Synergies with Other Facilities:

Near-IR and/or mid-IR may be necessary to obtain H recombination lines for distant galaxies in order to measure ionic and elemental abundances.

### Description of Observations:

Observations would consist of:

- Sample of z∼0 galaxies for self-consistent measurements of abundances and metallicity, comparison with optical determinations. From Huα (HI 7-6 at 12.37 μm) in the Dwarf Galaxy Survey with Spitzer (Cormier et al. 2015), we estimate HI (9-8) around ∼2×10$^{-18}$ W m$^{-2}$ in that sample, definitely within PRIMA's capabilities. With PRIMA it becomes possible to access quiescent/low-surface brightness galaxies as well.

- Sample of bright galaxies at least until Cosmic Noon. High spectral resolution mode may be necessary to overcome confusion noise.

- Maps will be useful to examine the spatial distribution of abundances/metallicity in the Magellanic Clouds, e.g., within/around the 30 Doradus star-forming region.

Main tracers:

- H: HI 9-8 27.8 μm.

- N: [NII] 122, 205 μm, [NIII] 57 μm.

- O: [OI] 63, 145 μm, [OIII] 52, 88 μm, [OIV] 25.9 μm.

- Ne: [NeIII] 36 μm, [NeV] 24 μm.

- Si: [SiII] 34.8 μm.

- P: [PII] 32.9, 60.6 μm.

- S: [SIII] 33.5 μm.

- Cl: [Cl II] 33.3 μm.

- Fe: [FeI] 34.71 μm, [FeII] 25.98, 26.55, 35.33 μm, [FeIII] 33.03 μm, [FeV] 25.9 μm.

- In faint galaxies: lines from F and Co...

- Depending on redshift: many mid-IR lines (e.g., Pfα, Huα, [NeII], [ArII], [ArIII], [SIV]...).

## 32. Feedback in primitive galaxies: how do internal processes and feedback drive their transformation into the present day galaxy population?

Vianney Lebouteiller (Université Paris-Saclay, Université Paris Cité, CEA, CNRS, AIM, 91191, Gif-sur-Yvette, France), Pierre Guillard (Institut d'Astrophysique de Paris), Frédéric Galliano, Suzanne Madden (Université Paris-Saclay, Université Paris Cité, CEA, CNRS, AIM)

Feedback mechanisms are crucial in shaping and energizing the interstellar medium (ISM). In turn, star formation is regulated by a complex, non-linear, coupling between chemical processes, thermodynamical, as well as turbulent processes in the ISM. Heating mechanisms are generally not observable directly and our knowledge relies on cooling processes. While HII regions are relatively well understood, the star-forming, neutral, gas reservoir of galaxies is subjected to various mechanisms (cosmic rays, shocks, X-ray photoionization etc…) that need to be identified, constrained, and disentangled. Heating mechanisms are especially little known in metal-poor galaxies where fundamental differences regarding ISM conditions and the nature and properties of energetic sources are expected. At stake is the understanding of the formation of the first stellar generations.

### General Scientific Area:

Growth of Galaxies and Black Holes Through Cosmic Time

### Science Justification:

Feedback mechanisms (shocks, ionization by UV-X-rays photons or by cosmic rays, outflows...) are crucial in shaping and energizing the interstellar medium (ISM). In turn, star formation is regulated by a complex, non-linear, coupling between chemical processes, thermodynamical, as well as turbulent processes in the ISM, which ought to be examined in various environments, especially in metal-poor galaxies where fundamental differences regarding ISM conditions and the nature and properties of energetic sources are expected. At stake is the understanding of the formation of the first stellar generations.

We need to understand how the gas cools down, i.e., how the energy injected through various feedback mechanisms is dissipated, so that gas can condense and eventually form stars. The phases in which mechanical energy is dissipated are, for instance, particularly uncertain and difficult to constrain. The cooling lines accessible with PRIMA not only trace the dissipation of feedback energy, but they also contribute to the gas cooling, providing an invaluable tool to link spectral emission to dissipation.

The far-infrared domain provides access to many fine-structure cooling lines in the ionized (e.g., [OIII] 52,88 µm), neutral atomic (e.g., [CII] 157 µm, [SiII] 34.8 µm, [OI] 63,145 µm), and neutral molecular phase (e.g., $H_2$ S(0) 28 µm, high-J CO Jupper>15 lines). Of particular note, [SiII] has received relatively little attention in external galaxies because it could only be observed at z∼0 with Spitzer with a relatively poor sensitivity and it is not observed by JWST. It must be noted, however, that [SiII] can be an important coolant in the low-metallicity ISM where the silicon gas-phase abundance is enhanced due to the lower dust-to-gas mass ratio (e.g., Kaufman et al. 2006, Chevance et al. 2016, Lebouteiller et al. 2017).

Infrared cooling lines are observed at very high-redshift with ALMA and NOEMA (e.g., Wong et al. 2022) and hold tremendous diagnostic power on ISM physical conditions for star-formation, and more generally on the nature of energetic sources at work (mechanical, massive stars, cosmic rays, compact objects...) and consequently on many diagnostics involving, for instance, the star-formation history or the mass of molecular gas. Such diagnostics, however, are possible and reliable only if the relevant heating mechanisms are identified. While it will remain difficult to access multi-phase cooling lines in extremely metal-poor sources at high-redshift, star-forming dwarf galaxies in the nearby Universe are ideal laboratories to understand the interplay between feedback, ISM properties, and star formation in primitive environments.

## Why PRIMA?

Spitzer and Herschel have barely explored such mechanisms as shocks, turbulence, and ionization by cosmic rays in moderately metal-poor (few times below solar) galaxies notably thanks to the Dwarf Galaxy Survey; Madden et al. 2013, Cormier et al. 2015, Lebouteiller et al. 2017, Cormier et al. 2019). Moreover, only the brightest star-forming galaxies have been examined and the study of heating and cooling mechanisms could not be performed in relatively quiescent galaxies.

PRIMA will access multiple diagnostics involving fainter lines and better constraints for the various feedback mechanisms, through a large sample of metal-poor galaxies potentially up to Cosmic Noon. Thanks to intricate ISM models involving multiple phases and multiple components (e.g., Lebouteiller et al. 2022), theoretical predictions will finally be put to the test, in particular for galaxies with metallicities lower than 10%, solar for which fundamentally different energetics are expected as compared to even moderately metal-poor sources, with an expected lower importance of the photoelectric effect on dust grains in favor of ionization by cosmic ray particles and/or X-rays from compact objects (e.g., Bialy & Sternberg 2019).

PRIMA can tackle several objectives to understand how the evolution of metal-poor environments is regulated and identify the main triggers that transform them into chemically evolved galaxies:

- Determine the nature and properties of the physical sources that energize the low-metallicity ISM (e.g., compact objects, super-stellar clusters...).

- Infer heating mechanisms that may impact significantly the regulation of star formation and their predominance range as a function, for instance, of the ISM phase and metallicity.

- Probe outflows for the first time in the nearest extremely metal-poor galaxies.

- Revisit our understanding of classical cooling line diagnostics to trace the star-formation rate and extend metallicity calibration as compared to previous studies (de Looze et al. 2014, HerreraCamus et al. 2015).

## Instruments Required:

FIRESS pointed observations, at high res and low-res modes

## Program Size:

- Large (100 + hr)

## Special Capabilities Needed:

- None

## Synergies with Other Facilities:

- eROSITA and ATHENA will observe and characterize many X-ray binaries in nearby galaxies and PRIMA is the best mission to reveal their signatures in the neutral ISM.

- JWST should enable the detection of $H_3^+$ in a few objects, leading to a better description of the cosmic ray ionization rate and its possible variations as a function of the environment.

- Mid-IR photometry from Spitzer and WISE and spectroscopy from JWST and ELT may complement the mid-IR range below 24 µm and provide access to emission of polycyclic aromatic hydrocarbons for a full characterization of the photoelectric effect heating at z~0, although the latter is expected to contribute relatively less to the total heating as metallicity decreases.

- ALMA/NOEMA may access lower-J CO lines as well as molecular outflows in moderately metal-poor sources, while PRIMA can probe outflows through atomic tracers in XMP galaxies.

## Description of Observations:

Observations would consist of :

- Nearby star-forming metal-poor dwarf galaxies within ~200Mpc (including, e.g., DGS; Madden et al. 2013; Cormier et al. 2019). Time required: ~1h/object for the main lines (incl. relatively faint lines such as [NII] 122, 205 µm, [PII], [FeII], [FeIII], [OIV]...) in ~50 objects, with additional deep observations up to ~20h/object for the faintest lines due to high-mass X-ray binaries (e.g., [NeV]) for a selected few objects.

- Nearby low surface brightness metal-poor dwarf galaxies within 1 to ~10Mpc (including, e.g., LITTLE THINGS; Hunter et al. 2012). Time required: ~20h/object to access main lines including faint lines such as [NII] 122, 205 µm in ~10 objects.

- Recently discovered extremely metal-poor (XMP) galaxies with SDSS in the redshift range 0 to ~0.3 (e.g., Izotov et al. 2021). Required time: ~1h/object for the brightest lines ([CII], [OI], [OIII], [SiII], [FeII]) in ~100 objects.

Most objects will be unresolved spatially. High spectral resolution mode is required to probe atomic and molecular outflows through asymmetric and wide profiles as done with R~600 with Spitzer/IRS (e.g., Spoon et al. 2009a, 2009b).

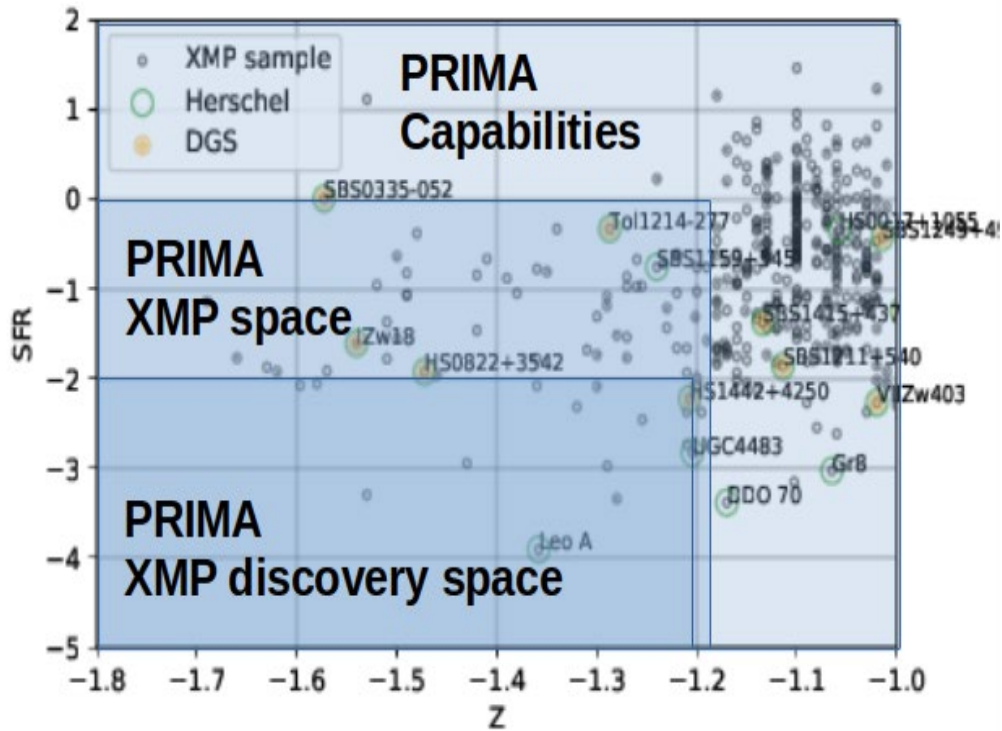

Figure 1: SFR vs. metallicity for a sample of extremely metal-poor galaxies. Most observed XMP galaxies are actively star-forming, and PRIMA will access for the first time numerous quiescent XMP galaxies.

**Main tracers:**

- Photoionization by X-rays: ionized gas (e.g., [OIV], [NeV]) and enhanced line fluxes near the ionization front ([SiII], [FeII], [Cl II]...).

- Shocks: [OI] 63 μm, high-J CO lines, $H_2O$, as well as [SiII], [FeII], and [FeIII] lines.

- Outflows: the low $H_2$ fraction in metal-poor galaxies dictates that outflows should be mostly probed through ionized gas (e.g., [OIII]) and neutral atomic gas (e.g., [CII]) tracers.

# 33. Dissecting the dust composition and magnetic topology of AGN tori with FIR polarimetry

Frédéric Marin (Université de Strasbourg, CNRS, Observatoire Astronomique de Strasbourg), Enrique Lopez-Rodriguez (Kavli Institute for Particle Astrophysics and Cosmology (KIPAC), Stanford University)

Active galactic nuclei (AGN) are powered by mass accretion from a surrounding circumnuclear reservoir of dust and molecular gas. This circumnuclear structure, so-called 'torus', is spatially complicated to resolve because it has a size of a few parsecs for objects situated at megaparsecs from us. They are key structures to understand the activity of supermassive black holes, from accretion to mass ejection. To explain the transfer of energy and mass to the potential well, magnetic fields are often suggested. Far-infrared (FIR) polarimetry has recently proven to be a powerful tool to probe the magnetic field orientation and turbulence in AGN tori, well-below the angular resolution of the observations thanks to the signature of magnetically aligned dust grains. The FIR observations showed that radio-loud AGN are highly polarized, ∼5–11%, while radio-quiet are lowly polarized, <1%. The interpretation is that the magnetic field may be the cause of the radio jet production or, reversely, it could be the effect of having a powerful radio jet. Yet, only a handful of very bright and nearby targets have been observed so far, due to instrumental constraints. The Band 2 of PRIMAger, being able to measure with high statistical significance the 80-265 µm polarization will open a new observational window for AGNs. PRIMAger will be able to provide statistical results of whether the magnetic fields surrounding AGNs are intrinsically related to the strength of the jet power (the stronger the jet power is, the larger and stronger the toroidal magnetic field). This study will provide a solution to the long-live problem of the radio loudness dichotomy in AGN. In addition, PRIMAger will also construct panchromatic polarized spectral energy distribution of AGNs and thus check for the presence of an active core in the most obscured targets. Such goals are within PRIMAger capabilities within a few tens of hours of observation.

## General Scientific Area:

Magnetic Fields

## Science Justification:

It is now widely accepted that a supermassive black hole (SMBH) lies at the center of virtually every massive galaxy. The large majority of black holes are quiescent, i.e., they accrete at low luminosity (e.g., $Lx \lesssim 10^{-6}$ LEdd, see Soria et al. 2006; Ho 2009; Gallo et al. 2010; Pellegrini 2010). However, once fed with a continuous flow of gas through accretion disks, they become active and are coined Active Galactic Nuclei (AGNs). The accreted material radiates across a wide range of wavelengths, part of this radiation being absorbed and then re-emitted as thermal emission at infrared wavelengths by the dust surrounding the SMBH.

According to the AGN unified model (Antonucci 1993; Urry & Padovani 1995), AGNs are believed to be surrounded by a circumnuclear reservoir of dust and molecular gas, often called a 'torus', that absorbs the ultraviolet and optical emission from the accretion disk and reradiate it, like a calorimeter, from the near-infrared to the sub-mm bands. The exact composition and geometry of this dusty region are poorly known, but the torus is likely expanding from the sublimation radius (where temperatures are larger than 1500 K, at a fraction of a parsec) to a few tens of parsecs. This structure is thought to act as the reservoir of SMBHs, i.e., part of the accretion flow, and/or the outflowing material supporting the loss of angular momentum of the accretion disk and collimation of the jets.

To explain the transfer of energy surrounding SMBHs, most of the theoretical models, if not all, must incorporate magnetic fields (B-fields). Matter rotating around a SMBH must have a B-field to account for the loss of angular momentum (Blandford & Znajek 1977, Blandford & Payne 1982). Although this process can explain the dynamics of matter at sub-parsec scales, there are still no clear observations of how the matter from the host galaxy, which contains kiloparsec-scale B-fields in close equipartition with the diffuse ISM (Beck et al. 2019), feeds SMBHs at 10-100 parsec-scales. The main reason resides in that B-fields remain difficult to characterize, until recently (Lopez-Rodriguez et al. 2018, 2021, 2022), due to the small size of the torus (~parsec) with respect to the cosmological distances of AGNs (~mega-parsecs).

A key point here is that dust grains will be oriented in a specific orientation with respect to the local B-field, and the emission, transmission, and absorption of light through magnetically aligned grains gives rise to (dichroic) polarization. Specifically, dust grains preferentially absorb radiation along one of their axes, making them spin up along their greatest moment of inertia (i.e., their short axis). Then, the dust grains acquire a magnetic moment, making them precess along the orientation of the local B-field. The final configuration is one where the short axis of the dust grains is parallel to the local B-field. Hence, by measuring the infrared polarization in AGNs, one can probe both the magnetic field orientation (for dichroic emission in the far-infrared the observed position angle of polarization is perpendicular to the orientation of the B-field), its intensity, and the chemistry of dust (different chemical compounds give rise to different wavelength-dependent polarization signatures).

The polarization capabilities of Band 2 of PRIMAger are thus the perfect opportunity to shed light on the orientation and intensity of B-fields in the tori of AGNs by probing the infrared polarization between 80 and 265 μm. FIR polarimetry is essentially an uncharted territory, only recently probed by SOFIA/HAWC+ (Vaillancourt et al. 2007; Dowell et al. 2010; Harper et al. 2018) at 53, 89, and 214 μm in a set of 15 nearby galaxies (Lopez-Rodriguez et al. 2022b) which gave extremely interesting results regarding the dichotomy between radio-loud and radio-quiet AGNs (Lopez-Rodriguez et al. 2022c).

Radio-loud and radio-quiet AGNs are often separated by the (overwhelming) presence of collimated jets of relativistic particles. However, the physical phenomena generating radio-loudness are still under debate. In the case of radio-loud objects, synchrotron emission from the jets often dominates the ultraviolet to near-infrared bands, while it is the emission from the accretion engine that dominates radio-quiet AGNs. However, in the infrared, dust reemission often prevails. The synchrotron polarization is negligible at far-infrared wavelengths, and the polarized far-infrared spectral energy distributions (SED) have similar shape as the infrared bump in total flux (Lopez-Rodriguez et al. 2018, 2022). This result is interesting because it makes FIR polarimetry wavelengths a new window of exploration into the physics of the environment surrounding the SMBH at the scales of tens of parsecs.

In particular, Lopez-Rodriguez et al. (2022c) found that the 89 μm dust emission cores of radio-loud AGNs are polarized in the range 5 – 11%, while radio-quiet AGN are essentially unpolarized (< 1%). This indicates that the size and strength of the B-fields surrounding AGNs are intrinsically related to the strength of the jet power (the stronger the jet power is, the larger and stronger the toroidal B-field). In other words, the measured B-field may be part of the accretion flow in radio-loud AGNs that ultimately supports the transfer of gas inwards and jet production. Alternatively, the measured B-field may be the signature of magnetized winds supporting the collimation of the jets in radio-loud AGN. However, the sample to reach final conclusion is very small (only 9 objects, and most of them with only a single measurement at 89 μm, see Figure 1).

PRIMA can provide a statistical sample by performing a larger survey of radio-loud and radio-quiet AGNs at redshifts 0–2. The AGNs in Lopez-Rodriguez et al. (2022c) were limited to objects with total intensity > 1 Jy integrated within the beamsize of SOFIA/HAWC+ (7.8" at 89 μm), and PRIMAger Band 2 is expected to go down as a few mJy (estimates of point source sensitivity including confusion reach). Based on the Photodetector Array Camera and Spectrometer (PACS) at the Herschel Space Observatory, hundreds of radio-loud and radio-quiet AGNs will be available for PRIMAger (Hardcastle et al. 2010, Melendez et al 2014). PRIMA is the only capability that can probe with great detail the radio-quiet and radio-loud AGN magnetic dichotomy by doing what other instruments cannot do: measure the B-fields in the densest region of the interface between the AGN and host galaxy.

As a by-product of such observations, PRIMAger will provide crucial data points to better reconstruct the infrared polarized SED of AGNs, to be related to ultraviolet, optical, near-infrared and radio polarized fluxes. As said, the polarized far-infrared SED of AGNs have the similar shape as the infrared bump in total flux, but the panchromatic polarized SED of AGNs has distinctive wavelength-dependent signatures that can be related to the AGN and host galaxy physics (Marin 2018, Marin et al. 2020), allowing us to check whether all type-2 AGNs observed with PRIMAger will contain type-1 cores, as expected from the unification model of AGNs.

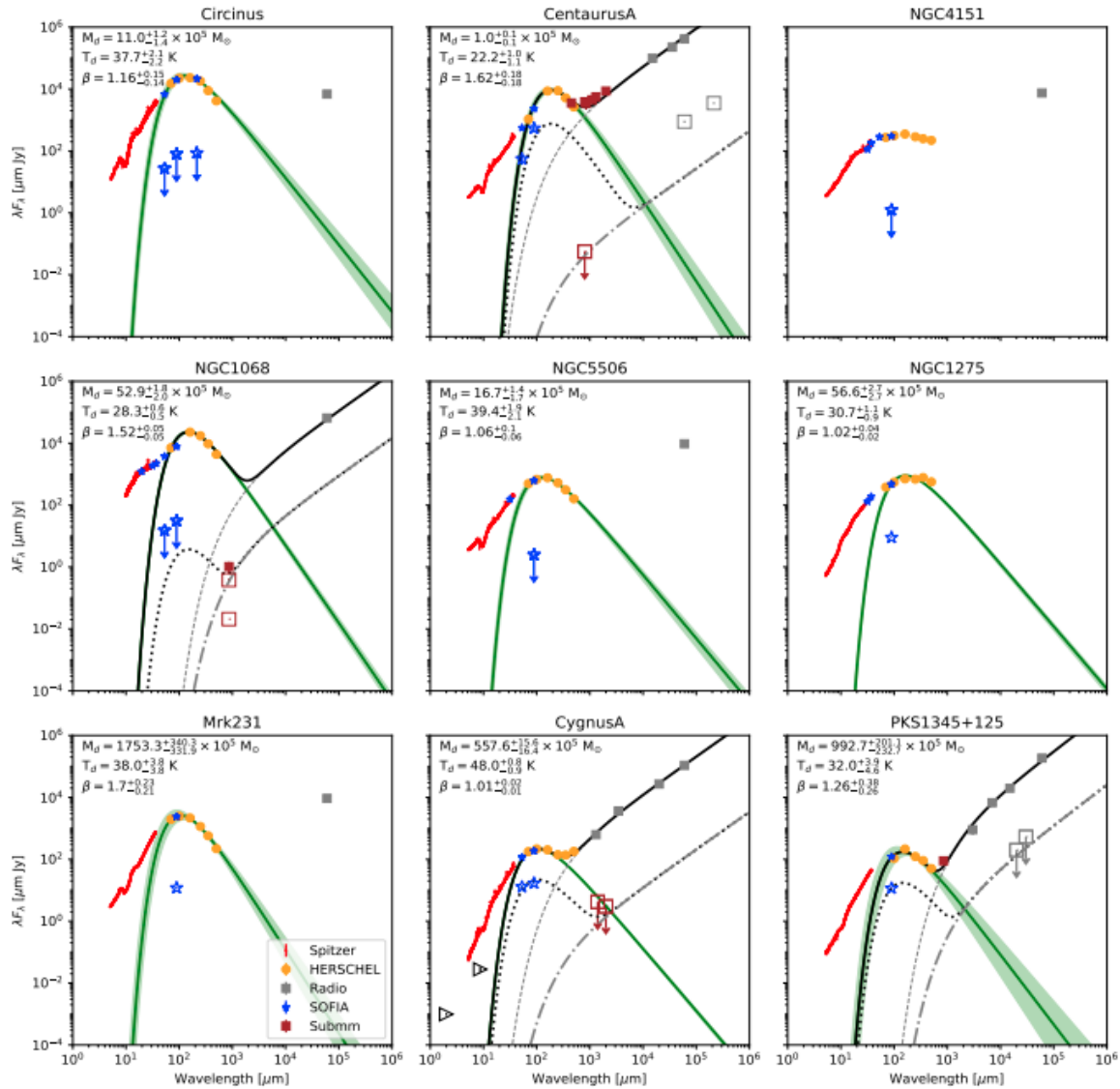

**Figure 1.** 5 μm - 5 GHz total and polarized nuclear SED of AGN in order of redshift. Total SED (filled symbols) using *Spitzer*, PACS/SPIRE/*Herschel*, FORCAST/HAWC+/SOFIA and radio data as shown in the legend of the bottom left panel. Besides the differences in the radio emission from these sources, the $10-1000$ μm modified blackbody function (green line, Eq. 1) shows a similar bump at $\lambda_d = 100 \pm 20$ μm arising from dust emission with a characteristic dust temperature of $T_d = 34.8 \pm 7.8$ K (Section 3.1.1). The best-fit parameters of the modified blackbody functions are shown in each panel. A fiducial synchrotron emission function (dashed grey line, Eq. 3) with a slope of $\alpha_{tk} = +0.21$ and $\alpha_{th} = -0.9$ in the optically thick and thin regimes, respectively with a turnover wavelength at 543 μm for Cygnus A and Centaurus A, and 2500 μm for PKS 1345+125 (Section 3.1.2) is shown. The total flux SED model (black solid line) is shown. Polarized flux SED (open symbols) using NIR (black triangle), SOFIA (open star), and sub-mm (open squares) are shown. The scaled synchrotron (green dashed dotted line) and thermal (black dotted line) polarized SEDs are shown for the polarized AGN (Section 3.3). The IR polarized SED of Centaurus A, Cygnus A, and PKS 1345+125 follow the shape of the IR bump in the total flux SED with negligible contribution of synchrotron emission.

## Instruments Required:

PRIMAGER small maps in polarimetry band

## Program Size:

- Medium (10–100 hr)

## Approximate Integration Time:

To reach a 5σ detection of quasar polarization in the PRIMAger Band 2, 10 hours of integration (in total for the 4 filters) are need for sources that are brighter than 0.8 mJy @ 96 µm (estimates of point source sensitivity including confusion for extragalactic science cases). Small maps (10'×10') resulting in a 4'×4' region covered by all four filters is enough for point-like sources such as quasars.

Table 1: Radio-loud and radio-quiet quasars FIR mean flux densities in the 100, 160, and 250 µm Herschel bandpasses (Kalfountzou et al. 2014). This catalog lists faint sources first detected in the radio band (but the FIR and radio luminosities are correlated).

| Class | z range | N per bin | F100µm (mJy) | F160µm (mJy) | F250µm (mJy) |
|---|---|---|---|---|---|
| Radio-quiet | 0.2–1.0 | 24 | 7.9 ± 1.9 | 7.6 ± 1.7 | 18.5 ± 2.2 |
| | 1.0–1.5 | 30 | 8.2 ± 1.8 | 16.7 ± 4.2 | 36.8 ± 2.1 |
| | 1.5–2.0 | 21 | 4.1 ± 1.4 | 7.1 ± 1.3 | 17.3 ± 2.3 |
| | 2.0–5.0 | 18 | 2.6 ± 1.4 | 5.6 ± 1.9 | 18.3 ± 2.3 |
| Radio-loud | 0.2–1.0 | 264 | 7.3 ± 0.5 | 9.9 ± 0.7 | 21.2 ± 1 |
| | 1.0–1.5 | 355 | 5.1 ± 0.6 | 8.2 ± 1 | 20.3 ± 1.6 |
| | 1.5–2.0 | 230 | 3.9 ± 0.4 | 9.5 ± 0.5 | 18.7 ± 0.8 |
| | 2.0–5.0 | 158 | 4.4 ± 0.6 | 7.4 ± 1.2 | 17.59 ± 1.7 |

Based on the Kalfountzou et al. (2014) list of faint sources, the approximate integration time to observe ~ 25 nearby radio-quiet AGNs and ~ 25 nearby radio-loud AGNs with the 4 filters of the PRIMAger Band 2 is ~ 50 hours (linear scaling P ∝ T) or ~ 160 hours (square-root scaling P ∝ √T) for 5σ detections in each filter. However, if one accounts for more standard, i.e., brighter, sources (80 mJy @ 100 µm), ~100 AGNs can be observed in ~ 10 hours (linear scaling) or ~100 hours (square-root scaling). This does not account for telescope dead-time (repositioning, calibration, Earth occultation...).

## Special Capabilities Needed:

Definition of stable polarized standards (planets can be used), 10% calibration uncertainty, precise estimation of the polarized instrumental background (~0.5–1%).

## Synergies with Other Facilities:

Ultraviolet, optical and radio telescopes that are sensitive to the polarization will be very useful for constructing the polarized SEDs of AGNs, in order to better determine the true contribution of dusty torus (with respect to the jet and host galaxy). Contemporaneous polarimetric observations will be enough, no need for coordinated observations. This program does not heavily rely on other facilities to achieve its prime goals.

## Description of Observations:

To reach the goals of this program, about 100 targets (an equal number of radio-quiet and radio-loud AGNs) will have to be selected. The list of targets is large and easily reaches thousands of sources. All sources should be situated at z<1 and with at least 80 mJy fluxes @ 100 µm.

We need to measure the polarization of discrete sources (point-like) with >5σ precision. To do so, we can use the minimal map in band 2, which is 10'×10'. It will result in a 4'×4' region covered by all four filters, more than enough for quasars.

Proper calibration of the instrument is crucial for reliable I, Q, and U measurements of the Stokes parameters, so that both a total flux and a polarized SED can be constructed. At each waveband of Band 2, we will be able to measure the B-fields in the densest region of the interface between the AGN and host galaxy. Crucial information can be obtained from the polarized SED if the PRIMAger observation is complemented with contemporaneous ultraviolet, optical, near-infrared, and radio observations.

Once the first sample is established, additional targets at 1 < z < 2 can be selected to determine if and how the AGN dichotomy evolves with time. This will require another medium-sized program (maybe a large one, depending on how the polarimetric capabilities of PRIMAger scale with the integration time).

# 34. Chemical evolution of dusty galaxies in the cosmic noon

Tohru Nagao (Ehime U.), Juan Antonio Fernández-Ontiveros (CEFCA), Takuya Hashimoto (U. of Tsukuba), Kohei Ichikawa (Tohoku U.), Hanae Inami (Hiroshima U.), Takuma Izumi (NAOJ), Borja Pérez-Díaz (IAA-CSIC), Enrique Pérez-Montero (IAA-CSIC), Luigi Spinoglio (INAF-IAPS), Tsutomu T. Takeuchi (Nagoya U.), Yoichi Tamura (Nagoya U.), Hideki Umehata (Nagoya U.), José Vílchez (IAA-CSIC), Takehiko Wada (NAOJ)

Metallicity is a key parameter characterizing the galaxy evolution, and its systematic measurements at various redshifts have been carried out mostly in the rest-frame optical bands. However, galaxies in the most actively evolving phase are deeply obscured by dust, and therefore beyond the reach of optical metallicity diagnostics. Since a large fraction of star-forming galaxies in the cosmic noon (1<z<4, where the cosmic star-formation rate density reached its peak) is dusty, new metallicity measurements of those galaxies with extinction-free diagnostics are crucial to understand the whole picture of the chemical evolution of galaxies in the cosmological time scale.

This GO program focuses on infrared luminous populations of galaxies such as ultra-luminous infrared galaxies (ULIRGs) and sub-millimeter galaxies (SMGs) in the cosmic noon, and proposes to measure their chemical abundances based on metallicity-sensitive fine-structure lines. More specifically, 75 infrared luminous galaxies in the range of 1<z<4 (25 objects in 3 redshift bins with delta_z = 1) are observed with PRIMA/FIRESS to detect fine-structure lines including [SIII]18.7,33.5, [SIV]10.5, [NeII]12.8, [NeIII]15.6,36, [NIII]57, and [OIII]52 (also [OIII]88 for galaxies at z<1.6). The combination of these emission lines enables us to determine independent O/H and N/O abundance ratios, the ionization parameter, and the gas density, without suffering from the dust extinction. This measurement will be a critical test to understand the metallicity evolution in the total population of galaxies, i.e., the whole picture of the cosmic chemical evolution.

## General Scientific Area:

Rise of Metals and Dust

## Science Justification:

**Broader context:**

Metallicity is a key parameter to understand the galaxy evolution because it is tightly connected to the processes that regulate the build-up of the stellar mass in galaxies. Hence, the content of heavy elements is highly sensitive to the past star-formation history, the gas inflows from the surrounding environment, and the gas outflows due to energetic events such as starburst and AGN activity (e.g., Maiolino & Mannucci 2019). So far, rest-frame optical diagnostics (e.g., Nagao et al. 2006) have been used to measure the ISM metallicity of many star-forming galaxies up to the cosmic noon, i.e., the redshift of z~1–4, and consequently the redshift evolution of the mass-metallicity relation has been reported (e.g., Maiolino et al. 2008). However, rest-frame optical diagnostics are powerful only for optically thin galaxies, whereas most of star-forming galaxies in the cosmic noon are heavily obscured by dust (e.g., Finkelstein et al. 2015). Interestingly, in the low-z Universe, rest-frame optical diagnostics for massive (>$10^{10.5}$ M$_\odot$) dusty galaxies (ultra-luminous infrared galaxies; ULIRGs) indicate significantly lower metallicity than the expectation from the mass-metallicity relation (Rupke et al. 2008; Caputi et al. 2008). This may suggest the chemically younger nature of dusty galaxies compared to optically thin (less dusty) galaxies. Another possible interpretation is that we observe only the optically thin part of ULIRGs selectively. These two possibilities cannot be distinguished by rest-frame optical diagnostics in principle.

**Science question:**

The true chemical nature of dusty galaxies, especially in the cosmic noon where dusty galaxies dominate the cosmic star-formation rate density, remains an open question. To address this, metallicity diagnostics in infrared wavelengths are required in order to avoid the dust-reddening effect. One of such infrared diagnostics is the N3O3 index, which consists of [OIII]52,88 (or only [OIII]52) and [NIII]57 lines (Nagao et al. 2011; Pereira-Santaella et al. 2017; Rigopoulou et al. 2018; Peng et al. 2021). The solar-like N/O abundances traced by the N3O3 index using Herschel and SOFIA measurements in low-z ULIRGs (z~0.1) suggest their metal-rich nature (i.e., consistent to the mass-metallicity relation), which strongly supports the idea that ULIRGs are already matured chemically at least in the local Universe (Rigopoulou et al. 2018; Herrera-Camus et al. 2018; Chartab et al. 2022; see also Inami et al. 2013). Nevertheless, deviations from the local O/H–N/O calibration, as those reported by Peng et al. (2021) and Spinoglio et al. (2022), point to a more chemically complex scenario with sub-solar metallicities –closer to the optical determinations– and high N/O abundances. In short, the chemical nature of ULIRGs is highly controversial even in the low-z Universe, and future studies require an independent determination of the O/H and N/O abundances (e.g., Fernández-Ontiveros et al. 2021, Pérez-Díaz et al. 2022). This makes the situation even more uncertain at the cosmic noon, because the infrared-based ISM metallicity (based only on the N3O3 index) has been reported for only a few gravitationally magnified galaxies such as SDP.11 at z=1.78 (Lamarche et al. 2018) and HLSW-01 at z=2.96 (Rigopoulou et al. 2018). As for unlensed galaxies in the cosmic noon, only a lower limit of the N3O3-based ISM metallicity is inferred even through the stacking analysis of Herschel data (Wardlow et al. 2017). As a conclusion, a super-sensitive infrared spectroscopic capability is needed to provide a complete picture of the chemical evolution of galaxies during the cosmic history.

**Need for PRIMA:**

To measure the metallicity of dusty galaxies in the cosmic noon, a cooled infrared space telescope with a high sensitivity and wide wavelength coverage is needed. PRIMA perfectly satisfies such requirements, thanks to its extremely high sensitivity (~1.5 dex deeper than Herschel/PACS) and wide enough wavelength coverage (~24–230 µm with only 2 exposures). The targets of this program are 25 dusty galaxies with $L_{IR} > 10^{12.2}$ $L_\odot$ in 1<z<2, 25 with $L_{IR} > 10^{12.5}$ $L_\odot$ in 2<z<3, and 25 with $L_{IR} > 10^{12.9}$ $L_\odot$ in 3<z<4 (i.e., 75 objects in total; see Figure 1), to achieve moderately high statistics in each redshift bin. With the pointing observation mode of PRIMA/FIRESS (in the low-resolution mode; R~130), the faintest line ([NIII]57) among useful emission lines can be detected with ~1–2 hr integration for each target (i.e., ~200 hr in total since 2 exposures are needed to cover the whole spectral range). Note that this program cannot be realized by JWST nor ALMA, because this program requires a high sensitivity in the whole wavelength range of ~24–230 µm, while JWST/MIRI covers only <30 µm (maximum redshift coverage of z<0.6 for inclusion of [SIII]18.7µm) and ALMA/Band10 covers only >300 µm.

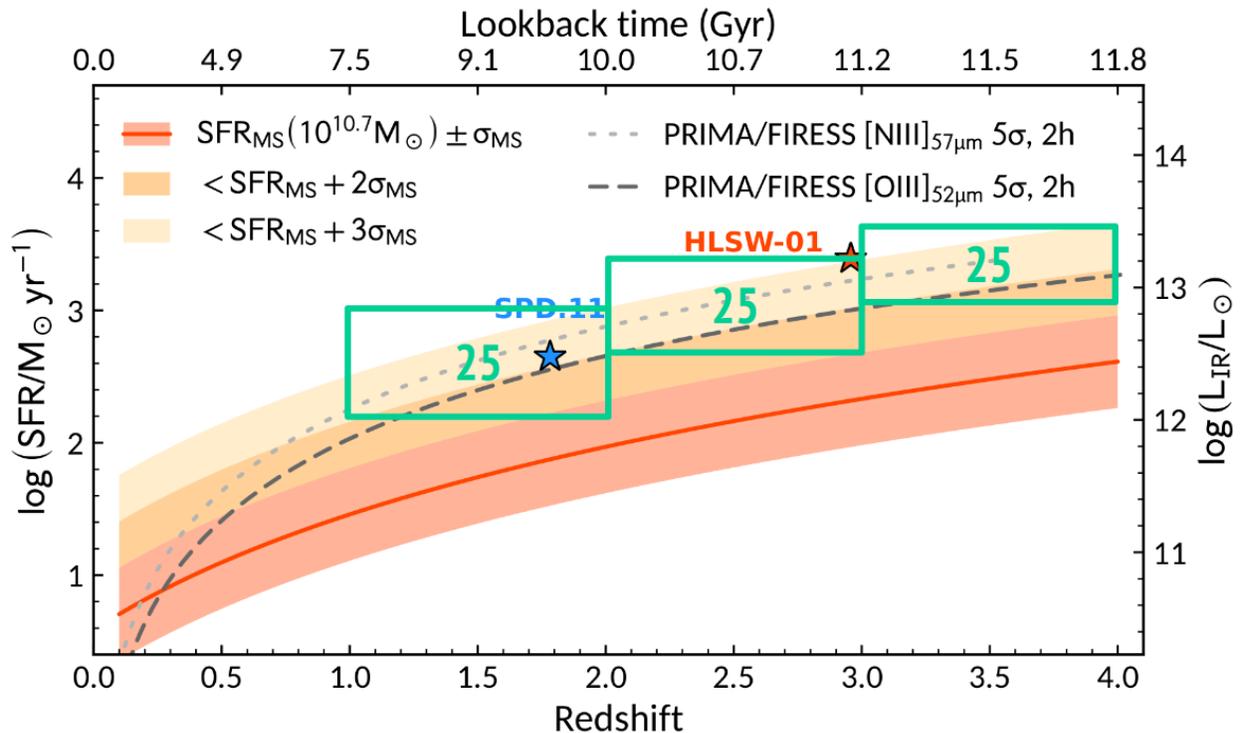

Figure 1: PRIMA/FIRESS detection limits of [OIII]52 (dashed line) and [NIII]57 (dotted line) on the plane of log SFR (or log LIR) versus redshift (5σ detection with 2 hr integration). The red solid line denotes the SFR of main-sequence galaxies at the knee of the mass function ($10^{10.7}$ $M_\odot$; Scoville et al. 2017), while the red, orange, and yellow shaded regions denote the ranges of SFR$_{MS}$ ± 1σ$_{MS}$, < SFR$_{MS}$±2σ$_{MS}$, and < SFR$_{MS}$±3σ$_{MS}$, respectively, with σ$_{MS}$ = 0.35 dex as the intrinsic scatter around the main sequence (Schreiber et al. 2015). Green boxes show the parameter range covered by this program. Blue and red stars denote lensed galaxies with a [NIII]57 detection (see the main text), but the intrinsic SFR (i.e., after the magnification correction) is plotted. These two data plots suggest that PRIMA/FIRESS can detect [NIII]57 without the lensing magnification effect with a reasonable integration time.

## Interpretation methods:

Thanks to the wide spectral coverage of PRIMA/FIRESS (~24–230 µm with only 2 integrations), various redshifted emission lines can be efficiently observed for each target. Although past metallicity diagnostics use 2 or 3 lines, the larger number of emission lines provide the more accurate metallicity estimates because the effect of the variance of some important parameters such as the gas density, ionization parameter, and N/O relative abundance ratio, can be considered. For such metallicity estimates based on infrared emission lines, a code HII-CHI-mistry-IR (https://home.iaa.csic.es/~epm/HII-CHI-mistry.html) can be used (Fernández-Ontiveros et al. 2021; Pérez-Díaz et al. 2022; see also Pérez-Montero 2014). In this code, the emission lines of [NeII]12.8, [NeIII]15.6, [SIII]18.7,33.5, [SIV]10.5, [OIII]52,88, and [NIII]57 are simultaneously used to diagnose the ISM metallicity, taking the effects of the gas density, ionization parameter, and N/O relative abundance ratio into account. Some lines are missed by FIRESS at a specific redshift range (e.g., [SIV]10.5 at z<1.3, [OIII]88 at z>1.6, [OIII]52 and [NIII]57 at z>3.4), but the best-effort metallicity is inferred by using available lines flexibly by HII-CHI-mistry-IR. Measuring the optical-based metallicity for dusty galaxies at 1<z<4 is challenging due to their faintness in optical due to the dust extinction, but 30 m-class ground-based telescopes such as ELT, TMT, and GMT will be available in the PRIMA era and thus extremely deep optical spectroscopic observations can be carried out to the targets in this program. As HII-CHI-Mistry (Perez-Montero 2014) will allow us to perform the same metallicity diagnostics but for optical emission lines, the comparison of the infrared-based and optical-based metallicity under the same prescriptions will reveal the chemical evolutionary stage of dusty galaxies in the cosmic noon, which is the main epoch of the galaxy growth.

## Instruments Required:

PRIMA FIRESS Spectrometer: Pointed Low-res R~130, 75 targets

## Program Size:

- Large (100 + hr)

## Approximate Integration Time:

To estimate the required integration time approximately, the empirical relations among $L_{IR}$ and $L_{line}$ (Mordini et al. 2021) are adopted. By targeting galaxies with $L_{IR} > 10^{12.2}$ $L_\odot$ in 1<z<2, $L_{IR} > 10^{12.5}$ $L_\odot$ in 2<z<3, and $L_{IR} > 10^{12.9}$ $L_\odot$ in 3<z<4, galaxies with a higher star-formation rate (SFR) than the main sequence plus ~1 σ are observed (see Figure 1). In this case, ~2 hr integration is needed to achieve the 5 σ detection of the faintest line ([NIII]57) among useful emission lines, assuming the FIRESS spectroscopic sensitivity of 2 x $10^{-19}$ W/m$^2$ (5 σ, 1 hr). However the line flux estimated here is underestimated (or equivalently, the required observing time is overestimated), because the Herschel [NIII] detections suggest systematically higher fluxes than our empirical estimates (based on the line fluxes reported by Lamarche et al. 2018; Rigopoulou et al. 2018). By considering also this fact, the estimated observing time required to detect all useful emission lines in the full spectral coverage is ~1–2 hr per spectral setting (i.e., ~2–4 hr per target) times 75 targets resulting in ~200 hr.

## Special Capabilities Needed:

- None

## Synergies with Other Facilities:

Comparisons between infrared-based metallicities (measured with PRIMA) and optical-based metallicities of dusty galaxies are interesting, but the rest-frame optical spectroscopy of dusty galaxies in the cosmic noon is very challenging for the existing ground-based telescope given their faintness (due to the dust extinction) in optical. Forthcoming 30m-class telescopes (ELT, TMT, and GMT) will be available in the PRIMA era, which makes such comparative studies feasible.

## Description of Observations:

This GO program targets 75 dusty star-forming galaxies, which consists of 25 objects at 1<z<2, 25 objects at 2<z<3, and 25 objects at 3<z<4, to cover the cosmic noon. Ideally the targets should be selected in an unbiased way, and one of such selections is to rely on FIRESS blind spectroscopic survey(s). Actually, one FIRESS blind spectroscopic survey is proposed by Spinoglio et al. (for 10x10 arcmin$^2$), but the achieved cosmic volume is insufficient to select luminous infrared galaxies in the cosmic noon. Thus, some targets will be selected from samples of known ULIRGs or SMGs, though PRIMAger wide-field deep imaging survey may be able to select the whole target sample for this program. Although some of the targets may contain powerful AGNs that can be identified by the obtained FIRESS spectrum ([NeV]24.3 and/or [OIV]25.9), the metallicity of AGNs, not only star-forming galaxies, can be inferred by using HII-CHI-mistry-IR (Pérez-Díaz et al. 2022). The observations are carried out in the standard manner with the pointed low-resolution mode of FIRESS. The full spectral coverage should be covered for all targets, so two spectral settings are needed for each target.

# 35.  A PRIMA "wedding cake" of blank-field surveys

Seb Oliver (University of Sussex, UK), James Donnellan (University of Sussex, UK), Luigi Spinoglio (IAPS INAF, Rome, Italy), Hanae Inami (Hiroshima University, Japan), Juan Antonio Fernández-Ontiveros (CEFCA, Teruel, Spain), Matt Malkan (UCLA, Los Angeles, USA), Tohru (Nagao, Ehime University, Japan), Amit Wishwas (Cornell University, USA)

Blank field surveys have been an essential tool for extragalactic studies, particularly surveys at Far Infrared (FIR) wavelengths which have a unique role in probing dust enshrouded star formation activity and measuring galactic nuclear activity from dusty tori. Previous space observatories and ground-based surveys have made groundbreaking discoveries in the field, including measuring the cosmic infrared background, constraining models of galaxy formation, and mapping clustering to constrain halo models. Despite their successes, the surveys still leave vast regions of parameter space unexplored. The proposal aims to fill some of the gaps in survey parameter space with an efficient survey design, and the key factors in survey design are discussed, including wavelength, area, depth, location, and geometry. We propose a specific a tiered set of surveys, a "wedding cake".

## General Scientific Area:

Growth of Galaxies and Black Holes Through Cosmic Time

## Science Justification:

### Introduction:

A fundamental challenge in astrophysics is to understand the evolution of galaxies and active galaxies over the history of the Universe. An essential tool is the statistical study of populations seen at different epochs (redshifts). Blank field surveys have thus been an essential and highly successful observational tool for extragalactic studies. Surveys at Far Infrared wavelengths have a particular role to play as this probe emission signposting and measuring dust enshrouded star formation activity and a relatively orientation independent measure of galactic nuclear activity from dusty tori.

Noteable surveys from space observatories in the FIR include IRAS (all-sky, Beichmann et al. 1986), COBE FIRAS, ISO (e.g., ELAIS, Oliver et al 2000), Spitzer (e.g., SWIRE, Lonsdale et al. 2004), Akari (all-sky), Planck (all-Sky), Herschel (HerMES, Oliver et al 2012, PEP, Lutz et al. 2010, HATLAS, Eales et al. 2009). These are complemented by surveys from the ground at longer wavelengths including with SCUBA-2 (e.g., Geach et al. 2020) and ALMA (e.g., Dunlop et al. 2017). These surveys have had profound influence on our statistical understanding of galaxies and AGN and have made ground-breaking serendipitous discoveries that were not widely expected.

Statistical success include: measuring the cosmic infrared background, demonstrating the roughly equal contribution of obscuration to the observed energy budget (Puget et al. 1996, Fixen et al); constraining models of galaxy formation with number counts (e.g., Oliver et al. 2010) or luminosity functions (e.g., Gruppinoi et al. 2012); tracing of the cosmic star formation history (e.g., Rowan-Robinson et al. 1997, Burgerella et al. 2013); mapping clustering to constrain halo models (e.g., Viero et al. 2010). Serendipitous discoveries include: IRAS F10214 (Rowan-Robinson et al. 1990), HDF 850.1 (Hughes et al. 1997); large samples of gravitational lenses (e.g., Negrello et al. 2010); HFLS 3 (Riechers et al. 2013).

However, despite their successes, the surveys have left unexplored vast regions of parameter space. E.g., Herschel surveys have covered large areas, but limited by confusion at longer wavelengths (≥ 250μm) and telescope temperature at shorter wavelengths (70<λ<250μm). Spitzer covered shorter wavelengths (λ<70μm) but with low sensitivity, while ALMA covers only very small areas. This means that, e.g., our current understanding of the cosmic star formation history relies on an extrapolation from statistical behaviour seen at low redshifts and high luminosities to higher redshift and lower luminosities and there remains many possibilities of theory-busting serendipitous discoveries. This proposal will fill some of the gaps in survey parameter space with an efficient survey design.

Moreover, a selection of galaxies detected by the PRIMAger large field survey based on the 12μm rest-frame band will be unbiased for both AGN, and Star forming galaxies (Spinoglio et al 1995) because of the linear relation between the 12 μm luminosity and the bolometric luminosity for both classes of objects, with a different bolometric correction. Moreover, the 12μm AGN selection is not biased for AGN type 1 or 2 sources (Spinoglio & Malkan 1989). So, the outcome of a wide and deep survey will be the definition of complete and unbiased samples of galaxies at different redshift, that can be further characterised with PRIMA-FIRESS spectroscopy. Fig. 1 illustrates the PRIMAger sensitivity for AGN and SF as a function of redshift.

## Survey Design:

The area affects the number of rare objects. This provides a limit at the high luminosity end of luminosity functions and constrains AGN studies. In addition, for statistical studies, including environmental dependency, one needs to cover a representative survey volume, likely to include enough of the most massive dark matter halos - this argues for areas of ~30 deg² (e.g., Jarvis et al.), smaller areas can probe mean properties in modest environments until about 1 deg² when sampling variance begins to be a limit.

A single survey of large area and deep depth would be ideal scientifically, but is costly observationally. So, a more efficient strategy is a tiered set of surveys, a "wedding cake". For the halo mass considerations above we would argue that the largest tier should be 30 deg², while for sampling variance considerations the smallest should 1 deg².

For the more detailed design we follow the methodology in Oliver et al. 2012. We focus on the objective of tracing the star formation history. We characterise this as probing the far infrared luminosity density, i.e., the product of luminosity and number density (mapping this to SFR density, shown as the contour plot in Figure 1). Thus, low luminosity galaxies with high number density and galaxies with high luminosity but low number density are less significant. Herschel surveys with SPIRE did not cover the peak of the luminosity (SFR) density.

We take the SFR/luminosity density from the SIDES models of Bethermin et al. 2017 as a function of redshift and Luminosity or Star formation rate. Any given survey can then constrain this parameter space limited on the low luminosity or SFR side by the depth and limited on the high luminosity side by numbers and thus number density and area.

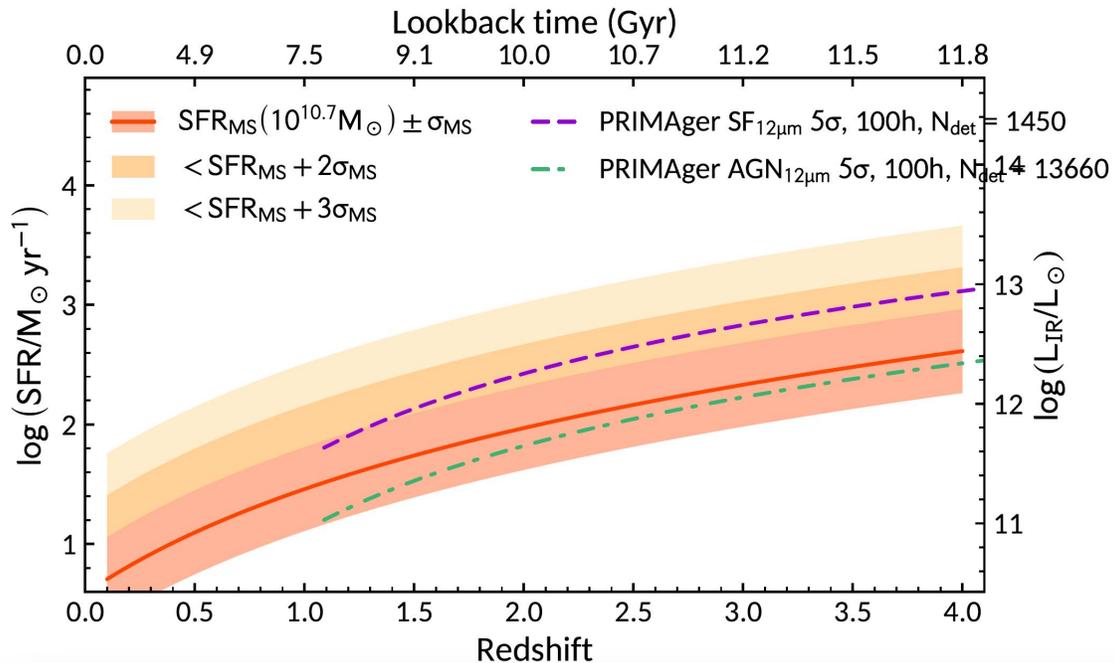

Figure 1: Using the Wang et al (2019) luminosity function and the 12μm to bolometric luminosity relations from Spinoglio et al. 1995 and 2023 (in prep.), we derive the following numbers of galaxies and AGN in a 1 deg² field at a 5σ sensitivity of 70 μJy in the 25–45 μm band integrating for 100 hours.

Our confusion noise estimates come from the photometric fluctuations in maps from the SIDES simulations of Bethermin et al. 2023 (Technical Notes). And our instrumental noise and integration times come from the PRIMAGER fact sheet. Our limit on numbers is defined to detect 25 galaxies in a dLogSFR=0.1, dz=0.1 bin for the number densities in the SIDES simulations.

We have considered a number of survey areas and the time taken to reach the confusion limit for each of the PRIMAger bands but consider that the combination of 3 surveys of size 1 deg², 5 deg² and 30 deg² each with 300 hr observation time provides a good compromise.

We show these parameter space plots from these three surveys in Figure 2.

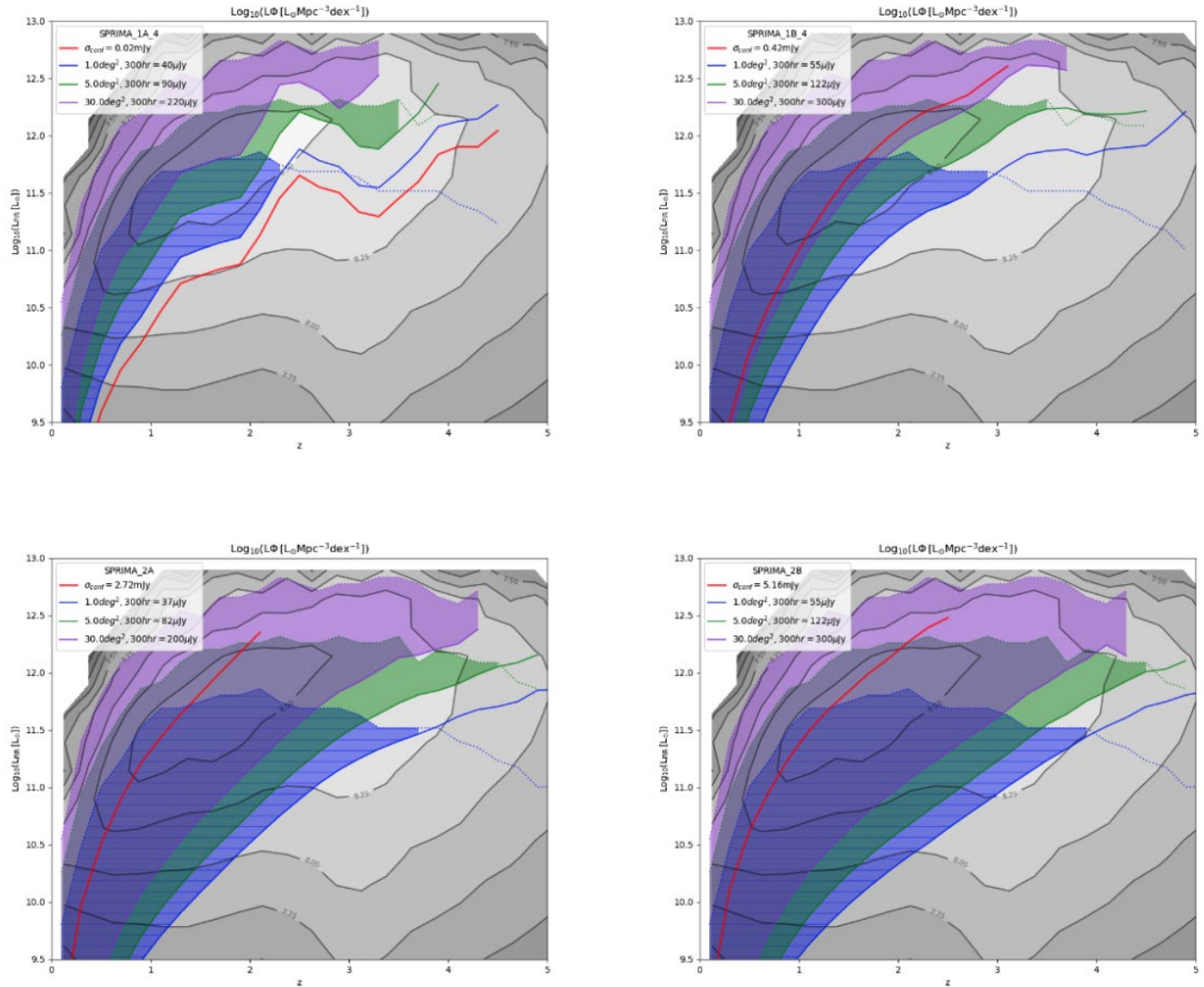

Figure 2: FIR luminosity density (grey-scale and contour diagram) as a function of FIR luminosity and redshift as calculated from the 2.0 sq. deg SIDES simulation. Overplotted are the regions of the luminosity-redshift plane covered by surveys of areas 1, 5 and 30 deg$^2$ in 300hr each. The solid red line shows the 5σ confusion limit for each PRIMAger band.

## Instruments Required:

PRIMAGER large map in both hyperspectral and polarimetry bands

## Program Size:

- Large (100 + hr)

## Approximate Integration Time:

We propose a 3-tiered set of surveys observing in both PRIMAger bands, with areas of 1 deg$^2$, 5 deg$^2$ and 30 deg$^2$. Each survey will require ~300hr. Since we have deep 1 deg$^2$, our 5 deg$^2$ thus just needs 4 deg$^2$ of new data (saving 300/5 =60 hr). Likewise, our 30 deg$^2$ only needs 25 deg$^2$ extra saving 300/6=50 hr. Therefore, we are requesting a total observation time of ~790 hr.

**Special Capabilities Needed:**

- None

**Synergies with Other Facilities:**

- JWST, possibly MATISSE at ESO (the 10 μm silicate feature).

**Description of Observations:**

We propose a 3-tiered set of surveys observing in both PRIMAger bands, with areas of 1 deg$^2$, 5 deg$^2$ and 30 deg$^2$. From Figure 2, in order to push towards the confusion limit in band 1A, and to cover as much of the luminosity density peak as possible, each of the surveys will require ~300 hr. The luminosity density peak will be entirely covered in band 1B-2D. For shorter observation times, the peak will still be covered for bands 2A-2D, however their respective confusion limits will severely impact the population of sources recovered. Moreover, such a survey will provide a complete and unbiased sample of AGN, and star-forming galaxies at different redshifts, as per Figure 1.

# 36. Identifying extreme star-forming galaxies with the PRIMA mission

Seb Oliver (University of Sussex, UK), James Donnellan (University of Sussex, UK)

One of the key challenges in studying galaxy formation and evolution is understanding the properties of the most extreme star-forming galaxies in the Universe. Herschel has observed extremely bright, hyper-luminous 250 μm sources which exhibit extreme star formation. If the star formation rates of these sources are confirmed, they pose a serious challenge for current galaxy formation models. PRIMA will allow us to target a sample of ∼ 100 Herschel 250 μm-selected, hyper-luminous candidates. FIRESS will obtain spectra of these sources to help determine their energy source or determine if they are composed of multiple components. PRIMAger will provide high resolution imaging of the candidates.

## General Scientific Area:

Growth of Galaxies and Black Holes Through Cosmic Time

## Science Justification:

### Introduction:

Galaxies are the basic building blocks of the Universe, and the formation and evolution of galaxies is one of the most fundamental topics in astrophysics. One of the key challenges in studying galaxy formation and evolution is understanding the properties of the most extreme star-forming galaxies in the Universe. In this proposal, we aim to identify and study these extreme objects using the planned PRIMA mission.

### Scientific Motivation:

There have been claims of observing extreme star-forming galaxies in the Universe, Rowan-Robinson et al. (2018) have reported the discovery of a population of extremely dusty galaxies that appear to be undergoing a massive burst of star formation. These objects exhibit star formation rates that are several hundred times higher than typical star-forming galaxies in the local universe. However, these extreme objects pose a significant challenge to theoretical models of galaxy evolution.

Current models of galaxy formation and evolution suggest that star formation is regulated by the interplay between gas inflows, outflows, and feedback from supernovae and black holes (e.g., Madau and Dickinson 2014). However, these models struggle to explain the existence of galaxies with such extreme star formation rates. Some models have proposed that these objects may be the result of major mergers between galaxies (e.g., Di-Matteo et al. 2008), while others have suggested that they may be powered by accretion onto massive black holes (e.g., Volonteri et al.

2015). However, these models are still not able to fully explain the observed properties of extreme star-forming galaxies.

One other explanation for these bright sources is that they are not actually one single source but instead are multiple component sources which have been blended. Several studies have found that some bright 250 μm Herschel sources are actually composed of multiple components (Yan et al. 2014, Scudder et al. 2016). Hodge et al. 2013 used the higher resolution of ALMA to identify that 35-50% of LESS sources resolved into multiple sub-mm sources.

## Objectives:

Our objectives are twofold: (1) To identify and study a sample of ~100 hyper-luminous far-infrared galaxy candidates selected from the Herschel mission, specifically from the HerMES survey and detected with the SPIRE instrument at 250 μm; and (2) To determine whether the energy source in these galaxies is star formation or AGN activity.

To achieve these objectives, we will use two instruments on the proposed PRIMA mission: PRIMAger, the PRIMA Imaging Instrument, and FIRESS, the PRIMA Far-Infrared Enhanced Survey Spectrometer. We will use the short wavelength PRIMAger bands (1A, 1B, 2A, 2B) from 25 to 141 μm to obtain higher resolution imaging of the selected hyper-luminous far-infrared galaxy candidates. This will allow us to distinguish between individual objects in the cases where the Herschel data leaves ambiguity about whether the detected emission comes from one object or more than one.

In addition, we will use the FIRESS to obtain spectra of these galaxies. The spectra will provide valuable diagnostic information about the metallicity and ionization strength of the underlying energy source. If a blended source contains objects at different redshifts, the spectra will also provide a spectral decomposition to distinguish between the different sources.

## Feasibility:

Our observation will require a total of ~ 50 hr telescope time, which is feasible given the planned capabilities of the PRIMA mission. The PRIMAger and FIRESS instruments will provide us with the necessary resolution and sensitivity to achieve our scientific goals.

## Conclusion:

Our proposed observation of a sample of hyper-luminous far-infrared galaxy candidates using PRIMAger and FIRESS on the proposed PRIMA mission will provide valuable insights into the physical properties and energy sources of extreme star-forming galaxies. By studying a large sample of these galaxies, we can place valuable constraints on theoretical models of galaxy formation and evolution. This proposal will make a significant contribution to our understanding of galaxy evolution and the most extreme environments in the Universe.

## Instruments Required:

- PRIMA FIRESS: Pointed Low-res R~130

- PRIMAger: Small Map (< 20'×20') ~100 targeted objects, Hyperspectral band (25–80 μm; 12 filters; R=10), Polarimeter band (96, 126, 172, 235 μm; R=4)

## Program Size:

- Medium (10–100 hr)

## Approximate Integration Time:

Using the FIRESS sensitivity curve, we expect these bright 250 μm sources to be detected at 10σ in ~20 minutes (including overheads). For the whole 100 target sample, we therefore request a total FIRESS observation time of ~ 33 hr. For the follow-up PRIMAger imaging of these sources, we expect them to be detected above 5σ in 10 mins (including overheads) and therefore require a total PRIMAger observation time of ~ 17 hr. As such, **for the entire program we request ~ 50 hr of observation time.**

## Special Capabilities Needed:

- None

## Synergies with Other Facilities:

- ALMA
- JWST

## Description of Observations:

The goal of this proposal is to obtain spectra of ~ 100 targeted hyper-luminous, Herschel selected sources using the FIRESS instrument in its pointed low-resolution mode. We have estimated that the spectra of these sources will be detected above 5σ in ~ 20 minutes (including overheads) and therefore will require a total observing time of ~ 33 hr for the whole ~ 100 target sample.

In addition, we will use the short wavelength PRIMAger bands (1A, 1B, 2A, 2B) from 25 to 141 μm to obtain higher resolution imaging of the selected targets. We expect these to be detected above 5σ in 10 mins (including overheads) and therefore require a total PRIMAger observation time of ~ 17 hr. As such, **for the entire program we request ~ 50 hr of observation time.**

# 37. Coevolution of star formation and black hole accretion at the Cosmic Noon

Luigi Spinoglio - IAPS INAF, Rome, Italy, Hanae Inami, Hiroshima University, Japan, Juan Antonio Fernández-Ontiveros, CEFCA, Teruel, Spain, Matt Malkan, UCLA, Los Angeles, USA, Tohru Nagao, Ehime University, Japan, Amit Wishwas, Cornell University, USA

PRIMA will be able to collect mid- and far-IR observations that simultaneously observe the star-formation and black hole activity in hundreds of galaxies, using detailed spectroscopy of dust features, ionized and neutral atomic gas lines and warm molecular gas emission. The spectroscopic mapping capability of PRIMA will provide an entirely unbiased estimate of the cosmic star formation rate density (SFRD) and black hole accretion rate density (BHARD) and allow us to explore both processes in samples of galaxies at the cosmic noon for the first time in the mid-infrared. This can be achieved through a deep spectroscopic survey using FIRESS in low resolution mode (R=130) in a blind field of $10 \times 10$ arcmin$^2$. To cover an area of $10 \times 10$ arcmin$^2$, we have estimated a total observing time (2 settings) of about 400 hours. The same area can be imaged at the required depth (20–30 μJy) in low resolution mode (R∼10) by the PRIMAGER in about 25 hours. So, the total observing time of this program is ∼ 425 hours.

The main fine structure lines to measure the BHAR are the high ionization lines of [OIV]26μm and [NeV]14.3 and 24.3 μm (see e.g., Spinoglio et al. 2022). The main features to measure the SFR are the PAH emission features at 6.2, 7.7 and 11.3 μm in dusty solar-like metallicity galaxies, and the [NeII]12.8μm, [NeIII]15.6 μm and [SIV]10.5 μm emission lines in low-metallicity star-forming galaxies, where these lines become extremely bright due to the strong radiation fields (e.g., Cormier et al. 2015, 2019).

## General Scientific Area:

Growth of Galaxies and Black Holes Through Cosmic Time

## Science Justification:

### Broader context

A complete study of galaxy evolution requires deep mid-IR spectroscopic surveys to locate in space and in time the presence of star-forming galaxies and AGN, and to assess their activity. During the peak epoch of star formation in galaxies, the UV and optical light is absorbed by dust grains and re-emitted in the infrared (IR), where ∼90% of the energy emitted by young stars emerges (e.g., Madau & Dickinson 2014, and references therein). Observations of the UV rest-frame continuum associated with young massive stars (1400–1700Å), in spite of being one of the most popular methods to measure the SFR density (Lilly et al. 1995; Madau et al. 1996; Cucciati et al. 2012; Bouwens et al. 2015), is highly unreliable in obscured environments such as those at the cosmic noon. Therefore, rest-frame ultraviolet (UV) and optical observations cannot access these crucial dust-obscured regions where massive stars form.

To understand galaxy evolution, we need to measure the rate at which stars form and black holes accrete matter as a function of time. The similar histories of star formation and black hole accretion with redshift, the fact that most active galaxies display an AGN and enhanced star formation, and the black hole–stellar mass correlation seen in the local Universe (Magorrian et al. 1998, Ferrarese & Merritt 2000), all strongly suggest that the two processes are intimately linked.

**Science questions**

Why do the cosmic star-formation rate density and cosmic black-hole accretion rate density follow a similar shape with redshift, peaking at z~2, yet they arise from processes occurring over at wildly different physical scales?

What are the roles of star formation, accretion onto and feedback from central black holes and supernovae in shaping galaxy evolution over cosmic time?

**Need for PRIMA**

PRIMA will quantify the contribution of AGN, measuring their accretion rate (through [OIV] and [NeV]) and hot, young stars, measuring the star formation rate (from the PAH and the brightest nebular fine structure lines, e.g., [NeII], [NeIII], [SIII], [SIV] and [OIII]), to the overall energy budget of samples of heavily dust-obscured galaxies. The power of these IR diagnostics was theoretically predicted and demonstrated with ISO, Spitzer and Herschel (Spinoglio & Malkan 1992; Genzel et al. 1998; Sturm et al. 2002; Armus et al. 2007; Veilleux et al. 2009; Fernández-Ontiveros et al. 2016). No other currently planned telescope will be able to perform such detailed spectroscopic investigations. PRIMA will fill the large spectral gap from 28 μm to 300 μm left between JWST and ALMA.

In Fig. 1 we show the potential of PRIMA-FIRESS to measure the SFR (left vertical axis) through the 11 μm PAH feature and the [NeIII] fine-structure line, and the luminosity of the [OIV] line (right vertical axis), as a proxy of the BHAR, in a "Main-Sequence" (M.-S.) galaxy at the knee of the mass function (M = $10^{10.7}$ M$_\odot$) during the cosmic noon (1 < z < 3). For the case assessment we refer to the preparation of the SPICA galaxy evolution science program (Spinoglio et al. 2017, 2021).

With this program we will be able to address the evolution of a M.-S. galaxy with redshift, however it is also important to study the role of the other main physical parameters, i.e., the environment, the chemical evolution and the stellar masses. Therefore, a much larger survey (of order of thousand galaxies) will be required in order to obtain a thorough view of the galaxy evolution from cosmic noon to our times.

## Interpretation Methods

Fig. 1 highlights the region in the redshift-luminosity space that will be covered by PRIMA.

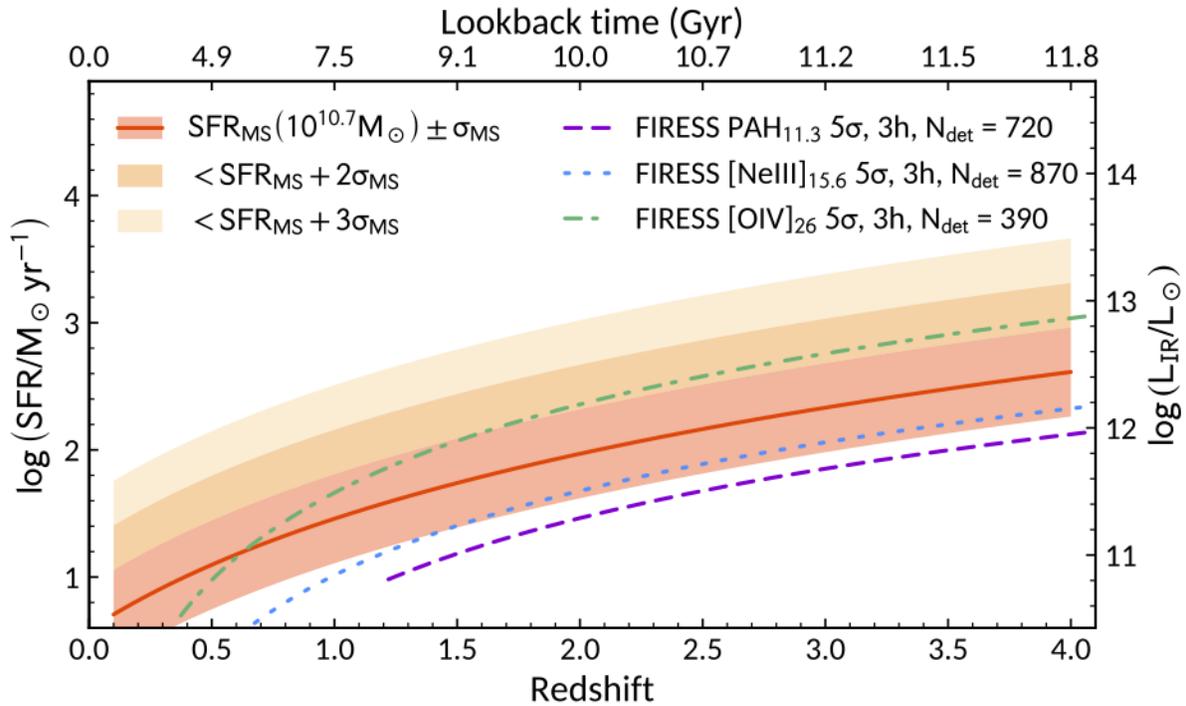

Figure 1: Star Formation Rate (left y-axis, SFR in $M_\odot$ yr−1) and total IR luminosity (right y-axis) as a function of redshift for a $10^{10.7}$ $M_\odot$ galaxy in the Main Sequence (MS, red solid line; Scoville et al. 2017). The red-shaded area shows the 0.35 dex intrinsic scatter around the MS indicated with a red line (Schreiber et al. 2015), while the dark and light orange-shaded areas indicate the location of galaxies whose SFR is +2σ and +3σ above the MS, respectively. The blue dotted line shows the SFR (left y-axis) and luminosity (right y-axis) detectable with the [NeIII]15.6 μm at 5σ in 3 hours by FIRESS for a low-metallicity galaxy. The purple long-dashed line shows the same with the 11.3 μm PAH emission feature for a dusty star-forming galaxy. The green dot-dashed line indicates the [OIV]25.9μm luminosity as a function of redshift detectable with FIRESS. PRIMA would measure the BHAR in these galaxies through the [OIV] emission.

In order to estimate the number of galaxies that can be detected with the FIRESS spectrometer deep spectroscopic $10 \times 10$ arcmin$^2$ blind survey in the features and lines, we have followed up the work of Spinoglio et al. (2021) for the preparation of the SPICA mission. The methodology is as follows: we have used the far-IR luminosity functions (Wang et al. 2019) that were derived using all available data (Spitzer, Herschel etc.) to obtain a complete distribution of galaxies in the luminosity-redshift diagram. The results are presented in the following Table 1, showing that a total of 400–800 detected galaxies are expected within a $10 \times 10$ arcmin$^2$ field at a limiting flux of $1 \times 10^{-19}$ W/m$^2$ at 5σ. The objects have been divided in three classes, star forming galaxies (SF), detected in the PAH@11.3 μm feature, dwarf (low-metallicity) galaxies, detected through the [NeIII]15.6 μm line, and AGN, detected through the [OIV]26 μm line. The number of galaxies given in the table has been computed assuming that all galaxies belong to the chosen class, and therefore are upper limits.

Table 1. Predictions of the number of detections vs redshift, using the PAH@11.3 µm for Star Forming galaxies (SF), the [NeIII]15.6 µm line for Dwarf galaxies and [OIV]26µm for AGN. Note that the total number of detections is computed from the Wang et al. (2019) luminosity function assuming, in each case, that all galaxies belong to the same class.

| z | < 0.5 | 1.0 | 1.5 | 2.0 | 2.5 | 3.0 | 3.5 | 4.2 | total |
|---|-------|-----|-----|-----|-----|-----|-----|-----|-------|
| SF | — | — | 236 | 248 | 108 | 79 | 35 | 17 | ~720 |
| Dwarf | — | 254 | 329 | 156 | 64 | 44 | 18 | 6 | ~870 |
| AGN | 231 | 96 | 41 | 18 | 6 | 1 | 0 | 0 | ~390 |

## Instruments Required:

- FIRESS large map in low res (R~130), 400 hours
- PRIMAGer small map in hyperspectral band down to 20–30 uJy, 25 hours

## Program Size:

- Large (100 + hr)

## Approximate Integration Time:

For a limiting sensitivity of $1 \times 10^{-19}$ W/m$^2$ at 5σ, we estimate about 400 hours to cover an area of $10 \times 10$ arcmin$^2$, with the FIRESS spectrometer (200h x 2 settings). The same area can be imaged at the required depth (20–30 µJy) in low resolution mode (R ~ 10) by PRIMAger in about 25 hours. So, the total observing time of this program is ~ 425 hours.

## Special Capabilities Needed:

- None

## Synergies with Other Facilities:

The study of AGN and SF coevolution and of the metallicity evolution using IR tracers is under way with both JWST at low redshift (z < 1) and at high redshift with ALMA (z > 4). PRIMA will fill the gap on these studies at the cosmic noon (1 < z < 4), when ~50% of the stellar mass in galaxies was formed. In the future, the large surveys performed by, e.g., the Nancy Grace Roman telescope and Euclid can be exploited to identify the best field to perform our blind spectroscopic survey, while facilities such as SKA and Athena will provide complementary multi-wavelength information for the detected sources.

## Synergies with other PRIMA proposals:

The blind spectroscopic survey of this proposed program will benefit from other larger field (of the order of square degrees) surveys planned to be done with the PRIMAger. In particular the FIRESS imaging spectroscopy can follow up targets and limited sky regions for which PRIMAger observations have revealed the presence of SF galaxies through PAH features and of AGN through their SEDs. Moreover, a 12µm rest-frame selection of galaxies with a PRIMAger large field survey will be able to discover AGN (see, e.g., Spinoglio & Malkan 1989) and longer wavelength selections can discover preferentially SF galaxies. We will exploit such wide surveys to follow up the discovered galaxies with FIRESS.

# 38. Detecting FIR emission from the circumgalactic medium

Elizabeth Tarantino (Space Telescope Science Institute), Alberto Bolatto (University of Maryland)

The CGM is the complex, multiphase gas that exists outside of galaxies. It contains the material that is released through feedback from galaxies and can subsequently cool and reaccrete back onto its host, providing an integral role in the ecosystem of galaxies and their evolution. We propose to use PRIMA to detect and map the circumgalactic medium (CGM) in far-infrared (FIR) fine-structure emission lines in order to better understand the CGM. The FIR fine structure lines, in particular [CII] and [OI], are some of the most important coolants in the ISM and can be used to constrain the cooling rate in the CGM. We construct simple Cloudy photoionization models that predict the surface brightness of the FIR lines for a range of densities. With the exquisite sensitivity and spectral capabilities of PRIMA, our simulations show that we will be able to detect the CGM through the [CII] line. Our proposed observations will open up a new discovery space to study the CGM in emission.

## General Scientific Area:

Galaxy Ecosystems and Interstellar Medium

## Science Justification:

The circumgalactic medium (CGM) is the gas that surrounds the outskirts of galaxies. It contains the material that outflows from galaxies that can subsequently cool and re-accrete onto galaxies, stimulating the formation of stars (see Tumlinson et al. 2017 for a recent review). The CGM provides an integral role in galactic evolution, but the diffuse, low surface brightness nature of its structure makes it difficult to study. Most advances in understanding the CGM come from pencil beam absorption line studies that use background quasars as an illumination source. These measurements can probe the low densities in the CGM and find a variety of atoms and ions that span in ionization state, such as HI, CII, CIV, and OVI. The presence of a mixture of species at varying ionization states and critical densities implies that the CGM has a complex multiphase structure. Since the absorption line studies are usually limited to one sightline per galaxy, the detailed resolved structure of the CGM remains unknown. We propose to use the exquisite sensitivity and spectral capabilities of PRIMA to detect the CGM in bright, far-infrared cooling lines.

The fine-structure [CII] 158 µm and [OI] 63 µm transitions are two of the most important coolants in the ISM, especially at low densities and temperatures of the atomic gas (Wolfire et al. 1995, 2003). The [CII] line is also often the brightest emission line in the FIR from galaxies, amounting to about 0.1%–1% of the integrated FIR continuum emission (Crawford et al. 1985; Stacey et al. 1991). Like the CGM, the $C^+$ ion is multiphase and is present in the ionized, molecular, and atomic gas. Thus [CII] emission, and to a lesser degree [OI], have great potential to trace the extended CGM halos around galaxies. The FIR band also contains species that trace higher ionization states, such as [OIV] and [NeIII]. One pointing of PRIMA can therefore potentially cover a wide range of species, providing physical diagnostics to study the CGM in emission instead of the narrow spatial scale of absorption.

We develop a simple model to predict the surface brightness of a variety of FIR fine structure lines through the photoionization code Cloudy (Ferland et al. 2017). Photoionization equilibrium in the CGM is a safe assumption for the low and intermediate ions we are interested in studying (e.g., CII, OI, SIII, OIV, etc.), but we note that collisional ionization may contribute when examining higher ionization ions (Tumlinson et al. 2017). For the input ionizing SED, we use the metagalactic radiation field presented in Gnat & Sternberg 2004, which represents the radiation field directly outside the Milky Way and is constrained through optical/UV and x-ray observational data. Gnat & Sternberg 2004 report a "standard" metagalactic field as well as a "maximal" field that contains an enhancement of the Lyman limit to 0.25 keV, consistent with the upper limits from Hα non-detections. We run plane-parallel Cloudy models using isotropic radiation fields to estimate the predicted line intensities in the FIR. We also compare to 10% of the local interstellar radiation field defined in Black 1987. The metallicity is the average of the warm and cold gas phase abundance at the solar circle. We create a grid of models that range from hydrogen densities of $\log n_H = -4$ cm$^{-3}$ to $\log n_H = 0$ cm$^{-3}$ and stop the models at a hydrogen column density of $N_H = 3 \times 10^{18}$ cm$^2$. The Cloudy models output the predicted surface brightness of the selected FIR lines at each input density and the results for the three radiation fields used are presented below.

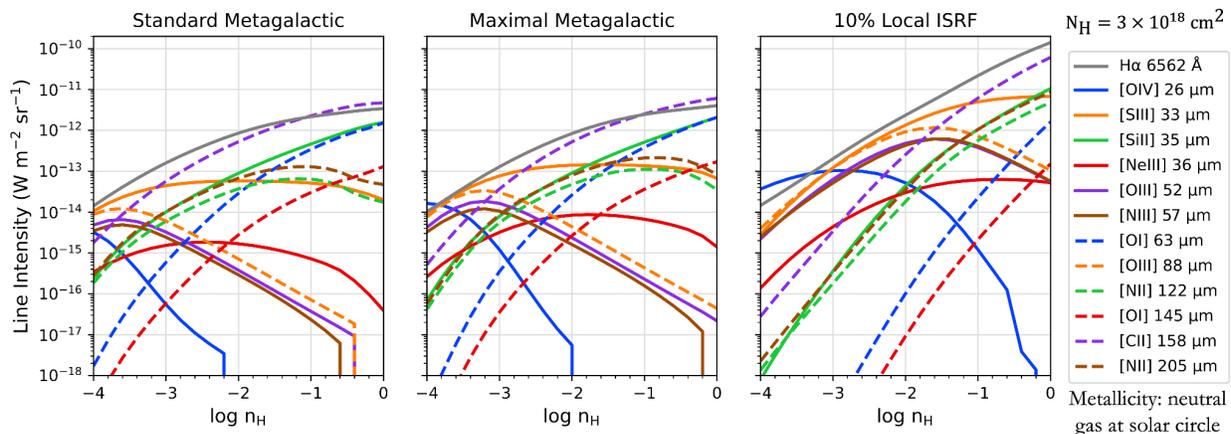

Figure 1: Predictions of the FIR surface brightness from simple Cloudy simulations of the CGM. The FIR lines PRIMA will observe are given by the solid and dash colors and the Hα line is presented in gray for comparison.

Of the FIR fine structure lines, [CII] is the brightest in our simple simulations, especially at the higher densities in the grid. At lower densities ($\log n_H < -3$ cm$^{-3}$), [SIII] and [OIII] are slightly brighter than the [CII] line. We also present the Hα line in gray for comparison which has a similar predictions as the [CII] line. Given the exquisite surface brightness sensitivity of PRIMA, it should be possible to detect the CGM in emission through the [CII] line in the higher density CGM. Mapping the CGM through the [CII] line would lead to a giant leap in our understanding of galaxy evolution. We will calculate the cooling rate of the CGM through the [CII] line and put constraints on how the gas from the CGM cools and falls back onto galaxies, fueling star formation. We will compare the [CII] emission maps to the pencil beam absorption studies to gain a more complete understanding of the multiphase structure of the CGM. If any other FIR line is serendipitously detected, we can model the resolved ionization and density structure of the CGM. Our proposed observations with PRIMA therefore have the potential to revolutionize our understanding of galactic ecosystems and the gas around galaxies.

## Instruments Required:

FIRESS low resolution observations, in pointed mode. Small map and medium map modes could also be used.

## Program Size:

- Medium (10–100 hr)

## Approximate Integration Time:

- 75 hours

## Special Capabilities Needed:

- None

## Synergies with Other Facilities:

- The Circumgalactic Hα Spectrograph (Melso et al. 2022), Dragonfly Telescope, ALMA

## Description of Observations:

Based on the results of the simple Cloudy photoionization models presented in Figure 1, the sensitivity of PRIMA will open a discovery space in detecting the CGM through FIR cooling lines. We use the PRIMA fact sheet to inform our integration time estimate which reports a $10^{-10}$ W $m^{-2}$ $sr^{-1}$ $5\sigma$ detection at the [CII] 158 μm wavelength with a 5-minute integration. We request an integration time of 75 hours for a source in pointed mode, which would yield a sensitivity of $3 \times 10^{-12}$ W $m^{-2}$ $sr^{-1}$ for the [CII] line. We will therefore be able to detect the CGM in emission for both the standard and maximal metagalactic radiation fields in a density range of log $n_H$ = 0 $cm^{-3}$ to log $n_H$ = −1 $cm^{-3}$. For the 10% of the local interstellar radiation field, we will be able to detect [CII] in a slightly larger range, down to log $nH$ = −1.5 $cm^{-3}$.

For this detection experiment, we will select a source near previous absorption measurements of the CGM that report a relatively high density close to our range of detectability and a high column density of $N_H$ = $10^{16}$ $cm^2$. This will increase our chances of detectability and allow for comparison between the CGM detected in emission and absorption.

# 39.  The buildup of galaxy clusters

Julie Wardlow (Lancaster University)

Galaxy clusters are crucial for understanding the build-up of large-scale structures and the evolution of stellar systems. However, identifying and studying (proto)clusters at z>~2 is challenging as most current search methods rely on the red-sequence or cluster gas already being in place. Instead, far-IR searches are proving fruitful, though they are expensive with current facilities. Additionally, understanding the processes driving the transformation of cluster members into quiescent galaxies requires detailed study of their ISM i.e., mid-IR spectroscopy of fine-structure lines and PAH features. PRIMAger colors (particularly hyperspectral colors) will enable the efficient selection of protoclusters from wide-field surveys and follow-up with FIRESS will be used to study the galaxy's ISM during transformation onto the red sequence.

## General Scientific Area:

Galaxy Ecosystems and Interstellar Medium

## Science Justification:

The dominant stellar systems in local galaxy clusters are passive early-type galaxies, whose properties contrast markedly with the star-forming late-type galaxies found in the surrounding low-density field. The elliptical galaxies in clusters are thought to have formed most of their stars in short bursts at z >~ 2 (Blakeslee et al. 2003), in contrast to the field where most of the activity occurs at z < 1 (Lilly et al. 1995). However, the progenitor population of these clusters and their local, massive early-type galaxies remain enigmatic and challenging to study.

Surveys for high-redshift protoclusters are challenging due to the wide areas that they are expected to cover (tens of Mpc at z~2; Muldrew et al. 2015) and the biases inherent in traditional selection methods. For example, searches focused on the red sequence have been exceptionally successful up to z~1.5 (e.g., Gladders & Yee 2000, Cerulo et al., 2016), but they require the member galaxies to be passive having already ceased star-formation, and they are therefore not suitable at higher-redshifts. Similarly, X-ray and searches using the Sunyaev–Zeldovich effect require a hot gas reservoir to have already formed and are also limited in redshift (e.g., Bleem et al. 2015, Koulouridis et al. 2021). QSOs and massive radio galaxies are thought to signpost overdensities and studies targeting sources around these objects have also been successful (e.g., Wylezalek et al. 2013), though they are inefficient and the requirement for a central QSO or radio galaxy may bias these searches to certain types of protoclusters.

Since local clusters are dominated by passive early-type galaxies it is expected that their high-redshift progenitors were dusty star-forming galaxies (DSFGs; see Casey, Narayanan & Cooray for a review). Follow up of individual DSFGs has also proved fruitful for identifying overdensities and protoclusters (e.g., Pope et al. 2005, Capak et al. 2011) though systematic surveys are needed to reliably put these structures in cosmological context. Thus, identifying groups of DSFGs is an effective way of identifying protoclusters. The power of far-IR surveys for systematically signposting overdensities and protoclusters has been demonstrated in recent years, with e.g., Herranz et al. (2013), Clements et al. (2014) and Cheng et al. (2020) (amongst others) using Planck and Herschel data to both find and measure the overdensity of such structures, and other studies using SPT (e.g., ZZZ) or Spitzer and Planck (e.g., Martinache et al. 2018) data. However, these studies are typically relatively small (tens of objects) and require extensive, expensive, multi-facility follow-up programs to confirm and characterise the candidate overdensities from initial far-IR selection.

With PRIMA, identification and confirmation of such structures can be performed efficiently with a single-facility: wide area surveys with PRIMAger will be able to robustly identify groups of DSFGs with similar (hyperspectral) colours (i.e., redshifts) in small areas of the sky. These groups will be strong candidates for being high-redshift protoclusters that can be confirmed and their ISM's studied with FIRESS spectroscopic mapping or targeted spectroscopy (e.g., PAH and/or fine-structure line observations). Furthermore, protocluster candidates identified with other means (e.g., the next generation of X-ray of SZ searches) can also be confirmed and properly characterised with FIRESS.

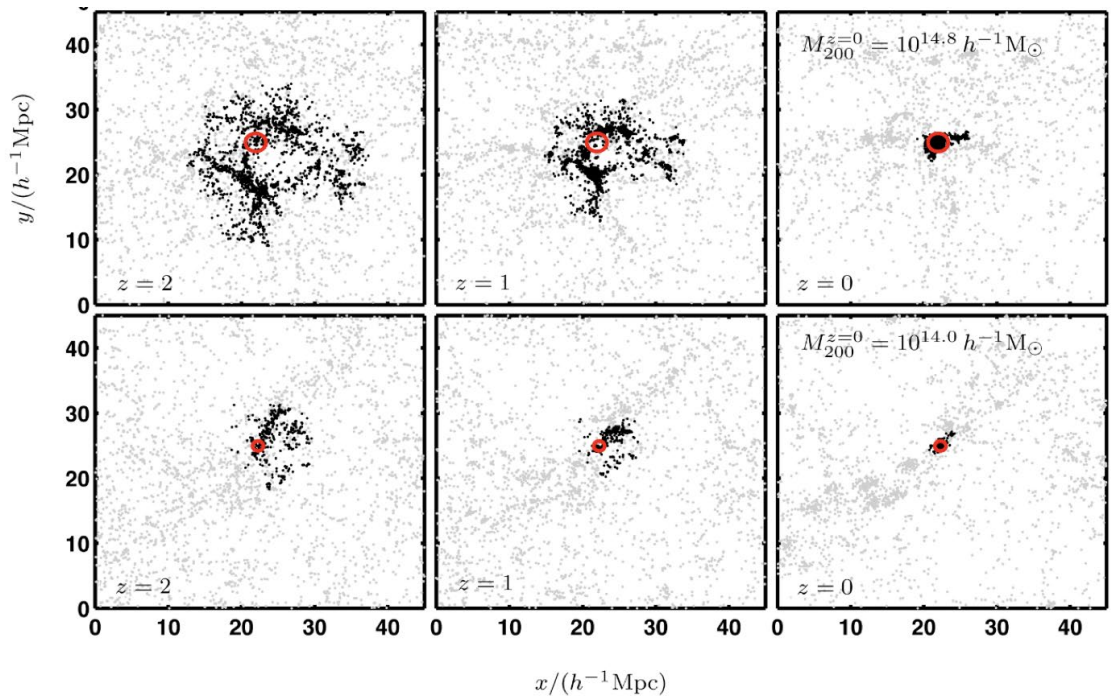

Figure 1: Simulation of the evolution of clusters of different mass from z=2 to z=0, showing that at high-z most of the member galaxies (black) are widely spaced, though those that are gravitationally bound (red circles) are in much smaller areas. The galaxies are still actively forming stars and therefore can't most high-z protoclusters can't be selected using red sequence techniques, and instead require searches for star-forming galaxies such as in the FIR. (Figure from Muldrew et al. 2015).

The mid-IR spectroscopic study is particularly important as (proto)cluster members and infalling galaxies transform onto the red sequence. The processes governing the shutdown of star-formation are poorly understood, and the role of environment (e.g., interactions vs ram-pressure stripping vs strangulation etc) are not well understood. PAH features and mid-IR fine-structure lines (e.g., [OIV]26 μm, [SiII]35 μm, [OIII]52 μm, [OI]63 μm) are excellent tracers of star-formation conditions in the ISM as well as of AGN activity (e.g., Spinoglio et al. 2019) and these will be used to identify galaxies with similar ISM conditions as well as those at different stages of their evolution, to be correlation with spatial location (core, filament, field etc.), mass and luminosity. For additional interpretation measurements of low-J CO lines (e.g., with ALMA or JVLA) would track the cold gas available for star-formation as well.

## Instruments Required:

- PRIMAger: Hyperspectral Imaging and Polarimetry, probably use deep extragalactic field surveys
- PRIMA FIRESS Spectrometer: Pointed Low-res R~130 or Small Map

## Program Size:

- Large (100 + hr)

## Approximate Integration Time:

Depends on existence/collaboration with extragalactic PRIMAger survey. Spectroscopy likely needs ~50 hour per protocluster (to cover ~10 arcmin$^2$ to ~$10^{-18}$ W/m$^2$), though smaller areas and shallower depths may be possible, or pointed spectroscopy may be more efficient depending on source distribution.

## Special Capabilities Needed:

- None

## Synergies with Other Facilities:

- Synergy with optical/near-IR colour-based, X-ray and SZ searches for high-z clusters.
- Synergy with ALMA and optical/near-IR for physical follow-up spectroscopy.
- Synergy with proposed wide-field ground-based far-IR facilities for searches and follow-up, with these vs PRIMA likely finding protoclusters at different redshifts dominated by galaxies with different dust temperatures and other properties.

## Description of Observations:

This science project can begin from other planned mid/wide-field PRIMAger blank-field extragalactic surveys, and doesn't require its own specific PRIMAger survey unless no other extragalactic surveys are undertaken. The FIRESS spectroscopy will need to be targeted follow-up, though the depth and areas required depend upon the targets found. Some protoclusters will be able to be confirmed with only the PRIMAager data (e.g., with photometric signatures of PAH's), though detailed study of the chemistry in these systems compared with the field and local galaxies will require spectroscopy. Depths will depend on the individual targets.

# 40. The ISM in dusty star-forming galaxies

Julie Wardlow (Lancaster University), Co-authors: George Bendo (University of Manchester), Andrew Blain (University of Leicester), David Clements (Imperial College)

Dusty star-forming galaxies (DFGs) are common at z~2 and are extreme star-forming galaxies with ULRIG-like infrared luminosities and star-formation rates (SFRs) of hundreds to thousands of solar masses per year. Locally, ULIRGs are predominantly merger-powered, with compact high-density regions where their star formation is focused. However, the nature of DSFGs at higher redshifts is poorly understood. In order to fully study them and get a good understanding of their nature and the mode of their star formation, spectroscopic observations of the star-forming regions and gas are required. In particular, PRIMA FIRESS will enable simultaneous spectroscopic coverage of tracers of PDRs, HII regions and AGN activity in a large sample (at least tens of objects) of DSFGs. Such a study requires space-based mid-IR spectroscopy and a sample large enough to investigate the role of stellar mass, IR luminosity and AGN presence on the ISM of DSFGs.

## General Scientific Area:

Galaxy Ecosystems and Interstellar Medium

## Science Justification:

High-luminosity dusty star-forming galaxies (DSFGs) are selected form their high flux densities at far-IR and submillimeter wavelengths and have IR luminosities in excess of $10^{12} L_\odot$ and star-formation rates (SFRs) of hundreds to thousands of solar masses per year. These galaxies contribute up to 20% of the cosmic star-formation rate density at z~2 where they are most populous and they are good candidates for being the progenitors of local massive elliptical galaxies (e.g., Chapman et al. 2005; Hickox et al. 2010; Wardlow et al. 2011).

These DSFGs appear to lie both on and above the stellar mass – SFR "main sequence" (e.g., da Cunha et al. 2015), suggesting that some of these galaxies are in a short starburst phase, whereas others may evolving more secularly. However, this inference is highly dependent on stellar masses which are challenging to reliably measure due to systematic uncertainties and the faintness of DSFGs in the rest-frame optical. It also requires that the distance of a galaxy from the main sequence is robust measure of its "burstiness", although this is still subject to debate (e.g., Wuyts et al. 2011, Sargent et al. 2014, Matthee, et al. 2019, Mountrichas et al. 2021). Conversely, the idea that DSFGs are dominated by starbursts is supported by their high SFRs and short gas depletion times (e.g., Birkin et al. 2020), but inconsistent with their location both on and above the main sequence.

To resolve this debate and determine the mode of star-formation in these galaxies requires study of the conditions in their ISM to measure the gas density and FUV field strength in the star-forming regions (i.e., with mid- and far-IR spectroscopy). Whilst some of the lines needed for this are covered by ground-based instruments (e.g., [NII]205 μm with ALMA) it is only with mid-IR spectroscopy that a range of conditions can be probed to allow for a full and complete analysis of the gas in DSFGs, particularly at z~2, where they are most common and where the Universe is at its most active.

Additionally, the role of AGN in the evolution of these galaxies is still poorly understood. Whilst only ~20% of DSFGs contain AGN that are detectable at X-ray or radio or near/mid-IR wavelengths (e.g., Alexander et al. 2005; Wang et al. 2013) the gold-standard for identifying deeply obscured AGN is mid-IR spectroscopy (e.g., Pope et al. 2008).

Thus, mid-IR spectroscopy is required both to measure the conditions in the ISM in DSFGs and to identify AGN and investigate whether the AGN affects the conditions in different galaxies. The latter scenario might be expected due to AGN feedback and scenarios in which a DSFG is triggered by an inflow of gas (e.g., galaxy mergers or interactions).

A handful (low tens) of DSFGs have been studied in the mid-IR using Spitzer/IRS to target PAHs (e.g., Pope et al. 2008) or Herschel/PACS for fine-structure lines (e.g., Wardlow et al. 2017). However, these studies cannot be expanded or deepened until another orbiting mid-IR spectroscopic facility is available. For fine-structure lines such observations were only available for the brightest objects (often lensed sources) and non-detections were common leading to low SNR stacking being the only option for analyses: deeper data on a wider variety of sources with both PAH and fine-structure line observations are required.

The low-resolution mode of PRIMA FIRESS will be able to observe the 12 μm PAH and its underlying continuum (tracing star-formation vs AGN contributions) as well as key fine structure lines including [OIV]26 μm (tracing AGN), [SiII]35 μm, [FeII]26 and 35 μm (tracing PDRs), and [OIII]52 μm, [OI]63 μm, [SIII]33 μm, and [NIII]57 μm among others (tracing HII regions). The sensitivity required will depend upon which lines are targeted, but sigma $<\sim4\times10^{-19}$ W/m$^2$ in the lines are required even for lensed z~2 DSFGs (Wardlow et al. 2017). Based on this experience and the sensitivity curve ~1 hour per target is likely to be the typical on-source time needed.

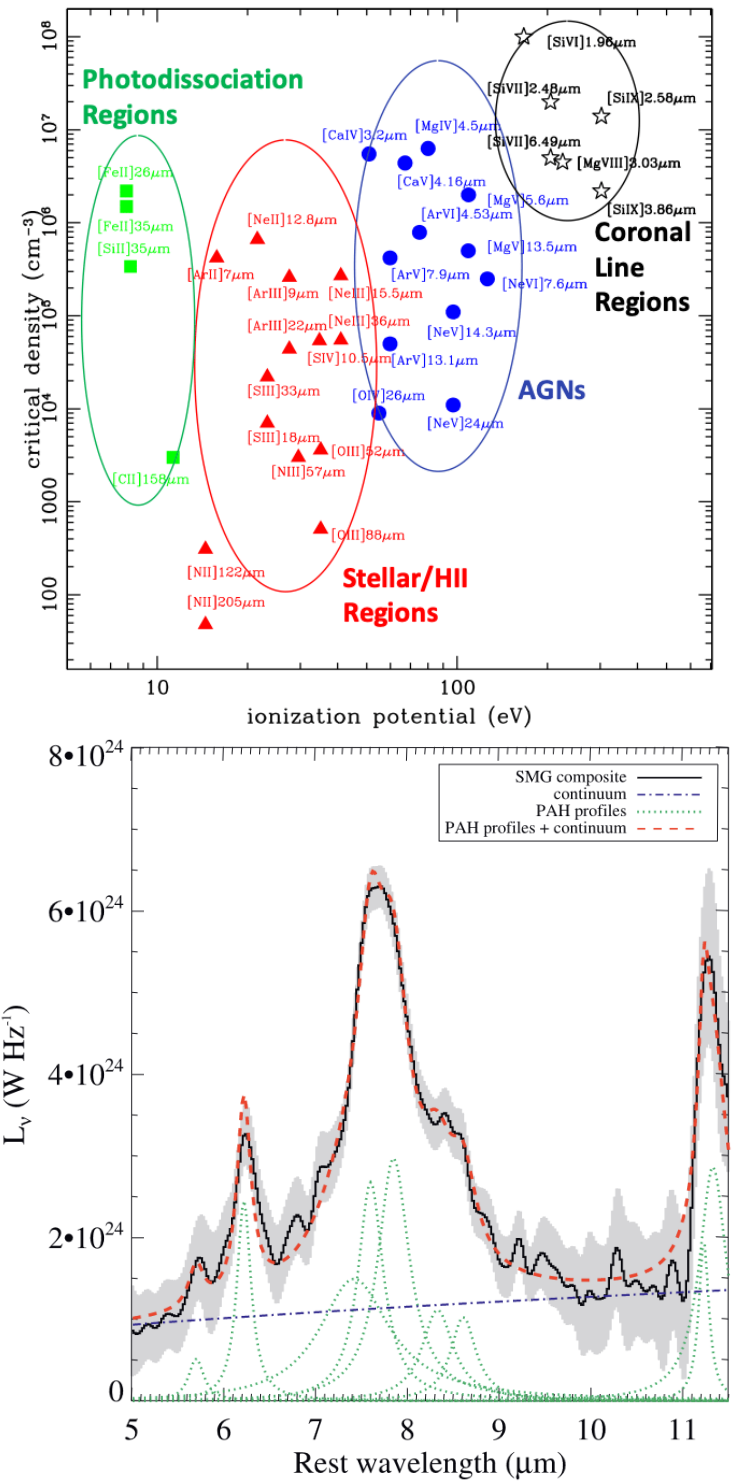

Figure 1: Left: Plot from Spinoglio et al. (2019) showing that different lines trace different regions of a galaxy. To probe star-forming regions and PDRs at z∼2 mid-IR observations are required, necessitating orbiting facilities. Right: The composite PAH spectrum of 12 DSFGs from Pope et al. (2008) showing the complexity of the PAH lines and the domination of star-formation over AGN powering (the latter would be dominated by power law emission at these wavelengths).

## Instruments Required:

- PRIMA FIRESS pointed observation, low resolution
- PRIMAGer Small maps

## Program Size:

- Medium (10–100 hr)

## Approximate Integration Time:

~1–2 hours per source. An initial sample of a few tens of sources (~40) would enable binning by redshift, stellar mass and SFR.

## Special Capabilities Needed:

- None

## Synergies with Other Facilities:

Synergy with ALMA for e.g., [CII] lines probing both PDRs and HII regions as well as the star forming gas (va CO) and continuum. JVLA will also be good for tracking cold gas (CO(1-0)).

## Description of Observations:

Targeted pointing of individual z~2 (and possibly beyond) DSFGs with FIRESS low-resolution mode. Each source will need to reach sigma <~4×10$^{-19}$ W/m$^2$ in the lines (based on stacking of mostly lensed sources with Herschel/PACS; Wardlow et al. 2017), but aiming for <10$^{-19}$ W/m$^2$ will be a more reliable strategy. Multi-band photometric mapping with PRIMAger would also be helpful for pinpointing the SED, but sample selection could be optimized to ensure that the targets are covered by prior PRIMAger survey programs.

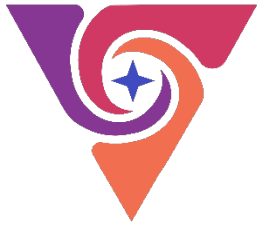

# Exoplanets, Astrobiology, and the Solar System

# 41.  Comets and the origin of Earth's water

Dariusz C. Lis (Jet Propulsion Laboratory, California Institute of Technology), Paul F. Goldsmith (JPL/Caltech), Paul von Allmen (JPL/Caltech), Youngmin Seo (JPL/Caltech), Arielle Moullet (NRAO)

Understanding how the Earth obtained its water and whether water-rich Earthlike planets are common in the Universe is one of the central themes in NASA's vision. The Astro2020 Decadal Survey speaks of the core concept of gas and dust particles in a circumstellar disk assembling into planets. Whether this process can frequently produce planets with Earthlike characteristics remains unknown. Our Solar System is currently the only laboratory where the conditions of habitability can be investigated in detail. Such investigations are crucial for placing potentially habitable planets in context. Understanding the trail of water from its interstellar reservoir to forming planetary systems can be best studied by following the isotopic composition of water. An important fingerprint is the D/H isotopic ratio, which is strongly dependent on the water formation temperature and exhibits large variations among interstellar and Solar System objects, making the required accuracy of individual measurements relatively modest and achievable to derive valuable information from remote sensing observations. Studies of the water content and isotopic composition of water-rich asteroids and comets are of key interest for understanding the late accretion stage of the Solar System cometary and chondritic materials. PRIMA can make an important contribution to solving this long-standing problem by carrying out measurements of the D/H ratio in a statistically significant sample of Oort cloud and Kuiper belt comets, sampling the isotopic composition of the present-day outer Solar System. This would allow comparisons between different comet reservoirs, and with inner Solar System measurements in meteorites, as well as searching for correlations with physical parameters, such as hyperactivity, providing quantitative constraints on the dynamical/chemical models of the early Solar System.

## General Scientific Area:

Solar System

## Science Justification:

The water mass fraction has been shown to increase with the distance from the Sun, from about 0.1% in the Earth to 50% in comets and small TNOs (van Dishoeck et al. 2014). In the textbook model of the protosolar disk, Earth formed inside the snow line meaning that the temperature in the terrestrial planet zone was too high for water ices to exist and water and organics had to be delivered later by impacts of water-rich comet- or asteroid-like bodies. However, the latest isotopic measurements in meteorites (Broadley et al. 2022) and dynamical models (e.g., Woo et al. 2023) suggest a much more complex, multi-stage process, in which volatiles on the Earth and other terrestrial planets have been heterogeneously sourced from different Solar System reservoirs. The late accretion of chondritic and cometary materials is an important, final stage in this complex process (Figure 1).

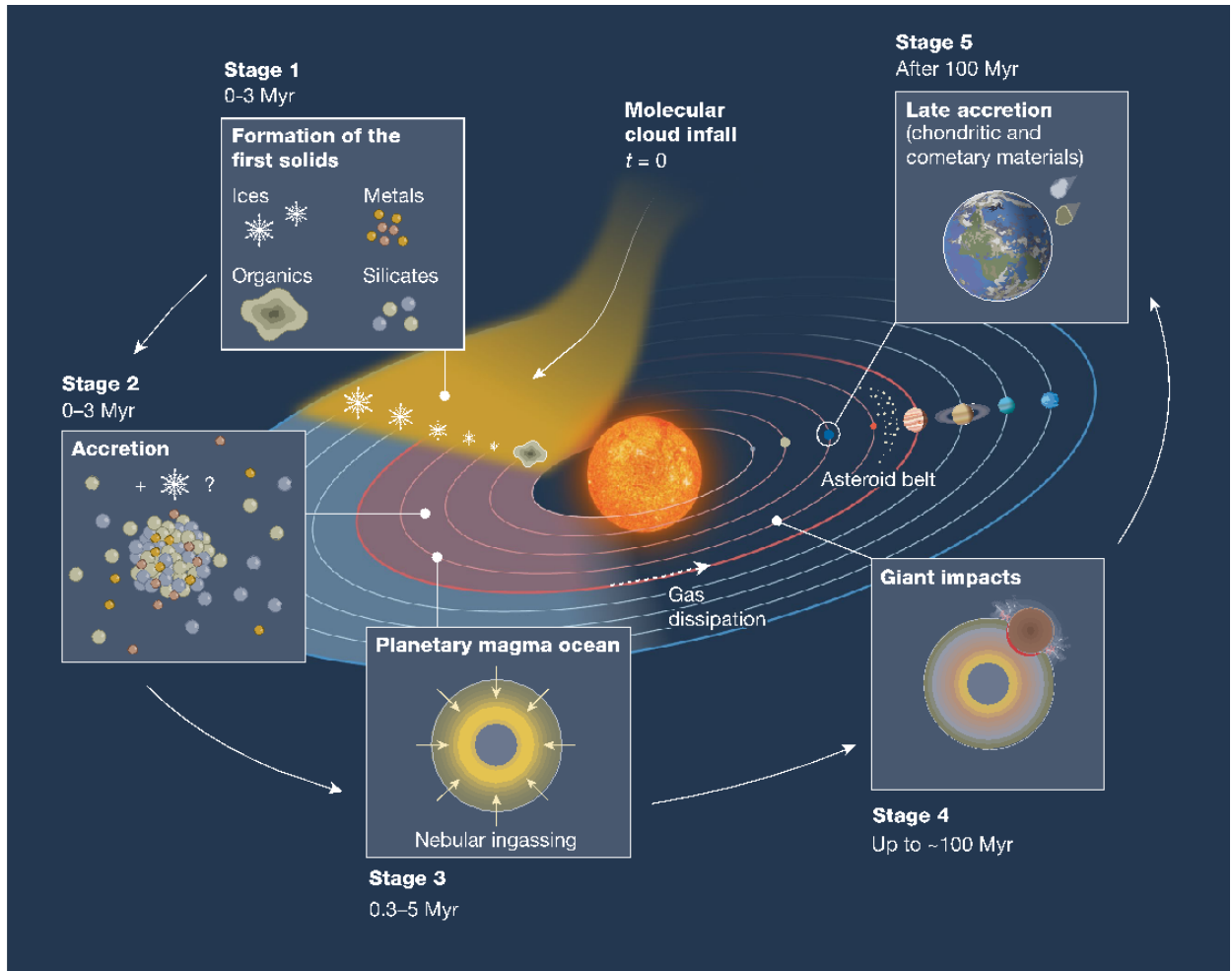

Figure 1: Multiple stages of volatile incorporation, with the late accretion of chondritic and cometary materials constituting the final stage (from Broadley et al. 2022).

Figure 2 summarizes D/H measurements in Solar System bodies. The primordial (Big Bang) D/H ratio is $(2.53 \pm 0.04) \times 10^{-5}$ (Cooke et al. 2014). The Solar Nebula ratio measured from the solar wind composition, as well as from $HD/H_2$ in the atmospheres of Jupiter and Saturn, is $(2.5 \pm 0.5) \times 10^{-5}$ (see Robert et al. 2000, and references therein), implying little processing in stars between the Big Bang and the formation of the Solar System. The D/H ratio of Earth's-ocean water is enhanced by a factor of 6 to $1.5576 \times 10^{-4}$ ("Vienna Standard Mean Ocean Water", VSMOW), with somewhat lower values reported in deep mantle materials (Hallis 2015).

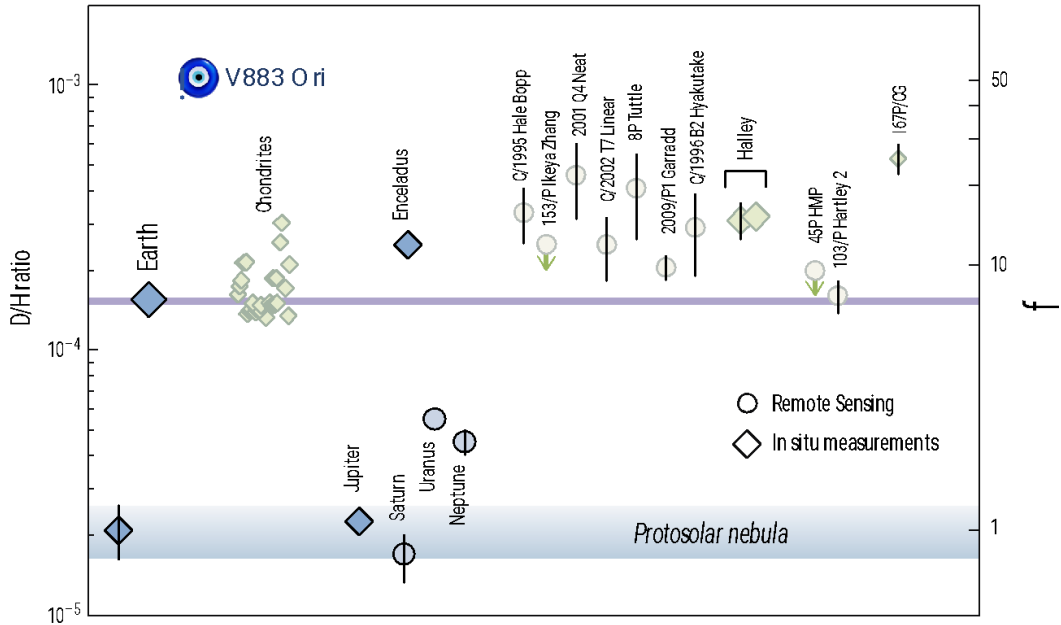

Figure 2: D/H ratio in water in the Solar System (from the Origins Mission Concept Study Report after Altwegg et al. 2015). Cometary D/H ratios vary between 1 and 3 times VSMOW, in both Oort Cloud and Jupiter Family comets. The blue disk marks the first ALMA measurement in the V883 Ori disk (Tobin et al. 2023).

The outer asteroid belt is a natural reservoir of icy bodies in the Solar System. Most C-type asteroids show hydrated silicate absorption features. Models suggest that buried ice can persist within the top few meters of the surface over billions of years (Schorghofer 2008), and Dawn observations (Raponi et al. 2018) showed that water ice can accumulate in permanently shaded areas of craters. Main-belt comets (Hsieh & Jewitt 2006) constitute a separate class of intrinsically icy bodies in the outer Asteroid belt, in which water outgassing was only recently reported using JWST (Kelley et al. 2023).

Comets have formed and remained for most of their lifetime at large heliocentric distances. Therefore, they contain some of the least-processed, pristine ices remnant from the Solar Nebula disk. Due to the presence of gaseous atmospheres, their composition can be studied through remote sensing. Numerous complex molecules have been identified in comet 67P by Rosetta and the derived composition of cometary ices is very similar to low-mass star-forming regions, suggesting formation in the pre-solar cloud or in the cold outskirts of the Solar Nebula (Drozdovskaya et al. 2019).

Cometary D/H ratios span a wide range (Figure 2) linked to the significant mixing of bodies from different regions in the Solar System caused by giant planet migration. It has also been suggested that a substantial fraction of Oort Cloud comets might have been captured from other stars in Sun's birth cluster (Levison et al. 2010). The variations in D/H ratio are currently interpreted as being the result of isotopic exchanges with D-poor gases ($H_2$, OH) in the hot, inner regions of the young solar nebula, combined with turbulent mixing (Ceccarelli et al. 2014). However, paradoxically, the high $D_2O/HDO$ ratio measured in comet 67P suggests that cometary water is largely unprocessed and was formed on grains in the presolar cloud (Altwegg et al. 2017).

While many comets show D/H ratios considerably in excess of the terrestrial value, a category of hyperactive comets has been identified in which the D/H ratio is consistent with the Earth's value (Lis et al. 2018). Hyperactive comets contain icy grains expelled from the nucleus that sublimate completely in the coma and may be more representative of the bulk isotopic composition of the nucleus than water sublimating from the surface, which may be subject to fractionation effects. Cryogenic sampling of the coma ice has been suggested as a better way of understanding the bulk isotopic composition of comets than direct nucleus sampling (Sunshine & Feaga 2021).

Observations to date, carried out over the past 35 years using ground-based and space facilities, have produced fewer than a dozen measurements of the D/H ratio in comets, of which only 4 space-based measurements (comets 1P, 103P, 67P, and 2009/P1) can be described as accurate (Figure 2). This compares to over 100 measurements in meteorites that probe the isotopic composition of the objects in the Asteroid belt (Robert 2003).

The long-term science objectives for studying the origin of the Earth's water with PRIMA are to: (a) constrain the shape of the D/H probability density function in cometary water with accuracy comparable to the meteoritic measurements; (b) study differences in the D/H distribution between different dynamical classes of comets; (c) study correlations with physical parameters, such as hyperactivity. Observations of a statistically significant sample of comets with PRIMA, compared with similar measurements in disks and meteorites, could help resolve these apparent contradictions and improve our understanding of the complex water cycle in the Solar System, and by proxy in other exoplanetary systems.

The statistical approach will allow for addressing various testable hypotheses and narratives to describe how, when, and in what quantity comets contributed to the Earth's water content. For example, if the Solar System small bodies that formed at different heliocentric distances in the protosolar disk were efficiently scattered due to the migration of giant planets, one may expect the same distribution of the D/H ratio in the Oort cloud and Kuiper belt objects. However, if a significant fraction of Oort cloud comets were captured from other stars in the Sun's birth cluster, one may expect different distributions in the two dynamical reservoirs. PRIMA observations will thus provide quantitative constraints for the state-of-the-art dynamical/chemical models of the early Solar System.

As was the case for past FIR observations of comets (e.g., Hartogh et al. 2011; Lis et al 2018), the D/H ratio will be determined from observations of optically thin water isotopologues, HDO and $H_2^{18}O$. Simultaneous observations of multiple HDO and $H_2^{18}O$ lines in a very similar field of view significantly simplify the analysis and decrease measurement uncertainties. Well-tested coma excitation models will be used to convert the observed line intensities to molecular production rates. The PRIMA data set is self-contained and does not rely on supporting observations from other telescopes. While the production rates depend on the assumed temperature profile in the coma, the broadband wavelength coverage of the PRIMA spectrometer will allow simultaneous observations of multiple lines of methanol, which will be used to constrain the temperature profile.

### Instruments Required:

PRIMA FIRESS: Pointed High-res R~4400@ 112 μm

**Program Size:**

- Large (100 + hr), 240 h per year (30 Oort cloud and Kuiper belt comets in a nominal 5-year PRIMA mission)

**Approximate Integration Time:**

The PRIMA FIRESS FTS CBE line sensitivity is $4.7\times10^{-19}$ Wm$^{-2}$ (5σ, 1h). We consider 20h as the longest observing time for a pointed observation of a single target. This gives a 5σ line flux detection limit of about $1\times10^{-19}$ Wm$^{-2}$. Detection of the HDO lines drives the required observing time, as lines of other water isotopologues will be brighter. Non-LTE cometary excitation models suggest that such flux level for the brightest 235-µm HDO line in the PRIMA wavelength range corresponds to the equivalent activity FOM=$2\times10^{28}$ (FOM = Q/Δ, where Q is the water production rate per second and Δ is the geocentric distance in au). Calculations carried out by the Origins Space Telescope Science and Technology Definition Team suggest that about 30 such comets can be expected in a nominal 5-year mission, or 6 comets per year. Three additional bright HDO lines at 126, 131, and 184 µm are accessible to PRIMA. When stacked, they give approximately the same SNR as the 235-µm line. To cover all 4 HDO lines, 2 independent observations will be required. The total observing time required is thus 40 h per target, or 240 h per year. If the FIRESS sensitivity is 4 times lower than the CBE value assumed here, a total of 20 D/H measurements can be expected in a nominal 5-year PRIMA mission, assuming the same total observing time.

**Special Capabilities Needed:**

- Non-sidereal tracking of Solar System objects.

**Synergies with Other Facilities:**

- ALMA, JWST, and large ground-based optical/UV telescopes have been shown capable of providing D/H measurements, but only in bright, nearby comets (Figure 2).

**Description of Observations:**

Observations will be carried out using the PRIMA FIRESS Spectrometer in the high-resolution FTS mode. The spectra will cover multiple lines of water isotopologues, including HDO, $H_2^{17}O$, $H_2^{18}O$, and $H_2^{16}O$. A combination of optically thin HDO and $H_2^{18}O$ lines was used in previous measurements of the D/H ratio, e.g., with Herschel and SOFIA (Hartogh et al. 2011; Lis et al. 2019). The 4 brightest HDO lines are covered by the 2 longest wavelength bands of the spectrometer, which cannot be observed simultaneously. Two independent observations will thus be required for each target. The nominal integration time for weak comets will be 20 h per target, per setting, as described above. For bright comets, multiple shorter observations will be carried out, spread in time to cover a range of heliocentric distances pre- and post-perihelion.

## Acknowledgement:

A portion of this research was carried out at the Jet Propulsion Laboratory, California Institute of Technology, under a contract with the National Aeronautics and Space Administration (80NM0018D0004).

# 42. Size distribution in the Solar System Kuiper Belt

Arielle Moullet (National Radio Astronomy Observatory), Joseph Masiero (Infrared Processing and Analysis Center), Noemi Pinilla Alonso (University of Central Florida), Dariusz C. Lis (Jet Propulsion Laboratory, California Institute of Technology)

The size frequency distribution (SFD) of the Solar System's Kuiper-Belt objects reflects the formation history of the outer Solar System. Accessing the size regime below 100 km is important to constrain the respective role of destructive (collisions) and accretive processes that modified the initial planetesimal mass function. While optical surveys have the ability to detect such targets, only thermal infrared observations have the ability to disentangle the effect of the albedo in retrieving equivalent sizes.

We describe a large PRIMAger targeted thermal survey of KBOs at 100 µm, completing and significantly extending earlier thermal surveys by detecting objects in the 35–80 km diameter range, inaccessible to Herschel's (and Spitzer). The retrieved SFD, once un-biased, can be effectively matched to theoretical model, discriminating against hypotheses initial mass function and collisional grinding at the time of Kuiper-Belt formation.

## General Scientific Area:

Solar System

## Science Justification:

The Solar System's Kuiper Belt is considered to be both the most pristine remnant of the Early Solar nebula, and the most accessible example of the last stage of an evolved debris disk. As such, the composition of individual Kuiper Belt Objects (KBOs) can be linked to the original composition of the Outer Solar System. On the other hand, population-wide physical and dynamical characteristics are indicative of past and present processes controlling planetesimal formation and distribution across a planetary system. In particular, planetesimal accretion processes are important to understand timescales in the transition from the protoplanetary disk stage to a fully evolved planetary system.

Today's Kuiper Belt size-frequency distribution (SFD) is one of the main observables that can be compared against planetesimal evolution models in the Solar System. Theoretical models (e.g., Davis & Farinella, 1997; Kenyon and Luu, 1999; Pan & Sari, 2005) can include catastrophic and non-catastrophic collisions, collisional grinding cascades and different accretion modes and stages (e.g., streaming instability, runaway growth, pebble accretion). Note that based on the surface history of large KBOs, the leading theory is that the size distribution has not been modified by collisional and accretional effects since the surfaces of these bodies formed (Spencer et al., 2020), and the SFD must hence represent the integrated effect of historical processes up to the time of Kuiper Belt formation.

The outcome of such models is usually a theoretical SFD displaying a break ('knee') in its slope at a specific size, a distinct typical size over which objects are considered of primordial size (hence representative of the initial planetesimal mass function), as well as a characteristic size under which objects are considered collisional fragments. The location of the SFD knee, and the slope of the SFD on either side of the slope break, are indicative of the efficiency of accreting and destructive processes over time, which are themselves linked to fundamental properties such as relative velocity (orbital inclination and eccentricity distribution) and solid bulk strength (Kenyon & Bromley, 2020).

Currently, the characterization of the SFD is mostly based on very sensitive optical photometric surveys such as the Outer Solar System Origins Survey CFHT (OSOSS, Bannister et al., 2018), which has characterized about 1000 objects with sizes down to ~20 km diameter. However, such surveys only directly measure absolute magnitude, not size. Magnitude is only a proxy of an equivalent size when the object's albedo is assumed, but such assumption may not be solid as albedos are known to vary between 2–70% amongst KBOs. Assuming the typical median albedo value of 10% for a given object can hence translate in a factor of 2 error in the derived diameter.

Albedo-unbiased size surveys are much more limited than photometric surveys, as they typically rely on thermal measurements (infrared or mm-wave) which are challenging due to limited sensitivity in the thermal regime. The most extensive albedo-unbiased survey was achieved with Herschel through the 'TNOS are cool' large program (Vilenius et al., 2012). Herschel's PACS sensitivity enabled access to thermal detection on ~130 KBOs of diameters down to 150 km diameter.

The SFD below 100 km, inaccessible by Herschel, would however be much more discriminant and informative in terms of comparison to theoretical models, as it covers the 40–100 km region where models display a slope break in the distribution (e.g., Abedin et al., 2022).

Using the exquisite far-IR point-source sensitivity offered by PRIMAger at 100 μm, approximately 100 times better than Herschel-PACS, one can design a pointed survey of 100–200 known objects across several dynamical classes, which will focus on the size regime of interest in regard to theoretical models. Specifically, with 1–4 hours of on time integration per object, one can access thermal detections for objects down to 35 km diameter.

The survey's immediate outcome will be photometry. Fluxes can be converted to equivalent sizes using available optical magnitudes and thermal emission models. There is some plausible range in the parameters for thermal models (emissivity, thermal inertia), which brings additional uncertainty to the derived size, but in a much more moderate manner (±20%) than the uncertainty from albedo bias. The derived size distribution will then be de-biased for observational limits (see Vilenius et al., 2014), and analyzed separately for distinct dynamical classes.

We expect that such a survey, directly complementing the size regime investigated by Herschel and commensurate to the size regime addressed by OSSOS, can be used to:

- provide an independent constraint to the Kuiper Belt SFD in the 35–100 km range.

- verify or improve the appropriateness of albedo-assumptions used in optical surveys, making the optically derived SFDs much more robust.

Together those constraints will facilitate a more meaningful comparison of the Kuiper Belt SFDs with formation models will help to understand the respective historical roles of accretion and collisions.

A separate science question that can be obtained on the largest targets is the presence of inhomogeneous surface features, which are known to be present on large KBOs, but whose presence of targets <100 km is not confirmed, possibly revealing geologically linked thermal structures and anomalies. Rotational monitoring with optical photometry provides similar accuracy but does not allow to disentangle size projection effects from surface features.

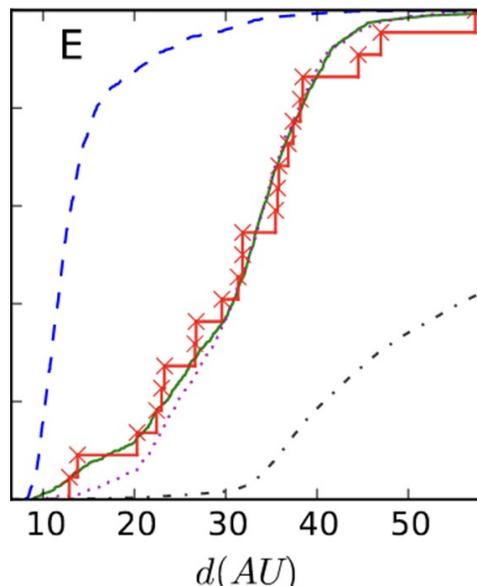

Figure 1: Adapted from Shankman et al., 2016: Cumulative distribution of KBOs scattered population as a function size (red - based on OSOSS observations) compared to three modeled distributions. Albedo assumption is 5% for all bodies.

## Instruments Required:

PRIMAGer small maps in hyperspectral band 2A.

## Program Size:

- Large (100 + hr)

## Approximate Integration Time:

Kuiper-Belt Objects as observed by PRIMA are point-sources, well below 0.1" in equivalent apparent angular size, and we require a SNR of ~10 on the detection to be able to constrain size to a meaningful level, especially considering modeling uncertainty on the top of measurement noise.

On the warmer side for KBOs, assuming a blackbody emission with a 50 K brightness temperature, a target with angular size 10 mas (~240 km diameter at 34 AU from Earth) would emit ~4.7 mJy at 100 μm. On the colder side, with a 30 K brightness temperature, a target with angular size 10 mas (~350 km radius at 50 AU from Earth) would emit ~2.8 mJy at 100 μm.

For a minimal 10'×10' PRIMAGER map at 100 μm, 1 hour of integration would yield a 10 σ detection for a 200 μJy target, and 4 hours of integration would yield at 10 σ detection for a 100 μJy target.

Based on these starting points, we estimate that to observe sources of sizes 35–95 km, integration times between 1–4h for each source are needed. Deeper observations could be warranted for specific targets, and targets on which we aim to perform monitoring may need 5–10h continuous observations.

With the considered sample of KBO targets, this leads to a total observing time of ~500h.

### Special Capabilities Needed:

All sources are Solar System sources with slow apparent motion: non-sidereal tracking is needed (less than 2"/h). The science case relies on high accuracy photometry, and special attention should be put towards flux reference accuracy. Continuous observations of several hours (5–10h) can be needed on some sources.

### Synergies with Other Facilities:

- JWST, possibly MATISSE at ESO (the 10 μm silicate feature).

### Description of Observations:

We propose a pointed photometry survey at 100 μm (PRIMAGER Band 2A) of ~200 point sources across the ecliptic. On 10–20 large sources, additional rotational monitoring could be performed.

## Acknowledgement:

A portion of this research was carried out at the Jet Propulsion Laboratory, California Institute of Technology, under a contract with the National Aeronautics and Space Administration (80NM0018D0004).

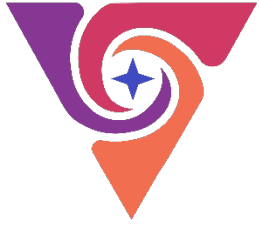

# ISM and Star and Planet Formation

## 43. How stars assemble their mass: uncovering the primary mode of stellar accretion with PRIMA

Cara Battersby (UConn), Henrik Beuther (MPIA), Will Fischer (STScI), Adam Ginsburg (University of Florida), Doug Johnstone (NRC-Herzberg), Rachel Lee (UConn), Klaus Pontoppidan (STScI)

The mass of a star predominately determines its physical characteristics, life cycle, and ultimate fate. Decades of research have uncovered a common initial stellar mass distribution, however, the processes by which stars assemble their mass remain largely undetermined. Several lines of evidence suggest that forming stars can undergo substantial episodic mass accretion events that are short compared to the formation time; however, the predominance of this episodic mode of accretion is unknown. Do episodic accretion events dominate the buildup of a star's mass ($\sum M_{burst} \gtrsim 50\% \, M_{star}$) or are they less consequential? With PRIMA, we can unequivocally answer this question.

Only in the Far-IR is a protostar's luminosity directly linked with its mass accretion rate. While mid-IR and submm observations have demonstrated widespread and substantial luminosity variations indicative of both large and small accretion events towards protostars, emission at these wavelengths does not directly correlate with the magnitude of the mass accretion events. Through Monte Carlo simulations, we demonstrate that repeated observations of 2,000 Galactic protostars with PRIMA allows us to detect enough large bursts to unambiguously resolve the primary mode of stellar mass accretion. Since protostars are highly clustered, and assuming a minimum map size of 3° x 1°, mapping 2,000 known Galactic protostars within 1.5 kpc can be achieved through observing 60 sq deg. of the sky. We propose to regularly monitor these protostars on cadences from two weeks to the entire five year PRIMA mission lifetime, and compare over 25 years against archival Herschel data.

### General Scientific Area:

Time Domain

### Science Justification:

**The mass of a star is important but how this mass is accumulated is largely unknown.**

The mass of a star determines nearly every aspect of its life and death. Its mass determines what elements it will produce in its lifetime, how those elements will ultimately be re-distributed, and the feedback effect that the star will have on its environment. The lifetime and final fate of a star is primarily dependent upon its mass.

Countless studies have focused on determining the initial mass function (IMF) of stars and whether this distribution varies across environments in the universe. However, the detailed processes by which a star assembles its mass are largely shrouded in mystery. What is the role of steady-state accretion versus episodic accretion bursts? How common are small (≲ factor of 2) versus larger (> factor of 10 or 100) accretion bursts and how long does each last? How do such episodic bursts affect the chemical composition of forming planets? An entire field of research lies largely unharvested because it has not yet been possible to answer these questions. PRIMA will enable us to plow these fields and unambiguously determine the dominant accretion processes for forming stars.

**Mass accretion events are important but parameter space is massively undersampled.**

Classically, it is thought that protostars smoothly acquire mass through their protostellar disks via steady-state accretion from the surrounding molecular cloud (Stahler et al. 1980, Pineda et al. 2022). However, several lines of evidence suggest that episodic mass accretion events play an important role in their mass assembly. A decade of observations from the mid-IR (e.g., Park et al. 2021; Zakari et al. 2022) to sub-mm (e.g., Lee et al. 2021) have established that the emission from protostars varies a lot. These protostellar luminosity variations (ranging from about 2x–100x) are the result of mass accretion events (Audard et al. 2014; Fischer et al. 2022).

The idea that episodic accretion events may be highly consequential was originally motivated by the luminosity problem (e.g., Kenyon et al. 1990), where mass accretion rates inferred from protostellar luminosities and integrated over the protostellar lifetime yield final stellar masses that are incompatible with the observed IMF. Therefore, mass accretion onto a protostar must change with time, which combined with the observed protostellar variability suggests a solution requiring periods of rapid mass accretion. The highly variable knotted morphology of protostellar outflows (e.g., Plunkett et al. 2015, Frank et al. 2014) independently suggests time-variable protostellar accretion. More directly, astronomers have been lucky enough to see several dozen extreme mass accretion events (e.g., Strom & Strom 1993; Caratti o Garatti et al. 2011; Safron et al. 2015, Audard et al. 2014) by investigating protostars in the time domain.

The state of the field and observations to date are summarized in the review by Fischer et al. (2022). We know that protostellar luminosities vary on every timescale that has been observed, from a few weeks (Billot et al. 2012) to many decades (Kenyon et al. 2000) with luminosity outbursts ranging from a factor of 2 to over 100 and everything in between. The most dramatic outbursts (FU Ori events) can increase a protostar's luminosity by a factor of ∼100x, but are thought to be quite rare and last for a quarter of a century or longer. Much more common are smaller burst-like variations, down to factors of 2-5x which may only last for weeks. Thus far, observations are consistent with a continuum between these two extremes, but parameter space is massively undersampled. Current data defy any simple classification system.

**Far-IR is the only wavelength that reliably traces mass accretion. We need PRIMA to understand how stars get their mass.**

Far-IR is the only wavelength capable of accurately tracing mass accretion onto forming protostars. While outbursts can and have been tracked in mid-IR and in the sub-mm, the measured luminosity at these wavelengths only loosely correlates with the amount of mass accreted. In Figure 1 and the analysis from Fischer et al. (in prep.), we demonstrate that by measuring the far-IR luminosity changes over time you can reliably and with relatively little scatter infer the true mass accretion magnitude of the event. At mid-IR wavelengths optical depth (extinction) causes high uncertainty and depending on the orientation of the disk, outflow, and envelope, the mass accretion estimate can be off by many factors. At sub-mm wavelengths, the optically thin envelope emits freely but the brightness depends instead on the dust temperature which varies with distance from the protostar and responds in a strongly non-linear way to changes in the source brightness.

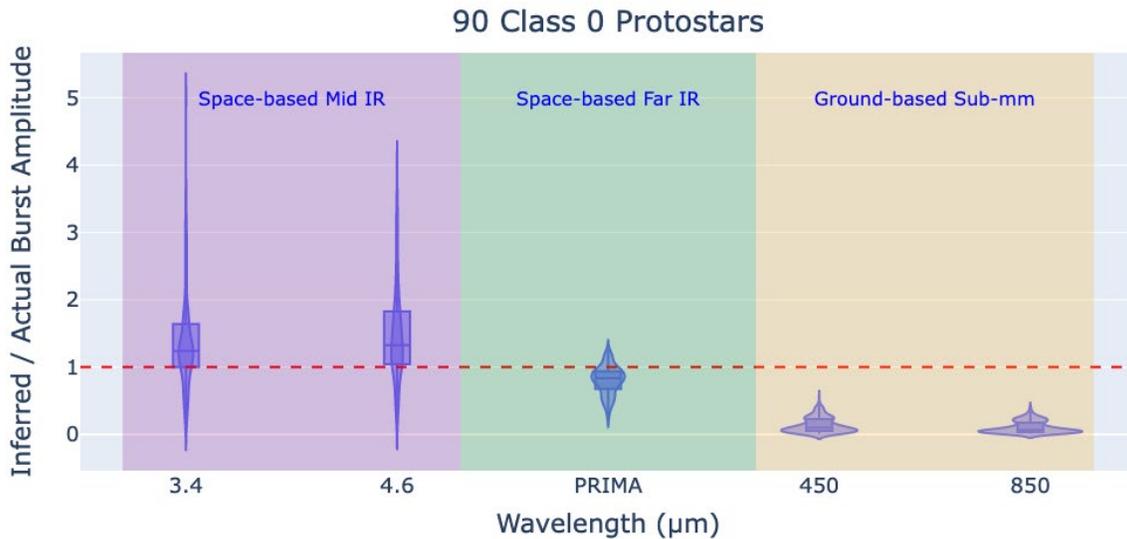

Figure 1: Far-IR is the only wavelength sensitive to the amount of mass being accreted onto a protostar during an accretion event (figure from Fischer et al., in prep.). This figure uses detailed protostellar models towards 90 Class 0 protostars in Orion (Furlan et al. 2016) with a range of burst amplitudes (from 10 to 100) to plot the inferred burst amplitude versus the true (modeled) burst amplitude. In the mid-IR, there is a high amount of uncertainty due to extinction and disk, envelope, and outflow orientation and it is not possible to infer the true burst amplitude. In the sub-mm the optically thin envelope emits freely but the brightness depends on the dust temperature which varies with distance from the protostar and responds in a non-linear way to changes in the source brightness. However, in the Far-IR with PRIMA, we can unambiguously trace the protostellar mass being accreted through the Far-IR luminosity.

**We need to observe 2,000 protostars with PRIMA in order to unambiguously determine the primary mode of protostellar mass assembly.**

With PRIMA, we aim to unambiguously determine whether or not the majority of a protostar's mass is assembled via episodic mass accretion events or via steady-state accretion, or in other words, is $\sum M_{burst} \gtrsim 50\% \ M_{star}$? In the limiting case, we can assume that a protostar gains its mass primarily in rare but powerful episodic accretion events, which sets the limit of the number of protostars required for observations. Specifically, we take the case of extreme (FU Ori) bursts whereby the brightness and accretion rate increases by a factor of 100 over steady-state and lasts for a quarter of a century or longer. In order for such bursts to account for 50% of the final stellar mass, the protostar must undergo such bursts 1% of the time. We consider this the limiting case, because it is the most stringent. In any other case, where smaller more common bursts play such an important role in mass assembly, we will automatically detect a sufficient number of bursts due the necessity of their occurring significantly more often.

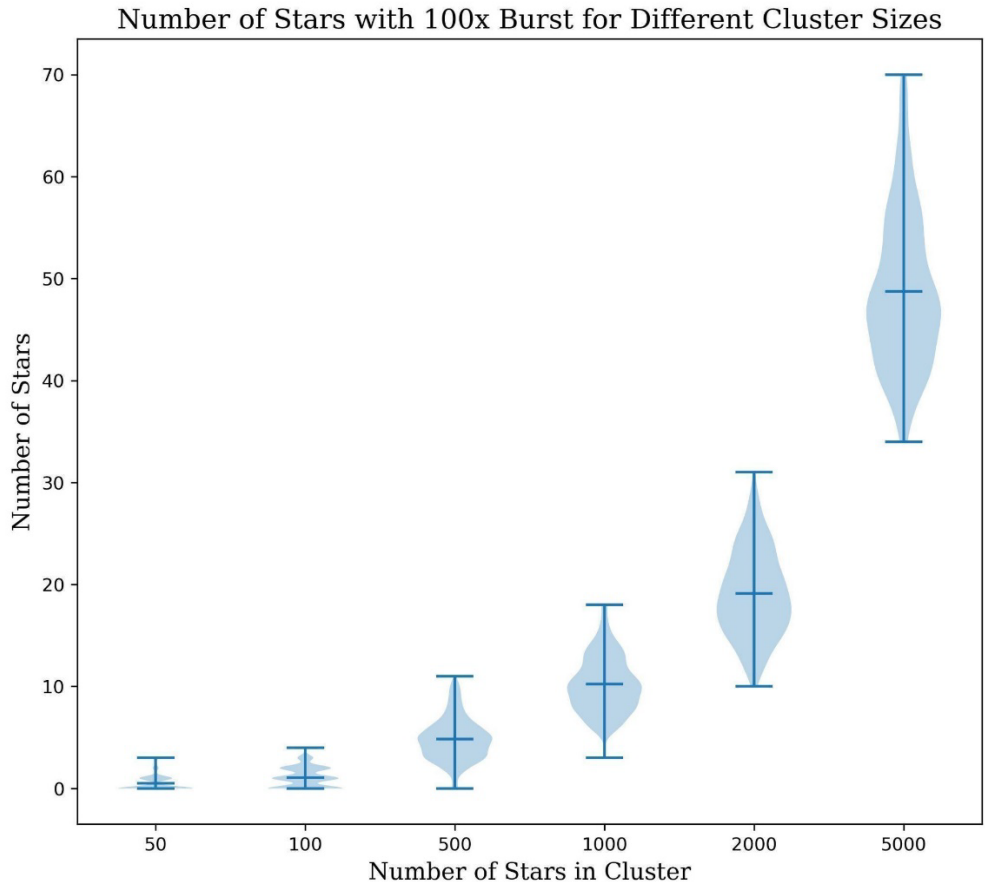

Figure 2: Monte Carlo simulations from Lee et al. (in prep.) show that we require observations of 2000 protostars to reliably and unambiguously determine the primary mode of protostellar mass assembly. This figure shows the number of stars that we would detect undergoing a 100x burst in each "PRIMA lifetime survey" for different protostellar cluster sizes. The burst frequency was set at 1% as described in the text and we determine that the cluster size of 2000 is necessary and sufficient to reliably detect sufficient bursts in our repeated simulations.

We run simple Monte Carlo simulations of protostellar clusters with various burst magnitudes, lifetimes, and probabilities that are then "observed" with a PRIMA-type facility (Lee et al. in prep.). We assume a PRIMA lifetime of 5 years and archival Herschel observations 20 years previous. We find that in this limiting case, where a protostar gains most of its mass through large episodic accretion bursts, we need to observe 2,000 protostars in order to be able to detect sufficiently rare, large bursts. We demonstrate this in Figure 2, which shows for a fixed input burst fraction, 1% as calculated above, the expected number of large (x100) burst detections with duration 25 years in a PRIMA lifetime survey ranging from 50–5000 objects. We determine that observing 2,000 protostars is necessary and sufficient to unambiguously decide the primary mass assembly mode for protostars.

**Understanding how stars assemble their mass is of fundamental importance in astrophysics. This field of research is largely unexplored due to previous observational limitations, yet it is teeming with possibility. With our ambitious large survey on PRIMA, we can unambiguously determine the main mode of stellar mass assembly. This survey is planned as a community collaboration to optimize the science return, including Target of Opportunity Observations**

**(ToOs), a complementary spectral line survey and post-outburst follow-up, as well as a complementary polarization survey.**

## Instruments Required:

- PRIMAger: Hyperspectral Imaging and Polarimetry, Large Map (> 20'×20')
- Followup with: PRIMA FIRESS Spectrometer: Large MapLow-res R∼130 and/or Pointed High-res R∼4400@ 112 μm

## Program Size:

- Large (100 + hr)

## Approximate Integration Time:

Assuming that the mapping speed is about 1 sq. deg. per hour, then each target epoch of 60 sq. degs will require 60 hours. We plan to obtain 10 epochs per target for a total of about 600 hours of science time. The sensitivity should not be the limiting factor since protostars are relatively bright, rather we would be limited by mapping speed.

## Special Capabilities Needed:

- We need to be able to observe all the protostars in the sample on cadences from 2 weeks to 5 years (start and end of mission). The short cadence observations should occur at the start of the mission so they can be repeated if necessary.
- We require a wide dynamic range to capture both dim protostars and bright outbursts in the same field.
- We require the ability to perform target of opportunity observations (ToOs) followup towards active protostellar bursts.

## Synergies with Other Facilities:

We plan to coordinate with ground-based and space-based facilities for complementary protostellar outburst monitoring. We will trigger a ToO towards an active burst in multiple facilities to more fully map out the protostellar mass assembly process. The time-domain is becoming ever more important for astronomy and several up-coming facilities anticipate dedicated monitoring campaigns.

## Description of Observations:

We will observe the nearest ∼2,000 protostars, within 2 kpc, over the five-year PRIMA mission lifetime with PRIMAGER. We plan to repeat the protostar survey ten times over the mission lifetime, with the shortest cadence being two weeks and the longest being the full five-year mission lifetime. Since the protostars are highly clustered, we will observe them in small maps of 1-3 sq. deg. in size.

In addition to our main observing project with PRIMAGER, we will coordinate:

1) target of opportunity observations (ToOs) with other facilities and with PRIMA FIRESS when a protostellar outburst occurs,

2) an initial spectral scan of all our protostellar targets with FIRESS, so that we can evaluate spectral changes during or after an outburst event, and

3) "bonus" polarimeter observations with PRIMAGER for each protostar observation. While a single map alone will not have sufficient sensitivity to detect dust polarization toward these star-forming regions, integrating all ten region visits will detect dust polarization.

## Cadence:

The shortest timescale on which we expect protostellar variation is about two weeks, the time it takes light to travel across a 6" beam at the distance of Orion, a typical region in our proposed survey (Johnstone et al. 2013). We plan to observe all the protostars in small maps as soon as possible when PRIMA begins science observations, then to follow up on the targets again two weeks later. Additional observations, save the final one, can happen on a somewhat flexible schedule, with the caveat that we aim to have the shortest cadence observations at the start of the mission. The first observations may yield unexpected variability which would lend key insights into the accretion process onto stars and may result in an altered observing program.

The final PRIMA protostellar observations should occur near the end of the mission lifetime, about five years after the first observations. We also plan to compare our observations with archival Herschel data to look for protostellar variability on timescales up to about 25 years. The rarest type of protostellar outbursts are also the most powerful, increasing the protostars luminosity by a factor of about 100 and lasting for about a century. These are extremely rare and in order to unambiguously determine the importance of these outbursts, we need to maximize our time baseline as much as possible.

## Mapping strategy:

We plan to map each region in the smallest map possible (down to about 1° × 1° to be the most efficient), but for the purpose of this proposal, we assume a minimum map size of 3° × 1°) with PRIMAGER. The protostars are bright so the limiting factor is the mapping speed and overheads, not the sensitivity required.

As described elsewhere, we also plan to simultaneously map with polarization and low-spectral resolution sensitivity with PRIMAGER. Additionally, we plan to do an initial protostellar survey with FIRESS so we can monitor for spectral changes during and after a protostellar outburst.

## Target of Opportunity Observations (ToOs):

During the first major (TBD based on modeling efforts in the years to come what is meant by major and when we should trigger additional observations, but likely at least a 3–5x increase in luminosity) detected outburst(s), we will trigger additional observations of that protostar / those protostars to observe: 1) how the outburst luminosity changes with time and 2) how the spectral features change with time. The ToO will be with PRIMA FIRESS as well as being coordinated with relevant ground- and space-based facilities. For these follow-up ToO observations to be most useful, we require an initial spectral scan of all the targets with FIRESS at the start of the PRIMA mission to establish the pre-burst spectrum.

## "Bonus Observations:"

In addition to the ToOs, community members rightly suggested that we include PRIMAGER hyper-spectral band as well as polarization observations. By integrating all ten region visits, we expect to detect substantial dust polarization signal toward each of the molecular clouds, fueling new and interesting studies of magnetic field properties towards these regions, e.g., determining whether ISM-scale magnetic fields play a role in regulating burst timescales.

## Targets:

We plan to observe the following regions in individual maps (the smallest we can make, 1° × 1° is the most efficient, but our estimates assume a minimum map size of 3° × 1°) with PRIMAGER. We have two categories of objects, those within 1 kpc and those out to about 2 kpc. Within 1 kpc, we have about 1000 protostars over approximately 44 sq. deg. The farther out targets yield about 1500 additional protostars over between $10^{-16}$ sq deg, for a total estimated observing area of about 60 sq. deg to yield at least 2000 Class 0 protostars.

Table 1: Targets within 1 kpc.

| Region | # of protostars | Reference | Map size (sq deg) | Typical source separation (pc)[1] | Distance (kpc) | Physical PRIMA resolution (pc)[2] |
|---|---|---|---|---|---|---|
| Perseus | 111 | Dunham et al. (2015) | 6 (one 2×1 + one 2×2) | 0.05 | 0.25 | 0.009 |
| Auriga/CMC | 43 | Dunham et al. (2015) | 5 (5×1) | 0.05 | 0.45 | 0.016 |
| Orion | 319 | Furlan et al. (2016) | 10 (two 3×1 + one 4×1) | 0.05 | 0.42 | 0.015 |
| Mon R2 | 188 | Gutermuth et al. (2011) | 6 (3×2) | 0.05 | 0.83 | 0.029 |
| Ophiuchus | 71 | Dunham et al. (2015) | 6 (3×2) | 0.05 | 0.13 | 0.005 |
| Aquila | 148 | Dunham et al. (2015) | 6 (3×2) | 0.05 | 0.26 | 0.009 |
| Serpens | 52 | Dunham et al. (2015) | 0 (included in Aquila) | 0.05 | 0.43 | 0.015 |
| Cepheus | 29 | Dunham et al. (2015) | 3 (three 1×1) | 0.05 | 0.32 | 0.011 |
| IC 5146 | 38 | Dunham et al. (2015) | 2 (2×1) | 0.05 | 0.95 | 0.033 |
| **Total** | **999 YSOs** | | **within about 44 sq. degs.** | | | |

[1] *Figure 2 of Gutermuth et al. (2009) suggests that adopting a 0.05 pc source separation for all regions would be a reasonable way to proceed.*

[2] *at 70 μm*

Table 2: Targets within 2 kpc.

| Region | # of protostars | Reference | Map size (sq deg) | Typical source separation (pc) | Distance (kpc) | Physical PRIMA resolution (pc) |
|---|---|---|---|---|---|---|
| Cygnus X | 200 class 0 670 class I | Beerer et al. 2010 | 4 | clustered | 1.5 kpc | 0.052 pc |
| Carina | 247 class 0/I | Povich et al. 2011 | 2.25 | clustered | 2.3 kpc | 0.081 pc |
| M17 | 406 YSOs | Povich et al. 2009 | 1.7 | clustered | 2.1 kpc | 0.074pc |
| W3 | 592 class 0/I | Rivera-Ingraham et al. 2011 | 1.5 | clustered | 2.0 kpc | 0.070 pc |
| **Total** | **1445 - 2115 YSOs** | | **about 10-16 sq. deg. when accounting for map shape** | | | |

# 44.  Ammonia chemistry in protoplanetary disks with PRIMA

Jenny Bergner (UC Berkeley), Romane Le Gal (Universite Grenoble Alpes and IRAM), Nicole Arulanantham (STScI), Klaus Pontoppidan (STScI)

Nitrogen is a major volatile element and a key ingredient for forming habitable worlds, yet the nitrogen chemistry in protoplanetary disks remains poorly constrained. $NH_3$ and semi-refractory $NH_4^+$ salts are likely a major reservoir for nitrogen in protoplanetary disks, but cannot be detected in disks with existing observational facilities. Only a far-IR facility with broad wavelength coverage and high sensitivity can provide access to both the cool $NH_3$ reservoir in the bulk outer disk, and the warm $NH_3$ reservoir tracing $NH_3$ ice and $NH_4^+$ salt sublimation around the water snowline. We show that 10 ortho-$NH_3$ and 14 para-$NH_3$ lines should be detectable with PRIMA for nearby disks, with upper-state energies spanning tens to hundreds of Kelvin. Characterizing the $NH_3$ chemistry in nearby disks will provide key new insights into the nitrogen budget available to forming planets.

## General Scientific Area:

Planet Formation

## Science Justification:

Planets form within disks composed of dust, ice, and gas in orbit around young stars. Observations of molecular lines towards protoplanetary disks provide a window into the chemistry and physics at play during planet formation. In the past decade, ALMA has revolutionized our view of disk chemistry in the millimeter, and JWST is poised to do so at near/mid-IR wavelengths (e.g., Tabone et al. 2023). Crucially, the far-IR fills in the gap between these facilities and provides access to so-called 'light hydride' molecules, including $NH_3$, $NH_2$, and NH; $H_2O$ and OH; CH and CH+; and HS and $H_2S$. Among the most exciting use cases is the detection of $NH_3$ lines towards protoplanetary disks.

Nitrogen is one of the most abundant volatile elements and likely a key ingredient for forming habitable worlds (e.g., Krijt et al. 2022). The main nitrogen carrier in disks, $N_2$, does not produce detectable emission at the temperature of the cold bulk gas, and as a result the N budget in disks is very poorly constrained. Trace N carriers like HCN, CN, and $N_2H^+$ provide some constraints on disk nitrogen chemistry (Bergner et al. 2019, van Terwisga et al. 2019, Qi et al. 2019), but we are missing a key carrier: $NH_3$ (e.g., Pontoppidan et al. 2019). $NH_3$ is an important constituent of protostellar ices (~5% wrt water; Oberg et al. 2011) but has only been detected towards two protoplanetary disks to date (Salinas et al. 2016, Najita et al. 2019). $NH_3$ may be present in disks either in the gas-phase, the ice-phase, or as semi-refractory ammonium salts. These salts can sublimate at somewhat hotter temperatures than $H_2O$, and may be a major reservoir of nitrogen in disks (Altwegg et al. 2020). Accessing the full $NH_3$ inventory therefore requires sensitivity to cool emission from the bulk outer disk, and warm emission tracing sublimation of $NH_3$ ice and $NH_4^+$ salts around or interior to the $H_2O$ snowline. This can only be achieved with a far-IR facility, which uniquely provides coverage of numerous $NH_3$ lines with upper-state energies spanning tens to hundreds of K. In addition to providing key constraints on the N budget, far-IR observations of multiple $NH_3$ lines will enable the ortho-to-para ratio of NH3 to be determined, which encodes information on its formation chemistry (e.g., Faure et al. 2013). Moreover, detections of the N hydride radicals NH and $NH_2$ can further constrain the nitrogen chemistry (Le Gal et al. 2014).

As a sensitive far-IR facility with broad wavelength coverage, PRIMA will be capable of revealing the $NH_3$ chemistry in disks. Most of the lines of interest are at relatively long wavelengths (>~100 um) and will therefore not be spectrally resolvable with a resolution R of a few thousand. Nonetheless, coverage of 10–15 lines each of ortho and para NH3 will enable retrievals of the molecular excitation temperatures, which can be used to infer the origin of the line emission within the disk. Also, line fluxes can be used in conjunction with disk chemistry models to determine molecular abundances, particularly in the case of nearby and well-characterized disks with auxiliary chemical and physical constraints (e.g., Salinas et al. 2016, Pontoppidan et al. 2019).

The sensitivity required to detect far-IR $NH_3$ lines with PRIMA is estimated based on the detection of $NH_3$ towards the nearby disk TW Hya with Herschel (Salinas et al. 2016). A model that can reproduce those observations (Bergner et al. 2022) was used to predict $NH_3$ line fluxes observable from 28–239 μm, shown in Figure 1. The predictions assume a statistical ortho-to-para ratio (OPR) of 1 and a source distance of 120 pc, typical for nearby star-forming regions. The PRIMA 5σ sensitivity is based on a 12-hour integration time, and accounts for a moderate far-IR disk continuum flux of 1 Jy. 10 o- $NH_3$ lines and 14 p-$NH_3$ lines are expected to be detectable. The upper-state energies of detectable lines span ~80-600K, providing access to cool-to-warm emission as well as leverage for determining the molecular excitation temperatures and column densities.

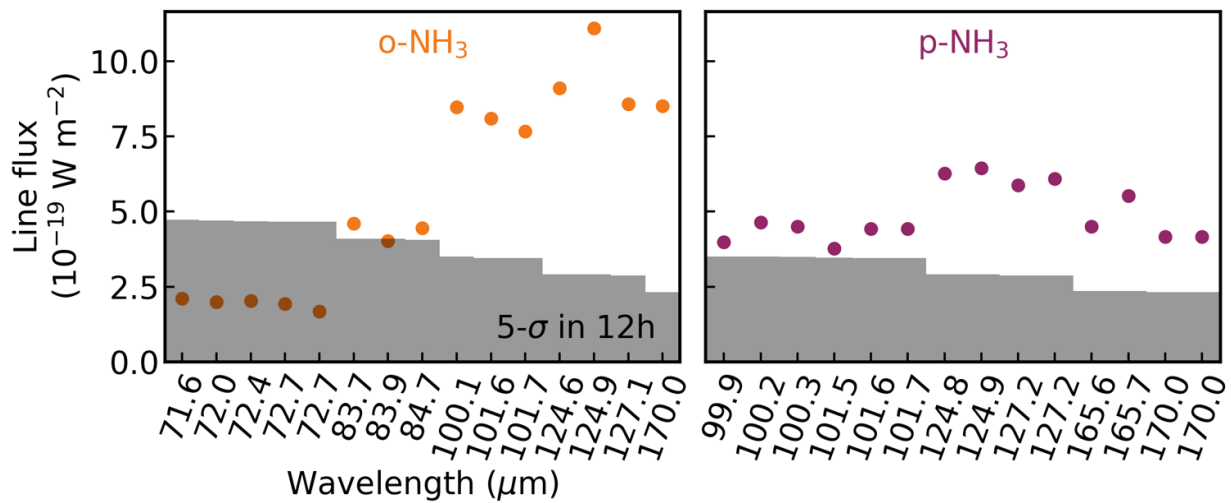

Figure 1: Predicted $NH_3$ line fluxes across the FIRESS band for a disk at 120 pc. Grey bars show the 5σ sensitivity for 12-hour integrations, given a 1 Jy continuum flux.

In summary, $NH_3$ (and $NH_4^+$ salts) may be a major reservoir of nitrogen in protoplanetary disks. The inability to detect cool-to-warm $NH_3$ lines with existing observing facilities is a major obstacle in our understanding of disk nitrogen chemistry. With broad wavelength coverage and high sensitivity, PRIMA will provide novel constraints on the N budget and chemistry in disks, key to our understanding of the formation of habitable planets.

*Other enabled science:* In addition to $NH_3$, the far-IR provides access to numerous other unique probes of disk chemistry and physics. Observations of common gas-phase molecules like CO, CN, and HCN in the far-IR will access emission from intermediate radii between the hot inner disk probed by JWST and the cold outer disk probed by ALMA. And, coverage of small radicals/ions like OH, NH, and $CH^+$ will help reveal the mechanisms of radical/ion-driven chemistry in disk gas, as well as the radiation environment in photochemical disk layers. These lines will be covered 'for free' as part of other disk spectral line programs targeting e.g., $H_2O$ and HD that cover the full FIRESS band.

## Instruments Required:

- PRIMA FIRESS: Pointed High-res R~4400@ 112 µm.

## Program Size:

- Medium (10–100 hr)

## Approximate Integration Time:

- 12h per setting

## Special Capabilities Needed:

- None

## Synergies with Other Facilities:

- Auxiliary constraints on disk chemical/physical structures from ALMA. Future coverage of low-frequency $NH_3$ lines with ngVLA.

## Description of Observations:

These observations use the FIRESS instrument in high-resolution mode with maximum spectral resolving power, achieving velocity resolutions of ~20–70 km/s from 28–112 µm. Disks will be spatially unresolved within the PRIMA beam and require a single pointing. The whole FIRESS band is required to cover all target lines, requiring two executions per disk. Figure 1 shows that for a typical nearby (120 pc) disk with a moderate continuum flux (1 Jy), numerous ortho and para $NH_3$ lines should be detected >5σ with a 12-hour integration per setting. For disks with lower $NH_3$ abundances, further distances, and/or brighter continuum fluxes, longer integration times will be necessary. Promising nearby disk targets for proof-of-concept studies include TW Hya, V4046 Sgr, and HD 163296.

# 45. Probing dust in the extremely low star forming intergalactic clouds

Caroline Bot (Université de Strasbourg, CNRS, Observatoire Astronomique de Strasbourg), Lucie Correia (Université de Strasbourg, CNRS, Observatoire Astronomique de Strasbourg), Pierre-Alain Duc (Université de Strasbourg, CNRS, Observatoire Astronomique de Strasbourg), Frédéric Galliano (AIM, CEA, Saclay); Suzanne Madden (AIM, CEA, Saclay); Nathalie Ysard (Institut d'Astrophysique Spatiale)

What are the properties of dust in very low density and low star formation rate environments? Is the life cycle of metals and the interstellar medium the same inside and outside galaxies? Thanks to the high sensitivity of PRIMA in the 24-260 μm wavelength range, we will be able to observe and study the properties of dust associated to diffuse gas in outer disks of galaxies or even in extragalactic clouds and tidal streams. Probing these extreme environments will shed light on dust survival and growth processes, as well as bring new elements on the evolution of dust and metal properties with environment. We propose to observe dust emission in total intensity with PRIMager in the Leo intergalactic Cloud. This cloud is a unique occasion to open the field of very low-density dust emission detection: the cloud it is at the right distance to be detected and has a large amount of ancillary observations from which the existence of dust has been inferred and which will provide a good characterization of the environment for the interpretation (star formation rate, gas density, heating sources, metallicity). The challenge for such detection, beyond the faintness of the signal, will be to mitigate confusion effects. We propose to first observe star forming clumps in a few arcminutes fields in all PRIMAger hyperspectral bands to detect the mid-FIR peak of the intergalactic dust emission. The partial band coverage in the surroundings will allow to explore the mapping strategy for even lower density gas. These observations of two star-forming clumps in the Leo Intergalactic cloud will open the way to further studies of low-density features.

## General Scientific Area:

Rise of Metals and Dust

## Science Justification:

Studies of dust in nearby galaxies have enabled us to probe wider range of environments (e.g., star formation rates, metallicity) than what is available in our Milky Way. Such studies provide important constraints not only on the dust and metal life cycles in galaxies, but also on the dust itself. Conversely, dust emission has been used to trace gas, including « dark » phases like $H_2$ without CO or diffuse gas thanks to mid-IR JWST observation [1], unveiling a bubbly structure only seen in simulations before. Since dust emission in the local universe is brightest in the FIR, it might be surprising that dust associated to diffuse gas is only observed thanks to the fainter mid-IR PAH emission. Yet previous missions like Herschel had low sensitivities and spatial filtering, biasing the detection of mid- to far-IR dust emission, even in very nearby galaxies [2]. This situation is going to drastically change with the high sensitivity of PRIMA in the 24–260 μm wavelength range. This wavelength range will sample the peak of the dust emission from small and large grains which represent most of the dust mass. We will hence be able to detect and study the intensity and shape of dust emission in previously inaccessible environments like the outer disks of galaxies or even in the extragalactic clouds and tidal streams far away from galaxies. By studying dust properties (dust-to-gas mass ratio, dust-to-metal ratio, dust temperature, size distribution) in these diffuse environments we will sample extreme cases of the parameter space (star formation, densities) and shed lights on dust survival and/or growth processes, as well as on the metal and interstellar medium life cycles in and outside galaxies.

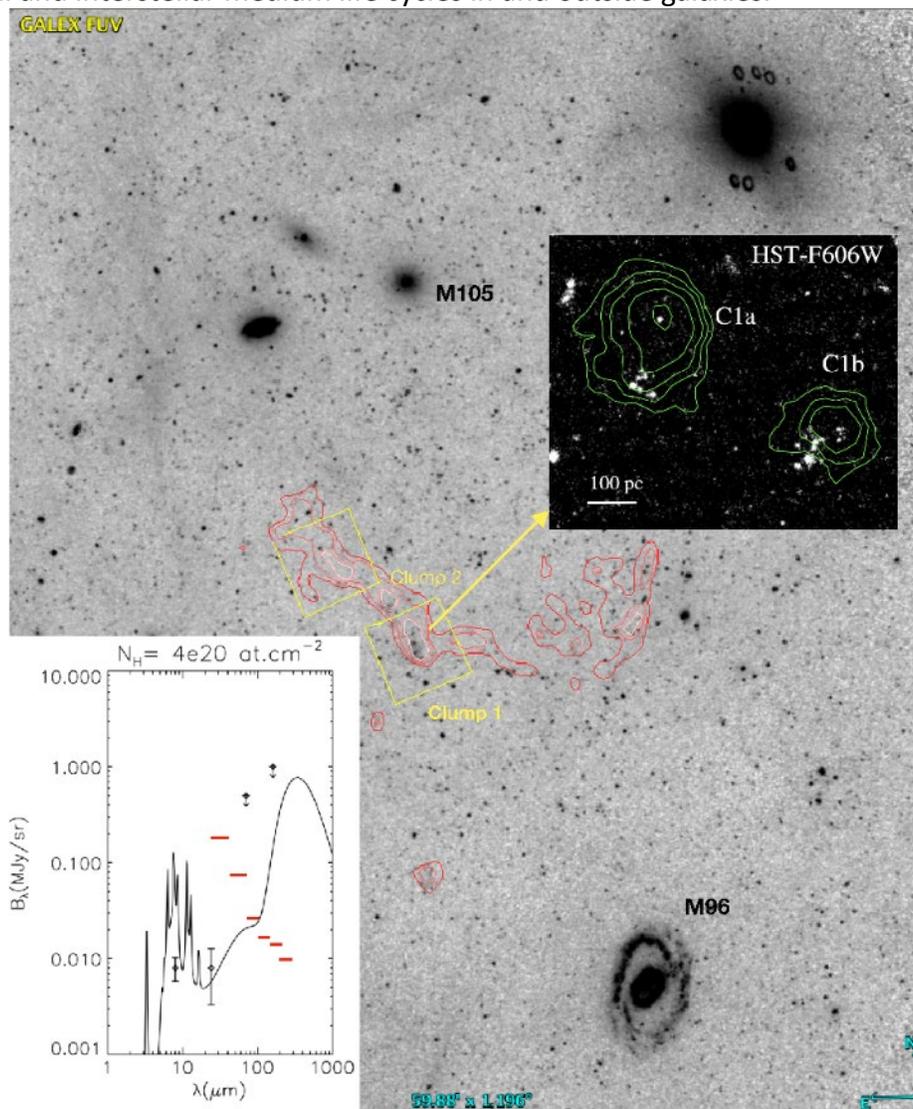

Figure 1: FUV mosaic of the M96 group with HI contours of the Leo cloud in red and Spitzer IRAC fields in yellow. The bottom left corner shows the expected dust emission together with Spitzer detections and upper limits in the clumps. 5σ sensitivity estimates for 10h of observations are overlaid. MUSE Hα emission in clump 1 is shown with green contours on the HST image in the excerpt on the top right corner (taken from Fig 6 of [8]).

We propose to observe dust emission in total intensity with PRIMAger in the Leo Intergalactic Cloud. The Leo Intergalactic Cloud was discovered in HI in the M96 group [3] and has a ring like shape. Higher resolution (45'') VLA observations showed the presence of HI clumps with densities reaching a few $10^{20}$ at.cm$^{-2}$ [4]. Dust emission associated to the highest density clumps was detected at 8 and 24 μm with Spitzer [5] but only upper limits to the emission were obtained at 70 and 160 μm. GALEX observations showed recent in-situ star formation in the highest density clumps [6]. This was confirmed and characterized by three 1'×1' MUSE observations [7], [8]: 4 HII regions powered by massive O stars were identified and the analysis of their line ratios suggests close to solar metallicities and moderate visual dust extinctions. Given the extremely low star formation rate, the intergalactic cloud is probably made of pre-enriched gas removed from a galaxy disk by tidal forces [9] and is now forming a small number of stars in-situ.

The high sensitivity of PRIMA in the FIR will provide a new window on dust and gas emission in extremely diffuse and low star-forming environments. The Leo ring is a unique opportunity to study such environments (measured densities and star formation rate densities in clump 1 are $n_H$=0.4 at.cm$^{-3}$ and $10^{-4.3}$ M$_\odot$.yr$^{-1}$ kpc$^{-2}$ respectively) thanks to its near solar metallicity, relatively nearby location (D ≤ 20Mpc) and large amount of archival observations. Observations with PRIMAger of two different ~1' star forming clumps will be the first probe of dust properties in this unexplored part of environmental parameter space. This will open the way to further studies of low-density features (e.g., [10]) and bring general constraints on dust survival or (re-)formation in the diffuse ISM.

### Instruments Required:

- PRIMAger: Small Map (< 20'×20')
- Hyperspectral band

### Program Size:

- Medium (10–100 hr)

### Approximate Integration Time:

- 10h for each of the 2 fields targeting a star-forming clump

### Special Capabilities Needed:

- None

### Synergies with Other Facilities:

Obvious synergies in the Leo cloud are with ALMA to trace CO and mm dust continuum, as well as with JWST to probe PAH dust emission.

### Description of Observations:

To mitigate confusion effects and because we expect the emission to be extremely faint, we aim to observe 2 HI VLA clumps also observed with MUSE. For each, we want to observe a small field (a few arcminutes) in all PRIMAger hyperspectral bands (in practice, the smallest map possible with PRIMAger is about 10'×10'). The hyperspectral observations of the 2 central regions from 25 to 235 μm will enable us to detect and constrain the dust mass, dust-to-gas, dust-to-metal ratio and dust size distribution in these two different clumps. The mapping strategy to obtain all bands in the central positions will naturally provide us with images of the surroundings (although in fewer bands) that will sample even lower density gas further away from heating sources. Targeted spectral observations on the two detected HII regions in clump 1 may be used to complement MUSE observations with mid-IR lines diagnostics to further probe the state of the gas, excitation, and elemental abundances. While we expect the emission in the 2 HII regions to be faint (given the Hα luminosity), we expect the lines to be fairly contrasted with respect to the continuum.

## 46. Resolving the mystery of dust emissivity index variations in the Andromeda galaxy

Ilse De Looze (Universiteit Gent, Belgium), Jeremy Chastenet (Universiteit Gent, Belgium), Monica Relano-Pastor (Granada University, Spain), Matthew Smith (Cardiff University, UK), Steve Eales (Cardiff University, UK), Maarten Baes (Universiteit Gent, Belgium), Jacopo Fritz (UNAM Morelia, Mexico), Frédéric Galliano (CEA Paris-Saclay, France), Suzanne C. Madden (CEA Paris-Saclay, France), Anthony Jones (IAS, France), Simone Bianchi (INAF, Italy), Viviana Casasola (INAF, Italy), Angelos Nersesian (Universiteit Gent, Belgium), Manolis Xilouris (IAASARS, Greece)

Interstellar dust grains play a vital role in the shielding of dense interstellar clouds from radiation where complex molecules can form, and new stars are born. Dust grains also impact the heating and cooling balance in the interstellar medium and obscure a non-negligible fraction (30–50% on average, up to 99% in extreme cases) of starlight. To understand these physical processes, and recover the starlight obscured by dust, we need to know how dust grain sizes and properties vary from one region to another. The observed strong variations in dust emissivity index on spatially resolved galaxy scales suggest strongly varying dust properties on interstellar cloud scales. To understand how grain properties vary from one environment to another is important to correctly recover the total dust mass and to correct for the dust obscuration of starlight, especially in star-forming regions where young stars remain enshrouded by their dusty birth clouds for a long time.

The Andromeda galaxy is best suited for studying grain properties on resolved scales of <100 pc. The strong radial variations in the Rayleigh-Jeans slope in this galaxy remain unexplained and require (1) the full mid-infrared to sub-millimetre wavelength coverage to disentangle temperature mixing effects from intrinsic dust property variations and (2) dust polarimetric observations to constrain variations in the grain composition and structure on resolved galaxy scales. We propose to use the PRIMAger instrument onboard PRIMA to map the entire Andromeda galaxy with the hyperspectral and polarimeter bands from 25 to 265 μm. This will allow us to test whether the radial variations in the Rayleigh-Jeans slope are mostly sensitive to the width of the SED peak (hence, dust temperature mixing effects) or to intrinsic variations in the dust opacity law (and, thus, genuine changes in the dust properties). These changes in dust properties can be constrained independently through modelling the observed polarisation fractions. This information will be crucial to understand how dust evolves from one ISM phase to another, how the formation and processing of dust grains affect their properties and how the mid-infrared-to-submm SEDs of more distant galaxies should be interpreted.

## General Scientific Area:

Galaxy Ecosystems and Interstellar Medium

## Science Justification:

Although interstellar dust accounts only for roughly 1% of the interstellar mass, dust grains are indispensable in galaxies (1.) to cool the gas and initiate star formation in dense clouds; (2.) to drive galactic outflows through radiation pressure on dust grains; (3.) to catalyse the formation of molecular hydrogen from which new stars will eventually form. Interstellar dust is also responsible for the obscuration of 30 to 50% of all stellar light, and thus obstructs our view and inhibits our understanding of how star formation events unfold in galaxy disks if dust obscuration is not properly corrected for. While we have come to realise the importance of dust during the last decade — as evidenced by the inclusion of dust in cosmological simulations, the suite of laboratory studies aiming to quantify dust properties, and the plethora of chemical evolution studies for galaxies nearby and far away — *we have not reached a consensus about the properties of interstellar dust*.

Studies of dust extinction curves in the Milky Way, Small and Large Magellanic Clouds have revealed a large diversity in the shape of their dust extinction curves — characterised by their ultraviolet slopes and the 2175 Angstrom bump strengths. These variations suggest that the grain size distributions and dust composition not only vary from one galaxy to another, but also strongly vary on resolved galaxy scales. As an example, the famous SMC-bar dust extinction curve characterised by a flat UV slope and the absence of the 2175 Angstrom bump differs from the average $R_V$=3.1 Milky Way dust extinction curve with a strong 2175 Angstrom bump. However, SMC-like dust extinction curves have also been seen along some sight-lines in the Milky Way which suggests that internal dust processing within galaxies may dominate over global galaxy variations.

A complementary view on these dust property variations is provided by the far-infrared and sub-millimetre dust emission detected with Herschel, where the dust emission $F_\lambda$ is typically modelled with a blackbody of a given temperature T, modified by a power-law dust emissivity with emissivity index β:

$$F_\lambda = \kappa_\lambda \, B_\lambda \, (T) \ \text{ with } \ k_\lambda = \kappa_0$$

Intriguingly, the dust emissivity index β — which dictates the behaviour in the Rayleigh-Jeans regime of the dust emission and is sensitive to dust property variations — shows strong variations with local galaxy environment. Specifically, the dust emissivity index has been shown to peak in the central regions of well-resolved nearby galaxies (M31 and M33, see Fig 1 for the former) with beta values of around 2.5, and drops off to beta values of 1.5 in the outer galaxy regions (Smith et al. 2012, Tabatabaei et al. 2014). Radial variations in beta were also witnessed in galaxies outside the Local Group, where the radial trends (positive or negative gradients) differ from one galaxy to another (Hunt et al. 2015). Dust emissivity index variations are also observed on global galaxy scales and seem to correlate with galaxy properties such as stellar mass (surface density), metallicity, HI mass fraction, star formation rate (SFR) surface density, and the ratio of SFR over dust mass in these galaxies (Lamperti et al. 2019, Smith et al. 2019). These radial trends remain unexplained up to now and hold crucial information about the variations of dust properties with the local environment.

The best environment to study dust emissivity index variations is our nearest neighbour, the Andromeda galaxy, located at a distance of 785 kpc, where we can study resolved structures such as molecular clouds, HII regions and more diffuse ISM media at unprecedented scales of <100 pc (see Figure). In fact, a recent study in M31 has gone beyond studying the radial trends in M31, and has studied and compared dust emissivity index variations inside and outside molecular clouds (Athikkat-Eknath et al. 2022). While we would expect significant dust processing to take place inside molecular clouds - where accretion and coagulation processes can alter the grain structure and shift the grain size distribution to larger grains, resulting in larger beta values - this recent study did not find any significant variations between the dust emissivity index inside and outside molecular clouds in M31. This suggests that the radial dust emissivity index variations must have another origin.

An important aspect, that requires verification, is that the effective dust emissivity index inferred from spectral energy distribution (SED) modelling may not be unambiguously linked to the intrinsic dust opacity law of the dust grain population. The dust emissivity index is also sensitive to the temperature distribution within a galaxy with a wide temperature distribution leading to low beta values (Remy-Ruyer et al. 2015). To disentangle dust opacity variations from dust temperature mixing, we require full coverage of the mid-infrared to sub-millimetre wavelength range at a similar resolution as probed with Herschel and SCUBA-2. The 70 to 850 µm wavelength range was covered with Herschel and SCUBA-2, but the limited wavelength coverage and sensitivity at the shortest wavelengths has made it impossible to tackle degeneracies between dust temperature mixing and dust property variations. For this purpose, we require deep observations with PRIMA to constrain the full distribution of dust temperatures and to infer the origin of the dust emissivity index variations in M31. Moreover, with PRIMAger we will map the polarized far-infrared emission which will give us an important independent constraint to characterise the dust properties. In a recent paper, we have used SOFIA dust polarised fractions at mid-infrared wavelengths to constrain the grain composition (parametrised through a carbon-to-silicate grain ratio) in the Galactic supernova remnant the Crab Nebula (Chastenet et al. 2022). The PRIMAger polarimetric capabilities will provide vital constraints on the grain properties that can be compared with the dust emissivity variations.

In summary, we will use the PRIMAger instrument onboard PRIMA to map the entire disk of the Andromeda galaxy in the hyperspectral and polarimeter bands from 25 to 265 µm. These data will be complemented with the existing Herschel (Fritz et al. 2012) and SCUBA-2 (Smith et al. 2021) observations with 6–18" resolutions to do spatially resolved SED fitting of >100 000 of pixels on <100 pc scales with physically motivated models that include a mixture of grain sizes, grain compositions and radiation field strengths (e.g., Relano et al. 2022, Galliano et al. 2021) and polarised dust emission (Chastenet et al. 2022). This will allow us to

(1.) constrain the multi-temperature components contributing to the mid-infrared to sub-mm dust emission;

(2.) infer unbiased estimates of the dust emissivity index β characterising the Rayleigh-Jeans slope of the dust SED;

(3.) link variations in dust polarisation fractions to dust property variations;

(4.) get precious constraints on how dust properties vary with local conditions on <100 pc scales which will be vital to guide dust emission model fitting of more distant galaxies where uncertainties on the assumed dust emissivity index β and dust opacity κ currently make dust mass estimates uncertain easily up to an order of magnitude.

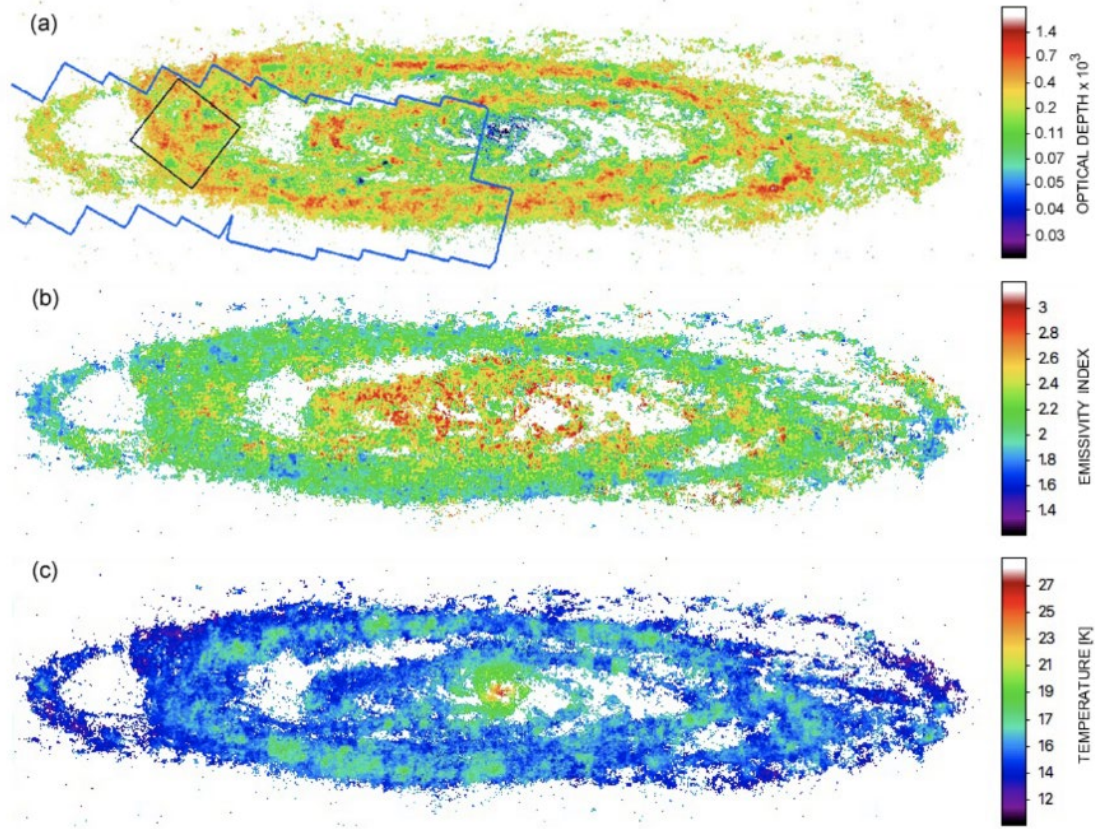

Figure 1: Maps of the dust optical depth (top), dust emissivity index β (middle) and dust temperature (bottom) in the Andromeda galaxy, as inferred from SED fitting (Fig. adopted from Whitworth et al. 2019, see also Smith et al. 2012). The origin of the radial trend in beta is unclear and seems to point to radial dust property variations but could also be linked to temperature mixing effects. PRIMA, with its excellent imaging and polarisation capabilities, will allow us to disentangle both effects.

## Instruments Required:

- PRIMAger: Large Map (> 20'×20') in Hyperspectral band and Polarimeter band (100 μm)

## Program Size:

- Medium (10–100 hr)

## Approximate Integration Time:

- ~90 hours.

## Special Capabilities Needed:

- None

## Synergies with Other Facilities:

- Herschel and SCUBA-2

## Description of Observations:

We request PRIMAger mapping in the hyperspectral and polarimeter bands. For an integration time of 30h/deg$^2$, we will reach surface brightness sensitivities of 260 kJy/sr and 635 kJy/sr in the hyperspectral imaging bands between 45–80 and 25–45 µm, respectively. With the same observing time we achieve surface brightness sensitivities ranging between 30-70 kJy/sr in the polarimeter bands. This observing strategy will allow us to go 2x times compared to existing Herschel observations (with a typical noise level of 72 kJy/sr in the SPIRE 250 µm band). Given the typical surface brightness levels of the dust emission in the spiral arms of the Andromeda galaxy, and assuming a conservative 1% of polarisation (cf. SOFIA observations of nearby galaxies have recovered average polarisation levels above 1%, Kim et al. 2023), these deep PRIMA observations will allow us to study the typical dust polarisation levels in the spiral arms of Andromeda. To cover Andromeda's disk out to a galactocentric radius of 20 kpc, we require a total observing time of 90h.

# 47.  Self-consistently probing the dust properties and the ISM structure of nearby galaxies

Frédéric Galliano (DAp, CEA Paris-Saclay, France), Maarten Baes (Ghent University, Belgium), Simone Bianchi (INAF, Italy), Caroline Bot (Observatoire astronomique de Strasbourg, France), Viviana Casasola (INAF, Italy), Jérémy Chastenet (Ghent University, Belgium), Christopher Clark (STScI, USA), Stavroula Katsioli (Athens Observatory, Greece), Ilse De Looze (Ghent University, Belgium), Mika Juvela (Helsinki University, Finland), Vianney Lebouteiller (DAp, CEA Paris-Saclay, France), Suzanne Madden (DAp, CEA Paris-Saclay, France), Takashi Onaka (Tokyo University, Japan), Lara Pantoni (DAp, CEA Paris-Saclay, France), Monica Relaño Pastor (Granada University, Spain), Matthew Smith (Cardiff University, UK), Tsutomu Takeuchi (Nagoya University, Japan) Manolis Xilouris (Athens Observatory, Greece)

The idea is to exploit the fact that PRIMA will allow us to self-consistently obtain, at the same time, the dust continuum and FIR lines in nearby galaxies. In a sample of nearby galaxies, this would allow us to solve some of the degeneracies one often encounters modeling dust SEDs. The information provided by the FIR lines can indeed be used to constrain the ISM/star topology. With this knowledge, we can thus solve the radiative transfer, at subpixel scales, using the information provided by the gas lines to more accurately model the dust.

## General Scientific Area:

Galaxy Ecosystems and Interstellar Medium

## Science Justification:

One of the main challenges when interpreting the Spectral Energy Distribution (SED) of an object is to separate the effects due to the variation of the microscopic grain properties (composition, size distribution) and those due to the macroscopic distribution of the InterStellar Medium (ISM). This problem has been addressed, concerning the gas, by state-of-the-art models, where a large number of lines can be used to constrain both the physical conditions of the ISM and its topology in galaxies (e.g., Cormier et al. 2012). This has been possible because atomic physics is more precisely known than dust physics. The fact that different ions have a wide range of critical densities allows us to use a few well-chosen lines to characterize the main phases of the ISM. The knowledge of the average structure of the ISM and of the stellar distribution, provided by the lines, could thus be used as an a priori to model the dust properties. Yet, obtaining at the same time a well-sampled SED and the intensity of the main IR gas lines is observationally challenging and has been achieved in only a handful of sources, most of the time for a single pointing.

The combined Mid-InfraRed to Far-InfraRed (MIR-to-FIR) photometric and spectroscopic capability of PRIMA opens the window to obtaining consistent spatially resolved maps of the total dust and multiphase gas properties in extragalactic regions. Such spectra were only obtained in the past by combining ISOSWS and ISOLWS (e.g., Peeters et al. 2002), but with a poor sensitivity, a poor spatial resolution and stitching problems. If we have only a few broadbands, scattered over the whole IR domain, as it is usually the case, we are unable to solve the degeneracy discussed above (e.g., Galliano et al. 2018, for a review). The unique feature brought by PRIMA will be consistent maps of the brightest far-IR lines, with the well-sampled dust continuum.

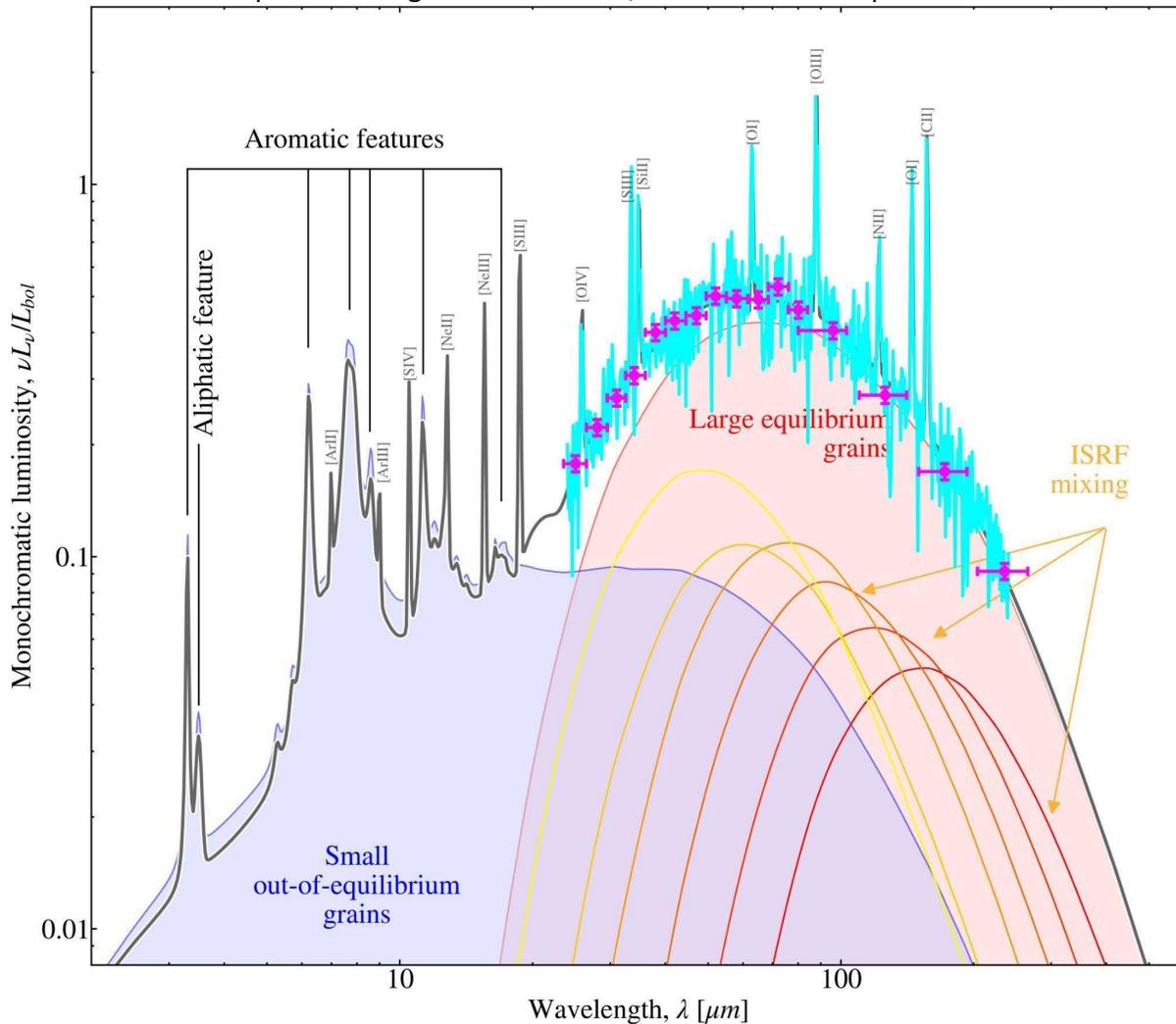

Figure 1: Typical SED of a star-forming region with the brightest gas lines. The magenta error bars correspond to broad-band spectrophotometry with R≈3.5 (10σ), and the cyan line is a R≈200 spectrum (5σ).

## Instruments Required:

PRIMA FIRESS: Medium Map Low-res R~130, PRIMAger: Small Map (< 20'x20')

## Program Size:

- Large (100 + hr)

**Approximate Integration Time:**

- 200 hr

**Special Capabilities Needed:**

- None

**Synergies with Other Facilities:**

NIR and MIR lines observed with JWST could advantageously be combined with the FIR lines observed with PRIMA.

**Description of Observations:**

We will observe a sample of nearby galaxies, both in narrow-band imaging and low-resolution spectroscopy. We do not need to reach the lowest emission of these objects, but we would need to have contiguous maps, in order to understand the spatial variations. Such combined dust-gas maps could be modeled using Cloudy (Ferland et al. 2017) and the Meudon PDR code (Le Petit et al. 2006) for the gas, a dust evolution model such as THEMIS (Jones et al. 2017) for the dust and a radiative transfer code such as SOC (Juvela 2019). We could observe 30 disk galaxies (20'×20') and 100 low-metallicity galaxies (2'×2'; 1/10 to 1/3 $Z_\odot$) in order to be able to build an evolutionary sequence of the dust properties.

- We will observe this galaxy sample with all PRIMAger bands, and with PRIMA-S at low spectral resolution.

- The continuum intensity is based on the Jones et al. (2017) model, at $U$=3, and N(HI)=5×$10^{20}$ H/$cm^2$. This corresponds to the typical extended emission of disk galaxies found in the DustPedia sample. This is however not the most diffuse emission of the galaxy, but going down to N(HI)=5×$10^{20}$ H/$cm^2$ is sufficient for the present science case.

- These fine-structure lines are all bright. In addition, we do not need to spectrally resolve them. Low-resolution spectroscopy should thus be sufficient to measure their intensity. We have estimated the line intensity by assuming they are proportional to the total IR luminosity. The proportionality factor is the average of the sample in Cormier et al. (2019). Sensitivity is not an issue for this science case, as we are not aiming at the faintest regions.

# 48. Surveying the mineralogical diversity of the ISM

Frédéric Galliano (DAp, CEA Paris-Saclay, France), Maarten Baes (Ghent University, Belgium), Simone Bianchi (INAF, Italy), Caroline Bot (Observatoire astronomique de Strasbourg, France), Jérémy Chastenet (Ghent University, Belgium), Christopher Clark (STScI, USA), Ilse De Looze (Ghent University, Belgium), Mika Juvela (Helsinki University, Finland), Hidehiro Kaneda (Nagoya University, Japan), Suzanne Madden (DAp, CEA Paris-Saclay, France), Mikako Matsuura (Cardiff University, UK), Takashi ONAKA (Tokyo University, Japan), Lara Pantoni (DAp, CEA Paris-Saclay, France), Monica Relaño Pastor (Granada University, Spain), Tsutomu Takeuchi (Nagoya University, Japan) Manolis Xilouris (Athens Observatory, Greece), Nathalie Ysard (IAS, Orsay, France)

PRIMA will provide continuous medium-resolution spectra in the FIR, for the first time since ISO. With its exceptional sensitivity, we should be able to detect numerous solid-state features in the ISM. We plan to obtain medium resolution spectral maps of star forming regions in the Milky Way (MW), Large Magellanic Cloud (LMC) and Small Magellanic Cloud (SMC). Star-forming regions will provide strong enough background sources to accurately characterize the bands, and the temperature gradient as a function of distance from the central cluster will allow us to perform better decomposition. We will thus better constrain the dust composition in the nearby Universe.

## General Scientific Area:

Galaxy Ecosystems and Interstellar Medium

## Science Justification:

One of the most challenging open questions concerning the physics of the InterStellar Medium (ISM) is the chemical composition and structure of dust grains. We have approximate indirect constraints coming from the elemental depletions and from the detection of a few broad features. The large uncertainty of dust models however is a consequence of our ignorance of this composition and structure. For instance, Zubko et al. (2004) assume it is made of 4.6% Polycyclic Aromatic Hydrocarbons (PAHs), 16.4% of graphite and 79 % of amorphous silicates; Compiègne et al. (2011) assume it is made of 7.7% of PAHs, 15.8% of amorphous carbon and 76.5% of amorphous silicates; Jones et al. (2017) assume it is made of 31% of partially hydrogenated amorphous carbon and 68.9% of amorphous silicates (forsterite and enstatite in equal proportions) with hydrogenated amorphous carbon mantles and Fe/FeS inclusions; Hensley & Draine (2022) assume it is made of 5% PAHs and 95% of a mash-up of silicate, hydrocarbons, iron and various oxides. This uncertainty is the main limitation of dust studies and of dust-based diagnostics of the physical conditions. This is also the limitation in the precision of gas-physics simulations, such as PhotoDissociation Region (PDR) models Röllig et al. (2007).

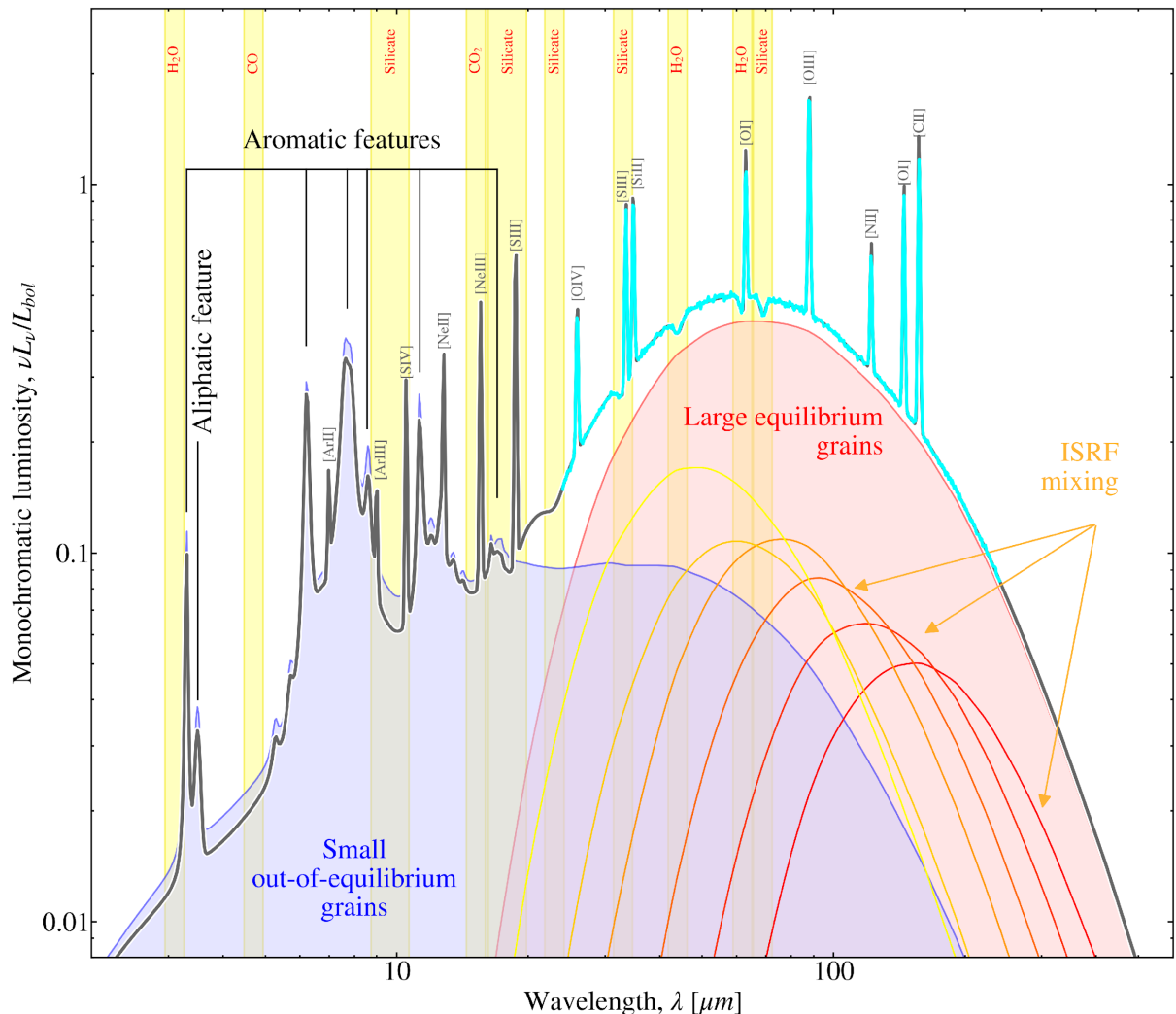

Figure 1: Typical SED of a star-forming region. The cyan line is a R=200 spectrum (100σ). We show a few absorption features from crystalline silicates and ices, and the brightest gas lines.

The most straightforward way to more precisely constrain the dust composition would be to look for solid-state features. This is the only way to unambiguously identify a particular chemical composition. The Mid-InfraRed (MIR) range is potentially the richest domain as it is where the vibrational modes of chemical bonds are located. For instance, we know that interstellar silicates are mostly amorphous (Spoon et al., 2022). However, precisely knowing this fraction and the way it varies with the physical conditions (radiation field, gas density and metallicity) would provide invaluable constraints on dust evolution. Another puzzle is that there is too much depleted oxygen in the ISM, compared to what we can put in silicates. It is thus possible that a fraction of the dust is in the form of various oxides (e.g., $Al_2O_3$, $CaCO_3$, etc.) or organic carbonates (Jones & Ysard, 2019). The bands of these compounds may have eluded previous spectroscopic surveys. Finally, recent X-ray investigations of the dust chemical composition in our Galaxy suggest that Mg-rich amorphous pyroxene represents the largest fraction of dust (about 70 % on average; Psaradaki et al. 2023). This fraction may change with environments and amorphous pyroxenes can be studied through their MIR-to-FIR features as well.

## Instruments Required:

FIRESS medium maps at low resolution

## Program Size:

- Large (>200 h)

## Approximate Integration Time:

- 980 hours

## Special Capabilities Needed:

- None

## Synergies with Other Facilities:

The spectroscopic information about X-ray absorption edges, provided by future X-ray satellites, such as ATHENA, will be complementary.

## Description of Observations:

Low spectral resolution (R), high signal-to-noise ratio (S/N) observations over the whole MIR-to-FIR range would give us access to a wealth of features (Fig. 1). However, such spectra were only obtained in the past by combining ISOSWS and ISOLWS (e.g., Peeters et al. 2002), with a poor sensitivity, a poor spatial resolution and stitching problems. This is a domain where PRIMA will particularly innovate. We could map star-forming regions of different metallicities, such as in the LMC and SMC. It would be important to obtain maps in order to understand how these features vary with the distance from the central cluster. This would highlight the role of photo-processing as well as the role of mantle growth in dust evolution. We could be looking for both emission and absorption features. However, the emission features come primarily from small grains or large hot grains. They might not represent the bulk of the dust mass. This is why absorption features are preferred. This is also why we would need to target star-forming regions, so that we have a strong MIR background.

We will use FIRESS to perform R=200 spectral maps of a sample of 8 nearby extragalactic star-forming regions (≈20'×20').

Table 1: Spectroscopic time estimates.

| Band [μm] | Fields | Signal [W/m²] | Integration Time [hours] |
|---|---|---|---|
| 32 | 8×20'×20' | $2.0 \times 10^{-17}$ | 976 |
| 57 | 8×20'×20' | $2.3 \times 10^{-16}$ | 2 |
| 101 | 8×20'×20' | $5.7 \times 10^{-16}$ | <0.5 |
| 179 | 8×20'×20' | $4.4 \times 10^{-16}$ | <0.5 |

8 regions could be mapped in 980 hours (Table 1).

- The flux sensitivity has been estimated using the dust model of Jones et al. (2017) with a radiation field intensity of U=35 and for N(HI)=$3\times10^{21}$ H/cm$^2$. This corresponds approximately to the extended parts of the star-forming regions of the LMC.

- We require a high S/N (≈100), because the goal is not so much to detect the intensity of a feature, but to unambiguously measure it. These features are usually weak and broad, we thus need to make sure that they are not diluted in the continuum.

- The sizes are those of typical star-forming regions in the LMC/SMC (N11, N66, etc.).

## 49.  Unveiling the elusive dust properties of the diffuse ISM of nearby galaxies

Frédéric Galliano (DAp, CEA Paris-Saclay, France), Maarten Baes (Ghent University, Belgium), Simone Bianchi (INAF, Italy), Caroline Bot (Observatoire astronomique de Strasbourg, France), Viviana Casasola (INAF, Italy), Jérémy Chastenet (Ghent University, Belgium), Christopher Clark (STScI, USA), Stavroula Katsioli (Athens Observatory, Greece), Ilse De Looze (Ghent University, Belgium), Suzanne Madden (DAp, CEA Paris-Saclay, France), Mikako Matsuura (Cardiff University, UK), Takashi Onaka (Tokyo University, Japan), Lara Pantoni (DAp, CEA Paris-Saclay, France), Monica Relaño Pastor (Granada University, Spain), Marc Sauvage (DAp, CEA Paris-Saclay, France), Matthew Smith (Cardiff University, UK), Tsutomu Takeuchi (Nagoya University, Japan), Manolis Xilouris (Athens Observatory, Greece), Nathalie Ysard (IAS, Orsay, France)

We plan to use the exceptional sensitivity of PRIMA to map the LMC and SMC with PRIMAger, down to the confusion limit. The goal is to obtain a reliable characterization of the SED of the diffuse ISM in these two systems. Combined with already existing estimates of the extinction curves and elemental depletions, this will allow us to build extragalactic dust models for these two galaxies, and thus understand the impact of metallicity on the dust properties.

### General Scientific Area:

Galaxy Ecosystems and Interstellar Medium

### Science Justification:

The constitution of interstellar dust (i.e., its composition, size distribution and abundance) in external galaxies is believed to differ significantly from the Milky Way (MW). Dust build-up and evolution indeed depend on the specific Star Formation History (SFH) of the galaxy. In particular, the metallicity, Z, which is the mass fraction of elements heavier than He, appears to be one of the most important factors (e.g., Galliano, 2021). This parameter, Z, quantifies the cumulated elemental enrichment of a system. To properly interpret observations of galaxies, we thus need to understand how the dust properties vary as a function of Z. Yet, contemporary dust models (e.g., Jones et al., 2017; Hensley & Draine, 2022), that are used to provide such an interpretation, are exclusively constrained by observations of the MW, a system with a narrow Z range around the Solar value, $Z_\odot$. We are therefore biased by the particular properties of the MW when modeling the dust Spectral Energy Distribution (SED) of other galaxies. This bias especially questions our ability to accurately understand nearby dwarf galaxies and early Universe galaxies when the metal enrichment is expected to be low.

Arguably, no properly constrained dust model of external galaxies currently exists. This is because there is a deficit of observational constraints. The mixing of physical conditions along the sightline and within the telescope beam indeed renders the SED degenerate. For instance, if we are observing a region where there is a gradient of InterStellar Radiation Field (ISRF), we will not be able to distinguish an overabundance of small grains from the spread due to the variation of the equilibrium temperatures of large grains (e.g., Fig. 3 of Galliano et al., 2018). This is why dust models are calibrated on observations of the diffuse InterStellar Medium (ISM) of the MW. The low optical depth of this medium (A(V)≈0.1 for N(H)=2×10$^{20}$ H/cm$^2$, at Z$_\odot$) ensures that the grains will be uniformly illuminated by the average ISRF. It is possible to build the observed SED of the diffuse ISM of the MW, by averaging the high Galactic latitude fluxes given by IRAS, COBE and Planck. This is however not yet possible in external galaxies. IRAS, COBE and Planck did not sufficiently resolve galaxies to allow the extraction of their diffuse ISM emission. And other observatories with a finer angular resolution, such as Spitzer and Herschel, were not sensitive enough (Herschel could barely go below N(H)=10$^{22}$ H/cm$^2$).

The good angular resolution and the exceptional sensitivity of PRIMA over the whole Mid-InfraRed-to-Far-InfraRed (MIR-to-FIR) window gives us a way to palliate this problem, without having to resort to feathering (e.g., Clark et al., 2023). The two closest galaxies, the Large and Small Magellanic Clouds (LMC & SMC; d$_{LMC}$=50 kpc and d$_{SMC}$=60 kpc; Z$_{LMC}$=1/2 Z$_\odot$ and Z$_{SMC}$=1/5 Z$_\odot$; Fig. 1) are ideal targets. We will be able to resolve regions of ≈10 pc size at λ=250 μm. With the addition of already estimated elemental depletions and extinction curves (e.g., Gordon et al., 2003; Tchernyshyov et al., 2015), the well characterized broad-band SED of their diffuse emission per H atom will allow us to build the first properly-constrained extragalactic dust models.

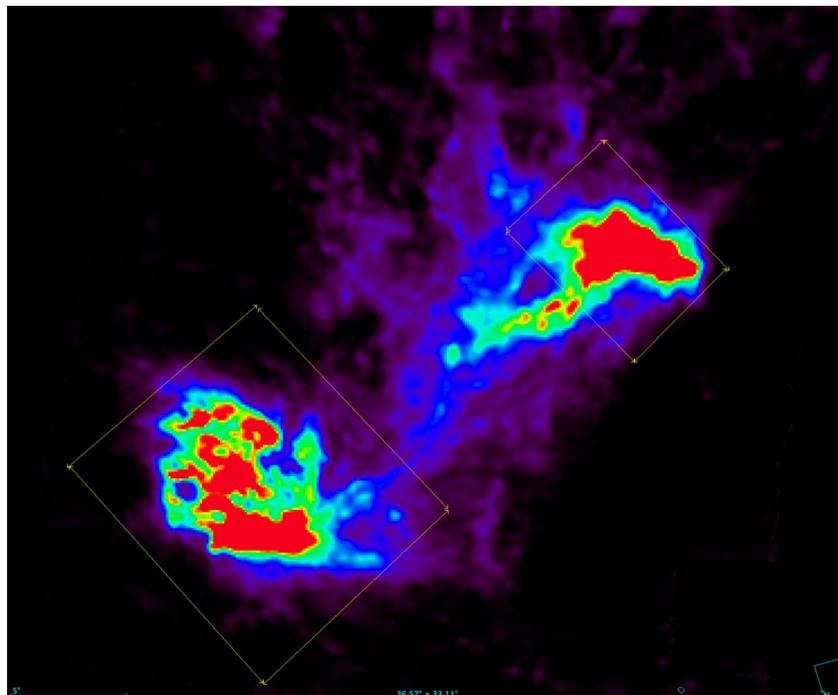

Figure 1: HI map of the LMC/SMC (Brüns et al., 2005). The contours correspond to N(H)=10$^{20}$ H/cm$^2$ and N(H)=2×10$^{20}$ H/cm$^2$. The yellow squares are areas covering most of the HI emission of both sources (LMC: 12°×12° SMC: 8°×6°).

**Instruments Required:**

- PRIMAger: Large Map (> 20'×20')
- Hyperspectral band
- Polarimeter band

**Program Size:**

- Large (100 + hr)

**Approximate Integration Time:**

- 500 hr

**Special Capabilities Needed:**

- None

**Description of Observations:**

To measure the MIR-to-FIR SED of the diffuse ISM of the LMC and SMC, we need to make deep maps of these two galaxies in all available bands. In addition, Guillet et al. (2018) showed how important FIR measures of the polarization fraction were to constrain the grain composition.

We will use PRIMAger, to measure both the total power and the polarization fraction (Table 1).

- We will map most of the 2 galaxies with PRIMAger at the 4 long-wavelength bands.

- We will make smaller maps of a diffuse region in each galaxy, in each of the short-wavelength bands, as confusion is less problematic in the MIR.

We will make large maps of the LMC (12°×12°) and SMC (8°×6°) through all the PRIMAger bands.

- The flux sensitivity has been estimated using the dust model of Jones et al. (2017) with a ISRF intensity of U=1 and for N(H)=2×10$^{20}$ H/cm$^2$. At first approximation, the emission and the column density both scale with Z. We should thus scale the emission of the model by Z/Z$_\odot$. However, we are interested in the emission of an optically thin medium (A(V)≈0.1). This A(V) will be reached at a Z$_\odot$/Z times higher column density than in the MW. Z therefore cancels out in this estimate.

- The Spitzer and Herschel maps of the Magellanic clouds are good (Meixner et al. 2006; Gordon et al., 2011; Meixner et al. 2013). However, they are not deep enough for this science case. Besides, PRIMA will provide a finer spectral sampling.

- The main challenge of this science case is the confusion with the diffuse dust emission of the MW and with the cosmic infrared background. Redundancy is the key to subtract these contaminations. This is why we need to map large areas.

Table 1: Sensitivity and observing time estimates. We used the PRIMAger Python time estimator
(Feb. 2023 version; Marc Sauvage, priv. comm.).

| Band [μm] | Field [deg²] | Sensitivity [MJy/sr] | Photometry [hours] | Polarimetry [hours] |
|---|---|---|---|---|
| **LMC** | | | | |
| 35 | 0.01 | 0.07 | 250 | ... |
| 63 | 4 | 0.24 | 190 | ... |
| 92 | 144 | 0.99 | 120 | 830 |
| 126 | 144 | 1.9 | 120 | 120 |
| 172 | 144 | 2.1 | 120 | 120 |
| 235 | 144 | 1.7 | 120 | 120 |
| **SMC** | | | | |
| 35 | 0.01 | 0.07 | 250 | ... |
| 63 | 4 | 0.24 | 190 | ... |
| 92 | 48 | 0.99 | 42 | 290 |
| 126 | 48 | 1.9 | 42 | 42 |
| 172 | 48 | 2.1 | 42 | 42 |
| 235 | 48 | 1.7 | 42 | 42 |

## 50.  PRIMA study of a new phase of the ISM: the Dense Warm Interstellar Medium (D-WIM)

Paul Goldsmith (Jet Propulsion Laboratory, California Institute of Technology), Jorge Pineda (Jet Propulsion Laboratory, California Institute of Technology), Dariusz Lis (Jet Propulsion Laboratory, California Institute of Technology), Youngmin Seo (Jet Propulsion Laboratory, California Institute of Technology)

We propose a study of the Dense Warm Interstellar Medium (D-WIM) via the two fine structure lines of ionized nitrogen ([N II]). This new phase of the interstellar medium (ISM) is fully ionized gas at density $10 \text{ cm}^{-3} \leqq n(e) \leqq 100 \text{ cm}^{-3}$, and has effectively been identified throughout much of the Milky Way and in nearby galaxies. The unexpected conditions in the D-WIM immediately raise important questions both about this phase of the ISM and its relationship to the ISM as a whole. These include the structure of the D-WIM - what is its filling factor, distribution, and morphology? A second is - what maintains its ionization; is this a new source or can it be explained by radiation from stars? A third question is the origin of the D-WIM - what is its relationship to stars - is it the remnants of very old HII regions that were produced by clusters of massive stars? And fourth, what is the evolution of this new phase of the ISM – does the overpressure result in expansion leading to dissipation? The PRIMA survey requires a coverage of a large fraction of the plane, plus selected regions at high galactic latitudes. A sensitivity of $10^{-9} \text{ W m}^{-2} \text{ sr}^{-1}$ would be highly desirable. A spectral resolution $R = 10^2$ would be acceptable to measure the total intensity. A conservative estimate based on the expected PRIMA sensitivity is that a survey covering 760 square degrees can be carried out in 400 hours. This PRIMA survey could be part of a broader survey of the ISM, and include a number of other important ISM tracers including [CII] 158 µm, [OI] 63 µm and [OI] 146 µm, depending on wavelength coverage of the instrument. The sensitivity requirement for study of the D-WIM is likely more stringent than for the other lines, and may require special consideration for successful implementation as well as for determining the area that can be observed.

### General Scientific Area:

Milky Way Interstellar Medium

### Science Justification:

Regions of ionized gas are a major component of the interstellar medium (ISM). These range from the relatively dense (with electron densities $n(e)$ as high as $1000 \text{ cm}^{-3}$) ionized gas in the immediate vicinity of massive young stars (HII regions) to the diffuse warm ionized medium (WIM) having a density $n(e) \sim 0.1 \text{ cm}^{-3}$ (Reynolds 1991, Hill et al. 2008, Ferrière 2001, and references therein). Determining the properties of these regions has been an ongoing task, with the question of the relationship of these different components of the ISM being an especially difficult challenge. While a great deal of information has been gained by the use of radio techniques, major questions remain, especially about the lower density ionized ISM. Densities 1–10 cm$^{-3}$, an order of magnitude higher than those of the WIM, have been suggested by observations of cm-wavelength recombination lines (Shaver 1976) and by comparison of low-frequency and high-frequency recombination lines (Anantharamaiah 1985, 1986). These regions may be the extended highly evolved HII regions, and are sometimes referred to as Extended Low Density or ELD HII regions (Mezger 1978). A quite different technique to study ionized regions is to employ the fine structure lines of ionized nitrogen; there are two [NII] lines at wavelengths of 205 μm and 122 μm (e.g., Bennett et al. 1994, Oberst et al. 2006). The 14.53 eV ionization potential of atomic nitrogen means that this element will be ionized only in regions where hydrogen is also ionized and **the ratio of the two line intensities is an excellent probe of electron densities for 5 ≦ n(e) ≦ 300 cm$^{-3}$** (Oberst et al. 2011; Goldsmith et al. 2015).

## The Dense Warm Interstellar Medium (D-WIM)

The [NII] line ratio was used by Goldsmith et al. (2015) to determine the electron density in an unbiased survey of the Galactic plane. For 116 positions in the range −60° ≦ $l$ ≦ +60°, the average electron density was 29 cm$^{-3}$, with almost all lines of sight falling in the range of 10 cm$^{-3}$ to 50 cm$^{-3}$ with a few regions having n(e) up to 100 cm$^{-3}$. The presence of widespread, dense, ionized gas was a very surprising result, as its density is an order of magnitude higher than the densities of the ELD HII regions and its widespread nature makes any intimate connection with stellar--driven ionization unlikely. And if the D-WIM does not ultimately have a stellar energy source, what is providing its ionization? Compared to the WIM, the D-WIM is overpressured by 2 orders of magnitude, so will expand. Is it being continually recreated as it diffuses and merges with the WIM? At this density, recombination is relatively rapid, and another major question is – what is the source of ionization?

Figure 1 shows the observed [NII] fine structure line ratio and derived electron density as a function of Galactic longitude. All of the above questions point to a new phase of the ISM that could be a very large-scale impact of massive star formation, but could also be indicative of a widespread additional ionization source. What is needed to pin down the properties of the D-WIM is to obtain 2D images of the [NII] lines and derived electron densities and column densities. This is what the proposed PRIMA survey will uniquely provide in two dimensions throughout the entire disk of the Milky Way.

The extensive, n(e) ~30 cm$^{-3}$ ionized gas was called the Dense Warm Ionized Medium (or D-WIM) by Geyer & Walker (2019). These authors modeled the D-WIM as being ionized only by collisions with electrons, and derived an unusually high temperature of ~19,000 K for this component of the ISM. Observations were carried out by Langer et al. (2021), and their modeling, including collisional ionization of nitrogen by proton charge exchange with protons from ionized hydrogen, suggested a much more reasonable temperature range of 3,400 K to 8,500 K, consistent with the line widths measured from velocity-resolved spectroscopy using the SOFIA/GREAT instrument. Observations of the same pair of [NII] fine structure lines in 21 galaxies from the KINGFISH survey yielded a range of densities from 1–300 cm$^{-3}$ (Herrera-Camus et al. 2016). This is similar to the range found in the Milky Way survey mentioned above, and even more striking, the extragalactic mean value, <n(e)> = 30 cm$^{-3}$, is essentially identical to that found in the Milky Way. While there are certainly variations and gradients, this confirms on a much broader scale the existence of the D-WIM phase of the ISM. But the origin of the D-WIM remains a major puzzle. What keeps it ionized, and what is its volume filling factor? As mentioned above, obtaining additional information on the morphology of the D-WIM is thus of critical importance for understanding the ISM of a wide range of galaxies.

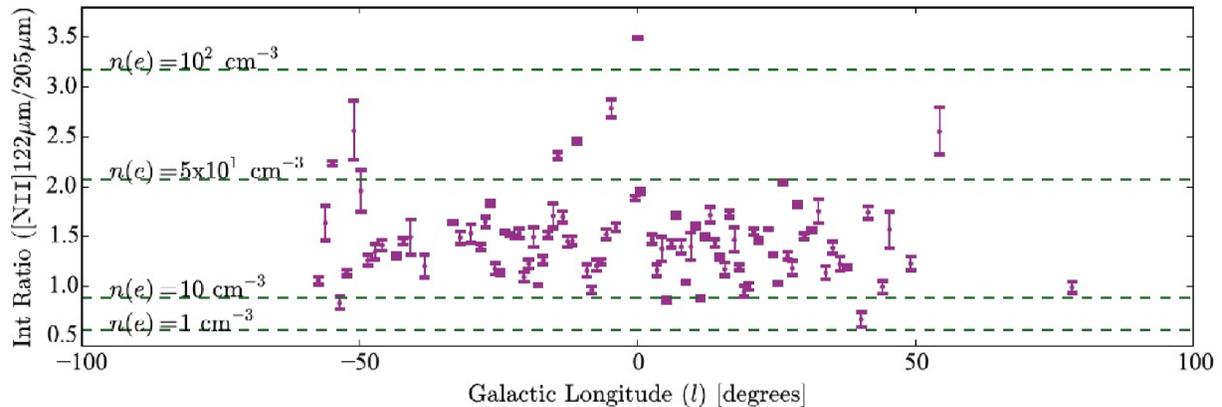

Figure 1: [NII] fine structure line ratios and derived electron densities from Herschel survey (Goldsmith et al. 2015).

Pineda et al. (2019) developed a hybrid technique in which the 205 [NII] line was combined with radio recombination line (RRL) data. As shown by these authors, the ratio of the integrated antenna temperature of the [NII] 205 μm line to that of a RRL is dependent on the electron density (Pineda et al. 2019, equation 9). They employed this technique to obtain electron densities along 11 lines of sight using the 205 μm line data from SOFIA and RRL data from the NASA Deep Space Network and the GBT. They found electron densities between 8 and 170 cm$^{-3}$, again generally consistent with the D-WIM densities determined by Goldsmith et al. (2015). A critical point is that these LOS were chosen **to avoid known HII regions.**

This project will definitively confirm the existence of a new phase of the interstellar medium. This may be connected with very late stages of evolution of HII regions produced by massive stars, but may well point to new processes that are required to sustain this surprisingly dense, warm, ionized gas (D-WIM) filling a significant fraction of the Galactic disk. A very significant additional result will be complete imaging of the [CII] 158 μm line and the ability to determine the fraction of its emission arising in ionized regions. An extension of this project to cover a selection of nearby galaxies would be extremely valuable, but the data required may well be obtained as part of other imaging projects.

### Instruments Required:

FIRESS Large map at R~130

### Program Size:

- Large (100 + hr)

## Approximate Integration Time:

- 400 hr

- If the two [NII] lines cannot be observed simultaneously, the integration time would be doubled. This would bring in some interesting shorter-wavelength lines including [OIII] 52 µm and lines of S and Si in different ionization states.

## Special Capabilities Needed:

- None

## Synergies with Other Facilities:

- Balloon Missions (GUSTO, ASTHROS & reflights thereof)

## Description of Observations:

PRIMA with a 2-m diameter telescope will have FWHM beam widths ~25" at [NII] 205 µm and ~15" at [NII] 122 µm. These are comparable to or smaller than Herschel PACS footprint, so there should not be a problem with confusion between the more diffuse ionized gas and HII regions. What is needed is well-sampled, extensive maps of a significant fraction of the Galactic plane. Exploration of off-plane emission is also important as the very large (~7°) beams employed by Wright et al. (1991) and Bennett et al. (1994), while suggestive of extended emission, did not have adequate angular resolution or sampling to obtain definitive results. The intensity distribution of the [NII] lines observed is shown in Figure 2 (from Goldsmith et al. 2015). The mean intensity is <log I> = $10^{-7.7}$ Wm$^{-2}$sr$^{-1}$. Given the desire to get more complete sampling of the outer galaxy as well as the high-latitude regions, a 1σ sensitivity of $10^{-9}$ Wm$^{-2}$ sr$^{-1}$ would be highly desirable. With a 2 m diameter telescope, PRIMA's diffraction-limited solid angle is ~5.4×$10^{-9}$ sr, yielding a desired 5σ flux density of 2.7×$10^{-17}$ Wm$^{-2}$.

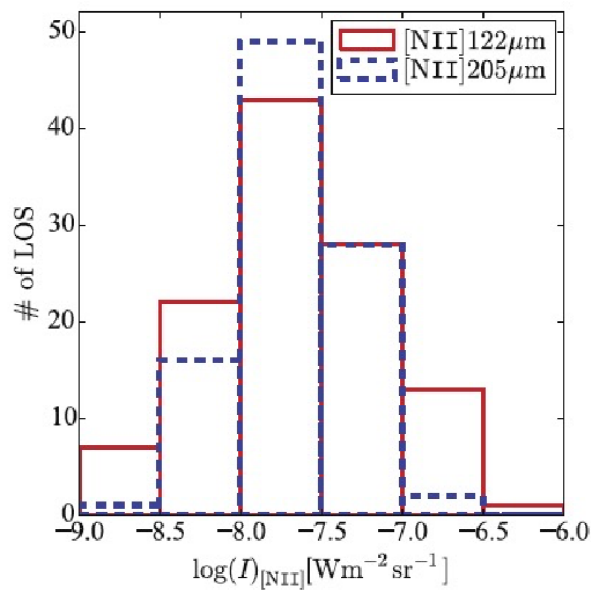

Figure 2: From Goldsmith et al., 2015: Distribution of intensities obtained from the line-of-sight positions detected on Herschel/PACS observations of the Galactic plane, shown by histograms of [N II] 122 µm (blue) and 205 µm (red).

For low resolution spectroscopy (R~100) PRIMA FIRESS will map a 900-arcminute square area to 5σ sensitivity of 3×10$^{-19}$ Wm$^{-2}$ in 1960 hours for [NII] 205 μm and in 2118 hr for [NII] 122 μm. We require 90 times lower sensitivity, so that the mapping speed should be 8100 times greater, requiring 0.97 hr and 1.05 hr per square degree for 205 μm and 122 μm, respectively. The coverage of this study should as a minimum include (1) Galactic plane −90° ≦ *l* ≦ 90°; −1° ≦ *b* ≦ +1°: 360 square degrees total, and (2) Strips in the outer Galaxy and strips including high latitudes; the extent is TBD, but we want to sample at least a few degree-wide strip including the Central Molecular Zone, and another such strip at moderate longitude: 40 square degrees total. The total area to be mapped is thus 400 square degrees. This would thus take 400 hours plus overhead and calibration time. Less areal coverage would still be acceptable and provide valuable information.

If it is possible and appropriate, the integration time could be adjusted to permit deeper integrations at larger longitudes and higher latitudes. The intensity of the [NII] lines is approximately 1/10 of that of the [CII] 158 μm line, as discussed by Goldsmith et al. (2015), making this a challenging project and one that is unlikely to be carried out by any other facility, e.g., a balloon or other suborbital mission. There is limited information on the spectral characteristics of the [NII] emission from the spectra observed in the Galactic plane by Goldsmith et al. (2015), Pineda et al. (2019), and Langer et al. (2021). In most cases, the spectral lines are single--peaked features, with line width 20 km s$^{-1}$ to 50 km s$^{-1}$. In some cases, there are multiple peaks, but generally, the [NII] emission is dominated by a single feature. Thus, low resolution PRIMA FIRESS mapping will be appropriate for this project.

## Acknowledgement:

A portion of this research was carried out at the Jet Propulsion Laboratory, California Institute of Technology, under a contract with the National Aeronautics and Space Administration (80NM0018D0004).

## 51. Starbursts under the microscope: how the feedback from a super star cluster impacts its surroundings and its host galaxy

Paul Goldsmith (Jet Propulsion Laboratory, California Institute of Technology), Cara Battersby (U Conn), Lars Bonne (SOFIA), Mélanie Chevance (Univ. Heidelberg), James Di Francesco (NRC, Canada), Perry Hatchfield (JPL), Volker Ossenkopf-Okada (Univ. Koeln)

Massive stars have an impact on the structure and evolution of galaxies far greater than would be expected purely based on their numbers. Their huge luminosities and mass losses result in major impacts on surrounding interstellar material during their lifetimes. When they explode as supernovae, they dramatically enrich the interstellar medium with heavy elements. The impact of massive stars is extremely nonuniform, however, because they are almost exclusively formed in clusters, of which the most extreme are Super Star Clusters (SSCs) that can contain 100s of O, B and WR stars. These SSCs highlight the major question of how massive stars are formed. While there are a number of competing theories, it is clear that the process is highly dependent on their environments, with the presence of massive stars promoting the formation of additional stars. Thus, studying SSCs probes the effects of massive stars on the interstellar medium, including how this feedback itself contributes to controlling the rate of star formation, and also how massive stars are formed. Competitive accretion plays a crucial role in their formation but the geometry of this process is basically unknown because it is deeply enshrouded in the parental concentrations of dust and gas. The extreme star formation characteristic of SSCs may well have been a critical mode of star formation in starburst galaxies and in the early evolution of today's galaxies. This General Observer project will observe key fine structure lines throughout the entire wavelength range accessible with PRIMA to characterize and spatially resolve the impact of stars in SSCs in the Milky Way and nearby galaxies on surrounding material. By imaging spectral lines emitted by neutral and ionized gas, and multiple lines of carbon monoxide molecules, we will derive a much-improved picture of the structure of all ISM phases surrounding SSCs. We will also develop templates for characterizing and analyzing SSC FIR emission that can be used in a broad range of extragalactic observations.

### General Scientific Area:

Galaxy Ecosystems and Interstellar Medium

### Science Justification:

Understanding the structure and evolution of galaxies requires understanding massive star formation, as a few massive stars dominate the feedback to the structure of the surrounding interstellar medium (ISM), the overall rate of star formation, and on metal enrichment of the interstellar medium. Some of these processes are intertwined, as additional heavy elements result in increased cooling of the ISM, which can accelerate the process of cloud contraction and star formation while reducing the fraction of massive stars formed. The transfer of radiative and mechanical energy from a massive young star to its surroundings can disperse clouds, reducing the rate of star formation, or can compress and radiatively implode clouds, increasing the rate of star formation.

Diagnosing the energy transfer from a massive young star (or cluster thereof) to the surrounding ISM can be achieved by studying the emission from the heated and excited ions, atoms, and molecules therein. At temperatures characteristic of material so impacted, the emission will be primarily at far-infrared and millimeter wavelengths. The stars themselves are surrounded by a region of ionized hydrogen, which itself lies onion-like, within a larger, hot photon dominated region (PDR) in which most material is atomic but where carbon and many polycyclic aromatic hydrocarbons (PAHs) are still observed in their ionized form. As the temperature and the hardness and magnitude of the radiation field decrease with increasing distance from the star, the composition of the surroundings gradually changes, with hydrogen becoming molecular, followed by carbon becoming first neutral and then largely bound up as CO. In fact, the onion structure contains many more layers. When looking at species with different ionization energies we find those with the highest ionization threshold only closest to the massive stars while a network of chemical reactions allows for more and more complex molecules with increasing distance.

This layering description, however, is already a vast oversimplification. In a radially symmetric configuration, massive stars could not even form as their radiation would shut off any further growth. It is clear that mass accretion occurs instead through disk structures but their nature is basically unknown because the process is deeply embedded in the parental clouds so that only far-infrared emission can reach an external observer. We can use the diagnostics of those lines that arise from gas that requires different ionization energies to perform a "tomography of the geometry of massive star formation," by measuring what fraction of gas is illuminated by different levels of hardness of UV radiation and thereby resolving the geometry and kinematics of the star-forming environment around the massive stars. The work by Chevance et al. (2016), employing multiple far-infrared lines, indeed found that the HII region around R136 has a porous and inhomogeneous structure instead of the spherical onion-like layers.

A full accounting of the energy deposited requires observations of ionized, atomic, and molecular phases. This can be accomplished by far-infrared observations of young star forming regions employing the entire PRIMA wavelength range. Some of the most important fine structure and representative molecule lines are included in Table 1. The rotational lines of CO add significantly to the regions probed by fine structure lines, and are vital for determining overall energetics of the surroundings of SSCs, as described by Lee et al. (2016). In addition, several of these fine structure lines are important diagnostics of the star formation activity in galaxies at different redshifts in the universe. Large maps of their emission around the prominent star formation regions in the Milky Way and nearby galaxies will provide important templates to understand the origin of this unresolved emission in more distant galaxies.

Table 1: Parameters of Fine Structure and Representative Molecular Lines to be Observed.

| Species | U | L | λ(µm) | ΔE(K) | $A_{ul}(s^{-1})$ | $n_{cr}(cm^{-3})$ | | IP(eV) |
|---|---|---|---|---|---|---|---|---|
| S⁺⁺ | $^3P_1$ | $^3P_0$ | 33.5 | 430 | 4.7e-4 | 1400 | (e-) | 35.12 |
| Si⁺ | $^2P_{3/2}$ | $^2P_{1/2}$ | 34.8 | 414 | 2.2e-4 | 3.4e5 | (H) | 8.15 |
| O⁺⁺ | $^3P_2$ | $^3P_1$ | 51.8 | 280 | 9.8e-5 | 3400 | (e-) | 35.12 |
| S | $^3P_0$ | $^3P_1$ | 56.6 | 255 | 3.0e-4 | 7.1e5 | (H) | 0.0 |
| O | $^3P_1$ | $^3P_2$ | 63.2 | 228 | 9.0e-5 | 9.7e5 | (H) | 0.0 |
| Si | $^3P_2$ | $^3P_1$ | 68.5 | 210 | 4.2e-5 | | | 0.0 |
| O⁺⁺ | $^3P_1$ | $^3P_0$ | 88.4 | 160 | 2.7e-5 | 500 | (e-) | 35.12 |
| CO | 28 | 27 | 96.8 | 149 | 1.8e-3 | 1.2e7 | (H2) | - |
| N⁺ | $^3P_2$ | $^3P_1$ | 121.9 | 120 | 7.5e-6 | 280 | (e-) | 14.53 |
| Si | $^3P_1$ | $^3P_0$ | 129.7 | 111 | 8.2e-6 | | | 0.0 |
| O | $^3P_0$ | $^3P_1$ | 145.6 | 98 | 1.7e-5 | 1.5e5 | (H) | 0.0 |
| C⁺ | $^2P_{3/2}$ | $^2P_{1/2}$ | 157.7 | 91 | 2.4e-6 | 2700 | (H) | 11.26 |
| N⁺ | $^3P_1$ | $^3P_0$ | 205.2 | 70 | 2.1e-6 | 45 | (e-) | 14.53 |
| CO | 9 | 8 | 325.5 | 44 | 5.1e-5 | 6.1e5 | (H2) | - |

For each ion, atom or molecule the columns are the upper and lower level (rotational quantum numbers for molecules) of the transition, the wavelength of the transition, the energy difference, the spontaneous decay rate, the critical density for relevant collision partner, and the ionization potential to produce ion in question.

Table 2 gives some parameters of the most prominent SSCs in the Milky Way and in neighboring galaxies.

Table 2: Super star clusters and regions of massive star formation.

| Name | l (deg) | b (deg) | D (kpc) | Age (Myr) | Log (Msun) | Radius (pc) | # O & WR stars | Total stellar mass (M_sun) | Refs |
|---|---|---|---|---|---|---|---|---|---|
| RCW38 | 268.03 | -0.98 | 1.7 | <1.0 | | 0.8 | | | |
| Westerlund 2 | 284.25 | -0.40 | 5.4 | 2.0 | 4.0 | 0.8 | >14 | 1-3×10⁴ | Furukawa+ 2009 |
| Trumpler 14 | 286.41 | -0.58 | 2.6 | 2.0 | 4.0 | 0.5 | | | |
| NGC 3603 | 291.62 | -0.52 | 6-7 | 2.0 | 4.1 | 0.7 | >50 | 1-1.8×10⁴ | Brandl+ 1999, Fukui+ 2014 |
| Westerlund 1 | 339.55 | -0.40 | 5.2 | 3.5 | 4.5 | 1.0 | ~140 | 1×10⁵ | Clark+ 2005 |
| [DBS2003] vb179 | 347.58 | 0.19 | 7.9 | 3.5 | 3.8 | 1.2 | | | |
| G305 | 305.506 | 0.085 | 3.8 | 1.5-3.0 | | 8.3 | 29 | 1.1×10⁴ | Borissova+ 2019 |
| R136 / 30Dor(LMC) | 05h38m42s.4 -69o06'03.4" (J2000) | | 48.5 | 1.5 | | 1.0 | 70 | 5×10⁴ | Chevance+ 2016 Crowther+ 2016, Schneider+ 2018 |
| N159 (LMC) | 05h39m50s -69o44'45" (J2000) | | 48.5 | | | | 24 (YSO) | | Nayak+ 2018, Lee+ 2016, Okada+ 2016 |
| H72.97-69.39 (LMC) | 04h51m53s -69o23'26" (J2000) | | | | | | | | Nayak+ 2019 |
| NGC 604 (M33) | 01h34m33s.2 +30o47'06" (J2000) | | 840 | 3.5 | | 1.5 | 200 | 1×10⁵ | Maíz-Apellániz+ 2004 |

## The Need for PRIMA

To accomplish the goal of understanding the structure of SSCs and their energy return to the interstellar medium requires observation of a large number of fine structure and molecular lines with good angular resolution. PRIMA will cover all lines of interest in only two setups. SOFIA did not have this option. As a result, the studies that have been carried out generally focus on a very limited number of lines (e.g., Simpson et al. 2012). The sensitivity of PRIMA is orders of magnitude better than Herschel or SOFIA and will thus strongly reduce the required time per region while being able to detect the fainter extended emission further away from the star clusters. This will result in a (large) sample of regions completely mapped with uniform sensitivity. Due to sensitivity limitations, SOFIA observations were restricted to relatively small regions when observing multiple lines of interest (see Klein et al. 2023). Herschel/PACS observed fine structure lines one-by-one with impressive results (e.g., Chevance et al. 2016), but a great deal of time would be required to map the ($\sim$15') regions that have to be covered to trace an individual SSC. Because of the clumpiness of the medium, we see emission of high ionization lines such as [OIII] tens of pc away from a SSC (e.g., Kawada et al. 2011 for 30Dor). Therefore, we need to map a large region ($\sim$15') to sample the 'sphere of influence' of an SSC. Balloon missions do not have sufficient angular resolution (GUSTO) and are not able to reach the shorter wavelength portion of the far-infrared wavelength range.

The spectral line integrated intensity images for each region will be modeled using a combination of CLOUDY (for ionized gas) and Meudon PDR (for PDR/neutral gas) codes. While still limited by geometry, the exceptionally broad wavelength coverage of PRIMA gives us a large number of tracers that span a wide range of physical conditions. These together will allow convergence on best-fit models for the total radiative input of each SSC. The large number of lines available will be key to avoid the degeneracy that exists when only 2 or 3 lines are available (Chevance et al. 2016; Okada et al. 2019). We will obtain values for the total luminosity including gas and dust (which is modeled in the two codes along with the gas). We expect to obtain complementary lower-J CO emission from ALMA that will give information on the kinematics of the gas. The critical output of this project will be calculation of the total energy input to the material surrounding Super Star Clusters at different stages of evolution, determination of the total mass reservoir that was available to form a SSC, and how the extreme star formation affects both subsequent formation of stars and the evolution of material in the region.

### Instruments Required:

FIRESS low-res small and medium maps

### Program Size:

● Medium (10–100 hr)

### Approximate Integration Time:

The "nominal" FIRSS sensitivity used to derive the 1960 hr required to map 900 sq arcmin region (as required for SSC in the Milky Way) is 3e-19 W/m², or 100 times lower flux. Achieving the R136 level (described below) for a 30'≥30' region would nominally then require .196 hr, or 706s. This has to be doubled to cover the entire FIRSS wavelength range. It does ignore the additional noise produced by the photon flux of the background, which should be calculated. It indicates that ~1hr observing time per typical SSC will be required with a few of the largest Milky Way SSCs requiring approximately a factor 5 more time.

## Special Capabilities Needed:

- None

## Synergies with Other Facilities:

We will be using images of CO and high- density tracers such as CS and HCN from ALMA and the LMT to determine the extent and kinematics of the molecular gas in the surroundings of SSCs. We will use ammonia observations from the GBT to determine temperatures of the dense gas. We will be using JWST to explore some infrared lines from ionized species and search for small scale structures such as voids such as those found in NGC628 by Barnes et al. (2023).

## Description of Observations:

We have eliminated SSCs in the central portion of the Galactic Plane (including the Arches and the Quintuplet) as the R~130 resolution of PRIMA FIRSS will not allow distinguishing between gas in these regions and elsewhere along these complex lines of sight. The SSCs selected span a range of distances, from ~2 kpc to 8 kpc in the Milky Way, 48 kpc in the LMC, and 840 kpc in M33. The SSC sizes given in Table 2 refer to the stellar cluster. Typically, this is surrounded at some distance by a PDR region and molecular cloud, which are clearly being affected strongly by the radiative and mechanical input from the SSC. We wish to trace the full spatial extent of this feedback. The observations of R136 in the LMC (Chevance et al. 2016) indicate that 5' is a reasonable angular sized region to include this at a distance of 48 kpc, implying a physical extent of 70 pc. This corresponds to angular sizes of $140' - 30'$ for sources within the Milky Way. The PRIMA slit lengths range from 6' to 17' (increasing with longer wavelengths). Development of a detailed observing strategy will include optimizing the scan directions and offsets. We can imagine a first pass across the region on the order of 10' in length followed by a second slightly offset pass to fill in gaps in the detector arrays for the extragalactic regions. This could be accomplished entirely with the steering mirror. For the Galactic sources, we may want to map regions as large as 4 square degrees; this would require a combination of steering mirror and spacecraft pointing.

We are fortunate to have reference for many of the fine structure lines we wish to observe from the data presented by Chevance et al. (2016) for the R136 Super Star Cluster in the LMC. The weakest line is the [NII] 205 µm line which has intensity ~1e-8 W/m$^2$/sr. With a nominal beam solid angle of 3e-9 sr, this corresponds to a flux of 3e-17 W/m$^2$. This sets the nominal required sensitivity level used above to calculate time requirement.

## Acknowledgement:

A portion of this research was carried out at the Jet Propulsion Laboratory, California Institute of Technology, under a contract with the National Aeronautics and Space Administration (80NM0018D0004).

## 52. HIGGLES (Hot Ionized Gas in the Galaxy's Local Extremes Survey): Exploring the Galactic Center with PRIMA's FIRESS instrument

H Perry Hatchfield (Jet Propulsion Laboratory, California Institute of Technology), Paul F. Goldsmith (Jet Propulsion Laboratory, California Institute of Technology), Cara Battersby (University of Connecticut), Mark Morris (UCLA), Jennifer Wallace (University of Connecticut), Dani Lipman (University of Connecticut), Brian Svoboda (NRAO)

The Milky Way's Galactic Center provides a unique observational opportunity to enhance our understanding of the impact of high-mass stellar feedback within a region of extreme physical parameters. The Galaxy's Central Molecular Zone (CMZ) can be used as a local analog to high-redshift environments, with its extremely dense molecular cloud complexes, globally high interstellar radiation field, turbulent cloud kinematics, and widespread star cluster formation. We propose the complete, contiguous mapping of the Galactic Center with PRIMA's FIRESS instrument, opening up a previously sparse window into the distribution, chemical makeup, and kinematic properties of the CMZ's hot ionized gas phase. Observations of the diversity of fine structure lines of various species within FIRESS's bandwidth towards the CMZ's many HII and photodissociation regions will enable us to probe the 3D structure of feedback regions and understand their role in the evolution of the extreme interstellar medium. A strategically placed supplemental grid of high-resolution-mode pointings will provide crucial kinematic information to resolve line-of-sight degeneracy in key locations. Previous instruments have lacked the spectral bandwidth to simultaneously cover a sufficiently wide variety of lines to study the ionization shells in depth, while also lacking the sensitivity to enable a highly complete mapping strategy, including less bright feedback remnants and cloud-cloud collision sites. The diffuse, hot phase of gas in the Galactic Center remains relatively underexplored outside of a handful of well-studied clouds. Understanding this elusive phase will provide a previously unrealized window into the role of feedback and cloud-cloud collisions in feeding the supermassive black hole, Sgr A*, driving Galactic outflows, and modulating the evolution of our Galaxy's most physically extreme environment.

### General Scientific Area:

Galaxy Ecosystems and Interstellar Medium

### Science Justification:

High-mass stars are known to be crucially important in governing the evolution of the interstellar medium (ISM) through their stellar winds, ionizing radiation, and energetic finales as supernovae. These mechanisms factor heavily into models for interpreting distant stellar populations and constraining the evolution of galaxies where the ISM is largely unresolved as well as the development of simulations where the relevant physical models are treated as sub-grid prescriptions. Such models are often tuned using observations of high-mass star-forming regions in local Galactic sources, though this greatly limits their extrapolation to regions of the universe where the interstellar environment differs greatly from the relatively calm disk of the Milky Way. The dynamically complex and physically extreme conditions within the Milky Way's Galactic Center provide a unique and comparatively nearby (8.2 kpc, The GRAVITY Collaboration 2019) window into a parameter space that is comparable to that characteristic of high-redshift galaxies (e.g., Kruijssen and Longmore 2013).

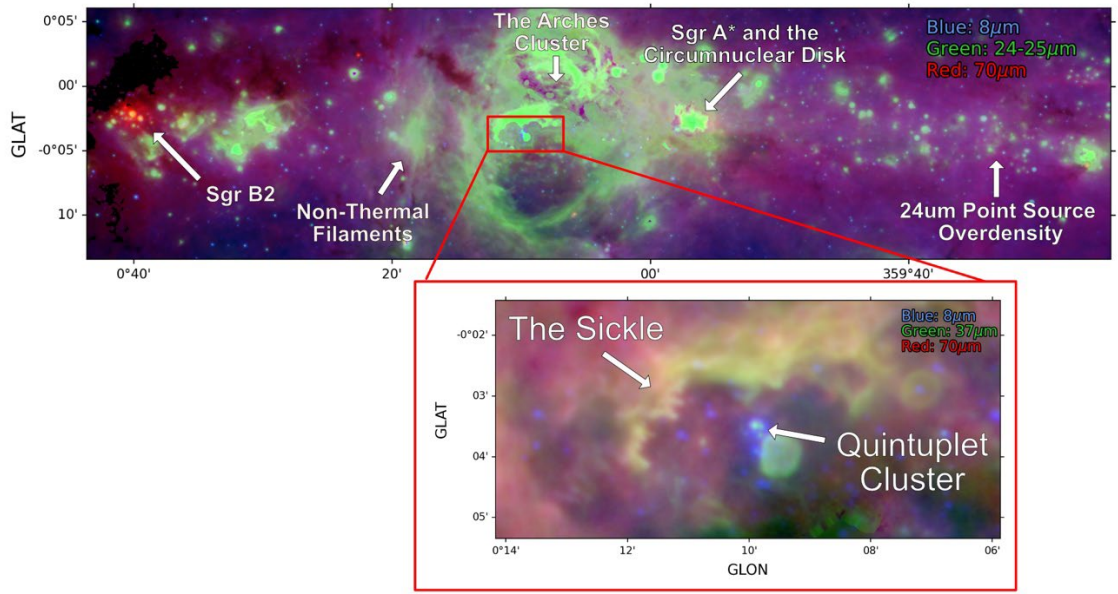

Figure 1: A multi-wavelength infrared mosaic of the innermost ~1 degree of the Galaxy, with a zoom-in showing the far-infrared landscape surrounding the Quintuplet cluster and Sickle Region.

The innermost ~3-degree region of the Galaxy is known as the Central Molecular Zone (CMZ), as it contains a large sample of the densest molecular clouds in the Milky Way, which orbit within the Galactic Center's potential. The CMZ hosts a population of high-mass star-forming regions as well as copious signposts of ongoing, past, and possible future bursts of star formation (e.g., Nogueras-Lara+2020, Schödel+2023). Numerous radio-bright filaments of mysterious origin and diverse morphology, a plethora of supernova remnants, turbulent gas flows, and widely distributed clusters of [HII] and photodissociation regions make this one of the most complex and rich regions for study within the Milky Way. The infrared and submillimeter continuum landscape of the Galactic Center is relatively well explored on ~30'' scales by Herschel, Spitzer, and SOFIA (e.g., Benjamin+2003, Molinari+2010, 2011, Hankins+2017, 2020), laying a solid foundation for understanding the CMZ's cold dust and molecular gas structures. The warm, ionized phase of gas in the CMZ remains relatively underexplored on similar spatial scales. Previous work has characterized populations of compact HII regions within sites of recent star formation (e.g., García+2016, Harris+2021), though relatively little has been done to deeply survey the hot (though not so hot as to be categorized as a hot X-Ray plasma) ionized phase on more global scales. There appear to be numerous diffuse HII regions throughout the Galactic Center, suggesting that other such signposts of long-past high-mass star formation might be discovered with more complete and higher sensitivity observations of fine-structure lines from singly ionized carbon ($C^+$) and nitrogen ($N^+$), and neutral ($O^0$) and doubly ionized oxygen ($O^{++}$), denoted ([CII], [NII], [OI], and [OIII], respectively). Some past studies have explored the distribution of ionized species such as $C^+$ towards notable sources like the Sgr B complex, Sgr A* and the circumnuclear disk, the Arches and Sickle regions, and the dense molecular cloud known as the Brick (e.g., Garcia+2016, Langer+2017, Harris+2021). The [CII] emission in this region appears extended and reveals an interconnected structure of star forming regions, photodissociation regions (PDRs) and other radio-continuum bright sources that have been well studied over the past several decades (e.g., Mehringer+1992).

Table 1: A summary of several regions of interest in the Galactic Center, including [CII] flux estimates, where available. The proposed survey will include all of these regions including the diffuse medium surrounding and connecting them.

| Source Name | Coordinates [deg] | Notable Features | [CII] intensity | Reference |
|---|---|---|---|---|
| 20 km s$^{-1}$ cloud, 50 km s$^{-1}$ | G359.889–0.093, G0.001–0.058 | Warm Gas Streamers, deeply embedded protostellar population | | García+2016 |
| Sgr B2 | G0.699–0.028 | High-mass protocluster | ~20–200 [K km s$^{-1}$] | Harris+2021 |
| Sgr B1 | G0.530-0.055 | HII regions, PDR regions, cloud-cloud collision sites, extended C+ observations | ~100–600 [K km s$^{-1}$] | Harris+2021, Langer+2017 |
| Sgr A*, CND | G359.948–0.052 | Supermassive black hole, streamers, nuclear stellar cluster | | García+2021, Langer+2017 |
| Quintuplet Cluster, Sickle Region, Arches Region | G0.162-0.057, G0.187-0.055 | Super star cluster site, stellar wind interactions | ~300 [K km s$^{-1}$] | Hankins+2020, García+2016, Rodrígez-Fernández+2004, Langer+2017 |
| Sgr E | G358.720+0.011 | 24um source overdensity, isolated HII regions, proximity to far dust lane | ~40–60 [K km s$^{-1}$] | Anderson+2020, Langer+2015 |

The spectral lines accessible to PRIMA's FIRESS instrument will provide a previously unrealized window into an observationally elusive phase of our Galaxy's most extreme environment. In particular, access to the numerous spectral lines of ionized species will permit a sort of ultraviolet tomography- a detailed study of how the boundaries and structures in various line ratios relate to the radiation field from nearby high-mass stellar feedback and radiation. This project aims to address several of the many questions about the properties and life cycle of the Galactic Center and the structures therein:

- **Are the winds and supernovae from high-mass stars responsible for driving the large-scale outflows seen in x-ray observations, or are these outflows a signature of past activity of the supermassive black hole Sgr A*?** By studying the broader distribution of warm, ionized gas components, we can understand how past sites of clustered high-mass star formation and the Milky Way's core are linked to signatures of outflowing gas.

- **What is the structure of the Galactic Center, and how can we learn about sites of past star formation from today's distribution of stellar nurseries and feedback regions?** [CII], [NII] and other fine structure lines are crucial for identifying connections between the CMZ's component clouds, many of which have been well studied individually. A fully-sampled map of the ionized structures that are widespread throughout the CMZ opens the door to discovering new feedback remnants in the warmer phase and is crucial for placing existing observations of Galactic Center sources in their environmental context.

- **How do cloud-cloud collisions impact the evolution of the multiphase ISM?** Recent observations and contemporary models of the Galactic Center suggest that clouds in this region are frequently buffeted by violent collisions, making this an excellent laboratory for studying the influence of these events on cloud properties. The fine-structure line emission from cloud-cloud collisions in the CMZ seems to be near or below the sensitivities of observatories like SOFIA and Herschel (Langer+2017, Harris+2021), and will tell us about the energetics of such collisions and their impact on the evolution of the Galactic Center's ISM.

- **Is the feedback from nearby high-mass stars responsible for the intermittent or episodic feeding of the circumnuclear disk and possibly Sgr A\*?** While past observations of the circumnuclear region revealed complex emission morphologies and characterized their relationships to Sgr A\* and nearby molecular clouds, more complete and higher sensitivity observations of a wider range of transition lines will open the door to new discoveries regarding the interaction of high-mass star formation and central supermassive black holes.

Each of these questions intersects with multiple subfields' interests in understanding of the Galactic Center's evolution. Answering these questions will demand a more complete observational understanding of the interconnected distribution and kinematics of feedback-driven bubbles, large-scale gas flows, and multi-phase dynamics of this unique Galactic ecosystem- an understanding that PRIMA will ultimately be capable of providing.

## Instruments Required:

- PRIMA FIRESS Spectrometer: Pointed High-res R~4400@ 112 μm.; Large Map Low-res R~130

## Program Size:

- Medium (10–100 hr)

## Approximate Integration Time:

Low-Resolution Mode: ~12.4 hr to reach $5\sigma \sim 3.0 \times 10^{-17}$ W m$^{-2}$ over ~2 square degrees, doubled to account for the entire FIRESS wavelength range, though this will likely be greatly increased by overhead for contiguous mapping of such a large region.

## Special Capabilities Needed:

- None

## Synergies with Other Facilities:

The region we intend to map with PRIMA coincides with numerous large programs with virtually every modern observatory, though there are several past and upcoming observational efforts that will synergize very well with the observations proposed here. In particular:

- The ALMA CMZ Exploration Survey (ACES) is an ongoing ALMA large program which surveys the entire CMZ in a contiguous mosaic, resolving gas and dust with ALMA's Band 3, with a variety of dense gas tracers and shock tracers. The global understanding of dense-phase gas kinematics and dust continuum properties for the CMZ's entire cold phase will be crucial in interpreting the ionized gas landscape.

- SOFIA's FORCAST instrument recently completed a large program toward the Galactic Center, the SOFIA Forcast Galactic Center survey, which observed many of the key regions of interest at 25.2 µm and 37.6 µm. Additionally, SOFIA's HAWC+ instrument completed a survey of the magnetic field of the CMZ at 214 µm with tens of thousands of dust polarization vectors, soon to be published. Understanding the dust continuum and polarization landscape of the CMZ at far infrared wavelengths with these surveys will be crucial for interpreting the hotter component of the Galactic Center.

- The MeerKAT observations of the Galactic Center have revealed a menagerie of non-thermal emission at 1.28 GHz, including a diverse population of bubbles and rings and mysteriously long and coherent filamentary structures both parallel and orthogonal to the Disk. These features being characterized in the radio allow for highly complementary studies with PRIMA to understand how the hot phase permeates and connects these structures.

- Telescopes sensitive to the x-ray bright plasma throughout the CMZ's hottest material provide another window into the multiphase composition of the Galactic Center. Telescopes like Chandra and XMM may provide highly complementary surveys in the coming years.

## Description of Observations:

The aim of this project is to contiguously map the Milky Way's entire Galactic Center in the wealth of spectral lines provided by FIRESS. In order to contain the majority of CMZ targets and a portion of the dust lanes, an ultimate mosaic of order 30' × 240' will be required. The shape and exact extent of this mapping strategy will be determined with the help of the FIRESS instrumentation team in order to optimize observations using the observatory's slit shapes and mapping speed. Based on previous observations from Harris+2021, some typical extended CII intensities towards Sgr B1 are on the order of $6 \times 10^{-15}$ W m$^{-2}$ sr$^{-1}$. Using this and the nominal beam size we estimate that the low-resolution mode mapping of the intended target area will require less than 1 hr, likely limited by the rate at which PRIMA can map a region of ~2 square degrees given its ultimate mirror steering speed and spacecraft pointing adjustments. Other lines are sure to be significantly weaker (García+2021 shows that, towards several Galactic Center sources, [NII] line intensities are a factor of 10-100 lower than those of [CII] lines), and if we estimate using the Harris+2021 [CII] intensities, we estimate a total mapping time in low-resolution mode (double to account for both sides of the wavelength range) of ~6 hours to cover the 2 square degrees.

There are significant observational challenges to observing towards the Galactic Center, but perhaps most notable for a low spectral resolution (R~130) mapping using PRIMA's FIRESS instrument is confusion along the line of sight. Given the substantial kinematic complexity of this region, any given line of sight is almost certain to contain multiple components attributed to distinct regions of the CMZ, as well as material in the Galactic Disk. Past work has shown that [CII] emission peaks coincide neatly with observed continuum sources at 70 um and 20 cm, though we cannot rely on this coincidence to eliminate the line-of-sight degeneracy. This is especially true in regions where the fine structure emission is fainter, which are also the areas most ripe for new discoveries.

We intend to work with the PRIMA team to develop a strategy that combines a 30' × 240' map of the Galactic Center using the FIRESS low resolution mapping mode, observed twice to include FIRESS's entire spectral coverage, with strategic pointings in high-resolution mode (R~4400) in order to distinguish between line-of-sight degenerate velocity components. The typical velocity of Galactic Center clouds is about a factor of 5 greater than the spectral resolution with R=4400, so these pointings will provide crucial kinematic information for interpreting the low-resolution map. The choice of pointings and integration times for the high-resolution component of this survey will be largely dependent on the region being observed, and will need to be developed carefully along-side the mapping strategy for the low-resolution mode. We also note that the morphology of other gas tracers can be helpful in interpreting whether certain gas structures may be associated. For instance if a streamer is detected in [CII] with FIRESS's low-resolution mode and follows closely the morphology of a filament detected in 20 cm continuum observations, this may suggest their correlation in velocity as well as position, although these situations will need to be handled carefully on a case-by-case basis.

## Acknowledgement:

A portion of this research was carried out at the Jet Propulsion Laboratory, California Institute of Technology, under a contract with the National Aeronautics and Space Administration (80NM0018D0004).

## 53. Calibrating the nature of CO-dark gas as traced by C$^+$ in nearby galaxies

Arshia Maria Jacob (Johns Hopkins University), Dariusz C. Lis (Jet Propulsion Laboratory, California Institute of Technology), Andreea Petric (STScI), Darshan Kakkad (STScI)

While CO is often assumed to trace the bulk of the molecular material in the interstellar medium (ISM), in diffuse and translucent regions, it is less efficiently self-shielded than H$_2$ and thus readily photodissociated by the surrounding interstellar radiation field (ISRF), particularly in low metallicity environments. Therefore, in such regions, a substantial amount of molecular hydrogen is left undetected and often referred to as 'CO-dark' molecular gas. Since knowledge of the amount, and physical and excitation properties of the molecular gas is vital for our understanding of the processes leading to star formation and the consequent chemical evolution, it is of great importance to investigate the composition of this dark gas. By taking advantage of the simultaneous coverage between 40–75 µm and 135–240 µm provided by Bands 2 and 4 of the FIRESS spectrometer on board PRIMA, we will be able to characterize the properties of CO-dark gas using multiple tracers including [OI] 63 µm, 145 µm, CH 149 µm, [CII] 158 µm and the [NII] 205 µm. Such observations will allow us to calibrate the contributions of the widely used CO-dark gas tracer – [CII] 158 µm, which traces multiple gas phases, across a wide range of sources. Furthermore, by sampling a variety of extra-galactic environments with varying degrees of star formation, nuclear activity and metallicities, we will be able to provide global statistics on the dependence (if any) of CO-dark gas on these properties.

### General Scientific Area:

Galaxy Ecosystems and Interstellar Medium

### Science Justification:

One of the fundamental questions in modern astronomy concerns the life cycle of molecular material in the universe and addressing how molecular clouds – the sites of star formation, are formed. Forming the major reservoir of material that fuels star formation, it is critical to investigate the properties of the neutral phases of the interstellar medium (ISM), which is mainly composed of atomic (HI) and molecular hydrogen (H$_2$). However, despite being the most abundant molecule in space, observations of H$_2$ tracing the bulk reservoir of cold molecular material are challenging, partly due to its widely spaced energy levels and lack of a permanent dipole moment. Fortunately, astronomers have overcome this hurdle by using other associated chemical species (including dust) as surrogates for H$_2$, with the ubiquitous CO (1–0) emission being the most extensively used proxy, today.

However, the reliability of CO as a cold molecular gas tracer has been debated over the recent years. This was based on comparisons of the total neutral gas column traced by the combination of HI and CO with those other tracers including, the diffuse gamma-ray flux (EGRET; Grenier et al, 2005) and the sub-mm dust optical depth (Planck; Planck Collaboration et al, 2011). These comparisons revealed an excess of gas invisible in emission from HI and CO (see for example Fig. 1 of Grenier et al, 2005). This excess gas can be attributed to either saturation in the HI emission line profile (Li and Goldsmith, 2003), or the inability of the standard proxy CO to trace $H_2$, in regions of low dust column densities (for example, low-metallicity galaxies; Schruba et al, 2012). While the fraction (which can vary from 30% to 100%) and composition of this excess gas are sightline dependent, both models and observations have revealed that the bulk of the missing material is molecular (Liszt et al, 2018, Liszt and Gerin, 2023b). The presence of CO-dark molecular gas then leads to uncertainties in the total molecular gas mass, limiting our understanding of the efficiency with which stars are formed.

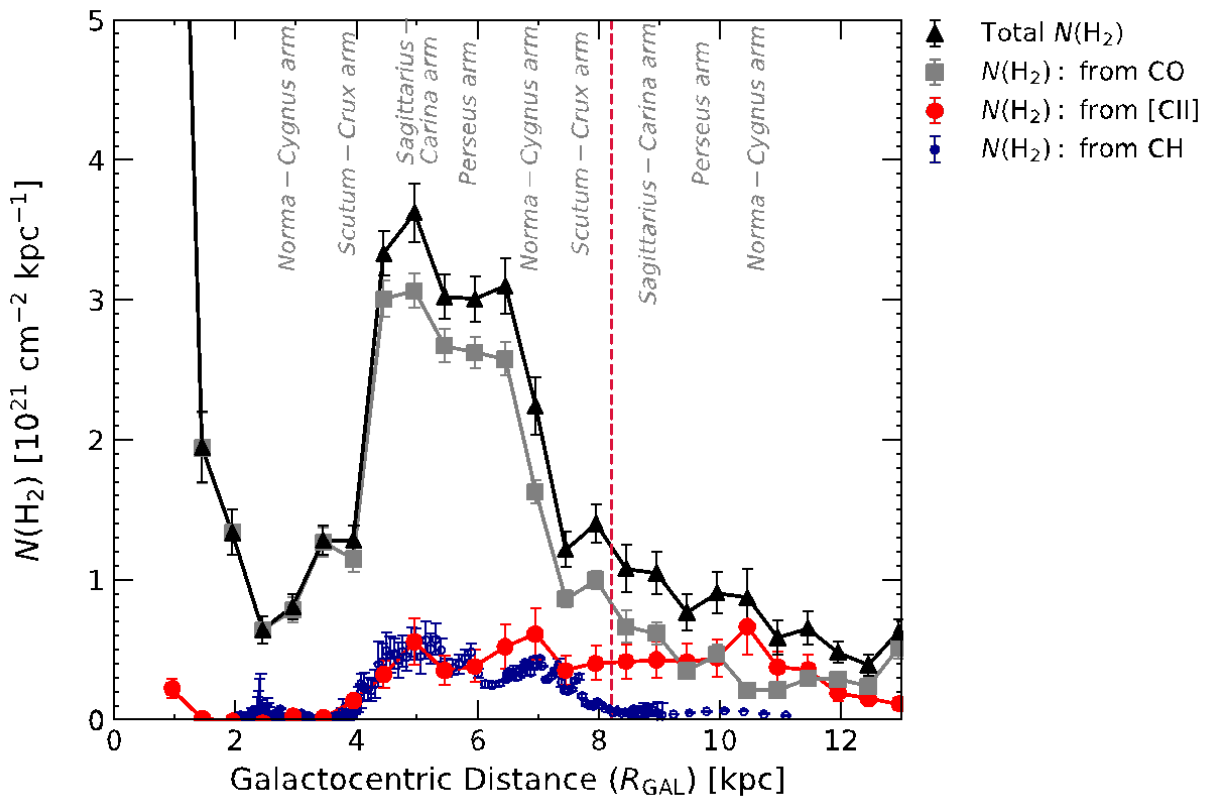

Figure 1: Radial distribution of the total $H_2$ column density in the plane of the Milky Way alongside the relative contributions of the $H_2$ column density traced by CO (grey squares) and CO-dark gas tracers like [CII] (red circles) and CH (blue dots). Taken from Jacob et al. 2023b in review.

As an important coolant in the ISM and chemical precursor to CO, the [CII] 158μm line has been widely used to study the properties of CO dark gas both in the Milky Way (GOTC+; Pineda et al, 2013; Langer et al, 2014) as well as external galaxies near and far (Bigiel et al, 2020, Chevance et al, 2020b, Madden et al, 2020, Schaerer et al, 2020, Tacconi et al,2020). While the [CII] line emission is in principle a good tracer of CO-dark molecular gas, it traces multiple phases of the ISM and includes contributions from both the warm, and cold neutral media. Therefore, it is necessary to decompose the contributions of the [CII] emission arising from molecular gas from that associated with ionized gas. This requires observations of other molecular and ionized gas tracers such as the CH, and [NII] lines, respectively. This is illustrated in Fig.1, which displays the distribution of molecular gas as traced by [CII], CO and CH, across Galactocentric distances in the Milky Way. The similarities between the $H_2$ column density profiles obtained from decomposed column densities of [CII] and CH – a purely molecular gas tracer, lends credence to their use as CO-dark gas tracers. Moreover, while detailed studies have been carried out within the galaxy and toward a handful of nearby galaxies, extending such an analysis to a broader galactic parameter space is crucial.

The capabilities of the high-resolution spectroscopic mode (necessary to avoid blending between the two key tracers targeted) provided by FIRESS onboard PRIMA can facilitate such observations toward external galaxies. A statistical study, the targets are selected based on their infrared emission in excess of 30 Jy at 100 μm based on the IRAS Revised Bright Galaxy Sample (Sanders et al. 2003) with far-infrared luminosities ranging between ∼$10^7$ and $10^{12}$ $L_\odot$ and varying nuclear activity (AGN, starburst, …). In addition, the sample will also consist of a few of the brightest low-metallicity dwarf galaxies observed under the Herschel Dwarf Galaxy Survey (DGS; Madden et al. 2013). These observations will not only be synergetic to previous ones carried out with the Herschel Space Observatory and the IRAM 30-m telescope but will also provide a unique exploration space, spectroscopically, for the identification of other neutral and ionized species be it atoms or molecules.

By carefully modelling the properties of the observed lines, we will be able to make independent measurements of the physical and excitation conditions probed. This will allow us to quantify what percentage of the molecular gas is associated with a warm gas reservoir.

### Instruments Required:

FIRESS Pointed high-res observations

### Program Size:

- Large (100 + hr)

### Approximate Integration Time:

The key lines of interest (CH 149 μm, CII 158 μm, and NII 205 μm) are all in PRIMA FTS Band 4. The PRIMA FIRESS FTS CBE line sensitivity is $4.7 \times 10^{-19}$ Wm$^{-2}$ (5σ, 1h). The required sensitivity of $2.16 \times 10^{-18}$ Wm$^{-2}$ (5σ) can thus be reached in 170 s. The number of FTS spatial pixels available in Band 4 is 8, which corresponds to a solid angle $7.06 \times 10^{-5}$ sr. To map an 5' × 5' area, we need 98 instrument footprints. The total time per target is thus 4.66 h, or 326 h for the full sample of 70 sources. This puts the program in the Large Program category.

A uniform mapping area of 5' × 5' was selected in order to cover both the compact dense gas associated with the star-forming regions of the targeted galaxies (∼1') as well as the more diffuse and extended gas fueling this process. The sensitivity criteria for this sample of sources are derived based on the previous detections made using Herschel toward a sub-set of the sources in this study.

## Special Capabilities Needed:

- None

## Description of Observations:

Observations will be carried out using the PRIMA FIRESS Spectrometer in the high-resolution FTS mode. As discussed above the high-resolution mode is necessary in order to disentangle contributions from the main targets of this study—CH 149 μm and [CII] 158 μm.

## Acknowledgement:

A portion of this research was carried out at the Jet Propulsion Laboratory, California Institute of Technology, under a contract with the National Aeronautics and Space Administration (80NM0018D0004).

# 54. What are the recipes for star formation and stellar evolution in metal-poor environments?

Vianney Lebouteiller (Université Paris-Saclay, Université Paris Cité, CEA, CNRS, AIM), Suzanne Madden, Frédéric Galliano (Université Paris-Saclay, Université Paris Cité, CEA, CNRS, AIM)

Nearby star-forming dwarf galaxies provide the opportunity to probe the metal-poor interstellar medium (ISM) – as low as ~2% the Milky Way metallicity – and extend our knowledge of macroscopic star formation laws to better understand the role and importance of molecular gas in the star formation process. The quest to detect dense cold molecular gas in metal-poor galaxies faces a challenge due to the current lack of sensitivity and spatial resolution to resolve small molecular clouds. While the warm layers of these clouds may be probed in the near-IR with, e.g., JWST, the large-scale molecular gas reservoir available for future star formation mostly lies in a "CO-dark" molecular phase that can be constrained through specific tracers and state-of-the-art models. However, the evolution of the molecular gas and its various tracers throughout a galaxy star formation history is poorly known.

The molecular inventory of metal-poor galaxies has been restricted to CO and warm $H_2$, but PRIMA will enable deep observations of various molecular gas tracers including high-J CO ($J_{upper}$>15), HD, $H_2O$, or even $HeH^+$, as well as transitions from atoms that co-exist with $H_2$ such as [CII], [OI], [FeII], or [SiII]. Furthermore, the existence of detectable molecular gas is contingent upon its consumption and destruction timescales, and PRIMA will access for the first time not only actively star-forming galaxies but also more quiescent ones. Together with metal-poor resolved galaxies such as the Magellanic Clouds, we now have the opportunity to study low-metallicity star-formation in a comprehensive way.

## General Scientific Area:

Galaxy Ecosystems and Interstellar Medium

## Science Justification:

Theoretical models predict a metallicity quenching for low-mass galaxies at high redshift (e.g., Krumholz & Dekel 2012). Nearby star-forming dwarf galaxies allow us to isolate metallicity effects in a time when gas infall is less dominant. Furthermore, such galaxies and their elusive molecular component (e.g., Leroy et al. 2012A, Schruba et al. 2012, Cormier et al. 2014, Hunt et al. 2015, Grossi et al. 2016) represent important cosmological laboratories to examine star-formation laws in primitive environments and provide a stringent test for the current paradigm that molecular gas should be the primary fuel for star formation (primarily based on the observed correlation between star-formation rate (SFR) and H2; e.g., Kennicutt 1998, Bigiel et al. 2008, Leroy et al. 2013).

The cycle of matter for star formation may proceed differently in low-metallicity environments as compared to our metal-rich Galaxy with the atomic phase participating in – or even possibly dominating – the star-formation process (e.g., Glover & Clark 2012, Krumholz 2013). The extremely metal-poor galaxies are the best candidates to witness potential star formation associated with neutral atomic gas.

The CO molecule has been widely used in the nearby and high-z universe to estimate the mass of molecular gas through the use of a CO-to-$H_2$ conversion factor (e.g., Bolatto et al. 2013). However, this factor becomes impractically large for metallicities <1/10$Z_\odot$ (e.g., Schruba et al. 2012), and, although we need to understand its dependency with metallicity, it is even more important to seek other means of inferring the total molecular gas mass reservoir, most of which is expected to lie in a "CO-dark" phase (where CO is selectively photodissociated due to the low dust-to-gas mass ratio (D/G); Röllig et al. 2006, Wolfire et al. 2010), especially in the metal-poor ISM (e.g., Madden et al. 1997). Only recently with intricate ISM models have we been able to quantify the relatively lower volume filling factor of CO-bright clouds / larger filling factor of ionized gas in metal-poor galaxies (Cormier et al. 2015, Lebouteiller et al. 2017, Cormier et al. 2019).

## Why PRIMA?

PRIMA provides access to the main CO-dark phase tracers, namely 1) [CII] and [OI] (e.g., Cormier et al. 2019, Madden et al. 2020) from multiphase and multicomponent ISM models that require a wealth of tracers that PRIMA can provide, and 2) the dust emission through the use of D/G (e.g., Sandstrom et al. 2013, Groves et al. 2015) which PRIMA can explore through observations of a large sample of metal-poor galaxies, including relatively quiescent ones and as a function of the ISM density (Rémy-Ruyer et al. 2014, Galliano et al. 2021). PRIMA is thus ideally placed to answer the following questions:

- Is $H_2$ abundant in metal-poor galaxies but in the CO-dark $H_2$ phase?

- Is $H_2$ not abundant and star formation proceeds in the atomic gas?

- Is $H_2$ not abundant and star formation more efficient?

- Is $H_2$ mostly consumed in the star-formation process and/or dissociated in the star-formation aftermath?

Thanks to its increased sensitivity compared to Herschel, PRIMA can observe relatively quiescent galaxies to examine whether pre-starburst low-metallicity galaxies do contain initially a significant $H_2$ reservoir. Star-formation timescales are also impacted by the coupling efficiency between the stellar feedback and the gas (Kruijssen et al. 2018) and, with this in mind, a lower metallicity environment might be affected differently by the feedback processes and by the lower dust abundance.

PRIMA will observe many tracers enabling multi-phase models of the ISM including in the diffuse medium where Herschel was severely limited. More generally, PRIMA is well adapted to perform a molecular inventory in metal-poor galaxies, with CO-dark $H_2$ tracers [CII], [OI] but also HD. For moderately metal-poor sources, high-J CO lines ($J_{upper}$>15) and $H_2O$ (with likely enhanced formation rate for the latter due to warmer gas temperature; Bialy & Sternberg 2015) should be accessible as well, and possibly the "primordial'" molecule $HeH^+$ in a few objects.

With an accurate description of the molecular content in metal-poor galaxies, it will be possible to finally provide a consistent picture of star-formation scaling laws down to metallicities on the order of ~1% solar, i.e., a regime where our knowledge is currently very limited. Apart from the potentially fundamental role of atomic gas in the star-formation process, one of the main questions at stake is to know whether star-formation efficiency is higher at low metallicity (e.g., Dib et al. 2011, Turner et al. 2015).

Finally, observations of the Magellanic Clouds will offer a unique close-up view of the complete star-formation process by characterizing the gas and dust reservoirs at different stages and by connecting the properties of the wide-scale star-forming gas reservoir to the population of young stellar objects and their environment.

## Instruments Required:

- FIRESS low res pointed observations and small maps
- FIRESS point high-res observations

## Program Size:

- Large (100 + hr)

## Approximate Integration Time:

1 hour per object can be sufficient for most metal-poor galaxies, and up to 10h for nearby low surface brightness metal-poor dwarf galaxies.

## Special Capabilities Needed:

- None

## Synergies with Other Facilities:

- ALMA and NOEMA may also observe the CO-dark gas through OH or $HCO^+$ in absorption, but only for the most metal-rich sources. Sub-mm observatories are useful to perform deep CO and [CI] observations for moderately metal-poor sources.

- Radio observations (eVLA and SKA) will be important to measure the mass of H0 in order to obtain the total ISM gas mass, while HI absorption can reveal the fraction of optically thick HI. γ-ray observatories will also play an important role to get to the total gas mass, including in the dark neutral medium.

- JWST observations are useful in detecting the integrated emission of the warm molecular layers in the most metal-poor sources (see, e.g., Vanzi et al. 2011).

## Description of Observations:

Observations would consist of:

- Strips in the Magellanic Clouds, covering, e.g., the Molecular ridge and 30 Doradus in the LMC or N66 in the SMC.

- Nearby star-forming metal-poor dwarf galaxies within ~200 Mpc (including, e.g., DGS; Madden et al. 2013; Cormier et al .2019). Time required: ~1h up to a few hours/object for $H_2$ in ~50 objects, with additional deep follow-up observations for specific molecular tracers such as OH, $H_2O$...

- Nearby low surface brightness metal-poor dwarf galaxies within 1 to ~10 Mpc (including, e.g., LITTLE THINGS; Hunter et al. 2012). Time required: ~10h/object to access main lines including most CO-dark $H_2$ lines for ~10 objects, with additional deep follow-up observations for specific molecular tracers such as $H_2$ S(0) 28µm...

- Recently discovered extremely metal-poor (XMP) galaxies with SDSS in the redshift range 0 to ~0.3 (e.g., Izotov et al. 2021). Required time: ~1h/object for the main CO-dark $H_2$ tracers ([CII], [OI], [SiII], [FeII]) in ~100 objects.

Most objects will be unresolved spatially apart from the Magellanic Clouds for which large maps would be invaluable to probe large-scale [CII], [OI], $H_2$, and high-J CO emission. The most nearby, brightest, objects should be observed in high-resolution mode to resolve the [CII], [SiII] line profile and distinguish the PDR emission from the ionized gas emission and to associate [CII] and [OI] to the molecular gas.

Main tracers:

- CO-dark molecular gas: [CII], [OI], OH, HD...

- Molecular inventory in moderately metal-poor galaxies: $H_2$ S(0) 28 µm, high-J CO ($J_{upper}$>15), $H_2O$, CH, $CO^+$, $CO_2$, $HeH^+$...

## 55. Revealing the role of magnetic fields in the formation of giant molecular clouds and star formation in local galaxies

Enrique Lopez Rodriguez (KIPAC/Stanford University), Mehrnoosh Tahani (KIPAC/Stanford University), Brandon Hensley (Princeton University), Susan Clark (KIPAC/Stanford University), Julia Roman-Duval (STScI), Kate Pattle (University College London), Paulo Cortes (ALMA), Matthieu Bethermin (Strasbourg Observatory/LAM)., Valentin LeGouellec (SOFIA), Sergio Martin-Alvarez (KIPAC/Stanford University), Darren Dowell (JPL)

The formation of stars and planets ultimately depends on the balance between the gravitational, turbulence, and magnetic properties of the giant molecular clouds (GMCs). These GMCs are the interface between the kpc-scale ordered B-fields driven by galaxy dynamics and the pc-scale turbulent B-fields driven by stellar feedback. Thus, a multi-scale analysis of the B-fields in the GMCs of nearby galaxies spanning different evolutionary stages and star-forming activities is required. PRIMA will provide the measurements of the alignment between the magnetic field and GMC orientation for a sample of few 1,000 GMCs. Based on PRIMA-like mock observations using numerical MHD simulations, this alignment can provide information about the difference between the magnetic and gravitational collapse in GMCs ultimately unveiling the processes that regulate the star formation history in galaxies. We propose a study via the characterization of the gas density and B-field structures by mapping entire, local disk galaxies at GMC spatial resolutions (<100 pc) using 100 μm imaging polarimetric observations.

### General Scientific Area:

Magnetic Fields

### Science Justification:

Stars and planets form in dense, dusty, and molecular clouds distributed in the interstellar medium (ISM) of galaxies. Giant molecular clouds (GMCs, ~50–100 pc) form in a compressed, stratified, and magnetized ISM (e.g., Inutsuka et al, 2015, Hennebelle & Inutsuka 2019). On the one hand, further compression causes fragmentation, which can enhance star formation. On the other hand, strong B-fields can stop fragmentation and decrease star formation. In addition, turbulence injects kinetic energy at large scales and cascades down to the smallest scales, which generates eddies perturbing the cloud. The size of a molecular cloud correlates with the velocity dispersion within it (Larson 1981). This kinetic energy can be available to amplify the B-field in the cloud. Thus, the formation of stars and planets ultimately depends on the balance between the gravitational, turbulence, and magnetic properties of the GMCs.

GMCs are the interface between the kpc-scale ordered B-fields driven by galaxy dynamics and the pc-scale turbulent B-fields driven by stellar feedback. At pc-scales, observations of molecular clouds within the Milky Way suggest that B-fields may control the molecular cloud formation by directing the large-scale gas flows onto star-forming filaments and hubs (e.g., Planck intermediate results XXXV 2016, Planck intermediate results XXXIII 2016, Pillai et al. 2020, Pattle et al. 2022). Within star formation complexes (sub-pc to pc scales), low-density gas structures preferentially align parallel to the B-field, while high-density gas structures tend to be perpendicular to the B-field (Hennebelle & Inutsuka 2019). These results indicate that the low-density gas may be magnetically supported, while the high-density gas is governed by the gravitational collapse driving star formation. The (mis-)alignment between B-fields and gas density structures provides a proxy of the relative contribution of the B-field energy with respect to gravity. In addition, if the B-field is strong, then it impedes the gravitational collapse of the cloud. The critical density and B-field strength at which the ISM is magnetically or gravitationally dominated remains unknown. Characterizing other energy sources, such as the galactic scale B-field and gravitational instabilities during the formation and evolution of molecular clouds, is therefore critical.

At kpc scales, all observed galaxies have large-scale ordered B-fields with typical strengths of $\sim 10^{-6}$ G in their multi-phase ISM (e.g., Beck et al. 2019, Lopez-Rodriguez, et al. 2022b). These large-scale scale B-fields have been found to be dynamically important in the dynamics of the large-scale gas flows of the ISM due to the close equipartition with the turbulent kinetic energy in the ISM of nearby galaxies (Beck et al. 2019). The Parker instability, triggered by magnetic buoyancy driving loops above and below the disk, can control how the gas is collected in the disk of the galaxy (Mouschovias 1974; Mouschovias, Kunz & Christie 2009). In this picture, gas flows along the B-field lines, which are parallel to the plane of the galaxy and vertically confined due to the galaxy's gravitational weight. These B-field lines twist and create valleys, fragmenting the ISM, creating molecular clouds, and enhancing star formation regions (Kortgen et al. 2018). This scenario and the aforementioned observations invite the question: **how do galactic B-fields influence the assembly of GMCs and the star formation rate in galaxies?**

Figure 1 demonstrates how PRIMA will differentiate between the magnetic and gravitational collapse in molecular clouds. Mock polarimetric observations using PRIMA for a face-on view of a MW-like galaxy with low and high magnetization at a spatial resolution of 50 pc were generated using RTCRMHD (Radiative Transfer Cosmic Rays MagnetoHydroDynamical) models from Martin-Alvarez et al. (2022). PRIMA will observe the B-field orientation and the total dust thermal emission of local disk galaxies at a resolution <100 pc (middle panels). This enables the determination of the alignment ($\zeta$) between the B-field orientation and the density distribution from GMCs across the galaxy.

We show that the B-fields strongly affect the shape of the interstellar gas and the rate of star formation in a galaxy. Specifically, MHD models with strong B-fields reduce the star formation (mean in the entire galaxy: $<SFR_{100Myr}> = 3 \times 10^{-6}$ M$_\odot$ yr$^{-1}$ kpc$^{-3}$), and generate smoother and larger filamentary structures parallel to the B-field ($\zeta=0.6\pm0.2$). Those MHD models with weak B-fields have higher star formation activity ($<SFR_{100Myr}> = 3 \times 10^{-2}$ M$_\odot$ yr$^{-1}$ kpc$^{-3}$) and thinner filamentary structures increasing the number of star-forming regions. For weak B-field, the B-field orientation is isotropically distributed with the gas density (ζ=0.0 ± 0.2), except at high densities where it tends to be perpendicular.

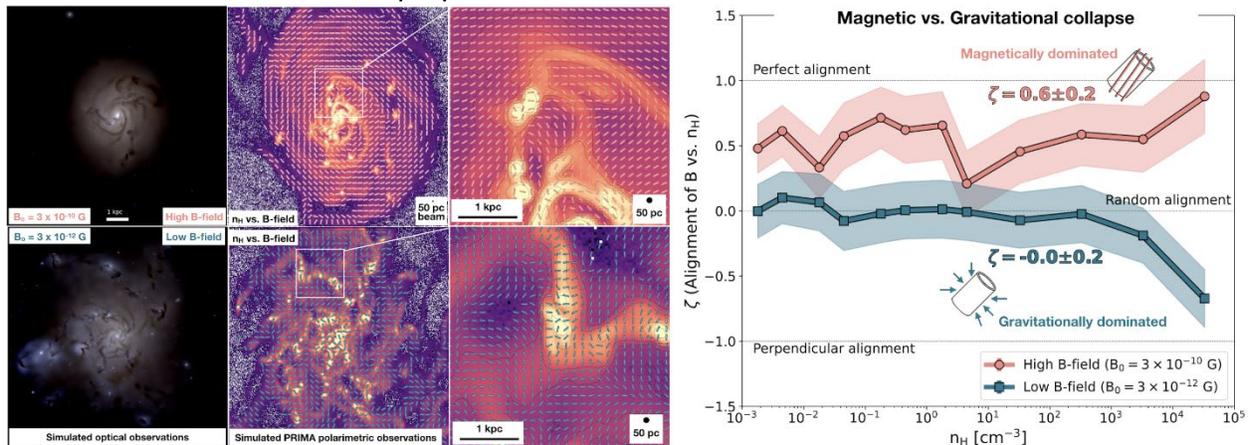

Figure 1: PRIMA will differentiate between the magnetic and gravitational collapse in giant molecular clouds unveiling the processes that regulate the star formation history in galaxies. Simulated optical (1st column) and FIR polarimetric observations (2nd column) of a face-on view of a MW-like galaxy generated using RTCRMHD models from Martin-Alvarez et al. (2022). We show high (top) and low (bottom) magnetization at a spatial resolution of 50 pc. A 4×4$ kpc2 zoom-in of a region of a galaxy is shown (3rd column). For the middle panels, the background image shows the volume density, and the lines show the B-field orientations. Note that our sample will provide observations with a spatial resolution in the range of 2.3-84.7 pc (Table 1).

PRIMA will observe local (D<2 Mpc) dwarf and spiral galaxies at a spatial resolution <100 pc. Spiral galaxies have well-defined large-scale B-fields generated by the differential rotation of the disk, while dwarf galaxies have a low (or non) rotation but show active star formation activity due to galaxy interaction. Local dwarf galaxies have weaker B-field strengths, ∼1 µG, than spiral galaxies, ∼10 µG (Chyzy et al. 2011, Beck et al. 2019). We identified six local galaxies (Table 1) that will be sampled at spatial resolutions of 2.3–85.5 pc at 100 µm. These observations will provide us with a few hundred resolved GMCs to analyze. Our sample enables us to characterize the SFR, B-field vs. density at different stages of magnetization of the galaxy, and definitely determine the origin and formation of GMCs.

## Instruments Required:

- PRIMAger: Small Map (< 20'×20')
- PRIMAger: Large Map (> 20'×20')
- Polarimetry band (100 µm)

## Program Size:

- Medium (10–100 hr)

## Special Capabilities Needed:

- None

## Synergies with Other Facilities:

- ALMA to obtain the velocity dispersion of the molecular gas as a proxy of the turbulent kinetic energy.

## Description of Observations:

PRIMA is the only observatory that can obtain FIR polarimetric observations at <100 pc spatial resolutions in local galaxies. The current sensitivity of FIR polarimeters (i.e., SOFIA/HAWC+, BLAST-TNG, JCMT/POL-2) is insufficient to trace B-fields in local galaxies at spatial resolutions <50–100 pc. In addition, coarser resolution causes depolarization and changes in the B-field orientation due to the vector quantity nature of the polarization measurements. We estimate that polarization fraction decreases a few percent from 50 pc to 250 pc spatial resolution observations. Observations with a 250 pc spatial resolution have B-field orientations dominated by the large-scale ordered B-fields driven by differential rotation of the galaxy.

PRIMA has to map nearby galaxies down to a volume density of $n_H = 10^{-3}$ cm$^{-3}$ to disentangle between a magnetically dominated or gravitationally dominated environment forming form giant molecular clouds (Fig. 1). From our MHD simulations, this $n_H$ corresponds to $N_H = 3 \times 10^{19}$ cm$^{-2}$. To convert to flux density, we assume a dust temperature of $T_d = 20$ K and an emissivity index of $\beta = 1.5$, yielding $I_{150\mu m} \sim 0.3$ mJy sqarcsec$^{-1}$ FIR (150 µm) polarimetric observations of spiral galaxies measured a median polarization fraction of $3.3 \pm 0.9\%$ (Lopez-Rodriguez et al. 2022b). We assume a conservative polarization of 2%, which corresponds to 6 µJy sqarcsec$^{-1}$ in polarized flux, and a 5σ detection. PRIMA's sensitivity of 150 µJy sqarcsec-1 (beam area $\sim 9.4'' \times 9.4'' = 88.36$ sqarcsec, 16 µJy sqarcsec$^{-1}$) can map 1 sqdeg in 10h with a 5σ in I. Thus, PRIMA can reach a sensitivity of 6 µJy sqarcsec in polarized flux (4.2 µJy sqarcsec-1 in I) at a 5σ detection mapping a 1 sqdeg in 1h of observing time. Table 1 shows the on-source time per object in our sample. **We request a total of 105h of on-source time observations.**

Table 1: List of Targets to be observed with Band 2A (100 µm) with a beam size of 9.4"

| Name | Distance | Scale (pc/") | PRIMA scale (pc/beam) | Size | Obs. Time (h) |
|---|---|---|---|---|---|
| LMC | 50 kpc | 0.24 | 2.3 | 8°× 8° | 70 |
| SMC | 61 kpc | 0.29 | 2.7 | 5° × 5° | 25 |
| IC1613 | 0.7 Mpc | 3.39 | 31.9 | 20' × 20' | 2 |
| M33 | 0.84 Mpc | 4.07 | 38.3 | 1° × 1° | 1 |
| IC10 | 0.86 Mpc | 4.17 | 39.1 | 25' × 25' | 0.5 |
| M31 | 0.89 Mpc | 4.31 | 40.5 | 3° × 2° | 6 |
| NGC300 | 1.86 | 9.01 | 84.7 | 15' × 25' | 0.5 |
| | | | | Total Time (h) | 105 |

## 56. Use the magnetic fields to link the cloud formation process and their resulting core and stellar mass functions (CMFs and IMFs)

F. Louvet (Univ. Grenoble Alpes, CNRS, IPAG), V. J. M. LeGoullec (Laboratoire AIM, Paris-Saclay, CEA/IRFU/SAp - CNRS - Université Paris Diderot), F. Motte (Univ. Grenoble Alpes, CNRS, IPAG), S. Bontemps (OASU/LAB, Univ. de Bordeaux - CNRS/INSU), I. Ristorcelli (Université de Toulouse, UPS-OMP, IRAP), P. Hennebelle (Laboratoire AIM, Paris-Saclay, CEA/IRFU/SAp - CNRS - Université Paris Diderot), F. Herpin (OASU/LAB, Univ. de Bordeaux - CNRS/INSU), F. Tabatabaei (Centre for Astrochemical Studies, Max-Planck-Institut für extraterrestrische Physik), A. Gusdorf (Laboratoire de Physique de l'Ecole Normale Supérieure, ENS, Université PSL, CNRS), N. Peretto (Cardiff University, School of Physics & Astronomy), M. Juvela (Department of Physics, University of Helsinki), F. Poidevin (Instituto de Astrofisíca de Canarias; Departamento de Astrofísica, Universidad de La Laguna (ULL)), P. Cortes (Joint ALMA Observatory), K. Pattle (Centre for Astronomy, Department of Physics, National University of Ireland Galway), E. Lopez-Rodriguez (Kavli Institute for Particle Astrophysics & Cosmology (KIPAC))

We aim at discovering the formation process of high-mass star-formation regions. All high-mass star-forming regions show sign of collapse and are prone to strong feedback mechanisms that destroy the trace of their formation process. Therefore, instead of studying the high Av (>100 mag) material of high-mass star-forming regions we propose observing the low and intermediate Av material (1-100 mag) that surround them. We will use the unique capacities of PRIMA to map the magnetic field morphology around a sample of intermediate- and high-mass star-forming regions. We postulate that high-mass star-forming regions result from large-scale gas-flow interaction that will disorganize the large-scale magnetic field morphology, whereas intermediate-mass star-forming regions (as well as low-mass) form from large-scale global collapse that will organize the large-scale magnetic-field morphology. In synergy with ground-based telescopes we will compare the cloud formation mechanism with the core and stellar mass function they generate.

### General Scientific Area:

Star Formation in the Milky Way

## Science Justification:

### Context

There is a growing body of evidence that the large-scale gas dynamic impacts the generation of massive proto-clusters. These large-scale gas motions are best exposed by the large-scale low-velocity shocks they generate. These shocks can be characterized by using multi-transition molecular line observations together with shock modeling. Nevertheless, such an approach cannot point toward the very origin of the shocks, commonly attributed either to cloud-cloud collision (e.g., [1], [2], [3]) or to the global collapse of the putative precursors of massive proto-clusters (IRDCs and/or GMCs). In parallel with their formation, the molecular clouds fragment into clumps of a few parsecs, with a fragmentation process that is highly impacted by the intensity of the magnetic field at intermediate scale (a few tens of pc). In detail, strong fields tend to prevent the formation of low/intermediate mass clumps, and could quench the formation of massive clumps (e.g., [4], [5], [6]). Inside these clumps, several interferometric studies reaching the core scale (≈ 0.01 pc) have reported top-heavy (dominated by high-mass cores) core mass functions (CMFs) — whose high-mass slope disagrees with the one of the canonical initial mass function (IMF, [7], [8], [9], [10]). That suggests a tight relation between the IMF and the physical properties of a cloud.

We aim at linking the atypical CMFs we observe in massive proto-clusters with the intermediate-scale magnetic field morphology and strength, and the latter with the cloud formation process by conducting deep observations with the polarimeters of PRIMA, in association with its spectrometers.

Whatever the process of cloud formation, the high-density fraction (Av > 100 mag) — that can be easily observed from the ground — displays signature of collapse. Therefore such observations will not be able to distinguish between a large-scale global collapse and large-scale gas-flow interaction. Instead we propose to use the unique capabilities of PRIMA to probe the magnetic field all the way from low to high Av (1 to 100 mag). The low-density medium, affected neither by local collapse nor by protostellar feedbacks, keeps the signature of the cloud formation process. The magnetic field will have an hour-glass shape magnetic field morphology associated with low-Av accretion streams if the cloud is formed through a global collapse or a disorganized magnetic field morphology not associated with the accretion streams in the case of large-scale gas-flow interaction (see the Fig. 1 and numerical simulations by e.g., [11]).

### Need for PRIMA:

Only PRIMA will be sensitive enough to probe the magnetic fields in the low-Av material around the massive proto-clusters and inform us on how these dense structures have formed. We will further use the dispersion of the polarization angles at ≈10 pc scale to derive the mean magnetic field intensity using the Davis-Chandrasekhar-Fermi technique and compare with predictions of cloud fragmentation (e.g., [12], [13]).

## Instruments Required:

PRIMAger: Large Map (> 20'×20'); Polarimeter band (96, 126, 172, 235 μm; R=4)

## Program Size:

- Medium (10–100 hr)

## Approximate Integration Time:

- 90 minutes per field times 20 fields = 30 hours

## Special Capabilities Needed:

- None

## Synergies with Other Facilities:

A synergy with two high-resolution LP is planned: with an ALMA and a NOEMA projects that probe the CMF in the high and intermediate mass star-forming regimes, respectively. The gas dynamic can be assessed from the (on-going) LP at APEX in CO(3-2) which has an angular resolution of 14". Noteworthy, by the time PRIMA will be operating ancillary polarimetric data (from the JCMT/Pol2, NIKA2-Pol, SOFIA/Hawc, ALMA) will be available at the clump-scale (Av > 100 mag). But, we stress that *only* PRIMA will be able to perform polarimetric observations at low Av, allowing quantitative comparison between the large-scale magnetic field orientation and strength to the core and stellar content inside molecular clouds at small-scale.

## Description of Observations:

We request a 10σ detection in polarized intensity for material at 1 mag in Av. Taking Hershel observations of G012.80 as a reference (see Fig. 1), it corresponds to ≈2000 MJy/sr fluxes in Stokes I at 250 μm. Assuming a polarization fraction of 1%, we thus need a sensitivity of 2 MJy/sr in the band 2D of PRIMA. We can achieve this sensitivity with one single scan of PRIMA over one squared degree (5400 seconds). We would get sensitivities to polarized power of 2.85, 2.82, 2.40 and 1.92 MJy/sr in band 2A, 2B, 2C and 2D, respectively. Therefore, at one mag in Av we will obtain magnetic field orientation with 10σ detection with 29" angular resolution from Band 2D, and 3σ detection with 11" angular resolution from Band 2A.

We plan to observe 10 intermediate and 10 high-mass star-forming regions with single maps of one squared degree each, resulting in a total observing time of 30 hours with PRIMA. An example toward the massive star forming region G012.80 is shown in Fig. 1.

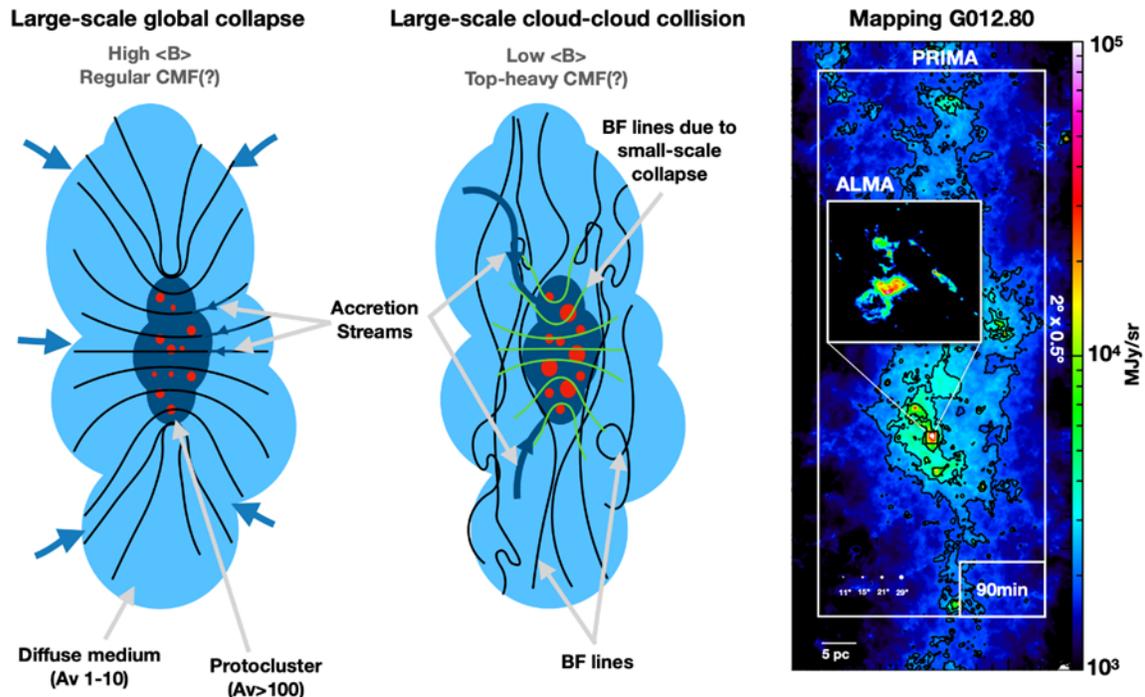

Figure 1: Left: Schematic view of the two competing scenarios for the formation of intermediate/massive molecular clouds. The left panel displays a large-scale global collapse that induces a coherent magnetic field morphology from large-scale to small scale. The middle panel displays a large-scale cloud-cloud collision where only the densest material is under gravitational collapse. Right: HERSCHEL/250 µm observations toward the massive star forming region G012.80. The contours are at 2000, 3000 and 5000 MJy/sr. The insert displays the 1.3mm continuum emission as seen with ALMA. The white rectangle shows the coverage that is proposed in the present science case. It has a size of 1 deg² that permits to observe a similar fraction of low and high Av material. The beams of PRIMA at 96, 126, 172 and 235 µm are shown with white circles.

# 57. PRIMA to test the properties of dust grains in the ISM and star-forming cores

Anaëlle Maury (CEA Astrophysics department, France), V. Le Gouellec (NASA AMES, USA), V. Guillet (IAS, France), L. Testi (Univ. Bologna, Italy) & N. Ysard (IRAP, France)

Dust plays a key role in the chemistry and the physics of the interstellar medium (ISM) but the dust composition and how grains grow during the early phases of the star formation process. Observing the total and polarized submm dust emission, Planck, BLASTPOL and ALMA have shown that the standard dust models clearly need to be revised. Further multi-wavelength observations are crucial to sample the polarized spectral energy distribution (SED) in the Wien regime where dust models dramatically evolve in order to robustly constrain the dust models and study how grain sizes evolve from the irradiated regions of the ISM to planet-forming disk scales.

We propose to sample the dust polarized emission in a handful of both the diffuse ISM and dense environments of protostars, with PRIMAger, allowing to test state-of-the-art dust models and obtain the first constraints on dust grain sizes at the onset of the star and planet formation process. PRIMAger will allow to get access to local variations of the total and polarized SED, which can now be confronted to predictions from synthetic observations produced from state-of-the-art MHD models combined with radiative transfer models, to solve some long-standing issues regarding the nature and evolution of dust grains in the ISM and in star-forming cores.

## General Scientific Area:

Magnetic fields

## Science Justification:

Dust is the key ingredient to form planetary systems around young stellar objects through its evolution from sub-micronic ISM dust to macroscopic objects. Moreover, dust is responsible for coupling the magnetic field to star-forming material, and the grain size distribution plays a role in setting the efficiency of magnetic fields to regulate, for example, the properties of disks around solar-type protostars. Detailed knowledge on dust properties, and its evolution from the diffuse ISM to the densest phases where stars and planets form, is thus essential. Planck observations of the Galactic diffuse and cold ISM have revealed discrepancies between the dust masses predicted from dust models and those derived from the optical, as well as local variations in the diffuse ISM where empirical dust models as the Draine & Li (2007) models are usually calibrated (Planck Collaboration Int. XI 2014).

While dust thermal emission is used to estimate masses in dense astrophysical structures, its polarized emission is routinely used to trace the magnetic fields in astrophysical structures. In the current paradigm to produce the dust polarization, grains are aligned with local magnetic fields but their initial momentum is set by the local radiation field (Radiative Alignment Torques, RATs, Dolginov & Mitrofanov 1976, Draine & Weingartner 1997, Hoang & Lazarian 2009, 2016). Complementing the total dust emission, Planck and BLASTPOL sampled the polarized dust SEDs at submm wavelengths, providing constraints on the nature of grains (Planck collaboration XXI & XXII, 2015; Planck Collaboration XI 2018). If local variations of the polarization fraction can be attributed to fluctuations on the structure of the B-field along the line of sight (Planck Collaboration XII 2018) or in the grain alignment efficiency (Andersson et al. 2015), the shape of the polarized SED is a direct signature of the variations in the grain physics itself (Smith et al. 2000, Fanciullo et al. 2017). Using this approach, Guillet et al. (2018) showed in particular that the flat profile of **the polarization fraction depends on the chemical composition of the grains responsible for the observed polarization**. In particular, it depends on the amount of aligned carbon grains compared to silicate ones (Fig. 1) while a model with carbon grains aligned and silicates with carbon inclusions seems to better reproduce the polarization observed. In spite of these promising results, many questions remain on the nature of the various grain populations. In particular, Planck is not sensitive to the warm dust emission: the temperature of the aligned grain population thus remains mostly unconstrained. While new dust models and modeling tools are currently updated to include this polarized component of the emission (Jones et al. 2017, Hensley & Draine 2023, Guillet et al. 2018), they predict that **observing the polarization fraction of dust thermal emission in the Wien regime provides unique probes of the dust composition**. As ALMA, NOEMA and other telescopes now provide polarization datasets that are widely used to probe the geometry of B-fields, and hence their role, in dense environments forming stars, understanding the exact origin of the dust polarized emission, its dependency with wavelength and with the local radiation field becomes crucial to correctly interpret polarimetric observations.

Moreover, in recent years, observational works probing dust at millimeter wavelengths have also suggested that dust grains may have grown significantly in the dense structures forming stars (see the recent reviews by Tsukamoto et al. 2022 and Maury et al. 2022). For example, envelopes of extremely young (< 100.000 yrs old) protostars observed in the millimeter domain have revealed protostellar dust with **emissivity indices β < 1** (see Fig. 2, Galametz et al. 2019). Although dust emissivity is sensitive to dust composition, porosity and size, studies on various interstellar dust analogues suggest that mm β values <1, like those observed in these protostars, can **only be produced by grains larger than 10 μm** (see Ysard et al. 2019). <span style="color:red">Similarly, unexplained dust emissivities are measured in some star-forming filaments (Lowe et al. 2022), calling for new ways to constrain dust properties in the dense ISM.</span>

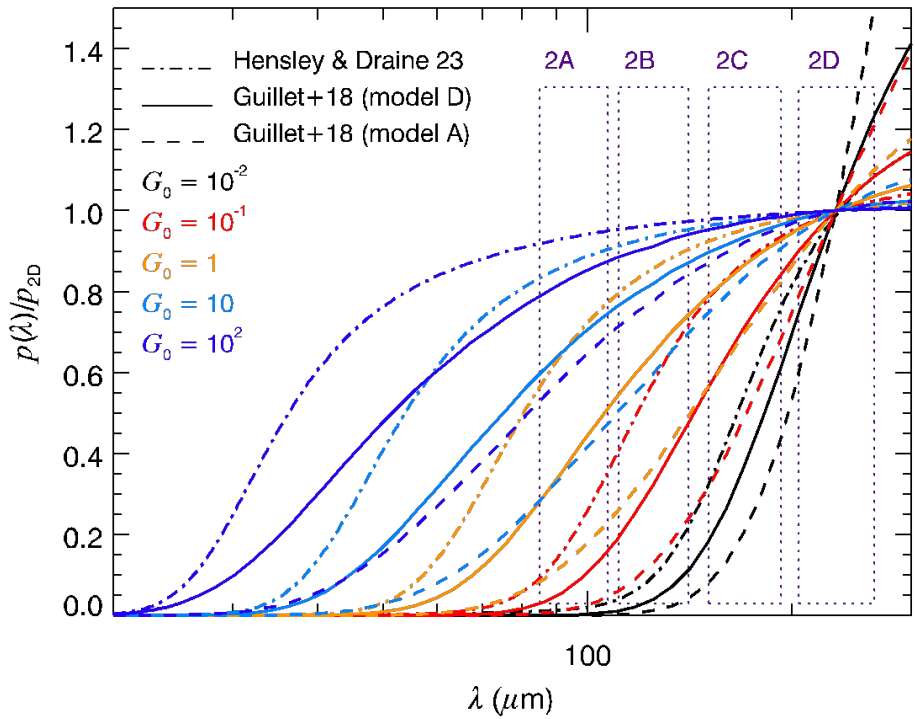

Figure 1: Dust polarization fraction, p = P/I (normalized to the P/I of PRIMAger band 2D), as a function of wavelength, for two dust models of Guillet et al. (2018) and the Astrodust model by Hensley & Draine (2023). Model A includes silicates grains that are aligned and carbon grains that are not. Both are aligned in model D, with carbon inclusions (6% in volume) in the silicate matrix. The vertical boxes show the PRIMAger bands sampling the Wien regime where the polarized emission evolves dramatically depending on the dust model considered. The figure also illustrates how the polarization fraction is also expected to evolve under a range of irradiation conditions, with values G0 typical from diffuse ISM to dense gas.

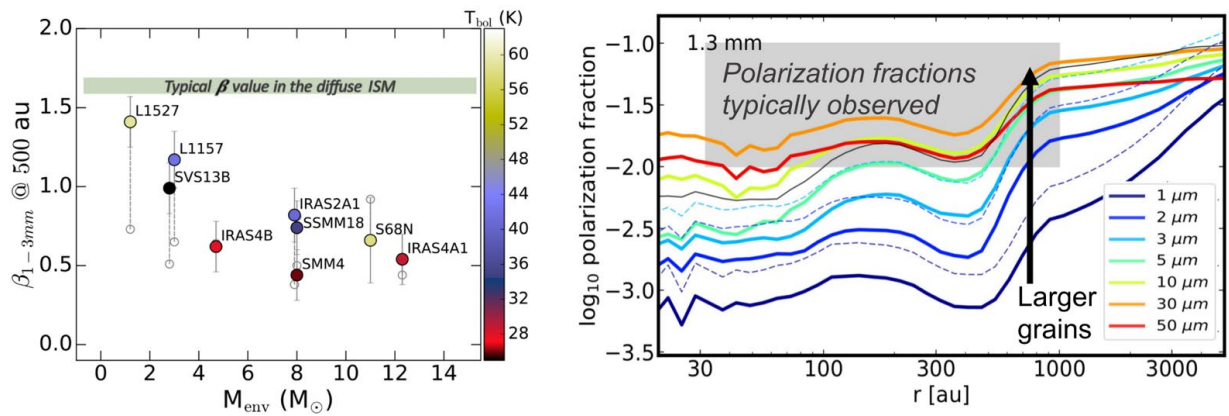

Figure 2: Left: Observed millimeter dust emissivity index in a sample of young protostars (Galametz et al. 2019): all protostars show lower values than the progenitor diffuse ISM at radii < 1000 au. Right: Modeled radial profiles of the polarization fraction from the 1.3 mm dust emission from RATs and MHD models of protostellar cores (Valdivia et al. 2019). Large (> 10 μm) dust grains are required to match the observations of the polarized emission (grey area), however, the lack of far-infrared wavelengths prevents a more refined characterization of the size of these dust grains.

Recently, it has been shown that **polarized dust emission can be a powerful tracer to measure the sizes of the largest dust grains in dense environments**: indeed, since the RAT mechanism only allows to align grains with sizes comparable to the wavelength of the incoming photons, in high-density environments where most of the protostellar radiation is quickly reprocessed in infrared / far-infrared photons, the resolved study of the dust polarized emission might be a unique window to put significant constraints on the underlying grain size distribution. Observations of the dust polarized fractions at millimeter wavelengths compared to models suggest protostellar environments may be favorable to develop, early on, a grain size distribution containing a significant fraction of large grains, with sizes 10µm (Valdivia et al. 2019, Le Gouellec et al. 2019, see the right panel of Fig. 2).

However, how large exactly are these grains cannot be constrained with current observations, because of the lack of observations of the polarization fractions at far-infrared wavelengths, needed to probe grains 50–200 µm in size. Hence, all the first steps of grain growth are missed. This is a key domain of grain sizes, where dust grains are expected to decouple from the gas thus promoting faster growth: we currently are blind to these dust grains in the pristine dense environments that are the cradle of star and planet formation processes, which is a significant shortcoming of our current understanding of planet formation.

The presence and formation mechanisms of these large dust grains in protostellar envelopes are vigorously debated in the community, and testing whether they are present already at large scales in filaments and in the outer layers of protostellar cores would also put stringent constraints on the scenarii and physical processes responsible for the first steps of grain coagulation. Not only this characterization of dust properties is key to set constraints on the first steps of dust evolution towards planetary systems in dense environments, but also because dust grain properties are key to couple the magnetic field to the circumstellar material and their sizes for example may directly influence the size of disks formed around protostars, for example.

## Observables

The spectral dependence of the polarization fraction (sampled with Planck and BLASTPOL) is relatively flat in the submm range (Fig. 1), which complicates its use to disentangle between the contributions of the various (silicate and carbon) dust populations. On the contrary, dust models predict measurable variations of the polarization fraction with the dust composition in the PRIMA wavelength regime (Band 2 at >80 µm onwards, see Fig. 1). Using PRIMA in polarization will allow to test the various models of dust grains by measuring the polarization fractions of dust grains in the Wien domain. For example, As shown in Figure 1, PRIMA could discriminate very different dust models with large grains (1-µm sized dark dust as in Siebenmorgen 2023 or the THEMIS-II model by Ysard et al. in prep.), models with two dust populations (Guillet et al. 2018) and new models with only one component of composite grains as in Hensley & Draine 2023.

Sampling the polarized SED in different FIR bands will allow to constrain the optical properties, temperature and composition of the dust population. Dust models also predict dramatic variations of the polarization fraction with the local radiation field intensity, which also depends on the dust mixture (Fig. 1): probing the polarization fraction in different $G_0$ regimes could be another additional test of dust models, from diffuse (high $G_0$) to dense (low $G_0$) environments.

Moreover, the percentage of polarized emission P at far-infrared wavelengths can be a powerful tool to investigate the very early stages of grain growth in dense, shielded environments where the local radiation field is well understood. Obtaining polarization fractions of the dust emission in several bands at far-infrared wavelengths with PRIMA would put unique constraints on the maximum size of dust grains ranging from a few dozens to a few hundreds of microns. This kind of work would not be possible without detailed modeling of the physical conditions in protostellar envelopes (temperature, density), which became possible only recently thanks to the development of radiative transfer tools and analytical models of aligned grains (see for example works by Valdivia et al. 2019, Le Gouellec et al. 2023, Chau Giang et al. 2023).

## Need for PRIMA

The good spatial resolution and sensitivity of PRIMA to image the continuum emission and its polarized component will allow to directly investigate the grain properties, both in regions of the ISM showing large $G_0$ variations, and in star-forming regions where dust temperatures and radiation fields are known from ancillary data (e.g., ALMA, VLA, JWST), thus providing invaluable information on the size of the largest grains present in these protostellar cores.

The high fluxes of star-forming structures in the far-IR bands make them easily detectable at the PRIMA sensitivities: PRIMAger will be a unique tool to test state-of-the-art dust models and firmly establish grain properties at the onset of the star and planet formation processes.

## Interpretation methods

Analyzing the local variations of the total and polarized dust SED in the far infrared obtained with PRIMA will allow direct tests constraining the nature of the aligned grains. The observational constraints can now be confronted to predictions from synthetic observations produced from state-of-the-art MHD models combined with radiative transfer models including different alignment mechanisms and dust models (see e.g., Le Gouellec et al. 2021, Valdivia et al. 2022 and Chau Giang et al. 2023 for example).

## Instruments Required:

PRIMAGER large maps in the polarimeter band

## Program Size:

- Medium (10–100 hr)

## Approximate Integration Time:

To compute a typical integration time needed to perform measurements of polarized SEDs in regions experiencing high irradiation, we used as prototype the NGC1579 (500–700 pc, Andrews & Wolk 2008; Fig. 2) Hii region, illuminated by the early-type B star LkHα101. NGC1579 also possesses a regular B-field morphology on large scale that is crucial if we want to disentangle the dust composition and alignment from variations linked with a non-uniformity in the B-field orientation. This region shows large range of local radiation field intensities, as $G_0$ varies from 200 to 7000. We used Herschel maps at 100, 160 and 250 μm to estimate the required sensitivities, since they are the closest to the PRIMAger bands we ought to use in this science case. Moreover, their spatial resolutions are similar (from 9" to 18" for 100 and 250 μm respectively). As an example, to detect the polarized dust emission at a few percent level one would require a typical 5σ sensitivity 0.2 mJy per squared arcsec in polarized emission in the Band 2a, which is critical for this science case. This makes about 9000 kJy/sr, while PRIMAger provides a typical 5σ in 10hours of 210 kJy/sr. Hence, a typical region such as NGC1579 would need around 2 hours of integration time for its polarized SED to be investigated with PRIMAger. Assuming about 10 targeted regions, a program testing dust models in highly irradiated regions would require around 20 hours.

The integration times to perform observations to measure polarized SEDs of the dust sitting in the dense structures that are protostellar cores have been estimated in a similar fashion, using the far-IR flux densities from the Herschel/SPIRE observations at 250 μm in the protostars of the Aquila/Serpens South cluster (see Maury et al. 2011), and assuming a conservative 3% polarized fraction. A non-detection at such level in Band 2a would allow to put stringent constraint on the maximum grain size producing the polarization observed at longer wavelengths. Obtaining the polarized dust SEDs in a complete sample of about 40 protostars in such a star-forming region would require about 80 hours with PRIMAger.

### Special Capabilities Needed:

- None

### Synergies with Other Facilities:

- Ancillary datasets: AKARI, Herschel and Planck. Synergy with ALMA, NOEMA, JWST and SOFIA/HAWC+.

# 58. Search for missing oxygen in the interstellar medium

Takashi Onaka (University of Tokyo), Itsuki Sakon (University of Tokyo), Takashi Shimonishi (Niigata University), Mitsuhiko Honda (Okayama University of Science)

The recent study of interstellar elemental depletion poses an important problem in the interstellar matter that at least a quarter of the total oxygen (~160 ppm relative to hydrogen) is not accounted for in any known form in the translucent or dense interstellar medium (ISM). The study of the 3 μm absorption feature of water ice suggests that one fifth of the missing oxygen may reside in 3 μm-size water ice grains. Recent near-infrared (NIR) spectroscopy indicates the ubiquitous presence of water ice in star-forming regions, further supporting a possibility that a significant fraction of oxygen may hide in even larger water ice grains in the dense ISM. However, the 3 μm feature becomes complex and weak for grains larger than 3 μm, and thus the NIR spectroscopy is not the best means to study large ice grains reliably. Here we show that sensitive observations of the far-infrared (FIR) features of water ice at 44 and 62 μm enable us to constrain the amount of crystalline water ice grains up to 5 μm or even larger accurately. Oxygen is one of the key elements of the ISM chemistry, and [O I] 63 μm is a dominant cooling line in the neutral ISM. The understanding of the actual form of the missing oxygen in the ISM is crucial for the study of the ISM and star-formation. To detect the FIR features of the crystalline water ice over the expected strong continuum, a sensitive FIR spectrograph represented by PRIMA/FIRESS is indispensable for this study. Since the feature is broad, the low resolution of R~130 is sufficient, but accurate relative calibration better than 1% is required.

## General Scientific Area:

Milky Way Interstellar Medium

## Science Justification:

The study of elemental depletion provides unique, quantitative information on the relative abundance of the major elements in gas and solid phases and strongly constrains the model of dust grains in the ISM (e.g., Draine 2003). The recent improvement of the analysis of the abundances of the Sun and B-stars removes the uncertainties in the reference abundance (Asplund et al. 2009; Nieva & Przybilla 2012). Together with the improved reference abundance, a detailed study of elemental depletion observations, however, poses a new, significant problem of the missing oxygen in the ISM. Jenkins (2009) and Whittet (2010) show that at least a quarter of the total oxygen (~160 ppm relative to hydrogen) is not accounted for in any known form in the translucent or dense ISM with the density larger than 7 cm$^{-3}$, while recent X-ray spectroscopy suggests that there may not be missing oxygen in the less dense, diffuse ISM (Psaradaki et al. 2023). Jenkins (2009) proposes that the missing oxygen may be hidden in large water ice grains because water ice is only oxygen-bearing species that can be in the solid phase without conflicting the abundance constraint and because the strong absorption band at 3 μm of water ice becomes weak and complex as the grain size increases and thus can be easily elusive in the 3 μm spectroscopy. Poteet et al. (2015) investigated this possibility in detail towards the well-studied line-of-sight to rho Oph using the 3 μm spectrum, concluding that even including contributions from 2.8 μm-size water ice they detected and other species, there remains two thirds of the missing oxygen (~100 ppm), which is not accounted for.

However, the 3 μm feature becomes weak and complex for grains larger than 3 μm, and it is difficult to study the presence of micron-sized grains in the NIR. High-energy resolution X-ray observations will offer unique, independent data on the elemental abundance both in gas and solid phases in the ISM (e.g., Psaradaki et al. 2023). However, large grains (> 1 μm) also become optically thick at the oxygen K-edge and the X-ray spectroscopy is not sensitive to large ice grains either. FIR features do not have this problem unless the grans are too large, e.g., larger than 10 μm. The FIR spectroscopy provides a unique means to study the presence of large ice grains and their amount in the dense ISM unambiguously.

Figure 1 shows the results of simple simulations. We assume a mixture of water ice and carbon dust and calculate the average refractive indexes using the effective medium theory of Maxwell Garnett (Bohren & Huffman 1983). We adopt the refractive indexes by Bertie et al. (1969) for crystalline ice and those for amorphous ice by Hudgins et al. (1993) at 10 K for 2.5–200 μm. We assume the refractive indexes of BE carbon for the carbon inclusion (Zubko et al. 1996). The results with the silicate inclusion do not differ significantly. We fix the volume fraction of carbon inclusion as 0.1, following the value used in Poteet et al. (2015) of 0.15. Figure 1a clearly shows that the 3 μm extinction feature of water ice per mass shows complicated features and becomes very weak for grains larger than 2 μm. On the contrary, Figure 1b indicates that the absorption cross-section per mass at 44 μm of crystalline water ice does not significantly change even for the grain of 5 μm and the feature is visible even for that of 10 μm, demonstrating the effectiveness the FIR feature in detecting large water ice grains compared to the NIR feature. Figure 1c shows the predicted emission from the ISM with and without ice grains (red and black) for the crystalline water ice of 5 μm under the radiation field intensity relative to the solar value, $U = 600$, using the DustEM code (Compiègne et al. 2011). We assume that the missing oxygen of 160 ppm all resides in the water ice grains. Note that the vertical scale is expanded to emphasize the difference in the spectral shape. The 44 μm feature is only ~1% over the linear baseline. For the amorphous ice, the feature becomes weaker even with $U = 1000$ (Figure 1d). The detectability of amorphous ice increases if we observe a target with much higher $U$.

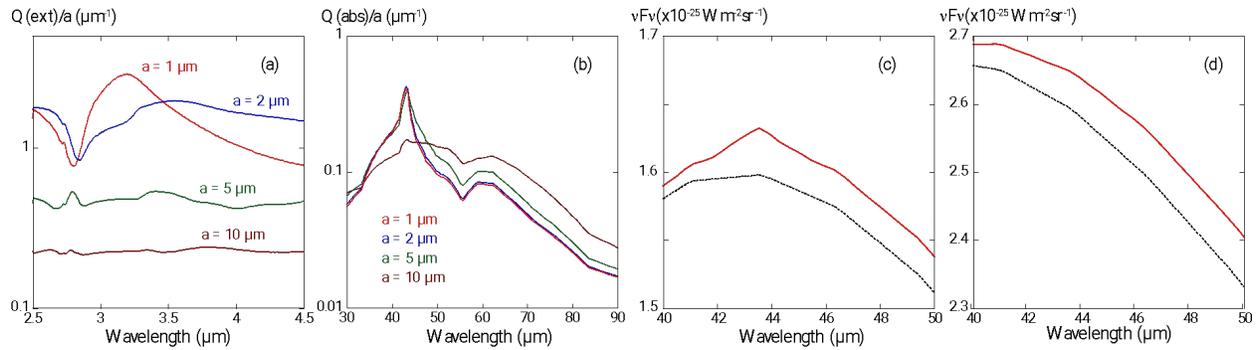

Figure 1: (a) Extinction cross-section per mass at 3 μm and (b) absorption cross-section per mass at 44 μm for a mixture of crystalline water ice and carbon dust for 1, 2, 5, and 10 μm size. (c) Expected FIR emission with and without (black) crystalline water ice grains of 5 μm with U = 600 per hydrogen. (d) Expected FIR emission with and without amorphous water ice grains of 2 μm with U = 1000 per hydrogen.

NIR spectroscopy of star-forming regions with AKARI reveals the relatively ubiquitous presence of the broad, shallow 3 μm absorption feature of water ice (Mori et al. 2014), supporting the possibility that the missing oxygen may in fact be sequestered in large water ice grains. The 3 μm spectroscopy also suggests the presence of crystalline water ice (Boogert et al. 2015) and large icy grains (Terada & Tokunaga 2012) in various objects. Formation mechanisms of crystalline water ice in the ISM have been discussed (Jenniskens & Blake 1996). A recent laboratory experiment of liquid-like behavior of amorphous ice at 65-150K under UV radiation suggests that crystallization of ice could proceed in a relatively short time scale in the ISM (Tachibana et al. 2017). A significant fraction of water ice could be crystallized in the ISM. If oxygen hides in large icy grains, FIR spectroscopy is the only means to put a stringent upper limit on the total mass oxygen in them unambiguously. It should also be added that [O I] 63 μm emission is a dominant coolant in the dense ISM and oxygen is one of the important elements in the ISM chemistry (Hollenbach & Tielens 1997). It is crucial to correctly understand the relative abundance in the gas and solid phase of oxygen. FIR spectroscopy offers a unique opportunity to study oxygen in both phases. Since the water ice feature is faint and the underlying continuum emission is strong, we need a very sensitive, well-calibrated FIR spectrometer for this study.

## Instruments Required:

PRIMA FIRESS: Small Map Low-res, R∼130

## Program Size:

- Medium (10–100 hr)

## Approximate Integration Time:

Figure 1 b suggests that the peak target flux will be $1.6 \times 10^{-25}$ W/m$^2$/sr per hydrogen. Assuming a conservative hydrogen column density of $10^{20}$ cm$^{-2}$, this corresponds to 10 MJy/sr at 40 μm. We would like to reach a S/N > 100, since the water ice feature is very weak. The PRIMAger Band 1B surface brightness sensitivity is 450 kJy/sr in 10 hr, 5σ in a 1 square degree area, which results in a S/N ratio above 100. We request a small area map (< 20'× 20'), thus the integration time per target would be about 1 hr. We would like to observe at least 10 targets in this program.

As described in the text, this program requires very accurate relative calibration, which would limit the detection rather than the standard S/N ratio. Therefore, each observation will be repeated 3 times to evaluate the quality of the calibration or improve the S/N ratio in weaker sources. This results in a 30-hr program.

### Special Capabilities Needed:

- High accuracy in the spectral response of FIRESS (better than 1%) is required.

### Synergies with Other Facilities:

High-energy resolution spectroscopy in the X-ray, e.g., XRISM and Athena, will provide complementary information on the depletion of elements. Radio observations of CO will also be important to estimate the total oxygen in the ISM.

### Description of Observations:

The crystalline water ice has three lattice vibration modes in the FIR at 44, 52, and 62 μm. The 44 μm band is the strongest. However, because of the lack of sensitive spectrometer in this spectral range, only a few detections have been reported so far; one in emission in an OH/IR star (Omont et al. 1990) and a few in absorption in embedded infrared sources (Dartois et al. 1998). The 62 μm band has been observed in several more sources in emission in planet-forming regions and in a young star (McClure et al. 2012, 2015; Min et al. 2016). It has also been detected in a shock region (Molinari et al. 1999). For the present study, we will observe dense, but warm ISM regions. At 44 μm, the dust emissivity is still proportional to the mass of the crystalline ice as far as the size is smaller than 5 μm and the fraction of the inclusions (silicate or carbon) is small. The size of ice grains larger than 5 μm can be estimated by the band profile (Figure 1b). Once the temperature of ice grains is well determined, the mass of the crystalline ice can be constrained accurately, and we can put a strict upper limit of oxygen hiding in crystalline grains up to ~5 μm size. The temperature of the dust grains that emit the continuum can be estimated from its shape. Although it may be similar, the ice grains could have a different temperature from the grains that emit continuum. The grain temperature depends on the size and the fraction and composition of the inclusions. We need to estimate the temperature of ice grains independently. Detection of both 44 and 62 μm features is thus crucial.

Figures 1c and d show the FIR emission of typical interstellar clouds per hydrogen atom. With U = 600, the temperature of the ice grain becomes T = 30.8 K. Since the features in question is at 44 and 62 μm, higher T and thus higher U environments will significantly enhance the detectability of the feature (e.g., Kamp et al. 2018, 2021). Figures 1c and d present rather conservative cases in this regard. We will select about 10 dedicated targets for this study carefully based on the past FIR surveys of Spitzer, AKARI, and Herschel observations, for which the feature at 62 μm can also be detected and a signal-to-noise ratio larger than 100 is expected within the integration time of several hours. Since the targets will be diffuse regions, a small map (< 15 arcmin) will be useful to remove any artifacts and to study spatial variations of the spectrum.

As indicated in Figures 1c and d, the expected feature is very weak relative to the underlying continuum even for the crystalline feature, which is the primary target of the present study. For amorphous water ice, the 44 μm feature becomes weak and broad, and the 62 μm feature is almost invisible. For the detection of amorphous water ice feature at 44 μm, therefore, we need accurate knowledge of the shape of the underlying continuum emission. For this purpose, FIR spectra taken in other programs for various environments, which include more diffuse regions, can be very beneficial for the accurate estimate of the shape of the underlying continuum emission. They will also be used to increase the reliability of the present study and the sample size. FIR spectra of planet-forming regions, where the ice features have been detected, will also provide the useful actual crystalline ice features in the ISM.

# 59.  Polarimetry of the warm dust in the Central Molecular Zone: magnetic fields and star formation in the Galactic Center

Dylan Michelson Paré (Villanova University); David Chuss (Villanova University); Natalie Butterfield (NRAO); Mark Morris (UCLA)

The Central Molecular Zone (CMZ) of the Galactic Center (GC) Milky Way contains a substantial fraction of the molecular mass of the Galaxy ($10^7$ $M_{sun}$) yet exhibits a far lower star formation rate (SFR) than expected given the high densities found in this region (<0.1 $M_{sun}$/yr). There are multiple possible explanations for the depressed SFR in the CMZ, like feedback, strong turbulence, free fall timescales, and magnetic fields. It is currently unclear which of these mechanisms is the dominant inhibitor of star formation in the CMZ. It is important to understand the star formation process in the extreme environment of the CMZ because it exhibits properties like those observed at redshifts of z~2, the time of peak cosmic star formation in the universe. One way to discriminate between the different SFR inhibiting mechanisms is through high-spatial resolution polarimetric observations of the CMZ that will provide insight into the magnetic field behavior at both sub-arcminute resolutions and over a range of dust temperatures. The PRobe far-Infrared Mission for Astrophysics (PRIMA) will be uniquely capable of providing both capabilities. The arcsecond-scale resolutions of PRIMA will yield insight into the magnetic field geometry on protostellar core spatial scales for the first time in the CMZ. Furthermore, polarimetric observations conducted at ~80 μm with PRIMA will provide a unique window into the warm dust pervading the CMZ region that coincides with active star-forming regions. Studying this distinct dust population will enhance our understanding of how the magnetic field varies with different local conditions within the CMZ. In addition, the warm dust serves as a unique connection between the field traced by radio synchrotron structures and the field traced by cool dust. Polarimetry of the warm dust is therefore a key measurement for understanding the magnetic field in the GC.

## General Scientific Area:

Star Formation in the Milky Way

## Science Justification:

The central region of the Milky Way, known as the Central Molecular Zone (CMZ), contains a substantial percentage of the molecular mass of the Milky Way (4% or $10^7$ $M_\odot$, Barnes et al. 2017). The CMZ possesses greater molecular temperatures and densities (e.g., Mills et al. 2018), elevated turbulence (Henshaw et al. 2019), and stronger magnetic field strengths (Pillai et al. 2015) than what is observed in the Galactic disk. The conditions of the CMZ, coupled with its relative proximity to Earth (~8 kpc away, Gravity Collaboration 2019) makes it an important local analog to the conditions of the universe that were present during the time of peak cosmic star formation (z∼2, Kruijssen et al. 2013). The CMZ can therefore be observed to obtain insight into the cosmic history of star formation to greater detail than is currently possible using extragalactic observations.

The fraction of the Galactic star formation rate (SFR) in the CMZ is roughly the same as the fraction of molecular mass of the Galaxy present in the region (3—6%, e.g., Rickert et al. 2019). However, this rate is far lower than would be expected given the high densities (e.g., $10^6$ cm$^{-3}$, Mills et al. 2018) found in the CMZ. There are multiple possible explanations for the low SFR in the CMZ, which include the strong magnetic field of the CMZ, the strength and compressibility of turbulence, and that the CMZ could be in a period of inactivity between episodic bursts of star formation (Krumholz & Kruijssen 2015). **Determining which of these factors is inhibiting star formation in the GC will expand our understanding of the properties of star formation in extreme environments like the CMZ and the universe during the time of peak cosmic star formation.**

One way to assess the impact of mechanisms like magnetic field strength and turbulence on star formation is to study the structure of magnetic fields over multiple spatial scales ranging from e.g., protostellar core scales (1000s of AU) to molecular cloud scales (100,000 AU). This multi-scale analysis has been conducted in the Galactic disk for numerous star-forming regions (see Liu et al. 2022 for a review). For example, Hull et al. (2017) studied the magnetic field of the star forming region Ser-emb-8 over a range of spatial scales and determined that the turbulence of the cloud was more significant in inhibiting star formation than the magnetic field. This kind of multi-scale magnetic field analysis has not been conducted in the CMZ, despite the CMZ possessing a significant reservoir of dense molecular gas.

Currently the magnetic field of the CMZ has been traced on arcminute angular scales (Mangilli et al. 2019; Guan et al. 2021), corresponding to the molecular cloud spatial scales in the CMZ. These observations reveal an ordered magnetic field that is largely oriented parallel to the Galactic plane (Mangilli et al. 2019). Higher resolution observations at an angular resolution of ∼18″ were obtained from the Stratospheric Observatory for Infrared Astrophysics (SOFIA) High-resolution Airborne Wideband Camera Plus (HAWC+) instrument (Butterfield et al. 2023; Paré+2023, in prep). The preliminary magnetic field derived from these observations is shown in the upper panel of Figure 1 overlaid on the Herschel 250 μm observations of the cool dust in the CMZ. This higher resolution magnetic field is more varied than what is observed by large-scale surveys with lower angular resolution (e.g., Mangilli et al. 2019; Guan et al. 2021), possibly revealing the dominance of the turbulence or gravitational/kinetic energy density over the magnetic field in regulating star formation in the CMZ.

**The PRobe far-Infrared Mission for Astrophysics (PRIMA) will be uniquely suited to continuing high resolution studies of the magnetic field in the CMZ. Studies of magnetic fields at high resolutions will enhance our understanding of the impact of the magnetic field on the process of star formation, a fundamental process in astronomy.** Furthermore, with the discontinuation of SOFIA, PRIMA will be the only far-infrared instrument capable of making these key measurements of the magnetic field of the warm dust phase of the CMZ.

PRIMA will also expand the discovery space of the CMZ by probing the warm dust population in the CMZ. The warm dust (observable at ~80 μm) appears to be a distinct component of the CMZ as seen from Herschel observations at 70 μm in the lower panel of Figure 1 (Molinari et al. 2011). However, there is currently no way to study the magnetic field distribution of the warm dust at sub-arcminute resolutions. The magnetic field of the warm dust probes the dust near proto-stellar cores in active star-forming regions, an important physical region of the CMZ that the cool dust cannot probe. **The high resolutions and sensitivities of PRIMA will provide a unique opportunity to conduct a multi-scale analysis of the warm dust component, yielding insight into how the impact of the magnetic field on regulating star formation varies at different temperatures and regions in the CMZ.** Studying the magnetic field over a range of temperatures in this way will make it possible to perform magnetic field tomography to assess how the magnetic field varies with local environment in the CMZ. Obtaining high resolution magnetic field measurements of the warm dust component will make it possible to compare the observed magnetic field distribution with MHD modeling of the large-scale CMZ field.

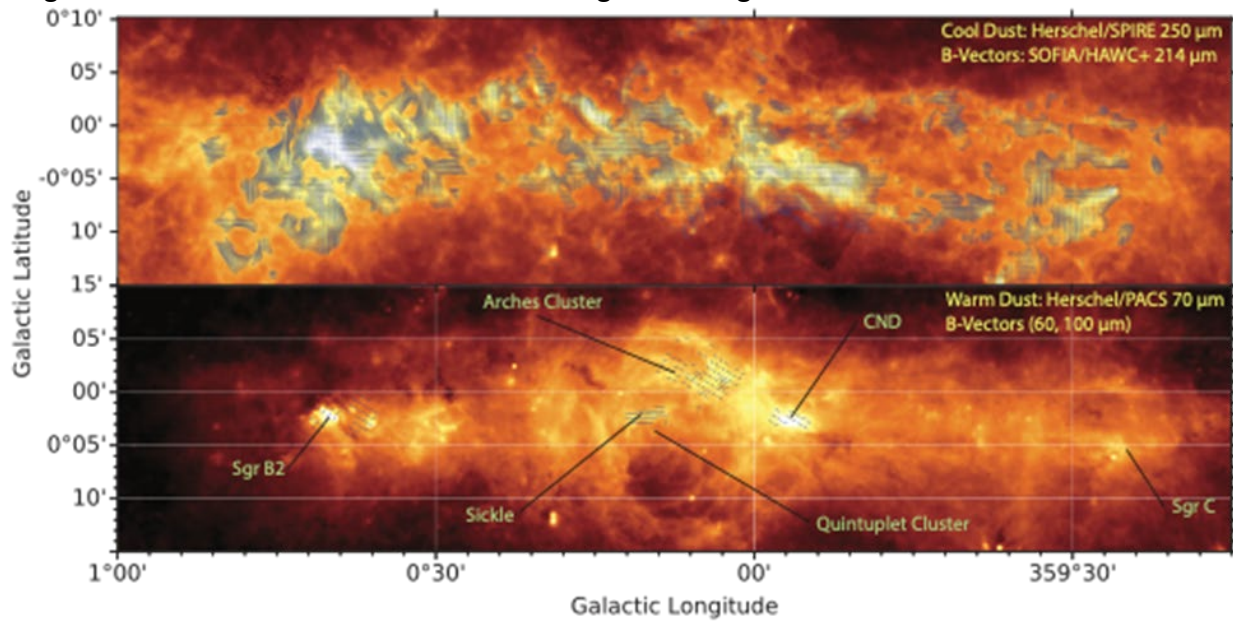

Figure 1: Upper panel: continuum emission from the cool dust in the CMZ as measured by Herschel at 250 μm (Molinari et al. 2011) with magnetic fields derived from 214 μm dust polarization from FIREPLACE overlayed. Lower panel: continuum emission from the warm dust in the CMZ as measured by Herschel at 70 μm (Molinari et al. 2011) with magnetic field vectors derived from 60, 100 μm polarimetry of peak warm dust emission regions only as measured by the Kuiper Airborne Observatory (Dotson et al. 2000).

The warm dust component within the CMZ is furthermore the dust population contained in structures that are co-spatial with prominent synchrotron radio structures known as the non-thermal filaments (NTFs). The Sickle and Arches structures marked in the lower panel of Figure 1, for example, coincide with portions of the prominent Radio Arc NTF. Studying the magnetic field traced by the warm dust at high resolutions may therefore reveal additional connections between the magnetic field derived from dust polarization and the field traced by the NTFs. **The arcsecond-scale observations provided by PRIMA would make it possible to address questions regarding the large-scale magnetic field distribution pervading the GC.**

## Instruments Required:

PRIMAger: Large Map (> 20′×20′); Polarimeter band (96, 126, 172, 235 μm; R=4) emphasizing shorter wavelengths

## Program Size:

- Medium (10–100 hr)

## Approximate Integration Time:

Given the 1.5-degree × 0.5-degree map size required, we anticipate an integration time of ~10 hours.

## Special Capabilities Needed:

If possible, we would like to extend PRIMA's polarimetric capabilities into the frequencies covered by band 1. The warm dust emission peaks at ~50 μm. The high resolution and sensitivity of PRIMA will provide an unprecedented view of the warm dust population in the CMZ in band 2A, and this would be further optimized by extending the polarimetric capabilities of PRIMA to lower wavelengths, if possible.

## Synergies with Other Facilities:

The observation specifications and science goals would synergize well with the Atacama Large millimeter/submillimeter Array (ALMA). The arcsecond and sub-arcsecond resolution capabilities of ALMA would further complement the goal of studying multi-scale magnetic fields over a range of temperatures. For the same reasons the proposed PRIMA project would also synergize well with next-generation radio instruments like the ngVLA and SKA, which will likely be operational during the lifetime of the PRIMA instrument.

## Description of Observations:

To achieve the scientific objectives of this proposal PRIMA would observe the entire CMZ using band 2A with full polarimetry. Full polarimetric capability at the frequency of band 2A will be critical to obtain polarimetric observations of the warm dust population. Without full polarimetric capability at these frequencies, there is no facility currently available or planned that will be able to study the magnetic field of the warm dust component in the CMZ.

Observations conducted using PRIMA band 2A will provide arcsecond-scale angular resolution (observations of the warm dust component in the CMZ). This resolution constitutes an order-of-magnitude improvement over previous observations of this molecular population. Observations of the entire CMZ would require a 1.5-degree × 0.5-degree size map. To have significant detections of the polarization in both bright and faint warm dust structures we would require a sensitivity of 0.3 Jy/arcsec$^2$. This number is derived from scaling the Herschel 70 μm emission and it assumes that the clouds observed are only 1% polarized. This percentage polarization limit is conservative since molecular clouds in the CMZ have typically been found to be a few % or more polarized.

## 60. Mapping magnetic fields in the star formation process from clouds to cores

Kate Pattle (University College London, UK), Simon Coudé (Worcester State University/CfA, USA), Valentin Le Gouellec (NASA Ames, USA), Paulo Cortes (NRAO, Chile), James Di Francesco (University of Victoria, Canada), Brandon Hensley (Princeton University, USA), Enrique Lopez-Rodriguez (Stanford University, USA), Fabien Louvet (Université Grenoble-Alpes, France)

A key unknown in studies of star formation and the physics of the interstellar medium (ISM) is that of the role and relative importance of magnetic fields. Stars form from the gravitational collapse of the densest regions within molecular clouds. These molecular clouds are threaded by magnetic fields on all size scales. High-resolution, wide-area mapping of magnetic field geometries within these clouds, and the star-forming filaments and cores within them, is thus crucial in order to understand the dynamic role that magnetic fields play in the star formation process.

We thus propose an unbiased survey of star-forming molecular clouds within 0.5 kpc of the Earth in polarized light with Primager. Such observations would map magnetic fields over entire molecular clouds at linear resolutions of $\sim 10^{-3}$–$10^{-2}$ pc ($\sim 10^3$–$10^4$ au) in Primager Bands 2A–2D. We will thus observe with sufficient resolution to resolve magnetic field structure within individual star-forming filaments and cores, and with sufficient sensitivity to map magnetic fields in the most diffuse regions of molecular clouds. This will allow us to systematically investigate the wide range of open questions about the role of magnetic fields in star formation and the evolution of the interstellar medium. We estimate that the time required to map a 160 square degree area (the area observed by the Herschel Gould Belt Survey) to the cirrus confusion limit in polarized light is $\sim 600$ hours for a $3\sigma$ detection of 2% polarized emission. We note that this time estimate can be simply scaled up or down in order to map a larger or smaller number of molecular clouds.

### General Scientific Area:

Magnetic fields

### Science Justification:

**Context:** Star formation is one of the most important unsolved problems in modern astrophysics, because how and where stars form has consequences across the discipline, from galaxy evolution to planet formation. A key unknown in studies of star formation and the physics of the interstellar medium (ISM) is that of the role and relative importance of magnetic fields. Ordered magnetic fields permeate the ISM on all size scales: stars form from the gravitational collapse of over-dense regions within molecular clouds, which are threaded by magnetic fields. High-resolution, wide-area mapping of magnetic field geometries within these clouds, filaments and cores is crucial, in order to understand the dynamic role that they play in the star formation process.

**Science Questions:**

The magnetic fields within molecular clouds appear to transition from dominating over gravity and being near equipartition with turbulence in low-density cloud envelopes, to being energetically sub-dominant in dense, gravitationally unstable star-forming cores (Pattle et al. 2022, and refs. therein). They may play an important role in the formation of cloud substructure, and can direct the accretion of material onto star-forming filaments and hubs. However, key questions remain to be answered, including:

- What is the role of magnetic fields in structure formation and fragmentation within molecular clouds?

- Do molecular clouds undergo a transition from sub- to super-criticality (from magnetically dominated to gravitationally collapsing), and if so, is there a characteristic size or density scale on which this occurs?

- Are there distinct weak- and strong-field modes of star formation, and if so, do they produce measurable differences in star formation efficiency or the Initial Mass Function?

- How do magnetic fields and stellar feedback interact, and what role does this interaction play in regulating star formation?

## Need for PRIMA:

There has never yet been a space-based telescope sensitive to far-infrared polarization. Magnetic fields in the interstellar medium can be traced through polarized emission from dust grains which are preferentially aligned with respect to the magnetic field (e.g., Andersson et al. 2015). 850 μm observations with Planck are low-resolution (Figure 1); while 850 μm JCMT POL-2 (e.g., Pattle et al. 2017) and 50-240 μm SOFIA HAWC+ observations (e.g., Pillai et al. 2020), with resolutions comparable to that of PRIMA, show the complexity of these fields. However, POL-2 and HAWC+ can only observe over very limited areas, and at the very highest surface brightnesses, due to atmospheric constraints. n unbiased survey of magnetic field morphologies in star-forming molecular clouds with both sufficient coverage to map entire molecular clouds and also sufficient resolution to resolve individual star-forming filaments and cores would allow us to systematically investigate the questions listed above.

## Interpretation Methods

We will measure the magnetic field strength using a modern implementation of the Davis-Chandrasekhar-Fermi (DCF) method (e.g., Liu et al. 2021) and, through comparison with existing spectroscopic observations, determine the energy balance between gravity, magnetic fields, non-thermal gas motions and stellar feedback in these molecular clouds. We will quantify the relationship between magnetic fields and density structure using the Histogram of Relative Orientations (HRO) analysis (Soler et al. 2013). We will compare the magnetic fields which we observe to predictions from simulations which have undergone radiative transfer post-processing using tools such as POLARIS (Reissl et al. 2016) in order to be directly comparable to observations. We will be able to perform these polarization-based analyses in a wide range of molecular cloud environments, and compare with star formation indicators such as star formation rate and efficiency, and with the mass distribution of dense cores, the Core Mass Function.

**Link to testable hypotheses:** (1) We will test molecular cloud formation mechanisms by comparing to models. Some models, notably cloud-cloud collision models (e.g., Inutsuka et al. 2015) already provide testable predictions of magnetic field geometry. By the time that PRIMA is operational, there will be clear predictions from other models, such as large-scale turbulence and Parker Instability models. (2) Magnetic fields are expected to be preferentially aligned parallel to gas structures at low column density, and perpendicular at high column density (e.g., Soler 2019): this transition is likely related to the onset of gravitational instability in the cloud. We will test this hypothesis by investigating if and how the column density of this transition varies with molecular cloud properties, magnetic field strengths and energetic balance. (3) We will investigate proposed relationships between magnetic fields, (proto)stellar feedback and star formation efficiency (Krumholz & Federrath 2019) through comparison with star formation rates measured from mid-IR source number counts, and with gas dynamics inferred from spectroscopic observations.

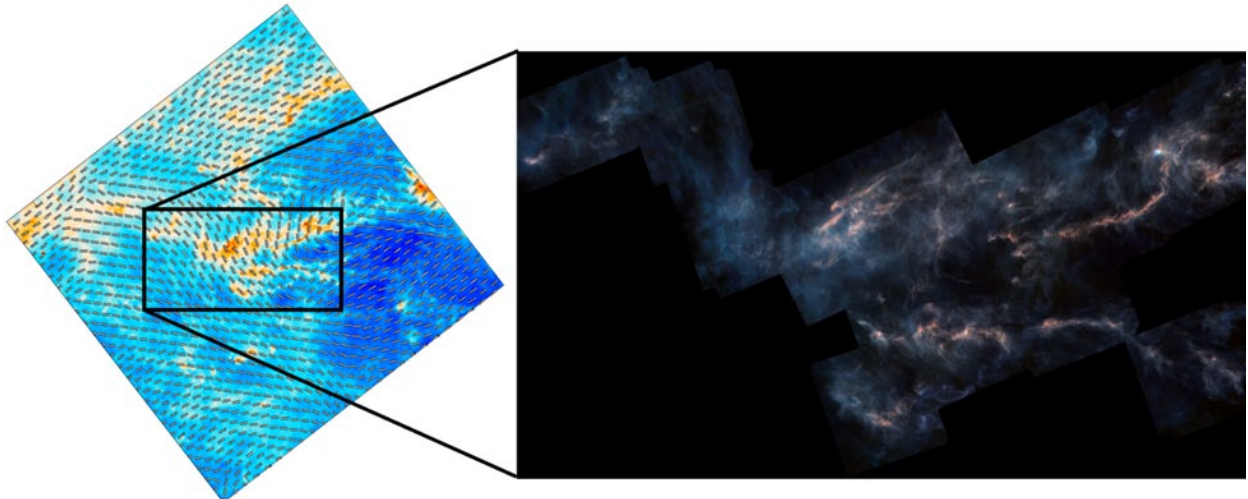

Figure 1: An illustration of the improvement in magnetic field mapping that PRIMA would produce. The Taurus Molecular Cloud, showing the magnetic field on ∼5 arcminute scales (left; Planck Collaboration 2015), and column density structure on ∼10 arcsecond scales, comparable to the resolution of PRIMA (right; ESA/Herschel/NASA/JPL-Caltech; R. Hurt).

## Instruments Required:

PRIMAger: Hyperspectral Imaging and Polarimetry, Large Map (> 20'×20'), Polarimeter band (96, 126, 172, 235 μm; R=4)

## Program Size:

- Large (100 + hr)

## Approximate Integration Time:

We base our estimates on Herschel Space Observatory observations of nearby molecular clouds. The Herschel Gould Belt Survey (André et al. 2010) found that Herschel SPIRE maps are confusion-limited in nearby star-forming regions. The cirrus confusion level is ~30 mJy/18" beam (André et al. 2010) = 5000 kJy/sr at 250 μm. In the following estimates we take Herschel SPIRE 250μm brightnesses to be comparable to PRIMA Band 2D (235μm) values. We perform our sensitivity calculations for the most stringent case, in which we wish to detect polarized emission down to the cirrus confusion limit.

The maximum polarization fraction in the diffuse ISM is ~20% (Planck Collaboration 2015). However, polarization fractions at high gas densities can be as low as 2-5% (e.g., Pattle et al. 2019). Assuming the worst-case scenario of 2% polarization, we would need to achieve a good detection of polarized emission of surface brightness 100 kJy/sr. We note that this is a conservative estimate, and that polarization fractions in the relatively diffuse parts of molecular clouds are likely to be greater than a few percent.

Following the PRIMAger fact sheet, in Band 2D, a surface brightness of 80 kJy/sr in polarized intensity is detectable at 5σ sensitivity over 1 square degree in 10 hours. Assuming that sensitivity integrates as roughly sqrt(t), we would thus require 6.4 hours to achieve a 5-σ detection of polarized emission of surface brightness 100 kJy/sr.

The Herschel Gould Belt Survey covered an area of ~160 sq. deg. To map the full 160 sq. deg. area observed by the Gould Belt Survey in polarized light to the confusion limit would thus require 1024 hours of telescope time for 5σ sensitivity, and 614 hours for 3σ sensitivity.

We note that this time estimate could easily be scaled up or down either in terms of number of clouds to be mapped, area to be mapped within molecular clouds, or desired polarization sensitivity.

## Special Capabilities Needed:

- None

## Synergies with Other Facilities:

We note that by the time PRIMA launches, the Square Kilometre Array (SKA) will be operational, and will be mapping non-thermal sources of polarization, notably synchrotron emission. There may be useful synergies between SKA and PRIMA polarization observations, particularly in higher-mass star-forming regions.

Spectroscopic observations from a range of ground-based large-aperture single-dish telescopes, such as the JCMT, IRAM and APEX will be used to provide information about gas kinematics and dynamics within the clouds observed by PRIMA.

## Description of Observations:

We propose to systematically map all of the star-forming clouds within 0.5 kpc in polarized light, in order to have a complete sample of magnetic field morphologies in star-forming regions, observed from cloud to core scale. Our preferred mapping strategy would involve making wide-area maps, using tiles of ~1 square degree or greater.

We have based our proposed observations on the areas mapped by the Herschel Gould Belt Survey (Andre et al. 2010), which mapped nearby molecular clouds down to the cirrus confusion limit. We describe the time required to detect 2% polarized cirrus emission in the integration time calculation above.

Our observations would have a linear resolution of 0.006 pc in Band 2A and 0.016 pc in Band 2D in the nearest clouds, at a distance of approximately 130 pc: Ophiuchus, Taurus, Corona Australis and the Pipe Nebula. We would have a linear resolution of 0.024 pc–0.060 pc at a distance of 500 pc. Thus, even in the higher-mass and more distant clouds, and particularly in the high-mass Orion A and B clouds, we would be able to resolve the magnetic field within individual star-forming filaments and cores, which have typical size scales ~0.1 pc (e.g., Pineda et al. 2022).

Although our sample would include some high-mass star-forming regions, we note that there is a lack of infrared-dark clouds (IRDCs) and other extremely massive star-forming regions within 0.5 kpc. This could easily be addressed by scaling this project up to also follow up more distant targets. In this case, targets could be selected from the Herschel HOBYS (Motte et al. 2010) and/or HiGAL (Molinari et al. 2016) surveys, or from surveys of large-scale Galactic filaments (Zucker et al. 2018).

# 61. Probing solid dynamics and evolution in protoplanetary disks using dust mineralogy

Youngmin Seo (Jet Propulsion Laboratory, California Institute of Technology), Karen Willacy (Jet Propulsion Laboratory, California Institute of Technology), Dariusz Lis (Jet Propulsion Laboratory, California Institute of Technology), Paul Goldsmith (Jet Propulsion Laboratory, California Institute of Technology), Geoff Bryden (Jet Propulsion Laboratory, California Institute of Technology), Neal Turner (Jet Propulsion Laboratory, California Institute of Technology), Klaus Pontoppidan (Jet Propulsion Laboratory, California Institute of Technology), Nicole Arulanantham (STSci)

We describe how PRIMA could be used to investigate the dynamics and the chemical evolution of solid materials in protoplanetary disks at different evolutionary stages and environments by a statistical survey of the far-IR spectral features of dust and ice species toward ~100 disks around T Tauri and Herbig Ae/Be stars. As addressed in the NASA Strategic Plan and the Astro2020 Decadal Survey, one of the key science themes in astronomy is the formation of planetary systems and the emergence of life on the planets. In past decades, planet formation has been intensively studied, and the overall evolutionary path from molecular clouds to protoplanetary disks and planets is well understood. However, details of how the solid materials assemble to form protoplanets and how the volatiles, including water and prebiotic ingredients, are supplied to terrestrial planets are largely unknown. The key challenge is that probing the dynamics and evolution of solid materials is not trivial; this is much easier for gas-phase constituents for which line profiles can be used to infer the dynamics. Also, emissions from key solid materials ($H_2O$, silicates, $CO_2$, and CO) are mostly in the infrared from 5 $\mu$m to 200 $\mu$m, which can only be observed from space. Key solid features such as water emission bands at 43 and 62 $\mu$m, $CO_2$ at 86 $\mu$m, $NH_3$ at 28 $\mu$m, forsterite at 69 $\mu$m, and fayalite at 93 and 110 $\mu$m provide critical insights into the dynamics and chemical characteristics of volatiles in protoplanetary disks based on their chemical structure (e.g., amorphous vs. crystalline; Maldoni et al. 1999) and the physical characteristics related to their formation (e.g., crystalline forsterite formation at 1000 K and its mixing in disks; Wooden et al. 2007). PRIMA offers a unique opportunity to observe these features in space and make a critical step forward in understanding planet formation and life emergence.

## General Scientific Area:

Planet formation

## Science Justification:

Protoplanetary disks are the birthplace of planets, and their properties are thought to determine exoplanet properties, including demographics, chemical composition, and habitability. The overall concept of planet formation in disks has been well-established with observations using large telescopes and theoretical models. However, as discussed in the Astro2020 Decadal Report, the details of planet formation, including the growth of planetesimals, the chemical evolution of disks, and the delivery of volatiles to terrestrial planets, are still largely unknown and are the next focus in the study of planet formation.

The critical problem is the lack of observations tracing the dynamics and evolution of solid materials in disks. To date, observations of disks have been heavily biased toward gaseous materials in the disks since their narrow emission/absorption lines at radio/millimeter wavelengths are easily accessible from the ground. On the other hand, solid features of specific molecules and minerals are mostly at infrared wavelengths, which are strongly attenuated by Earth's troposphere. Also, the solid spectral features are relatively broad and often overlap (e.g., Rocha et al. 2021), so decomposing and interpreting the solid spectral features requires low-noise data obtained from a highly sensitive telescope. PRIMA provides a unique opportunity to study solid features with high sensitivity and sufficient resolving power. Studies of key solid spectral features of forsterite, fayalite, water, $CO_2$, and $NH_3$ in far-IR using PRIMA will provide key missing information on the evolution of solids in protoplanetary disks. The key science questions related to solid materials in disks are:

1. Do solid materials undergo significant radial mixing?

2. Is vertical mixing of solids efficient in disks?

3. How do solid materials evolve in disks?

These questions can be answered by PRIMA observations of critical solid features closely related to the dynamics of disks. The solid spectral features are typically broad; thus, the dynamics of solids cannot be directly inferred using the shape of the spectral features as is done using the shape of narrow gas emission lines. On the other hand, the formation conditions of certain silicates and the solid structure of volatiles (amorphous vs. crystalline) are closely connected to the radial and vertical mixing in protoplanetary disks; thus, the detection of certain silicates and their temperature infers the existence and degree of disk dynamics. PRIMA can observe water emission at 43 and 62 μm, $CO_2$ at 86 μm, $NH_3$ at 28 μm, forsterite at 69 μm, and fayalite at 93 and 110 μm toward 100 protoplanetary disks spanning different evolutionary stages around various stars, including T Tauri and Herbig sources.

Forsterite is one of the key minerals probing the radial mixing in disks (e.g., Abraham et al. 2009; Maaskant et al. 2015, Kospal et al. 2023). Forsterite is a mineral in the olivine family, which can form at a temperature above 750 K and becomes crystalline above 1000 K. In a protoplanetary disk, a temperature above 1000 K corresponds to the innermost regions in the disks, where there is no volatiles (e.g., water, $CO_2$, and CO). However, crystalline forsterite has been found in comets (Brownlee et al. 2006a; Wooden et al. 2007), which are aggregates of volatiles and refractories. Two mechanisms have been suggested to account for crystalline forsterite in disks. One mechanism is that amorphous forsterite in the interstellar medium has gone through thermal annealing by nearby stars. While this might be a plausible scenario, observations of the forsterite features at 23 μm and 69 μm toward a few protoplanetary disks suggest that the size of crystalline forsterite grains is roughly 10 μm. This is significantly larger than the typical size of interstellar dust grains; therefore, the observed forsterite is unlikely to be of interstellar origin (e.g., Maaskant et al. 2015). The other mechanism is that the crystalline forsterite formed in the innermost region of protoplanetary disks (> 1000 K) and was transported outward through radial mixing. In this scenario, large crystalline forsterite grains can form, but radial mixing in the disks is required (Brownlee et al. 2006b, Wooden et al. 2007). Therefore, the presence of large grains of crystalline forsterite in the cold outer region of disks may be the "smoking gun" for radial mixing. The strength and shape of forsterite features at 23 μm and 69 μm are known to probe the crystallization of forsterite (Figure 1). The temperature of the dust grains can be measured using the forsterite feature at 69 μm (crystalline silicate), and the fayalite features at 93 μm and 110 μm (detectable only at low temperatures; Koike et al. 2006). Observations to date, carried out by ISO and Herschel, have revealed forsterite features only in a handful of disks (e.g., Molster et al. 1999; Sturm et al. 2013), which is inadequate to understand the physical processes of solid mixing in disks. PRIMA can uniquely observe the 69, 93, and 110 μm features, complementing JWST observations of the feature at 23 μm. Observation of these features toward disks at different evolutionary stages and environments will infer how the mixing varies with disk evolution and environments.

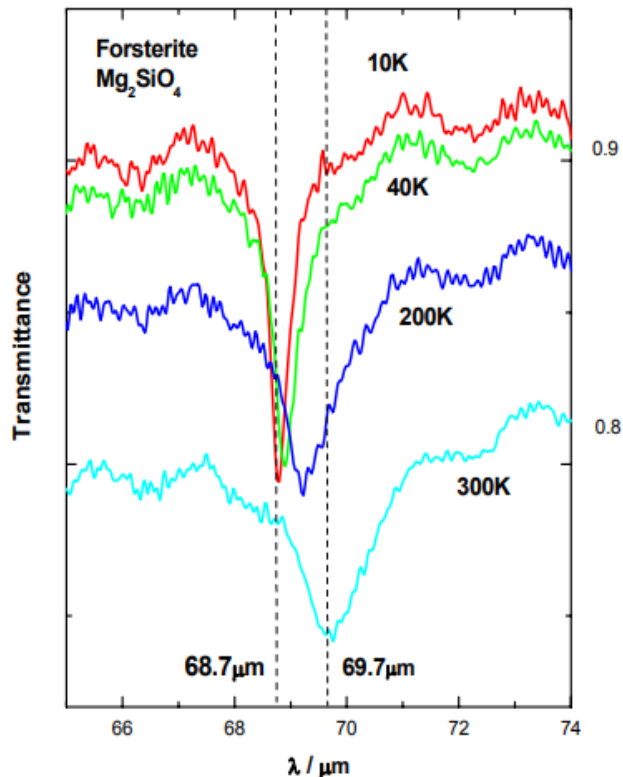

Figure 1: Transmittance of forsterite as a function of temperature. The opacity peak and the shape of the feature vary with temperature, which can be used to infer the physical properties of forsterite in disk observations using PRIMA. Credit: Franz Kerschbaum

Other key solids in disks are volatiles, including water, $CO_2$, $NH_3$, and CO ices. These volatiles are the catalyst of the formation of complex organic molecules (COMs) and the main reservoir of COMs that can be transported to terrestrial planets. Overall solid chemistry of volatiles, including COM formation, is sensitive to temperature and the abundance ratios between ices (e.g., Tabone et al. 2023). The temperature of volatiles is mainly determined by the location within the disks. However, the location of dust grains may not be fixed in time. They may move around due to turbulent mixing within protoplanetary disks. While a direct temperature measurement of these volatiles is not trivial, we may probe the thermal history of the volatiles through their solid structure (amorphous vs. crystalline; Maldoni et al. 1999). If volatiles have gone through a significant temperature fluctuation by turbulent mixing within the disk, crystallization may occur if dust grains are stirred up into the atmosphere of disks (right panel of Figure 2), which will affect the solid chemistry as well. Water and $CO_2$ ice have unique crystalline signatures in the far-IR at 62 µm and 85 µm, respectively. Also, the abundance ratios among water, $NH_3$, CO, and $CO_2$ and refractory materials may vary with the temperature fluctuation (left panel of Figure 2), which affects COM chemistry significantly. ISO and Herschel made a handful of measurements probing these ice features in the far-IR. However, the sensitivities of the two observatories were not sufficient to make a clear detection of the ice features. With its high sensitivity compared to earlier FIR missions, PRIMA will provide a unique window to far-IR solid features of volatiles, water at 45 and 62 µm, NH3 at 28 µm, and $CO_2$ at 87 µm to trace the history of volatile evolution in protoplanetary disks.

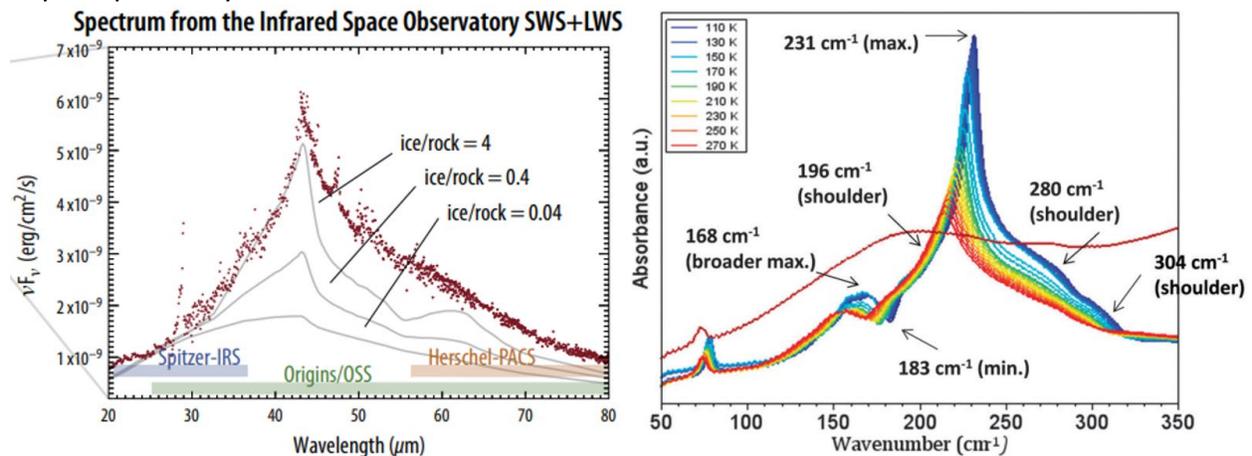

Figure 2: Iso observation of water ice feature at 43 µm (dots, HD 142527, Malfait et al. 1998) and the models of 43 µm feature with different ice/rock ratios (gray lines, McClure et al. 2015) (left: OST mission concept study report, 2019) and with different temperatures (right: Caer et al. 2013). The peak wavelength and the shape of the water feature at 43 µm vary with the ice/dust ratio and ice temperature.

PRIMA observations of these solid features toward protoplanetary disks at different evolutionary stages (Class 0/I to Class III) will enable the evolution of solids to be tracked, while their observations towards different stars (M, K, G, F, A, and B) will test our understanding of solid chemistry in different environments.

## Instruments Required:

FIRESS pointed observations, high-res mode

## Program Size:

- Large (100 + hr)

## Approximate Integration Time:

The FIRESS instrument at a high-resolution setting will be required to resolve the shapes and peaks of solid features properly ($\Delta\lambda < 0.05$ μm). The PRIMA FTS weak line sensitivity estimate is $4.7 \times 10^{-19}$ W m$^{-2}$ (5σ, 1h), which is approximately 5 mJy with a resolving power of R ∼ 400 at 60 μm. The previous studies of forsterite using Herschel showed that the bright forsterite features at 69 μm toward intermediate mass disks (Sturm et al. 2013) are at 0.1–10 Jy, with 0.1 Jy being close to the sensitivity limit. Therefore, 5 mJy will be sufficient enough to survey the disk populations around a wide range of central stars.

To cover the entire spectral range is required since solid features span from 28 μm to slightly over 100 μm. This will require 2 pointings per source since PRIMA may cover half of the spectral range per pointing. We may observe 100 protoplanetary disks to cover different evolutionary stages and central stars, which will provide statistically meaningful numbers per evolutionary stage (Class I, II, III of T Tauri stars) and per stellar mass (M, K, G, F, A, and B). To achieve the target sensitivity and cover the required wavelength range, 2 hours per source are needed and 200 hours in total.

## Special Capabilities Needed:

- None

## Synergies with Other Facilities:

The study described here will have a strong synergy with JWST observations. While many critical solid features for the disk mineralogy can be observed using PRIMA, there are also key solid features that are complementary in near- and mid-IR, for example, CO and amorphous NH$_3$ features. There are also amorphous silicate features at 10 μm, which are complementary to forsterite features in far-IR. To have a complete picture of solid evolution, JWST NIRSPEC and MIRI observations are complementary to PRIMA observations.

## Description of Observations:

To answer the questions listed in the Science Justification would require statistical surveys of disks at different evolutionary stages to probe the evolution of solid materials. Including Class 0/I disks (or protostar) allows the solid features to be studied at the earliest possible time. The observations obtained from Class 0/I sources may be degenerate in that they may include information from both disks and envelopes. However, it is very hard to observe the solid features from molecular cloud cores directly using background stars since the brightness of background main sequence stars falls below the PRIMA sensitivity in far-IR wavelengths. The observations from Class 0/I to Class III will show how the chemistry of solids evolves through time and at what stage of disks we see significant solid mixing disks.

Observations of Herbig Ae/Be disks will also be required to understand how the properties of solids in the disks around low-mass stars are different from the properties of solids in disks around high-mass stars.

We target a total of 100 disks to have a meaningful number of disks per evolutionary stage and stellar mass. The total observation time required is 200 hours.

## Acknowledgement:

A portion of this research was carried out at the Jet Propulsion Laboratory, California Institute of Technology, under a contract with the National Aeronautics and Space Administration (80NM0018D0004).

# 62. Far infrared fine structure lines in local galaxies

Jessica Sutter (UCSD), Karin Sandstrom (UCSD), Thomas Williams (University of Oxford), Elizabeth Tarantino (Space Telescope Science Institute)

The PRIMA FIRESS instrument provides an exciting opportunity to explore the interstellar medium (ISM) of local universe galaxies through measurements of far-infrared fine structure emission lines. These lines, which include the [CII] 158 μm line, the [OI] 63 μm line, and the [Si II] 34.8 μm line, are responsible for cooling the majority of the ISM and have therefore been proposed as tracers of energetic processes like star-formation. Their far-infrared wavelengths also make them ideal targets for millimeter observations of high redshift galaxies. Studies of these lines with the previous generations of far-infrared observatories have already revealed the interpretive power of these lines as tracers of star-formation rate, photoelectric heating efficiency, density, and radiation field hardness, but lacked the ability to efficiently map these lines in large areas or at low surface brightnesses. PRIMA FIRESS will be able to simultaneously map many of these lines along with the infrared continuum in the entire disks of nearby galaxies, exploring parts of the ISM previously unstudied at these wavelengths.

## General Scientific Area:

Galaxy Ecosystems and Interstellar Medium

## Science Justification:

### Broader Context

Far-infrared (FIR) fine structure lines are an important tool for probing interstellar medium (ISM) conditions in galaxies near and far. These lines include the 34.8 μm line of ionized silicon (the [SiII] line), the 63 μm and 145 μm lines of neutral oxygen (the [OI] lines), the 52 μm and 88 μm lines of doubly-ionized oxygen (the [OIII] lines), the 158 μm line of singly-ionized carbon (the [CII] line), and the 122 μm and 205 μm lines of singly-ionized nitrogen (the [NII] lines). Unlike many optical and UV lines, these FIR fine structure lines have minimal attenuation, making them valuable tracers of embedded star forming regions and other heavily obscured sources. The wavelengths of these lines also make them accessible to millimeter arrays like ALMA at redshifts greater than 4. As many of these lines are also important cooling channels for photodissociation regions (Wolfire, Vallini, and Chevance, 2022), they can be related back to the star formation rate and energy input into the ISM (see, for example, Herrera-Camus 2015, de Looze 2014). Additionally, these fine structure lines can be used to explore the full range of environments in a multiphase ISM, with tracers of both the neutral medium ([CII], [OI] and [SiII]) and ionized gas ([OIII], [NII]). This makes them the ideal tools for creating a full picture of the diverse environments within a galaxy (see, e.g., Madden 2013, Ramos-Padilla 2022).

Previous instruments able to detect these lines (Herschel PACS and SPIRE, SOFIA FIFI-LS and GREAT), had limited field of views and sensitivities, and there are no currently operational observatories that can detect these lines in sources below a redshift of ~2. The available archival data in local galaxies are therefore primarily focused on star-forming regions, with few complete maps of nearby sources at any of these lines. The lack of data in nearby galaxies has limited the interpretation of detections of these lines in the high redshift universe. The improved sensitivity of the PRIMA FIRESS instrument, along with its ability to rapidly map large areas covering multiple FIR fine-structure lines, would allow us to fill in the gaps left by previous instruments and address many of the outstanding questions surrounding the interpretation of these lines. Additionally, these data would provide a wealth of data on the ISM conditions of local sources, specifically illuminating the efficiency of cooling in galaxy environments previously unstudied.

In order to build on the legacy of Herschel and SOFIA while providing context for high-z FIR fine-structure line detections, we propose to use PRIMA FIRESS to survey the full disk of a sample of nearby, star-forming, face-on galaxies. The galaxies in this sample were selected from the Physics at High Angular resolution in Nearby Galaxies (PHANGS) dataset, as these galaxies have a compendium of multi-wavelength maps for comparisons and constraining galaxy properties. Mapping the entirety of the star-forming disk of these galaxies with the higher sensitivity of PRIMA FIRESS in all four spectral bands will provide maps of all the important cooling lines as well as the total infrared luminosity in diffuse gas and extended halos inaccessible to previous instruments.

## Science Questions

- How do the conditions of the ISM vary as a function of galaxy environment, and how does this affect photoelectric heating efficiency?

Photoelectric heating is the dominant source of heating in a variety of ISM environments, and will likely vary with conditions like metallicity, dust to gas ratios, and properties of the smallest dust grains, the polycyclic aromatic hydrocarbons (PAHs). JWST is currently providing a novel view of the PAHs, shedding light on how these small dust grains heat the ISM of local universe galaxies (see e.g., Sandstrom 2023). While JWST can provide measurements of the sources of photoelectric heating, FIR fine-structure lines responsible for a bulk of the cooling in multiple phases of the ISM are not observable with any currently operational observatory (Wolfire, Vallini, and Chevance, 2022). Comparing the strength of these lines to traditional tracers of heating like total infrared luminosity and PAH emission maps available from JWST will allow for measurements of how photoelectric heating efficiency varies as a function of ISM phase, galactic environment, and metallicity. PRIMA FIRESS will provide an improvement from previous studies by covering larger areas in multiple lines and continuum simultaneously, providing measurements of full disks as well as low-surface brightness diffuse gas. By comparing maps of multiple FIR fine-structure lines to the available ISM models (Cloudy, PDR toolkit, etc.) properties like ISM density and interstellar radiation field (G0) can be estimated. These lines provide an attenuation-free measurement of electron density and metallicity (Chen 2023, Herrera-Camus 2016), which could then be compared to optical and UV indicators. These data would also illuminate what conditions lead to different FIR fine structure lines becoming the dominant cooling channel.

- What is the cause (or are the causes) of the [CII] deficit?

Studies of the [CII] line have shown that the proportional strength of this line to total infrared luminosity decreases as a function of galaxy activity (see e.g., Smith 2017, Sutter 2019). Understanding the cause of this deficit will provide the context necessary for interpretation of the multitudes of [CII] detections observed through surveys like ALPINE (z~4–7, Le Fèvre 2020) and REBELS (z~7-9, Fudamoto 2022). Studies of [CII] emission in well-studied local universe galaxies where specific ISM environments can be resolved will provide the data needed to determine the conditions that can lead to either a deficit or enhancement of [CII] emission. This information will be essential for determining when [CII] can be used as a tracer of important indicators of galaxy evolution, like star-formation rate.

- How spatially extended is FIR fine-structure line emission in local universe galaxies?

Recent detections of [CII] in z~5 galaxies have shown extended halos of [CII] surrounding the host galaxy (e.g., Herrera-Camus 2022). Meanwhile, JWST maps of nearby U/LIRGS have discovered extended halos of 7.7 μm emission, tracing the polycyclic aromatic hydrocarbons (PAHs) at large radii. As PAHs provide a bulk of the photo-ejected electrons in the neutral ISM, PAH emission is often considered a tracer of photoelectric heating. Similar extended maps of these local universe galaxies at the [CII] line would determine whether the extended halos observed at high redshift are also present in local sources, which would further clarify what mechanisms are creating the [CII] halos early in universal history. By comparing any observed extended [CII] emission to the extended PAH emission, the photoelectric heating efficiency could be measured in these extended halos. As the shortest wavelength FIRESS band also covers the H2 28 μm, which in combination with the [CII] and HI data will provide measurements of the pressure in these extended halos (Herrera-Camus 2017).

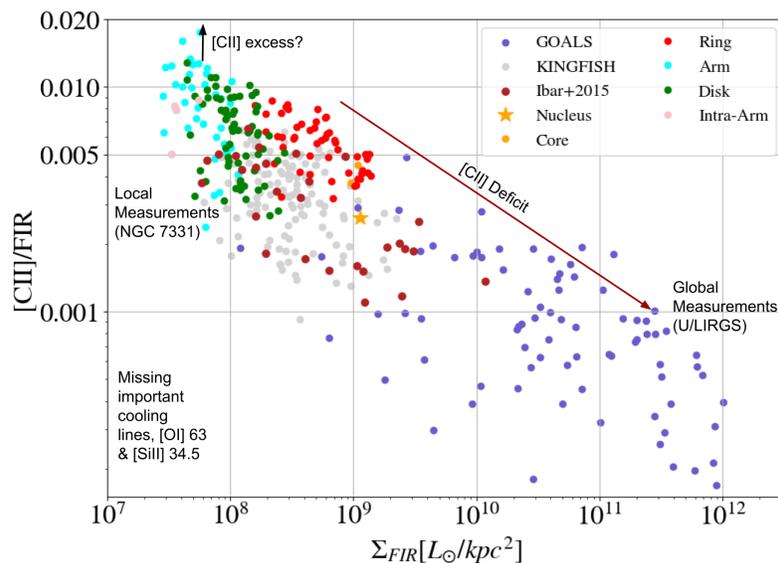

Figure 1: The [CII] deficit, connecting local and global measurements (Sutter and Fadda, 2022). PRIMA will be able to further determine what galaxy environments lead to either enhanced or limited [CII] emission. By providing additional cooling lines ([OI] 63 and [Si II] 34.5) PRIMA observations will better quantify the total photoelectric heating efficiency, clarifying whether this trend is exclusive to [CII] or seen in all FIR fine structure lines, supplementing work done on global scales (McKinney+2020, Malhotra+2001).

## Need for PRIMA

As there are currently no operational observatories that cover these FIR fine structure lines in galaxies below z~2, all studies of these lines rely on archival data. PRIMA FIRESS will provide the necessary sensitivity and mapping capabilities to efficiently study these lines across the entire star-forming disk of nearby galaxies. Additionally, the sensitivity of FIRESS will allow for studies of FIR fine structure emission in extended galaxy halos and the diffuse ISM. As archival datasets do not include these regions or are not sensitive enough to detect emission lines in low-density ISM, PRIMA would allow for these studies for the first time. PRIMA will also be able to cover the full FIR spectral energy distribution, providing FIR fine structure line and total infrared luminosity measurements simultaneously. This unprecedented data acquisition technique will allow for immediate comparison of gas and dust properties, testing models of photoelectric heating in diverse galactic environments.

## Interpretation methods

- Emission line ratios like [NII] 122/205, [OI] 63/145, [OIII] 52/88 to determine electron density
- Measurements of photoelectric efficiency with ([CII] 158 + [OI] 63 + [Si II] 34.8)/TIR or PAH emission
- Comparisons to PDR models using PDR toolbox (Pound & Wolfire 2023)
- Isolation of [CII] emission from ionized and neutral ISM with the [CII]/[NII]205 values
- Measurements of ISM pressure using [CII], HI, and CO (or H2) emission

## Instruments Required:

PRIMA FIRESS: Medium Map Low-res R~130

## Program Size:

- Medium (10–100 hr)

## Approximate Integration Time:

To cover the full extent of 19 nearby galaxies selected based on available multi-wavelength data, we would need ~30 hours of integration time.

## Special Capabilities Needed:

- None

## Synergies with Other Facilities:

ALMA maps of CO will provide data on molecular gas content, JWST maps of PAH emission to trace heating within the ISM, and MUSE IFU data will provide optical emission line diagnostics.

## Description of Observations:

To complete this project, we propose mapping 19 nearby galaxies with FIRESS in all four bands to a sensitivity of $2.5\times10^{-18}$ W m$^{-2}$. The galaxies were chosen from the PHANGS sample, and have available JWST, ALMA-CO, and MUSE maps for constraining the ISM properties of each environment. The sensitivity was chosen to detect the [NII] 205 μm line in diffuse ISM, based on measurements of resolved regions in local galaxies with the SPIRE-FTS instrument as well as using the relationship between the 24 μm flux and the [NII] 205 μm line provided in Hughes et al, 2017. Observations at all bands are required to cover all FIR fine-structure lines, so all times are multiplied by two to complete the wavelength coverage. The galaxies, areas (determined using the WISE3 maps), and required observing times are listed in the following table. To completely observe all 19 galaxies in all four FIRESS bands, the total observing time required is 31 hours.

| Galaxy | Area (square arcmin) | Required integration time (hr) |
|---|---|---|
| IC5332 | 22.3 | 0.72 |
| NGC0628 | 49.2 | 1.57 |
| NGC1087 | 5.94 | 1.9 |
| NGC1300 | 12.25 | 0.39 |
| NGC1365 | 72.92 | 2.33 |
| NGC1385 | 13.14 | 0.42 |
| NGC1433 | 27.62 | 0.88 |
| NGC1512 | 29.75 | 0.95 |
| NGC1566 | 44.23 | 1.42 |
| NGC1672 | 25.34 | 0.81 |
| NGC2835 | 2.66 | 0.1 |
| NGC3351 | 27.48 | 0.88 |
| NGC3627 | 35.31 | 1.13 |
| NGC4254 | 15.56 | 0.50 |
| NGC4303 | 30.9 | 0.99 |
| NGC4321 | 26.85 | 0.86 |
| NGC4535 | 9.92 | 0.32 |
| NGC5068 | 26.17 | 0.84 |
| NGC7496 | 3.38 | 0.11 |

# 63. Investigating the role of magnetic fields in the formation and evolution of striations in interstellar clouds with PRIMA

Konstantinos Tassis (University of Crete and IA-FORTH, Greece), Aris Tritsis (EPFL, Switzerland), Raphael Skalidis (Caltech, USA), Paul F. Goldsmith (Jet Propulsion Laboratory, California Institute of Technology, USA)

We propose to use the PRIMAger instrument abord the PRIMA space observatory to study the largely underexplored diffuse parts of interstellar molecular clouds. We plan to spend ~5h to map the striation regions in three nearby molecular clouds. Striations are diffuse, linear, and magnetized structures located in the outskirts of interstellar medium clouds. Striations hold important clues about the formation and evolution of dense molecular clouds and are thus critical for understanding the early phases of star formation. Recent evidence suggests that striations are formed by magnetohydrodynamic (MHD) wave-modes. PRIMAger gives us the high dynamic range required to explore the magnetic field properties of striations down to 0.1 pc scales, which have been inaccessible so far. We particularly aim to search for the passage of small-scale wave modes that would be a unique probe of the MHD nature of striations and would open up a new window to access the local conditions in molecular clouds. We also plan to combine the polarization data that we will get from PRIMAger with optical stellar polarization data in order to explore the properties of dust grains in the ISM.

## General Scientific Area:

Magnetic Fields

## Science Justification:

### Studying Striations

In the last several decades the community studying the interstellar medium has shifted its focus from mapping the large-scale medium to resolving the smallest, densest structures where protostars form with spectacular results (such as the amazing diversity of protoplanetary disks revealed by ALMA). This has come to the expense of the study of the lower density, diffuse medium that requires large-dynamical--range observations. But it is in the diffuse parts of the interstellar medium that we hope to find clues about the open questions on the origins of the dense molecular clouds; the sites of star formation. The dense parts where stars have already formed are affected by the feedback of the star formation process that dominates the local conditions, making it difficult to deduce the physical conditions at the origins from observations of the actively star-forming sites. If we want to learn how it all begins, we have to study the early phases that take place in the lower density parts of clouds.

Elongated hair-like structures, called striations, are observed in the low-column--density outskirts of molecular clouds. These striations are remarkably similar to wispy structures seen in IRAS images of cirrus clouds (Boulanger & Perault 1988). Striations were first observed in $^{12}$CO, and $^{13}$CO by Goldsmith et al. (2008) in the northwest part of the Taurus molecular cloud where they appear to be unassociated with denser structures. These features arise over a velocity range of 2 km s$^{-1}$ but exhibit substructure over narrower velocity intervals. The width of the striations was marginally resolved by the 45″ resolution of the $^{12}$CO data. The surface brightness increments are 15-25% above the background $^{12}$CO emission over the full velocity interval but larger (50–100%) over small velocity ranges that kinematically isolate individual features. Striations were also observed by Herschel in dust emission. One of the most representative examples is the Polaris flare where well-ordered, low-density elongations are seen throughout the cloud (Miville-Deschênes et al. 2010). Hennemann et al. (2012), Palmeirim et al. (2013), and Alves de Oliveira et al. (2014) found striations in Herschel dust emission maps of various molecular clouds. In some clouds, striations are connected to dense filaments within which protostellar objects are observed.

Striations are found to be aligned with the plane-of-sky component of the local magnetic field revealed by polarization observations. The alignment between these structures and the magnetic field has been pointed out in all clouds by all relevant studies in the literature (Goldsmith et al. 2008; Chapman et al. 2011; Hennemann et al. 2012; Palmeirim et al. 2013; Alves de Oliveira et al. 2014).

While the exact physical origin of low-density striations (Heyer et al. 2016; Tritsis & Tassis 2016, 2018; Chen et al. 2017) is not well understood and remains debated in the literature, there is little doubt that magnetic fields are involved. Tritsis & Tassis (2016) modelled striations as density fluctuations associated with magnetosonic waves in the linear regime (the column-density contrast of observed striations does not exceed 25%). These waves are excited as a result of the passage of Alfvén waves, which couple to other magnetohydrodynamic (MHD) modes through phase mixing. In contrast, Chen et al. (2017) proposed that striations do not represent real density fluctuations but are rather a line-of-sight column-density effect in a corrugated layer forming in the dense post-shock region of an oblique MHD shock. While others (e.g., Li et al. 2013) interpreted striations as sub-Alfvénic streamlines in which material flows into or out from more dense filaments and/or clumps guided by strong magnetic fields. While recent evidence of enhanced CO formation on striations (Skalidis et al. 2023) indicate that they represent real density enhancements, high-resolution and sensitivity polarimetric imaging data would be of great interest to set direct observational constraints and discriminate between these possible models. Specifically, the magnetosonic wave model predicts that a zoo of MHD wave effects should be observable in these regions. One of them, that linear waves in an isolated cloud should establish standing waves (normal modes) imprinted in the striations pattern, has been confirmed in the case of the Musca cloud (Tritsis & Tassis 2018). Other such effects include the ``sausage'' and ``kink'' modes (Figure 1), which are studied extensively in the context of heliophysics (e.g., Nakariakov et al. 2016) and could open a new window to probe the local conditions in ISM clouds (Tritsis et al. 2018).

Only an instrument like PRIMAger, with its polarimetric capabilities and sensitivity, will be able to shed light on the physics that governs the origin and evolution of striations. The target angular resolution of PRIMAger will allow us to study, for the first time, the small-scale fluctuations in the density and the magnetic field of striations. Such a study was not feasible yet because past observatories, such as Planck, could not reach the 0.1 pc resolution (15″ at Taurus) that is necessary in order to probe the properties of striations (see Planck beam size in Figure 1).

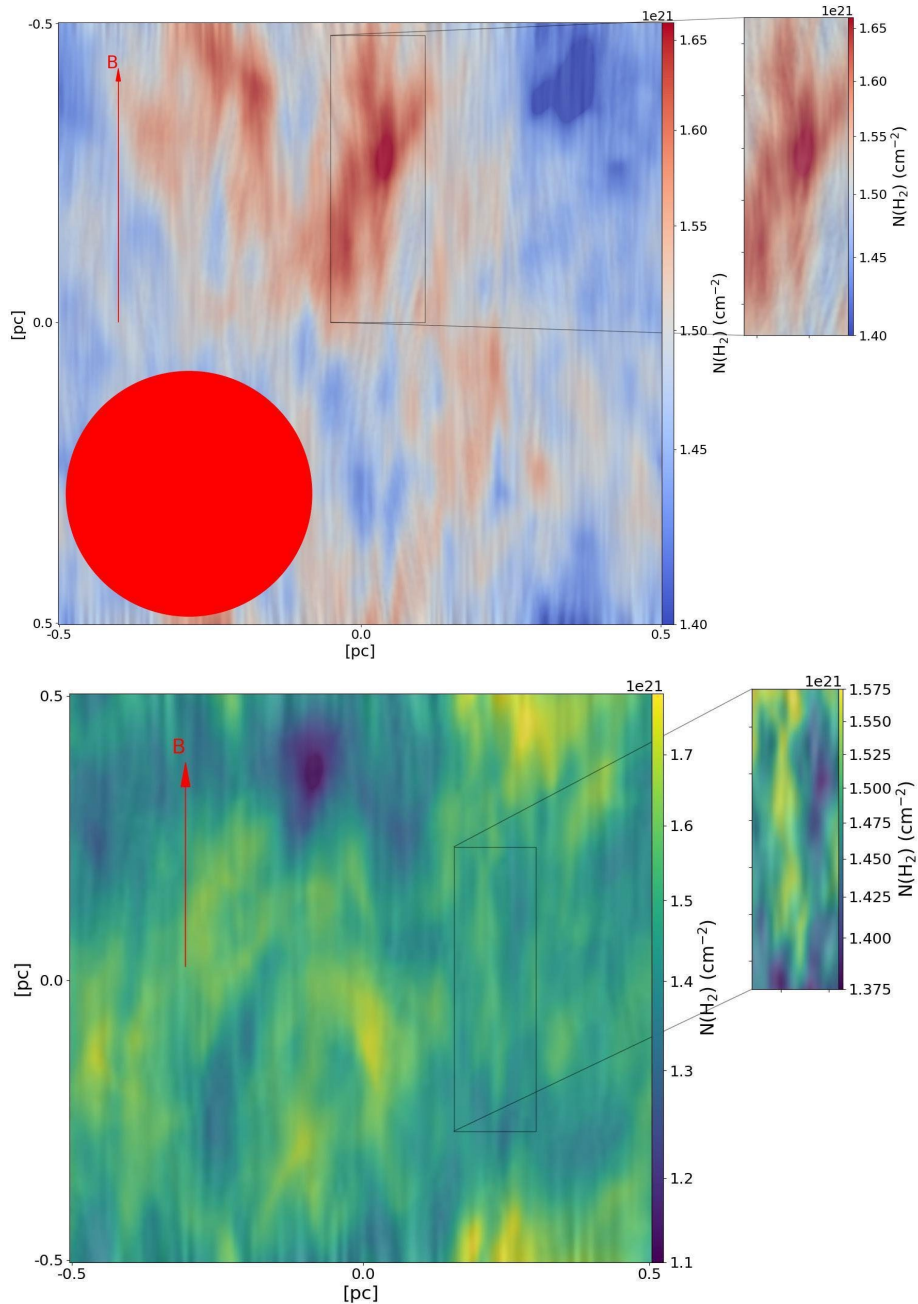

Figure 1: Striations in a 3D MHD simulation of a supersonic (MS=2) but sub-Alfvenic (MA=0.75) cloud (Tritsis & Tassis 2016). The background colormap shows the column density of the simulated cloud and the drapery pattern shows the magnetic field lines, both convolved to PRIMAger's resolution at 235 μm. The polarization map was calculated following King (2018) and convolved to PRIMA's resolution. Top: The zoomed-in panel shows the passage of a small-scale MHD mode ("kink mode") that propagates along the mean field orientation. The high angular resolution of PRIMAger will allow us to see the imprint of these modes in the column density maps. For comparison the Planck beam is shown at the lower left corner. Bottom: The zoomed-in panel shows the passage of a "sausage mode".

We aim to explore the correlation of the column density and polarization maps of striations to search for the unique imprint of the passage of the various wave modes. We propose to target three ISM clouds with prominent striations: Polaris Flare, Taurus, and Musca. The striations of these clouds have been studied extensively in the past and there is already a wealth of archival data, such as emission lines of CO (and CO isotopologues), and optical polarization data, that will allow us to build a comprehensive picture of striations and how they shape the early stages of star formation.

## Studying the physics of the ISM grains

The proposed observational plan will also be beneficial for studying the properties of the ISM grains. The degree of dust-induced polarization scales with the wavelength following the so-called Serkowski relation (Serkowski 1973). This is an empirical relation that provides a unique way for constraining the dust properties, e.g., dust grain size and the alignment efficiency between the magnetic field and the ISM grains associated to dust polarization.

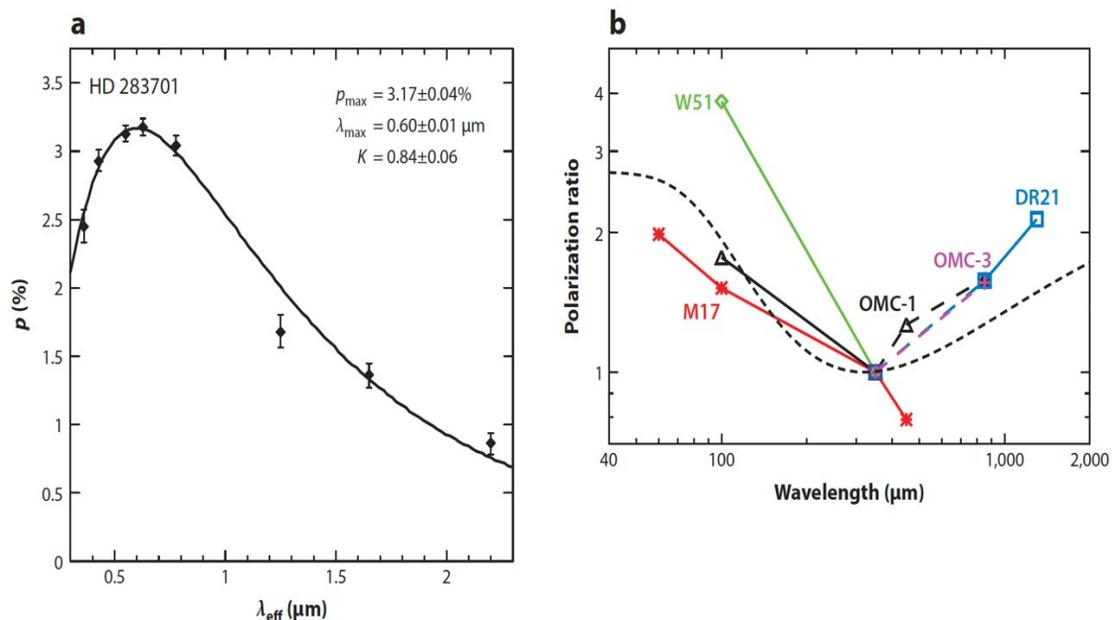

Figure 2: The Serkowski curve (Andersson et al. 2015) at optical, UV wavelengths (left panel), and at far-infrared wavelengths (right panel). The curve seems to have a local minimum, indicative of the ISM dust grain properties, at far-infrared wavelengths. The PRIMAger wavelength coverage will gives us polarization data at wavelengths that have been inaccessible so far.

The observational constraints of the Serkowski relation show that the degree of polarization is maximum at optical wavelengths and seems to decrease non-linearly elsewhere. However, the Serkowski relation observed towards star-forming clouds has a minimum at far-infrared wavelengths, ~300 μm. It was suggested that a two-component dust population model is required to explain this minimum, while the rise of dust fractional polarization at wavelengths larger than 300 μm might be due to variations of the dust grain emissivity, and grain alignment.

*The Serkowski relation is poorly constrained at far-infrared wavelengths because there has been no far-infrared polarimeter with high sensitivity and angular resolution at these wavelengths.* For example, the sensitivity of the Planck satellite is low, and the signal needs to be convolved over

a coarser-than-the-native resolution (at least 10 arcmins) for getting a detection. This approach is problematic when studying the Serkowski relation, given that there is a large beam difference between optical (where a pencil beam applies) and far-infrared wavelengths, and hence the two observations do not probe the same material.

The high angular resolution makes PRIMAger the ideal instrument for minimizing the beam size difference between optical and far-infrared polarization. There is a wealth of archival optical polarization data towards the three target clouds (Polaris Flare, Taurus, Musca) but we also plan to fully map those regions which are included within the Pasiphae (www.pasiphae.science) optopolarimetric survey (Tassis et al. 2018). We aim to combine the stellar polarization data with data that we will get from PRIMAger in order to constrain the Serkowski curve from optical to far-infrared wavelengths. The synergy between optical and far-infrared polarization is largely unexplored – the far-infrared polarimeter on-board the SOFIA observatory has mainly targeted massive clouds where optical polarization is undetectable. We can make a significant contribution with such a synergistic study, which will provide valuable information about the ISM dust grain properties in the diffuse part of the molecular clouds, using the far infrared polarimetric data that we will get towards the striation regions of the Polaris Flare, Taurus, and Musca. Better understanding dust polarization will advance our knowledge of dust grain physics. This will allow improved modeling of grain induced polarization and thus: i) better use of polarization data to determine the role of the magnetic field in critical processes such as cloud, core, and star formation, and ii) more accurate dust polarization foreground subtraction from observations seeking a cosmological B-mode signal in the polarization of the cosmic microwave background.

## Instruments Required:

PRIMAGER: Large Map (> 20'×20') for 3 maps of 1sqdeg each and Polarimeter band (96, 126, 172, 235 μm; R=4).

## Program Size:

- Small (<10 hr)

## Approximate Integration Time:

We estimate that approximately 5h will be required for our program assuming that PRIMAger bands 2A, 2B, 2C and 2D can be observed simultaneously.

## Special Capabilities Needed:

- None

## Description of Observations:

We propose to image the polarized thermal dust emission of three 1 deg$^2$ areas towards the Taurus cloud, the Musca cloud, and the Polaris Flare cloud, using PRIMAger to investigate the nature of striations for understanding the MHD turbulence in molecular clouds. In Fig. 3, we show the proposed target region of each cloud with the green rectangle. The target regions have been selected because they show prominent striation-like (linear) structures. The dynamics of these regions have been studied extensively and bare interesting features, which we summarize below: 1) In the target region towards the Polaris Flare (left panel in Fig. 3), compressible wave modes propagating along the magnetic field lines tend to enhance the formation of CO gas (Skalidis et al. 2023); 2) In the striations of Taurus (middle panel in Fig. 3), the power spectra of the column density and velocity maps are strongly correlated perpendicular to the magnetic field lines, which is unique evidence of the theory that striations are formed by compressible wave modes (Tritsis & Tassis 2016); 3) Musca (right panel in Fig. 3) is an isolated cloud pulsating at the characteristic frequencies of the compressible modes that form the striations (Tritsis & Tassis 2018). All these features are characteristic of the dynamic properties of striations, rendering the proposed target regions ideal testbeds for studying the MHD nature of striations at small scales.

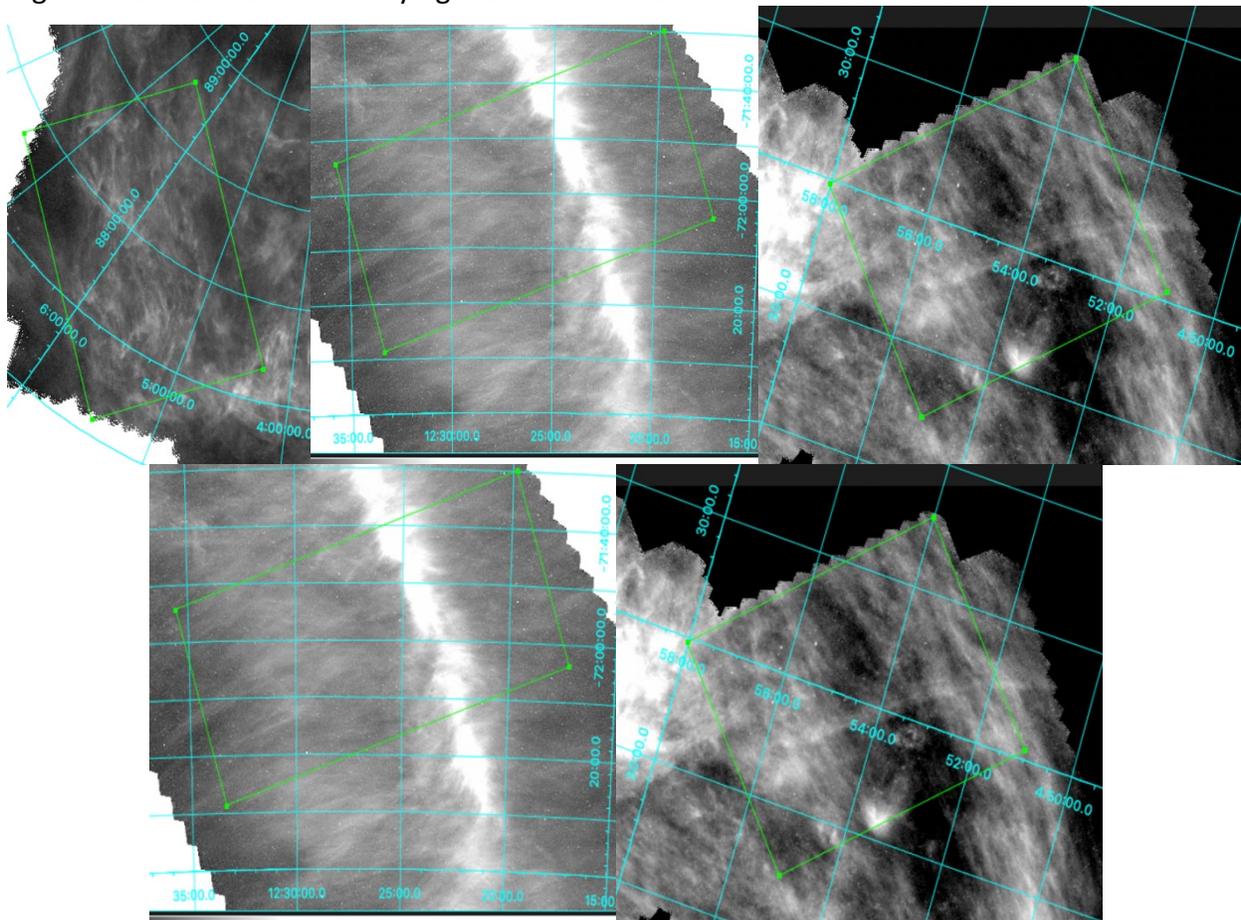

Figure 3: The green rectangles correspond to the proposed target region towards the three clouds with prominent striations. The background maps correspond to the total dust intensity at 350 μm observed with the SPIRE instrument onboard the Herschel observatory. The yellow segments show the orientation of the striated structures, which are aligned with the plane-of-the-sky magnetic field morphology. The center coordinates of each region are shown in the table below. The area of each region is approximately 1 square degree.

## Exposure time estimation:

In the table below, we summarize the coordinates of the proposed target regions. From the Planck data we find that the polarized intensities (P) at 353 GHz towards the three target regions are: 1) Polaris Flare, P = 0.08 MJy/sr, 2) Musca, P = 0.27 MJy/sr, and 3) Taurus, P = 0.20 MJy/sr.

We consider these reference values for estimating the integration times; polarized intensity at far-infrared wavelength might vary with frequency but such variations are below the detection limit of the Planck satellite (Planck Collaboration Int. XXII. 2015) and could only be detected with the high sensitivity of PRIMager. Based on the sensitivities quoted in the fact sheet, for each band a 5σ detection in the polarized intensity towards the three proposed regions would require the integration times shown in the table. If PRIMager could observe simultaneously all four filters of the polarimeter band (Band 2), then the total integration time for the Polaris Flare would be ~3 hr, for Musca 1 hr, and for Taurus 1 hr. This yields a total observing time of ~5 hr. We estimated the dust intensity fluxes using a modified black body with the dust temperature constraints from Planck. For the Polaris Flare, and Musca the mean dust temperature is 15 K, while for Taurus it is 17 K. We assumed that the dust spectral index is β=1.6 and estimated the fluxes towards the proposed regions for each filter of the Band 2 (Table). The target sensitivity of PRIMager will allow us to obtain high signal to noise measurements (>>5σ) in total intensity for the estimated integration times.

| Cloud | Ra | Dec | Integration time for Band 2 | | | | Flux(Stokes I) estimates (MJy/sr) | | | |
|---|---|---|---|---|---|---|---|---|---|---|
| | | | 2A | 2B | 2C | 2D | 96 μm | 126 μm | 172 μm | 235 μm |
| Polaris Flare | 05:46:19 | +87:57:12 | 2.5 hr | 2.5 hr | 1.5 hr | 1 hr | 11 | 33 | 62 | 67 |
| Musca | 12:26:24 | -71:59:48 | 1 hr | 1 hr | <1 hr | <1 hr | 21 | 67 | 124 | 135 |
| Taurus | 04:54:31 | +27:00:44 | 1 hr | 1 hr | <1 hr | <1 hr | 87 | 205 | 300 | 276 |

## Acknowledgement:

A portion of this research was carried out at the Jet Propulsion Laboratory, California Institute of Technology, under a contract with the National Aeronautics and Space Administration (80NM0018D0004).

## 64.  Protoplanetary disk properties over a wide range of star-forming environments

Marion Villenave (Jet Propulsion Laboratory, California Institute of Technology), Karl Stapelfeldt (Jet Propulsion Laboratory, California Institute of Technology), Romane Le Gal (IPAG, IRAM), Youngmin Seo (Jet Propulsion Laboratory, California Institute of Technology), Ke Zhang (University of Wisconsin-Madison)

Determining how protoplanetary disks evolve and the different paths that disk evolution can take is critical to better understand planet formation. Analyzing the spectral energy distribution of protoplanetary disks provides a way to obtain statistical trends about their properties but is currently limited by the number of disks observed at far infrared wavelengths. This program aims to double the number of disks observed at these wavelengths, allowing to obtain disk properties over a wider range of star-forming environments, and to identify the dependence of disk properties with environmental parameters.

### General Scientific Area:

Planet formation

### Science Justification:

Protoplanetary disks are the birthplace of planets, which are thought to form from micron-sized dust particles, initially present in the interstellar medium. Characterizing the dust properties and evolutionary mechanisms of protoplanetary disks is thus essential to understand planet formation. The analysis of spectral energy distribution (SED) of disks has long been one of the most important ways to study their properties. The near and mid infrared emission characterizes the innermost regions of disks, the far-infrared is most sensitive to the disk's vertical structure, while millimeter and centimeter observations help characterize the maximum grain size and dust mass of disks.

While analyzing SEDs is known to be a highly degenerate problem, the recent implementation of Bayesian frameworks coupled to machine learning techniques allowed to obtain parameter uncertainties and degeneracies. Fast and complex modeling can now be performed for samples of tens to hundreds of disks (Ribas et al. 2020, Kaeufer et al. 2023, Rilinger et al. 2023). These studies are however currently limited by the number of systems with complete SEDs from the optical to the millimeter. In particular, the far infrared range, which is most sensitive the disk's vertical structure and critically determines the disk's temperature structure, is missing for many systems, limiting the statistical analysis of disk properties to a subsample of disks (e.g., Ballering & Eisner 2019).

The high sensitivity of the PRIMAger instrument can obtain far infrared photometric observations of all star-forming regions up to 1.5 kpc and to complete the spectral energy distributions for thousands of disks. Such survey would provide access to a wider range of star-forming environments by observing more older regions (such as Upper Sco, 5-10Myr), to trace the evolution of disks, and more high-mass star-forming regions (e.g., Lambda Orionis), to trace different UV environments. This would allow the community to determine disk properties (for example grain sizes, vertical extent, settling efficiency) over a sample more than twice larger what is currently available (see Figure 1), and to study their dependence to various parameters such as the stellar spectral type, age, mass, the binarity of the system, or the disk location within a high mass or low mass star forming-region. While the main goal of this survey would be to study the properties of Class II protoplanetary disks, this program would map entire star-forming regions, such that younger stellar objects (in the Class 0 or I) could also be studied.

In addition to obtaining more complete spectral energy distributions for the protoplanetary disks already identified by previous missions, this survey would discover new candidate disks that were too faint for previous observations. In particular in nearby star-forming regions, faint disks such as disks around brown dwarfs, edge-on disks, or disks transitioning between the protoplanetary disk and the debris disk phase could be discovered and followed up with resolved imaging. Moreover, as this program would target the star-forming regions that are currently missing mid/far-infrared photometry, the combined analysis of this PRIMA survey and Gaia membership or spectroscopy data could identify a large number of new disks. Such large statistical study will address the question of how do disk evolve and assess the different paths for disk evolution.

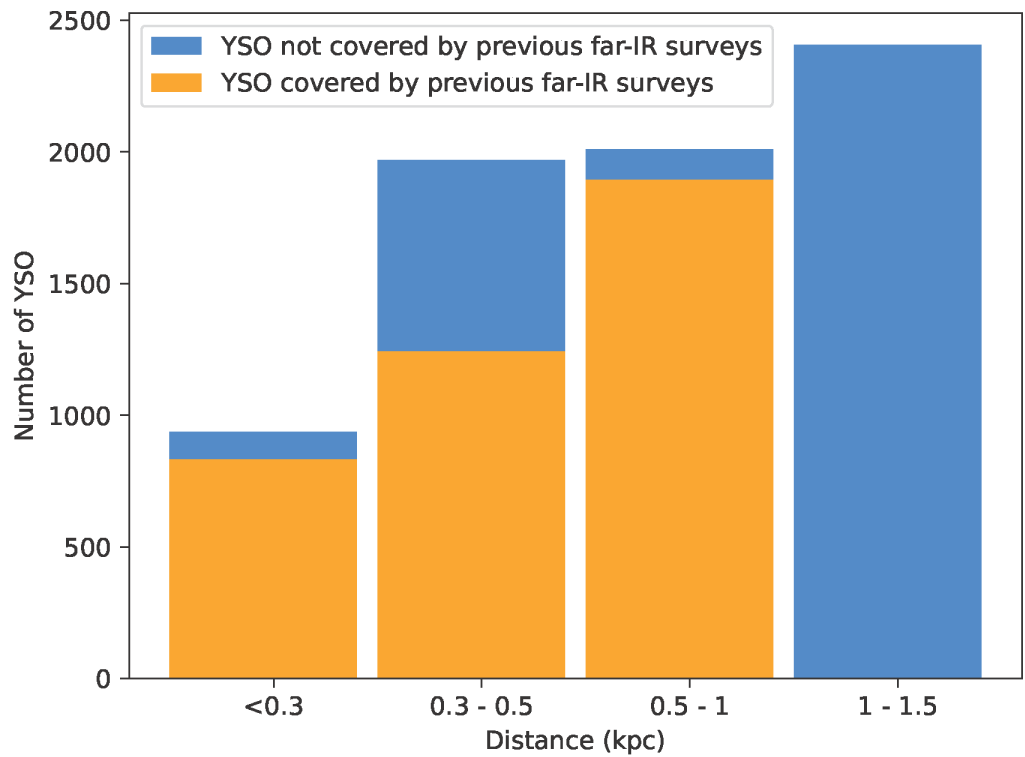

Figure 1: Estimate of the number of young stellar objects (YSO) of all evolutionary stages included in regions previously covered or not covered by far-IR surveys. PRIMA will allow to double the number of YSO with far-IR fluxes and permit to expand statistics on disk properties over a wider range of stellar ages and external environments.

- Previously covered: Perseus, Taurus, Chameleon, Lupus, Ophiuchus, CrA, Aquila, Orion OB1, Lynds 1641, Sigma Orionis, Cepheus, Cassiopeia, Orion A&B, Monoceros R2, NGC 2264, Vela C, Vela D, Serpens

- Partially or not covered: Upper Sco, NGC 1333, IC348, Lambda Orionis, Canis Major, Cygnus, S106, NGC 2362, Lagoon Nebulae

## Instruments Required:

PRIMAger: Large Map (> 20'x20'), Hyperspectral band (25–80 μm; 12 filters; R=10) and Polarimeter band (96, 126, 172, 235 μm; R=4)

## Program Size:

- Large (100 + hr)

## Approximate Integration Time:

This science case aims to complete the far-infrared mapping of star-forming regions within 1.5kpc. We identified 9 regions of interest, covering sky areas between 1 deg$^2$ and 100 deg$^2$. We propose to observe these regions with all PRIMA bands with a sensitivity similar to previous far-infrared surveys, leading to integration times ranging between 2 hours and 200 hours depending on the size of the region.

## Special Capabilities Needed:

- None

## Synergies with Other Facilities:

This survey would discover new candidate disks that were too faint for previous surveys, such as disks around brown dwarfs, edge-on disks, or disks transitioning between the protoplanetary disk and the debris disk phase. It then will be possible to follow up these new disks discovered in nearby star-forming regions, with resolved images with ALMA, NOEMA, JWST, VLT/SPHERE, HST, GPI, or Keck/NIRC2 to directly study their morphologies and properties.

## Description of Observations:

We propose a large continuum survey of star-forming regions aiming to detect thousands of protoplanetary disks in the far-infrared to study disk evolution. This survey would complete the far-infrared mapping of all star-forming regions within 1.5 kpc, distance at which confusion becomes a major difficulty.

Previous far infrared surveys have targeted protoplanetary disks and young stellar objects. In particular, the Gould Belt, HOPS and HOBYS Herschel surveys (André et al. 2010, 2009, Motte et al. 2010, Fisher et al. 2013) obtained observations between 70 and 500 μm of large cloud areas in 23 star forming regions, located up to 3 kpc. These surveys mostly aimed to study very young stellar objects (in the Class 0/I phases) and did not cover some older star-forming regions within 1.5 kpc. To complete the spectral energy distribution of over three thousand young stellar objects, we propose that several PRIMA programs would observe these additional star-forming regions with a similar sensitivity to the Gould Belt survey. Regions of interest are Upper Scorpius

(140 pc), NGC 1333 (300 pc), IC348 (310 pc), Lambda Orionis (400 pc), Canis Major OB1 (CMA OB1, 570 pc–1650 pc), Cygnus (1.5kpc), S106 (1.5 kpc), NGC 2362 (1.5 kpc), Lagoon Nebula (1.5 kpc). These regions cover approximately sky areas ranging between 1 deg² and 100 deg². These regions increase the range of star-forming environments with far-infrared fluxes to both older star-forming regions (e.g., Upper Sco, Lambda Orionis, 5–10 Myr) and more high-mass star forming regions (e.g., Cygnus, Lagoon Nebula), allowing to trace the evolution of disks with age and for different UV environments.

The Gould Belt survey achieved a sensitivity of ∼10 mJy at 250 μm. Given the point source sensitivity of PRIMA, it would require only 30 minutes on a 1 deg² field to reach this sensitivity. To obtain photometry information across all the wavelength range available with PRIMA (25 μm–235 μm), and especially the far-infrared range (bands 2A to 2D), this program will observe each region four times. Each observing sequence will pair observations with one hyperspectral mode and one polarimetric mode (2A-1A, 2B-1A, 2C-1B, 2D-1B). The observing time for such survey would be between 2 hours for the smallest star-forming regions, and up to 200 hours for the most extended star-forming regions. However, careful analysis of existing maps needs to be performed to design the fields of views.

## Acknowledgement:

A portion of this research was carried out at the Jet Propulsion Laboratory, California Institute of Technology, under a contract with the National Aeronautics and Space Administration (80NM0018D0004).

# 65.  Linking magnetic fields from cloud to galactic scales

Thomas G. Williams (Oxford), Simon Coudé (Worcester State University/CfA), Stephen Eales (Cardiff), Eric Koch (Harvard/CfA) Kate Pattle (UCL), Frédéric Poidevin (IAC,ULL), Matthew W. L. Smith (Cardiff)

We propose a survey linking the large-scale magnetic structure of the galactic disc to the small-scale structure around giant molecular clouds (GMCs) at ~50pc resolution in the two most nearby spiral galaxies: M31 (Andromeda) and M33 (the Triangulum). These observations are enabled by the huge increase in PRIMA's mapping speed over previous observatories, and will produce exquisite maps of the cold dust continuum and polarization for these two iconic galaxies. These results will transform our view of magnetic fields at the small scale – for instance, previous results in M33 have shown tentative correlations for the B-field orientation in GMCs preferentially along the spiral arms; these magnetic fields are coherent and linked. There may also be a link between polarization fraction and star formation, but to verify this requires observations that localize individual star forming sites. With the hyperspectral imaging capabilities of PRIMA, we will also be able to study the importance of local processes in setting the balance of dust heating. By studying the entire disc of two nearby galaxies, we can link the magnetic fields across different dynamical structures (e.g., ring, spiral arm, interarm) in two very different systems (one massive, solar metallicity, mostly quiescent and one much less massive, at half solar metallicity). We will produce large polarization and hyperspectral maps, requiring a total of 24 hr observing time. We will produce polarization fraction and magnetic field direction maps in an analogous way to e.g., the SALSA survey on SOFIA, and with the wealth of high-quality data that already exist for these galaxies a large number of studies will be enabled. These observations will increase the number of B-field measurements within GMCs by at least two orders of magnitude, and will provide the definitive resource linking the magnetic fields in clouds with large-scale galactic structure.

## General Scientific Area:

Magnetic fields

## Science Justification:

External galaxies offer an ideal, 'top-down' perspective on galactic structures, without the confusion of Milky Way observations from our own location within it (for the difficulty in mapping GMCs in the Milky Way, see e.g., Zucker et al., 2018). The more nearby a galaxy, the better, as for a fixed-diameter dish this maximizes our physical resolution. For the most nearby, we can resolve the molecular clouds - the sites and fundamental unit of star formation - whilst placing them within the context of the larger galactic environment. We propose a survey of the two closest spiral galaxies, M31 (Andromeda) and M33 (the Triangulum) with hyperspectral imaging and polarimetry. These galaxies offer a wide variety of internal environments (e.g., flocculent and more grand-design spiral arms, rings), and are very different systems, with M33 at half-solar metallicity and M31 more similar to MW metallicity. At similar distances (~800 kpc), PRIMA will be able to observe them with 10s of parsec resolution, resolving the molecular clouds and offering us a new perspective on the cold dust continuum and, for the first time, large-scale mapping of the dust polarization across these extended galactic discs. These are the only two galaxies where we will be able to link the dust polarization in molecular clouds to the large-scale B-fields in spiral galaxies.

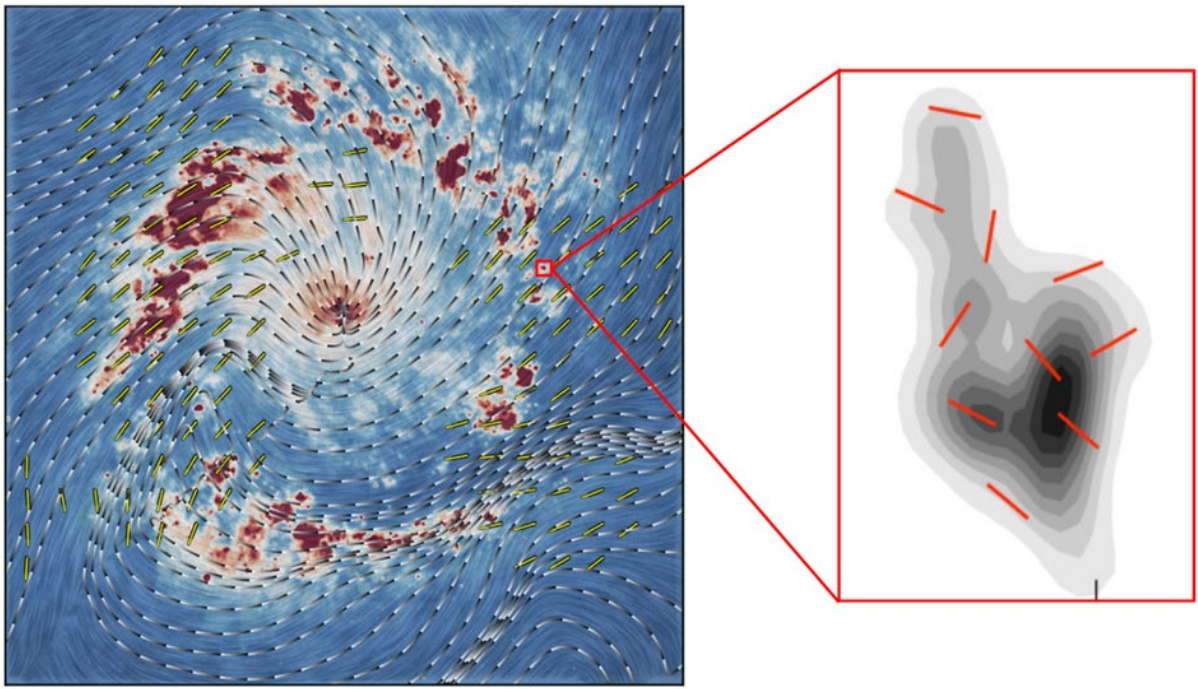

Figure 1: Left: The large-scale magnetic field around NGC1097 (Lopez-Rodriguez et al., 2021). Right: The small-scale magnetic field within a single GMC (Li & Henning, 2011). Our proposed survey of M31 and M33 will fully map these galaxies (left) at the resolution of individual clouds (right).

A survey of these two galaxies will offer huge legacy value, superseding existing Herschel maps with much deeper dust continuum, as well as adding more datapoints around the peak of the dust spectral energy distribution (SED). Combining this with Herschel and existing SCUBA-2 imaging at 450 and 850 µm will provide the best sampling of any galaxies to date, and with advanced image processing techniques such as PPMAP (Marsh, Whitworth & Lomax, 2015) we will be well-placed to study the variation in dust properties across two very different spiral galaxies. For instance, these data will allow for studies of the local variation in the hot-cold dust balance, linking to sources of ionizing radiation such as HII regions, where we may expect elevated dust temperatures relative to the diffuse background, and, motivating the need for a large-area survey, differences between different HII region morphologies (e.g., shells and filled regions, Relaño et al. 2016)

The game-changer in using PRIMA, however, will be the far-infrared polarimetry. The sum total of polarized GMC measurements in external galaxies is six (Li & Henning, 2011). These new observations will map hundreds of GMCs in both of our galaxies, opening up a whole new parameter space to explore. The results from Li & Henning (2011) showed a preferential alignment of the GMCs with the spiral arms, indicating that the spiral arms are intimately linked to GMC formation. However, this sample of six can only draw preliminary conclusions, and is not robust or representative. By surveying the entire disc of two galaxies, we will measure B-fields for hundreds of clouds, and we can test both inter- and intra-galaxy variation - how do the B-fields look for clouds outside of spiral arms, for example, and between the more flocculent arms of M33 and better-defined arms of M31? These observations will clearly distinguish between different cloud formation models, and inform better the links between spiral arms and cloud formation. Such measurements will be compared with magnetic field structures derived from synchrotron polarized measurements probed at radio wavelengths in the hot diffuse ISM of M31 (Fletcher et al. 2004) and of M33 (Tabatabaei et al. 2008), hence allowing to study the role of the magnetic fields in different phases of the ISM.

PRIMA is the only instrument that can map the large angular areas these galaxies subtend on the sky in a reasonable time. The only previous study of B-fields in extragalactic clouds is that of Li and Henning (2011), a sample of six clouds that each required ~3 hours of observing time. We plan our observations to detect every cloud in M31 and M33, an increase of at least two orders of magnitude in the number of clouds, and will enable detailed study of our science questions. Ground-based sub-mm and mm observatories require prohibitive integration times given the required sensitivities. As such, PRIMA is a unique instrument, the only one capable of carrying out these large sky-area surveys, in order to answer these important and outstanding questions about the interstellar medium.

For imaging, we will construct dust spectral energy distributions which are well-explored within the literature (e.g., Smith et al., 2012; Lamperti et al., 2019; Williams et al., 2019a), or more complex radiative transfer models, linking dust properties to the stellar components heating them (e.g., Viaene et al., 2017; Williams et al. 2019b). For the polarimetry, we will construct polarization fraction and orientation maps using tested methods from e.g., the SALSA large survey on SOFIA (e.g., Borlaff et al., 2021), and measure orientations using techniques such as the Histogram of Relative Orientations (HRO; Soler et al., 2013). Dust-selected cloud catalogues exist for both galaxies (Kirk et al., 2015; Williams et al., 2019a), and we will use these to distinguish clouds in the data. The resolution of our observations will allow for a cross-spatial analysis, to link the lower resolution observations in the literature to the high-resolution observations we propose here.

These observations will be a legacy dataset that will provide new insights into the cold and warm dust in the nearby Universe. As such, these are a non-exhaustive list of the possible studies that are only enabled by this dataset:

- Local variations in the cold and warm dust balance in galaxies:

    o How do dust temperatures vary between different types of HII region?

    o How does the dust temperature vary between the dense and diffuse ISM?

  o How do stellar populations heat the surrounding dust, and how homogeneous is this?

● Links between large and small-scale magnetic fields

  o Do GMC B-fields align with spiral arm B-fields?

  o Are the orientations consistent between GMCs in different environments?

  o Do the B-fields probed from polarized dust emission in cold or neutral phases of the ISM align with the B-fields probed from the polarized synchrotron emission in hot plasma regions?

● Linking GMC B-fields to star formation

  o Do magnetic fields enhance or suppress star formation?

  o Are there variations within the GMC population?

## Instruments Required:

PRIMAger large map, in both hyperspectral and polarimeter Band (polarimetry is the main focus)

## Program Size:

● Medium (10–100 hr)

## Approximate Integration Time:

● 24hr

## Special Capabilities Needed:

● None

## Synergies with Other Facilities:

M31 and M33 are two iconic, well-studied galaxies. They have been observed with almost every facility that can target them. In particular, we will have access to all phases of the gas via new ALMA observations (M33) for the cold, molecular gas, and 21cm VLA observations (both galaxies) as part of the Local Group L-Band Survey (LGLBS). With robust star formation rate measures (from a variety of sources, see e.g., Viaene et al., 2014 and Williams et al., 2018) we will be able to study the complex links between magnetic fields, gas, and star formation, to ascertain whether the magnetic fields enhance or suppress star formation. In addition, deep Herschel and JCMT SCUBA-2 observations well-cover the entire cold dust continuum, allowing for cloud-scale measurements of dust mass and properties. We also have HST from the PHAT/PHATTER surveys, and JWST observations for a portion of M33, which will provide the exposed and embedded stars, respectively.

## Description of Observations:

We set a sensitivity requirement to detect a 1% polarization signal at 5$\sigma$ from every GMC in M31 and M33 (M>10$^4$ Msol). PRIMAger's polarimetry capabilities will provide a census of >500 GMCs in each galaxy complete to an Orion-like molecular cloud (Lada et al. 2010), and spanning all star forming stages across a variety of galactic environments. To achieve this goal, we require a 5σ point-source sensitivity of 900 μJy in the 238 μm band. In an equivalent observing time, the other 3 bands will be ~2 times more sensitive. For the shorter wavelength hyperspectral bands, we will have sensitivity of ~60 μJy (1A) and ~150uJy at worst (1B), i.e., close to the confusion limit and significantly deeper than previous images at similar wavelengths.

Our sensitivity requirement is calculated assuming a dust temperature of 20 K, β=2, and gas-to-dust ratio of 80 based on previous M31 and M33 dust studies (Kirk et al. 2015; Williams et al. 2019a). We assume a dust opacity following Chiang et al. (2018).

M31 and M33 cover a large area and will require a large mapping mode. We estimate the mapping areas based on the complete dust-detected regions in archival Herschel SPIRE-250 maps. To calculate the mapping time, we have assumed the total mapping time scales linearly with area. **PRIMAger can map M33 (0.8 deg²) in 5 hr and M31 (2.8 deg²) in 17 hr.** To account for polarization of the Galactic foreground, we can moderately expand the observed fields to enable a direct comparison with the Planck polarization measurements, adding approximately 2 hr to the total time. Accounting for this, we estimate that PRIMAger can map both galaxies in just 24 hr of observations.

# 66. The PRIMA All-Sky Survey

Jason Wright (Penn State), Randolf Klein (SOFIA/USRA), Margaret Meixner (Jet Propulsion Laboratory, California Institute of Technology), Kate Pattle (UCL), Marc Sauvage (CEA), Stanislav (George) Djorgovski (Caltech)

The long-term legacy of PRIMA will be greatly enhanced with a FIR all-sky survey. The state-of the art for FIR fluxes for most sources on the sky are due to IRAS and AKARI, which achieved similar point source sensitivities. These fluxes (and upper limits) remain one of IRAS's primary legacies and enable science far beyond its stated science case. PRIMA's legacy will similarly benefit, and perhaps be defined, by the foundational science provided by an all-sky survey that improves by two orders of magnitude on the earlier IRAS and AKARI surveys.

A PRIMA FIR all-sky survey as part of its primary mission would:

1. Enable new GO science by discovering new sources for follow-up investigations, enhancing PRIMA's pointed science.

2. Make other GO science more efficient by providing flux anchor points for previously undetected sources, allowing optimal exposure time selection, and preventing wasted time on non-detections.

3. Enable novel and unanticipated science for decades into the future.

4. Provide a "deep field" of the Ecliptic Poles, which overlaps with similar fields in many other bands.

Strategically, an all-sky map is uniquely enabled by PRIMA's fast mapping capabilities, and would provide a strong science case that distinguishes PRIMA from competing FIR probe concepts.

## General Scientific Area:

Multiple areas including Galaxy Ecosystems and Interstellar Medium, Star Formation in the Milky Way, Magnetic Fields, Milky Way Interstellar Medium, Planet Formation.

## Science Justification:

### Broader Context

The IRAS Point Source Catalog (PSC) is now almost 40 years old, but still highly referenced and a reliable source for all-sky far infrared point source fluxes. The IRAS satellite provided a spatial resolution of about 1' and the PSC reaches a sensitivity of about 1 Jy (Beichman et al 1988). In 2010, AKARI provided a mid- and far- infrared all-sky point source catalog with a factor of a few (at best) sensitivity improvement and slightly higher spatial resolution than IRAS. But the AKARI catalogs still contain only 1.7 and 3.5 times more sources than IRAS at 60 and 100 μm, hardly a transformational improvement.

## The Need for PRIMA

It is high time for a new all-sky infrared survey and catalogs. PRIMA provides a once-in-a-generation opportunity to provide such a survey, being the only proposed FIR probe concept that has the high mapping speed required to contemplate an all-sky map. Not only is PRIMA the natural vehicle for this, but it may be the last chance to do this transformational science for decades to come.

We are proposing that PRIMA provide an all-sky survey that will supersede the IRAS PSC catalog by providing three times better spatial resolution and 100 times better sensitivity. The resulting all-sky images and catalogs will become the new reference sources for infrared fluxes and will support the research of a wide range of astronomers.

To estimate how many sources a PRIMA survey may detect with good precision (we chose $50\sigma$ = 2% photometry), we extrapolated the source counts in the AKARI Far Infrared Bright source catalog (using only sources with FQUAL=3) down to 10 mJy. This estimate is also conservative because PRIMA will offer better spatial resolution than AKARI. Thus, we can reasonably expect to find more than 10 to 100 million sources, which is 100 times more than the IRAS PSC.

## Science Addressed

If done early in the mission, the all-sky map will support all PRIMA GO science in two ways. First, it will detect many new sources for GO programs to consider for PRIMA. Second, by providing flux anchors for new and known sources in the FIR, the PRIMA all-sky survey will allow accurate exposure times to be calculated. This will prevent under- and over-exposure of targets, optimizing GO time requests and science return. It will thus at least partially "pay for itself."

A PRIMA all-sky survey will also allow for a huge number of survey and demographic studies to be performed concurrently with other PRIMA GO science. This will put pointed GO targets in context and raise new questions that can be addressed with currently unanticipated PRIMA science.

The all-sky map itself will be a lasting legacy of PRIMA across astronomy in ways difficult to enumerate or estimate. The IRAS Faint Source, Point Source, and nearby galaxy catalog papers have over 1,000 citations, and those papers have over 50,000 citations. Even this is an underestimate of the IRAS legacy, in that IRAS photometry is so foundational to astronomy that many authors do not even bother to cite it when using the fluxes. An all-sky survey is also important because we cannot anticipate which sources the astronomers of tomorrow will need fluxes for. PRIMA will set SED points or upper limits for every persistent source in the sky, which will be used for decades to come, much like the successful MIR WISE all sky survey.

Just as IRAS revolutionized our understanding of the obscured star formation and AGN, by going 100 times deeper and generating at least 100 times as many sources as IRAS, the PRIMA all sky survey may reveal previously unanticipated populations of obscured sources, in addition to expanding the numbers of the known ones, by going deeper down the luminosity function and out to higher redshifts.

The all-sky survey would surpass the Spitzer/Herschel 70 μm sample size of debris disk stars by roughly a factor of 10, and allow better investigation of disk luminosity & temperatures as a function of age, host star type, host star metallicity, environment, binarity, presence of exoplanets, etc.

PRIMA's optimal scan strategy will also revisit the Ecliptic Poles every rotation, providing a "deep field" there, overlapping with the JWST Time Domain Field, the Hubble Deep Fields, and deep stares by Planck, eROSITA, and WISE, enhancing the science of those efforts.

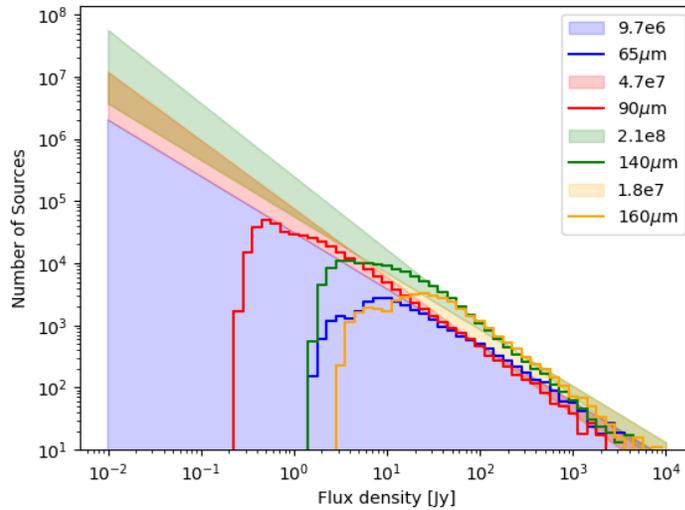

Figure 1: Figure 1:AKARI FIS source counts extrapolated to 10 mJy.

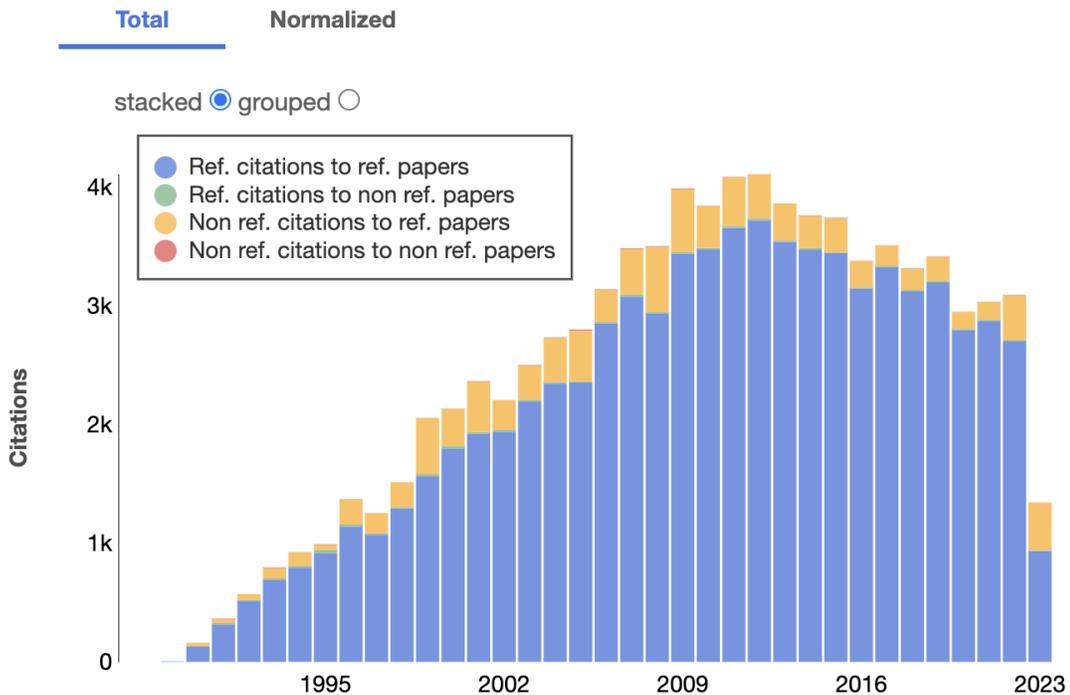

Figure 2: Citations to papers citing four of the IRAS All Sky Catalogs, from ADS

## Instruments Required:

PRIMAger: Large Map using polarimeter bands (All-sky survey broken up into 10˚×10˚ tiles scanned at 240"/s).

## Program Size:

- Large (100 + hr)

## Approximate Integration Time:

5000 hours based on mapping 10˚×10˚ maps at 240"/s in 11.8h. With this mapping speed the following sensitivities (5σ) can be reached based on estimates by Marc Sauvage.

| Wavelength (µm) | 92 | 126 | 172 | 235 |
|---|---|---|---|---|
| 5σ Sensitivity (mJy) | 0.84 | 1.20 | 1.63 | 2.30 |
| 5σ Sen. for pol. (mJy) | 1.18 | 1.69 | 2.31 | 3.25 |

## Special Capabilities Needed:

Scanning at 240"/s and tilt in two axes up to ±5° to cover a 10˚×10˚ field in one go.

## Synergies with Other Facilities:

Innumerable, but especially the ROSAT, GALEX, Gaia, 2MASS, SphereX, WISE, and VLA all sky surveys. Provides a key FIR SED point for tens of millions of sources, and weak detections or upper limits for hundreds of millions more.

## Description of Observations:

The most natural strategy is for PRIMA to spin at its maximum survey rate between the Ecliptic Poles. Over the course of a year, this would enable PRIMA to survey the entire sky. Pointing constraints may require the survey to be extended into additional years to complete the survey.

The optimal band for the map is 92 µm, because it can be cross-calibrated most easily with the IRAS 100 µm maps, because it provides the best sensitivity, and because it will not be confusion limited in most fields at this depth. However, other PRIMAger bands will be observed simultaneously.

## Acknowledgement:

A portion of this research was carried out at the Jet Propulsion Laboratory, California Institute of Technology, under a contract with the National Aeronautics and Space Administration (80NM0018D0004).

## 67. Feedback from high-mass stars: Impact on young star formation properties

Zavagno, Annie (LAM, Marseille, France), Russeil, Delphine (LAM, France), Battersby, Cara (University of Connecticut, USA), Di Francesco, James (University of Victoria, Canada), Schneider, Nicola (University of Cologne, Germany), André, Philippe (CEA-Saclay, France), Arzoumanian, Doris (NAO, Japan), Zhang, Siju (PKU, China), Suin, Paolo (LAM, France), Figueira, Miguel (NCNR, Poland), Loren Anderson (WVU Astrophysics, USA), Michael Mattern (CEA, Saclay, France), Manash Samal (PRL, India)

We propose to use PRIMA to study the impacts of high-mass star early feedback (photoionization, wind) on star formation properties in our Galaxy. We propose to map the surrounding of a series of Galactic ionized (HII) regions in different evolution stages to characterize the star formation ongoing in their associated photo-dissociation regions (PDRs). The high-sensitivity data in the far infrared obtained with PRIMA will allow to sample down to low mass the star formation activity. Combined with multi wavelength data and state-of-the art high-resolution numerical simulations, the aim of this program is to study the impacts of early feedback from high-mass stars on star formation properties and star formation laws (SFR, SFE versus the gas surface density, $\Sigma_{gas}$).

### General Scientific Area:

Star Formation in the Milky Way

### Science Justification:

Star formation occurs in dense and cold molecular clouds in galaxies. The far-infrared domain is particularly well suited to study the earliest stages of star formation.

Massive O and B stars ($M_{star} > 8\ M_\odot$) have a profound impact on their surrounding through their radiative feedback. Recent JWST-MIRI images of nearby galaxies, like the Phantom galaxy (NGC628), show how important this impact can be in shaping the surrounding molecular medium (Kim+2023). However, little is known about the physics of this feedback. In particular, the impact this feedback might have on future formed star properties is highly debated. Is this impact constructive (favoring the formation of new stars) or destructive (dispersing the gas and halting further star formation)? Numerical simulations tend to conclude that this feedback is destructive (Walch+2013, Geen+2015) while observations indicate the opposite, even favoring the formation of a new generation of high-mass stars (Deharveng+2010).

In many Galactic ionized (HII) regions, the photodissociation region (PDR) that surrounds the HII region is observed as a dense layer of gas and dust where new stars form, including high-mass stars. The supersonic expansion of the ionized gas in the surrounding medium creates a shock that travels in front of the ionization front, allowing the formation of a dense layer between the two fronts. Under its own gravity this layer becomes unstable and fragments, forming dense clumps located in the PDR where a new generation of stars form.

Star formation occurs in dense clumps (of 0.1 pc in size) that fragment into cores (of about 0.01 pc in size). The clumps and their associated cores are hosted by filaments. The star formation process that starts with the assembly of the interstellar medium into filamentary structures up to the formation of cores is still highly debated.

The way the star formation process is impacted by the physical conditions (level of turbulence, magnetic field) of its original medium also remains puzzling, even if recent works show that these conditions clearly impact the properties of the future star formation (Zhang+2020, 2021). In particular, the compression exerted on the surrounding medium by an expanding HII region increases the local density and could favor the formation of a new generation of stars (Zavagno+2020). This phenomenon is observed in many Galactic (Palmerim+2017) and extragalactic (Bernard+2016) star forming regions. In this process, young stars are observed at the edges of the HII regions, towards their associated dense PDR, in filamentary structures that compose it.

We propose to use PRIMA and its suite of instruments to study the feedback from high-mass stars on star formation in our Galaxy. The questions we want to address are:

- How does the compression from the incoming stellar radiation (ionizing star that form the HII region) influence the properties of filaments and clumps in the PDR around HII regions?
- How different are the properties of filaments and clumps in directly irradiated (facing the incoming radiation) and less-irradiated zones of the PDR?
- How the evolution of the physical conditions (density, temperature, turbulence, magnetic field) impact the evolution of star formation in the PDR?

We will compare the results of PRIMA observations with state-of-the art high resolution magneto-hydrodynamical simulations of star formation (Verliat+2022) and dedicated PDR models (Tiwari+2022).

### Instruments Required:

- PRIMA FIRESS: Small Map Low-res, R~130
- PRIMAger: Large Map (1 square degree)
- Polarimeter band

### Program Size:

- Medium (10–100 hr)

### Approximate Integration Time:

The regions we plan to observe are located in the Galactic plane and are bright in the infrared (from 30 to 2000 MJy/sr at 8 μm with Spitzer and from 500 to 3500 at 250 μm with Herschel-SPIRE).

We plan to observe with PRIMAger several Galactic HII regions to well-sample the ionizing stars properties, the evolution stage and the HII region's geometry (spherical and bipolar regions).

We will do maps of maximum 1 square degree to well sample the surrounding interstellar medium impacted by the HII regions. Polarimetric observations will be obtained to characterize the role of magnetic field combined with the feedback towards the PDRs.

We will also use the PRIMA FIRESS Spectrometer to characterize the physical conditions towards different sites of star formation, from quiescent clumps to the densest and most active regions of star formation.

We plan to use PRIMAger to map 8 (1 square degree) regions of the Galactic plane selected to contain well-studied HII regions that sample different physical conditions (column density, temperature, level of turbulence and magnetic field) in their surroundings, known from previous studies obtained by the proposing team from near to far infrared, including Herschel (70–670 µm) imaging (see Figure 1) and spectroscopic data (see Figure 2). We will reach a surface brightness sensitivity (total power I) in the polarimetry bands between 210 and 60 kJy/sr (5σ, 10 hr, 1 sq deg) and the surface brightness sensitivity in P will be between 210 and 80 kJy/sr. These sensitivities will allow us to detect the lowest mass/most embedded young stellar objects present in the field to obtain an unbiased census of the young stellar content for an estimate of the Star Formation Rate and Star Formation Efficiency. The PRIMA measurements will allow to study both the continuum emission and the point sources (mainly young stellar objects in different evolutionary stages) down to low mass sources (0.1 M$_\odot$). The high sensitivity of PRIMA will be key to study the Star Formation Rate in the regions by avoiding the strong effect of the sensitivity bias recently discussed by Suin+2023 using high-resolution numerical simulations and comparing with estimates from Spitzer and Herschel observations where the sensitivity limits clearly lead to an underestimate of the SFR. The imaging part of the proposal leads to a time estimate of 80 hours.

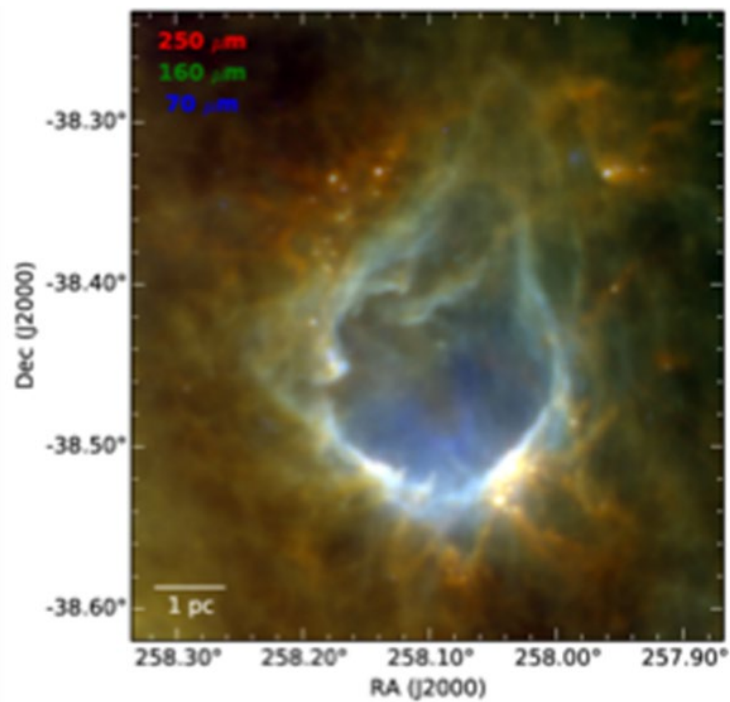

Figure 1: Herschel PACS and SPIRE images of the Galactic HII region RCW 120. Herschel-PACS 70 µm (blue), 160 µm (green) and Herschel-SPIRE 250 µm (red). The field size is 21.8' × 24.5'. North is up, east is left. (from Figueira+2017)

To better characterize the physical conditions in the observed regions, we propose to complete the imaging part of the program by obtaining spectra with FIRESS in the low-resolution mode for 20 positions. These positions will be selected to sample the variety of conditions observed in the regions. We are particularly interested in looking at evolution in star formation properties, from low to high density regions, affected (and not) by ionization feedback. The FIRESS spectra will nicely complement the Herschel PACS and FTS-SPIRE spectra obtained (Rodon+2015, Zavagno priv. comm, see Figure 2). The regions we observed with Herschel exhibit bright lines and continuum (both in the PACS and SPIRE range, see Figure 2), and using the current estimate of 2x10-19 W/m2 (5σ, 1 hour sensitivity), we plan to spend 30 minutes (at maximum) on each position to obtain the FIRESS spectrum. The spectroscopic part of the proposal represents 10 hours of observing time.

The total amount of time required for this program, including imaging (80 hours) and spectroscopy (10 hours), is 90 hours.

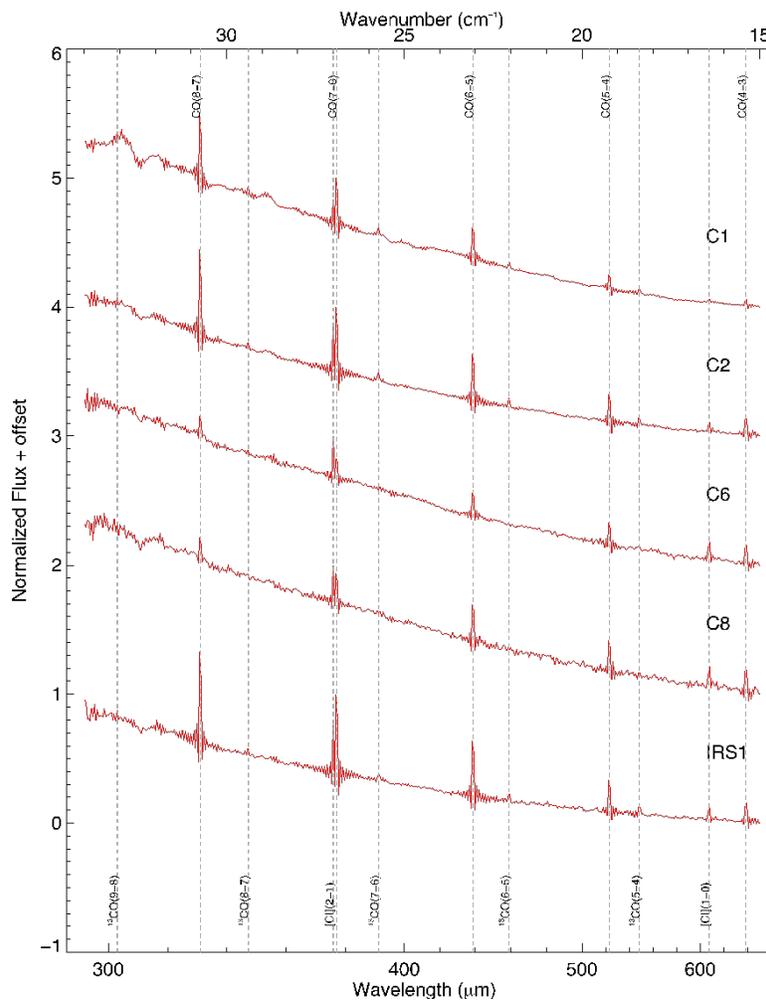

Figure 2: Herschel SPIRE-FTS spectra obtained at different locations of star formation towards the Galactic HII region RCW 120. (from Rodon+2015)

## Special Capabilities Needed:

- None

## Synergies with Other Facilities:

- ALMA
- JWST
- Existing IR data (from near to far-infrared) and submillimeter data (continuum and spectroscopy)

The idea is to lead a multi wavelength study and compare the results with high-resolution MHD numerical simulations (Verliat +2022, Suin+2023).

## Description of Observations:

Capitalizing on the results obtained in this research field with the Spitzer, AKARI and Herschel Space Observatories, we propose to benefit from the unprecedented PRIMA sensitivity in the far-infrared to characterize the physics of feedback from high-mass stars. We propose to map a sample of carefully selected Galactic star forming regions that contains HII regions at different evolution stage (from compact to extended) and with different geometries (spherical, bipolar, hub systems). The regions are bright (from 30 to 2000 MJy/sr at 8 μm) but we want to access the faintest population of young stars and characterize their distribution and properties at the early stages of their birth. Because we are interested in understanding the impact of the physical conditions set by the feedback (higher temperature, higher density, higher level of turbulence) on star formation, we plan to use both continuum and hyperspectral imaging. The polarimetric imaging will be used to study the impact of the magnetic field on the dust and young stars observed in the PDR.

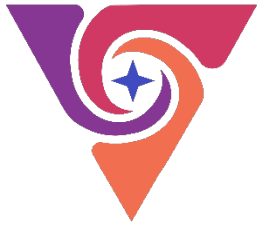

# Stars, the Sun and Stellar Populations

# 68. Determination of forsterite grain temperatures in circumstellar environments using narrow emission bands at 49 µm and 69 µm

Janet E. Bowey (Cardiff University)

Infrared emission bands due to forsterite (the regularly identified crystalline silicate in the mid- to far-infrared) are observed in circumstellar regions around evolved stars, young stellar objects, protoplanetary discs and debris discs in the Milky-Way and beyond. Systematic changes in the peak wavelengths and widths of forsterite bands near 49 and 69 µm with temperature can be used to estimate grain temperatures from emission features in circumstellar environments.

Grain temperatures of the optically thin component are indicative of their location relative to optically thick emission from dust in a circumstellar disc, envelope or outflow. This information is used to constrain the system morphology, to aid the interpretation of images and could constrain chemical studies. Time-series observations in rapidly evolving environments including SN, AGB stars and eruptive variables might also prove instructive. PRIMA will be the first observatory able to provide data for the 49-µm peak and will detect the 69-µm peak in more environments than was possible with Herschel or ISO.

## General Scientific Area:

Milky Way Stars and Stellar Evolution

## Science Justification:

### Context

Infrared emission bands due to forsterite (the most regularly identified crystalline silicate in the mid- to far-infrared) are observed in circumstellar regions around evolved stars, young stellar objects, protoplanetary discs and debris discs in the Milky-Way and beyond. Laboratory data by Bowey et al. (2001) showed bands near 49 and 69 µm in the spectrum of crystalline forsterite, $Mg_2SiO_4$, sharpened and shifted to shorter wavelengths when the temperature was reduced from 295 K to 4K. Systematic changes in the peak wavelengths and widths of features in the laboratory data with temperature (Fig. 1) have been utilized to estimate the grain temperatures of forsterite grains responsible for optically thin 69 µm emission in circumstellar environments (Bowey et al. 2002, Sturm et al. 2013, Blommaert et al 2014). PRIMA will be the first observatory able to provide data for the 49-µm peak.

These temperature shifts constrain radiative transfer models (e.g., Maaskant et al. 2015) because dust temperatures are usually estimated from the broader continuum or the ratios of the 23 µm and 69 µm forsterite band strengths (I23/I69 ratio). Grain temperatures of the optically thin component are indicative of their location relative to optically thick emission from dust in a circumstellar disc, envelope or outflow. This information is used to constrain the system morphology and may aid the interpretation of images. For example, Maaskant et al. found that forsterite in flat disks surrounding Herbig Ae/Be stars is located in the inner few AU and that the detection rate for the 69 μm feature is higher for objects with lower millimetre luminosity. This may indicate that as disks evolve toward lower masses, optical-depth effects or increased production of forsterite or transportation of forsterite into cold regions of the outer disk enhance the strength of the 69 μm feature. Temperature data can also utilized in chemical studies of grain condensation from the gas-phase and dust evolution. Time-series observations in rapidly evolving environments including SN, AGB stars and eruptive variables might also prove instructive because grain temperatures will vary as the dust is transported from hotter to colder regions, or vice versa.

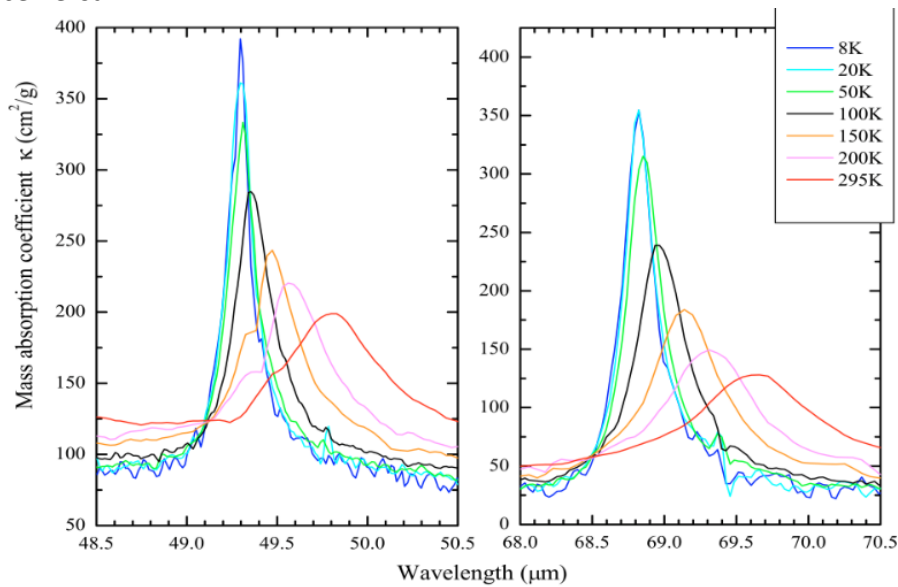

Figure 1: Effect of temperature on the 49 and 69 μm bands of forsterite (figure from Koike et al. 2006).

## Science Questions:

- How frequently are the features observed?

- Where are they most common?

- Is there any correlation with the orientation of the source?

- Is there any correlation with its location – metal-rich environment, metal-poor environment in the MW or beyond.

- What is the temperature of the forsterite dust?

- Are there several optically thin temperature components (indicated by broadened bands)?

- Are temperatures deduced from the 49- and 69-μm bands in each source the same?

## Need for PRIMA:

PRIMA is required to conduct the study because these wavelengths are inaccessible from ground-based observatories. Infrared features at shorter wavelengths exhibit smaller temperature changes which cannot be disentangled from the effect of grain composition or other molecular bands within circumstellar envelopes. At the lowest temperatures the 49- and 69-μm bands are 0.15 and 0.3μm wide, respectively indicating they could be detected at R~770, and R~230 but these limit the accuracy of width measurements and confusion with gas-phase emission. Hence, we require the high-resolution (R~4400 at 112μm) spectrometer.

The other constraint is the contrast between the continuum flux and the optically thin emission because source lists would be tailored to match the overall flux sensitivity. Detected PACs/Herschel (e.g., the PNe in Fig. 2) and LWS/ISO peak strengths were typically 3% to 8% of the continuum with outliers at 1% and 29% so we request a contrast sensitivity of 1%. Blommaert et al. detected the band in 8/15 oxygen-rich AGB OH/IR stars, 11/18 post-AGB and low-mass evolved stars and 3/8 massive evolved stars. Maaskant et al. (2015) found the band in 6/23 Herbig YSOs.

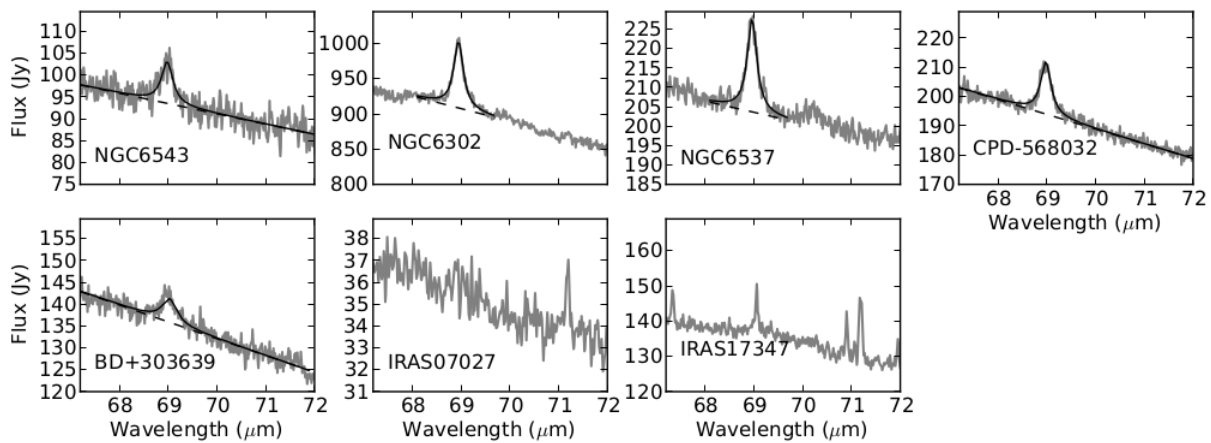

Figure 2: Spectra of PNe from Blommaert et al. 2014. Detected 69 μm features are fitted with Lorentzian profiles. IRAS 07027 and IRAS 17347 are OH maser sources which contain no forsterite band at 69 μm but might contain gas lines.

## Interpretation Methods

The peaks are fitted with Lorentzian profiles superimposed with low order polynomials to represent the continua. Results are then compared with those for the laboratory data (see Fig. 3, reproduced from Blommaert et al.).

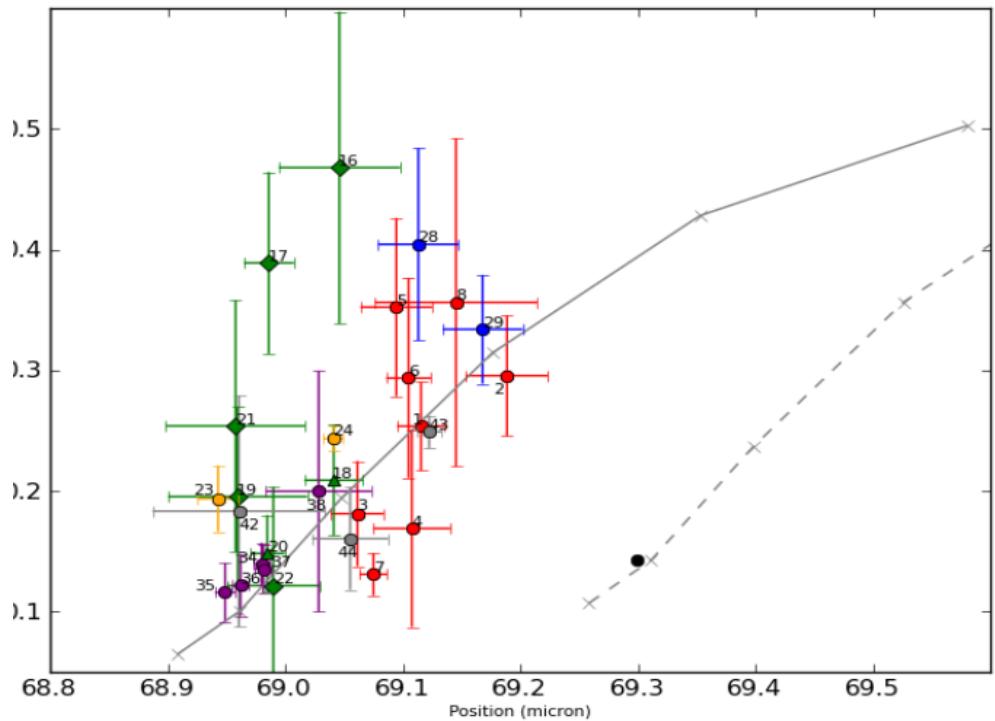

Figure 3: Position and width of all detected 69 um bands. In red are the OH/IR stars, in green the post-AGB stars, in plus the post_AGB stars with circumbinary disks, in yellow two other likely disk sources, in purple the PNs, and in grey the massive evolved stars. In contrast to the 69 um bands of the evolved stars, the position and width of the 69 μm band of the disk around the young main-sequence $\beta$ Pictoris is shown in black. In solid and in dashed gray are the width and position of the 69 μm bands using the laboratory measurements of Koike et al. (2003) and Suto et al. (2006).

## Instruments Required:

FIRESS Pointed High-resolution mode (R~4400 @ 112 μm)

## Program Size:

- Medium (10-100 hr)

## Approximate Integration Time:

On the faintest sources, a 1% contrast line at 69 μm can be detected in 3 hours of integration. About 30 hours of observations could be sufficient to observe ~30 targets of different brightnesses.

## Special Capabilities Needed:

- None

## Synergies with Other Facilities:

- JWST, possibly MATISSE at ESO (the 10 μm silicate feature).

# 69. Understanding the mineralogy in evolved stars

Franciska Kemper (ICE-CSIC / ICREA / IEEC)

We propose to do a spectroscopic survey of dust producing evolved stars to quantify the mineralogy of the stellar dust production, in order to compare it with the interstellar dust properties, in our own Milky Way and external galaxies. The survey will be done with the FIRESS instrument on PRIMA, using the pointing mode observations at R~130. We will target dusty evolved stars in the Local Group, covering a range of metallicities, galaxy properties, but also stellar dust mass-loss rates and ZAMS masses. A sample of a few hundred objects will be targeted in order to obtain statistically relevant results.

Specifically, we will be looking to measure the crystalline silicate content in the stellar ejecta, which may be used as a tracer of recent star formation in external galaxies, and hope to detect more instances of circumstellar diopside, water ice and carbonates. This proposal complements the proposal submitted by Janet Bowey who is targeting nearby evolved stars and planetary nebulae (such as NGC 6302) in order to study the dust condensation sequence in those objects in detail.

## General Scientific Area:

Milky Way Stars and Stellar Evolution

## Science Justification:

The interstellar medium of galaxies is continuously replenished with the ejecta from post-main-sequence stars. These ejecta are often enriched with the products of nucleosynthesis and as such are an important driver to galactic chemical evolution. The conditions in the stellar winds are also often conducive for the condensation of solid-state particles (dust), and, in the case of Asymptotic Giant Branch (AGB) stars, the condensation of dust is even required to accelerate the wind and drive the (enhanced) mass loss. Thus, post-main-sequence stars can also directly contribute dust particles to the interstellar dust reservoir, rather than just contributing the elements for later dust assembly in molecular clouds. Indeed, the presence in meteorites of presolar grains with an isotopic signature pointing to an AGB or other non-solar origin (Clayton & Nittler, 2004), proves that dust grains formed by post-main-sequence stars not only get deposited into the ISM directly, but some even survive up until the point of incorporation in a planetary system in formation.

There is some question however to what extent the dust in the interstellar medium is being processed after being ejected by post-main-sequence stars. Considering that the residence timescale of dust in the ISM, based on elemental depletions, is $3 \times 10^9$ yr, and that the erosion life time of dust under supernova shock processing is $4 \times 10^8$ yr for refractory elements like Si, Mg and Fe, and $2 \times 10^8$ yr for C, we expect that around 11% of the silicate and 6% of the carbon dust in the ISM is of stellar origin (Jones & Nuth, 2011). The rest is recondensed or regrown in the interstellar medium. For the Large Magellanic Cloud (LMC), the erosion life times due to supernova processing are a factor of ~10 shorter (Temim et al. 2015), and consequently, even smaller fractions of original stellar dust in the ISM remain. Indeed, the composition of freshly produced stellar dust does not match with the composition of interstellar dust in the LMC (Kemper 2015), and it has also been established that the total dust production by evolved stars does not match the dust mass measured in the interstellar medium (Srinivasan et al. 2016), in this case for the Small Magellanic Cloud (SMC).

Specifically, the crystalline fraction of silicates in the interstellar medium is observed to be very low (<2%; Kemper et al. 2004), while scattered measurements for the crystalline fraction of silicates in evolved stars seem to suggest a fraction of ~10%, but a proper measurement is lacking. This is in line with our understanding of the residence and dust destruction time scales in the ISM.

However, we now have the capacity of detecting crystalline silicates in external galaxies, with Spoon et al. (2022) reporting them in around 800 galaxies, almost a quarter of the sample observed with Spitzer. Because crystalline silicates require temperatures higher than what is available in the ISM to form, it is thought that these crystalline silicates must be of stellar origin, and can trace recent star formation activity. In order to put a strong constraint on the importance of this process, we need to have a better handle on the mineralogy of dust around post-main-sequence stars, including its composition and crystalline fraction.

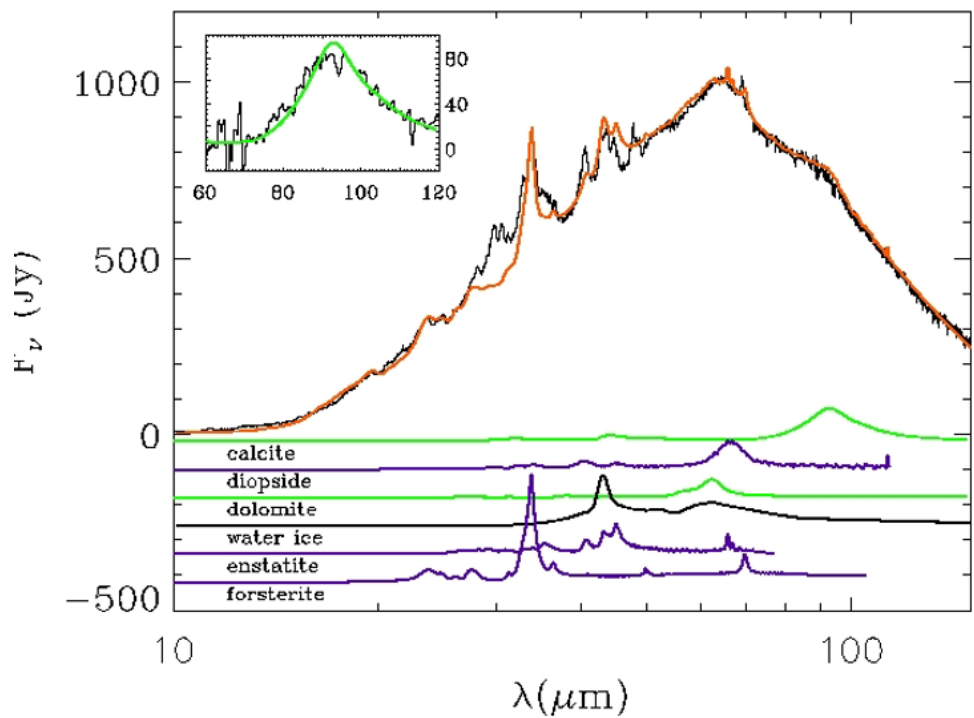

Figure 1: ISO SWS/LWS spectrum of Planetary Nebulae NGC 6302, along with a fit showing thermal dust emission from crystalline silicates, diopside, water ice and carbonates (Kemper et al. 2002).

The PRIMA Far-Infrared Enhanced Survey Spectrometer (FIRESSS) will operate from ~28–240 µm. In this wavelength range the crystalline silicates forsterite and enstatite have strong resonances at 33 and 43 µm, respectively, with forsterite having an additional, weaker, at 69 µm (Molster et al. 2010). Assuming the continuum at these wavelengths is due to amorphous silicates at the same temperature, and the dust is optically thin, the crystalline fractions can be calculated using the feature over continuum ratio. The 69 µm feature is found to be a sensitive tracer for the Fe-content in the lattice, thus further probing the conditions at the time of dust formation and crystallization. Additional species with resonances in the FIRESS spectral range include Ca-bearing crystalline silicates, such as diopside (Koike et al. 2000); carbonates (see Fig. 1); and crystalline water ice (Hoogzaad et al. 2002); all of these previously detected around at least one evolved star. Due to their specific mineralogy, these species further trace physical conditions in the stellar outflows.

The FIRESS spectral resolution of R=130 matches very well with the spectral resolution of the available optical properties.

We propose to do a SAGE-Spec-like survey (Kemper et al. 2010) of evolved stars in the Large and Small Magellanic Clouds, and other Local Group galaxies, sampling a range of metallicities, and galactic properties. We will also include targets in the Milky Way. We aim to observe a few hundred objects, to get a statistical inventory of the dust production and crystallinity of the produced silicates across a range of physical conditions. Due to the exploratory nature of this experiment, it is not possible to estimate the expected strength of the spectral features, but it is clear that PRIMA will be uniquely capable to do a spectroscopic survey of this kind, and it will greatly expand our understanding of dust formation and provide a tool to use silicate crystallinity as a probe for star formation activity in external galaxies.

## Instruments Required:

- FIRESS pointed observations at low-res.
- Possible additional FIRESS Small maps at low-res

## Program Size:

- Large (100 + hr)

## Approximate Integration Time:

- For a legacy type program with statistically relevant numbers of objects, we estimate about ~200 hours in total

## Special Capabilities Needed:

- None

## Synergies with Other Facilities:

- JWST, possibly MATISSE at ESO (the 10 µm silicate feature).

## Description of Observations:

Targets will be selected from the IRAC and MIPS point source catalogues currently available for the LMC (Meixner et al. 2006), SMC (Gordon et al. 2011), and other Local Group galaxies (Khan et al. 2015; Khan 2017). Within the Milky Way, the GLIMPSE survey may be used (Benjamin et al.

2003). It has been demonstrated that the reddest AGB stars are the most dust-producing ones (Srinivasan et al. 2016) and therefore the most relevant to quantify the properties of dust injected into the ISM, but we will ensure also to include a number of lower mass-loss rate objects. Furthermore, we will include higher-mass post-main-sequence stars which are also thought to contribute significantly, such as LBVs (Agliozzo et al. 2021). For most non-Galactic AGB stars, a single point source observation will be sufficient, but it may be possible to resolve some of the Galactic AGB stars with a small map observing mode.

## 70. Exploring the magnetized circumstellar environment of evolved stars through dust polarization analysis

Laurence Sabin (Institute of Astronomy-UNAM, Mexico), Edgar Ramirez (Tecnologico de Monterrey, Guadalajara, Mexico), Abraham Luna (INAOE, Mexico), Omar Serrano (INAOE, Mexico), Jesús Toalá (IRyA-UNAM, Mexico), Alejandro Garcia (INAOE, Mexico), Margaret Meixner (USRA)

The study of magnetic fields in evolved stars originating from both low and high mass precursors is still a relatively unexplored research area. Magnetic fields are known to play a critical role in the formation and evolution of stars, and can have a significant influence on their circumstellar environment. Most evolved objects, namely (pre)Planetary nebulae, Symbiotic Stars (Symb), Wolf-Rayet (WR), and Supernovae Remnants (SNRs) are characterized by their high mass-loss, producing large amounts of dust (and gas), which is dispersed into the Interstellar Medium (ISM). How do magnetic fields interact with the stellar wind and other surrounding material, how do the magnetic fields influence the evolution of these stars, what is the strength and structure of the magnetic fields in these stars or how do the magnetic fields affect the shape and distribution of the stellar material in these systems are amongst the open questions that still need to be answered.

Dust polarization measurements are therefore a powerful tool for tracing and studying the magnetic fields of evolved stars, as they can provide detailed information about their strength and structure but also, and importantly, about the dust characteristics. PRIMA will be an ideal tool to perform sensitive multi-wavelength and multi-scale polarization analysis in the far-infrared and therefore providing a wealth of new information on the late stage of stellar evolution.

### General Scientific Area:

Milky Way Stars and Stellar Evolution

### Science Justification:

#### Broader context:

Unlike the star formation phase, the presence and influence of magnetism in the later stages of a stellar evolution is not as well known or even studied in some cases. This is especially true for the descendants of low-mass stars (1–8 solar masses) such as (pre-)Planetary Nebulae and the binary systems leading to Symbiotic stars or Type Ia supernovae, as well as the descendants of very high-mass stars (>8 solar masses) leading to WR nebulae and supernova remnants with a core-collapsed origin. All these astrophysical objects are characterized by a relatively strong mass loss induced by stellar winds or more explosive events that inject energy, momentum, and processed material into the interstellar medium (ISM). This material includes gas but also a non-negligible amount of carbonaceous-type and/or silicate-type dust (e,g, ∼0.001–0.2 solar masses in young PNe and WR nebulae (Rubio et al., 2020; Jiménez-Hernández et al. 2020)). The importance of these evolved stars in the Galaxy is therefore uncontested.

In order to advance our understanding of the role of magnetism in the formation and evolution of evolved stars and their interaction with the ISM, the best solution is to explore magnetic fields through dust polarization analysis. This technique is based on the principle of alignment of spinning, non-spherical paramagnetic dust grains in the presence of a magnetic field. The dust emission is then polarized as all the grains have their long axis perpendicular to the field lines (Lazarian 2007). The most popular mechanism suggested as responsible for the alignment of dust grains is the one via radiative torques (Dolginov & Mytrophanov 1976; Draine & Weingartner 1996) which is seen as a more efficient process as it involves (isotropic) radiation and hence allows the grains (with sizes similar to the wavelength) sufficient time to rotate and become fully aligned with the field. We note that even in the case of mechanical alignment, the dust grains will be aligned with respect to the magnetic field (Lazarian 2007).

PRIMA and its polarimetric imaging capability, will be a unique tool to trace the magnetic field topology and geometry as well as to quantify its strength in the MIR-FIR. It will therefore be possible to perform a sensitive multi-wavelength and multi-scale polarization analysis. Such observations will allow not only the characterization of the dust (shape, size, type) but also, in synergy with other ground-based interferometers such as ALMA or the future ngVLA, it will be possible to have a full picture of the magnetic field distribution from the inner region to the most external part of the shells. Furthermore, it is important to remark that the results of such studies can provide valuable input to stellar models and MHD simulations. The project will therefore use a combination of theoretical and observational methods to interpret the data.

## Science questions

This project principally seeks to understand the role played by magnetic fields in evolved stars issued from both low and high mass progenitors via the study of polarized dust emission. Specifically, the project will address the following questions:

- How do magnetic fields interact with the stellar wind and other surrounding material? (e.g., Magnetic field energy vs. thermal energy, effect on the mass loss)

- How do the magnetic fields influence the evolution of these stars?

- What is the strength and structure of the magnetic fields in these stars?

- How do the magnetic fields affect the production of X-ray emission from these stars?

- How do the magnetic fields (and magnetized accretion disks) affect the shape and distribution of the nebular material?

- What is the information on the dust obtained from scattering, absorption and emission processes?

- How do the dust grain properties and magnetic field strengths vary between stars that originated from low and high mass progenitors?

- How does the presence of a magnetic field affect the dust grain alignment and the resulting emission?

By gathering and analyzing observational data from a variety of evolved stars, the project will provide a better understanding of the properties of dust and magnetic fields in evolved stars. This will provide first-hand valuable insights into the late stage of stellar evolution that could be compared to existing (MHD) models.

## Need for PRIMA:

With SOFIA being no longer an option, the MIR-FIR wavelength range is poorly constrained from the ground. The reasons of the need for PRIMA are described by the following:

1) **Typical nebulae around evolved stars are extended:** In case of low mass stars (below 8 solar masses), during the AGB phase, the evolved stars develop an extended dusty circumstellar structure, which grows as the time goes by. The 'superwind' phase causes the shift from spherical to axisymmetric structure (Gledhill et al. 2001), driving the circumstellar structure far away from the central star and when the shift from proto-planetary nebulae (PPNe) to planetary nebulae (PNe) happens, the extended envelope have an extension with an average radius from 30 arcseconds to 5 arcminutes in some cases. In the case of the supernovae (SNe) and supernova remnants (SNRs) the extended nebula (which contains mainly dust and hot SN-gas (Chastenet et al. 2022)) could reach an average radius from 2 to 7 arcminutes and has (in some cases) an irregular morphology.

2) **Need of large field of view:** Because of the characteristics of the aforementioned targets, the need of a large field of view is crucial in order to unveil the morphology of the envelope that hosts the evolved stars and their interaction with the interstellar medium. Also, a large field of view will allow us to study the interaction of the circumstellar envelope with the stellar or galactic magnetic field.

3) **Interferometric instruments would filter out most of their emission:** Despite the small angular resolution that can provide, the use of interferometric instruments could filter out most of their extended emission in the data processing pipeline.

## Interpretation methods:

As aforementioned the determination of the magnetic field geometry will rely on dust polarization analysis via the principle of alignment of spinning, non-spherical paramagnetic dust grains due to radiative torques. However, the presence of scattering and absorption components cannot be discarded. In the circumstellar and interstellar medium, various physical processes occur that polarize electromagnetic waves. We can group these phenomena into two: polarization of origin and polarization acquired by transport. Among those of origin are the processes of synchrotron emission, Zeeman effect, Goldreich-Kalafis effect, maser emission and thermal emission of aligned non-spherical dust grains. Transport methods include Faraday rotation, scattering, reflection, and dichroism (selective absorption). The polarization observed in scenarios associated with evolved stars is caused mainly by synchrotron, by dust emission or absorption (dichroic) and by scattering, the latter being the physical process that does not depend on an external magnetic field, but in the first three it is essential, being the magnetic field who is in charge of aligning the paramagnetic grains of non-spherical dust. For the analysis of the data produced in this research, we will use techniques such as those followed by Chastenet et al. 2022 to properly separate the emission due to each of these polarization mechanisms in M1 observational data with SOFIA/HAWC+, in which the different efficiency that these mechanisms have in different bands from NIR to FIR is taken advantage of; in combination with typical distribution patterns observed for the different mechanisms and models, such as those used by Jones, T. (2000) in the analysis of M82.

## Link to testable hypotheses

Dust emission and polarimetry results will be the driver for modern magneto-hydrodynamical simulations, tailored to the diversity of nebulae around evolved stars.

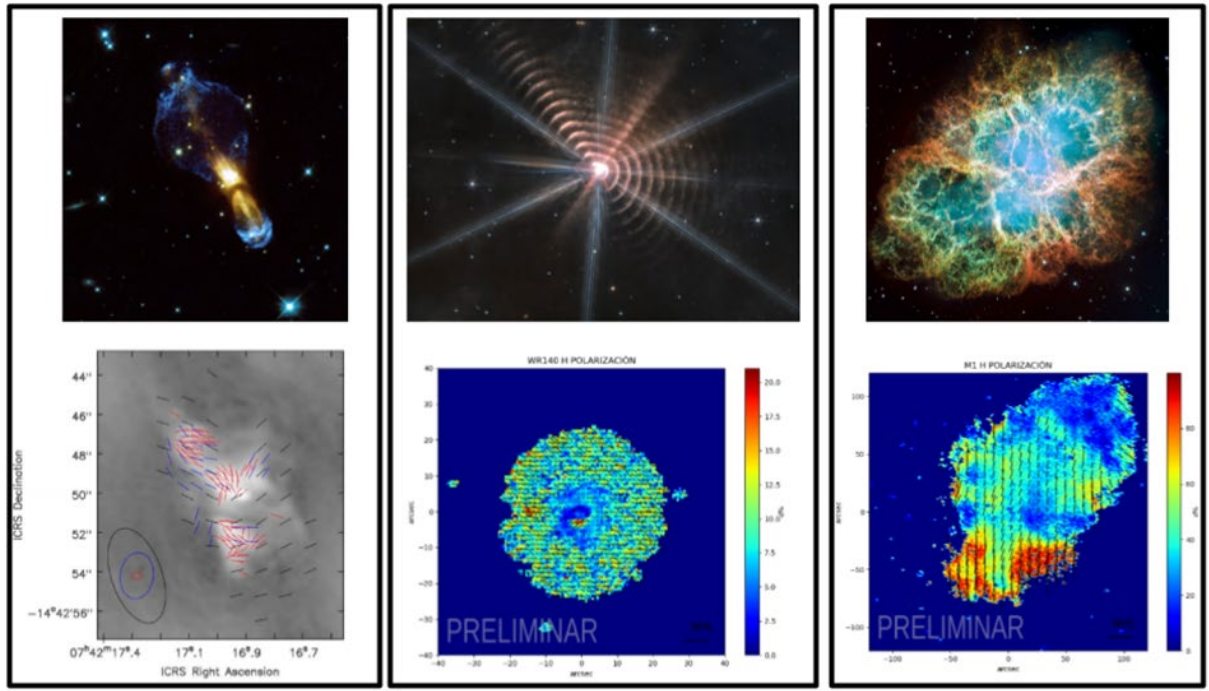

Figure 1: Column 1 to 3 shows examples of the targets in the science case (top row) and the polarization information (bottom row). First a) The pre-PN OH231.8+4.2 (HST) and the associated magnetic field distribution a multiple wavelength and scales (CARMA: 2.5" to ALMA: 0.5") an external helicoidal magnetic field is suspected (Sabin+2020), b) The WR shell WR 140 (JWST) has been recently observed in near IR with the instrument POLICAN and the polarization vectors in H band are shown here for the first time. The magnetic field distribution will be better traced at larger wavelengths and the data will complement the NIR information (Luna+in prep), c) The SNR M1 (HST) and the polarization map obtained with POLICAN in the NIR is also the type of object expected to be better constrained at larger wavelengths (Luna+in prep) (see text).

## Subsections:

**(pre-)PNe/Symb:** Magnetic fields have been invoked as an agent responsible for the drastic departure from spherical symmetry of a large number of evolved intermediate mass stars such as pre-Planetary Nebulae (pre-PNe) and Planetary Nebulae (PNe) via the generation of collimation mechanism(s). It has now been (theoretically) established that magnetic fields are likely to arise within a close binary system (thought to be the main shaping agent). Hence, Balick, Frank & Liu (2019,2020) showed that toroidal magnetic fields, which could result from the presence of accretion disks, could play an important role in the shaping of pPNe displaying jets along the outflow symmetry axis and Garcia-Segura et al. (2020,2021,2022) analyzed how a magnetized envelope was developing after a common envelope (CE) episode and they found that magnetic outflows could be launched from a $\sim 10^4$ G (toroidal) magnetic field compressed into the circumbinary disk. Therefore, the very latest theoretical investigations not only pointed out the crucial role of magnetic fields in generating particular morphological structures but also described in detail their jets/outflows launching ability. But can we actually probe or test these predictions with observations? Does the actual distribution of the magnetic fields in these evolved objects match the theoretical prognosis? There were few observational discoveries which suggested the dynamical importance of magnetic fields and their possible active and efficient role in launching, collimating and shaping the bipolar jets/outflows. Sabin et al. (2015a,2020) worked on dust polarization to map the magnetic field in the circumstellar envelopes of these evolved objects. The imaging polarimetry revealed organized toroidal and polar large scale magnetic field configuration. All the studies pointed out the role of magnetic fields in the shaping of the circumstellar envelopes of pre-PNe/PNe via outflow launching and collimation.

**WRs:** WR nebulae form by mass ejections from very massive stars (>25 Solar mass) in a very similar process as that of PNe. However, WR nebulae are short lived and of the 600 WR stars in our Galaxy only about 20 of them exhibit the clear presence of a nebula. Consequently, the role of magnetic field in the formation and evolution in WR nebulae are less understood than in the case of PNe, if not completely unknown. Characterizing the magnetic field through dust polarization with PRIMA will help us assess the characteristics of the previous evolutive phase (which can be a red supergiant or a luminous blue variable star) and peer into the formation and evolution of WR nebulae. This will be done by comparing PRIMA results with modern magneto-hydrodynamical simulations (i.e., Meyer 2021).

**SNRs:** Supernovae produce a large mass of dust and were important sources of dust in the early universe (Laporte et al. 2017). Far-infrared polarization observations improve the contrast of the SNR with respect to the ISM, allowing cleaner detection of dust emission from the SNR. In addition, the composition of the grains can be determined, being able to identify the shapes, sizes and types of dust grains as well as their relationship with the magnetic fields of the region, finding that they are predominantly silicates, which are large and elongated since they were formed and that dust grains are capable of surviving the energetic environment of SNRs (Fry et al. 2020).

## Instruments Required:

PRIMAGER small maps (pointed observations) in the polarimeter band

## Program Size:

- Small (<10 hr)

## Approximate Integration Time:

The worst case of point-source sensitivity in polarized flux at Band 2A (96μm) is 300 μJy for a 10 hour observation over one square degree, or for 15 minutes observations over the minimal map area of 15'×15'. An estimation of time using this "worst" case for the different kind of sources sums a total time for our proposal of **approximately 5 hr**, divided as follows:

**SNRs:** The brightest case for SNRs is M1, it is 7×5 square arcmin, with local polarized flux density @ 1500 GHz of approximately 6 Jy (Ritacco et al. 2018). This is smaller than the minimal map size observed by PRIMAger. Just a few minutes are needed to reach a sufficient signal to noise.

We select small size SNRs at different types (II, IIb, Ia-Ib, Ic) and with polarized fluxes larger than 10mJy. The finest and smallest case requires a map of (5×5) sq.arcmin and needs approximately 15 min each. Four selected targets sum a subtotal of 1 hr of observing time.

**Low-Intermediate mass evolved stars:** We will use the sample from the THROES catalog (Ramos-Medina et al. 2018) which includes various objects already detected and measured with HERSCHEL/PACS which leads to 121 targets. The total flux ranges from ∼2 Jy to ∼0.5 Jy (96–250 μm) for the faintest targets up to ∼1000 Jy to ∼250 Jy (96–250 μm) for the brightest. Assuming polarized fluxes of 250 mJy, our estimation in time for these cases sum a subtotal of 3 hr of observing time.

**WRs:** These sources are bright (>1 Jy) in polarized flux and small (<2×2) arcmin², for five selected cases (NGC 6888, NGC 2359, M1-67, WR16, WR140), we estimated a subtotal of observing time of 1 hr.

## Special Capabilities Needed:

- None

## Synergies with Other Facilities:

- The TolTEC camera has a total of 7716 detectors distributed in the 1.1, 1.4, and 2.0 mm bands. Each element is a Lumped Element Kinetic Inductance Detector (LEKID), sensitive to the linear polarization of the incoming radiation and additionally has a half-wave plate to recover the polarization in all three bands. Four general science areas focused by TolTEC are coincident in science objectives with PRIMA, especially the called "The Fields in Filaments Legacy Survey", designed to probe the distribution of magnetic fields in filaments as traced by the polarization of dust. Just now TolTEC is in the commissioning stage (http://toltec.astro.umass.edu/).

- The ALMA interferometer completes PRIMA's wavelength coverage in the submm and mm range with sensitive imaging from 84 to 950 GHz (3 mm to 320 μm), allows an angular resolution down to 0.015" (at 300 GHz) and full polarization capabilities. It will therefore be possible to map the sources near their central engine and retrieve the magnetic field's information at this location.

- The Next Generation Very Large Array (ngVLA), while still being a project, will be a useful complement to the PRIMA data in the 1.2–116 GHz range. Set to be 10 times more sensitive than ALMA and to reach mas resolution. Its polarization capabilities will allow to trace the magnetic field even closer to the central structure and then perform multi-scale/resolution analysis of the magnetic field and dust.

- The full intensity high-resolution images obtained from the JWST are a good supplement to the PRIMA data in order first to clearly identify the topology of the magnetic field with respect to the nebula and to complete the dust analysis (using models).

## Description of Observations:

Polarimetry requires strict control and precision of the angles, so the recommendation is to use the raster observation mode at different angles to eliminate correlation effects per trace in the coadded map. Mappings must consider observing a "sky" at least equivalent in area to the observed object. For example, for M1 (5×5 sqr.arcmin), the minimal PRIMAGer map of 15x15sqr.arcmin is sufficient to get enough data from the object and sky.

Several calibrators are needed such as a gain calibrator:(e.g., 0730-116), a passband calibrator (e.g., 3C 84) and a flux calibrator: generally, a planet (e.g., Mars/Uranus). But these can be adjusted as in all cases we will use the closest calibrators from the sources. Those can be shared with other programs.

# 71. The extraordinary deaths of ordinary stars

Raghvendra Sahai (Jet Propulsion Laboratory, California Institute of Technology)

Most stars in the Universe that evolve in a Hubble time or less (i.e., with main-sequence masses, m(MS) ~1–8 $M_{sun}$) go through their death throes as Asymptotic Giant Branch (AGB) stars, ejecting half or more of their total mass in the form of nucleosynthetically-enriched material into the interstellar medium (ISM) via dusty, isotropic, slowly-expanding winds. The heavy mass-loss dramatically alters the course of stellar evolution, seeds the ISM for future generations of stars and habitable planets, and drives galactic chemical evolution. Yet this crucial stellar evolutionary process is very poorly understood because the full mass-loss histories of such stars over long time scale is very poorly understood; observations that can probe mass-loss from intermediate to long time-scales (~$10^3$ to $10^5$ yr), for a statistical sample of AGB stars, are needed. In addition, during the final death throes of these stars which convert them to planetary nebulae, a dramatic transformation of the geometry of mass-ejection from roughly isotropic to a rich diversity of bipolar, multipolar and elliptical shapes is observed. The current consensus is that a variety of strong binary interactions are responsible, but in order to pin down theoretical models for this transformation, a knowledge of the mass ejected during this transition phase (~$10^3$ yr) as a function of morphology is needed. We propose a medium-sized program comprised of two sensitive far-IR surveys, one for AGB stars and another for young planetary nebulae, using the PRIMAger and FIRESS instruments, to address these two goals.

## General Scientific Area:

Milky Way Stars and Stellar Evolution

## Science Justification:

### Broader context:

Most stars in the Universe that evolve in a Hubble time or less (i.e., with main-sequence masses, m(MS) ~1–8 $M_\odot$) go through their death throes as Asymptotic Giant Branch (AGB) stars, ejecting half or more of their total mass in the form of nucleosynthetically-enriched material into the interstellar medium (ISM) via dusty, isotropic, slowly-expanding (Vexp~10–20 km/s) winds at rates of dM/dt~1e-7 to 1e-4 $M_\odot$/yr (e.g., Olofsson 2008). The heavy mass-loss dramatically alters the course of stellar evolution, seeds the ISM for future generations of stars and habitable planets, and drives galactic chemical evolution (e.g., increase in metallicity in the early Universe). Yet this crucial stellar evolutionary process is very poorly understood because the full mass-loss histories of stars, with m(MS)~1–8 $M_\odot$ as a function of m(MS) and metallicity, are very poorly known. In addition, there is increasing observational evidence that binary interactions may play a crucial role in the late evolution and deaths of such stars (Sahai & Trauger 1998, Sahai+2011 [SMV11], Decin+2020). For example, HST imaging surveys of planetary nebulae or PNe (the end-products of AGB evolution) show that although AGB stars eject mass roughly isotropically over most of their lives, most young PNe have extreme and varied shapes – very few are round (SMV11). Theoretical studies show that strong binary interactions during the late AGB phase – such as common envelope (CE) evolution (e.g., Blackman & Lucchini 2014), are likely needed to produce high-velocity jets that can sculpt, from the inside-out, the mass ejected during the final death throes of AGB stars and produce the observed departures from spherical symmetry (Sahai & Trauger 1998).

*There are two important mass-loss time-scales that remain poorly investigated -- (a) the relatively long times-scales* (>~1e5 yr for an average dM/dt ~ 1e-5 M⊙/yr) of mass-loss in AGB stars required to lose most of the star's envelope, and (b) relatively short (~1e3 yr) time-scales that probe the very last death throes of AGB stars, sometimes referred to as the "superwind" phase. This phase is best probed by studying young PNe, and thus the physics of strong binary interactions (a process of fundamental and wide-ranging importance for stellar astrophysics) which result in the varied shapes of PNe. A fundamental question that such a study of young PNe can address is if binarity also impacts the very onset of the star's death phase. We describe below how PRIMA's unprecedented far-IR capabilities, especially its sensitivity, in order to probe the above mass-loss times-scales.

## Science questions

**(1) Probing the long-term mass-loss history of AGB stars:** Determining the full history of heavy mass-loss in AGB stars remains a major observational challenge. *We do not know, in general, how long AGB stars have been undergoing dense mass-loss (i.e, at rates >~$10^{-7}$ M⊙/yr), if the mass-loss rates change by significant factors, and if so, whether this happens in a periodic, episodic or irregular manner over long time-scales.* A dramatic example of such a change is the strong, episodic increase that has been found (from observations of millimeter-wave CO line emission) for a few C-rich (C/O>1) stars by the detection of a detached, geometrically-thin expanding circumstellar shell (Olofsson+1996), inferred to have resulted from a thermal pulse associated with He-burning and 3rd dredge-up nucleosynthesis around the stellar core (this process produces much of the C in the Universe and converts the normally O-rich (C/O<1) stellar atmosphere to one that is C-rich (C/O>1). Millimeter-wave CO line emission from gas and scattered light from dust in the circumstellar envelope (CSE) – the standard tracers of the mass-loss history from ground-based observations – are limited typically to time-scales of <6,500 yr for a typical expansion velocity of 10 km/s and mass-loss rate <$10^{-5}$ M⊙/yr, because, at relatively large radii (> $3.5×10^{17}$ cm) in the CSE, CO is photodissociated by the interstellar UV field (e.g., Ramstedt et al. 2020), and dust scattered light becomes undetectable because it becomes too faint. Observations of atomic hydrogen (HI) from the wind, generally resulting from the photodissociation of H2 in the molecular wind, are usually strongly confused by Galactic emission (e.g., Matthews et al. 2015).

*Hence, the total amount of mass ejected into the ISM, M(ejecta) (which depends on the envelope's outer extent -- e.g., M(ejecta)~r(out) (for a constant mass-loss rate at a constant expansion velocity) remains unknown;* as a result the progenitor mass is also unknown. For example, in the

case of IRC+10216, the best-studied mass-losing AGB star, the CSE seen in CO or scattered light extends to about 200'' ($3.5 \times 10^{17}$ cm), and the inferred M(ejecta) is ~0.15 $M_\odot$, a small fraction of what this star has had to have lost, given its late evolutionary phase. The unexpected discovery of a very extended astrosphere (resulting from the shock interaction of the stellar wind with the ISM) in far-UV emission around IRC+10216, led to a first estimate of the total duration of heavy mass-loss, t(hml)~69,000 yr, and total ejected mass, M(ej,tot)~1.4 $M_\odot$, a value consistent with the theoretical expectation that C-rich stars have progenitor masses of ~2 $M_\odot$ (and that IRC+10216 is close to transitioning to the post-AGB phase).

*Far-IR emission from dust in the CSE provides a method of tracing the mass-loss history over very long time-scales.* Previous far-IR missions, beginning with IRAS, followed by ISO, Akari and Herschel detected very extended emission from CSEs, at radii much larger than probed by standard tracers in a few objects, either from detached shells, and/or from the astrospheres (e.g., Waters+1994, Izumiura+1996, Ueta+2006). A dedicated far-IR survey at 70 and 160 µ of a sample of 78 evolved stars (mostly AGB stars and a few red supergiants), using Herschel's PACS instrument revealed astrospheres in ~40% of the sample (Cox et al. 2012: Cox12; Fig. 1). About 20% of the sample, comprising of C-rich or S-type (C/O~1) stars (which experience thermal pulses), showed circular rings. The astrosphere radius, Rc, and width provide a natural estimate of the time-scale for the full duration of AGB mass-loss.

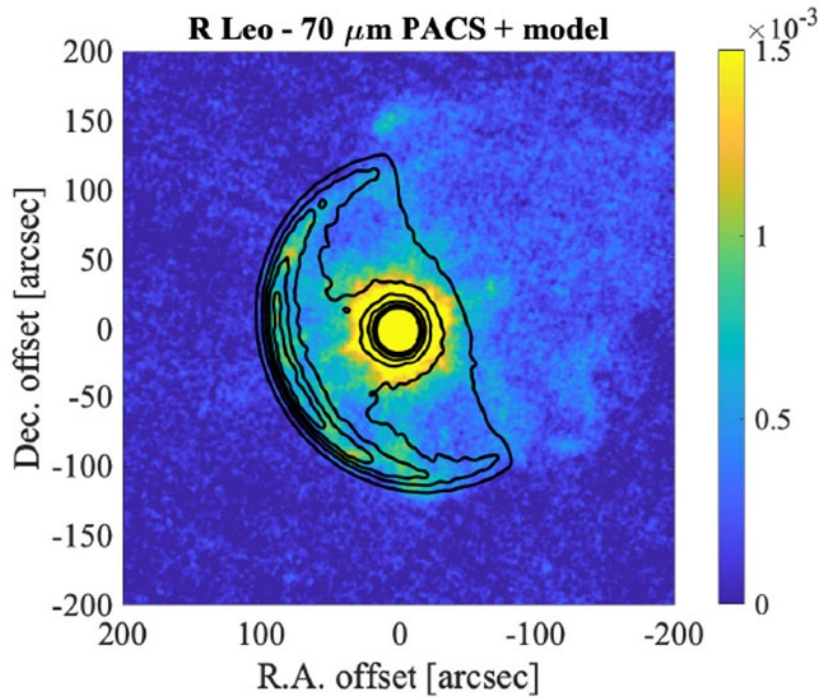

Figure 1: The astrosphere of the O-rich AGB star, R Leo, as seen at 70 µm with PACS/Herschel (color scale), together with a model (contours) for analysing the shock interaction of the AGB wind with the ISM (adapted form Mercker+2022) to estimate the long-term mass-loss history. A PRIMA far-IR imaging survey will enable us to similarly estimate the mass-loss histories of a statistical population of AGB stars on ~10^5 yr time-scales.

Even *for intermediate mass-loss time-scales* that can be probed in CO mm-wave lines, far-IR imaging may offer a distinct advantage. For CO, the combination of interferometric and total-power (to recover interferometric flux losses) observations needed to image areas of (say) 5 × 5 sq. arcmin, with high (~few arcsec) angular resolution, is very time-consuming (several hours/target) even for relatively bright sources and prohibitive for fainter ones. In contrast, striking discoveries of unexpected morphological structures (inferred to result from the presence of a binary companion) via far-IR imaging could be made with PACS imaging at 70 and 160 μm, e.g., the presence of 2 perfectly diametrically opposed spiral-shaped arms in the symbiotic star R Aqr and possible spiral structures in the S-star W Aql (Mayer+2013) with significantly smaller exposure times.

*In summary, in spite of the above studies, many questions remain unanswered.* For example, for most AGB stars, the values for M(ej,tot) derived from the far-IR studies (as well as the far-UV) are generally far below those required to bring these stars close to the post-AGB evolutionary phase (Maercker+2022, Sahai & Stenger 2023). Interestingly, within the small sample of stars for which M(ej,tot) could be estimated (~30), the C-rich and S-type stars typically show significantly higher values of t(hml) and M(ej,tot), which is expected since O-rich stars evolve into C-rich stars. So it is a puzzle as to why there are also O-rich (C-rich) stars with M(ej,tot) comparable to the highest (lowest) ones found for C-rich (O-rich) stars. In addition, some stars show shock structures that do not look like bow-shocks, but have "eye" shapes – the origin of such is not understood (the simplest explanation, that the mass-loss ejection properties are different in two opposing hemispheres, is unlikely.) Others, like W Hya, show exceedingly complex mass-loss histories via their FUV emission (Fig. 2); hints of these complexities are seen in the PACS images of W Hya, and in many other stars in the Cox12 study. Finally, the potential contribution of line emission (e.g., [OI] 63 μm, [CII] 158 μm) from shocked gas in the astrospheres remains completely unknown, but may be significant as indicated by numerical simulations (Mohamed+2012). In the current sample of AGB stars with extended rings, attributed to a thermal pulse, surprisingly only one or two show two rings (Mecina+2014), instead of just one, although AGB stars are expected to undergo a large number before their post-AGB journey. This may be due to sensitivity (the outer ring is much fainter than the inner one) and/or because of the star's initial mass (since the interpulse timescale is smaller for stars with higher initial masses, e.g., ~2.8e4 yr for 3.6 M⊙) – a more sensitive survey of a larger sample (e.g., with PRIMA) can help address this issue.

**(2) Mass-loss during the Final Death Throes:** The very last death throes of AGB stars that transform (most) of them into aspherical PNe (with WDs at their centers), hold the key to understanding the binary interaction mechanisms that are believed to either cause (as, e.g., Common Envelope Evolution or CEE, Grazing Envelope Evolution or GEE, resulting in the ejection of most or part of the stellar envelope) and/or influence (as, e.g., formation of a jet-powering accretion-disk due to Bondi-Hoyle or Roche-lobe Overflow) this transformation. The diversity of morphologies (Fig. 3), which mostly includes bipolar, multipolar and elliptical shapes (with a ubiquitous presence of point-symmetry) and only a small fraction of round ones, is well cataloged. A knowledge of the mass ejected during this transition phase as a function of morphology in young PNe is crucial for pinning down the nature of the binary interactions (e.g., in contrast to the other binary interactions, CEE will eject a lot of mass, ~few × 0.1 M⊙, in a relatively short time), but is lacking. This is because most of this ejecta mass is expected to be neutral and warm, for which the primary gas cooling lines (e.g., [OI]63 μm, [CII]158 μm) lie in the far-IR region – but only very few young PNe have been observed in these lines (e.g., with ISO: Liu+2001, or PACS: Ramos-Medina+2018; Fig. 4).

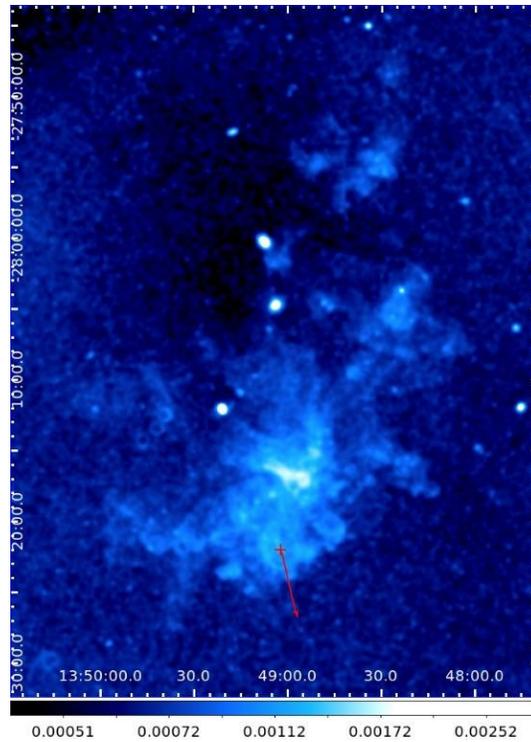

Figure 2: The astrosphere of the O-rich AGB star, W Hya (red cross), as seen in FUV emission with GALEX, showing a complex morphology that includes multiple arcs and collimated-outflow like features, some of these are tentatively seen in the far-IR PACS image as well. Red vector shows proper motion of star thru the local ISM (adapted from Sahai+2023). A PRIMA far-IR imaging survey will reveal the fraction of AGB stars that undergo similar complex mass-loss histories.

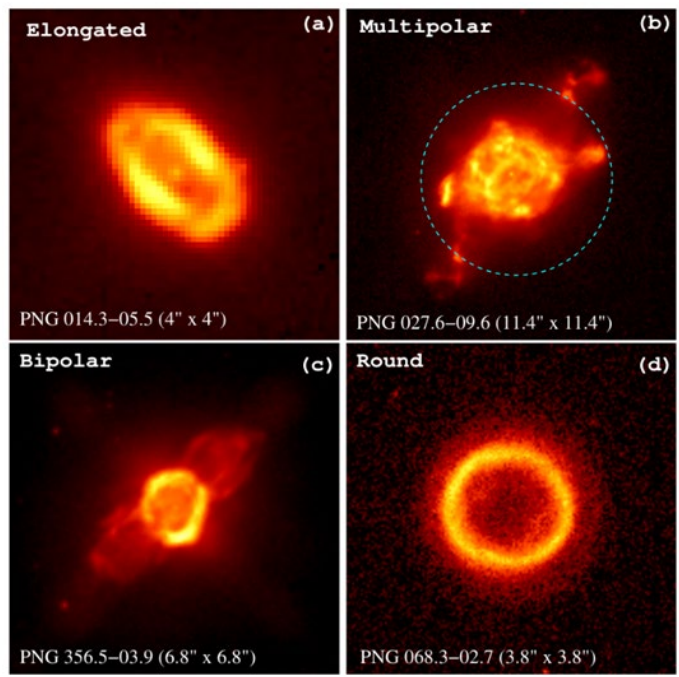

Figure 3: HST H-alpha images of PNe belonging to 4 major primary morphological classes. Dashed circles in panel b shows a 6.5 arcsec diameter circle for scale (adapted from SMV11).

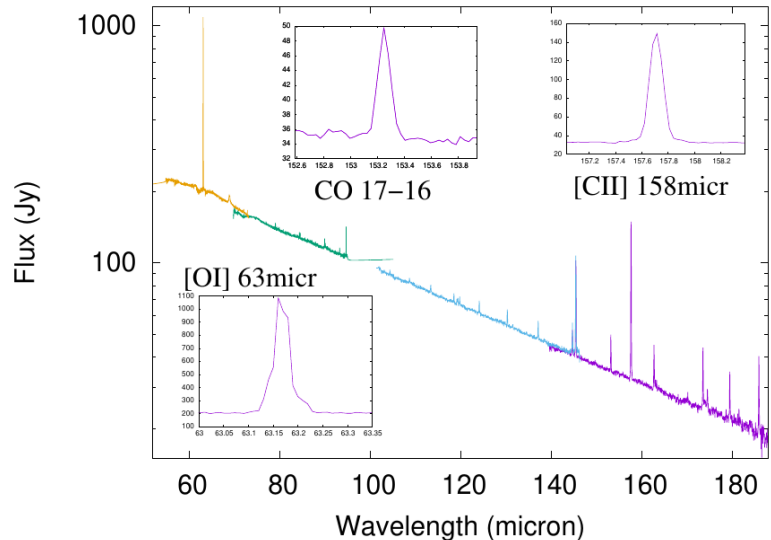

Figure 4: Typical spectrum of a young PN (CPD-56d8032) showing fine-structure and CO lines, taken using PACS/Herschel (data from Ramos-Medina+2018). A PACS survey of a statistical sample of such PNe, as a function of morphological type, will enable the determination of the mass ejected during the final death throes of AGB stars as they transition into PNe, and help constrain theoretical models of binary interactions that are believed to cause this transition.

## Instruments Required:

- FIRESS pointed high-resolution observations
- FIRESS small maps at low resolution
- PRIMAGER small maps, at both hyperspectral and polarimeter bands

## Program Size:

- Medium (10–100 hr)

## Approximate Integration Time:

- 95 hr

## Special Capabilities Needed:

- None

## Synergies with Other Facilities:

- ALMA, NOEMA, SMA, JCMT

## Description of Observations:

We will carry out PRIMA surveys of a statistical sample of AGB stars and young PNe, that are a factor >10 deeper in sensitivity than what has been achieved so far. Our survey has strong synergies with surveys of mass-loss from AGB stars that have been (or are being, or will be) carried out with ALMA (e.g., DEATHSTAR: Ramstedt+2020, ATOMIUM: Decin+2020) and JCMT (NESS: Scicluna+2022).

*(a) Long-term Mass-Loss History:* The dust temperature in the shocked wind in astrospheres is found to be typically 30-50K (assuming no line contribution), thus PRIMAger bands 1B and 2A-2C are the most important ones for probing these, while Band 1A is most important for tracing the unshocked wind closer to the star. We will aim for detecting a continuum surface brightness of ~1 MJy/sr in the 50–200 μm wavelength range, which is a factor 10 deeper than the faintest sources detected in Cox12. We will first determine the contribution of any line emission in different bands down to 10% of the wide-band emission seen in the PACS 70 and 160 μm filters (i.e., line flux ~8.5e-15 W/m$^2$ for a line-width of ~1 km/s emitting over a 0.1 arcmin$^2$ area) by making FIRESS/FTS observations using single-pointings centered on the brightest regions of the astrospheres for a small representative sample of 8 of the brightest targets in the Cox12 survey in both bands, which will include 2 O-rich stars, 2 C-rich stars, 2 S-type stars and 2 RSGs – this can be completed in a few hours. We will use the PRIMAger to make small maps (~15 × 15 arcmin$^2$) in both bands of the far-IR emission (since the FOVs of the two bands are separated by about ~15 arcmin, each target will have to be mapped twice). To achieve a 5σ detection in Band 1B, we will need 0.25 hr/source, or 75 hr for a survey of the nearest 300 AGB stars (extracted from the compilation of 18,381 AGB stars by Suh 2021). The sample size is dictated by the requirement of obtaining statistical samples of the 4 different morphological classes of extended emission sources (fermata, eyes, rings, irregular: Cox12), and the 3 different abundance types, and conservatively assuming a detection rate that, although expected to be higher, may be the same as in Cox12, i.e., 40%.

*(b) The Final Death Throes:* We will use FIRESS to observe multiple far-infrared lines that are diagnostic of the density and temperature in the PDR (e.g., [OI]63, 146 μm & [CII]158 μm) and molecular region (two CO lines). Young PNe generally have small-angular sizes (typically < 10 arcsec) (Fig. 3), so a single-pointing per target will be adequate to capture all of the flux, as well as provide information on the emission extent (in the unlikely case that the emission is extended). The faintest lines of importance observed in a few targets observed with PACS (ref) are ~1e-16 W/m$^2$; we therefore aim for a 5σ sensitivity of 1e-17 W/m$^2$ in a 1 arcmin$^2$ FOV, which, using the stated speed for mapping (small-map mode) will take ~3 min to cover 2/4 bands. Thus, our target list of ~150 young PNe taken from compilations of HST-imaged young PNe (SMV11, Guzman-Ramirez+2011, Stanghellini+2016), most of which have Spitzer spectra (Stanghellini+2012) revealing their dust chemistry, will require 15 hr of integration time. The sample size is dictated by the requirement of obtaining statistical samples of the 7 different primary morphological classes (bipolar, collimated-lobe-pair, multipolar, elliptical, spiral, round, and irregular: SMV11).

## Acknowledgements:

This research was carried out at the Jet Propulsion Laboratory, California Institute of Technology, under a contract with the National Aeronautics and Space Administration.

## 72. Spatially and spectrally resolved distributions of cold dust/gas toward the interface between the circumstellar shells and the interstellar medium in a sample of solar-neighbor evolved stars

Toshiya Ueta (University of Denver), Masaaki Otsuka (Kyoto University/ Okayama Observatory)

The life cycle of matter in the Universe begins with stellar nucleosynthesis. Various chemical elements synthesized deep within stars must be brought up to the surface and dispersed into the circumstellar environments via the process of stellar mass loss. The circumstellar shell forms as the direct product of this process, retaining the pristine records of mass loss and subsequent physico-chemical processes that may take place. Our studies are aimed at providing empirical quantitative evidence for this process that has not yet been deciphered from first principles of physics. The past far-IR opportunities enabled us to glimpse the interface region between the circumstellar shell and the interstellar medium, where the bulk of the circumstellar mass was found at the coldest temperatures.

We will use PRIMA to gather quantitative evidence that is directly comparable with existing theories for the first time. Specifically, we will spatially resolve multiple shell structures resulting from stellar wind thermal pulses during the latest stages of post-main sequence evolution to directly measure the inter-pulse timescale as a model constraint. Also, we will spatially and spectrally resolve the shell-ISM interface region to perform full-fledged atomic and molecular line diagnostics so that the empirically derived physico-chemical conditions are directly comparable with predictions made by existing PDR models. These investigations will be done with a set of about two dozen evolved stars with extended shells in the Solar neighborhood – all resolvable with PRIMA - that were detected in the previous far-IR opportunities.

### General Scientific Area:

Milky Way Stars and Stellar Evolution

### Science Justification:

### Broader Context

Stellar mass loss is one of the critical processes of stellar evolution that has significant implications for the rest of cosmos, from individual stellar chemical yield to the ISM to the metallicity and dust yield in the cosmological context. While existing stellar evolution codes are equipped with various mass loss prescriptions that would yield results that are generally consistent with observations in a statistical sense, specifics are still unknown as mass loss has not been physically/mathematically formulated from first principles and observational cues of the physically important junctures of mass loss have rarely been established (e.g., when mass loss begins/terminates are not yet specifiable by observables).

## Science Questions

We aim at addressing the following two specific issues surrounding stellar mass loss in terms of providing quantitative observational constraints on existing theories/models:

### (1)   The timescale and geometry of mass loss during the extreme AGB phase

During the latest stages of evolution during the asymptotic giant branch phase, the rate of mass loss is known to be enhanced by thermal pulses so that the bulk of the stellar envelope can be lost to the circumstellar space by the end of the AGB phase. Also, the geometry of mass loss is expected to change dramatically around this phase. These thermal pulses are theoretically predicted and mass loss ejecta from a single thermal pulse event have been observationally seen. However, in no case multiple shells of thermal pulse origin has been spatially resolved while unresolved cases have been detected [4,8,10,11; Figure 1]. We will spatially resolve these shells to observationally constrain the time interval between such shells of thermal pulse origin to be compared with existing model predictions. Also, hyperspectral mapping in thermal dust continuum can probe time-lapse of the geometrical transformation as bands at longer wavelengths see the spatial distribution in further past [7,8,10,11].

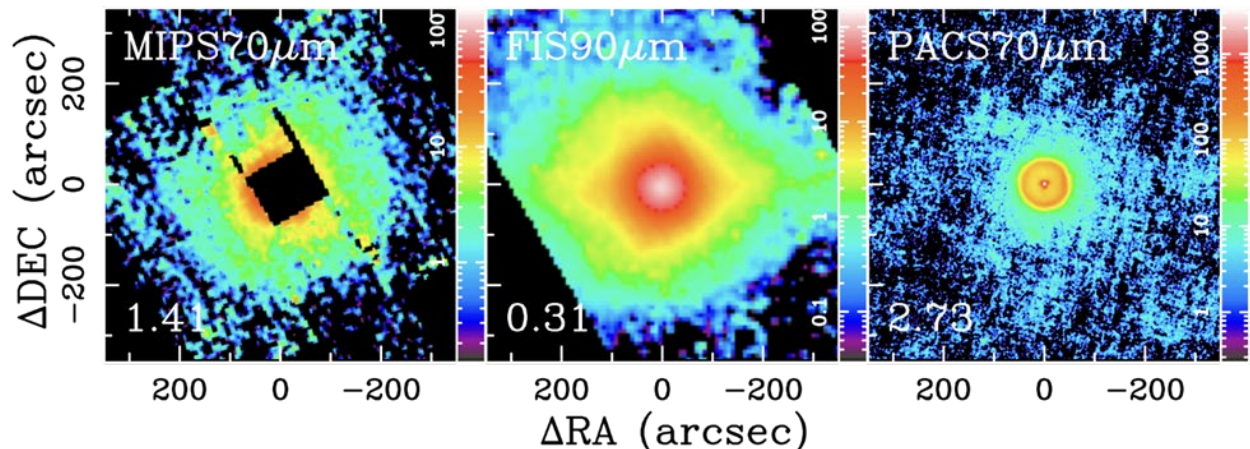

Figure 1: Far-IR maps of the extended circumstellar shell of U Ant taken with Spitzer at 70 μm (left; [8]), AKARI at 90 μm (middle; [10,11]), and Herschel at 70 μm (right; [4]). Each color wedge shows the surface brightness in MJy sr$^{-1}$. The detection limit in MJy sr$^{-1}$ is indicated by the value given at the lower left corner. The AKARI map (middle) achieved the sensitivity of 300 kJy sr$^{-1}$ which is enough to detect the extended shell cause by the thermal before last. However, the spatial resolution was not enough to resolve the extended shell cause by the previous thermal pulse, which is resolved in the Herschel map which does not have enough sensitivity to detect the extended shell. PRIMAger will provide both spatial resolution and sensitivity to detect both thermal pulse shells, allowing the direct measurement of the inter-pulse time scale.

### (2)   Spatially resolved physico-chemical conditions in the shell-ISM interface regions

Our previous attempts with Herschel found that the bulk of the circumstellar mass resided beyond the ionized region of planetary nebulae in the form of atomic/molecular gas [1,2,5,9] but did not reach sufficient sensitivities to probe this colder wind-ISM interface region [5,9; Figure 2]. With spectral images at various diagnostic lines in the PRIMA range augmented with other line maps, the physico-chemical conditions will be spatially resolved in a sample of extended planetary nebulae [9] and shell-ISM interaction cases [3,6,7] in the Solar neighborhood. We can then understand how emission arises especially in some exotic molecular radials and at the shell-ISM interface as well as individual and bulk/statistical abundances to be compared with predictions from existing PDR models.

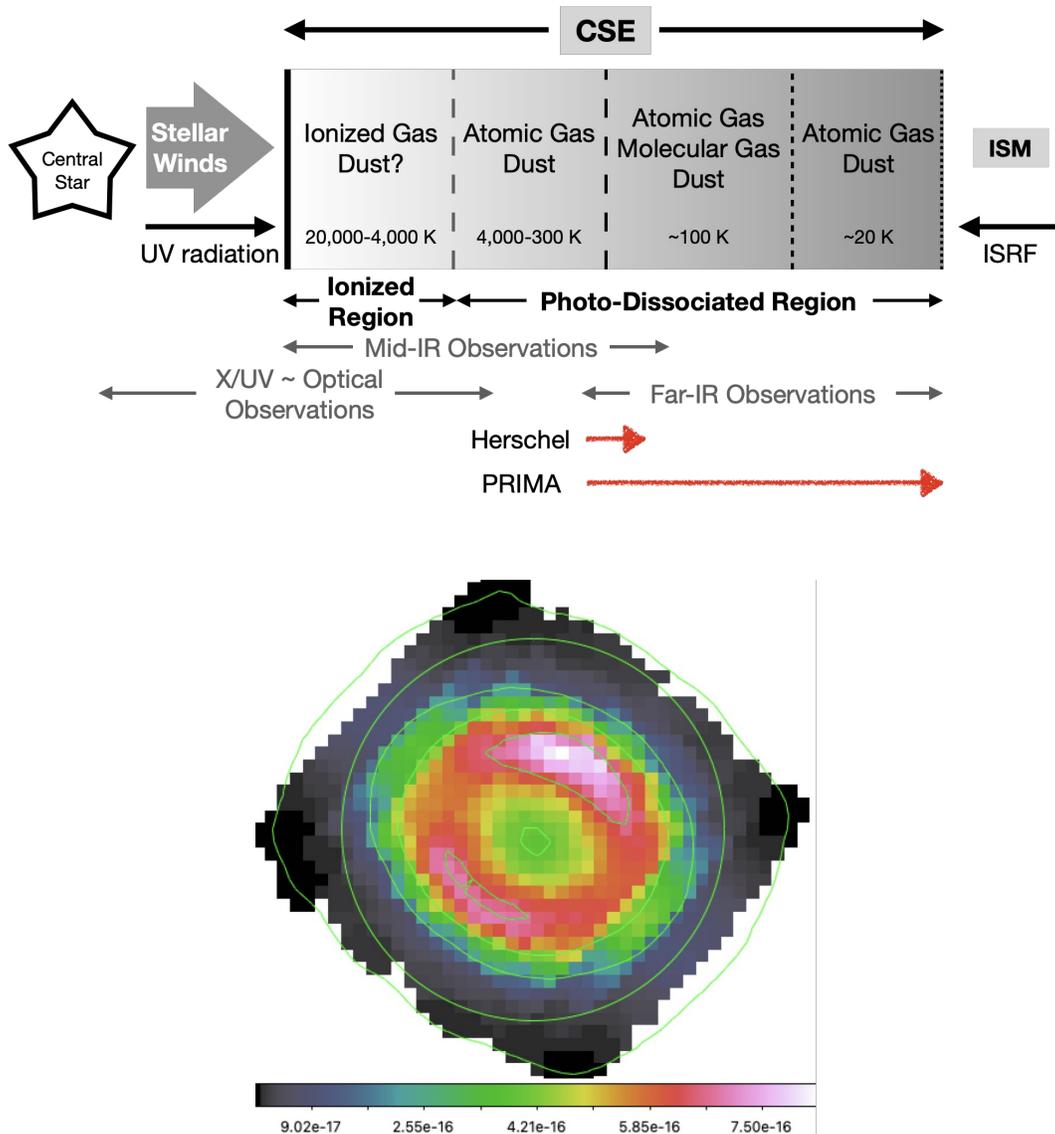

Figure 2: [Left] A schematic of a planetary nebula showing its construction and physico-chemical conditions of the circumstellar envelope (CSE). Herschel/PACS far-IR spectral mapping was just sensitive to begin to probe regions slightly beyond the ionized region. PRIMA/FIRESS is sensitive enough to probe the truly atomic/molecular regions toward the shell-ISM interface. [Right] Herschel/PACS [O III] line map of NGC 6720 (Ring Nebula; about 50" radius at 3" pix$^{-1}$; Ueta t al. in prep) with the color wedge showing the line flux distribution in W m$^{-2}$. The atomic/molecular region extends even further beyond the high excitation region seen via the [O III] line emission.

## Need for PRIMA

PRIMA would have sufficient resolution both spatially and spectrally, and especially would have sufficient extended-source sensitivities, all of which were not afforded by the previous far-IR facilities. To probe the coldest extended shell-ISM interface regions, access to the far-IR spectral range is essential.

### Instruments Required:

- PRIMA FIRESS: Small Map Low-res, R~130
- PRIMAger: Small Map (< 20'×20')
    - Hyperspectral band
    - Polarimetry band
- FIRESS small maps at least 5' × 5'.

### Program Size:

- Medium (10–100 hr): 60 hr for PRIMAger spectral mapping; 30 hr for FIRESS spectral mapping

### Approximate Integration Time:

#### PRIMAger spectral imaging:

Based on the existing most sensitive far-IR images of evolved star dust shells taken with Spitzer and AKARI, we need to achieve about 0.3 MJy sr$^{-1}$ at 90 μm. Using the values given in the 2023 March PRIMAger fact-sheet (surface brightness sensitivity for 5σ 10 hr mapping), we would need 2.5 hr per target. For a sample of about two dozen extended targets, the whole spectral mapping would take about 60 hr.

#### FIRESS spectral imaging:

Based on the existing Herschel/PACS spectral images of planetary nebulae in the [O III] 88 μm line, the detection limit is about $3 \times 10^{-17}$ W m$^{-2}$. If we want to aim at detecting 10 times fainter to probe the surrounding regions of a 5' x 5' map, using the formula provided in the March 2023 FIRESS presentation (2000 hours × [area / 900 sq arcmin] × [ 3×10$^{-19}$ W m$^{-2}$/ depth]$^2$), a single two-band scan mapping would take 0.6 hr. So, the 4-band coverage would require the total of 1.2 hr. For a sample of about two dozen extended targets, the whole spectral mapping would take about 30 hr.

### Special Capabilities Needed:

- None

### Synergies with Other Facilities:

- Ground-based optical IFU observations. ALMA spectral mapping.

## Description of Observations:

In both thermal dust emission hyperspectral (PRIMAger) and spectral (FIRESS) imaging, simple scan-mapping of a at least 5' × 5' field will do for most of the cases. Because of the nature of our target sources (emission is typically centrally concentrated), if there is a mapping strategy to increase redundancy (hence S/N) in the outer part of the extended structure would be nice.

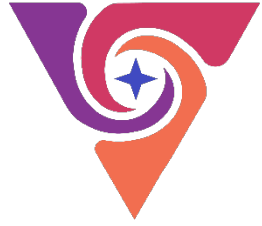

# Technical Notes

# 73. PRIMA confusion limit from extragalactic sources

Matthieu Béthermin (Strasbourg observatory/LAM), Guilaine Lagache (LAM)

Due to the limited size of telescopes in space, combined with the long wavelength, the angular resolution of space telescopes in the far-infrared reaches only a few tens of arc seconds. Even in the regions with low Galactic emission, the numerous faint galaxies blended in the beam of the instrument are causing map called confusion noise. Only the brightest sources emerge above these fluctuations even in absence of instrument noise. The lowest flux at which an individual point-like source can be extracted is called the confusion limit (see, e.g., Condon et al. 1974, Dole et al. 2004). This is an important parameter to design future observations, since confusion acts as a fundamental photometric noise, which does not decrease with integration time. However, advanced extraction techniques (e.g., Hurley et al. 2017) or statistical methods (stacking as in Dole et al. 2006, P(D) analysis as in Glenn et al. 2010) can be used to probe galaxy populations below the confusion limit.

In this technical note, we use the latest version of the SIDES simulation (Simulated Infrared Dusty Extragalactic Sky, Béthermin et al. 2022) to forecast the PRIMA confusion limit in photometry for both intensity and polarization. The confusion limit is computed both in term of fluctuations from faint sources and density limit. We find that the confusion will be a limitation for PRIMAger in intensity, especially in band 2. For instance, for a 10h survey covering 1 deg$^2$, wavelengths above 60 µm will be confusion limited, while shorter wavelengths will remain noise limited. In this case, advanced extraction methods using priors from shorter wavelengths will be necessary. In polarization, the confusion will not be a problem, expect for ultra-deep surveys (100h per deg$^2$).

## General Scientific Area:

Growth of Galaxies and Black Holes Through Cosmic Time

## Science Justification:

## Technical description:

At first order, the map variance caused by the confusion depends on the beam size and the galaxy flux distribution in the sky (e.g., Lagache et al. 2003):

$$\sigma_{\mathrm{conf}}^2 = \int \int b^2 d\Omega \times \int_0^{S_{\mathrm{lim}}} S^2 \frac{dN}{dS} dS,$$

where b is the beam function, Slim is the flux limit at which sources can be extracted, dN/dS are the differential galaxy counts, i.e., the number of galaxies per flux interval and per solid angle, as a function of the flux S. Usually, we define Slim as 5 σconf, and the equation above is solved iteratively. This formula assumes Poisson spatial galaxy distribution. The galaxy clustering can slightly increase this variance (e.g., Béthermin et al. 2017, Lagache et al. 2020). This is why we use a simulation-based estimate of the confusion limit in this document. Hereafter, this confusion limit is called photometric.

At shorter wavelength, the beam is smaller and photometric confusion is less problematic. In contrast, the density of sources usually limits how deep the source extraction can go. Historically, a density of 40 beams per source was used, but modern source extractor can go much deeper. In this note, we will use 15 beams per source, and we will refer to it as density limit. This is not an absolute limit, especially for source extractors using priors from shorter wavelengths.

We use the latest version of the SIDES simulation (Béthermin et al. 2022). This simulation starts from a dark-matter halo light cone and generates galaxy properties using a semi-empirical model. SIDES can reproduce a large set of observed galaxy properties such as the source number counts at various angular resolution, the redshift distributions, and the clustering observed in the cosmic infrared background.

For PRIMA, we integrated the spectral energy distribution of the SIDES galaxies in the corresponding filter. For simplicity, each hyper-spectral band is represented by 6 representative narrow filters. We then produced simulated SIDES maps corresponding to all the filters assuming a Gaussian beam (see Fig. 1). The 1-σ confusion noise is computed as the standard deviation of the map in Jy/beam after masking the pixels 5 σ above the confusion noise corresponding to bright sources. We iterate this process to converge on the final confusion noise. In Table 1, we list the 5-σ confusion limit obtained for all the PRIMAger filters. We also provide the confusion limit in density (15 beam/source) directly computed from the SIDES catalog. The confusion limit is the highest of these two values.

In addition to its capability in intensity, PRIMAger will also measure the linear polarization in band 2. The polarization data can also be affected by the confusion limit, but this limit is much lower if galaxies have a modest degree of polarization (e.g., Lagache et al. 2020). So far, we do not have constraints on the dust polarization of high-redshift galaxies. The SALSA SOFIA/HAWC+ program (Lopez-Rodriguez et al. 2022) showed that the polarization fraction of local galaxies is in average 1.3% with no obvious dependance on the starburst nature and a mild dependance on the inclination of the system.

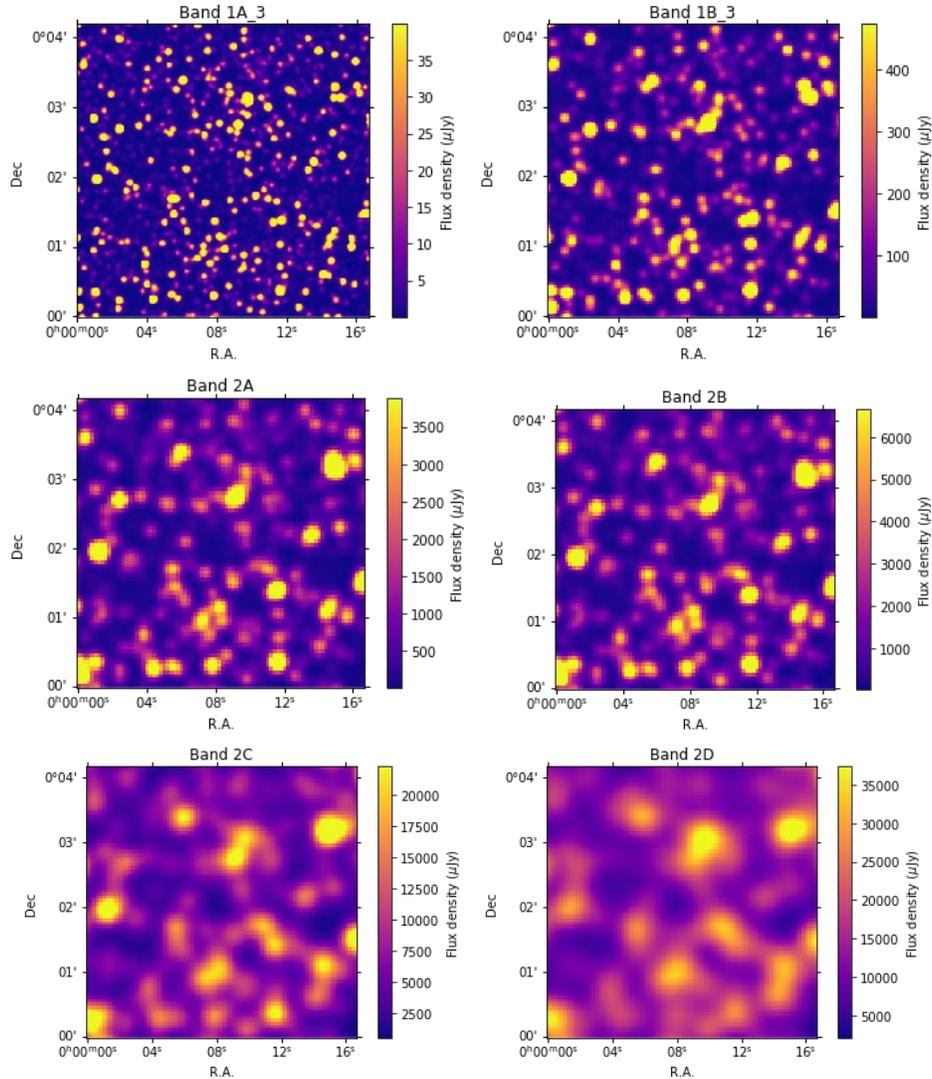

Figure 1: Cutouts of simulated noiseless PRIMA observations at 30 μm (upper left), 58 μm (upper right), 96 μm (center left), 126 μm (center right), 172 μm (lower left), and 235 μm (lower right) based on the SIDES semi-empirical model.

We thus produced polarized SIDES maps assuming a 1.3% polarization fraction p for all objects and drawing randomly the polarization angle α. For each galaxy, we compute the Stokes vectors Q and U in each band, together with the polarized flux P:

$$Q = Ip\cos(2\alpha),$$

$$U = Ip\sin(2\alpha).$$

where I is the flux density of the source in intensity in a given filter. We then produced Q and U maps ($m_U$ and $m_Q$) and combined them to derive the P maps ($m_P$):

$$m_P = \sqrt{m_U^2 + m_Q^2}.$$

Finally, we measured the confusion noise from the P maps in the same way as for the intensity maps, since in band 2 the confusion is driven by the photometric noise. We summarize our results in Table 2.

While the confusion limit for PRIMA will be high especially in band 2, the confusion in polarization will be reached only for the deepest surveys. For instance, for a 10 h/deg$^2$ survey, the confusion will be reached in intensity around 60 μm (at 58 μm, the confusion limit is 320 μJy and the instrument limit is 300 μJy), but even the longest-wavelength band (2D, 235 μm) will be limited by the instrument noise (detection limit of 700 μJy) and not the confusion (confusion limit of 252 μJy) in polarization. The polarized confusion limit in band 2D will be reached only by an ultra-deep 80h/deg2 survey (instrumental limit of 247 μJy versus a confusion limit of 252 μJy). At this depth, wavelength below 45 μm will still be dominated by the instrument noise (at 42.3 μm, the confusion limit is 77 μJy and the instrument limit is 78 μJy). There is thus a strong synergy between polarization observations close to the confusion limit in polarization at the longest wavelength, and ultra-deep intensity data at shorter wavelength.

Table 1: Confusion limit in intensity in the various PRIMA sub-bands predicted by the SIDES simulation. The photometric confusion limit is defined as 5 times the confusion noise, and the confusion limit in density is computed for 15 beam/source. The highest value is highlighted and should be used as confusion limit. Our estimates of the confusion are conservative and future extraction algorithm may allow us to extract fainter source.

| Band | Wavelength μm | FWHM arcsec | Photometric limit μJy | Density limit μJy |
|---|---|---|---|---|
| 1A | 25.0 | 2.94 | 6.4 | 8.3 |
| 1A | 27.8 | 3.27 | 10.4 | 13.3 |
| 1A | 30.9 | 3.63 | 16.8 | 21.0 |
| 1A | 34.3 | 4.03 | 26.5 | 32.4 |
| 1A | 38.1 | 4.49 | 42.3 | 49.8 |
| 1A | 42.3 | 4.99 | 68.0 | 77.1 |
| 1B | 47.0 | 5.53 | 111 | 121 |
| 1B | 52.3 | 6.16 | 186 | 195 |
| 1B | 58.1 | 6.83 | 313 | 320 |
| 1B | 64.5 | 7.58 | 534 | 530 |
| 1B | 71.7 | 8.43 | 925 | 892 |
| 1B | 79.7 | 9.36 | 1572 | 1462 |
| 2A | 96.0 | 10.3 | 3414 | 2827 |
| 2B | 126 | 14.8 | 6352 | 7976 |
| 2C | 172 | 19.0 | 22747 | 14503 |
| 2D | 235 | 26.7 | 30743 | 20123 |

Table 2: Confusion limit in polarization predicted by the SIDES simulation. The confusion limit is defined as 5 times the confusion noise in the polarized flux P map.

| Band | 2A | 2B | 2C | 2D |
|---|---|---|---|---|
| Confusion limit μJy | 32 | 57 | 176 | 252 |

## Special Capabilities Needed:

- None

## Synergies with Other Facilities:

Obtaining deep prior information from other wide-survey facilities (Euclid, Rubin, Roman, SKA) with a higher angular resolution could help to beat the confusion using advanced extraction methods as XID+ (Hurley et al. 2017).

# 74. What is the mineralogy of astronomical dust?

Janet E. Bowey (Cardiff University), Frédéric Galliano (DAp, CEA Paris-Saclay), Hendrik Linz (MPIA)

We aim to determine the abundance and macromolecular content of rock-forming materials observed in dusty environments beyond the Solar System. The study would utilize data from surveys of the physical properties of evolved stars, protoplanetary disks, and debris disks because observed dust features are highly dependent on source orientation, temperature and opacity. Some bands will be seen in emission and others in absorption. Each observation will sample a unique chemical environment but will give clues to dust properties in sources where the target alignment is less useful.

The project will address the following questions:

1. Which meteoritic materials can be found in astronomical dust?

2. On what spatial scales within a circumstellar environment does the mineralogy of the dust vary?

3. Do metallicity differences between stellar populations affect the observed mineralogy?

4. How does the observed mineralogy compare with theoretical and experimental predictions?

## General Scientific Area:

Rise of Metals and Dust

## Science Justification:

Refractory (rock-forming) astronomical dust is a repository of information on the evolution of metallicity in galaxies, and of stellar evolution, because many of the chemical elements condense into dust grains once they escape stellar interiors. Grains are observed in circumstellar regions around evolved stars, the dense and diffuse interstellar medium (ISM), young stellar objects, protoplanetary discs and debris discs, brown dwarfs and exo-planetary atmospheres. Space-dust is of public interest because nearly all terrestrial solids and living organisms contain atoms derived from it. Unlike carbon-bearing / organic materials and ices, which have characteristic spectral features in radio and mm and submm bands accessible to facilities like ALMA and radio arrays, most rock-forming oxygen-rich dust components do not have known features outside the 2 to 100µm range and their weak optical bands are obscured by the stronger atomic and molecular features of H and He. Consequently, astrophysical studies simplify dust properties to size-ranges and a handful of chemical compositions even though hundreds of oxygen-rich solid materials are found in the pre-planetary materials collected within the Solar system (meteorites and Interplanetary Dust Particles). PRIMA and JWST could revolutionize astronomical measurements of these properties and compare them with those of the Solar System because infrared spectra contain observable features at wavelengths specific to the chemical composition and lattice structures (crystallinity) of rock-forming dust.

Bands accessible to ground-based observatories (within the telluric transmission windows between 0.9 and 13 μm) and JWST (0.9 to 28 μm) are used to identify sub-micron-sized silicate dust due to measurements of bands at 10 and 18 μm associated with 'silicate', $SiO_4^{4-}$ . However, these bands are distorted and become opaque when column densities are high and grain sizes are large (≥ 0.3 μm; Hofmeister, Keppel and Speck 2003). The bands of mixed crystalline silicates can blend to produce broad features with little fine structure (see Bowey & Adamson 2002) to indicate anything but amorphous dust (see everybody else, e.g., Kemper et al. 2004) beyond the Solar System. Detailed mineralogy in these features is relatively difficult so that the appearance of narrow bands between 20 and 45 μm due to ordered crystalline olivines and pyroxenes in ISO/SWS spectra of evolved stars was quite a surprise to astronomers (Waters et al. 1996) with many more in the ISO/LWS 43 to 100 μm range. FIR identifications include: hydrous silicates in YSOs and in NGC 6302 (Malfait et al. 1998 and Hofmeister & Bowey 2006, respectively); crystalline olivines and pyroxenes (e.g., Cohen et al. 1999, Molster et al. 1999, Molster et al. 2002a, b, c); Ca and Al-bearing oxides and corundum (Hofmeister et al. 2004) and carbonates (Kemper et al. 2002). Most identifications were relatively tentative due to an absence of laboratory data (little existed because the mineralogical variety was unexpected) and the low number of targets sampled. PRIMA has the potential to revolutionize this area of cross-disciplinary research. Paradigm shifts would be enabled particularly if PRIMA and JWST were flying at the same time because whole SEDs could be observed at similar epochs. The only observatory to have explored this parameter space to date was ISO and the best observed target was NGC 6302 (Fig. 1) due to its use as a FIR calibrator.

In the PRIMA era, the discovery space will be larger due to the existence of more laboratory data (Fig. 1a) following those early discoveries and its extremely high sensitivity (the ISO mirror had a diameter 60 cm, poless-sensitive detectors and a non-ideal orbit). Tentative mineral identifications based on the new laboratory are in Fig. 1b. Broad 25 to 100 μm mineralogical features have rarely been explored since. This wavelength range is inaccessible from the ground. Regarding airborne and space missions, the AKARI instrumentation was unsuitable for this task, and the anticipated HERMES spectrograph did not arrive before SOFIA's demise. The wavelength coverage of Herschel was only longwards of 52 μm, and mineralogy work was limited to a few studies of the 69 μm Forsterite feature in bright targets (Herbig stars or evolved stars, see e.g., Sturm et al. 2013, de Vries et al. 2010).

The PRIMA range (≥25 μm) is key to sampling crystalline silicates because diagnostic features are more widely spaced in wavelength and overlap less than narrow bands within the broad 10 and 18 μm bands. These resonances, known as translational bands, sample a wider region of the crystal lattice and are due to movements of the cations (e.g., $Mg^{2+}$, $Fe^{2+}$) relative to the $SiO_4^{4-}$ groups. Minerals are more transparent in the FIR enabling the sampling of higher column densities and larger grain sizes (> 1 μm) and emission features are due to cooler grains: it is highly likely that these crystalline grains are a distinct population from the classic amorphous silicates identified in most 10 and 18 μm astronomical spectra. They are more like meteoritic samples than are classic astronomical silicates. PRIMA/FIRESS will enable astronomical mineralogy and meteoritic and planetary studies to explore mutually beneficial science in a way that has only been touched on in the past.

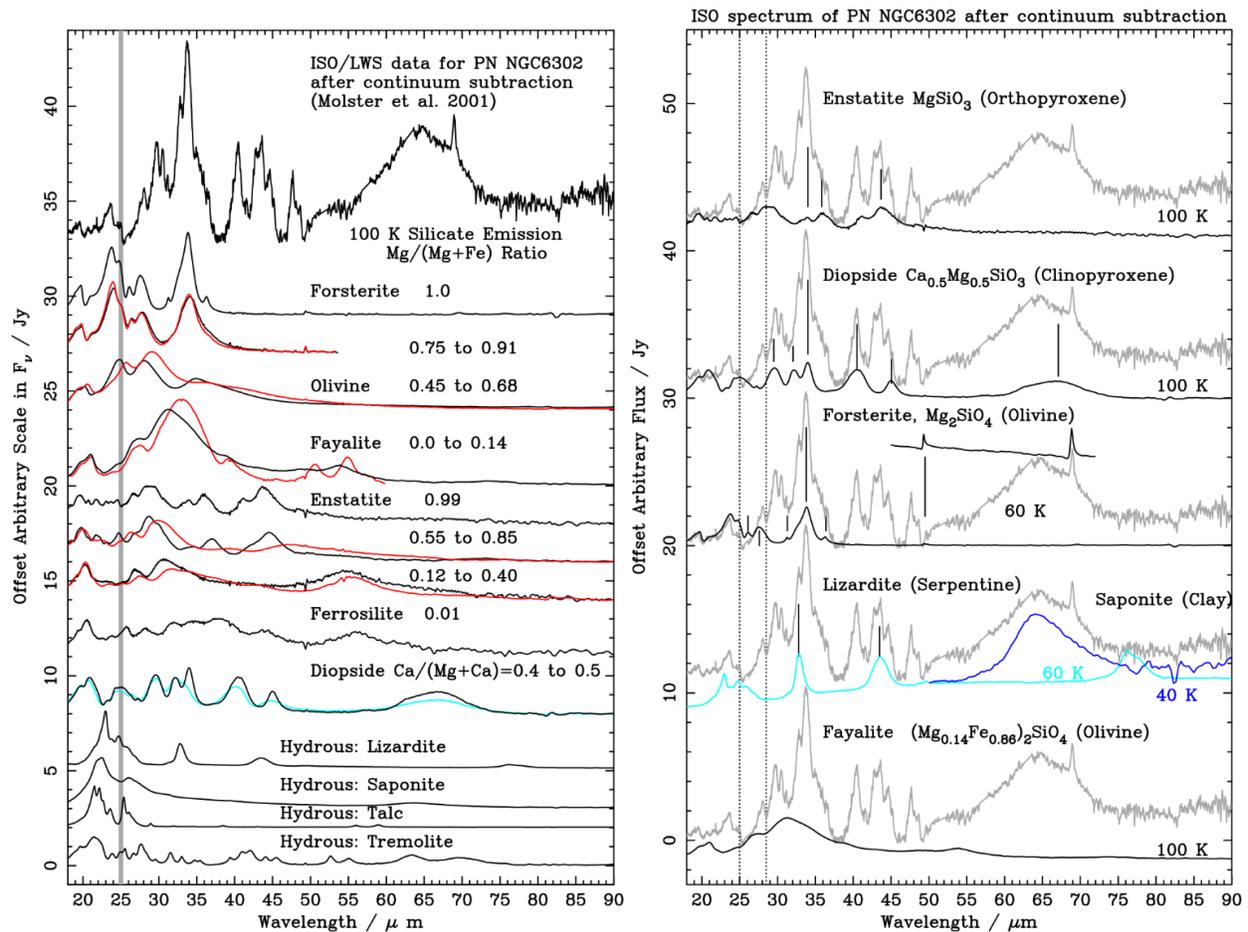

Figure 1: (a) Room temperature laboratory spectra of crystalline silicates: olivines (forsterite, olivine, fayalite)- Pitman et al. 2010; pyroxenes (enstatie, ferrosilite, diopside)- Bowey, Hofmeister and Keppel 2020; hydrous silicates: Hofmeister & Bowey 2006. Red curves indicate the more Fe-rich silicates in overplotted pairs, the cyan curve indicates the Ca-poor diopside.  (b) Continuum-subtracted ISO spectrum of PN NGC 6302 (grey curves; Molster et al. 2001) compared with simulated optically thin emission from crystalline silicates; emission temperatures are indicated. Laboratory data were obtained at room temperature, except the 49 and 69 μm peaks of forsterite which were measured at ∼ 4 Kelvin. Models are eyeball matches and non-definitive. Hydrated silicates are indicated in blue and cyan. Pairs of dotted lines indicate the overlap region between JWST and PRIMA.

## Instruments Required:

PRIMA FIRESS: Pointed Low-res, R∼130

## Program Size:

- Small (<10 hr)
- Medium (10–100 hr)
- Large (100 + hr)

## Special Capabilities Needed:

- None

## Synergies with Other Facilities:

- Spectral Energy Distributions (SEDs) obtained with MIRI/JWST at wavelengths below PRIMA range and submm SEDs obtained with ground-based facilities.

## Description of Observations:

- This program requires a good s/n (ideally ~ 100 or better) over a continuous wavelength range because many mineral bands are several μm wide. Integration times are highly dependent on the source characteristics which would include circumstellar and interstellar environments. Circumstellar dust studies would include evolved stars, protoplanetary disks and debris disks. Interstellar environments could include dusty galaxies, AGN and quasar absorbers as well as Milky-Way molecular clouds and the diffuse interstellar medium. Observations of unique object types and larger samples will benefit the program. 20 or more high s/n observations of each object type will provide useful statistics.

# 75. Understanding the formation and evolution of exoplanetary systems with PRIMA

Tiffany Kataria (Jet Propulsion Laboratory, California Institute of Technology), David Ciardi (Caltech/Infrared Processing and Analysis Center), Ian Crossfield (University of Kansas), Ty Robinson (University of Alabama)

We discuss three potential science investigations that could be pursued with PRIMA in the areas of exoplanet and brown dwarf science. The study of the atmospheres of exoplanets and their substellar cousins, brown dwarfs, allows us to understand how planets have formed and evolved, as the composition of those atmospheres can be tied to the volatiles in protoplanetary disks. Using time-resolved observations, we can learn about three-dimensional processes (i.e., winds and weather) on these objects. The far-IR could also enable a look into a planet's internal heating, providing new insights into the energy budgets of giant planets. Each of these potential science investigations would advance our knowledge of the formation and evolution of giant planets and brown dwarfs beyond the state-of-the-art and would be highly complementary to the comparative exoplanetology studies conducted by JWST and Ariel.

## General Scientific Area:

Exoplanets

## Science Justification:

Forming a better understanding of how planetary systems form and evolve is critical to comprehending how unique our own Solar System may be. Other than the central host stars, giant planets likely play a strong and influential role in sculpting the architecture and composition of the forming exoplanetary systems. However, the formation and evolution of the giant planets at and beyond the snow line themselves are still not well understood, while having profound influence on the formation and composition of the inner rocky planets, where life has formed in our own Solar System.

At present, the formation and evolutionary boundary between giant planets and brown dwarfs is not well understood. Traditionally, giant planets have a mass of 13 $M_{Jup}$ or less with brown dwarfs being more massive, up to about 0.1 $M_\odot$. But there is a plethora of examples of single brown dwarfs with mass estimates well-below 13 $M_{Jup}$ and giant exoplanets with masses above that limit. The formation and evolutionary processes of giant planets and brown dwarfs likely play a role in the final mass and orbital location of the sub-stellar objects and, as a result, in how those objects shape the rest of the exoplanetary systems.

PRIMA offers the opportunity to understand three critical areas regarding the formation and evolution of giant planets and brown dwarfs and, thus, cast insight on their influence on the planets in the inner solar systems where rocky, water-hosting, planets might exist.

1. Measuring the hydrogen – deuterium isotopic ratio in brown dwarfs and giant planets.

2. Measuring the rotational rates and atmospheric compositions of brown dwarfs.

3. Measuring the internal energy sources of brown dwarfs and giant planets.

## 1. The D/H Isotopic Ratio

Accurate determinations of the deuterium to hydrogen (D/H) ratio in the giant planets provide fundamental constraints for modeling planetary formation and evolution (Hersant et al. 2001). The abundance ratio of deuterium to hydrogen on the Earth compared to the same abundance ratio of comets and outer solar system planets like Jupiter can inform us on how and when water was delivered to the Earth during the early Solar System. Morley et al. 2019 showed that the D/H ratio could be calculated with JWST using signatures of methane in exoplanets and brown dwarfs, but not using water. PRIMA provides access to several water lines that can be used to determine the D/H ratio in isolated brown dwarfs and widely separated brown dwarfs in orbit around stars.

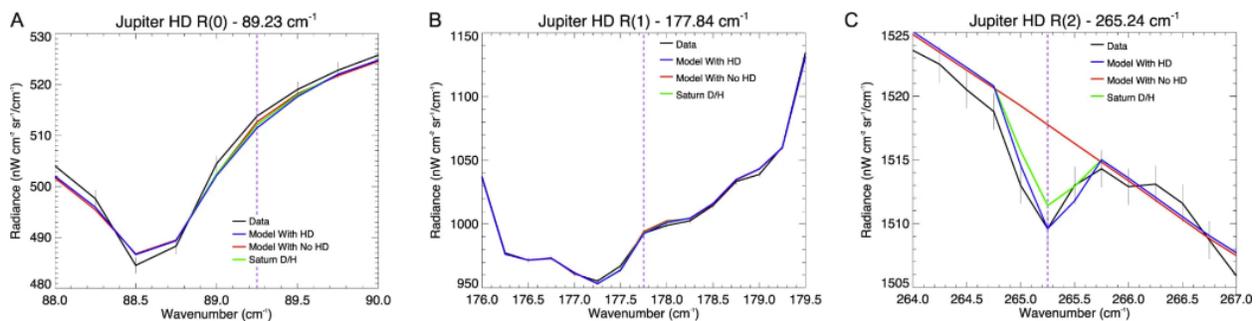

Figure 1: Far-infrared spectra of Jupiter modeled with and without deuterium (from Pierel et al. 2017). The results show that Saturn has relatively less deuterium than Jupiter, but more in line with the protostellar nebula. Similar observations could be made of brown dwarfs and giant exoplanets with PRIMA and its far-infrared spectrometers.

## 2. Brown Dwarf Atmospheric Composition and Rotational Properties

Determining the global atmospheric properties of substellar objects is critical to understanding how those objects formed and evolved. Spectroscopy for atmospheric characterization coupled to rotation monitoring of brown dwarfs can inform us about chemical abundances, cloud formation, and angular momentum evolution. In brown dwarfs, unlike in hotter stars, lithium exists in molecular form with lines from molecules such as LiF, LiCl, and LiH accessible in the far-infrared (Gharib-Nezhad et al. 2021). The presence and strength of lithium lines is related to the age and mass of the brown dwarf. By measuring the abundances of Li-bearing molecules in the atmosphere and measuring the rotation rates of the brown dwarfs, PRIMA can yield insight into the masses, angular momentum evolution, and atmospheric cloud formation processes in brown dwarfs and giant planets.

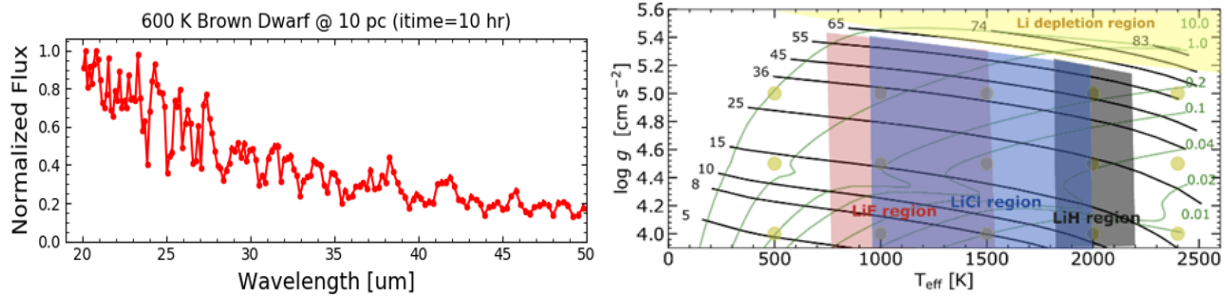

Figure 2: Far-infrared model spectrum of a 600K brown dwarf (left) and the expected regions of Li atomic and molecular line formation as a function of brown dwarf mass, temperature, and age. Objects in the yellow area are massive and old enough to burn protosolar Li atoms. LiH, LiCl, and LiF spectral features become apparent in the black, blue, and red areas, respectively.

## 3. Internal Heat Sources of Giant Planets

The radiant energy budget and internal heat are fundamental properties of giant planets and brown dwarfs. PRIMA offers the opportunity to obtain direct measurements of the energy output of brown dwarfs and giant planets, placing constraints on, and improving the current evolutionary theories and models for substellar objects across the planet-brown dwarf boundary. It is well-known that Jupiter and Saturn both contain internal heat sources and thus the radiant energy of these planets is greater than what would be expected from just absorption of solar energy. The transfer and distribution of radiant energy within the atmospheric systems of the giant planets modify the thermal structure and generate mechanical energy driving atmospheric circulation. The same is also likely true of all giant planets; thus, measuring the internal heat sources can place significant constraints on the formation and evolution of giant planets and their influence on the general architectures of exoplanetary systems.

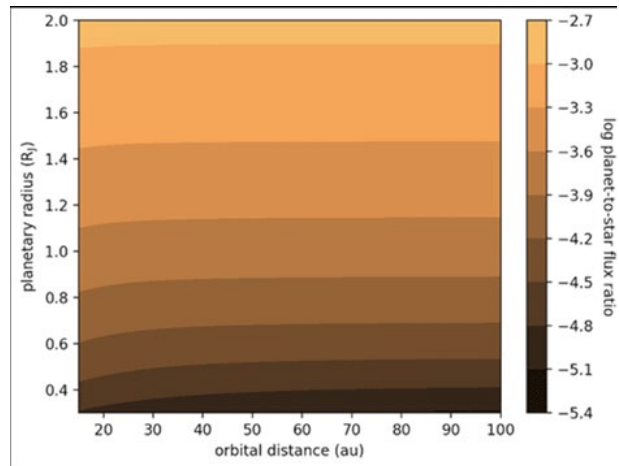

Figure 3: Estimates of the planet-to-star flux ratio of 100K giant planets measurable with PRIMA.

## Instruments Used:

We choose not to indicate a specific instrument mode or time estimate since this warrants further investigation (see Description of Observations).

## Special Capabilities Needed:

- Precise pointing would be needed during transit events (~hr).

## Synergies with Other Facilities:

While JWST will perform the lion's share of spectroscopic reconnaissance of the atmospheres of brown dwarfs and giant exoplanets, there is a wealth of discovery space offered within the far-IR. The HD and HDO lines, for example, have many of the strongest lines in the far-infrared. ESA's Ariel mission, which plans a comprehensive survey of ~1,000 transiting exoplanets, will only access vis-to-mid-infrared wavelengths, and will also be complementary to PRIMA far-infrared observations.

### Description of Observations:

As stated above, precise pointing would be needed for transit events on the order of hours. Additionally, the amplitude of chemical signatures may be small (~10s ppm or less) and therefore it may be challenging for PRIMA to achieve the sensitivity needed for identifying deuterium-bearing or other chemical species. Still, if broadband observations are sufficient for the science use case (e.g., the internal energy science case illustrated above), using PRIMAger hyperspectral mode might be a more suitable choice. Overall, while the FIRESS Fourier transformation spectrometer (FTS) would be the ideal instrument to use for the science cases outlined above, some of the science could possibly be achieved with PRIMAger. However, further quantitative investigation is needed.

### Acknowledgement:

A portion of this research was carried out at the Jet Propulsion Laboratory, California Institute of Technology, under a contract with the National Aeronautics and Space Administration (80NM0018D0004).

# 76. Unraveling the spectral energy distributions of beam-confused galaxies with PRIMAger

Howard A. Smith (CfA), J. Rafael Martínez-Galarza (CfA)

PRIMAger will image fields containing full of galaxies, many of them at high-z, in up to sixteen IR and FIR wavelength bands from 25–260 μm across the SED. The factor of ten in wavelength sampling of the SED also spans a factor of ten in spatial resolution, meaning that every set of images will contain galaxies that are resolved at the short wavelengths but blurred together at the long wavelengths. Reconstructing the SEDs of individual blended galaxies will be challenging. We propose a Bayesian algorithm to extract from these sets of images the most likely SEDs for each of the confused galaxies in the blurred, long-wavelength images. A similar technique can unravel the SEDs of infrared sources in confused galactic star clusters.

## General Scientific Area:

Growth of Galaxies and Black Holes Through Cosmic Time

## Science Justification:

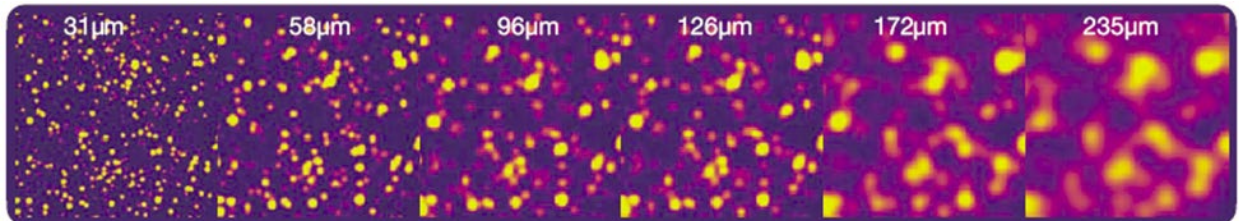

Figure 1: Simulated PRIMAger images demonstrating that one can beat the confusion by bootstrapping the position and sources staring at the shortest wavelengths, and going to the longest wavelengths. In dense confused fields like this one, in which long-wave beams encompass multiple sources, Bayesian algorithms can help unravel the most-probable single-object SEDs.

Our current algorithm is based on successful routines we have applied to YSOs in clustered star formation regions using 2MASS, Spitzer IRAC and MIPS, and Herschel PACS and SPIRE, covering 2 to 500μm. (Martinez-Galarza, Protopapas, Smith & Morales 2018).

Our method uses measured and simulated SEDs to create informative priors for the photometry of unresolved multiple systems. We then perform Bayesian inference to obtain the most likely physical parameters of individual objects given the set of images and blended photometry. The results yield most-likely physical parameters in the context of different models of galaxy evolution formation. Our SEDs would be based on our extensive sets of observed FIR galaxy SEDs in the nearby universe, our sets of simulated SEDs in merging galaxy pairs, and other methods. We have used CIGALE in much of our SED analysis papers, and we would work with the CIGALE team on this project for extracting physical parameters from the resultant most-likely SEDs.

The PRIMAger images present both new opportunities and challenges. The 16-band photometry covers more datapoints than we used in the published YSO analyses, and they span the peak of the galaxy SED emission with the shape of the peak offering key diagnostic capabilities. However, the galaxies clustered in an unresolved FIR image may lie at different redshifts, and statistical tools need to be included to infer the likelihood that galaxies seen close together are physically associated and at the same or similar z's, with photometric redshift analysis incorporated. Ancillary datasets (optical/near-IR) would be readily applied to facilitate determining these priors. We suggest developing a working demonstration using simulated PRIMAger images and our set of observed galaxy SEDs. At some point we would update the code with newer software and strategies.

The results of this program will enable PRIMA scientists to obtain more reliable galaxy SEDs for confused objects from the epoch of peak star formation; this translates into clearer evolutionary histories of galaxies in this epoch, better characterization of the dust including mass and heating via AGN activity and SF rates, and a better understanding of the development of the ISM over cosmic time. A similar technique can also be applied to unravel the SEDs of infrared sources in confused galactic star clusters.

## Instruments Required:

Pointed High-res R~4400 @ 112 μm - comment

## Special Capabilities Needed:

- None

## Synergies with Other Facilities:

- JWST, possibly MATISSE at ESO (the 10 μm silicate feature).

## References:

Martinez-Galarza, Protopapas, Smith & Morales 2018; https://doi.org/10.3847/1538-4357/aad503